# THESE

Présentée en vue de l'obtention du titre de

## Docteur de l'INSTITUT NATIONAL POLYTECHNIQUE DE TOULOUSE

### ECOLE DOCTORALE SYSTEMES

*Spécialité : Systèmes Industriels*

Par

## Leoncio JIMÉNEZ CANDIA

*Licencié en informatique et sciences humaines de l'Université Libre de Bruxelles*
*Maître de conférences à l'Université Catholique de Maulé*

---

## *Gestion des connaissances imparfaites dans les organisations industrielles : cas d'une industrie manufacturière en Amérique Latine*

---

Soutenu le 22 février 2005 à l'INP de Toulouse devant le jury composé de :

| | | |
|---|---|---|
| **M. Germain LACOSTE** | Directeur | *Professeur à l'INP de Toulouse* |
| **M. Jean-Michel RUIZ** | Rapporteur | *Professeur à l'Ecole Généraliste d'Ingénieurs de Marseille* |
| **M. Jean-Claude BOCQUET** | Rapporteur | *Professeur à l'Ecole Centrale de Paris* |
| **M. Emmanuel CALLAUD** | Rapporteur | *Professeur à l'Université Louis Pasteur de Strasbourg* |
| **M. Jean Marc LE LANN** | Examinateur | *Professeur à l'INP de Toulouse* |
| **Mme. Pascale ZARATE** | Examinateur | *Maître de conférences à l'IRIT / INP de Toulouse* |
| **Mme. Angélica URRUTIA** | Invitée | *Prof. associé à l'Université Catholique de Maulé (Chili)* |
| **M. Aquiles LIMONE** | Invité | *Professeur à l'Université Catholique de Valparaiso (Chili)* |

# Remerciements

Une thèse est une sorte d'égoïsme en soi, eh oui, la plume transformé en clavier d'ordinateur de nos jours, demande silence et solitude pour la faire danser, sur en papier blanche, qu'on ne touche jamais, mais on a l'illusion de le voir monter et descendre dans l'écran, je remercie donc les personnes qui ont compris le besoin de ma solitude, et non rien obtenu en échange.

Un grand merci ma chère Christine pour le délicat travail de la correction, je remercie aussi sa co-équipier de bureau, qu'a pris le relais de la retouche dans multiples occasion.

Une thèse est un projet de long aileine ou personne peut garantir a coup sur le succés de cette aventure, bien que cette aventure est guidé Directeur.

Je remercie à Germain Lacoste pour sa qualité de bon vivant, je vois en lui le mélange juste entre un bricoleur qui maîtrises ses outils et se mains, et un lieder qui maîtrise l'avenir et les hommes qui doivent le forger dans sa tête. Et oui, Germain est la réussite d'un modèle sociotechnique bien maîtrise à l'avance pour maîtriser l'avenir, il peut être fière, que sa trace intellectuel a construit de chemins de vie, On dit souvent, que derrière tout grand homme il y a une grand femme (l'inverse est toujours si vrai), ma chère Annie je te remercie aussi dans cette thèse, non seulement pour avoir transformer mon manuscrit initial un français propre, mais aussi pour ta générosité et amabilité chaque fois que j'ai sonné a sa porte.

Je voudrais remercie aussi à mes amis et collègues Aquiles Limone et Luis Bastias, qui m'ont initié dans l'autopoïèse et la connaissance, néanmoins la porté de mon effort dans cette thèse ne peut être que très modeste.

Je pense qu'une thèse se construit à travers des émotions avec soi même et avec autrui. Ma plus belle émotion est Angelica Urrutia, sans elle difficilement j'aurais arrivé à la fin. Un grand merci pour ton soutient permanente et ta joie de vivre.

Enfin, je remercie à tous ce qui ont partage avec moi une belle arrosage, atour d'une belle table dans et par la conversation entre bons amis et amies.

# Table des Matières

## Chapitre 4    Le système (connaissance et opérationnel) dans un environnement imprécis et incertain    204

## Chapitre 5    Cas d'étude: application à l'industrie du carton en Amérique Latine    247

# Table des Figures

Depuis les années 80, la gestion de projet, outil indispensable de la gestion de ressources : (humaines, temporelles, financières,…) de l'entreprise, évolue vers une utilisation prédictive de résultats[1] à partir de scénarios. Si l'on emploie le langage courant, pour comprendre le principe, on peut faire une analogie avec des films, constitués d'un nombre très important d'images, que notre système cognitif garde précieusement au fond de nous-mêmes avec nos expériences de tous les jours, depuis notre naissance (voire même avant). Nous utilisons ces films consciemment ou inconsciemment pour guider nos actions et nos comportements quotidiens. Ils peuvent alors décider de notre chemin de vie. Pour l'entreprise, les scénarios sont constitués, d'une part, par des menaces ou risques, et d'autre part, par des opportunités ou performances, stockées dans le système de connaissance de l'entreprise. Ils sont liés à des éléments politiques et stratégiques de gestion du passé, et ceci est vrai à tous les niveaux de l'entreprise (stratégique, tactique, opérationnel). Il faut comprendre que l'alimentation de ces scénarios se fait à partir d'informations (images), mieux : de connaissances, liées à des savoirs et de savoir-faire de l'entreprise, pour lesquels l'information ou la connaissance imparfaite, matérialisée, d'une part, par l'incertain lié aux événements, et d'autre part, l'imprécis lié aux données (inconnus, mal connus,…), ajoute finalement une complexité supplémentaire à la gestion de projet.

D'après le modèle intégratif du cycle de vie d'un projet, selon Lacoste[2], quatre phases apparaissent dans la gestion de projet :

- une première phase est la réponse à un appel de création, fabrication, production appelé habituellement appel d'offres ("ce qu'on veut"). Elle met directement en œuvre la base de connaissance[3] de l'entreprise (savoir-faire "ce qu'on peut") ; elle est liée à la décision (GO/NO GO) du projet induisant une connaissance approfondie des savoirs et savoirs-faire

---

[1] Nous validons cette argumentation à partir de notre Mémoire de Licence en Informatique et Sciences Humaines, Université Libre de Bruxelles, Belgique, intitulée *Contribution au lissage exponentiel généralisé*. Où nous avons développé un modèle prédictif à partir de données sur les ventes passées d'un produit [Jiménez, 95].

[2] Pierre Bonnal dans sa thèse intitulé *Planification possibiliste d'un grand projet industriel s'appuyant sur l'approche de la chaîne critique* (publié en 2002), détaille la logique de cette mécanique, qui permet de situer dans le temps (avant-projet, projet) tous les éléments décisionnels de l'entreprise, par exemple : veille, intelligence économique, bases d'informations (bases de données, bases de connaissances, base de règles, base de cas, etc.),… [Bonnal, 02].

[3] C'est l'ensemble de connaissances tacites (source d'innovation), explicites (source de compétences), émotionnelles (source de lidership).

de l'entreprise. Cette étape traduit également le degré de prétention de la sollicitation externe.

- une deuxième phase, dite de scénarios illustrant les divers chemins du futur projet, les réponses et les conditions de réussite (analyse des risques et du niveau de succès espéré) ;

- une troisième phase classique d'exécution et de gestion du projet coût/délais/performance ;

- une quatrième phase dite de retour d'expérience qui alimente, rectifie et modifie la base de connaissance initiale. Dans cette logique, les phases 1, 2 et 3 utilisent directement la base de connaissances.

Ces quatre phases témoignent que la gestion de projet en Génie industriel est une science d'action permettant de concevoir, construire et de maintenir des organisations industrielles.

Nous avons bien perçu que la plupart des recherches de l'équipe de Génie industriel se sont centrés dans les problématiques relatives aux phases 1, 2 et 3. Illustrons ceci par trois recherches.

Dans le domaine du génie des systèmes industriels de l'INPT, le projet européen DECIDE[4] qui a démarré en 1993 avait pour objectif « de mettre au point une méthodologie et un outil associé permettant d'aider les industriels dans leur processus global de réponse à un appel d'offres». La méthodologie de DECIDE proposé est un canevas de pensées pour la capitalisation de connaissances centré sur la conception de solutions techniques. Comme a dit Anne-Marie Alquier, DECIDE « cherche à améliorer la qualité et l'efficience de la procédure globale en leur permettant notamment de réutiliser le savoir-faire existant dans l'entreprise au travers des connaissances acquises à l'occasion de réponses à des appels d'offres précédents ». Ainsi, l'outil informatique de DECIDE est structuré autour de quatre modules afin d'aider le gestionnaire à proposer au client une solution technique pertinente. Le premier module permet de faire la gestion de l'analyse de la valeur de la solution technique. Le deuxième module permet de gérer l'estimation des coûts d'une solution technique. Le troisième module permet de gérer la fixation du prix de la solution technique. Le quatrième module permet de faire la gestion du cycle de vie de la solution technique à partir la description de modèles utilisés dans la conception de la solution technique (produits, process, production, maintenance). Il s'agit bien, donc, d'un véritable module de capitalisation de connaissances autour des produits, des process, de production et de maintenance de la solution technique. Le cinquième module permet de gérer le tableau de bord de la solution technique. Les informations à gérer concernent « le prix maximum conseillé pour une proposition, le taux de couverture des exigences du client pour une solution technique, son coût et ses chances de succès sont mises en perspective afin de permettre aux décideurs de répondre ou non à l'appel d'offre et/ou de choisir une solution technique et de coupler le niveau financier de la réponse ».

---

[4] Pour plus de détails sur le projet voir "DECIDE : Decision Support for Optimal Bidding in a Competitive Business Environment - Esprit Project 22298", Ed P. Zaraté, Journal of Decision Systems, vol 8 N°1, 1999.

Dans le cadre de la thèse de Sophie Bougaret, intitulée « Prise en compte de l'incertitude dans la valorisation des projets de recherche et développement : la valeur de l'information nouvelle [Bougaret, 02], se trouve explicitée une dynamique pour gérer un projet de Recherche et Développement[5] dans le domaine pharmaceutique. Cette démarche prend en compte l'attitude du décideur face à l'optimisation du choix de scénario de développement du projet par la valorisation de l'information nouvelle liée à chaque scénario du projet. L'originalité de cette démarche est la prise en charge de l'information imparfaite car traditionnellement les méthodes d'évaluation des projets incorporent dans ces modèles prévisionnels de l'information parfaite venue du passé et non pas des paramètres incertains de l'avenir. Comme a dit Bougaret « l'incertitude suppose l'inconnu, l'absence de référentiel antérieur qui pourrait permettre une probabilisation de ce qui va arriver » [Bougaret, 02]. Pourtant, l'avenir du projet peut et doit être considéré comme une situation incertaine[6] …qui ne permet pas ou peu de prévision à partir de données du passé.

Dans le même esprit de générations des outils pour la gestion de projet, se trouve la thèse de Pierre Bonnal, intitulée Planification possibiliste d'un grand projet industriel s'appuyant sur l'approche de la chaîne critique [Bonnal, 02]. Pour apprécier la différence avec l'incertain et l'imprécis, il déclare que « l'incertain est lié au caractère aléatoire d'un résultat, et à ce titre revêt une dimension probabiliste ; l'imprécis est lié au caractère imparfaitement défini d'un résultat, et il est de nature déterministe ». Puis il ajoute dans le cadre de la planification d'un projet industriel « on a fait aussi trop souvent l'amalgame entre deux concepts qui sont vraiment distincts : l'imprécis et l'incertain. Le premier revêt une dimension purement déterministe pour laquelle les acteurs disposent de moyens d'actions, le second renvoie à une dimension aléatoire, donc effectivement probabiliste, pour laquelle les acteurs n'ont aucun moyen d'action ». Il ajoute aussi « il y a unanimité pour considérer qu'entre incertitude, notion foncièrement probabiliste, et imprécision, notion déterministe, c'est plutôt cette dernière qui s'applique au contexte de la planification de projet » [Bonnal, 02].

---

[5] Pour Bougaret un projet de R&D peut être défini à partir d'un critère d'information parfaite comme « un processus d'acquisition de connaissances visant à réduire l'incertitude pour aboutir à une solution innovante (produit ou procédé) qui permettra d'améliorer la rentabilité de l'entreprise ». Ou bien, à travers d'un critère d'information imparfait comme « un processus d'implémentation cognitive (acquisition de connaissances), qui porte en lui un degré d'innovation et donc d'incertitude. Ses spécifications finales ne sont pas définies clairement au début du projet. La démarche projet aura justement pour objectif de réduire l'incertitude afin de préciser les spécifications par la production de connaissances livrée par le R&D. La finalité du projet de R&D est bien évidemment de contribuer à la rentabilité de l'entreprise » [Bougaret, 02].
[6] Pour Bougaret « une situation est incertaine dès lors que les paramètres qui la définissent sont inconnus ou mal connus, au sens non encore révélés » elle ajoute « le qualificatif "incertain" désigne quelque chose qui n'est pas déterminé, que cette chose soit bonne ou mauvaise, pour le futur » [Bougaret, 02].

Enfin, plusieurs autres contributions ont vu le jour au sein de cette équipe de recherche de génie industriels de l'INPT afin de mettre au point des outils performants pour la gestion de projet, d'une part liés à la conception de produit-procédé-processus innovants[7], et d'autre part, liés à la prise en compte de risques dans la gestion et le management de projets industriels. Néanmoins, peu d'efforts ont été engagés dans la spécification et la conception de bases de connaissances, capables de capitaliser les savoirs-faire, en laissant ainsi ouvert le champ à la capitalisation du retour d'expérience sous forme de bases de règles, bases de cas ou bases d'analogies ou …. Nous avons donc choisi, dans le cadre de cette thèse, de réfléchir aux Bases de Connaissances non pas appliquées à la gestion des appels d'offres mais plutôt liées à l'évolution des connaissances.

A partir de besoins industriels et de contraintes culturelles, la gestion des connaissances a fait l'objet de nombreux travaux de recherche [Charlet et al, 00]. Nous avons constaté que ces recherches sont organisées sur deux champs : l'un est la mise en place de méthodes et d'outils pour savoir localiser les gens qui ont la compétence requise dans l'entreprise (GRH : "qui fait quoi") [Hermosillo, 03], l'autre est la mise en place de méthodes et d'outils pour pouvoir et savoir évoluer la connaissance pour faire de l'innovation en continue [Longueville et al, 01].

Le contexte de cette thèse est la gestion de projets et non pas la GRH, c'est-à-dire la gestion des connaissances est liée à la capacité de l'entreprise d'intégrer (1) ce qu'elle veut faire (sa stratégie, ses objectifs) ; (2) ce qu'elle peut faire (ses ressources, ses moyens) ; et (3) ce qu'elle sait faire (ses compétences, ses modes d'action) pour faire de l'innovation en continue.

Nous plaçons notre recherche dans une perspective d'évolution des connaissances de l'entreprise à partir de modèles. Le concept de modèle est pris dans le sens d'un langage de représentation et communication symbolique du pourquoi et du comment de la réalité selon une certaine approche (c'est-à-dire un certain angle avec lequel l'observateur observe le phénomène). En ces termes, l'observateur à travers le modèle essaie de saisir et de représenter un phénomène afin de faire émerger quelque chose d'opérationnel et de capable qui assurerait la viabilité du business de l'entreprise[8]. Et pourtant, ce modèle est simplement un outil, ce qui importe c'est la réalité physique. A cet égard, Limone conforte cette position lorsqu'il dit « c'est en travaillant sur la réalité à l'aide d'un modèle et non exclusivement sur le modèle qu'on peut avancer petit à petit dans l'éclaircissement des phénomènes réels » [Limone, 77]. La phénoménologie existe simultanément et nécessairement sur deux domaines, l'un est le modèle (c'est-à-dire l'abstraction symbolique sous un certain angle), l'autre est la réalité physique.

---

[7] A ce sujet il serait utile de consulter les travaux de [Córdova, 99], [Jiménez, 00a], [Jiménez, 00b], [Córdova, 02], [Cortes, 03].

Cette thèse s'intéresse à une réalité de gestion des connaissances imparfaites, réalité dans laquelle la connaissance existe simultanément et nécessairement dans un domaine : flou et non flou à travers un système de connaissances[9] et un système opérationnel.

Cette réalité nous la percevons dans le système de connaissances à travers la "théorie" de la gestion des connaissances et la théorie de l'imprécis et de l'incertain. Pour Lotfi Zadeh[10] cette théorie est relative à « l'aptitude du cerveau humain (une aptitude que les actuels ordinateurs ne possèdent pas) à penser et à raisonner en termes imprécis, non qualitatifs, "flous" » [Kaufmann, 77]. C'est-à-dire que si nous faisons l'hypothèse que la connaissance existe dans le cerveau humain, alors le flou et le non flou font partie du système cognitif (l'individu : il fait assez beau aujourd'hui), et du système de connaissance (l'entreprise : nos clients nous adressent des demandes floues pas toujours satisfaisantes). En revanche, dans le système opérationnel nous percevons la réalité à travers des données imprécises et incertaines.

Nous avons centré ce travail dans le domaine industriel de la fabrication de carton. Ceci nous a permis d'observer une réalité imprécise et incertaine et de rester dans un cadre générique. La problématique de gestion des connaissances imparfaites se réfère à la prise en charge du flou, c'est-à-dire à l'imperfection de la connaissance sous l'angle des bases de données relationnelles floues par le biais des objets flous, des événements flous, des requêtes floues et le traitement de données imprécises et incertaines. La gestion des connaissances imparfaites signifie pour nous l'extraction des connaissances de données étendue aux données imprécises et incertaines d'une base de données relationnelles floues.

---

[8] Le terme "business" signifie pour nous l'énergie qui fait fonctionner l'organisation du système. Notre réflexion n'est pas si loin de la pensée du businessman Donald Trump, lors qu'il dit « business is to make money », et il ajoute « business is also about sex », dans un récent interview donné pour lui à la chaîne CNN (LateEditon) lors de la sortie de son livre *Trump : How to get rich*.
[9] Le terme *système de bases de connaissances* est un héritage du terme *bases de données*. Dans cette thèse un *système de bases de connaissances* de type *base de règles* ou *base de cas*, correspond plutôt à un *système opérationnel* et non pas à un *système de connaissances*. D'ailleurs, il faut souligner aussi qu'à certains moments nous préférons utiliser ce terme sans "s", afin d'emphatiser davantage l'unité, l'identité et l'autonomie d'un tel système.
[10] Nous matérialisons la contribution de Zadeh dans les concepts suivants : Fuzzy Sets (1965), Fuzzy Logic (1973), Berkely Initiative in Soft Computing (BISC, 1990), Human-Machine Perception (2000). Pour en savoir plus http://www.cs.berkeley.edu/~zadeh/

Ce travail comporte cinq chapitres et deux annexes.

Le chapitre 1, permet de décrire l'état de l'art de la gestion des connaissances et d'introduire notre proposition sur la base d'un modèle sociotechnique[11] selon (1) l'approche organisationnelle de Nonaka et Takeuchi fondée sur le concept knowledge creating-company ; (2) l'approche biologique de Maturana et Varela fondée sur le concept de l'arbre de connaissance ; et (3) l'approche managériale de Jean-Louis Ermine fondée sur le concept de la marguerite. Le chapitre 2, permet de présenter les approches système, cybernétique et autopoïétique de Santiago et de Valparaiso et leurs modèles associés (OID de Jean-Louis Le Moigne, OIDC de Jean-Louis Ermine, AMS de Jacques Mélèse, MSV de Stafford Beer, et CIBORGA de Aquiles Limone et Luis Bastias) qui, pour nous, sont à la base de l'approche managériale de la gestion des connaissances. Le chapitre 3, permet de concrétiser notre approche et de présenter nos réflexions. Le chapitre 4, permet de donner le cadre conceptuel de l'aspect social et technique de la gestion des connaissances que nous avons étendue aux données imprécises et incertaines, dans l'espoir de faire une extraction des connaissances à partir de ces données, ainsi que le cadre conceptuel de la représentation et l'interrogation des données dans une base de données relationnelles floues de type FSQL. Le chapitre 5, véritable cas d'étude et d'application à l'industrie du carton en Amérique Latine. Il permet de dégager une problématique industrielle capable d'être traitée par un filtre flou qui interroge une base de cas de type FSQL.

Nous terminons ce document par des conclusions générales sur les apports de notre recherche et nous en dégagerons des perspectives dans la même direction : gestion des connaissances, avec la particularité de management et retour d'expériences.

---

[11] Il s'agit d'un concept développé au début des années 50 par de chercheurs britanniques, visant à impliquer les individus et les groupes dans l'organisation des tâches. Aujourd'hui, on parle de l'école Tavistock de Londres, pour décrire ce mouvement de pensée systémique [Herbst, 74]. Enfin, Nous parlons indifféremment, de modèle, d'approche, de système ou de concept sociotechnique dans le cadre de cette thèse.

<table>
<tr><td>

# Chapitre 1

</td><td>

## Les feuilles de l'arbre de la gestion des connaissances : parties organisationnelle, biologique, managérial et NTIC du KM

</td></tr>
</table>

*Le monde du vivant, la logique de l'autoréférence et toute l'histoire naturelle de la circularité devrait nous dire que la tolérance et le pluralisme sont le véritable fondement de la connaissance. Ici, les actes valent mieux que les mots.*
*Francisco Varela, biologiste et informaticien chilien*

Dans l'introduction générale nous avons positionné la problématique de notre recherche dans un cadre général, c'est-à-dire, la gestion des connaissances par rapport à la gestion de projets innovants, la base de connaissance et le retour d'expérience dans un domaine industriel. Dans ce cadre, la gestion des connaissances est le moteur entre le business (l'activité), la mémoire collective de l'entreprise, c'est-à-dire l'ensemble des savoirs, savoir-produit, savoir-client, savoir-financier, etc., et la dynamique continu d'apprentissage organisationnel qui permet d'améliorer la rentabilité de l'entreprise. Nous constatons bien que les enjeux pour l'entreprise d'aujourd'hui ne sont plus la gestion des données, ni la gestion de l'information, mais la gestion des connaissances.

Dans ce chapitre nous ferons l'état de l'art de la gestion des connaissances, pour cela nous avons choisi un style de présentation différent de ce que nous avons trouvé dans trois thèses récentes sur ce sujet, nous parlons de la thèse de Pachulski, intitulée *Le repérage des connaissances cruciales pour l'entreprise : concepts, méthode et outils* (publié en 2001) [Pachulsky, 01], de la thèse Tounkara, intitulée *Gestion des connaissances et veille : vers un guide méthodologique pour améliorer la collecte d'informations* (publié en 2002) [Tounkara, 02], et de la thèse et de la thèse de Barthelme-Trapp, intitulée *Une approche constructiviste des connaissances : contribution à la gestion dynamique des connaissances* (publié en 2003) [Barthelme-Trapp, 03]. Le modèle choisi pour Tounkara afin de représenter la connaissance est le modèle de la marguerite de Jean-Louis Ermine, tandis que Pachulski et Barthelme-Trapp font appel au modèle des objectifs managériaux de Michel Grundstein. Pour nous, Ermine et Grundstein ont orienté leurs modèles vers une approche managériale de la gestion des connaissances. Néanmoins, pour Ermine la connaissance est liée à une activité il est donc possible de la modéliser par des outils descriptifs de la connaissance que nous trouvons dans le domaine de l'ingénierie des connaissances[1]. En revanche, pour

---

[1] Pour Charlet l'ingénierie des connaissances correspond à « l'étude de concepts, méthodes et techniques permettant de modéliser et/ou acquérir les connaissances pour des systèmes réalisant ou aidant des humains à réaliser des tâches se formalisant a priori, peu ou pas » [Charlet *et al*, 00]. Ainsi, comme le souligne Barthelme-Trapp « l'ingénierie des

Grundstein la connaissance est liée aux acteurs (individu, groupe, entreprise), et donc comme le souligne Barthelme-Trapp « il y a impossibilité d'isoler la connaissance et les acteurs ». Cela signifie que dans un modèle de gestion des connaissances où la connaissance est liée à l'activité, ce que l'on trouve dans le modèle est une relation entre concepts qui ont un sens pour l'activité elle-même. Tandis, que si la connaissance est liée à l'acteur, le modèle décrit une relation entre acteurs, ce que Barthelme-Trapp appelle « la dynamique des connaissances ». Pour elle, cette "dynamique" est une condition nécessaire à la production de nouvelles connaissances.

Et à ce propos, Pachulski, Tounkara et Bartheme-Trapp ont choisi de faire leur état de l'art par rapport aux approches classiques de la gestion des connaissances. Pourtant, dans ces thèses nous trouvons une liste de modèles descriptifs de la connaissance, tel que CommonKADS, MASK et autres, c'est pour cette raison que nous n'avons pas trouvé nécessaire de rentrer dans le détail de chaque modèle, pour établir notre état de l'art de la gestion des connaissances, en plus l'équipe de recherche ACACIA à l'INRIA Sophia Antiopolis, dirigé par Rose Dieng, a publié récemment deux ouvrages avec l'ensemble des modèles. L'un est *Méthodes et outils pour la gestion des connaissances* (publié en 2000), l'autre est *Méthodes et outils pour la gestion des connaissances : une approche pluridisciplinaire du Knowledge Management* (publié en 2001). Néanmoins, ce que nous avons trouvé intéressant dans l'état de l'art chez Pachulski, Tounkara et Bartheme-Trapp a été l'ordre et le choix des modèles, par rapport au sujet de thèse. En effet, pour Pachulski il s'agissait de repérer des connaissances cruciales pour l'entreprise en regard de la performance de celle-ci. Dans cette approche la connaissance est vue comme un processus de décision, bâti autour de deux dimensions : la dimension de "l'être", qui permet à un individu de construire sa "vision du monde" (terme emprunté de Watzlawick), et la dimension du "faire", qui permet à un individu d'agir sur le monde [Pachulsky, 01]. Par contre, pour Tounkara il s'agissait de repérer des concepts pour mettre en place un modèle de gestion des connaissances qui prend en charge une problématique : environnement et connaissance dans un cadre de veille scientifique et technologique[2]. Dans ce contexte le modèle de gestion des connaissances proposé par lui permet d'expliquer (analyser et optimiser) le processus d'interaction du patrimoine de connaissances de l'entreprise avec son environnement extérieur en vue d'acquérir, d'intégrer, de créer de nouvelles connaissances pour connaissances propose des méthodes de conception et de modélisations de systèmes à base de connaissances à travers des méthodes et techniques permettant l'acquisition et/ou la représentation des connaissances » [Barthelme-Trapp, 03].
[2] Pour Tounkara l'environnement concerne « l'ensemble des acteurs susceptibles d'avoir une influence sur l'entreprise ». Tandis que pour lui la connaissance concerne « l'ensemble des savoirs et savoir-faire mobilisés par les acteurs dans le cadre de leurs activités » [Tounkara, 02]. Puis il ajoute « cette définition implique que la connaissance n'est véritablement connaissance que si elle est prise dans l'action et elle n'a de sens que pour ceux qui la produisent et pour ceux qui l'utilisent » [Tounkara, 02].

---

l'entreprise [Tounkara, 02]. Enfin, l'état de l'art de Barthelme-Trapp a permis d'en dégager des concepts afin de formuler une approche dynamique de la gestion des connaissances sur la base d'un modèle de gestion dynamique de la connaissance [Barthelme-Trapp, 03].

Dans le cadre de cette thèse, nous plaçons notre problématique de gestion des connaissances dans un cadre sociotechnique relatif à la connaissance imparfaite dans une activité industrielle. C'est pour cette raison que nous avons décidé d'organiser notre état de l'art de la gestion des connaissances par rapport à un *modèle sociotechnique*[3].

Avant de continuer, nous avons trouvé nécessaire ici d'ouvrir une parenthèse pour expliquer ce modèle. Il a été développé au début des années 50 par de chercheurs britanniques, comme un outil conceptuel pour l'organisation du travail, visant à impliquer les individus et les groupes dans l'organisation des tâches. Depuis son apparition cette approche a été appliqué principalement dans l'industrie (par exemple Volvo, Philips, etc.) [Gousty, 98]. Dans ce contexte industriel, d'après Gousty, deux traits sont à la base de l'approche sociotechnique :

- l'organisation est divisée en segments fonctionnant en unités de production relativement autonomes ;
- les responsables des méthodes ne donnent qu'un minimum d'instructions indispensables au fonctionnement global de l'organisation, laissant aux unités autonomes le soin de prendre les autres décisions.

Ainsi, selon Gousty, « toute opération technique peut être considérée selon deux points de vue : un volet technique où prédomine la machine, un volet humain où les opérateurs tiennent une place essentielle dans le succès de la production ». Pour Gousty, cette approche sociotechnique est « particulièrement efficace lorsqu'il s'agit de mettre en œuvre une technologie complexe, dépendant des variables affectées de nombreuses incertitudes. L'inconvénient majeur réside dans la difficulté à maintenir la cohésion du groupe » [Gousty, 98].

Aujourd'hui, le concept sociotechnique traduit un chemin de pensée plus proche du constructivisme que du positivisme, pour faire intégrer dans une même démarche d'organisation du système (entrées/transformation/sorties) deux composants en interrelation permanente. L'un est la composante sociale, comprenant d'une part des acteurs (individu, groupe, entreprise), et d'autre part, la structure de l'organisation. L'autre est la composante technique, composé, d'une part, par des tâches, et d'autre part, par la technologie. Nous attirons l'attention, que la richesse de l'approche réside dans le fait que les éléments de ces composants sont aussi en relation avec les autres, dans différentes niveaux systémiques, en générant ainsi plusieurs niveaux explicatifs d'une même réalité.

---

[3] Dans cette thèse nous parlons indifféremment de modèle, système ou approche sociotechnique.

Enfin, Herbst, dans son livre intitulé *Socio-Technical Design* (publié en 1974) parle de l'école Tavistock de Londres, pour décrire ce mouvement [Herbst, 74].

Nous fermons cette parenthèse, afin de revenir à notre explication du cadre de la thèse. Nous utilisons alors le modèle sociotechnique comme un chemin explicatif d'une réalité (la gestion des connaissances). Ainsi, nous décrivons d'abord l'aspect social de la gestion des connaissances puis son aspect technique. En plus, nous introduisons l'adjectif "récursif", pour impliquer davantage l'aspect « dynamique des connaissances » entre la composante technique et la composante sociale de la gestion des connaissances à partir la dualité[4] social/technique et la dualité organisation/structure. Nous parlons, alors de *modèle sociotechnique récursif*.

Ce choix de l'approche sociotechnique dans notre démarche, nous le justifions aussi par le fait suivant. En effet, d'après l'article d'Aquiles Limone et Luis Bastias, intitulé : *Autopoiésis y Conocimento en la Organización. Fundamento Conceptual para una Auténtica Gestión del Conocimiento*, ils disent « tradicionalmente, la Gestión del Conocimiento (GC) se ha sustentado principalmente sobre la implantación de tecnologías informáticas, proporcionándole mayor relevancia a la tecnología que a la información o al conocimiento, entendidos como entidades epistemológicas. Esto ha sido grave ya que se ha traducido en una simple moda y la GC en un mero nombre comercial destinado a vender determinadas herramientas de software y hardware a las empresas » [Limone et Bastias, 02b]. Ce point de vue, que nous avons entendu l'année 2002 à l'Université de Talca au Chili à l'occasion du XIX conférence de l'ENEFA, nous laisse penser qu'il y a un double discours par rapport à la gestion des connaissances, d'abord un discours autour de la technologie, c'est-à-dire quelque chose que l'on peut voir, et un autre discours, que nous ne pouvons pas voir, lequel se construit autour de la question fondamentale : ça sert à quoi ? Et pour certains, la réponse se trouve simplement dans la gestion de données ou dans la gestion de l'information, mais gérer les connaissances, ce n'est pas une problématique de gestion de stocks, pour plus complexe que soit cette gestion.

---

[4] Comme nous verrons dans le chapitre 2, la dualité caractérise une complémentarité plutôt qu'une opposition, à partir le concept de *processus de causalité circulaire défini dans une dualité cause/effet*. Ce processus est attaché à la cognition, alors si le processus est relatif à l'approche cognitiviste de la cognition, ce que l'on cherche à construire est une chaîne logique de cause à effet, tandis que si le processus est associé à l'approche enactiviste de la cognition, alors la relation ne se construit pas nécessairement d'une représentation vraie ou logique de causes à effets, puisque le problème de l'enaction n'apparaît pas lié à la représentation symbolique d'une réalité, sinon qu'au maintient du système en vie (organisation/structure) et viable (sens).

Cette observation chez Limone et Bastias nous a permis de faire une analogie[5] avec un *arbre*, où d'une part, la partie visible de l'arbre symbolise la technologie[6], et d'autre part, les racines symbolisent la partie cachée de la gestion des connaissances. En plus, cette analogie est récursive car la connaissance a aussi une partie qui est visible et une partie qui est cachée dans nous-mêmes. Nous reviendrons plus loin sur cette argumentation à partir de l'arbre des connaissances de Maturana et Varela et l'arbre du Club Gestion des Connaissances de Jean-Louis Ermine. Sans oublier que cette analogie a donné le nom de ce chapitre pour présenter, selon notre point de vue, état de l'art de la gestion des connaissances.

Ainsi, nous décrivons les enjeux de la complexité[7] de l'aspect social et technique de la gestion des connaissances. Nous essayons alors de comprendre les mécanismes de création des connaissances nouvelles et d'apprentissage organisationnel afin d'établir, d'une part, des mécanismes génériques de la gestion des connaissances (ce que nous ferons dans ce chapitre 1 à l'aide des approches organisationnelles, biologiques, managériales, et NTIC du KM, sous l'hypothèse qu'un "pont" existe entre eux), et d'autre part, de dégager un modèle[8] de la gestion des connaissances, appelé *modèle autopoïétique de la gestion des connaissances imparfaites*[9] (ce que nous ferons dans le chapitre 3 à l'aide de modèles système, cybernétique et autopoïétique que nous présentons au chapitre 2). Dans ce contexte, la connaissance et la création de connaissance collective nouvelle (tacite ou explicite)[10] et d'apprentissage organisationnel apparaissent comme la dynamique de faire émerger la connaissance collective (tacite ou explicite) à travers l'apprentissage organisationnel et l'apprentissage émotionnel collectif.

---

[5] Pour Bourdieu « raisonner par analogie, c'est former un raisonnement fondé sur les ressemblances ou les rapports d'une chose avec une autre » [Bourdieu *et al*, 73]. De plus, Limone dit « une analogie n'est valable et féconde que si elle est guidée et soutenue par une réflexion théorique et par une démarche méthodologique » [Limone, 77].

[6] D'après Gousty, la technologie, en citant à Nollet *et al.*, est « un ensemble de méthodes, de procédures, d'équipements et même d'approches utilisées pour fournir un service ou produire un bien » [Gousty, 98].

[7] La complexité fait preuve des relations de causalité circulaire entre l'aspect social et l'aspect technique, et donc, la complexité n'a rien avoir avec la taille du système, le nombre de relations, ou les phénomènes compliqués du système. Une excellente démarche pour dialoguer avec la complexité (la méthodologie MIAPRO [Muñoz, 04]) a été développée par José Muñoz et appliquée dans une entreprise du secteur énergétique (dans le cadre de sa thèse, dirigée par notre grand ami Edmundo Leiva).

[8] Nous n'avons pas l'intention de proposer ici un modèle pour "faire-évoluer" et "faire-émerger" la connaissance, en fait nous sommes dans une phase de réflexion sur l'utilité et la faisabilité d'un tel projet. Néanmoins, notre démarche et canevas de pensée sera guidé par l'approche de l'enaction de Maturana et Varela [Varela, 96] et leur modèle autopoïétique. Ainsi que, les contributions de Limone et Bastias sur la relation connaissance et autopoïèse.

[9] Le fondement théorique de notre modèle est l'approche de l'enaction de Maturana et Varela [Varela, 96] et leur modèle autopoïétique, qui enrichi davantage l'approche de l'enaction de Karl Weick [Weick, 79], comme nous le verrons dans le chapitre 3. D'ailleurs, c'est l'approche de l'enaction de Karl Weick, le fondement théorique du modèle de la marguerite de Jean-Louis Ermine, et ses évolutions, par exemple chez Tounkara [Tounkara, 02].

Dans la littérature il est possible de trouver de multiples contributions dans le domaine de l'apprentissage organisationnel, notamment à travers les travaux de Drucker, Senge, Argyris, Nonaka et Takeuchi, Davenport, Prusak, Prax, etc., qui ont permis l'émergence et ont forgé les concepts de *knowledge worker* (ou *knowledge society*), *learning organization* (ou *systems thinking*), *actionable knowledge*, *knowledge-creating company*, *information ecology* (ou *information age*), *knowledge based economy*, *corporate knowledge* sans oublier les travaux de Michel Grundstein sur la capitalisation des connaissances de l'entreprise et Jean-Louis Ermine, plus orientés sur des modèles de capitalisation de connaissances métiers, tels que MKSM et MASK. Le point commun de ces auteurs est qu'ils regardent la connaissance d'un point de vue social (savoir-faire humain), alors que pour René-Charles Tisseyre et autres, la connaissance existe davantage d'un point de vue technique (savoir-faire encapsulé dans un système à base de connaissances). Dans ce cas l'offre technologique des nouvelles technologies de l'information et de la communication du Knowledge Management (les NTIC du KM) autour de l'Internet, l'Intranet, et l'Extranet est capitale.

En revanche, au niveau des contributions de l'apprentissage émotionnel dans l'entreprise, la littérature est un peu limitée. Le livre de Daniel Goleman, intitulé *L'intelligence émotionnelle 2* (publié en 1999), introduit l'hypothèse que l'intelligence émotionnelle peut être cultivé dans les relations de travail. Le modèle de Goleman a été pensé autour de deux sortes de compétences chez l'individu. L'un est la compétence personnelle (la connaissance de soi, la maîtrise de soi et la motivation), l'autre et la compétence sociale (l'empathie et les aptitudes sociales) [Goleman, 99]. Dans l'apprentissage collectif c'est justement la compétence sociale qu'il faut développer dans et par l'entreprise. Nous pensons que l'approche de Barnard sur la "connaissance comportementale" peut être approché à l'intelligence émotionnelle.

L'objectif de ce chapitre est d'une part, de faire l'état de l'art sur la gestion des connaissances, et d'autre part, d'introduire notre proposition sur la base d'un modèle sociotechnique récursif, qui permet, d'une part, d'établir des mécanismes génériques de la gestion des connaissances (que nous présentons dans la section 1.4, ainsi qu'un *framework* pour approcher la question fondamentale : qu'est-ce que la connaissance ?), et d'autre part, de dégager la problématique essentiel d'un modèle de la gestion des connaissances (que nous ferons dans la conclusion de ce chapitre 1).

---

[10] En général, la connaissance collective et la connaissance individuelle peuvent être tacites (idées, métaphores, créances, concepts, hypothèses, modèles mentaux, analogies, etc.) ou explicite (une source d'information, par exemple un document, un email, un modèle, une maquette, un plan, etc.). La représentation "matérielle" de la connaissance collective dans de nouveaux produits ou services, apparaît de nos jours comme un sujet de recherche important.

Ce chapitre est organisé en cinq parties. La première partie appelée « qu'est-ce que la gestion des connaissances ? » permet d'explorer la pensée de plusieurs auteurs contemporains qui ont investi du temps et de l'énergie sur la question, sans oublier de donner notre point de vue sur ce sujet aussi. La deuxième partie appelée « aspect social de la gestion des connaissances » essaie de dégager les mécanismes de création de connaissances nouvelles et d'apprentissage organisationnel d'un point de vue social, selon (1) l'approche organisationnelle[11] de Nonaka et Takeuchi fondée sur le concept de *knowledge creating-company* ; (2) l'approche biologique[12] de Maturana et Varela fondée sur le concept de l'*enaction*[13] ou l'*arbre des connaissances* ; et (3) l'approche managériale[14] de Jean-Louis Ermine fondée sur le concept de la *marguerite*. La troisième partie appelée « aspect technique de la gestion des connaissances » essaie de dégager les mécanismes de création de connaissances nouvelles et d'apprentissage organisationnel selon l'approche des nouvelles technologies de l'information et de la communication du Knowledge Management (l'approche NTIC du KM). La quatrième partie appelée « les mécanismes de généralisation de la gestion des connaissances selon l'aspect social et l'aspect technique » permet de généraliser les mécanismes de la gestion des connaissances à partir les mécanismes de création de connaissances nouvelles et d'apprentissage collective des approches organisationnelle, biologique, managériale, et NTIC du KM (sous l'hypothèse qu'un "pont" existe entre ces approches-là). Ces mécanismes sont représentés par des verbes à l'infinitif. La cinquième partie appelée « les origines de la connaissance industrielle » permet d'introduire l'importance de la connaissance dans notre histoire industrielle, d'abord comme un levier de productivité, puis comme un levier d'avantage concurrentiel ou compétitif et enfin, comme un levier d'avantage coopératif. Pour cela nous faisons appel à Taylor, Ford, Mayo, Simon et Penrose, entre autres pour illustrer les différents systèmes d'organisation du travail pour une économie de production (1930), une économie de service (1960) et une économie globalisée (1990).

---

[11] Dans ce contexte, il s'agit des mécanismes de création de connaissance et d'apprentissage social d'un point de vue organisationnel (les relations sont au niveau acteur : individu, groupe, entreprise).
[12] Dans ce contexte, il s'agit des mécanismes de création de connaissance et d'apprentissage social d'un point de vue biologique (les relations sont au niveau cellulaire).
[13] L'approche de l'enaction de la cognition correspond au fait de faire émerger un comportement intelligent. Ici l'intelligence ne se définit plus comme la faculté de résoudre un problème mais comme celle de pénétrer un monde partagé. Cette approche nous le verrons davantage dans le chapitre 2, ainsi que l'approche cognitiviste et l'approche connexionniste de la connaissance. Pour nous l'*arbre des connaissances* symbolise l'approche de l'enaction de Maturana et Varela, cette analogie nous la faisons à partir du site de Jean-Louis Ermine *Le Club de Gestion des Connaissances* http://www.club-gc.asso.fr, où un "arbre" est le logo du Club. Nous parlons indifféremment dans cette thèse d'enaction ou d'arbre des connaissances. Néanmoins, pour nous, cette approche est beaucoup plus riche que l'approche de l'enaction de Karl Weick, que dans son livre, intitulé *The Social Psychology of Organizing* (publié en 1979) [Weick, 79], décrit la relation organisation-environnement comme un processus d'interaction, où chacun se construit elle-même par cette interaction (et non pas comme une relation d'inclusion unilatéral entre l'organisation et leur environnement). D'ailleurs se justement l'approche de l'enaction de Weick qui est le fondement théorique du modèle de la marguerite de Jean-Louis Ermine (et ses évolutions), par exemple chez Tounkara [Tounkara, 02].
[14] Dans ce contexte, il s'agit des mécanismes de création de connaissance et d'apprentissage social d'un point de vue managérial (les relations sont entre l'entreprise et son environnement).

*« Descartes, dans l'histoire de la pensée, ce sera toujours ce cavalier français qui partit d'un si bon pas »*
*Péguy*

## 1.1.    Qu'est-ce que la gestion des connaissances ?[15]

Voilà, une question qui pour sa complexité nous pouvons être tentés de l'approcher séparément pour essayer de l'analyser et la comprendre, tel que l'a fait René Descartes pour étudier l'être humain au XVIIème siècle, lorsqu'il dit dans son ouvrage fondamental *Discours de la méthode* « ces hommes sont composés, comme nous, d'une Ame et d'un Corps. Et il faut que je vous décrive, premièrement, le corps à part, puis après l'âme aussi à part ; et enfin, que je vous montre comment ces deux natures doivent être jointes et unies, pour composer des hommes qui nous ressemblent » [Descartes, 66].

En réalité ce "si bon pas" a été le chemin suivi par René-Charles Tisseyre[16] pour approcher cette question. En effet, dans son livre intitulé *Knowledge Management* (publié en 1999) il a dit « définir le Knowledge Management ou sa traduction littérale, la "gestion des connaissances", est un art difficile car il fait appel à deux notions abstraites : la "gestion" et les "connaissances" qui ne peuvent donner naissance qu'à un concept lui-même abstrait dont l'existence même pourrait paraître étonnante » [Tisseyre, 99]. Pour lui « "gérer" n'est pas produire et correspond à de tâches d'organisation et la "connaissance" n'est pas l'information mais se rapproche plus du savoir-faire au sens classique » [Tisseyre, 99].

Il nous semble, que cette dualité cartésienne est un peu plus complexe, car (1) la "connaissance", apparaît, d'une part, liée à l'individu, tel que l'apprentissage, l'intelligence, la compétence, le savoir, le savoir-faire, le savoir-faire opératoire, le savoir-être, le savoir-partagé, le savoir-vivre, ou autre savoir qu'il faut mobiliser au sein de l'entreprise pour créer une avantage concurrentiel, compétitif ou coopératif durable[17], et d'autre part, la connaissance est sous-jacente à l'activité productive, tels que les données et les informations qui sont produits par cette activité et consommés par le processus de prise de décision de l'entreprise (élaboration de stratégies à tous les niveaux de l'entreprise) ; et (2) "gérer" apparaît aussi liée à l'individu et à l'activité productive qui lui réalise dans l'entreprise, d'une part, il s'agit de *capitaliser*, *partager* et *créer* des connaissances, et d'autre part, il s'agit de *créer*, *produire* et *offrir* de nouveaux produits ou services. D'une manière générale ce premier trait de la gestion nous l'appelons *gestion des connaissances*, tandis que le deuxième trait nous l'appelons *gestion de l'innovation*[18]. Ainsi, pour nous "créer" doit être lié à la *gestion des connaissances* pour décrire la création des connaissances nouvelles, mais aussi doit être lié à la *capacité de l'entreprise*[19] pour transformer cette idée en produits ou services innovants, c'est-à-dire pour créer une avantage concurrentiel, compétitif ou coopératif durable pour l'entreprise. Par conséquent, "créer" est, d'une part, le dernier échelon de la *gestion des connaissances* : *capitaliser*, *partager* et *créer*, et d'autre part, le premier échelon de la capacité de l'entreprise pour la *gestion de l'innovation* : *créer*, *produire* et *offrir* de nouveaux produits ou services.

---

[15] Dans la plupart de livres (Tisseyre, Prax), thèses (Pachulsky, Tounkara, Batherme-Trapp), articles, et différents sites web que nous avons répertoriés et qui abordent le sujet de la *gestion des* connaissances, le point de départ est justement cette question, c'est pour cette raison que nous ne voulons pas échapper au déjà établi.
[16] Il a été l'ex-directeur de l'offre de gestion des connaissances et ex-membre de l'équipe internationale *Applied Knowledge Management* (AKM) de la société Cap Gemini Ernst & Young, qui depuis le 15/04/04 le groupe s'appelle Capgemini http://www.fr.capgemini.com/home. Aujourd'hui Tisseyre est l'EFQM Project Advisor chez Capgemini.

Pourtant, la complexité de la gestion des connaissances est sans doute plus complexe que l'a vécu pour la gestion de l'information à partir des années 70, la méthode MERISE en témoigne. En plus nous pensons que cette complexité doit être intégrée dans toute tentative de réflexion sur la *gestion des connaissances*, tel comme nous la percevons chez :

a) Pour Wendi Bukowitz et Ruth Williams de la société PricewaterhouseCoopers, la complexité de la *gestion des connaissances* est perçue à travers le capital intellectuel, lorsqu'elle disent « gestion des connaissances : démarche selon laquelle l'entreprise génère de la richesse à partir de son savoir ou de son capital intellectuel » [Bukowitz et Williams, 00][20].

b) Pour Karl Sveiby de la société Sveiby, en citant à Leif Edvinsson, la complexité de la *gestion des connaissances* se trouve aussi dans une relation avec la valeur du capital intellectuel d'une entreprise, lorsqu'il dit « Knowledge Management, la nouvelle richesse des entreprises … vous aidera à développer une stratégie fondée sur le savoir et à approfondir vos intuitions de dirigeants sur les pièges d'un contrôle des coûts trop rigide. En vous montrant comment les actifs intangibles peuvent créer de la valeur, il vous permettra de découvrir des stratégies à mettre en œuvre pour améliorer la rentabilité de votre entreprise » [Sveiby, 00][21].

c) Pour René-Charles Tisseyre de la société Capgemini, la complexité de la *gestion des connaissances* se trouve dans une dimension humaine, que nous percevons dans cette définition « le Knowledge Management est à la fois une démarche, une nouvelle organisation, une nouvelle approche du rôle des acteurs par rapport à cette organisation, un nouvel effet de levier pour le développement de ces organisations et un nouvel usage des technologies » [Tisseyre, 99]. En revanche, pour lui la *gestion des connaissances* n'est « ni une mode ; car elle correspond à une besoin fort et concret de la part des organisations ; ni un nouveau concept, mais plutôt la redécouverte, à l'aide des nouvelles technologies, que les informations structurées ne sont pas tout ; ni de la réorganisation d'entreprise, il n'a pas cette ambition ; ni de l'intelligence artificielle, car il a une vocation plus large mais moins profonde ; ni uniquement de la technologie, même s'il s'appuie sur cette dernière pour se développer » [Tisseyre, 99].

---

[17] Dans l'économie de service depuis l'émergence de l'analyse stratégique de ressources, où ès deux outils d'analyse stratégique le plus répandus sont les deux modèles de Porter. L'un est le modèle de "cinq forces" qui permet l'analyse d'avantages concurrentiels dans l'industrie (concurrents, clients, entrants, substituts, fournisseurs) présenté dans son livre intitulé *Competive Strategy* (publié en 1980), et l'autre est le modèle de la "chaîne de valeur" qui permet l'analyse d'avantages compétitifs dans l'entreprise (ressources, capacités, compétences), présenté dans son livre intitulé *Competive Advantage* (publié en 1985). L'avantage coopératif est associé au concept de l'entreprise élargie. Selon Prax « l'idée de "l'entreprise élargie" est d'améliorer le fonctionnement d'un réseau professionnel par une transparence totale d'information, de savoirs et de savoir-faire entre les différents acteurs de la chaîne de conception-production … même s'ils sont concurrents » [Prax, 00].

[18] Ici nous parlons dans un sens large de l'innovation industrielle, c'est-à-dire de l'innovation d'un point de vu technique (un nouveau produit, un nouveau process ou procédée, etc.), mais aussi de l'innovation organisationnelle, c'est-à-dire de l'innovation d'un point de vu social (un nouveau méthode de travail, un nouveau mode de management, une nouvelle structure organisationnelle, un nouveau processus, un nouveau service, etc.). Néanmoins, dans les études de cas du livre de Nonaka et Takeuchi *The Knowledge-Creating Company. How Japanese Create the Dynamics of Innovation* (publié en 1995) l'innovation est relative à l'innovation industrielle, car les études de cas qui illustrent leur modèle, ils sont relatifs au développement de nouveaux produits. Nous ne pesons pas que cela reste généralité à leur modèle.

[19] Pour Jean-Claude Tarondeau « les capacités sont définies comme des routines de mise en œuvre d'actifs pour créer, produire et/ou offrir des produits ou services sur un marché » [Tarondeau, 98]. Ainsi que ressources, capacités, savoir et compétences sont liées, lorsqu'il dit « les ressources, y compris le savoir, deviennent capacités quand elles sont combinées, intégrées et coordonnées dans le contexte d'une activité » [Tarondeau, 98]. La liaison entre capacité et compétence nous la trouvons davantage chez Mélissa Saadoun, pour elle « compétence est une aptitude permettant de transformer une connaissance en action et d'obtenir le rendement escompté. Les compétences représentent les aptitudes et les capacités essentielles en milieu professionnel. L'aptitude correspond à la faculté d'apprentissage, tandis que la capacité est la faculté d'accomplir les tâches inhérentes à un emploi donnée » [Saadoun, 96].

---

[20] Pour Bukowitz et Williams « capital intellectuel est tout élément qui, détenu par des personnes ou dérivé de processus, de systèmes ou de la culture d'une organisation, présente une valeur pour cette dernière : compétences et qualifications individuelles, normes et valeurs, bases de données, méthodes, programmes informatiques, savoir-faire, brevets, marques, secrets de fabrication, pour n'en citer que quelques-uns » [Bukowitz et Williams, 00].
[21] Cette définition nous l'avons adaptée de l'introduction faite par Leif Edvinsson au livre de Karl Sveiby intitulé *Knowledge Management. La nouvelle richesse des entreprises. Savoir tirer profit des actifs immatériels de sa société* (publié en 2000).

d) Pour Jean-Yves Prax de la société CorEdge, la complexité de la *gestion des connaissances* se trouve aussi dans une dimension humaine, nous la percevons lorsqu'il se pose la question « pourquoi s'intéresser au management de la connaissance ? Les témoignages sont clairs : l'agent contemporain est littéralement submergé par un raz-de-marée informationnel, conséquence de la révolution bureautique. Son rêve pourrait se résumer en un phrase : apporte-moi l'information dont j'ai besoin, au moment ou j'en ai besoin, et si possible sans que j'en fasse la demande » [Prax, 00]. Voilà, un nouveau paradigme[22] organisationnel dans le sens que « l'organisation n'est pas tant un système de "traitement de l'information" mais bien de "création de connaissance" » [Prax, 00], ce qu'implique un nouveau besoin d'entreprise qui d'après lui et d'autres, fait appel, d'une part, à de nouveaux "métiers du savoir" en empruntent le mot de Prax, tels que le Knowledge Manager, Chief Knowledge Officier, Chief Learnig Officier, etc., et d'autre part, à de nouveaux "modèles du savoir" (mot inventé par nous)[23], pour la mise en place de systèmes de connaissances et la *gestion des connaissances* dans l'entreprise. Dans la pluspart de cas, il s'agit de mots "made in USA" qu'il faut adapter à la culture de l'entreprise, ou carrément de mots "made in France", comme nous le voyons chez Prax. En effet, dans *Le guide du knowledge management* (publié en 2000)[24], il a dit « nous y présentons un modèle original, baptisé *Corporate Knowledge*, qui tente de rendre compte de la problématique du management de la connaissance collective à l'échelle de l'entreprise » [Prax, 00]. Pour Prax la complexité de la *gestion des connaissances* peut être exprimé comme suit « (1) une approche qui tente de manager des items aussi divers que pensées, idées, intuitions, pratiques, expériences émis par des gens dans l'exercice de leur profession ; (2) un processus de création,

---

[22] Un paradigme, tel que le terme fut introduit par Kunhn dans son livre, intitulé *La structure des révolutions scientifiques* (publié en 1982) est la manière usuelle de référer à l'ensemble cohérent des idées scientifiques qui correspond à l'explication populairement admise d'un corpus phénoménologique. Cité par Varela dans [Varela, 96].

[23] Nous avons recherché sur le moteur google le terme "modèles du savoir" et nous avons repéré d'autres deux termes associées "modèles de diffusion du savoir", "modèles de structure du savoir", mais en dehors du domaine de la *gestion des connaissances*.

[24] C'est livre est la continuation de *Manager la connaissance dans l'entreprise* (publié en 1997), et plus récemment l'on trouve chez Prax (a) *La gestion électronique documentaire* (publié en 2004) pour lui la GED est le "chantier" (mot utilisé par Tisseyre) organisationnel et technologique plus répandu de la *gestion des connaissances* dans l'entreprise, c'est livre est une version enrichi par des études de cas du même livre publié en 1998, cependant ici le GED est perçu comme un véritable système de gestion de documents électroniques bureautiques, avec de capacités de transversalité, collaboration, accélération et capitalisation des documents ; (b) *Le manuel du Knowledge Management : Une approche de 2e génération* (publié en 2003), si le GED a si bien marché pour la *gestion des connaissances*, ce n'est pas vrai pour les autres projets, l'enjeu se trouve toujours dans le partage des connaissances et la coordination des acteurs (individu, groupe, entreprise) sur un lieu virtuel de travail. En fait, le passage d'une division vertical du travail à la coopération transversal et au réseaux de coopération des acteurs pour former une entreprise élargie (entreprise, clients, concurrents, fournisseurs, etc.) est une illusion et une résistance au changement pour beaucoup d'entreprises ; (c) *Le Management territorial à l'ère des réseaux* (publié en 2002), selon Prax le management (ou intelligence) territorial est un autre champ d'application de la *gestion des connaissances*, c'est livre est la prolongation du livre *Le guide du knowledge management* (publié en 2000), car à la fin de ceci l'on dit « à l'instar des grandes entreprises, qui ont su construire de nouveaux avantages "coopétitifs" dans la coopération avec leurs fournisseurs, voir avec leurs concurrents, nous avons la conviction que les principes du knowledge management s'appliquent au *territoire et à ses acteurs* : petites, entreprises, administrations, collectivités, tissu associatif, et qu'ils ouvrent la voie à un "être économique nouveau" » [Prax, 00].

---

d'enrichissement, de capitalisation et de diffusion des savoirs qui implique tous les acteurs de l'organisation, en tant que consommateurs et producteurs ; et (3) suppose que la connaissance soit capturée là où elle est créée, partagée par les hommes et finalement appliquée à un processus de l'entreprise » [Prax, 00].

e) Pour Jean-Louis Ermine, la *gestion des connaissances* « est revenue d'outre-atlantique, sous le vocable de Knowledge Management, avec une vigueur et une force qui en font un des mots-clés des entreprises actuellement » [Ermine, 03]. C'est dire que pour lui il s'agit aussi d'un mot "made in USA" pour décrire des approches, démarches, méthodes, outils, etc. Dans ce contexte, la complexité de la *gestion des connaissances* est perçue à travers le système de connaissances (un autre "si bon pas", mais cette fois-ci à partir de la dualité cartésienne cognition/action). Il s'agit à l'opposé de la généralisation des démarches pragmatiques, tels que les contributions Tisseyre et Prax, d'un modèle conceptuel : un macroscope dans le sens du macroscope (vers une vision globale) de Joël de Rosnay, pour observer la complexité de la connaissance dans un domaine social et technique au travers de l'information (données, traitements) qui prend une certaine signification (concepts, tâches) dans un contexte (domaine, activité) donné. Pour le domaine technique nous avons le modèle OIDC de Jean-Louis Ermine. Ce modèle est fondé sur l'hypothèse que le système de connaissances de l'entreprise peut être (1) contrôlé à travers des flux de connaissances (ou cognition en termes de Ermine) ; et (2) régulé à travers des flux de compétences, formant ainsi une machine cybernétique (de deuxième ordre). Ce model vient à enrichir le modèle OID de Jean-Louis Le Moigne, en posant sur lui une couche supplémentaire. A ce propos, Ermine a dit « la connaissance n'est donc pas un attribut propre à un des sous-systèmes, elle existe cependant en tant que telle, comme un patrimoine propre au système. Ceci justifie l'hypothèse de l'existence d'un quatrième système qu'on appellera "Système de (ou des) connaissance(s)", ou, pour reprendre l'expression d'Umberto Eco : "Patrimoine de connaissances". Ce sous-système est clairement un système actif. Il possède les deux activités fondamentales que lui prête Edgar Morin : *l'activité d'acquisition* des connaissances produites, et l'*activité de cognition*, relative à la transmission de ces connaissances » [Ermine, 96]. D'ailleurs, nous soulignons que le fondement théorique du modèle OID de Le Moigne est l'approche par niveaux utilisé par Von Bertalanffy pour décrire l'organisation du système. Pour le domaine social, nous avons le modèle de la marguerite, qui permet de guider, d'après nous, l'analyse de la complexité du couplage du système de connaissances (patrimoine de connaissances) et l'environnement (interne et externe) pour approcher les mécanismes d'adaptation et d'évolution entre eux, afin de garantir le maintient de l'organisation comme un tout, c'est-à-dire dans une vision globale. Nous pensons que ce modèle permet de structurer les racines de l'arbre de connaissances du Club Gestion des Connaissances, où son objectif est de rassembler « des entreprises et organismes de tout secteur, a pour vocation de développer les attitudes, la culture et les actions de gestion des connaissances, en tant que facteur de progrès des organisations ». Dans ce contexte les connaissances tacites de l'arbre de connaissances sont autour, pour le moment, de cinq démarches (1) le transfert intergénérationnel des savoirs, pour préserver la structure de l'évolution du savoir, dans ce sens la connaissance se construit pas à pas, tel le cerveau humain, ou la première couche correspond au période de l'interaction de l'organisme avec l'océan, la deuxième couche est relative au période de l'interaction de l'organisme avec le continent, et la troisième couche correspond à la période d'évolution moderne, où nous somme aujourd'hui. Dans ce contexte l'océan, la terre, la culture forment l'environnement ; (2) les technologies, pour la maîtrise de nouvelles technologies de l'information et de la communication du Knowledge Management (que nous simplifions par NTIC du KM) ; (3) la cartographie des savoirs, pour la mise au point des outils de représentation des connaissances dans l'entreprise ; (4) la mise en pratique des savoirs, afin de changer les mentalités et les habitudes de travail (le partage n'est pas une habitude humaine en soi, car elle est contraint au champ social et culturel dans lequel nous évoluons), c'est ce que nous appelons, le *système d'organisation du travail coopératif*[25] à travers de méthodes de travail et de modes de management ; (5) la rédaction des savoirs, pour établir un référentiel de bons pratiques afin de partager des connaissances de forme effective et éviter les pertes de temps et de ressources. Et donc, de (1), (2), (3), (4), (5), … doivent émerger le fruit (connaissance explicite) de l'arbre de connaissances du Club Gestion des Connaissances[26].

f) Pour nous[27], la complexité de la gestion des connaissances vient du faite que la connaissance est perçue pour l'entreprise comme un *objet de gestion* qu'elle doit gérer et cette gestion est soumise aux contraintes de temps, coûts et objectifs. Dans ce sens, le terme gestion relève du sens classique de l'administration des organisations mise en évidence par Henri Fayol : planifier, organiser, diriger et contrôler. Dans cette démarche qui est encore valable aujourd'hui pour gérer l'entreprise à tous ses niveaux (stratégique, tactique, opérationnel), l'on voit bien que la gestion est composée par des verbes qui agissent sur une « chose », ainsi, par exemple, le résultat d'organiser la « chose » est son organisation, ce résultat se justifie encore par les travaux d'Edgar Morin dans son livre *La méthode 3. La connaissance de la connaissance*. Ainsi, dans la gestion des connaissances, la « chose » est bien la connaissance qu'il faut gérer. Et donc, la performance du système d'organisation du travail coopératif, ou simplement le *business*, explique la capacité de l'entreprise pour être organisée dans le temps, c'est-à-dire la capacité de l'entreprise pour intégrer ce qu'elle veut faire (sa stratégie, ses objectifs), ce qu'elle peut faire (ses ressources, ses moyens) et ce qu'elle sait faire (ses compétences, ses modes d'action). Or du fait que pour nous le système d'organisation du travail coopératif, est composé d'un système cognitif (l'individu), et d'un système de connaissances (l'organisation), la performance de l'entreprise passe alors à travers le système de données et la gestion de données[28], le système d'information et la gestion de l'information, le système de connaissances[29] et la *gestion des connaissances*, le système de compétences et la gestion des compétences, le système d'innovation et la gestion de l'innovation, etc. Dans ce sens les connaissances, telles que les données, les informations, et les compétences, sont une ressource porteuse d'une valeur stratégique, tactique, ou opérationnel, et sa mise en action permet l'élaboration de compétences et l'apprentissage organisationnel à tous les niveaux de l'entreprise (stratégique, tactique, opérationnel). Cette idée nous le voyons aussi refléter chez Farey et Prusak lorsqu'ils disent « l'ambition du knowledge management réside dans la dissémination des savoirs pour permettre à de nouvelles idées de germer, de réduire le temps de développement des nouveaux produits et d'engendrer de meilleures décisions. La connaissance est créée et développée par des hommes ; le système de knowledge management doit donc savoir connecter les items de savoir avec les hommes qui savent l'utiliser » cité par [Prax, 00].

---

[25] L'organisation du travail correspond à la répartition du travail entre la force de travail. Historiquement nous avons deux types de répartitions. L'un est la division sociale du travail, dans ce cas l'on parle de répartition des activités à partir de fonctions spécialisées, par exemple dans une ville du moyen âge, la séparation des métiers (guerriers, prêtres, paysans, artisans, etc.) nous renvoie à une position sociale déterminée par le métier. L'autre est la division technique du travail, dans ce cas l'on parle de segmentation des tâches à partir de la production d'un même type de produit, par exemple dans la fabrication d'automobiles, plusieurs tâches sont ordonnancées dans une ligne d'assemblage de travail à la chaîne pour la fabriquer, certaines de ces tâches seront manuelles et d'autres automatisées. Ici, le terme coopératif implique davantage les mécanismes de la gestion des connaissances (capitaliser, partager, créer), la gestion de l'innovation (créer, produire, offrir), et la gestion des compétences (apprentissage, action) à travers un réseau des savoirs.
[26] http://www.club-gc.asso.fr/
[27] Maître de conférences en management de l'information à l'Université Catholique de Maulé, Licencié en informatique et sciences humaines à la Faculté des sciences sociales, politiques et économiques de l'Université Libre de Bruxelles.

En résumé, il faut adopter et adapter la gestion des connaissances à la culture et l'espace social de l'entreprise. Peut-être pour cette raison on parle davantage de *management des connaissances* [Boughzala et Ermine, 04]. Ceci implique l'émergence de trois paradigmes :

- paradigme organisationnel, la gestion des connaissances n'est pas un système de "traitement de l'information" mais bien un système de "création des connaissances nouvelles" et "d'apprentissage organisationnel". Ce paradigme introduit une problématique des *mécanismes* de création des connaissances nouvelles et d'apprentissage organisationnel ;

---

[28] En jargon informatique on prale plutôt de bases de données, et administration de bases de données.
[29] Voir aussi les systèmes experts, les systèmes à base de connaissances, et l'administration des bases de connaissances.

- paradigme stratégique, la gestion de connaissances permet de développer une stratégie fondée sur le savoir et non par à travers un contrôle des coûts. On parle de la gestion des connaissances comme la "nouvelle richesse" des entreprises à partir des "actifs immatériels" ("intangibles", "patrimoine des connaissances", etc.). Ce paradigme introduit une problématique des mécanismes d'*évaluation* du patrimoine des connaissances ;

- paradigme épistémologique. La gestion de connaissances permet la mise un place d'un système de connaissances opérationnel bâti sur l'idée que la connaissance peut être définie au travers de l'information (données, traitements) qui prend une certaine signification (concepts, tâches) dans un contexte (domaine, activité) donné. Ce paradigme introduit une problématique d'*évolution* de la connaissance, c'est-à-dire de créer un nouveau domaine de connaissance.

## 1.2.    Aspect social de la gestion des connaissances

Dans l'aspect social de la gestion des connaissances, la connaissance est perçue comme un système, d'une part organisé à travers un système de connaissances, et d'autre part structuré à travers la gestion des connaissances. Pour explorer cet aspect nous nous attacherons à retrouver la signification de la connaissance dans le travail et à justifier l'importance de cette connaissance comme un facteur clé dans l'organisation du travail coopératif.

Dés l'époque de Taylor, Ford et Mayo dans une économie de production, la connaissance a été considérée comme un levier de productivité puis, à l'époque de Nonaka et Takeuchi dans une économie de service, la connaissance a été considérée comme un levier d'avantage concurrentiel ou compétitif durable enfin, de nos jours dans une économie globalisée - en liaison d'une part aux nouvelles technologies de l'information et de la communication (NTIC, Internet, Intranet et Extranet, et d'autre part au concept de l'entreprise élargie - la connaissance est finalement considérée comme un levier d'avantage coopératif durable.

Pour organiser cette section nous avons repéré dans la littérature trois approches : l'approche organisationnelle de Nonaka et Takeuchi, l'approche biologique de Maturana et Varela, dans lequel la vraie connaissance se cache au fond de nous même ; et enfin, l'approche managériale de Jean-Louis Ermine. L'approche de Nonaka et Takeuchi est intéressante du fait que pour eux, la création de connaissance tacite (individuelle) est un facteur d'avantage compétitif lié au développement de nouveaux produits (pour l'industrie japonaise). Selon Nonaka et Takeuchi, la mécanique de création de cette connaissance, expliquée dans leur modèle, explique la sortie de la crise économique du japon des années 70 et le succès économique des années 80 des entreprises japonaises. Pour Maturana et Varela la vraie connaissance se cache au fond de nous mêmes (ils utilisent l'analogie avec un "arbre" pour expliquer que dans ses racines se cache cette connaissance). L'approche de Jean-Louis Ermine est intéressante du fait qu'il enrichit le modèle OID de Jean-Louis Le Moigne. Modèle largement éprouvé en France et ailleurs dans la conceptualisation de systèmes d'information dans l'entreprise.

### 1.2.1. Approche organisationnelle de la gestion des connaissances

Nous avons décidé d'approcher la gestion des connaissances d'un point de vue organisationnel à travers, l'ouvrage de Nonaka et Takeuchi, intitulé *La connaissance créatrice, la dynamique de l'entreprise apprenante* (publié en 1997)[30], étant donné que pour eux *l'aspect humain : l'individu, le groupe, l'entreprise* est à la base de la création des connaissances nouvelles et d'apprentissage organisationnel pour l'entreprise. Autrement dit, la connaissance est à rapprocher d'un système social[31] qui existe dans un domaine humain au travers du dialogue et du langage des acteurs (individu, groupe, entreprise). Au niveau individuel, la connaissance est crée de façon isolée et l'apprentissage se développe de façon autonome. Au niveau du groupe, la connaissance est crée en interaction de dialogues puis partagée avec le groupe, alors que l'apprentissage se développe de façon collective par création et intégration de données, d'informations et de connaissances. Au niveau de l'entreprise la connaissance est créée au travers d'un avantage concurrentiel ou compétitif (nouveaux produits ou services) alors que l'apprentissage se développe de forme organisationnelle. Cette hypothèse nous parait fondamentale pour la gestion des connaissances, car pour nous l'acteur est au centre de - la gestion des connaissances (capitaliser, partager, créer…) - la gestion de l'innovation (créer, produire, offrir…) - la gestion des compétences (apprentissage, action…). Dans

---

[30] L'ouvrage original s'appelle *The Knowledge-Creating Company. How Japanese Create the Dynamics of Innovation* (publié en 1995), et donc le mot *knowledge* a été traduit en français comme *connaissance*. En revanche, Jean-Claude Tarondeau, dans son livre *Le management des savoirs* (publié en 1998) traduit le modèle de création des connaissances nouvelles et d'apprentissage organisationnel de Nonaka et Takeuchi comme « modes de création des savoirs », alors que dans l'ouvrage origienl « contents of knowledge created by the four modes », Tarondeau traduit le mot *knowledge* comme *savoir*. Nous justifierons ceci plus loin. D'autre part, nous préférons utiliser le nom *création des connaissances nouvelles et d'apprentissage organisationnel*, pour désigner le modèle de Nonaka et Takeuchi, plutôt que *création de connaissances organisationnelles*, afin de bien préciser les deux dimensions du modèle : épistémologique et ontologique.
[31] Nous tenons a souligner que nous restons dans une approche constructiviste de la connaissance comme l'on fait Edgar Morin (*La Méthode, La connaissance de la connaissance*, publié en 1986), Francisco Varela, et Joël de Rosnay (*Le macroscope, Vers une vision globale*, publié en 1975), entre autres. Néanmoins, notre objectif reste les systèmes industriels, c'est-à-dire explorer davantage la relation connaissance et travail dans un système social. C'est pour cette raison que l'approche de Nonaka et Takeuchi nous parait plus appropriée pour le faire.

---------------------------------------------------------------------------------------------------

cette section nous allons analyser les différents modèles de création de connaissance nouvelle et d'apprentissage organisationnel à travers une approche philosophique (dimension épistémologique du modèle) et une approche managériale (dimension ontologique du modèle). De plus, nous allons étudier la connaissance tacite et la connaissance explicite proposée pour Polanyi, et la connaissance comportementale proposée par Barnard au travers de la relation entre connaissance et acteur (individu, groupe, entreprise), autrement dit entre l'individu et son espace de travail (l'environnement : le travail et ses outils de production). Cet aspect a été négligé par Polanyi et Barnard, alors que Nonaka et Takeuchi l'ont utilisé pour établir leur modèle générique de création de connaissance nouvelle et d'apprentissage organisationnel. Enfin, l'ouvrage de Nonaka et Takeuchi a popularisé aux Etats-Unis et ailleurs la pratique managériale de la gestion des connaissances dans les entreprises liée principalement au développement de nouveaux produits ou services.

De manière générale, le fait que la gestion des connaissances est devenu un vrai problème industriel et que l'entreprise doit y faire face pour disposer d'un avantage compétitif durable, (et non un simple thème à la mode popularisé aux Etats-Unis par Nonaka et Takeuchi), il a été possible d'œuvrer dans ce sens grâce aux contributions de Laurence Prusak dans son livre *Knowledge in Organization. Resources for the Knowledge based Economy* (publié en 1997) et Thomas Davenport dans son livre *Information Ecology: Mastering the Information and Knowledge Environmenet. Why Technology is not enough for success in the Information Age* (publié aussi en 1997). Pour ces auteurs, la gestion des connaissances est un nouveau style de management et de comportement organisationnel et l'entreprise doit y faire face pour survivre. Une réflexion d'ensemble autour des problématiques associées à la gestion des connaissances, se trouve dans leur ouvrage intitulé *Working Knowledge. How Organizations Manage what they Know* (publié en 2000).

Malgré cela, nous avons choisi l'approche de Nonaka et Takeuchi car elle reste basée sur un modèle de création des connaissances nouvelles et d'apprentissage organisationnel, tandis que pour Prusak et Davenport la gestion des connaissances est plutôt un nouvel élément culturel de l'entreprise qui pour les grands groupes industriels a donné lieu à des nouveaux métiers du savoir, appelés Knowledge Manager, Chief Knowledge Officier, etc. D'ailleurs pour Prusak et Davenport la gestion des connaissances est une activité courante parmi les cadres responsables des systèmes d'information de l'entreprise. Pour eux, l'enjeu de la gestion des connaissances passe tout d'abord par les ressources humaines et leurs capacités de créer, partager et appliquer la connaissance, et après par les NTIC du KM, afin de générer un avantage compétitif durable pour l'entreprise[32].

> ➢ **Qu'est-ce que la connaissance ? Selon l'approche de Nonaka et Takeuchi**

Les vrais problèmes de cette question sont autour de trois voies de réflexions (1) l'origine de la connaissance ; (2) la nature de la connaissance ; et (3) la validité de la connaissance. Nous n'avons pas l'intention ici d'ouvrir un nouveau chemin, ni de remettre en question les différents points de vues de philosophes depuis la Grèce antique à nos jours sur ces trois voies là. Mais plutôt de rechercher le pilier fondamental de la gestion des connaissances dans le modèle de création des connaissances nouvelles et d'apprentissage organisationnel[33] de Nonaka et Takeuchi. Pour eux le terme "création de connaissances organisationnelles", est une démarche qui traduit « la capacité d'une entreprise considérée dans son ensemble, de créer de nouvelles connaissances, de les disséminer au sein de l'organisation et de leur faire prendre corps dans les différents produits, services du système »[34] [Nonaka et Takeuchi, 97]. Dans cette logique, les mécanismes de création de connaissances nouvelles et d'apprentissage organisationnel se trouvent caractérisés par un processus dynamique que nous matérialiserons à travers les verbes "créer" et "disséminer" des connaissances. Ces expressions, nous les utiliserons pour caractériser les mécanismes de la gestion des connaissances afin de créer de nouvelles connaissances selon l'aspect social de l'approche organisationnelle de la gestion des connaissances.

Ainsi, dans ce processus dynamique de la gestion des connaissances, l'origine de la connaissance nécessaire pour innover (produits, services) et produire un avantage compétitif durable, se trouve dans la capacité de l'entreprise de pouvoir créer et pouvoir disséminer ses

---

[32] Un grand nombre d'articles de Nonaka, Takeuchi, Prusak, et Davenport se trouvent dans le site de la Harvard Business Review http://www /harvardbusinessonline.hbsp.harvard.edu/.

[33] Dans la thèse de Tounkara, intitulé *Gestion des Connaissances et Veille : vers un guide méthodologique pour améliorer la collecte d'informations*, au chapitre 1 (la connaissance : un facteur de performance, de stabilité, d'adaptabilité et un critère de positionnement stratégique), il est précisé « que lorsque nous utilisons le terme "connaissance", nous entendons par là "les connaissances dans l'entreprise" ». Pour nous la *connaissance organisationnelle* est le cœur du système d'organisation du travail de l'entreprise. Ce système est décrit à travers des méthodes de travail et des modes de management, et donc le terme connaissance est un système qui englobe les connaissances organisationnelles qui sont nécessaires au *business* de l'entreprise. Or, l'origine, la nature et la validité de ces connaissances sont propres au *business* et à sa survie en termes des coûts, temps, et objectifs.

[34] Nous constatons que cette définition enrichit la définition de *capacité de l'entreprise* de Tarondeau « les capacités sont définies comme des routines de mise en œuvre d'actifs pour créer, produire et/ou offrir des produits ou services sur un marché » [Tarondeau, 98] en conformité avec les définitions que nous avons données dans l'introduction. En plus, le terme "système" se réfère ici à l'ensemble des processus ou process (procédés) de l'entreprise.

---------------------------------------------------------------------------------------------------

connaissances de façon "circulante et dynamique"[35]. Le terme "circulante" signifie, pour eux, en interaction entre trois acteurs génériques (individu, groupe, entreprise), alors que "dynamique" signifie évoluant entre deux formes de connaissance : la *tacite* et l'*explicite* qui forment la nature de la connaissance (individuelle et collective) à tous les niveaux de l'entreprise (stratégique, tactique, opérationnel). La différence, se trouve alors dans le fait que la connaissance tacite est enracinée dans nos idées, métaphores, créances, concepts, hypothèses, modèles mentaux, analogies, etc., alors que la connaissance explicite est enracinée dans une source d'information (par exemple un document, un email, un modèle, une maquette, un plan, etc.). Ainsi, la connaissance explicite est la matérialisation de la connaissance tacite.

Dans cet esprit, la création des connaissances nouvelles et d'apprentissage organisationnel est possible grâce à un processus en "spirale ascendante" de "conversion" de connaissances. Le terme "spirale" implique un mouvement entre ces trois acteurs et les deux formes de connaissances à travers le dialogue et l'apprentissage. Le terme "ascendante" signifie que la connaissance est enrichie à chaque moment. En revanche, le terme "conversion" signifie que le transfert met en oeuvre quatre mécanismes de création de connaissances organisationnelles, individuelle et collective, dans l'espace de travail (l'environnement : le travail et ses outils de production) : (1) extériorisation : de tacite à explicite, le partage de la connaissance et de l'apprentissage se fait à travers le langage (discours ou écrit), alors que la création se matérialise à travers la formation des idées, concepts, analogies, métaphores, hypothèses, modèles mentaux, etc. ; (2) combinaison : d'explicite à explicite, la création de la connaissance et de l'apprentissage se fait par la mise en commun de la connaissance explicite à travers des réunions, de changements d'informations, données, etc. ; (3) internalisation : d'explicite à tacite, la création de la connaissance et de l'apprentissage se fait par l'expérimentation (apprentissage) de la connaissance qui a été explicitée sur un support (document, diagramme, modèle, email, etc.) ; et (4) socialisation : de tacite à tacite, la création de la connaissance et de l'apprentissage se fait par le partage de l'expérience mais sans le recours au langage sans un support écrit, simplement à travers l'observation de l'autre, et la réflexion ou l'imitation[36]. La figure 1.1 montre la dynamique du modèle.

---

[35] Comme nous verrons dans l'approche biologique de Maturana et Varela, la connaissance est aussi un mécanisme "circulant" afin de "faire-émerger" la signification. Ce mécanisme va être à l'origine de l'approche de l'enaction que ces auteurs ont popularisé.
[36] Nous n'avons pas eu besoin d'appliquer ce modèle dans cette thèse. Notre intérêt est resté plutôt théorique que pratique. Néanmoins, pour plus de précision pour la mise au point d'un tel modèle il est préférable de se référer à l'ouvrage de Nonaka et Takeuchi dans lequel sont discutés des cas d'application dans l'industrie japonaise.

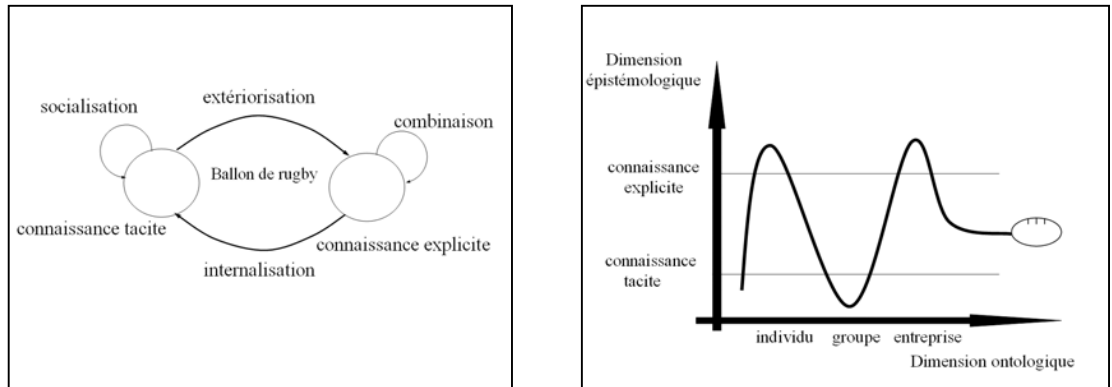

**Figure 1.1 :** Modèle du ballon de rugby (source propre)

Nonaka et Takeuchi font une analogie avec un "ballon de rugby" pour indiquer que la connaissance collective se trouve au sein de l'équipe (ceci est montrée par la partie gauche de la figure 1.1). En revanche, le processus en "spirale ascendante" de "conversion" de connaissances est montré dans la partie droite de cette figure.

Or, la validité de la connaissance, c'est-à-dire la validité du modèle de conversion de connaissances de Nonaka et Takeuchi (que nous préférons appeler *modèle de création des connaissances nouvelles et d'apprentissage organisationnel*) se traduit par des résultats opérationnels à travers de nouveaux produits, services, processus, ou procédés (innovation[37]). Cela signifie aussi que l'entreprise doit avoir la capacité suffisante pour mobiliser ou disséminer ses connaissances et ses compétences afin de les forger.

En résumé, le modèle de Nonaka et Takeuchi est structuré autour d'un processus de conversion de connaissances (que nous préférons aussi appeler *mécanismes de création et dissémination des connaissances nouvelles et d'apprentissage organisationnele*) lié à la capacité de l'entreprise de mettre en place l'extériorisation, la combinaison, l'internalisation, et la socialisation de la connaissance tacite et de la connaissance explicite à tous les niveaux de l'entreprise (stratégique, tactique, opérationnel) et entre ses trois acteurs génériques (individu, groupe, entreprise) afin d'innover (produits, services) et produire finalement un avantage compétitif durable pour l'entreprise. Le but recherché selon Nonaka et Takeuchi est la création de cet avantage et donc l'effort essentiel se trouve principalement, d'une part, dans la connaissance collective, et d'autre part, dans le mécanisme d'extériorisation de conversion de la connaissance tacite vers la connaissance explicite par génération d'analogies, de métaphores, de concepts, etc. La complexité se cache alors dans la mécanique "circulante et dynamique" du processus ou mécanisme de conversion de connaissances (extériorisation, combinaison, internalisation, socialisation), car il y a une implication mutuelle entre l'un par rapport à l'autre, et ceci se fait par le partage et le dialogue entre les individus.

---

[37] Chez Nonaka et Takeuchi l'innovation est relative à l'innovation industrielle, c'est-à-dire à l'innovation prise d'un point de vu technique (un nouveau produit, un nouveau process ou procédée, etc.), car les études de cas qui illustrent leur modèle sont relatifs au développement de nouveaux produits. Nous ne pensons pas que cela soit possible d'utiliser leur modèle dans le cadre de l'innovation organisationnelle, c'est-à-dire à de l'innovation prise d'un point de vu social (un nouveau méthode de travail, un nouveau mode de management, une nouvelle structure organisationnelle, un nouveau processus, un nouveau service, etc.).

Ce modèle, s'appuie sur une étude d'entreprises japonaises (Honda, Matsushita, etc.) qui ont innové par la mise au point de nouveaux produits dans le marché. Du fait qu'il s'agisse d'un modèle descriptif et prescriptif, ce modèle peut être généralisés (c'est pour cette raison que Nonaka et Takeuchi l'appellent aussi modèle générique de la création de connaissances organisationnelles). Néanmoins, rien ne garantit son succès en dehors du Japon, et donc sa validité pour générer une avantage compétitif durable reste à ce jour spécifique.

Notre idée maintenant, est d'utiliser ce modèle pour explorer les traces de la connaissance organisationnelle. Nous pensons que le point de départ dans leur construction a été la prise en considération des approches *cartésienne* et *sartrienne* de la connaissance. Dans l'approche cartésienne l'origine de la connaissance a été formulée, par René Descartes, à partir de la dualité philosophique : âme/corps, esprit/matière, sujet/objet, etc., comme il dit lui-même « ces hommes sont composés, comme nous, d'une Ame et d'un Corps, il faut donc que je vous décrive, premièrement, le corps à part, puis après l'âme à part ; et enfin, que je vous montre comment ces deux natures doivent être jointes et unies, pour composer des hommes qui nous ressemblent ». Dans l'approche sartrienne, l'origine de la connaissance se trouve dans la relation entre connaissance et action, comme le dit Jean-Paul Sartre « être c'est agir ; si nous voulons connaître le monde, nous devons agir, poursuivre une finalité ». Cette argumentation que prend ses sources dans la tradition épistémologique de la philosophie chez Sartre, nous pouvons la matérialiser dans une dualité : moi/autres. L'intérêt pour nous est que cette dualité peut très bien exister dans le domaine technique. Ainsi *agir*, peut très bien signifier, par exemple : *prends un marteau*.

Les dualités de ces approches cartésienne et sartrienne de la connaissance ont forgé, d'un point de vue philosophique, trois sortes de natures de la connaissance : la conception *rationaliste*, la conception *empiriste*, et la conception *existentialiste*.

Le courant rationaliste (la plus courante) prétend qu'à l'origine, la connaissance peut être obtenue par la réflexion de l'individu, c'est-à-dire sur la base d'un processus purement intellectuel, aussi appelé méthode déductive de raisonnement ou modèle mental. Dans ce courant de pensée, la dualité entre sujet/objet est interprétée comme le fait que l'individu crée la connaissance en analysant l'objet de forme externe. Ainsi la perception de l'objet doit correspondre à un modèle mental (pour avoir du sens, pour pouvoir formuler les concepts, les lois, théories, etc.,) de forme rationnel et logique. Un exemple d'utilisation de ce courant rationalisme est la *modélisation analytique*.

Dans le courant empiriste, la connaissance est obtenue sur la base d'une méthode inductive d'expériences sensorielles, c'est-à-dire que l'origine de la connaissance est l'expérience sensorielle de l'individu. La dualité sujet/objet est interprétée par le fait que la connaissance peut être créé par la implication de l'individu dans l'objet. Un exemple d'utilisation de ce courant empiriste est la *modélisation systémique*.

Enfin, dans le courant existentialiste, l'origine de la connaissance est l'acteur (individu, groupe, entreprise). Dans ce sens la connaissance est là pour accomplir un projet. La connaissance est créée par un comportement pratique de l'acteur (individu, groupe, entreprise) qu'induit une action (par exemple dire ou agir) afin de réaliser un projet.

Ces trois courants philosophiques de la connaissance (rationaliste, empiriste, existentialiste) ont permis d'organiser le modèle de conversion de connaissances de Nonaka et Takeuchi en deux dimensions : la dimension *épistémologique* et la dimension *ontologique* de la connaissance. Ainsi, en ce qui concerne la validité de la connaissance, elle est prise en charge par la validité de la méthode, soit déductive (conception rationaliste), soit inductive (conception empiriste), soit existentialiste des dimensions épistémologique et ontologique de la connaissance.

Avant de continuer nous avons trouvé nécessaire ici d'ouvrir une parenthèse pour faire une différence entre savoir, connaissance, compétence, et intelligence.

Le savoir

Pour Robert Tremblay, dans son livre *Vers une écologie humaine* (publié en 1990) « le savoir d'un individu ou d'une société est l'ensemble des représentations que cette personne ou cette collectivité tiennent pour vraies à propos de la réalité. Le concept de connaissance est plus restrictif, et recouvre l'ensemble du savoir pratique, des techniques et des sciences qui, d'une manière ou d'une autre, ont fait leurs preuves dans la pratique ». Le savoir est alors le champ existentiel de la connaissance, un espace culturelle où habite la connaissance, Tremblay symbolise le savoir comme une sphère de l'opinion, qui englobe la connaissance comme une sub-sphère de la raison, laquelle à la fois englobe la science comme une sub-sphère de la preuve, selon une validation logique ou basée sur les faits.

### La connaissance

Pour Jean-Claude Tarondeau, dans son livre *Le management des savoirs* (publié en 1998) « le savoir individuel est l'ensemble des croyances d'un individu sur les relations de cause à effet entre phénomènes ». Cette définition correspond à la définition de la connaissance chez Tremblay, nous pensons que ceci est cohérent car Tarondeau en se référant au terme de Peter Drucker *knowledge society*, dit « nous sommes entrées dans l'ère du savoir », puis traduit le modèle de création de connaissances organisationnelles de Nonaka et Takeuchi comme « modes de création des savoirs », (il s'appelle dans l'original « contents of knowledge created by the four modes »), et justifie que le mot *knowledge* soit traduit en français par connaissance ou savoir.

### La compétence

De plus, pour Jean-Claude Tarondeau « les compétences sont des capacités particulières de mise en œuvre d'actifs de façon organisée dans le but d'atteindre des objectifs », alors que pour Jean-Yves Prax « la compétence doit être définie en termes de savoirs opératoires et combinatoires : c'est-à-dire la capacité d'une personne ou d'un groupe à mobiliser les différentes ressources dont elle dispose : savoirs, savoir-faire, documents, réseaux relationnels, pour agir face à de situations professionnelles ». Enfin en citant Wittorski, Prax dit « une compétence est un "savoir-agir reconnu" : on ne se déclare pas soi-même compétent ; cela dépend d'une appréciation sociale ». Et donc, la compétence est une habilité qui permet à l'individu ou l'entreprise de mobiliser son savoir, savoir-faire ou autre ressource pour faire face aux situations organisationnelles.

<u>L'intelligence</u>

Pour Edgar Morin, dans son livre *La Méthode, La connaissance de la connaissance* (publié en 1986) « l'intelligence est une *métisse* qui mêle en elle des sangs très divers »[38]. Cela signifie que l'intelligence est aussi une habilité tel que comme la compétence est propre à l'individu ou à l'organisation, et non au patrimoine individuel ou organisationnel des connaissances.

En résumé, nous constatons à partir de ces quatre définitions et de ce qui a été dit antérieurement, que la différence entre savoir et connaissance est le propre de la dualité sujet/objet de l'approche rationaliste et de l'approche empiriste de la connaissance, alors que la compétence et l'intelligence est le propre de la dualité moi/autres exprimée par l'approche existentialiste de la connaissance.

Nous fermons cette parenthèses, afin de revenir aux dimensions épistémologique et ontologique de la connaissance du modèle de création de connaissances organisationnelles de Nonaka et Takeuchi. Pour eux, la dimension épistémologique permet de définir le *savoir* ou la *connaissance* de l'acteur (individu, groupe, entreprise) que nous avons matérialisés dans la dualité sujet/objet selon que l'on choisisse l'approche rationaliste ou l'approche empiriste de la connaissance. Cette dimension a été visualisée à travers des travaux de Polanyi, sur le champ philosophique, de la "connaissance tacite" et la "connaissance explicite" que Nonaka et Takeuchi ont généralisé dans le domaine industriel. En revanche, la dimension ontologique permet de définir la *compétence* et l'*intelligence* que nous avons matérialisé dans la dualité moi/autres selon l'approche existentialiste de la connaissance. Les travaux de Barnard sur la "connaissance comportementale" ont contribué à la formation de cette dimension.

Dans cette mécanique la création de connaissance organisationnelle est une source d'*apprentissage organisationnel*. Autrement dit, du processus de conversion des connaissances émerge l'*apprentissage* au niveau individuel, du groupe et /ou de l'entreprise.

A cet égard Ingham[39] commente et déclare dans l'introduction de l'édition française du livre de Nonaka et Takeuchi, que « les processus d'apprentissage concernent les "savoir quoi faire" et les "savoir pourquoi faire" il s'agira d'apprendre "comment faire" et le résultat prendra généralement la

---
[38] Le terme grec *métis*, correspond à la connaissance conjecturale, ruse, flair, etc.
[39] En fait, le professeur Marc Ingham de l'Université Catholique de Louvain est le troisième auteur de la version française du livre de Nonaka et Takeuchi.

-------------------------------------------------------------------------------------------------------------

forme d'un "savoir-faire". Mais ils pourront aussi entraîner une modification d'un comportement et avoir trait alors à un "savoir être" » [Nonaka et Takeuchi, 97].

En anglais "savoir-faire" correspond aussi au terme « know-how », et donc nous trouvons nécessaire de faire une distinction en français entre savoir, savoir-faire et savoir-être. Or, pour Jean-Yves Prax dans son livre *Manager la connaissance dans l'entreprise* (publié en 1997) « les Grecs avaient déjà défini plusieurs formes de connaissance (1) l'*épistémè*, connaissance abstraite généralisante ; (2) la *techné*, connaissance permettant l'accomplissement d'une tâche ; (3) la *phronesis*, sagesse sociale ; et (4) la *mètis*, connaissance conjecturale, ruse, flair, …, on voit déjà la place particulière occupée par la *mètis*, qui est une connaissance totalement tacite » [Prax, 00].

En utilisant ces quatre sources de connaissances nous pouvons généraliser afin de classifier quatre sources d'apprentissage basées sur les *savoir* (phronesis en grec), *savoir* (épistémè en grec), *savoir-faire* (techné en grec) et *savoir-être* (mètis en grec).

Dans ce cadre là, l'approche empiriste de la connaissance, forme la dimension épistémologique du modèle de Nonaka et Takeuchi, par le fait que la connaissance est décrite à travers la dualité sujet/objet et non pas comme dans l'approche rationaliste entre l'acteur (celui qui connaît l'individu, le groupe, ou l'entreprise) et l'objet (le connu). En fait l'acteur (individu, groupe, entreprise) crée de la connaissance en s'impliquant lui-même dans l'objet. En revanche, l'approche existentialiste forme la dimension ontologique de ce modèle par le fait que ce modèle est décrit à travers une dualité moi/autres, c'est-à-dire, entre l'individu et les autres (individu, groupe, entreprise).

En conclusion nous caractériserons l'apprentissage organisationnel à travers, d'une part, le *savoir-faire* (techné en grec, approche empiriste) qui correspond à la fois à la connaissance tacite et la connaissance explicite (identifie par Polanyi), et qui forme la dimension épistémologique du modèle de Nonaka et Takeuchi, et d'autre part, le *savoir-être* (mètis en grec, approche existentialiste) qui correspond à la connaissance comportementale (identifie par Barnard) et qui forme la dimension ontologique du même modèle entre l'individu, le groupe, et l'entreprise. Nous préférerons utiliser le terme *action* pour indiquer le *savoir-être* car ce terme est plus proche d'une compétence ou d'un comportement (connaissance comportementale chez Barnard). Enfin, la dimension épistémologique de la connaissance formée par l'approche empiriste peut se matérialiser dans la dualité connaissance tacite/connaissance explicite, alors que la dimension ontologique de la connaissance (formée par l'approche existentialiste) peut se matérialiser par la dualité moi(sujet) / autres(individu, groupe, entreprise).

Nous allons ouvrir maintenant une autre parenthèse pour expliquer davantage la connaissance tacite, explicite et comportementale chez Polanyi et Barnard.

Pour Polanyi[40], dans ses livres, intitulé *Personal Knowledge* (publié en 1958) et *The Tacit Dimension* (publié en 1966), « les êtres humains acquièrent la connaissance en créant et organisant activement leurs propres expériences », et donc, pour Polanyi, la connaissance ne peut pas être séparée de l'expérience, en plus la connaissance existe sur deux plans opposés, celui que l'on peut communiquer et celui que l'on ne peut pas communiquer aux autres. A cet égard, il déclare que « la connaissance qui peut être exprimée sous forme de mots et nombres ne représente que la partie visible de l'iceberg du corps complet des connaissances », puis il rajoute que « nous pouvons savoir plus que ce que nous pouvons exprimer ». Afin de faire la différence entre ces deux sortes de connaissances, Polanyi emploi les termes : *connaissance tacite* et *connaissance explicite*. La connaissance tacite « est personnelle, spécifique au contexte et de fait, il est difficile de la formaliser et de la communiquer ». La connaissance explicite, quant à elle « se réfère à la connaissance qui est transmissible dans un langage formel, systématique ». Pour Polanyi ces deux connaissances sont opposées. Tandis que selon Nonaka et Takeuchi, cette dualité entre connaissance tacite et connaissance explicite est plutôt une complémentarité qu'une opposition. La connaissance tacite « inclut des éléments cognitifs et techniques. Les éléments cognitifs se centrent sur ce que Johnson-Lair (1983) appellent "modèles mentaux" dans lesquels les êtres humains créent des modèles de fonctionnement du monde en recourant a des analogies et en les manipulant dans leurs esprits. Les modèles mentaux, comme les schémas, paradigmes, perspectives, croyances et points de vues aident les individus à percevoir et a définir leur monde ».

Voilà une nature de la connaissance tacite que si elle a bien ses origines dans un contexte philosophique, peut s'élargir au domaine industriel ou technique. Dans ce cas, l'élément technique de la connaissance tacite recouvre les savoir-faire concrets les habiletés et aptitudes concrètes. Elle est incrustée dans l'expérience individuelle et implique des facteurs intangibles tels que la croyance personnelle, la perspective et le système de valeurs … qui est difficile à articuler au moyen du langage formel ». Dans un même ordre d'idée, la connaissance explicite dans le domaine industriel

---

[40] Selon Nonaka et Takeuchi « Michael Polanyi fut un chimiste renommé très près d'obtenir le prix Nobel de chimie avant de se tourner vers la philosophie à l'age de 50 ans ».

« peut être articulée en langage formel comprenant les énoncés grammaticaux, les expressions mathématiques, les spécifications, les manuels… ». Nous pensons qu'il s'agit d'une approche très confortable et pratique de la connaissance que nous validons tous les jours dans notre vie professionnelle. C'est peut être pour cette raison que Nonaka et Takeuchi l'ont empruntée pour leur modèle de création de connaissances organisationnelles. Pour eux à l'inverse de l'approche de Polanyi « la connaissance tacite et la connaissance explicite ne sont pas totalement séparées mais sont des entités mutuellement complémentaires ».

D'autre part, nous constatons que la création de connaissance tacite dans le domaine industriel implique une relation profonde entre les produits et ses outils de production (process et processus), mais également entre connaissance et action selon l'approche existentialiste de la connaissance, ce qui donne naissance à la dimension ontologique de la connaissance dans le modèle de Nonaka et Takeuchi, à partir des travaux de Barnard. Cette relation entre la connaissance tacite et l'approche existentialiste de la connaissance peut se justifier davantage du fait que « les éléments cognitifs de la connaissance tacite font références aux images qu'un individu a de la réalité et à ses visions du futur ; à savoir ce qui "est" et ce qui "doit être" ». Cela signifie que la création de la connaissance tacite est le fruit de l'expérience et de l'action et que s'il n'y a pas d'action (dire ou agir) comme conséquence de l'expérience, de l'intuition, du jugement ou du savoir-être de l'individu, la création de connaissance tacite n'est pas possible. Comme le disent, Nonaka et Takeuchi « la connaissance tacite est crée ici et maintenant dans un contexte spécifique et pratique ».

En 1938, Barnard, dans son livre, intitulé *The Functions of the Executive*, a proposé une étude sur les fonctions managériales des dirigeants d'entreprises. A partir des recherches sur la théorie du management scientifique de Taylor (*The Principles of Scientific Management*, publié en 1911) et la théorie de relations humaines de Mayo (*The Human Problems of an Industrial Civilization*, publié en 1933). Selon lui les managers possèdent deux sortes de processus cognitifs pour résoudre leurs problèmes. L'un est le *processus logique* qui « se réfère à la pensée consciente ou au processus de raisonnement et qui peut être exprimé en termes de mots et signaux », l'autre est le *processus non logique* qui « comprend les processus mentaux inexprimables tels que le jugement, la décision et les actions pratiques ». Pour Barnard le processus logique est attaché à la connaissance scientifique que nous pouvons associer à l'approche rationaliste (savoir) ou à l'approche empiriste (savoir-faire), tandis que le processus non logique est attaché à la connaissance comportementale (l'approche existentialiste). Ainsi, les managers dans leurs activités de tous les jours « utilisent à la fois la connaissance scientifique obtenue par des processus mentaux logiques et la connaissance comportementale extraite des processus mentaux non logiques ». Cette vision de Barnard a contribué à la définition de la dimension ontologique de la connaissance du modèle de Nonaka et Takeuchi.

Si on généralise cette étude à la création de connaissances dans le domaine industriel chez l'acteur (individu, groupe, entreprise), le *processus logique* selon Nonaka et Takeuchi, permet la création de *connaissance scientifique* dans l'entreprise par le produit des aptitudes intellectuelles, de l'expertise et du savoir-faire technique d'un individu ou d'un groupe. Généralement, ce type de connaissance s'exprime en termes de mots, formules physiques ou chimiques, graphes, etc., que l'on peut stocker au fils du temps dans des documents papiers ou papiers électronique (GED[41]). Par contre, le *processus non logique* permet la création de *connaissance comportementale* dans l'entreprise. Cette connaissance ne peut pas s'exprimer en termes de mots pour la stocker dans un document, mais est par contre exprimable à travers les gestes, les émotions et l'action d'un individu ou d'un groupe, par exemple le jugement de l'état d'une activité en cours, la prise de décision face à la présence d'un événement redouté, la sensation de ce qu'il faut faire, le sens du pouvoir, etc. Ce type de connaissance peut s'exprimer en termes de *compétences*, c'est-à-dire à partir de qualités professionnelles de l'individu pour transformer la connaissance en action, et en *savoir-être*, ou à partir de qualités personnelles de l'individu pouvant se développer dans son espace de travail. Ce *savoir-être* peut s'exprimer au travers du phénomène de l'intelligence émotionnelle [Goleman, 99].

Nous fermons cette parenthèse en disant que la dimension ontologique de la connaissance, est construit à travers la dualité moi/autres, c'est-à-dire, entre l'individu, le groupe, l'entreprise. Dans cette logique, la connaissance scientifique et comportementale est créée et mobilisée par le partage et le dialogue entre individus.

En conclusion nous avons montré que le modèle générique de création de connaissance et d'apprentissage organisationnel de Nonaka et Takeuchi est centré principalement sur l'aspect humain : l'individu, le groupe, l'entreprise. Dans ce modèle, la connaissance (tacite/explicite) et l'apprentissage organisationnel est créée et disséminée (mobilisée) dans une dynamique circulante (implication mutuelle) entre la connaissance individuelle et collective (individu, groupe, entreprise) à travers d'un processus de conversion de connaissances (tacite/explicite) composé de quatre mécanismes (extériorisation, combinaison, internalisation, socialisation) et ceci à tous les niveaux de l'entreprise (stratégique, tactique, opérationnel) afin d'innover (produits, services) et produire un avantage concurrentiel ou compétitif durable pour l'entreprise basé principalement sur la connaissance collective et le mécanisme d'extériorisation (conversion de la connaissance tacite vers la connaissance explicite par la génération d'analogies, métaphores, concepts, etc.). Dans tout ce qui vient d'être explicité, la clé du modèle se trouve dans la relation directe entre l'acteur (individu, groupe, entreprise) et son espace de travail (l'environnement : le travail et ses outils de production). D'après Nonaka et Takeuchi ces aspects ont été négligés par Polanyi et Barnard, dans la configuration de la dimension épistémologique et la dimension ontologique de la connaissance.

---

[41] Gestion électronique de documents.

Et donc, les mécanismes de création de connaissances nouvelles et d'apprentissage organisationnel sont caractérisés par un processus dynamique, matérialisés à travers les verbes "créer", "disséminer" ou "mobiliser" la connaissance (tacite/explicite) individuelle ou collective et l'apprentissage organisationnel (individuelle/collective) dans l'entreprise. Ainsi, le processus dynamique de transfert de connaissance tacite à explicite et vice versa est un système récursif qui s'enrichit grâce au retour d'expérience de l'apprentissage organisationnel (individuelle/collective).

### 1.2.2. Approche biologique de la gestion des connaissances

Dans l'approche organisationnelle, nous avons caractérisé les mécanismes de création de connaissances nouvelles et d'apprentissage organisationnel par les verbes "créer", "disséminer" ou "mobiliser" la connaissance, dans le cadre d'un système social. L'objectif de cette section sera de retrouver ces mécanismes, mais cette fois dans le cadre d'un système naturel qui existe dans un domaine biologique (les relations sont au niveau cellulaire) au travers du "dialogue" et du "langage mais non pas au niveau humain mais au niveau cellulaire, organisé dans le système nerveux. Puis de faire l'hypothèse que ces mécanismes au niveau cellulaire peuvent être généralisés au niveau humain à travers un système social. La validation de cette hypothèse reste restreinte à ce jour au niveau des analogies.

Nous avons décidé d'approcher la gestion des connaissances par le point de vu biologique, au travers de l'ouvrage d'Humberto Maturana et Francisco Varela, intitulé *L'arbre des connaissances* [Maturana et Varela, 73][42], paru au Chili en 1973, et ceci pour trois raisons :

---

[42] Nous attirons ici l'attention qu'il ne s'agit pas d'établir un arbre de compétences collectives pour l'entreprise, tel qu'il a été développé en 1992, ou Pierre Lévy et Michel Authier proposèrent une solution de repérage des savoirs et savoir-faire, dans un ouvrage qu'ils baptisèrent *Les arbres de connaissances* », cité par Barthelme-Trapp, et qui se trouve implémenté dans un progiciel appelé GINGO.

La première raison est parce que nous avons eu le privilège de l'écouter à plusieurs reprises au Chili, et de constater la solidité de ses points de vues sur la relation entre connaissance et environnement à partir de l'étude du système nerveux comme un système clos[43]. Cette relation nous l'avons trouvée aussi dans l'approche organisationnelle de la gestion des connaissances caractérisée par le modèle de Nonaka et Takeuchi, qui décrit la relation entre l'acteur (individu, groupe, entreprise) et son espace de travail (l'environnement : le travail et ses outils de production) comme un système ouvert.

La deuxième raison est qu'une organisation vivante ne peut pas créer de la connaissance sans êtres vivants. Cette argumentation est tout à fait équivalente au fait que l'individu est au centre de la connaissance, de la compétence, et de l'innovation dans l'approche organisationnelle de la gestion des connaissances.

La troisième raison est que pour Maturana et Varela le paradigme simonien qui prétend que l'organisation est seulement une machine de traitement de l'information et non pas de création de connaissance n'est pas correct, et donc l'analogie entre le cerveau humain et l'ordinateur n'a pas lieu d'être. Comme l'a dit Varela « la métaphore populaire désignant le cerveau comme une machine de traitement de l'information n'est pas seulement ambiguë, elle est totalement fausse » [Varela, 89]. Dans l'approche organisationnelle de la gestion des connaissances Nonaka et Takeuchi, déclarent que « l'entreprise ne "traite" pas seulement de la connaissance mais la "crée" aussi » [Nonaka et Takeuchi, 97], et pour Prax « l'organisation n'est pas tant un système de "traitement de l'information" mais bien de "création de connaissance" » [Prax, 00].

> ➢ **Qu'est-ce que la connaissance ? Selon l'approche de Maturana et Varela**

Les origines du livre *L'arbre des connaissances* remonte à la fin des années 60, lorsque Francisco Varela était un jeune étudiant d'Humberto Maturana dans un cours appelé « Biologie de la connaissance », à la Faculté des Sciences de l'Université du Chili. Comme lui même a dit « *c'est d'Humberto Maturana que j'ai appris à concevoir le système nerveux comme un système opérationnellement clos. Il proposa explicitement cette idée dès 1969. Je considère aujourd'hui que cette intuition de Maturana est fondamentale. Il a établi un rapport essentiel entre les processus*

---

[43] Ce concept est introduit dans le chapitre 3. Dans un système clos le résultat d'une opération de transformation se situe à l'intérieur de la frontière du système lui-même, c'est-à-dire que les sorties d'un système clos ne font pas partie de l'environnement. En revanche, pour un système ouvert les sorties font parties de l'environnement.

*matériels et systématiques qui ont lieu à l'intérieur du système nerveux et propose une conception profonde et riche de la connaissance chez l'homme* » [Varela, 89], essence même de *L'arbre des connaissances*. Depuis 1969, ils se sont mis à réfléchir petit à petit sur la relation connaissance et environnement, et ceci d'un point de vue biologique (les relations sont au niveau cellulaire) à partir du système nerveux, afin de comprendre les mécanismes de création de connaissances et d'apprentissage social, et de conclure sur l'unicité de la nature humaine. Dans *L'arbre des connaissances* ils disent « la connaissance n'est pas armée comme un arbre avec un point de départ solide qui croît progressivement jusqu'à épuiser tout ce qu'il faut connaître, car la connaissance est un mécanisme "circulant" et "d'émergence de signification".

Pour eux ce mécanisme ou processus "circuler" et "faire-émerger" la signification est à l'origine de l'approche de l'*enaction*[44] dans lequel la connaissance est définie comme un système d'actions, qui prend en charge l'historique du *couplage structurel*, de la *détermination structurelle* et de la *clôture opérationnelle*[45] qui *enacte* (fait-émerger) un monde, et que nous symbolisons à travers du concept d'*arbre des connaissances*, pour insister sur le fait de créer. Néanmoins, à la différence de l'approche existentialiste de Sartre, où la connaissance dans un plan philosophique est là pour accomplir un projet téléologique, l'approche de l'enaction de Maturana et Varela met en relation la connaissance et l'environnement, et plus précisément entre la connaissance et l'action à travers un couplage structurel et non pas par un couplage par inputs[46]. Si l'on fait une analogie avec le système nerveux, dans l'approche de l'enaction[47], le système nerveux ne contient pas une fin en soi (une projection). Autrement dit un projet déjà défini en lui (un besoin nouveau), sinon qu'il y a un mécanisme d'équilibre en permanence avec l'environnement, c'est-à-dire il n'y a pas une "confrontation" sinon qu'un "dialogue" constant avec l'autre. D'autre part, dans l'approche rationaliste et l'approche empiriste la connaissance est un processus de représentation structurelle de ce que l'on voit ou de ce que l'on expérimente, et non pas le sens de faire émerger autre chose, que la juste vérité.

---

[44] Pour Varela, dans son livre intitulé *Invitation aux sciences cognitives* (publié en 1989), la cognition peut être classifiée selon trois catégories, à savoir : les symboles (l'hypothèse cognitiviste : l'approche cognitiviste) ; l'émergence (une alternative à la manipulation de symboles : l'approche connexionniste) ; et l'enaction (une alternative à la représentation : l'approche de l'enaction de la cognition). Nous reviendrons sur le sujet lorsque nous aborderons la question de l'autopoïèse (dans le chapitre 2) et de l'autopoïèse et la connaissance (dans le chapitre 3) afin de proposé notre modèle.

[45] Ces mécanismes sont expliqués dans le chapitre 2, section 2.2.3 (modèle autopoïétique).

[46] Le couplage par inputs est typique de la transformation : entrées/transformation/sorties de l'approche système pour représenter un système ouvert, tandis que le couplage structurel est typique pour un système clos (que n'a rien avoir avec un système fermé), comme nous verrons dans les chapitres 2 et 3.

[47] Nous utilisons davantage le terme approche *enactiviste* au lieu d'*enaction* afin d'aligner avec les deux autres approches (plutôt scientifique que philosophique) de la cognition de Maturana et Varela, et aussi pour ne pas le confondre avec l'approche de l'enaction de Karl Weick, qu'il a développé dans son livre, intitulé *The Social Psychology of Organizing* (publié en 1979) [Weick, 79]. Nous reviendrons au chapitre 3 au sujet de l'approche de l'enaction de Karl Weick.

En revanche, dans l'approche de l'enaction de Maturana et Varela « la nature de la connaissance humaine » n'a jamais été un fruit interdit ; au contraire, la connaissance est le propre de l'être vivant, la connaissance est « un grain que l'on sème dans le plus profond de nous mêmes ». Cela signifie, que la connaissance d'un point de vu biologique est quelque de chose commun à tous les êtres humains, et donc les mécanismes de création de connaissances le sont aussi (les racines biologiques de la connaissance sont les mêmes pour tous les êtres humains de la planète). A propos des mécanismes de création de connaissances et d'apprentissage social, ils affirment « ce qui nous unis à tous les hommes de tous les temps est la manière par laquelle nous faisons apparaître en nous nos significations existentielles, la manière par laquelle celles-ci sont créées, gérées, stabilisées, transformées. C'est justement, dans ce processus d'apprentissage social qu'émerge dans nous mêmes la signification du monde dans lequel nous vivons. Le fondement de la compréhension universelle de l'homme par l'homme »[48]. Cela signifie que, pour eux, les mécanismes de création de connaissances et d'apprentissage social d'un point de vue biologique (les relations sont au niveau cellulaire) sont caractérisés à travers un processus dynamique que nous matérialisons à travers les verbes "créer", "gérer", "stabiliser" et "transformer". Ces mécanismes nous les utiliserons pour caractériser les mécanismes de la gestion des connaissances afin de créer de nouvelles connaissances selon l'aspect social de l'approche biologique de la gestion des connaissances.

L'origine de cette caractérisation de la connaissance est bien entendu le domaine biologique, et il convient de l'élargir au domaine industriel, tel comme l'on fait Nonaka et Takeuchi avec la caractérisation de la connaissance de Polanyi et Barnard. Et donc, nous faisons l'hypothèse que les mécanismes de création de connaissance et d'apprentissage social selon l'approche biologique de l'aspect social caractérisé par un processus dynamique à travers les verbes "créer", "gérer", "stabiliser" et "transformer" la connaissance, complémentent les mécanismes de création de connaissances et d'apprentissage organisationnel du modèle de Nonaka et Takeuchi que nous matérialiserons par un processus dynamique au travers des verbes "créer", "disséminer" ou "mobiliser" la connaissance.

Nous pensons que ceci complète la relation entre connaissance et son environnement de l'approche organisationnelle identifiée dans le modèle de Nonaka et Takeuchi matérialisée à travers la dualité *connaissance tacite/connaissance explicite* pour indiquer que ces deux types de connaissances sont plutôt complémentaires qu'opposés. De même pour la dimension ontologique de la connaissance, que correspond à la connaissance comportementale de Barnard selon l'approche existentialiste et que nous avons matérialisée à travers la dualité moi (sujet)/autres (individu, groupe, entreprise), c'est-à-dire, entre la connaissance et l'action chez l'acteur (individu, groupe, entreprise).

---

[48] Les verbes qui caractérisent l'apprentissage social sont « crear, generar, estabilizar y transformar » que nous avons traduit par créer, gérer, stabiliser et transformer. Dans l'aspect social de la gestion des connaissances gérer et stabiliser nous les sont caractérisés par « organiser et stocker la connaissance ».

En conséquence, dans l'approche organisationnelle de l'aspect social de la gestion des connaissances à partir du modèle générique de création de connaissance et apprentissage organisationnel de Nonaka et Takeuchi, les mécanismes de la gestion des connaissances sont perçus au travers d'un processus dynamique de création et de dissémination (mobilisation) des connaissances. Dans l'approche biologique de l'aspect social de la gestion des connaissances, selon le point de vue de Maturana et Varela, les mécanismes de création de connaissance et apprentissage organisationnel dans un domaine biologique (les relations sont au niveau cellulaire) se trouvent matérialisés au travers d'un processus dynamique de création, gestion, stabilisation et transformation de connaissances, que nous matérialiserons à travers les verbes "créer", "gérer", "stabiliser" et "transformer" la connaissance afin de créer de nouvelles connaissances chez l'homme.

### 1.2.3. Approche managériale de la gestion des connaissances[49]

Nous avons vu que dans le modèle de création de connaissance et d'apprentissage organisationnel de Nonaka et Takeuchi, apparu au milieu des années 90 aux Etats-Unis, l'acteur (individu, groupe, entreprise) est au cœur de la connaissance, l'apprentissage et l'innovation. Pour eux la dimension humaine de l'organisation est comme un tout, le moteur de l'innovation et de l'avantage compétitif.

Dans ce modèle, nous semble qu'il y a une question fondamentale pour trouver la trace de l'approche managériale de la gestion des connaissances :

*How do Japanese companies bring about continuous innovation ?*
*Comment les entreprises Japonaise pratiquent-elles l'innovation continue ?*
*¿Cómo es que los japoneses se las arreglan para innovar todo el tiempo ?.*

---

Cette question, qui d'ailleurs reste toujours ouverte et applicable au reste du monde[50], a été abordée par ces auteurs à partir du fait « que les entreprises japonaises se sont continuellement tournées vers leurs fournisseurs, leurs clients, leurs distributeurs, leurs agences gouvernementales et même leurs concurrents pour rechercher toutes nouvelles idées ou indications qu'ils pourraient offrir »[51]. Ce changement d'attitudes est le fruit de conditions incertaines du marché, comme eux-mêmes l'indiquent « en période d'incertitudes, ces entreprises accumulent de façon désespérée les connaissances issues de l'extérieur ». D'ailleurs, à notre avis ces deux argumentations peuvent être considérées comme la base managériale de l'émergence du concept de "système de connaissances" dans l'entreprise, que nous attachons à Jean-Louis Ermine, puisque l'approche managériale spécifie davantage le rôle de la connaissance que le rôle de l'information dans un système social. Nous ne disons pas par là que les systèmes d'information ne sont pas nécessaires, mais que les systèmes d'information doivent créer de la connaissance et non seulement la traiter…en suivant le paradigme simonien de traitement de l'information. Comme l'ont écrit Nonaka et Takeuchi dans l'introduction de leur livre « ce livre part de la croyance que l'entreprise ne "traite", pas seulement de la connaissance mais la "crée" aussi ». De plus, la dimension ontologique du modèle de Nonaka et Takeuchi a été construit à partir l'étude de Barnard sur les fonctions managériales des dirigeants d'entreprises, dans son livre *The Functions of the Executive*, publié en 1938. Comme réponse à la question managériale qui englobe, selon nous, la philosophie de leur modèle (voir plus haut), ils écrivent « une façon de procéder est de se tourner vers l'extérieur et vers le futur en anticipant les changements dans les marchés, la technologie, la concurrence ou les produits ». Nous pensons que cette dynamique passe par la gestion des connaissances dans un sens plus pratique que théorique. Dans ce livre les études de cas ont été utilisées pour justifier le caractère générique du modèle de création de connaissance et d'apprentissage organisationnel. Néanmoins, la validité du modèle est purement théorique, car nous ne voyons pas dans ce modèle une démarche de gestion des connaissances, ni non plus une séparation entre l'aspect social et l'aspect technique de la gestion des connaissances.

L'avantage concurrentiel du Japon depuis la fin des années 80, en particulier dans les marchés de l'automobiles et de l'électronique avec l'émergence de nouveaux produits, a été un facteur décisif pour les autres pays, principalement les USA et l'Europe, pour valoriser la connaissance et l'apprentissage comme une ressource organisationnelle, stratégique et humaine qu'il faut gérer, en tant que porteuse d'une valeur économique par le biais de l'innovation et la créativité des acteurs (individu, groupe, entreprise) ; avec l'argumentation, que si cela a bien marché pour le Japon pourquoi pas ailleurs ? Nous pensons que le modèle de Nonaka et Takeuchi reste plutôt un lieu de réflexion plus qu'une pratique organisationnelle, car ce modèle est propre de la culture japonaise face au système d'organisation du travail coopératif caractérisé par des méthodes de travail et modes de management, ainsi que du contexte économique du moment au Japon. L'ouvrage de Nonaka et Takeuchi a, sur ce plan, le mérite (1) d'anticiper le terme *Knowledge Management*, car le chapitre 2 du livre s'intitule *Knowledge and Management* (connaissance et management) ; et (2) de valoriser les réflexions de Peter Drucker et Peter Senge sur la société de connaissance et de l'apprentissage organisationnel qui a pris petit à petit forme à travers la société industrielle de Taylor, Ford et Mayo, la société de service de Penrose, Nonaka et Takeuchi, et la société de l'information de Simon, March, Prusak, Davenport.

---

[50] C'est pour cette raison que nous avons préférer exprimer cette phrase en anglais, français et espagnol. *The Knowledge-Creating Company. How Japanese create the Dynamics of Innovation* (publié en 1995). *La connaissance créatrice, la dynamique de l'entreprise apprenante* (publié en 1997). *La organización creadora de conocimiento. Cómo las compañías japonesas crean la dinámica de la innovación* (publié en 1999).
[51] Cette définition de l'entreprise correspond bien à la définition de « l'entreprise élargie », mais nous avons préféré d'utiliser ce terme pour indiquer le paradigme de l'avantage coopératif que nous verrons plus loin. En effet, au milieu des années 90 la problématique des entreprises était orientée vers l'avantage compétitif (chaîne de valeur) ou concurrentiel (marché). Dans la dimension ontologique, Nonaka et Takeuchi parlent des ontologies : individu, groupe, organisation, inter-organisation, nous avons préféré utiliser le terme entreprise pour indiquer l'organisation et l'inter-organisation.

Depuis l'ouvrage de Nonaka et Takeuchi, plusieurs auteurs ont présenté la gestion des connaissances comme un facteur essentiel de croissance économique, l'approche managériale recouvre alors l'ensemble des démarches sur les applications de la gestion des connaissances dans l'entreprise. Néanmoins, il faut tenir compte que chaque démarche est apparue de façon pragmatique, et donc elle reste propre à une problématique particulière de gestion des connaissances. Dans cette perspective, la gestion des connaissances est perçue comme un problème de gestion, que pour comprendre et résoudre, il faut (1) diagnostiquer ; (2) évaluer ; et (3) recommander des actions :

- diagnostiquer signifie identifier ses causes, par exemple la perte de mémoire de l'entreprise, l'accumulation des systèmes experts qui ne sont pas intégrés, etc. ;

- évaluer signifie estimer d'un point de vue économique, social et technologique des solutions ;
- recommander des actions signifie établir une démarche, du point de vue social et technique, pour la mise un place d'un projet de gestion des connaissances dans l'entreprise.

➢ **Qu'est-ce que la connaissance ? Selon l'approche de Jean-Louis Ermine**

Nous avons décidé d'approcher cette question à travers l'ouvrage de Jean-Louis Ermine, intitulé *Les systèmes de connaissances* (publié en 1996)[52], étant donné que pour lui : (1) la connaissance[53] est un système selon la théorie générale des systèmes de Jean-Louis Le Moigne[54], et donc la structuration, la fonction, et l'évolution de la connaissance est fondamentale pour maintenir dans le temps la relation connaissance et environnement interne et externe du système dans lequel l'individu et la culture d'entreprise est au centre de l'organisation du système, d'une part, comme moteur de la dynamique de capitalisation et de partage de connaissances, et d'autre part, comme moteur de la dynamique d'apprentissage et de création de connaissances nouvelles pour l'entreprise ; (2) l'objectif de la gestion des connaissances est justement la capitalisation, le partage et la création de connaissances, et que ceci passe par un processus dynamique (capitaliser, partager, créer) d'alimentation permanente. Et donc, pour lui, les mécanismes de création de connaissance et d'apprentissage organisationnel d'un point de vue de l'approche managériale sont caractérisés à travers un processus dynamique circulante (implication mutuelle, c'est-à-dire une relation d'implication de l'un par rapport à l'autre), appelé gestion des connaissances, qu'il matérialise à travers les verbes "capitaliser", "partager" et "créer" la connaissance en collectivité (en fait un réseau de savoirs).

Dans son article *La gestion des connaissances, un levier stratégique pour les entreprises*, présenté dans les journées d'IC 2000 à Toulouse, et plus récemment dans son ouvrage *La gestion des connaissances* (publié en 2003), il écrit « la gestion des connaissances ("Knowledge Management") s'inscrit désormais dans la réalité de l'entreprise : la connaissance est un enjeu économique majeur de demain. Créer, capitaliser et partager son capital de connaissances est une préoccupation de toute organisation performante ». Ainsi, pour lui la connaissance est un capital économique, une ressource stratégique, un facteur de stabilité et apporte un avantage concurrentiel décisif. Autrement dit, le concept-clé de la gestion des connaissances rejoint celui de "capital" : la connaissance est considérée comme un capital qui a une valeur économique et un statut stratégique pour l'entreprise, au même titre (et souvent plus) que les actifs tangibles de cette entreprise. De ce fait, le « capital connaissance » est au centre de l'organisation et de la structuration du système de connaissances, ou, comme l'explique si bien Tounkara dans le cadre de sa thèse : (1) la connaissance est un facteur de performance et donc de croissance pour les entreprises ; (2) la connaissance est un facteur de stabilité et d'adaptabilité pour les entreprises ; et (3) la connaissance est un critère de positionnement stratégique pour les entreprises.

---

[52] Deux autres ouvrages sont venus après *La gestion des connaissances* (publié en 2003), et *Management des connaissances en entreprise* d'Imed Boughzala et Jean-Louis Ermine (publié en 2004).
[53] Nous parlons de connaissance, car d'après lui « le mot "connaissances" peut être mis au singulier, pour donner une couleur plus générique, ou faire symétrie avec les systèmes d'information ».
[54] Le Moigne utilise de préférence le terme *théorie du système général*, plutôt que le terme *théorie générale des systèmes* de Von Bertalanffy, peut être pour faire ressortir davantage l'organisation du système général, répandu en France et ailleurs, comme modèle OID.

Pour nous, ces trois traits de la connaissance vont constituer les objectifs de l'approche managériale de la gestion des connaissances qui doit guider toute démarche (méthodes et outils) systémique et constructiviste de la gestion des connaissances dans l'entreprise. Nous allons les explorer sommairement à l'aide du modèle de la marguerite de Jean-Louis Ermine, et dans la section que nous appelons *Les origines de la connaissance industrielle*, nous justifierons davantage la production de connaissances dans le domaine industriel à travers (1) la production des connaissances comme un levier de productivité, que nous matérialiserons dans le système d'organisation du travail à l'époque de Taylor, Ford et Mayo ; (2) la production des connaissances comme un levier d'avantage concurrentiel et compétitif que nous expliquerons par le système d'organisation du travail à l'époque de Simon et Penrose ; (3) la production des connaissances comme un levier d'avantage coopératif.

Dans l'introduction de son ouvrage *Les systèmes de connaissances* (édition 2000), Jean-Louis Ermine écrit que « l'objectif de cet ouvrage est d'aborder, à l'instar du système d'information, ce nouvel objet d'ingénierie qu'est le système de connaissances d'une organisation ». Pour lui la connaissance est carrément un objet d'ingénierie qu'il faut approcher à partir d'un projet d'ingénierie, tel que l'information a été approché par la méthode MERISE, largement utilisée et enseignée en France et ailleurs, pour le développement de systèmes d'information. Néanmoins, si dans le passé récent l'ingénierie des systèmes d'information a été développée pour le traitement de l'information, en suivant divers paradigme simonien et modèles, l'ingénierie des systèmes de connaissances[55] doit être développé, d'après nous, pour la capitalisation, le partage et la création de connaissances nouvelles sur la base de l'apprentissage organisationnel, la culture[56] de communication, de coordination et de coopération qui mobilise les acteurs (individu, groupe, entreprise) de la connaissance à tous les niveaux de l'entreprise (stratégique, tactique, opérationnel) tout ceci à travers un réseau de savoirs. Pour nous ces ingrédients donnent vie au système de connaissances de l'entreprise et plus largement au système d'organisation du travail coopératif structuré sur la basé de la gestion des connaissances, la gestion des compétences et la gestion de l'innovation. Et donc finalement, un projet d'ingénierie des systèmes de connaissances doit être développé dans un paradigme de gestion des connaissances sur un modèle systémique et constructiviste de la connaissance, (par exemple, le modèle OIDC de Jean-Louis Ermine). Dans ce sens, le système de connaissances est organisé, d'une part, par un patrimoine de connaissances, et d'autre part, structuré à travers un réseau de savoirs. L'enjeu pour le système de connaissances est alors l'évolution de ce patrimoine de connaissances pour maintenir dans le temps l'organisation. Ceci passe, selon nous, pour un renouvellement permanent de l'énergie du réseau des savoirs par l'apprentissage, l'innovation et la culture d'entreprise par principalement, un réseau (communication, coordination, coopération).

---

[55] Il ne faut pas la confondre avec l'ingénierie des connaissances ou les systèmes experts.
[56] Nous préférons de parler de culture d'entreprise au niveaux de 3 "C" du groupware, bien que René-Charles Tisseyre, introduit les termes culture de partage, et culture de diffusion, lorsqu'il dit « le type de culture d'entreprise est essentiel quant à l'adoption d'un programme global de Knowledge Management. S'il n'y a pas de culture de partage, 'il n'y a pas une culture de diffusion de l'information et donc une autonomie autour de cette information, il est clair que de mettre en place des messageries, des Intranets, des Internets ne changera rien à l'usage que l'ont fait de l'information et *a fortiori* des connaissances ».

Du fait qu'il y a dans cet ingénierie, un processus dynamique (capitaliser, partager, créer) en plus de l'aspect culturel et stratégique de la connaissance, le terme même "d'ingénierie" est un peu limité par cette complexité, d'où la nécessité de mettre le terme "gestion", ou bien le terme en anglais "management" qui amplifie davantage l'aspect humain de la connaissance dans l'approche managériale bien que à ce propos, Jean-Louis Ermine ayant dit que « l'ingénierie … ne se définissant pas à partir d'un objet, mais à partir d'un projet », il convient de retenir un objectif pour cerner ce qu'est un système de connaissances. Cet objectif peut être la *gestion des connaissances*, qu'on entendra au sens de *gestion des systèmes de connaissances* ».

Dans ce sens, le but de l'ingénierie des systèmes de connaissances ou la gestion des systèmes de connaissances est préférable à la gestion des connaissances pour insister sur le fait qu'il ne doit y avoir qu'un seul patrimoine de connaissance qu'il convient de gérer et de faire évoluer à l'aide de tous. C'est la réalisation d'un tel système opérationnel[57] (informatisé ou pas) dans un sens générique qui permettra d'enrichir la capacité de l'entreprise pour maintenir l'organisation comme un tout dans le temps et non pas pour ajouter une autre application au parc informatique de l'entreprise …sinon nous risquons de confondre la gestion des connaissances avec l'ingénierie des connaissances, les systèmes à base de connaissances ou les systèmes experts. Finalement, dans l'approche managériale de la gestion des connaissances les nouvelles méthodes et outils de l'ingénierie des systèmes de connaissances (gestions des connaissances) développés à partir des NTIC du KM, doivent être à la mesure de contribuer significativement à cet effort.

---

[57] Le bons sens indique que la solution informatique fait le tri entre les tâches qui seront automatisées et les tâches qui seront manuelles. Pour éviter le chaos d'une « usine à gaz », Ermine parle de « la conception d'un système opérationnel de gestion des connaissances » plutôt que de système informatisé ou système informatique.

Pour Jean-Louis Ermine les méthodes et outils de la gestion du système de connaissances doivent être orientés vers trois objectifs, trois verbes : capitaliser, partager et créer, qu'il emploie pour caractériser la gestion des connaissances, (1) capitaliser la connaissance « savoir d'où l'on vient, savoir où l'on est, pour mieux savoir où l'on va », cela signifie qu'il faut des méthodes et outils, pour structurer et gérer l'évolution interne du patrimoine des connaissances (système de connaissances) de l'entreprise. Ceci, d'une part, pour garder la trace du passé et présent du patrimoine des connaissances (systèmes de connaissances) de l'entreprise, et d'autre part, pour organiser l'évolution future du patrimoine des connaissances (système de connaissances) de l'entreprise comme un tout, c'est-à-dire dans sa globalité et dans sa complexité ; (2) partager la connaissance « passer de l'intelligence individuelle à l'intelligence collective », cela signifie qu'il faut des méthodes et outils, pour gérer l'évolution de l'interaction entre le patrimoine des connaissances (système de connaissances) et leur environnement interne et externe ; et (3) créer de la connaissance « créer, innover pour survivre » cela signifie qu'il faut des méthodes et outils, pour gérer l'évolution de la créativité et l'innovation du patrimoine des connaissances (système de connaissances).

Nous avons trouvé dans le livre *Management des connaissances en entreprise* d'Imed Boughzala et Jean-Louis Ermine (publié en 2004), mais aussi chez certains cabinets de conseils que nous verrons plus bas, un autre objectif pour les méthodes et outils de la gestion du système de connaissances, que nous pouvons caractériser par le verbe "évaluer", et qu'implique l'évaluation (qualitative, quantitative, financière, etc.) du patrimoine des connaissances de l'entreprise. Par exemple, l'approche IC-dVAL d'Amhed Bounfour permet de mesurer la valeur du capital immatériel de l'entreprise. Pour Bounfour le capital immatériel est constitué du capital humain, du capital structurel, du capital client et du capital innovation. Tandis, que Jean-François Tendron propose l'approche KMM (Knowledge Maturity Model, se définissant comme un modèle de maturité cognitive) pour évaluer le patrimoine de connaissances de l'entreprise. La gestion des connaissances (capitaliser, partager, créer) peut bien sur être élargie pour évaluer la connaissance, ce qui implique qu'au lieu d'avoir trois verbes pour caractériser la gestion des connaissances on peut en avoir quatre, et parler alors de gestion des connaissances dans le sens de capitaliser, partager, créer, évaluer, si l'on souhaite *faire-savoir* ce que l'on a.

Ces quatre objectifs de la gestion des connaissances peuvent être réalisés à partir de cinq processus dynamiques circulantes (c'est-à-dire une relation d'implication de l'un par rapport à l'autre) pour gérer le système de connaissances de l'entreprise. Ils ont été organisés à travers un modèle appelé *modèle de la marguerite* de Jean-Louis Ermine.

Dans ce modèle, le patrimoine de connaissances (système de connaissances) de l'entreprise est gravité par cinq processus : (1) le processus de capitalisation et de partage des connaissances ; (2) le processus d'interaction avec l'environnement ; (3) le processus de sélection par l'environnement ; (4) le processus d'apprentissage et de création de connaissances ; et (5) le processus d'évaluation du patrimoine de connaissances. En faite, ce dernier processus ne se trouve pas incorporé dans le modèle de la marguerite de Jean-Louis Ermine, tel que l'on trouve dans son livre, intitulé *Les systèmes de connaissances* (publié en 2000), mais nous l'avons trouvé ailleurs d'une façon implicite. D'abord, chez lui-même et Imed Boughzala, dans le livre *Management des connaissances en entreprise* (publié en 2004), puis chez autres auteurs, tels que : Wendi Bukowitz, Ruth Williams, Karl Sveiby et Leif Edvinsson que nous les avons présenté plus haut (voir section 1.2). Ainsi, nous ajoutons une complexité supplémentaire au modèle d'Ermine, que nous décrivons à continuation. Néanmoins, cette présentation nous la faisons dans un esprit de recherche des mécanismes création de connaissance et d'apprentissage organisationnel. Une description du modèle de la marguerite, ainsi que son schéma graphique, se trouve bien entendu dans [Ermine, 00].

Modèle de la marguerite[58]

Ce modèle se construit dans la conviction que l'entreprise avec son environnement interne et externe sont en dialogue permanent pour leur organisation et structuration, et donc qu'il y a une dynamique d'interaction à tout moment pour le développement des relations. Le point en commun, est logiquement l'individu, par le biais du système de connaissances (ou patrimoine de connaissances) de l'entreprise. Pour cette raison, nous pensons que cette approche peut être associée à l'approche de l'enaction de Maturana et Varela, du fait que l'individu est au centre de l'organisation. Néanmoins, il y a une différence de paradigme. L'approche de la marguerite est attachée à une problématique de "faire-évoluer" la connaissance, tandis que l'approche de l'enaction est attachée à une problématique de "faire-émerger" la connaissance.

Dans cette dynamique circulaire de processus du système de connaissances, se trouve :

---

[58] Dans ce modèle, symboliquement, la marguerite est formée d'un cœur et autour de celui-ci est composée de quatre pétales. Le cœur de la marguerite est appelé *patrimoine des connaissances* (ou *systèmes des connaissances*). Les quatre pétales ont pour noms : *repérer*, *préserver*, *valoriser* et *évoluer*.

a) <u>Le processus de capitalisation et de partage des connaissances</u>. L'objectif du processus est le développement des relations de conversion des connaissances (individuel et collectif selon le modèle Nonaka et Takeuchi afin de repérer les connaissances internes du business (connaissance métier) de l'entreprise, le mettre à la disposition de tous les acteurs (individu, groupe, entreprise) du système de connaissances et créer de nouvelles connaissances pour l'entreprise, ceci est possible par des outils de l'ingénierie de connaissances.

Les méthodes et outils proposés pour Jean-Louis Ermine afin de gérer un tel chantier du Knowledge Management, en empruntant le terme de René-Charles Tisseyre, ont été conçus à partir de travaux sur l'intelligence artificielle recensés dans deux livres écrits par lui. Le premier est *Systèmes experts, théorie et pratique* (publié en 1989), et le deuxième livre *Génie logiciel et génie cognitif pour les systèmes à base de connaissances* (publié en 1993 et relatant son expérience au Commissariat à l'Energie Atomique (CEA)). Ainsi sont nées deux méthodes[59] d'analyse des systèmes de connaissances (1) Method for Knowledge System Management (MKSM) ; et (2) Method for Analyzing and Structuring Knowledge (MASK), qui, au fils des années, ont été expérimentées et mis au point par lui et son équipe de chercheurs.

Pour Jean-Louis Ermine « MKSM est (ou du moins espère être) l'équivalent de MERISE pour les systèmes d'information, à savoir une méthode d'analyse de systèmes de connaissances pour aboutir à la conception d'un système opérationnel de gestion des connaissances ». Nous soulignons que la conception d'un tel système opérationnel n'implique pas directement la construction d'un système informatisé.

Et donc, comme MERISE qui a été utilisée comme une méthode d'analyse de l'information dans l'entreprise, MASK et MKSM sont deux méthodes d'analyse de la connaissance dans l'entreprise, et donc sont des outils conceptuels pour dialoguer avec la complexité afin de la comprendre comme un tout, et non pas comme une méthode réductrice qui aboutirait à une solution informatique. C'est justement ce que Jean-Louis Ermine a souligné pour MKSM (et nous pensons que ceci s'applique aussi à MASK), lorsqu'il dit « MKSM était au début le sigle de *Method for Knowledge System Management* ; il vaut mieux maintenant de le prendre comme un simple identifiant ! »[60]. Nous dirons plutôt qu'aujourd'hui MASK et MKSM et ses évolutions symbolisent un macroscope (en empruntant le terme de Joël de Rosnay) pour observer l'acteur (individu, groupe, entreprise) dans son poste de travail (connaissance métier) en tant que facteur clé pour améliorer et développer la connaissance collective et le travail un réseau (communication, coordination, coopération) de l'entreprise. Dans ces deux démarches les connaissances sont structurées par une approche métier avec des outils de l'ingénierie de connaissances (intelligence artificielle) et non pas par une approche des compétences. Nous n'irons pas plus loin dans l'explication de leurs modèles, leurs différences, et la comparaison avec d'autres méthodes et outils de l'ingénierie de connaissances, tels que CommonKADS/KADS, etc.

---

[59] Nous avons hésité à mettre une date de naissance, car ils ont été conçu au début pour le CEA et puis appliqués ailleurs. Par exemple, Ermine a dit « MKSM est utilisé au CEA depuis 1993 », alors que Prax a dit « la méthode MKSM inventée en 1996 par Jean-Louis Ermine, mathématicien au CEA ».

Dans les approches MASK et MKSM de la gestion des connaissances (conçu par Ermine et son équipe de recherche pour gestionnaire des connaissances au Commissariat à l'Energie Atomique (CEA)) « la connaissance se perçoit comme un signe, qui contient de l'information (quelle est alors la forme codée ou perçue du signe que je reçois ?), du sens (quelle représentation l'information engendre-t-elle dans mon esprit ?), et du contexte (quel environnement conditionne le sens que je mets sur l'information reçue). Dans ce cadre, la connaissance se perçoit comme un système, avec toujours trois points de vue : la structure, (comment se structurent les objets et les concepts de la connaissance ?) la fonction (dans quelle fonction, quelle activité s'inscrit la connaissance ?) et l'évolution (quel est l'historique de la connaissance ?) ».

La problématique de gestion des connaissances abordé par lui et son équipe au CEA a été relative aux « pertes de savoir et savoir-faire à la suite, souvent, de départ à la retraite, de transfert de personnes, de suppression d'effectifs »[61]. Dans ce sens pour le CEA « la gestion des connaissances vise à (1) rassembler le savoir et le savoir sur des supports facilement accessibles ; (2) faciliter leur transmission en temps réel à l'intérieur du CEA et en différé à ses successeurs ; (3) garder la trace de certaines activités ou actions sur lesquelles le CEA devra rendre des comptes dans l'avenir ». En plus « chaque direction opérationnelle, fonctionnelle et de centre est chargée de définir, dans son domaine de responsabilité, ce qui doit être écrit et conservé et d'organiser cette conservation, en s'appuyant, en tant que de besoin, sur les compétences de la Mission de l'Information Scientifique et Technique du CEA ».

---

[60] Nous avons rencontré deux articles, sur cette approche. L'un publié en 1996, intitulé *MKSM, méthode pour la gestion des connaissances*, l'autre publié en 2000, intitulé *Capitaliser et partager les connaissances avec la méthode MKSM*.
[61] Cette définition de la gestion des connaissances a été donnée dans le Manuel Qualité de cette institution, que nous s'avons repéré dans la thèse de Barthelme-Trapp et la thèse de Tounkara.

Dans cette démarche, les mécanismes de création de connaissance et d'apprentissage organisationnel sont caractérisés à travers un processus dynamique que nous matérialiserons à travers les verbes "rassembler", "faciliter", "garder". La validé de ce processus dynamique, et par suite de la démarche, est soumis aux résultats opérationnels de la mission fixée par le CEA[62]. Autrement dit, la connaissance qu'il faut « rassembler, faciliter, et garder », est la connaissance qui est au service de l'organisation.

Il existe d'autres verbes pour matérialiser ces mécanismes sur le plan social, ils sont relatifs aux processus de capitalisation et de partage des connaissances, notamment chez Michel Grundstein, dans les travaux des thèses de Barthelme-Trapp (2003) et Thierno Tounkara (2002). Néanmoins, nous avons gardé l'approche de Tounkara pour caractériser le processus d'interaction avec l'environnement. En effet, il utilise le même mécanisme de création de connaissance et d'apprentissage organisationnel que Grundstein mais en l'enrichissant par le verbe "faire évoluer". Pour Michel Grundstein dans son article *La capitalisation des connaissances de l'entreprise, une problématique de management* (publié en 1996), son modèle de gestion des connaissances est conçu à partir de cinq objectifs : repérer, préserver, valoriser, transférer et partager les connaissances cruciales de l'entreprise, car « le management des connaissances … couvre toutes les actions managériales visant à actionner le cycle de capitalisation des connaissances afin de repérer, préserver, valoriser, transférer et partager les connaissances cruciales de l'entreprise ». Nous constatons aussi que ces objectifs correspondent aux objectifs de la capitalisation des connaissances, puisqu'il dit dans son article *La capitalisation des connaissances de l'entreprise, système de production des connaissances* (publié en 1995) « repérer et rendre visible les connaissances de l'entreprise, pouvoir les conserver, les accéder et les actualiser, savoir les diffuser et mieux les utiliser, les mettre en synergie, deviennent des sujets de préoccupations actuels, que nous rassemblons sous l'expression générique "capitalisation des connaissances de l'entreprise" ». Finalement, pour Grundstein gestion et capitalisation reposent sur les mêmes mécanismes lorsqu'il s'agit de gérer ou capitaliser la mémoire d'entreprise (connaissances cruciales selon Grundstein, patrimoine de connaissances selon Ermine). D'après la démarche de Grundstein les mécanismes de création de connaissance et d'apprentissage organisationnel, sont caractérisés à travers un système de connaissances, qui englobe plus que la connaissance métier identifiée par MASK ou MKSM, organisée par un cycle de capitalisation des connaissances, et structurée par un processus dynamique que nous matérialiserons à travers les verbes "repérer", "préserver", "valoriser", "transférer" et "partager" les connaissances.

---

[62] Un exemple de capitalisation de connaissances au CEA est le livre de connaissance Silva. Il s'agit d'un support informatique qui regroupe quelques 5000 pages de documentation sur les connaissances mises en œuvre dans les procédés d'enrichissement de l'uranium par laser. Ce livre est destiné aux nouveaux intervenants sur le procédé.

En revanche, pour Barthelme-Trapp « la gestion des connaissances recouvre un ensemble de modèles ou méthodologies pouvant mettre en œuvre des outils de traitement de l'information et de communication visant a structurer, valoriser et permettre un accès à toute l'organisation aux connaissances développées et qui y ont été ou sont encore mises en pratique en son sein »[63]. Or, d'après cette démarche les mécanismes de création de connaissance et d'apprentissage organisationnel sont caractérisés à travers un système de connaissances, organisé à travers de modèles ou méthodologies de connaissances, et structuré par un processus dynamique que nous matérialiserons à travers les verbes "structurer", "valoriser", et "permettre" un accès aux connaissances. Ceci a été spécifié dans son modèle des « *objectifs managériaux des gestion des connaissances* », où les mécanismes de création de connaissance et d'apprentissage organisationnel sont caractérisés par les verbes "repérer", "identifier", "préserver", "contrôler", "actualiser", "initier", et "ouvrir", qui correspondent aux étapes génériques de la gestion des connaissances (1) repérer les connaissances cruciales ; (2) identifier les connaissances ; (3) préserver les connaissances ; (4) contrôler l'accessibilité au regard de la confidentialité des connaissances ; (5) actualiser et enrichir au fur et à mesure par l'introduction de nouvelles connaissances ; (6) initier l'innovation en permettant les combinaisons de connaissances ; et enfin (7) ouvrir le champ de connaissances sur l'environnement.

b) <u>Le processus d'interaction avec l'environnement</u>[64]. L'objectif de ce processus est le développement des relations d'adaptation avec l'environnement externe de l'entreprise afin de capitaliser les connaissances externes du business de l'entreprise, le mettre à la disposition de tous les acteurs (individu, groupe, entreprise) du système de connaissances, et enfin créer de nouvelles connaissances pour l'entreprise. Ceci est possible à partir d'outils de veille[65] (stratégique, concurrentielle, commerciale, scientifique, technologique, etc.) et d'intelligence économique[66], élargis à la gestion des connaissances.

---

[63] Cette démarche a été mise au point à l'occasion de sa thèse appelée *Une approche constructiviste des connaissances : contribution à la gestion dynamique des connaissances dans l'entreprise*, publiée en 2003 par l'Université Toulouse I, ainsi que dans son article *Analyse comparée de méthodes de gestion des connaissances pour une approche managériale* (publié en 2001).
[64] L'environnement est défini par Lesca, et cité par Tounkara comme « l'ensemble des acteurs susceptibles d'avoir une influence sur l'entreprise ». Et donc, l'environnement interne est constitué par exemple à partir des relations fournisseurs, tandis que l'environnement externe est constitué par exemple à partir des relations clients.
[65] La veille est défini par Lesca, et cité par Tounkara, comme « le processus informationnel, volontariste par lequel l'entreprise se met à l'écoute anticipative (ou prospective) des signaux précoces de son environnement socio-économique et technologique dans le but créatif de découvrir des opportunités et de réduire les risques liés à l'incertitude »et donc, dans cette optique, les actions de veille sont des outils différentes.

En effet, ces actions (stratégiques, concurrentielles, commerciales, scientifiques, technologiques, etc.) de veille et d'intelligence économique sont limités au traitement de l'information en suivant un paradigme simonien. L'enjeu maintenant est d'élargir ces approches pour la création des connaissances collectives nouvelles de l'entreprise et l'apprentissage organisationnel (valorisation des connaissances créées). Dans la thèse de Tounkara, intitulé *Gestion des Connaissances et Veille : vers un guide méthodologique pour améliorer la collecte d'informations*, on montre la synergie qui existe entre les mécanismes de gestion des connaissances et de veille scientifique et technologique dans la collecte d'information liée au processus d'interaction avec l'environnement selon le modèle de la marguerite de Jean-Louis Ermine. A ce propos Tounkara dit que « l'enjeu de ce travail de thèse est d'analyser et d'optimiser le processus d'interaction du patrimoine de connaissances avec l'environnement extérieur en vue d'acquérir, d'intégrer, de créer de nouvelles connaissances » [Tounkara, 02].

Pour Tounkara « gérer les connaissances de l'entreprise consiste à mettre en place des processus de capitalisation, d'apprentissage et de création, d'interaction dans le but de repérer les connaissances cruciales pour l'entreprise, de les préserver, de les valoriser et de les faire évoluer » [Tounkara, 02]. D'après cette démarche, les mécanismes de la gestion des connaissances sont caractérisés à travers un système de connaissances, organisé à travers la capitalisation, l'apprentissage, la création, et l'interaction des connaissances cruciales pour l'entreprise, et structuré par un processus dynamique que nous matérialiserons à travers les verbes "repérer", "préserver", "valoriser", et "faire évoluer" des connaissances.

c) Le processus de sélection par l'environnement. L'objectif de ce processus est le développement de relations d'évolution avec l'environnement afin de capitaliser les connaissances externes du business de l'entreprise, le mettre à disposition de tous les acteurs (individu, groupe, entreprise) du système de connaissances, et finalement de créer de nouvelles connaissances pour l'entreprise. Ceci est possible à travers d'outils comme CRM, ERP, marketing, etc., élargis à la gestion des connaissances.

---

[66] L'intelligence économique est définie par Martre, cité par Tounkara comme « l'ensemble des actions de recherche, de traitement et de diffusion (en vue de son exploitation) de l'information utile aux acteurs économiques ». Dans cette définition l'on voit nettement présente l'influence de Simon.

d) Le processus d'apprentissage et de création de connaissances. L'objectif de ce processus est le développement de relations de socialisation (transfert de connaissance tacite à connaissance non tacite selon le modèle de Nonaka et Takeuchi) entre tous les acteurs (individu, groupe, entreprise) du système de connaissances afin de capitaliser les connaissances internes et externes du business de l'entreprise et de son environnement, le partager et créer de nouvelles connaissances pour l'entreprise. Ceci est possible par des outils de e-learning, de créativité (par exemple TRIZ), de retour d'expérience, etc.

e) Le processus d'évaluation du patrimoine de connaissances (système de connaissances). L'objectif de ce processus est le développement des relations de valeur pour mesurer le capital connaissance[67] de l'entreprise. Ce processus, que nous avons trouvé intéressant de le rajouter au modèle ne se trouve pas inclus dans le modèle de la marguerite de Jean-Louis Ermine [Ermine, 00], mais nous l'avons trouvé par ailleurs, d'abord dans [Boughzala et Ermine, 04] et puis chez deux cabinets de conseil. L'un est PricewaterhouseCoopers[68] avec deux membres du groupe Asset Management, Wendi Bukowitz et Ruth Williams, qui dans leur livre *Gestion des connaissances en action* (publié en 2000), ont déclaré « le travail d'évaluation consiste, pour l'entreprise, à définir les connaissances d'importance stratégique et à comparer ses biens intellectuels présents à ceux dont elle aura besoin par la suite … Ces entreprises devront, de plus en plus à l'avenir, faire face à un défi supplémentaire : mettre au point un système de mesure traduisant le développement de sa base de connaissance ainsi que la rentabilité de leurs investissements en matière d'actifs intellectuels »[69]. Il s'agit donc de mettre en place un système de mesure du « capital connaissance du corps social » qu'est l'entreprise, c'est une tâche complexe car les relations de travail ne sont pas du cash-flow (information financière) que l'on peut associer à une valeur (ce l'on fait avec le système comptable de l'entreprise, ou bien avec le système de rémunérations de l'entreprise ou on peut associer individuellement une prime pour la compétence des salariés dans le cadre de la gestion de compétences individuelles). Finalement cette gestion des connaissances manipule des richesses invisibles qui se trouvent enracinées dans les têtes des individus et dans les relations coopératives du travail en réseau que nous appellerons système d'organisation du travail coopératif. L'autre groupe de conseil est la société Sveiby[70]. Dans le livre *Knowledge Management. La nouvelle richesse des entreprises. Savoir tirer profit des actifs immatériels de sa société* (publié en 2000), son PDG Karl Sveiby déclare « en 1986, j'ai écrit un premier livre sur le sujet (The Know-How Company). Depuis, je crois avoir acquis une compréhension plus profonde de ce que sont les entreprises basées sur le savoir … un gisement de ressources illimitées dans la capacité des hommes à créer des connaissances et dans le fait que, au contraire des actifs physiques traditionnels, comme le savoir s'accroît au fur et à mesure qu'il est partagé ». Dans cette logique, la problématique de la gestion des connaissances peut être résumée dans la question fondamentale « pourquoi les actions de certaines sociétés ont-elles une valeur boursière plus élevée que leur valeur comptable ? ». Le groupe Mazars[71] offrent des outils pour gérer et évaluer ce capital immatériel de l'entreprise (bien entendu avec ses relations avec leur environnement). Nous pensons que la validités des ces outils, restent toujours dans leur interprétation des résultats. Par exemple, si nous faisons une analogie avec l'information financière, une même valeur pour le certains indicateurs peut avoir différentes interprétations selon la méthode utilisée pour le calculer. En conséquence, dans le processus d'évaluation du système de connaissances, les méthodes et outils développés pour gérer et évaluer la connaissance sont basés dans des concepts *knowledge-based assets* (Karl Sveiby), *corporate longitude* (Leif Edvinsson), etc. dans lesquels « la gestion de l'immatériel, celle notamment du savoir, est au cœur de la création de valeur » pour l'entreprise[72]. Nous soulignerons enfin, que les outils d'évaluation (qualitative, quantitative, financière, etc.) du capital immatériel (patrimoine de connaissances, systèmes de connaissances, etc.), tels que l'approche IC-dVAL d'Amhed Bounfour, l'approche KMM de Jean-François Tendron, que nous avons mentionné plus haut, et les outils de PricewaterhouseCoopers, Mazars, Sveiby, etc., n'ont rien à voir avec les méthodes et outils pour la gestion des connaissances issus de l'ingénierie des connaissances, tel que MASK/MKSM, CommonKADS/KADS, etc. ou les outils d'évaluation et évolution de la mémoire d'entreprise. La figure 1.2 montre la dynamique du modèle.

---

[67] Pour nous, capital connaissance, patrimoine de connaissances, système de connaissances, actifs intangibles, actifs incorporels, capital intellectuel, capital immatériel sont des synonymes, bien que différents dans l'esprit de la construction du système de connaissances avec MASK, MKSM ou CommonKADS, il s'agit de la connaissance métier qu'il faut analyser, structurer pour être exploité par un système opérationnel (informatisé ou pas).
[68] http://www.pwcglobal.com/
[69] Pour Bukowitz et Williams « le capital intellectuel est tout élément qui est détenu par des personnes ou dérivé de processus, de systèmes ou de la culture d'une organisation et qui présente une valeur pour cette dernière des qualités telles que : compétences et qualifications individuelles, normes et valeurs, bases de données, méthodes, programmes informatiques, savoir-faire, brevets, marques, secrets de fabrication, pour n'en citer que quelques-unes » [Bukowitz et Williams, 00].

---

[70] http://www.sveiby.com/
[71] http://www.mazars.fr
[72] De nombres sites Web qui développent la relation entre gestion des connaissances et capital intellectuel se sont organisés, tel est le cas de http://www.unic.net/

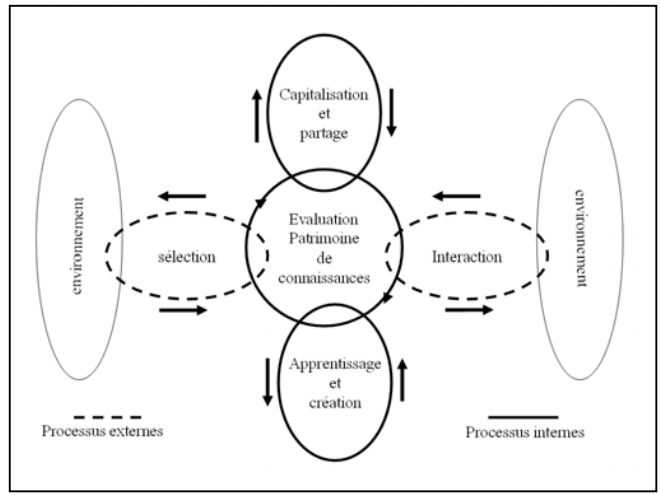

**Figure 1.2 :** Modèle de la marguerite[73] (source [Ermine, 96])

En conséquence, si les mécanismes création de connaissance et d'apprentissage organisationnel selon Nonaka et Takeuchi (approche organisationnelle de l'aspect social de la gestion des connaissances à partir du modèle générique de création de connaissance et apprentissage organisationnel) peuvent être perçus à travers un processus dynamique de création et de dissémination (mobilisation) des connaissances, selon Maturana et Varela (approche biologique de l'aspect social de la gestion des connaissances) peuvent être perçus à travers un processus dynamique de création, gestion, stabilisation et transformation de connaissances. Par contre, d'après le modèle de la marguerite de Jean-Louis Ermine (approche managériale de l'aspect social de la gestion des connaissances) les mécanismes création de connaissance et d'apprentissage organisationnel seront perçus à travers un processus dynamique de capitalisation, partage, création et évaluation de connaissances, que nous matérialiserons à travers les verbes "capitaliser", "partager", "créer" et "évaluer" la connaissance afin de créer de nouvelles connaissances dans l'entreprise. Pour nous, dans ces verbes se trouvent implicites les objectifs de la gestion des connaissances du CEA[74] que nous matérialisons à travers les verbes "rassembler", "faciliter", "garder".

Le tableau 1.1 montre une synthèse de ces trois approches

---

[73] Nous constatons que la marguerite est formée d'un cœur et autour de celui-ci se trouvent quatre pétales. Le maintien de la marguerite en vie se fait par un dialogue permanent avec son environnement (interne et externe) au travers de l'information.

[74] D'après, le Commissariat à l'Energie Atomique (CEA) « la gestion des connaissances vise à (1) rassembler le savoir et le savoir sur des supports facilement accessibles ; (2) faciliter leur transmission en temps réel à l'intérieur du CEA et en différé à ses successeurs ; (3) garder la trace de certaines activités ou actions sur lesquelles le CEA devra rendre des comptes dans l'avenir ». Où Ermine a joue un rôle important dans un grand nombre de projets liés à la gestion des connaissances dans cette institution.

---------------------------------------------------------------------------------------------------------------

| | Approche de Nonaka et Takeuchi | Approche de Maturana et Varela | Approche de Jean-Louis Ermine |
|---|---|---|---|
| Analogie | Ballon de rugby | Arbre | Marguerite |
| Fonctionnement | Processus de causalité circulaire cognitiviste | Emergence de signification | Sélection d'information |
| Mécanismes | - Socialisation<br>- Combinaison<br>- Extériorisation<br>- Internalisation | - Clôture opérationnelle<br>- Couplage structurel<br>- Détermination structurelle | - Capitalisation et partage<br>- Interaction avec l'environnement<br>- Sélection par l'environnement<br>- Apprentissage et création<br>- Evaluation |

**Tableau 1.1 :** Les trois approches de la gestion des connaissances

## 1.3.    Aspect technique de la gestion des connaissances

Nous avons fait l'hypothèse que la gestion des connaissances est un système, capable d'être décrit à travers un système sociotechnique récursif. La récursivité dans ce système implique que la connaissance est perçue par la composante sociale comme un système d'organisation du travail coopératif, organisé à travers un système de connaissances, et structuré à travers la gestion des connaissances. Par contre, la connaissance est perçue aussi par la composante technique, comme un objet de gestion, organisé à travers l'information, et structuré à travers de données. Ainsi, la récursivité se trouve dans la gestion des connaissances par une double existence, d'abord dans la dualité[75] organisation/structure, puis dans la dualité social/technique. Autrement dit, l'aspect social de la gestion des connaissances existe dans la dualité organisation/structure, tandis que l'aspect technique de la gestion des connaissances existe aussi dans cette même dualité.

Or, dans le système sociotechnique récursif, (1) la composante sociale qui est relative au système d'organisation du travail coopératif, est organisée par deux systèmes. L'un est le système cognitif (l'individu) géré dans une structure cognitive par un processus d'apprentissage et d'adaptation à l'environnement, et l'autre est le système de connaissances (l'organisation) géré dans une structure organisationnelle par un processus d'apprentissage organisationnel d'évolution avec l'environnement au travers de méthodes de travail et de modes de management ; et (2) la composante technique relative, d'une part à l'organisation dans une structure technique des NTIC du KM, et d'autre part, à la structuration de l'organisation. Ces composantes existent dans cette dualité organisation/structure. La composante sociale est relative à l'organisation et à la structuration de la connaissance à travers de méthodes de travail et de modes de management, tandis que la composante technique est relative à l'organisation et la structuration de la connaissance sur le plan technologique. Ainsi, selon l'aspect technique, la connaissance dans l'entreprise est perçue, d'une part, comme un *système*, organisé à travers un système opérationnel (informatisé ou pas) des NTIC du KM, et d'autre part, comme un *objet de gestion*, structuré à travers d'abord, la gestion de la communication, de la coordination et de la coopération entre les acteurs et les relations du système d'organisation du travail coopératif qui sont nécessaires pour créer des connaissances nouvelles, puis, la gestion de l'information et de données afin d'en extraire la connaissance et créer des connaissances nouvelles dans l'entreprise. Ici, comme pour la structure sociale, la structure technique de la gestion des connaissances est matérialisée aussi à travers de verbes.

---

[75] Au passage nous soulignons que pour nous la dualité caractérise une complémentarité plutôt qu'une opposition.

Néanmoins, dans la démarche de Grundstein nous constatons que seulement l'aspect social a été considéré, tandis que dans la démarche de Tounkara l'aspect technique de la gestion des connaissances est, d'après nous, implicitement mentionné à travers l'organisation et la structuration des NTIC qui sont à la base de la capitalisation et de l'interaction des connaissances lorsque l'on veut favoriser l'apprentissage collectif et la création des connaissances nouvelles dans l'entreprise. De plus, dans la démarche de Barthelme-Trapp l'aspect technique de la gestion des connaissances est organisé à travers des outils des NTIC, et la structure technique est matérialisée à travers les mêmes verbes que pour son aspect social de la gestion des connaissances. Enfin, avec les approches de Dominique Crié et René-Charles Tisseyre, les objectifs de l'aspect technique de la gestion des connaissances sont davantage caractérisés.

➢ **L'approche NTIC du KM**

Pour Dominique Crié dans son article intitulé *De l'extraction des connaissances au Knowledge Management* (publié en 2003), « la gestion des connaissances ou *Knowledge Management* (KM) se définit comme le processus de capture et d'enregistrement de l'expertise collective d'une entreprise quel que soit l'endroit où cette dernière réside (les bases de données internes ou externes, les documents de toute nature et format ainsi que dans la "tête des individus" puis de sa redistribution là où elle est susceptible de produire profits ». Pour cet auteur, les méthodes et outils de la gestion des connaissances doivent être développés à partir de trois axes (1) stockage de connaissances « vous savez ce que vous détenez » ; (2) partage de connaissances « vous savez ce que nous n'avez pas » ; et (3) extraction de connaissances « vous ne savez pas ce que vous détenez ». Elle propose des outils de fichier ou gestion des documents pour le stockage de connaissances, des outils de moteur de recherche traditionnelle pour le partage de connaissances, et des outils de text mining, web mining, et data mining pour l'extraction de connaissances.

D'après ce point de vue, les mécanismes de création de connaissances nouvelles et d'apprentissage organisationnel, selon l'aspect technique, sont caractérisés à travers un système opérationnel, organisé à travers le stockage et la diffusion des connaissances. Dans ce système, la structure technique est matérialisée par les verbes "stocker", "partager" et "extraire" des connaissances à l'aide des outils des NTIC. En comparaison, les mécanismes de création de connaissances nouvelles et d'apprentissage organisationnel, selon l'aspect social, sont caractérisés à travers un système de connaissances, organisé à travers la collecte, la diffusion et la production des connaissances, et structuré par un processus dynamique que nous matérialiserons à travers les verbes "capturer", "enregistrer", "redistribuer" et "produire" des connaissances.

Pour René-Charles Tisseyre, « il est utile d'analyser le processus de Knowledge Management autour de quatre actions : capter les connaissances, structurer les connaissances, diffuser les connaissances, et mettre à jour les connaissances (ce qui est le plus souvent oublié) » [Tisseyre, 99]. Il parle aussi de « faire une analyse de la nouvelle organisation pour alimenter, gérer et diffuser les connaissances » [Tisseyre, 99]. Au niveau des outils de la gestion des connaissances, « le chantiers technologique du KM définit les outils qui sont mis à la disposition des acteurs pour mieux acquérir, gérer et diffuser leurs connaissances ». Il parle des « outils de capitalisation qui permettent de recevoir les connaissances et de les structurer … des outils de diffusion permettant de restituer rapidement ces connaissances … des outils de travail qui permettent de les exploiter » [Tisseyre, 99]. Pour lui les outils de la création de connaissances nouvelles et d'apprentissage organisationnel peuvent jouer sur la transversalité (acquisition, stockage, diffusion) d'informations et de connaissances ou bien sur la verticalité (extraction de connaissances). Pour la transversalité par exemple, il peut exister un système à base de connaissances qui gère les connaissances et informations de plusieurs départements, et pour la verticalité on peut avoir un système expert pour l'extraction de connaissances à partir des données du département de ventes. Pour lui, l'infrastructure technologie de base des outils de la gestion des connaissances est possible grâce à l'Internet, l'Intranet, et l'Extranet autour de quatre phases « acquisition, structuration, recherche, diffusion » [Tisseyre, 99] d'informations et de connaissances.

Selon la démarche de Tisseyre, les mécanismes de création de connaissances nouvelles et d'apprentissage organisationnel, selon l'aspect technique peuvent être organisés par des outils des NTIC, dans lequel la structure technique est matérialisée à travers les verbes "acquérir", "structurer", "rechercher", "diffuser" et "extraire" des connaissances. D'ailleurs, nous constatons que dans cette démarche, ces objectifs sont implicites dans les objectifs de l'aspect social, qui sont caractérisés par un système de connaissances, organisé à travers la capitalisation, la structuration, la diffusion, et l'exploitation des connaissances, et structuré par un processus dynamique que nous matérialiserons à travers les verbes "capter" ou "alimenter" ou "acquérir", "structurer" ou "stocker" ou "gérer", "diffuser", et "mettre" à jour les connaissances.

En conséquence, sur la base des approches de Crié et de Tisseyre, pour nous les mécanismes de création de connaissances nouvelles et d'apprentissage organisationnel selon l'aspect technique sont centrés principalement dans la mise au point, d'une part, des méthodes et outils pour la collecte ou l'acquisition, l'extraction des connaissances,[76] la structuration des connaissances et la diffusion ou partage des connaissances, et d'autre part, l'utilisation des nouvelles technologies de l'information et de la communication du Knowledge Management (les NTIC du KM). Bien entendu, dans ces trois objectifs génériques de la gestion des connaissances, l'analogie avec la gestion de l'information est toujours possible, mais la complexité de la gestion des connaissances est fort différente, comme l'a souligné Tounkara à propos de la méthode MERISE « il n'existe pourtant pas beaucoup d'équivalent pour la connaissance, permettant une analyse et une structuration d'un patrimoine de connaissances, après la réflexion stratégique et avant l'implantation du système informatisé. C'est d'autant plus étonnant que le problème de la connaissance dans l'entreprise n'est pas réductible au problème de l'information, et semble intuitivement bien plus complexe » [Tounkara, 02].

En résumé, l'approche NTIC du KM, les mécanismes de création de connaissances nouvelles et d'apprentissage organisationnel, sont matérialisés à travers les verbes "acquérir", "structurer", "extraire" et "diffuser" des connaissances.

## 1.4. Les mécanismes de généralisation de la gestion des connaissances selon l'aspect social et l'aspect technique

Dans l'aspect social et technique de la gestion des connaissances, les mécanismes de création de connaissances nouvelles et d'apprentissage organisationnel des approches organisationnelles, biologiques, managériales, et NTIC du KM permettent de faire l'hypothèse qu'un "pont" existe entre elles et peut être caractérisé, à travers la relation qui existe entre connaissance et environnement. Cette relation peut être justifiée, d'abord par le modèle de Nonaka et Takeuchi, puis par l'approche de l'arbre des connaissances (ou l'enaction) de Maturana et Varela, ensuite par le modèle de la marguerite de Jean-Louis Ermine. Mais aussi elle peut être justifiée par d'autres modèles chez d'autres chercheurs ayant travaillé d'une part sur l'approche managériale, à travers le système cognitif (l'individu) et le système de connaissances (le patrimoine de connaissances), et d'autre part, sur la dualité organisation/structure (nous reviendrons largement sur cette dualité dans le chapitre 2, section 2.1) qui existe simultanément et nécessairement dans un modèle systémique et constructiviste de la connaissance. Ceci nous permet de faire une généralisation des mécanismes de la gestion des connaissances, en postulant que la relation connaissance/environnement est une dualité, dans le sens que l'un ne peut pas exister sans l'autre, ou l'un n'a pas de sens sans l'autre, car les deux sont construits dans un dialogue continu (qui permet de créer à tout instant la structure de l'organisation) capable de voir l'organisation comme un tout. L'adaptation et l'évolution du système de connaissance font alors partie d'un organisme intelligent, capable de décider sur les relations internes du réseau des savoirs.

---

[76] Les termes collecte ou acquisition sont utilisés pour la collecte ou acquisition des connaissances à partir d'informations, tandis que le terme extraction est utilisé pour l'extraction des connaissances à partir de données. Les technologies associées peuvent être du type GED pour la collecte ou acquisition d'informations, ou bien du type Text Mining, Web Mining ou Data Mining pour l'extraction des connaissances à partir de données.

Or, c'est justement dans cette relation entre connaissance et environnement que se trouve l'opportunité de définir les composantes sociales et techniques de la gestion des connaissances de notre modèle, c'est-à-dire les mécanismes génériques qui la créent. En effet, nous avons fait l'hypothèse que la gestion des connaissances peut avoir lieu (dans un système d'organisation du travail coopératif) à travers un système sociotechnique récursif qui existe simultanément et nécessairement dans une dualité organisation/structure. Nous constatons donc que "gérer" a un double sens :

- dans l'aspect social de l'organisation dans lequel, gestion est relié à produire ou créer des connaissances collectives nouvelles (en suivant le concept *knowledge creating-company* de Nonaka et Takeuchi, ou le concept de *la marguerite* de Jean-Louis Ermine, ou le concept de l'*arbre des connaissances* de Maturana et Varela[77]) ;

- dans l'aspect technique de l'organisation, ou gestion signifie aboutir à un système opérationnel qui matérialise la gestion des connaissances de l'aspect social à travers une structure de la gestion des connaissances.

---

[77] Non par hasard est le logo du Club Gestion des Connaissances dont Jean-Louis Ermine est président http://www.club-gc.asso.fr/

D'ailleurs, la dualité organisation/structure nous permet de constater, le point de vue erroné de René-Charles Tisseyre lorsqu'il dit « "gérer" n'est pas produire et correspond à des tâches d'organisation » [Tisseyre, 99]. En effet, comme nous l'avons montré, gérer selon l'aspect social signifie justement produire ou créer des connaissances collectives.

Finalement, dans l'aspect social la gestion des connaissances dans sa double existence, sera caractérisée d'abord à travers un système de connaissances organisé pour produire de la connaissance nouvelle et de l'apprentissage organisationnel. Ce système de connaissances est géré par la gestion des connaissances dans une structure organisationnelle composée par le capital humain en interaction permanente dans une relation de travail coopératif. Nous avons matérialisé cette observation à travers quatre mécanismes génériques identifiés par les verbes capitaliser, organiser, stocker, transformer des connaissances collectives[78]. Par contre, dans l'aspect technique, la gestion des connaissances est caractérisée au travers d'un système opérationnel[79]. Ce système sera gérés par la gestion des connaissances dans une structure technique que nous matérialiserons génériquement à travers les verbes acquérir, structurer, rechercher et diffuser, en relation avec l'acquisition, la structuration, la recherche et la diffusion de données ou informations.

Ces mécanismes liés à la structure technique seront traités avec des outils traditionnels des systèmes d'informations. Autrement dit, les NTIC du KM apparaissent comme une couche supplémentaire à la technologie des systèmes d'information et de bases de données. Dans cette logique, la composante sociale de la gestion des connaissances s'appuie sur la composante technique, et la composante technique de la gestion des connaissances matérialise la composante sociale (c'est le concept de récursivité pour nous).

En conséquence, dans la dualité organisation/structure nous rencontrons deux traits de la connaissance, d'une part la connaissance qui est à la fois un objet (une chose) de gestion, et d'autre part, la connaissance qui est un processus de gestion, un système à gérer si l'on veut.

De là, le concept d'*arbre des connaissances* de Maturana et Varela, Il représente pour nous dans cette thèse, le symbole de l'origine des mécanismes de la gestion des connaissances, qui permet de définir les composantes sociales et techniques de la gestion des connaissances, à savoir d'une part, l'organisation de la connaissance qui conforte le point de vue social de la connaissance (autrement dit la dynamique circulante[80] de créer, organiser, stocker, et transformer de la connaissance), et d'autre part la structure qui conforte le point de vue technique de la connaissance, autrement dit la représentation matérielle de la connaissance relative à l'acquisition, structuration, recherche et diffusion de la connaissance. Pour nous ces deux types de descriptions de la connaissance (sociale et technique) sont complémentaires, plutôt qu'opposés.

---

[78] Ces mécanismes génériques coïncident à peu près avec ceux de Jean-Louis Ermine dans l'approche managériale de la gestion des connaissances, identifiés par les verbes *capitaliser, partager, créer* des *connaissances*. Nous pouvons faire aussi une analogie avec le modèle de Simon, pour exprimer la capitalisation (création), le stockage et l'utilisation de la connaissance.
[79] Dans le cas d'un système informatisé, le système opérationnel peut être caractérisé par un système à base de connaissances (base de règle ou une base de cas, voir même un système expert), organisé par des règles de production, et pourquoi pas une combinaison de systèmes, organisé à travers les NTIC du KM.

Ainsi, la description du phénomène étudié à travers un modèle sociotechnique nous permet d'approcher la gestion des connaissances à travers l'aspect social et l'aspect technique de la connaissance. Ceci à l'avantage de classifier les différentes point de vue (l'approche organisationnelle, l'approche biologique, l'approche managériale, et l'approche NTIC du KM), en plus ce modèle nous a permis de faire une généralisation des mécanismes de création de connaissances nouvelles et d'apprentissage organisationnel à partir de la dualité organisation/structure.

L'aspect social de la gestion des connaissances

a) Pour l'approche organisationnelle de Nonaka et Takeuchi fondée sur le concept de *knowledge creating-company*, les mécanismes de création de connaissances nouvelles et d'apprentissage organisationnel sont caractérisés par un processus dynamique matérialisés à travers les verbes "créer", "disséminer" ou "mobiliser" la connaissance (tacite/explicite) individuelle ou collective et l'apprentissage organisationnel (individuelle/collective) dans l'entreprise. Ainsi, le processus dynamique de transfert de connaissance tacite à explicite et vice versa est un système récursif qui s'enrichit grâce au retour d'expérience de l'apprentissage organisationnel (individuelle/collective). Nous proposons alors d'utiliser ces mêmes verbes pour caractériser les mécanismes génériques de la gestion des connaissances, afin de créer de nouvelles connaissances et d'apprentissage collective (selon l'aspect social de l'approche organisationnelle).

b) Pour l'approche biologique de Maturana et Varela fondée sur le concept de l'*arbre des connaissances*, les mécanismes de création de connaissance et d'apprentissage social d'un point de vue biologique (les relations sont au niveau cellulaire) selon l'approche de l'enaction de Maturana et

---

[80] Ceci correspond à la dynamique circulaire de faire-émerger de la connaissance de Matura et Varela (domaine biologique), mais aussi de Nonaka et Takeuchi (domaine industriel).

--------------------------------------------------------------------------------------------------------------

Varela se trouvent caractérisés par un processus dynamique que nous matérialiserons à travers les verbes "créer", "gérer", "stabiliser" et "transformer" la connaissance afin de créer de nouvelles connaissances chez l'homme. Nous proposons alors d'utiliser ces mêmes verbes pour caractériser les mécanismes génériques de la gestion des connaissances, afin de créer de nouvelles connaissances et d'apprentissage collective (selon l'aspect social de l'approche biologique).

c) Pour l'approche managériale de Jean-Louis Ermine fondée sur le concept de la *marguerite*, les mécanismes de création de connaissance et d'apprentissage social d'un point de vue managérial (les relations sont entre l'entreprise et son environnement selon le modèle de la marguerite d'Ermine) se trouvent caractérisés par un processus dynamique, que nous matérialiserons à travers les verbes "capitaliser", "partager", "créer" et "évaluer" la connaissance afin de créer de nouvelles connaissances dans l'entreprise. Nous proposons alors d'utiliser ces mêmes verbes pour caractériser les mécanismes génériques de la gestion des connaissances, afin de créer de nouvelles connaissances et d'apprentissage collective (selon l'aspect social de l'approche managériale).

L'aspect technique de la gestion des connaissances

Pour l'approche NTIC du KM, les mécanismes de création de connaissances nouvelles et d'apprentissage organisationnel selon l'aspect technique sont matérialisés à travers les verbes "acquérir", "extraire", "structurer", et "diffuser" des connaissances, permettant, d'une part, des méthodes et outils pour la collecte ou l'acquisition, l'extraction des connaissances,[81] la structuration des connaissances et la diffusion ou partage des connaissances, et d'autre part, l'utilisation des nouvelles technologies de l'information et de la communication du Knowledge Management (les NTIC du KM). Nous proposons alors d'utiliser ces mêmes verbes pour caractériser les mécanismes génériques de la gestion des connaissances afin de créer de nouvelles connaissances et d'apprentissage collective (selon l'aspect technique de l'approche NTIC du KM).

Ainsi, l'aspect social et l'aspect technique de la gestion des connaissances nous permet d'approcher la question fondamentale : *qu'est-ce que la connaissance ?* :

---

[81] Les termes collecte et acquisition sont utilisés pour la collecte ou acquisition des connaissances à partir d'informations, tandis que le terme extraction est utilisé pour l'extraction des connaissances à partir de données. Les technologies associées peuvent être du type GED pour la collecte ou acquisition d'informations, ou bien du type Text Mining, Web Mining ou Data Mining pour l'extraction des connaissances à partir de données.

- pour l'approche organisationnelle, la réponse peut être résumée comme suit (1) *savoir* ou *savoir technique* pour exprimer la connaissance (tacite ou explicite) créée par le raisonnement de l'acteur (individu, groupe, entreprise), liés aux phénomènes d'intelligence humaine (individu, groupe), d'intelligence économique (entreprise), veille, etc. ; (2) *savoir-faire* ou *savoir-faire technique* pour exprimer la connaissance (tacite ou explicite) créée dans l'action (l'apprentissage) pour l'acteur (individu, groupe, entreprise) ; (3) *savoir-faire collectif* pour exprimer la connaissance organisationnelle ou collective, organisée comme un tout dans un système de connaissances et matérialisée par l'innovation (produits ou services) capable de produire un avantage concurrentiel ou compétitif durable pour l'entreprise ; et enfin (4) savoir-*comportemental* relatif d'une part à la compétence (transformation de la connaissance en action) des ressources humaines qui matérialisent les qualités professionnelles de l'individu dans son espace de travail (l'environnement : le travail et ses outils de production), et au savoir-être (les qualités personnelles de l'individu) exprimé au travers du phénomène de l'intelligence émotionnelle [Goleman, 99].

Dans cette approche, le modèle de gestion des connaissances de Nonaka et Takeuchi met en valeur la connaissance porteuse d'une avantage compétitif pour l'entreprise par le biais de l'innovation à travers deux mécanismes génériques de création de connaissance et d'apprentissage organisationnel, identifiés par des verbes à l'infinitif, à savoir : créer et mobiliser les connaissances à travers la dimension épistémologique (connaissance tacite et explicite) et la dimension ontologique (individu, groupe, entreprise) de la connaissance. Dans ce modèle nous constatons que la gestion des connaissances est décrite par rapport à l'aspect social (socialisation), et l'aspect technique (combinaison, extériorisation, internalisation) ;

- pour l'approche biologique, la réponse peut être résumée comme suit, il s'agit d'un mécanisme "circulant" et de "faire-émerger". C'est un mécanisme proche de l'enaction dans la relation entre l'individu (connaissance) et l'environnement (la connaissance en action). Nous reviendrons largement sur cette approche et les autres approches scientifiques de la connaissance (cognitiviste et connexionniste) dans le chapitre 2, afin de décrire la relation entre connaissance (individu) et action (environnement). Cette approche permet aussi d'enrichir la dualité moi/autres de la relation entre connaissance et action de l'approche existentialiste, et complète les autres approches philosophiques de la connaissance, telles que l'approche rationaliste et l'approche empiriste (que nous avons matérialisées dans la dualité sujet/objet). Dans ce contexte, nous pensons que "l'arbre des connaissances" permet aussi de représenter ces quatre courants cognitifs qui permettent de faire-émerger la connaissance (d'une part, les approches de l'enaction et existentialiste qui symbolisent les racines, et d'autre part, les approches rationaliste et empiriste qui symbolisent les feuilles).

Dans cette approche, le modèle de gestion des connaissances de Maturana et Varela[82] met en valeur la connaissance qui permet de maintenir l'organisme vivant (avec une structure et une organisation) et viable (avec un sens) à travers quatre mécanismes génériques de création de connaissance et d'apprentissage organisationnel, identifiés par des verbes à l'infinitif, à savoir : créer, gérer, stabiliser et transformer les connaissances pour "faire-émerger" la connaissance. Dans ce modèle nous constatons que la gestion des connaissances est décrite par rapport à l'aspect social (relations humaines), tandis que l'aspect technique est négligé ;

- pour l'approche managériale, la réponse peut être résumée par le fait que la connaissance est un système (dans le sens du macroscope de Joël de Rosnay) au travers de l'information (données, traitements) qui prend une certaine signification (concepts, tâches) dans un contexte (domaine, activité) donné. Cette définition peut être affinée, d'une part à partir du modèle OIDC au travers de flux de compétence et de flux de cognition entre la composante (C) et les autres composantes du modèle (O,I,D), et d'autre part, à partir du modèle de la marguerite au travers de la dynamique circulante entre de processus de capitalisation, de partage des connaissances, d'interaction avec l'environnement, de sélection par l'environnement, d'apprentissage, de création de connaissances, et d'évaluation du patrimoine de connaissances. Par exemple, Tounkara qui a été intéressé (dans le contexte de sa thèse) par le processus d'interaction avec l'environnement, élabore la définition suivante « nous proposons de définir les connaissances comme l'ensemble des savoirs et savoir-faire mobilisés par les acteurs dans le cadre de leurs activités. Cette définition implique que la connaissance n'est véritablement connaissance que si elle est prise dans l'action et elle n'a alors de sens que pour ceux qui la produisent et pour ceux qui l'utilisent ».

Dans cette approche, le modèle de gestion des connaissances de Grundstein met en valeur la connaissance cruciale de l'entreprise à travers quatre mécanismes génériques de création de connaissance et d'apprentissage organisationnel, identifiés par des verbes à l'infinitif, à savoir : repérer, préserver, valoriser, transférer et partager les connaissances cruciales de l'entreprise. Dans ce modèle nous constatons que la gestion des connaissances est décrite par rapport à l'aspect social (l'organisation, la stratégie, et le capital humain), tandis que l'aspect technique est négligé.

---

[82] Ce modèle est une mise en place de notre recherche, car ces auteurs n'utilisent pas le terme gestion des connaissances. De même que pour le modèle de gestion des connaissances de Nonaka et Takeuchi.

En revanche, le modèle de gestion des connaissances d'Ermine met en valeur la connaissance métier à travers le système de connaissances (modèle de la marguerite) et le système opérationnel (modèle OIDC) à travers cinq mécanismes génériques de création de connaissance et d'apprentissage organisationnel, identifiés par des verbes à l'infinitif, à savoir : capitaliser, partager, créer, évaluer et faire-évoluer les connaissances[83]. Dans ce modèle nous constatons que la gestion des connaissances est décrite par rapport à l'aspect social (domaines, activités), et l'aspect technique (concepts, tâches)[84].

Dans ce contexte, notre hypothèse est que l'approche managériale de la gestion des connaissances met l'accent, en plus du facteur humain, sur l'importance du facteur organisationnel et stratégique, (et ceci est vrai à tous les niveaux de l'entreprise, stratégique, tactique et opérationnel), au travers d'un réseau des savoirs. Autrement dit, dans l'approche managériale, la connaissance est perçue selon trois aspects qu'il faut considérer ensemble (1) l'aspect organisationnel qui indique que la connaissance est au service de l'entreprise à travers l'apprentissage, l'innovation et la culture d'entreprise de travail en réseau (communication, coordination, coopération) ; (2) l'aspect stratégique qui indique que la connaissance est une ressource utilisée pour atteindre un résultat stratégique, tactique ou opérationnel à travers le principe *l'union fait la force* ; et enfin (3) l'aspect humain qui indique que la connaissance est un patrimoine de l'individu, du groupe, et de l'entreprise à travers le dialogue ;

- enfin, pour l'approche NTIC du KM, la réponse peut être résumée comme suit : la connaissance est perçue par la composante technique, comme un objet de gestion, organisé à travers l'information, et structuré à travers des données.

Dans la suite nous reviendrons au fur et à mesure sur les penseurs qui ont impliqué la gestion des connaissances au cœur d'un système d'organisation du travail coopératif par la mise en valeur de méthodes de travail ou modes de management. Comme l'a si bien dit Jean-Louis Ermine « la productivité n'est plus seulement dans la force du travail et dans les outils de la production, mais passe désormais par la connaissance ».

---

[83] Les objectifs d'évaluer et de faire-évoluer le patrimoine des connaissances nous les avons repéré ailleurs.
[84] Ici l'interaction du système (connaissance et opérationnel) avec l'environnement permet la "sélection d'information".

## 1.5. Les origines de la connaissance industrielle[85]

Notre ambition dans cette section est de justifier l'importance de la création des connaissances à travers les différentes époques de notre histoire industrielle.

Regardons, avec un premier exemple la dynamique circulante de faire-émerger de la connaissance et l'apprentissage à travers sa représentation matérielle, c'est-à-dire suivant le point de vue social et technique de la connaissance. Selon les ethnologues des civilisations précolombiennes ou autres civilisations reposant sur la tradition orale, la sagesse d'un peuple était concentré dans les « anciens ». Cette sagesse était basée, principalement, sur le savoir et l'expérience des anciens. Ainsi les anciens, porteurs de la sagesse, distribuaient la connaissance aux plus jeunes, pour les conseiller et éviter de commettre les mêmes erreurs commises par eux dans le passé. Cela signifiait aussi que lorsqu'un ancien disparaît, c'était la mémoire d'un peuple qui disparaissait aussi. A cet égard Maturana et Varela disent « la tradition est une manière de voir et d'agir. Toute tradition se base sur ce qu'une histoire structurelle a accumulé comme évident, comme régulier, comme stable, …, mais aussi pour le non dit, l'interdit », c'est-à-dire que la tradition est aussi un lieu où habite l'histoire cachée.

Cette observation nous permet de justifier, d'une part, la dynamique circulante de faire-émerger la connaissance, et d'autre part, la représentation matérielle de la connaissance. Dans le premier cas, la connaissance est un patrimoine propre à la personne, ce patrimoine de connaissance est stocké dans une structure mentale et alimenté par le cycle de vie de la personne et de son action dans le monde. Or, cette distribution du savoir et de l'expérience, nous ne la voyons pas comme une volonté individuelle (la sagesse est à moi seul), mais comme une nécessité collective et propre de la nature humaine de survie dans un environnement en constante évolution. C'est justement dans cette survie en collectivité que la signification et le sens commun sont nécessaires pour communiquer.

Nous voyons la connaissance alors comme un outil pour organiser la vie collective en particulier le travail. En effet, la connaissance est là pour faire quelque chose. A cet égard, Maturana et Varela disent « la vie humaine émerge dans un univers de significations (perceptuelles), inter-agir socialement dans des domaines de significations mutuellement complémentaires, est une condition existentielle et reste la base de la communication ». Nous pensons que ceci a sens, seulement, dans une vie en communauté, c'est-à-dire organisée autour de la connaissance. Dans ce scénario, nous pensons, que l'argumentation de Maturana et Varela « le savoir est faire et le faire est savoir » prend de sens pour nous et plus particulièrement dans un système d'organisation du travail coopératif.

---

[85] D'après nous, ce détour fait partie d'une thèse en génie industriel.

Pour justifier cela, regardons, avec un autre exemple, la dynamique circulante de faire-émerger la connaissance et sa représentation matérielle. Dans le domaine industriel la substitution du travail à la main (énergie musculaire) par la machine (énergie), a été possible par la mise en service dans l'industrie à partir de la seconde moitié du XVIIIème siècle de la machine à vapeur de Watt. Or, la machine à vapeur de Watt n'est autre chose qu'une amélioration de ce qui déjà existait á l'époque, c'est-à-dire la machine de Savery et la machine de Newcomen[86]. En effet, tous les deux ont eu un problème en commun qui était le rendement et la continuité du mouvement propre d'une machine à vapeur « à simple effet ». L'intervention de Watt a été la mise au point d'une machine « à double effet » et a permis d'incorporer dans cette machine d'autres éléments tels que le volant, le régulateur à boules, et autres améliorations. La machine de Watt émerge par la transformation de ces éléments et l'intégration de nouveaux qui ont été inventé par lui (ses idées). La « circulation de la connaissance » prend fin en 1769 quand Watt dépose un brevet de son invention. Alors, la machine à vapeur de Watt devient un savoir-faire collectif qui pousse la réflexion humaine sur une autre échelle de la connaissance et du savoir, d'une part, dans sa fabrication en masse pour l'utilisation dans l'industrie, et d'autre part, sur une autre source d'énergie, la vapeur. Par la suite, la l'énergie électrique, a permis l'invention du moteur électrique (transformation de l'énergie électrique en force motrice), et d'autre part son utilisation dans l'industrie. Elle a bouleversé encore plus l'organisation du travail.

Cet bouleversement du travail est connu sous le nom de *Révolution Industrielle*, et est caractérisé par (1) le passage du « domestic system » au « factory system » ; et (2) l'incorporation de la vapeur dans l'industrie se substituant progressivement aux autres sources d'énergie existantes à ce moment, telles que l'énergie musculaire, animale, éolienne et hydraulique.

---

[86] Ces deux machines sont autant d'améliorations d'inventions, comme l'on dit dans le site *histoire de la machine à vapeur* « depuis l'Antiquité les hommes cherchèrent à utiliser les propriétés physiques de la vapeur d'eau. Héron d'Alexandrie inventa une machine appelée éolipyle (porte d'Eole). Cette machine était une chaudière hermétique remplie en partie d'eau que l'on plaçait sur un feu. De cette chaudière sortait un tube creux relié à une sphère pouvant tourner autour d'un axe horizontal. De cette sphère deux autres tubes perpendiculaires à l'axe laissaient sortir la vapeur qui par propulsion faisait tourner la sphère ». http://visite.artsetmetiers.free.fr/histoire_vapeur.html.

### 1.5.1. La production des connaissances comme un levier de productivité

> ➤ **Le système d'organisation du travail à l'époque de Taylor**

Dans la société industrielle, les premières réflexions et publications sur l'application de la connaissance comme un levier pour améliorer la productivité[87] des ouvriers due à l'expansion du marché et à la volonté de l'industrie de produire plus, datent, d'une part, de l'époque du taylorisme, et d'autre part de l'époque du fordisme. Dans cette époque, de nouveaux systèmes d'organisation du travail se sont mis en place.

Pour nous et d'autres auteurs que nous citons plus bas, Frederick Taylor[88] reste un humaniste, car il a essayé de comprendre l'homme, son macroscope pour l'observer a été le *travail*. Pour lui le travail est quelque chose qu'il faut organiser de façon systématique, autrement dit, le travail est un phénomène que l'on peut étudier de façon scientifique (mesurer les temps, déterminer les conditions efficientes d'exécution des tâches, etc.).

Ainsi, pour augmenter la productivité des ouvriers, les tâches doivent d'abord être conçues sous forme efficiente pour (1) éviter le gaspillage de matières primaires ; (2) éviter la perte du temps et les mouvements inutiles et fatigants des ouvriers pendant la journée de travail, pour ensuite (3) être exécutés selon "la méthode" conçue. Taylor a mis en place ainsi une standardisation des *façons de travailler* des ouvriers, due principalement à la séparation (division du travail) entre la conception et l'exécution des tâches, qui avec le temps est devenu le premièr système d'organisation du travail. Pour voir le rôle de la connaissance nous allons prendre un cas d'application de la méthode de Taylor.

En 1899 à l'usine BethIéem Steel Co., Taylor a étudié les temps, les mouvements et la structure des outils de production utilisés pour décharger des wagons de minerais de cette usine. D'après cette étude, il a constaté que (1) les méthodes pour faire ce travail étaient fort différentes entre les ouvriers ; (2) les temps de repos et les permis pour aller satisfaire leurs nécessités biologiques n'étaient sont pas contrôlés ; (3) les outils utilisés pour faire leur travail étaient des pelles de poids et de largeurs différentes. En observant les conditions d'exécution de la tâche de chaque ouvrier, il établit la valeur du temps et les mouvements les plus appropriés pour l'accomplir, ainsi que la forme et le poids de la pelle idéale. Ainsi, 140 ouvriers avec la même méthode de travail (temps, mouvements, etc.), et les mêmes largueur et poids des pelles (10 kg chacune) arrivaient à décharger chacun 59 tonnes par jour, alors que 600 ouvriers avec des pelles de largeurs et poids différents, chacun avec ses propres mouvements, qui dans pluspart des cas étaient des gestes inutiles, déchargeaient auparavant chacun 10 tonnes par jour [Bravo, 04].

---

[87] Rapport entre la production et le temps de travail consacré à l'obtenir.
[88] 1856 – 1915. Taylor est le fondateur de la théorie du management scientifique par l'apparition de son ouvrage *The Principles of Scientific Management*, publié en 1911. Il est connu aussi comme le père du génie industriel et de l'ergonomie, car il a été le premier homme à observer l'homme dans son poste de travail comme facteur clé pour améliorer la productivité.

Nous constatons que cette méthode de travail a été déterminée (1) par le savoir-faire individuel, autrement dit par les mouvements pour manoeuvrer la pelle des ouvriers les plus qualifiés ; (2) par le temps optimal de certains ouvriers ; et (3) par les formes des outils de production les plus appropriés pour accomplir la tâche (c'est-à-dire ce qui fatiguaient moins les ouvriers). Pour sa mise en œuvre, cette méthode de travail doit être (1) conceptualisée dans un savoir-faire collectif (les meilleurs mouvements et temps, et la pelle la plus performante); et (2) matérialisée dans un support physique (un schéma, une graphique, des règles, des formules, etc.), pour être enseignable et applicable à l'ensemble des ouvriers.

La période de formation permet la sélection des ouvriers les plus qualifiés (ceux qui ont les compétences) pour faire le travail, selon la méthode. L'efficacité de la méthode de Taylor est garantie par l'obtention d'une productivité maximale. La résistance au changement, due à la modification de la façon de travailler a été surmontée par une prime à la production. Pour Taylor la productivité des ouvriers ne s'améliore pas par le savoir-faire, l'expérience et la communication entre eux dans leurs postes de travail, mais par un système d'organisation du travail composé (1) d'une méthode (scientifique) de travail qui préconise une forte division du travail entre la conception et l'exécution des tâches (c'est-à-dire que les tâches sont définies et exécutés indépendant de talents et expertises individuelles des ouvriers) ; et (2) d'un mode de management (le *functional foremanship*) qui assure la supervision des tâches à partir de l'application et formation à ce nouveau méthode de travail.

Selon Taylor la seule motivation des ouvriers pour augmenter la productivité à l'usine est son salaire, ceci a donné naissance à la critique de Mayo comme nous verrons plus loin.

Revenons, à l'hypothèse de la méthode de Taylor qui établit la séparation entre la conception et l'exécution des tâches, pour justifier le rôle de connaissance dans l'élaboration de cette méthode. D'après l'exemple de l'usine BethIéem Steel Co. (voir plus haut) nous constatons que la connaissance se trouve dans la conception et l'exécution des tâches. Dans la conception, la connaissance est assimilée d'abord au savoir-faire individuel et puis au savoir-faire collectif. Le savoir-faire individuel se trouve dans la tête de chaque ouvrier, et varie selon l'expérience de l'ouvrier avec l'outil (de production), tandis que le savoir-faire collectif existe seulement dans un support physique. Nous allons commenter ceci à l'aide de certaines réflexions.

Pour Wilkins, « Taylor ne fait pas confiance au jugement individuel de l'ouvrier (qualifié). Il pense que celui-ci garde pour lui ses connaissances, qu'il ne tient pas à en faire profiter les autres, qu'il veut préserver ses "secrets professionnels", qu'il est fainéant et nécessite une surveillance » [Wilkins, 90]. Dans cette affirmation, la connaissance est associée au savoir-faire individuel, en plus Wilkins laisse comprendre que son jugement individuel ou savoir-faire individuel a une sorte de pouvoir différentiateur, qu'il convient de garder pour soi même et de ne pas le partager avec les autres.

Tandis que pour Nonaka et Takeuchi, Taylor « prescrit des méthodes et procédures "scientifiques" pour organiser et réaliser le travail. Les plus importantes furent l'étude des temps et des mouvements afin de trouver la "meilleure méthode" pour réaliser une tâche. Le "management scientifique" constituait une tentative de formalisation de l'expérience des travailleurs et des aptitudes tacites en connaissances objectives et scientifiques »[89]. Dans cette argumentation, la connaissance est associée, d'une part, à l'expérience et les aptitudes tacites, c'est-à-dire au savoir-faire individuel, et d'autre part, au savoir-faire collectif comme une sorte de connaissance objective ou scientifique. Nous voulons attirer l'attention ici que l'aspect "scientifique" du nom de la méthode de Taylor (le management scientifique) est relative à la séparation entre la conception et l'exécution des tâches plutôt qu'à l'émergence d'une nouvelle connaissance ou savoir scientifique. En effet, dans le cas de l'usine BethIéem Steel Co. (voir plus haut) nous ne voyons aucune loi scientifique à part la loi de la gravité.

---

[89] Nonaka et Takeuchi utilisent aussi les termes "connaissance tacite" pour indiquer les aptitudes tacites, et "connaissance explicite" pour indiquer la connaissance objective, nous allons revenir sur ce sujet, mais d'ors et déjà pour bien comprendre l'exemple de Taylor, la connaissance tacite est associée au savoir-faire individuel, et la connaissance explicite au savoir-faire collectif.

---------------------------------------------------------------------------------------------------

Pour nous l'apparition de la connaissance scientifique comme un facteur clé pour organiser le travail[90], voie le jour avec le fordisme, étant donné que le savoir scientifique doit être présent pour développer les outils pour le travail à la chaîne, le fordisme nous le présentons plus loin.

Or, pour Taylor le savoir-faire collectif est une source de productivité qu'il faut standardiser, et non pas le savoir-faire individuel. A cet égard, Nonaka et Takeuchi, affirment « il ne parvint pas à percevoir les expériences et les jugements des travailleurs comme étant des sources de nouvelles connaissances ». C'est vrai, la préoccupation de Taylor n'était pas la création de connaissances à partir de la connaissance individuelle (expériences, jugements, etc.), mais plutôt de l'augmentation de la productivité à partir de la modification du travail par une nouvelle méthode de travail. Là encore, Nonaka et Takeuchi, en citant Taylor, disent « la création de nouvelles méthodes de travail devint la responsabilité des seuls managers. Le managers furent chargés de classer et de réduire les connaissances à des règles et formules et de les appliquer au travail quotidien ».

Bref, nous pouvons dire que le facteur clé du taylorisme comme système d'organisation du travail afin de garantir l'augmentation de la production de l'usine pour satisfaire la demande du marché est, d'une part, la gestion des connaissances, caractérisée par la conversion du savoir-faire individuel au savoir-faire collectif (la "meilleure méthode"), et d'autre part, la gestion des compétences, caractérisée par la formation à la nouvelle méthode de travail. Ainsi, l'importance de la connaissance dans la méthode de Taylor se trouve d'une part, dans la conception de la tâche (en effet, la gestion des connaissances permet le passage du savoir-faire individuel au savoir-faire collectif), et d'autre part, dans l'exécution de la tâche (en effet, la gestion des compétences permet la mobilisation de l'ancien régime au nouveau régime). Cette modification de la façon de travailler, est assurée par une période de formation à la méthode de Taylor, et surtout par la volonté et la capacité des ouvriers et des managers (le personnel à charge de la conception et de l'exécution des tâches) de le faire.

Comme nous l'avons vu plus haut, pour Taylor le savoir-faire collectif (la connaissance explicite, en termes de Nonaka et Takeuchi) est une source de productivité qu'il faut standardiser, et non pas le savoir-faire individuel (la connaissance tacite, en termes de Nonaka et Takeuchi).

---

[90] Mais aussi, pour nous comme une "troisième révolution industrielle", les deux autres son relatives à l'incorporation de la machine à vapeur dans l'industrie, et à l'homme vivant en communauté autour de la sagesse comme un facteur d'organisation du travail, comme nous l'avons montré plus haut dans la section *Les origines de la connaissance industrielle*.

Avant de passer au deuxième système d'organisation du travail proposé par Ford, il nous parait intéressant, pour la suite de cette thèse, de laisser une trace sur le caractère humaniste de Taylor. A cet égard, Drucker dit « Taylor a été le premier homme dans l'histoire écrite qui a considéré le travail comme un sujet méritant d'être observé et étudié systématiquement. Après tout, de l'administration scientifique de Taylor repose l'énorme mouvement de bien-être des 65 dernières années qui a élevé les travailleurs des pays développés très au-dessus de tout niveau atteint précédemment ». Tandis, que pour Nonaka et Takeuchi « il convient de noter que Taylor portait un intérêt humaniste à l'attribution de salaires convenables et au développement productif. Dans la pratique cependant, les techniques que lui et ses suiveurs développèrent pour accroître la productivité du travail furent souvent mal utilisées ce qui entraîna des effets déshumanisants pour les travailleurs » [Nonaka et Takeuchi, 97]. Enfin, pour notre ami et collègue Juan Bravo[91], dans son ouvrage récent consacré à la vie de Taylor et au taylorisme comme système d'organisation du travail, passé, présent et futur, appelé justement *Frederick Winslow Taylor*, il dit que « Taylor avait toujours une manière personnalisée de traiter les travailleurs et veillait pour leurs conditions de vie, il voulait un partage juste des fruits de l'amélioration de la productivité, entre les chefs d'entreprise, les travailleurs et la société » [Bravo, 04]. Puis il dit aussi, « pourquoi le discours chez Taylor est tellement actuel ? Parce que son discours est orienté vers le dépassement de la pauvreté et parce que ses propositions, dûment mises à jour, pourraient produire de grands bénéfices dans l'économie de l'Amérique Latine » [Bravo, 04].

Ces observations chez Drucker, Nonaka, Takeuchi, et Bravo sur le personnage de Taylor, nous font sentir, que les critiques du taylorisme que l'on trouve dans la littérature ne sont pas fondées, puisque chez Taylor il y avait toujours un principe d'égalité face à la répartition du surplus de la productivité de l'entreprise.

Pour nous, il ne faut pas brûler Taylor[92], mais plutôt les applications déshumanisantes de sa méthode (*The Principles of Scientific Management*), et ses évolutions (par exemple, le taylorisme assisté par ordinateur), qu'en font certains technocrates de la répartition du travail.

---

[91] La contribution de Bravo à la formation des cadres au Chili et en Amérique Latine est digne de mentionner dans cette thèse. La liste des ouvrages de Juan Bravo peut être consulté dans http://www.evolucion.cl

[92] Dans le site web http://membres.lycos.fr/hconline/taylor.htm l'on pose la question *faut-il brûler Taylor ?*

➢ **Le système d'organisation du travail à l'époque de Ford**

L'autre système d'organisation du travail a été le travail à la chaîne, développé en 1909, par Henry Ford[93] dans son usine d'automobiles de Détroit[94]. Dans ce type d'organisation du travail nous avons d'une part, la segmentation des tâches conçues (gestion des connaissances) et exécutées (gestion des compétences) selon le taylorisme, et d'autre part, le déplacement mécanisé de l'objet travaillé d'un ouvrier à l'autre.

Nous pensons que c'est justement dans "le déplacement mécanisé de l'objet travaillé" qu'émerge le savoir ou la connaissance de type scientifique ou technologique. En effet, ce type de connaissance ou savoir permet de gérer la répartition des tâches manuelles (travail à la main) et automatisées (travail à la machine).

Dans ces deux types de systèmes d'organisation du travail (le tayloriste et fordisme), la connaissance joue un rôle primordial pour accroître la productivité des ouvriers. En effet, dans le taylorisme l'organisation du travail se fait par la standardisation du savoir-faire individuel en savoir-faire collectif[95].

Or, si nous employons le langage de Nonaka et Takeuchi, nous pouvons dire que la connaissance qui permet l'augmentation de la productivité est du type connaissance explicite et non tacite. Nous pensons, que la raison pour laquelle Taylor n'a pas utilisé la connaissance tacite comme levier de la productivité est purement historique. En effet, à cette époque de l'histoire économique et industrielle, l'offre attirait la demande (et non l'inverse) et pour cette raison l'industrie devait produire. On retrouve une même situation à l'époque du fordisme où la différenciation des produits n'était pas une argumentation nécessaire pour vendre. Cet effet de marché a encore plus été perçu avec la rentrée des Etats-Unis dans la première guerre mondiale.

Ainsi, les réflexions de Taylor et Ford autour de l'organisation du travail, ont été centrées, d'une part, sur la gestion des connaissances plus explicites que tacites, mais surtout, d'autre part, dans la gestion des compétences collectives plutôt qu'individuelles, pour augmenter la productivité de l'usine afin de satisfaire la demande forte du marché.

---

[93] 1863 – 1947.
[94] Depuis là, le fordisme a été répandu aux Etats-Unis et en Europe, dans d'autres domaines industriels. En France le fordisme a été amélioré par Louis Renault (1912) et adopté par les usines de fabrication de matériel de guerre entre 1914 et 1918.
[95] L'on passe de l'empirisme à la standardisation de méthodes et outils de production.

En fait, l'idée de présenter la connaissance tacite comme un levier de productivité dans l'industrie a eu lieu à l'époque d'Elton Mayo.

> ➢ **Le système d'organisation du travail à l'époque de Mayo**

A partir de l'ouvrage de Taylor (*The Principles of Scientific Management*, publié en 1911) et ses applications dans le monde industriel, Mayo a développé en 1933 une alternative à la théorie du management scientifique de Taylor, dans un livre intitulé *The Human Problems of an Industrial Civilization*, la *théorie des relations humaines*. Cette théorie suggérait que « les facteurs humains jouaient un rôle significatif dans l'augmentation de la productivité par l'amélioration des connaissances pratiques détenues par les travailleurs des ateliers ». Mayo a observé que la motivation des ouvriers et managers (le personnel à charge de la conception et de l'exécution des tâches) pour s'impliquer dans le système d'organisation du travail de Taylor, était garantie par un système de primes de rendement à partir de l'objectif de productivité de la journée de travail. Pour lui, penser que la seule motivation des ouvriers et managers pour faire le travail était le salaire et la prime de rendement, était une erreur. Il nie alors le principe de Taylor appelé *economic man* ou "hommes économiques".

De plus, il a observé l'importance de la communication dans les relations de travail. A cet égard, Mayo argumente : « les managers devraient développer des "talents humaines sociaux" pour faciliter la communication interpersonnelle au sein des groupes formels et informels de l'organisation du travail ».

A la différence de Taylor, Mayo ne dit pas comment il faut appliquer cette « théorie des relations humaines » pour augmenter la productivité de l'usine, cette théorie émerge simplement de la critique du principe *economic man* de la méthode de Taylor. Néanmoins, ce qu'il faut retenir chez Mayo est l'implication du savoir-faire dans l'organisation du travail, qui va prendre de l'importance quand il s'agira de l'obtention d'avantages compétitifs plutôt que de productivité.

## 1.5.2. La production des connaissances comme un levier d'avantage concurrentiel ou compétitif

L'idée ici, est de se focaliser sur la connaissance en tant que source de compétitivité dans l'industrie. Pour cela nous faisons appel aux réflexions de Drucker et Senge, sur le plan théorique, et de Nonaka et Takeuchi, sur le plan pratique. Pour introduire le sujet, regardons un article récent du

---

site web de Microsoft France, titré *Le Knowledge Management, composante de l'entreprise numérique*[96], où l'on peut lire que « dans le monde de l'entreprise, les premières réflexions et les premiers travaux sur les connaissances professionnelles datent de 1939. Peter Drucker, le "pape" américain incontesté de cette discipline invente cette expression en 1959. D'après lui, seule la connaissance permet la productivité (en l'appliquant à des tâches que nous savons faire) et l'innovation (en l'appliquant à des tâches nouvelles et différentes). L'objectif est donc finalement d'évaluer, de structurer et de développer le capital intellectuel de l'entreprise ». Dans ce commentaire de Drucker, que nous partageons, il y a deux termes qu'il convient de clarifier avant de poursuivre dans la production des connaissances comme levier d'avantage compétitif.

Le premier terme est la "connaissance professionnelle" qui dérive du terme "société de la connaissance" (*the knowledge society*), forgé par Nonaka et Takeuchi , Peter Drucker, Alvin Toffler, James Brian Quinn et Robert Reich. Ils annoncent, chacun à leur manière, l'avènement d'une nouvelle économie ou société, appelée "société de la connaissance" et comme l'affirme Drucker dans son dernier livre[97] « la connaissance n'est pas seulement une nouvelle ressource qui s'ajoute aux facteurs de production traditionnels – travail, capital, terre – mais la seule ressource qui ait une signification réelle aujourd'hui ». Autrement dit, dans la société post capitaliste d'aujourd'hui (au moins pour les pays du G8), les ressources de base pour survivre ne sont plus ni le travail, ni le capital, ni la terre (comme l'ont été au moyen age certains pays gouvernés par un régime féodal, ou après la deuxième guerre mondiale sous un régime communiste ou capitaliste), mais la connaissance. Or, pour cultiver cette connaissance, il faut la travailler et pour cela, il faut des travailleurs, d'où les autres termes si chèrs à Drucker "travail de connaissance" (*knowledge work*) et "travailleur de connaissance" (*knowledge worker*).

Nonaka et Takeuchi ont constaté que dans le langage de Drucker deux termes apparaissaient fortement : la connaissance et la productivité.

Le premier terme, la connaissance, permet d'augmenter la productivité des travailleurs de la connaissance, et sur laquelle est fondé la société de la connaissance est assimilée à une habilité ou savoir-faire (la connaissance « ne peut pas être expliquée sous forme de mots parlés ou écrits, elle peut seulement être montrée », puis il a ajouté « le seul moyen d'apprendre une *techné* est par l'apprentissage et l'expérience »). Et donc finalement, pour Nonaka et Takeuchi, « Drucker semble avoir reconnu l'importance de la connaissance tacite » en tant que source de compétitivité. La connaissance tacite peut être aussi associée au concept d'actif intangible. A cet égard, Nonaka et Takeuchi argumentent à propos des idées de Quinn, Drucker et Toffler que « le pouvoir économique et de production d'une entreprise moderne tient plus dans ses capacités intellectuelles et de service que dans ses actifs matériels ; terrains, usines et équipements ». Il franchit un pas de plus en mettant en exergue le fait que la valeur de la plupart des produits et des services dépend fondamentalement de la façon dont sont développés les actifs intangibles basés sur les connaissances ; tels que le savoir-faire technologique, le design des produits, la présentation marketing, la compréhension des clients, la créativité personnelle et l'innovation.

---

[96] Article apparu le 20/09/01 dans le site web de Microsoft France
http://www.microsoft.com/france/entrepreneur/Solutions/LeReseauInformatique/LeKnowledgeManagement.mspx
[97] Nonaka et Takeuchi se réfèrent au livre *Post-capitalist Society*, publié en 1992, et 1993 (pour la deuxième édition).

Le deuxième terme est "productivité". C'est vrai que depuis le premier ouvrage[98] de Drucker, intitulé *The End of Economic Man*, publié en 1939, on a valorisé l'importance de la connaissance comme un facteur clé dans un système d'organisation du travail, tel que l'ont fait Taylor et Mayo à leurs époques. Mais pour Drucker, il s'agit d'un facteur clé pour accroître la productivité des travailleurs dans une économie de service, et non pas la productivité de l'usine dans une économie de production. Plus précisément, si, pour Taylor la connaissance explicite est une source de productivité qu'il faut standardiser, pour Mayo c'est la connaissance tacite qu'il ne faut pas laisser de côté dans la définition du système d'organisation du travail, et pour Drucker, il y a un carrément un changement d'horizon, puisque dans le domaine industriel on est passé d'une économie de production à une économie de service (les produits doivent, en effet, être différencié pour satisfaire le besoins des clients), et donc finalement l'innovation et la compétitivité dans une économie de service ont pris le relais de la productivité dans une économie de production. Donc, la productivité, aujourd'hui ne peux pas être vue comme un problème (au niveau de la productivité du travail à la chaîne d'une usine, par exemple), mais plutôt un besoin au niveau de la productivité du travail de connaissance. La connaissance devient alors un levier de compétitivité. D'où l'importance de la standardisation du processus de création de connaissances et apprentissage organisationnel à travers de méthodes de travail et modes de management qui vont constituer en définitif le système d'organisation du travail coopératif de l'entreprise.

Pour Drucker le nouveau système d'organisation du travail doit être axé, au niveau de la méthode de travail, sur « (1) l'amélioration continue de chaque activité ; (2) le développement de nouvelles applications de ses propres succès ; (3) l'innovation continue en tant que processus organisé ». Au niveau du mode de management qu'il convient de mettre en place, Drucker a mis l'accent « sur le fait qu'une organisation doit augmenter la productivité des travailleurs au niveau de la connaissances et des services pour relever les nouveaux défis qui se présentent à elle ». Il justifie encore en disant que « ce défi, qui deviendra la préoccupation centrale en management durant les prochaines décennies, déterminera de façon ultime la performance compétitive des entreprises et, ce qui est encore plus important, déterminera le vrai façonnement de la société et la qualité de vie dans chaque nation industrialisée ».

---

[98] Il s'agit de son premier livre en anglais, car les deux antérieurs sont en allemand.

> **Le système d'organisation du travail à l'époque de Simon**

En 1945, Herbert Simon, a exploré plus en détails le rôle décideur du manager, distingué par Barnard comme l'une des fonctions du manager, dans son livre intitulé *Administrative Behavior*. Plus tard, avec la collaboration de March, ils ont développé toute une théorie sur la cognition, en supposant que les êtres humains ont une capacité limitée à traiter l'information dans une période de temps donné (en particulier les managers dans les processus de prise de décision). A partir de cette hypothèse de rationalité limitée, Simon et March affirment que l'entreprise peut être conçue comme une machine de traitement de l'information. De cette affirmation, beaucoup d'approches sur la conceptualisation de systèmes d'information dans l'entreprise sont apparues, comme le modèle OID de Le Moigne, et la société de l'information de Prusak et Davenport. Cette vision de l'information, comme une ressource organisationnelle (le système d'information) est intégrée dans un support technologique (le système informatique) a mis au point petit à petit des méthodes de travail et des modes de management de l'information, dans lesquelles le système d'organisation du travail coopératif est organisé autour de l'information Un exemple, de nos jours, d'un système d'organisation de travail coopératif focalisé sur l'information est l'ERP ou e-ERP (selon la plateforme client/serveur ou intranet que l'on utilise).

> **Le système d'organisation du travail à l'époque de Penrose**

Édith Penrose dans son livre intitulé *The Theory of the Growth of the Firm* (publié en 1959) déclare que « la croissance de la firme est à la fois encouragée et limitée par le processus véritablement dynamique et interactif qui apparaît lorsque le management recherche le meilleur usage possible des ressources disponibles ». L'hypothèse de Penrose montre que dans cette période de l'histoire économique américaine, le bon usage des ressources a été un facteur clé de croissance économique pour les entreprises, étant donné que d'une part, les ressources (productives, humaines et matérielles) sont limitées dans le temps et l'espace, et d'autre part, leur capacité de gestion est aussi limitée. Pour Penrose, les ressources sont au service des objectifs de l'entreprise, et donc il faut bien les gérer. Dans la pensée de Penrose, le service symbolise un mode d'action ou une capacité de l'entreprise pour mobiliser les ressources dans le but d'atteindre les objectifs. Une ressource sera considérée alors plus stratégique qu'une autre si elle est porteuse d'une avantage concurrentiel ou compétitif. D'ailleurs, ceci a été à la base de l'émergence de l'analyse stratégique de ressources, dans lequel les deux outils d'analyse stratégique les plus répandus sont les deux modèles de Porter. L'un est le modèle des "cinq forces" qui permet l'analyse d'avantages concurrentiels dans l'industrie (concurrents, clients, entrants, substituts, fournisseurs) présenté dans son livre intitulé *Competive Strategy* (publié en 1980), et l'autre, est le modèle de la "chaîne de valeur" qui permet l'analyse d'avantages compétitifs dans l'entreprise (ressources, capacités, compétences), présenté dans son livre intitulé *Competive Advantage* (publié en 1985). Dans un paradigme d'avantage concurrentiel ou compétitif, la ressource la plus stratégique est alors la connaissance, basée sur des savoirs et savoir-faire organisés au service d'objectifs. Ainsi, pour Penrose « la firme est un répertoire de connaissances ». Ceci a été enrichi par Tarondeau, en citant l'article de Grant *The Ressource-Based Theory of Competitive Advantage* (publié en 1991) « alors que les ressources sont la source des capacités d'une entreprise, ses capacités sont la source principale de son avantage concurrentiel ».

Dans cette logique, nous pouvons concevoir l'entreprise comme (1) un portefeuille de ressources qu'elle détient ; (2) un répertoire de connaissances et compétences qu'elle développe et déploie ; et (3) sa capacité de gestion pour gérer ses connaissances et ses compétences.

### 1.5.3. La production des connaissances comme un levier d'avantage coopératif

Nous avons vu à travers le modèle de Nonaka et Takeuchi que la création de connaissances nouvelles collectives et apprentissage organisationnel est possible, d'une part, dans une dimension épistémologique de la connaissance représentée par la dualité sujet/objet : le savoir et savoir-faire, et d'autre part, dans la dimension ontologique de la connaissance représentée par la dualité moi/autres : la relation avec les autres.

Nous avons vu aussi l'importance de la production des connaissances comme un levier de productivité dans l'entreprise, d'abord pour améliorer la productivité des ouvriers dans une économie de production (1930) caractérisée par l'émergence du taylorisme et fordisme, puis comme un levier d'avantage concurrentiel et compétitif afin d'améliorer la productivité des travailleurs de la connaissance dans (1) une économie de service (1960) caractérisée par l'émergence de l'analyse stratégique de ressources dans une société de l'information ; et (2) une économie globalisée (1990) caractérisée par l'émergence des nouvelles technologies de l'information et de la communication (NTIC) supportées à travers l'internet/intranet/externet avec le paradigme de l'entreprise élargie[99], l'outsourcing, l'e-entreprise ou l'e-business.

Un résultat évident de cette évolution du monde industriel est l'accumulation de données, d'informations, de connaissances et de compétences, que les entreprises doivent aujourd'hui gérer, mais non pas dans un esprit de compétition par la recherche d'avantage concurrentiel ou compétitif, mais plutôt dans un esprit de collaboration guidé par la recherche d'avantage coopératif ou "coopétitif", en employant le terme de Prax, qui ouvre l'ère de "l'être économique nouveau", selon le terme d'Hervé Sérieyx utilisé par Prax. A cet égard, nous percevons le terme "compétitif" comme le contraire de concurrentiel lorsqu'il dit que « l'entreprise a appris à collaborer avec tout son environnement professionnel : clients, fournisseurs, distributeurs, et même concurrents. La collaboration devient alors sa principale force, bien plus que la compétition : on parle alors d'*avantage coopératifs* ».

Pour Gérard Balantzian dans son livre *Après l'avantage concurrentiel, l'avantage coopératif, Le partenariat, la coopération, l'alliance stratégique* (publié en 1997) « le modèle d'organisation taylorienne fondée sur l'unité de temps, de lieu et d'action laisse progressivement la place au travail coopératif et à l'emploi juste-à-temps ». Il déclare également que « nous sommes entrés dans un type de société de la matière grise et du savoir déjà largement évoqué par des auteurs comme Alvin Toffler ou Peter Drucker. Le savoir et l'art de l'influence sont les nouveaux pouvoirs. Le XXe siècle fut le siècle de l'explosion, du développement et de l'épanouissement de la société industrielle. Le XXIe siècle sera celui de l'immatériel où les biens "invisibles" seront plus appréciés qu'ils ne le furent durant le siècle précédent ». Pour lui l'avantage coopératif est « un avantage décisif en s'appuyant sur le partenariat interne et externe. Cette coopération où l'on peut être allié mais différent (voir concurrent) permet des formes de rapprochements d'objectifs, d'activités et de ressources permettant de prendre une longueur d'avance. Chacun des partenaire peut garder son autonomie et son identité mais s'associer à l'autre autour d'une vision partagée du progrès ».

---

[99] Selon Prax « l'idée de "l'entreprise élargie" est d'améliorer le fonctionnement d'un réseau professionnel par une transparence totale d'information, de savoirs et de savoir-faire entre les différents acteurs de la chaîne de conception-production … même s'ils sont concurrents ».

Cela signifie, qu'aujourd'hui les enjeux pour les entreprises sont autour de la maîtrise de cet avantage coopératif, comme dans le passé récent il était autour de l'avantage concurrentiel ou compétitif durable. La performance passe désormais par la gestion des connaissances autour du *business* de l'entreprise, c'est-à-dire dans la capacité qu'a l'entreprise d'intégrer ce qu'elle veut faire (sa stratégie), ce qu'elle peut faire (ses ressources) et ce qu'elle sait faire (ses compétences). Maintenant la performance passe avec les autres, mais en gardant son autonomie et son identité. Et donc, dans un paradigme d'avantage coopératif, concurrentiel ou compétitif, la connaissance et l'apprentissage sont le principal capital de l'entreprise.

A ce sujet Prax déclare que « la création de connaissance devient un axe culturel majeur et la performance de l'entreprise est calculée en termes de capital intellectuel et de création d'idées ». Dans ce cadre, les enjeux sont autour de l'organisation (méthodes de travail), et de la gestion (modes de management) et donc des NTIC du KM, qu'il faut adopter pour survivre comme un tout dans une économie globalisée (moi, toi et les autres). Cette idée se reflète aussi chez Farey et Prusak « l'ambition du knowledge management réside dans la dissémination des savoirs pour permettre à de nouvelles idées de germer, réduire le temps de développement de nouveaux produits et engendrer de meilleures décisions. La connaissance est créée et développée par des hommes ; le système de knowledge management doit donc savoir connecter les items de savoir avec les hommes qui savent l'utiliser ».

En parcourant, ainsi l'histoire de notre société industrielle comme un système d'organisation du travail nous est apparu la source d'autre avantage, ce que nous appelons, l'*avantage collectif*. En effet, dans ces trois époques (production, économie de service, et économie globalisée) la connaissance est prise d'abord comme un levier de productivité, puis comme un levier d'avantage concurrentiel ou compétitif, et enfin comme un levier d'avantage coopératif. Ceci implique que la gestion des connaissances dans une économie globalisée (l'outsourcing du travail) exige la coopération de l'autre, ce qui n'est plus le cas dans une économie de service telle que nous la vivons aujourd'hui (modèle de Nonaka et Takeuchi). Autrement dit, serions-nous prêts à partager nos connaissances avec d'autres (voire une entreprise concurrente) ? Comme l'a si bien dit Gérard Balantzian « saurions-nous négocier le virage dans une démarche partenariale ? … la raison majeure de notre carence réside dans le fait que dans ces transformations, l'économie dite de service, est considérée comme une économie de l'intelligence ou une économie de la matière grise … tandis que l'économie globalisée exige de l'intelligence et de la connaissance coopérative ». D'ailleurs, l'aspect social de la gestion des connaissances amplifie le fait que, sans les individus il n'y a pas de création de connaissances et d'apprentissage organisationnel, alors que l'aspect technique de la gestion des connaissances indique que, sans les NTIC du KM, la connaissance créée ne pourra pas être mobilisée comme information à travers un réseau de communication, de coordination, et de coopération, d'une part de l'extérieur vers l'intérieur (pour accumuler le savoir et le capital innovateur de l'entreprise), et d'autre part de l'intérieur vers l'extérieur (afin de garantir la pérennité de l'entreprise à travers de nouveaux produits et services). Dans, ce contexte, la gestion des connaissances existe dans un plan moral et éthique[100] plus que social ou technique, c'est ce que nous appelons : *système d'organisation du travail* dans un paradigme d'*avantage collectif*[101].

Un exemple de ce système est le modèle autopoïétique proposé par Fernando Flores (le *réseau de compromis social*). Ce modèle (qui a été inspiré de l'approche de l'enaction de Maturana et Varela), offre un progiciel pour analyser les flux conversationnels d'une entreprise. Néanmoins, dans la pratique, le rejet social de la démarche de Flores (au Chili) nous renvoie aux mouvements syndicaux de l'époque de Taylor. Bien qu'il ne s'agisse pas d'une lutte de classe, mais plutôt d'une lutte de pouvoir entre celui qui sait et celui qui ne sait pas.

**Conclusion du chapitre**

Nous avons présenté dans ce chapitre, l'état des lieux bibliographique de la gestion des connaissances. En particulier nous avons mis en évidence :

- premièrement, les arguments "commerciaux" sur la question : *qu'est-ce que la gestion des connaissances ?* C'est ainsi que nous avons passé en revue les discours de Wendi Bukowitz, Ruth Williams, Karl Sveiby, Leif Edvinsson, René-Charles Tisseyre, Jean-Yves Prax, et Jean-Louis Ermine[102]. Cette mise en revue a permis de mettre en évidence ces enjeux autour de la gestion des connaissances, par le biais des NTIC du KM supportés par des sociétés de conseils en information, tels que : PricewaterhouseCoopers, Sveiby, Capgemini, CorEdge, et le Club Gestion des Connaissances, où ces auteurs sont concernés. Ces discours ont été symbolisés à travers la dualité cartésienne âme/corps, et la dualité cartésienne cognition/action (pour le Club Gestion des

---

[100] Ce plan moral et éthique de la gestion des connaissances est l'un de *4 Repères essentiels pour la gestion des connaissances dans l'entreprise* que nous développerons dans le chapitre 3.
[101] Le terme "collectif " correspond, d'une part, à la collaboration (point de vue social), et d'autre part, à la dynamique circulante de 3 "C" du groupware, à savoir : communication, coordination, et coopération (point de vue technique). Nous parlons aussi de système d'organisation du travail par avantage collectif.
[102] Nous avons donné aussi notre point de vue sur cette question.

Connaissances). Ces enjeux ont permis aussi de passer en revue les "métiers du savoir", tels que : Knowledge Manager, Chief Knowledge Officier, Chief Learnig Officier, etc.

- deuxièmement, l'aspect social et l'aspect technique de la gestion des connaissances nous ont permis de développer une approche sociotechnique (un *framework*) pour (1) approcher la question fondamentale : *qu'est-ce que la connaissance ?* ; (2) nos écarter d'un souci commercial autour de la gestion des connaissances ; et (3) faire le clivage des différentes contributions. C'est ainsi qu'ont été mis en évidence quatre perspectives, à savoir : l'approche organisationnelle de Nonaka et Takeuchi (fondée sur le concept de knowledge creating-company) ; l'approche biologique de Maturana et Varela (fondée sur le concept de l'enaction) ; l'approche managériale de Jean-Louis Ermine (fondée sur le concept de la marguerite) ; et l'approche de NTIC du KM. Cette démarche nous a permis de dégager les mécanismes de création de connaissances nouvelles et d'apprentissage organisationnel d'un point de vue social et technique.

- troisièmement, les approches organisationnelle, biologique, managériale, et NTIC du KM ont été caractérisées à travers des mécanismes de création de connaissances nouvelles et d'apprentissage organisationnel. Ces mécanismes, représentés au travers des verbes à l'infinitif, ont été utilisés pour généraliser les mécanismes de la gestion des connaissances selon l'aspect social et technique.

- quatrièmement, nous avons décrit la connaissance, à partir des origines de la connaissance industrielle. Ceci a permis de dégager trois grandes époques : l'économie de production (1930), l'économie de service (1960) et l'économie globalisée (1990). Dans ces trois époques la connaissance a été prise d'abord comme un possible levier de productivité, puis comme un levier d'avantage concurrentiel ou compétitif, et enfin comme un levier d'avantage coopératif. Cette vision historique, a permis aussi de passer en revue les "concepts du savoir", tels que : knowledge worker (1967), knowledge society (1969), learning organization (1990), systems thinking (1990), actionable knowledge (1996), knowledge-creating company (1995), information ecology (1997), information age (1997), knowledge-based economy (1997), corporate knowledge (2000), corporate longitude (2000), knowledge-based assets (2000), etc.

La conclusion générale de ce chapitre est qu'au carrefour de ces quatre approches, deux concepts sont le fondement de la problématique essentielle de la gestion des connaissances. L'un est le concept de "faire-évoluer" les connaissances, c'est-à-dire de créer de nouvelles connaissances là, où il n'y a pas de savoir. L'autre est le concept de "faire-émerger" la connaissance, c'est-à-dire de créer des connaissances nouvelles à partir d'une représentation non symbolique de la réalité[103]. En effet, tous les modèles de gestion des connaissances[104] de nos jours sont gérés à partir du passé (les bonnes pratiques, le retour d'expérience, la communauté de pratiques, etc.) et non pas à partir de l'avenir (l'inconnu, le chaos, le désordre, etc.). Nous pensons que les concepts "faire-évoluer" et "faire-émerger" la connaissance sont fort intéressants pour réfléchir sur la question.

C'est d'ailleurs ce deuxième concept que nous développerons davantage dans cette thèse[105]. Nous ferons ceci à travers la *connaissance imparfaite*, en plaçant notre modèle de gestion des connaissances, tout d'abord sur l'aspect social (union, identité, autonomie), puis sur l'aspect technique (communication, coordination, coopération). Dans ce modèle la capacité de l'entreprise pour créer, apprendre et mobiliser son savoir (ou connaissances) seront visualisées à travers un système d'organisation du travail collectif qui permet à la fois la gestion des compétences, des connaissances et l'innovation durable. Pour nous, ces trois conditions sont essentielles pour bâtir une vraie culture d'entreprise autour d'un *avantage collectif.*

---

[103] Dans le chapitre 3 nous étudierons, d'une part, le concept de "faire-évoluer" les connaissances selon l'approche de l'enaction de Karl Weick [Weick, 79] mis en évidence par Tounkara, dans sa thèse, intitulée *Gestion des Connaissances et Veille : vers un guide méthodologique pour améliorer la collecte d'informations* (publiée en 2002), et d'autre part, celui de "faire-émerger" la connaissance selon l'approche de l'enaction de Maturana et Varela [Varela, 96].

[104] Ici nous faisons référence au "catalogue" du KM fait par l'équipe de recherche ACACIA à l'INRIA Sophia Antiopolis, dirigée par Rose Dieng, à partir des ouvrages : *Méthodes et outils pour la gestion des connaissances* (publié en 2000), et *Méthodes et outils pour la gestion des connaissances : une approche pluridisciplinaire du Knowledge Management* (publié en 2001).

[105] Dans cette thèse nous sommes loin d'offrir un modèle pour "faire-évoluer" ou "faire-émerger" la connaissance. Néanmoins, cette thèse a le mérite de caractériser la problématique de la gestion des connaissances par ces deux concepts.

| **Chapitre 2** | **Les racines de l'arbre de la gestion des connaissances : partie système, cybernétique et autopoïétique** |
| --- | --- |

*La connaissance s'acquiert par l'expérience, tout le reste n'est que de l'information.*
*Albert Einstein, physicien américain*

Afin de positionner scientifiquement notre problématique de gestion des connaissances imparfaites, nous avons utilisé une analogie avec un "arbre"[1], que nous avons appelé : l'*arbre de la gestion des connaissances* (voir figure 2.1). Cet arbre (que symbolise le fait que la connaissance est un mécanisme "circulant" et "d'émergence de signification"), a deux parties :

- les feuilles de l'arbre de la gestion des connaissances : partie organisationnelle, biologique, managériale, et NTIC du KM ;

- les racines de l'arbre de la gestion des connaissances : partie système, cybernétique et autopoïétique.

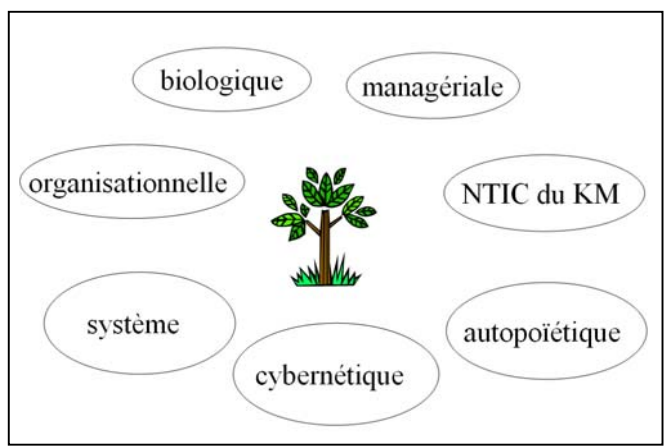

**Figure 2.1 :** Arbre de la gestion des connaissances (source propre)

Le chapitre 1 a permis d'explorer cette première partie, et de faire l'état de l'art de la gestion des connaissances, ensuite nous avons dégagé les mécanismes génériques de la gestion des connaissances à partir des mécanismes de création des connaissances nouvelles et d'apprentissage organisationnel d'un point de vue social et technique, ainsi que d'étudier la signification de la connaissance dans le monde du travail.

Dans ce chapitre 2, nous explorerons davantage la deuxième partie de cet arbre. Il nous a paru approprié d'interpréter et d'expliquer cette réalité à travers des modèles bâtis autour de trois approches plus au moins utilisées dans le domaine du *management* (en particulier, les systèmes d'information) et qui nous ont accompagnés et guidés tout au long de ce travail pour analyser et comprendre l'entreprise (lors de notre stage), à savoir : l'*approche système* qui permet d'observer l'entreprise comme un flux d'entrées/transformation/sorties ; l'*approche cybernétique* qui permet d'observer l'entreprise comme une organisation sous une contrainte de relations (variables d'état) entre les propriétés des composants ; et l'*approche autopoïétique de Santiago* et *de Valparaiso*[2] qui permet d'observer le système (l'entreprise) comme : (1) l'organisation de l'unité (dans un domaine de détermination interne, par une description organisationnelle de l'unité) à partir de trois processus génériques de production de ces mêmes composants génériques, à savoir : constitutives (primaires), spécification (structurelles), et d'ordre (décisionnelles) ; et simultanément et nécessairement, comme (2) le fonctionnement de l'unité (dans un domaine de composition, description structurelle de l'unité) qui opère sous trois contraintes génériques, à savoir : clôture opérationnelle, couplage structurel, et détermination structurelle. La richesse de l'approche est que le système (l'entreprise) est observé à travers trois phénomènes, à savoir : l'auto-organisation de l'unité (l'organisation), d'auto-maintient de l'identité (la structure), et l'auto-gestion de l'autonomie (la dynamique).

---

[1] Nous avons imaginé cette analogie à partir du site de Jean-Louis Ermine *Le Club de Gestion des Connaissances* http://www.club-gc.asso.fr, où un "arbre" est justement le logo du Club.

Nous avons repéré dans la littérature systémique, cybernétique, et autopoïétique cinq modèles, à savoir : le *modèle des systèmes organisés* (en abrégé modèle OID[3]) de Jean-Louis Le Moigne, le *modèle des systèmes de connaissances* (en abrégé modèle OIDC[4]) de Jean-Louis Ermine, le *modèle des systèmes de gestion* (en abrégé modèle AMS[5]) de Jacques Mélèse, le *modèle des systèmes de management* (en abrégé modèle MSV[6]) de Stafford Beer, et enfin le *modèle des organisations autopoïétiques* (en abrégé modèle CIBORGA[7]) d'Aquiles Limone et Luis Bastias.

Ces modèles ont vu le jour au cours du temps à travers des analogies[8] principalement avec le système nerveux [Maturana et Varela, 72], [Maturana et Varela, 73], [Varela, 89], [Beer, 72] et les entreprises [Mélèse, 72], [Limone, 77], [Limone et Bastias, 02a], [Le Moigne, 77], [Ermine, 96]. Ainsi, les modèles OID et OIDC trouvent leurs bases conceptuelles dans l'approche système, tandis que les modèles AMS et MSV en plus de considérer cette approche, considère l'approche cybernétique (de deuxième ordre), enfin le modèle CIBORGA a été fondé sur l'approche cybernétique (de deuxième ordre) et autopoïétique en plus de l'approche système. En d'autres termes, l'approche système est utilisée pour expliquer les modèles OID et OIDC, les approches système et cybernétique (de deuxième ordre) sont utilisées pour expliquer les modèles AMS et MSV, et les approches système, cybernétique (de deuxième ordre) et autopoïétique sont utilisées pour expliquer le modèle CIBORGA, sans oublier que dans la dualité modèle/réalité, la réalité que le modèle essaie de saisir et d'expliquer, représente beaucoup plus que le modèle, comme nous l'avons dit dans l'introduction de la thèse.

---

[2] La différence de l'un par rapport à l'autre se fait par rapport au phénomène décrit, dans l'approche autopoïétique de Santiago il s'agit des systèmes vivant (par exemple, le système nerveux, système immunitaire et les animaux multicellulaires), tandis que l'approche autopoïétique de Santiago et de Valparaiso, le phénomène observé est l'entreprise.
[3] Opération, Information, Décision.
[4] Opération, Information, Décision, Connaissance.
[5] Analyse Modulaire des Systèmes.
[6] Modèle des Systèmes Viables.
[7] Modèle Cybernétique de l'Organisation et de l'Apprentissage.
[8] Pour Bourdieu « raisonner par analogie, c'est former un raisonnement fondé sur les ressemblances ou les rapports d'une chose avec une autre » [Bourdieu *et al*, 73].

Pour justifier la validité des analogies, nous faisons appel à l'argumentation de Limone lorsqu'il dit « une analogie n'est valable et féconde que si elle est guidée et soutenue par une réflexion théorique et par une démarche méthodologique » [Limone, 77]. Nous n'avons aucun doute du caractère sérieux dans la réflexion et en démarche des auteurs concernés lors de la conception des modèles, ce qui donne un sens au cadre de référence de cette thèse. Néanmoins, nous pensons que dans tous les cas, il s'agit d'analogies superficielles. En effet, ces analogies ont été utilisées pour stimuler l'imagination afin de suggérer des idées ou voies possibles de recherche pour la description des phénomènes. Mais, la similitude entre les phénomènes n'a pas été faite au niveau de sa structure, fonctionnement et évolution, comme c'est le cas pour une analogie profonde[9].

Au fond, ce sont justement ces trois approches et leurs modèles associés qui nous ont servi pour fixer le cadre de référence de cette thèse. La démarche suivie (afin d'approcher une réalité de gestion des connaissances imparfaites) a été d'expliquer et distancier d'un point de vue système, cybernétique, et autopoïétique les notions de *données*, *information* et *connaissances*.

L'idée centrale de ces modèles, où certains sont plus connus et/ou utilisés que d'autres, est d'expliquer, à travers l'identification d'un *patron commun*[10] *d'organisation*, l'organisation du système, par la mise en évidence, d'une part, de relations entre composants (OID, OIDC, AMS, et MSV) et de relations entre processus de production de composants (CIBORGA). Ces composantes sont capables de contenir d'autres composantes et d'autres relations, de telle sorte que l'organisation de l'unité[11] des composantes et leurs relations appartiennent toujours à une même totalité, sans perdre son identité de classe.

L'objectif de ce chapitre 2 est de présenter les approches système, cybernétique et autopoïétique de Santiago et de Valparaiso et leurs modèles associés (OID de Jean-Louis Le Moigne, OIDC de Jean-Louis Ermine, AMS de Jacques Mélèse, MSV de Stafford Beer, et CIBORGA d'Aquiles Limone et Luis Bastias) qui sont à la base des approches organisationnelles, biologiques, managériales, et NTIC du KM de la gestion des connaissances, que nous avons développées au chapitre 1. Ces modèles sont considérés comme de modèles génériques de la gestion des connaissances.

---

[9] Dans [Limone, 77] est explique une demarche general a patir de travaux de Bertalanffy, Beer et autres auteurs sur la construction de modeles pour la descriptions des phenomenes, fondes sur les analogies superficielles et profondes, que nous avons utilise ailleurs.
[10] Nous avons trouvé aussi d'autres termes pour signaler l'existence de ce patron commun dans l'organisation (construction ou production) de la structure : l'invariant [Ashby, 72], [Limone, 77] ; principe d'organicité [Ashby, 72] ; principe organisateur [Morin, 77] ; patron organisationnel [Capra, 98], etc.
[11] Pour Varela, l'unité est « le fait d'être distinguable de son environnement et donc des autres unités ».

Ce chapitre est organisé dans deux parties. La première partie appelée « les dualités » permet de préciser d'abord ce que l'on entend, d'une part, par *dualité*, et d'autre part, par *domaine/unité*, *observateur/observé*, *unité/distinction*, *organisation/système* et *organisation/structure*. Ainsi, que la dualité *gestion/management*. En fait, ces expressions se rapprochent mais n'ont pas le même sens comme nous le verrons.

La deuxième partie appelée « les modèles de description de l'unité » essaie de dégager les mécanismes (ou fonctionnement) des modèles : OID, OIDC, AMS, MSV, et CIBORGA.

## 2.1.    Les dualités[12]

Les dualités, ont été le fondement de notre savoir, savoir-faire, savoir-technique, savoir-être, savoir-vivre, etc., depuis des siècles, puisque les dualités permettent de mettre en rapport des objets (abstraits et physiques)[13] qui nous entourent, c'est-à-dire les dualités donnent de sens à nos activités de tous les jours dans un monde réel et réfléchi. Et donc, les dualités sont un *mode d'interaction*.

Pour bien comprendre la richesse du concept de dualité, supposons que nous avons deux objets : *a* et *b* qui existent dans un domaine particulier, disons domaine *A* pour *a* et domaine *B* pour *b* alors la dualité *a/b*, veut dire que *a* et *b* ne sont pas liés dans leur domaine de définition (domaine A et domaine B). En revanche, dans la matérialisation de *a* dans le domaine *B*, et en même temps la matérialisation de *b* dans le domaine A, implique que *a* et *b* sont fortement liés. En conséquence, cette indépendance et (en même temps) dépendance, est réalisée (simultanément et nécessairement) grâce à ce que nous appelons : *processus de causalité circulaire[14]*.

---

[12] Le nom de cette section vient d'une argumentation de Varela que dit ceci « on redécouvre ici la dualité entre les descriptions organisationnelles et structurelles ». Dans ce contexte le terme "dualité" n'a rien à voir avec le sent cartésienne du terme, que nous avons introduit dans le chapitre 1 à l'aide des dualités âme/corps et cognition/action.

[13] Un "et" dans cette relation veut dire l'un ou l'autre avec un certain degré (une valeur réelle de l'intervalle [0,1]) d'appartenance à l'un ou l'autre. Dans un cas extrême, que cette valeur soit égale à 1, alors la relation est un "ou". Dans le contexte de cette thèse, c'est-à-dire dans une réalité de connaissances imparfaites, ce qui est bien entendu un "et" et non pas un "ou". En effet, comme nous verrons dans le chapitre 4, dédié à l'étude de la représentation floue et non floue de la connaissance (à l'aide de la théorie de l'incertain et de l'imprécis), et donc le "ou" n'est qu'un cas particulier du "et". Cela nous renvoie aussi à clarifier un peu le titre de cette thèse, car le nom correct serait "connaissances parfaites et imparfaites", mais par un souci de simplification (dans la pure tradition cartésienne) nous avons préféré laisser seulement "connaissances imparfaites", bien que cela peut être une source de confusion pour le lecteur non averti de notre démarche.

[14] Comme nous verrons dans le chapitre 3, ce processus est attaché à la cognition, alors si le processus est relatif à l'approche cognitiviste de la cognition, alors ce que l'on cherche à construire est une chaîne logique de cause à effet, tandis que si le processus est associé à l'approche enactiviste de la cognition, alors la relation ne se construit pas nécessairement d'une représentation vraie ou logique de causes et d'effets, puisque le problème de l'enaction n'apparît pas lié à la représentation symbolique d'une réalité, sinon qu'au maintient du système en vie et viable.

D'après cette argumentation, le processus de causalité circulaire définit alors une dualité cause/effet. Cela veut dire, que les causes et les effets sont *distinguables* dans des espaces fort différents, mais par le pouvoir du processus de causalité circulaire les causes auront une influence sur les effets, et simultanément et nécessairement, les effets auront une influence sur les causes, créant ainsi un autre domaine de définition de façon récursive.

En conséquence, les dualités ne doivent pas être perçues comme seulement une opposition mais également comme une complémentarité récursive, dans le sens : *L'union fait la force*[15].

D'ailleurs, nous avons trouvé dans la littérature, qu'on peut remplacer le terme dualité par un adjectif afin de faire ressortir certains aspects, par exemple : boucle cause/effet pour faire ressortir davantage l'aspect cybernétique (de premier ordre) entre la cause et l'effet, ou bien dialectique cause/effet pour faire ressortir davantage la dialectique observateur/observé (le projet politique de l'observateur dans sa distinction) selon un aspect cybernétique (de deuxième ordre), etc.

Dans cette section nous focalisons notre intérêt sur les dualités. D'abord, deux exemples nous permettront d'illustrer davantage le concept de dualité. Le premier exemple est la dualité *savoir/faire* de Ballay imbriquée avec la dualité *tête/mains* de Bernard. Le deuxième exemple est la dualité *figure/fond* de Valéry imbriquée avec la dualité *apparence/mécanisme* de Varela. Ensuite, cette dualité chez Varela est utilisée pour introduire cinq concepts, à savoir : domaine (conceptuel, physique), unité, observateur, observé, et distinction (opération de distinction). Ces concepts sont à la base de trois dualités : *domaine/unité*, *observateur/observé* et *unité/distinction*, qui seront détaillés également ici. Puis, sont introduites les dualités : *organisation/système* et *organisation/structure*. Nous soulignons que ces dualités seront expliquées à partir des approches systèmes, cybernétique et autopoïétique de Santiago et de Valparaiso. Néanmoins, ces approches sont explicitées dans la section 2.2.3, mais par la force des choses nous sommes obligés de les introduire ici (pour la clarté de l'exposé). Enfin, nous présentons la dualité *gestion/management*. Cette dualité prend ses racines dans deux ouvrages récents chez Jean-Louis Ermine. L'un est *La gestion des connaissances* (publié en 2003), l'autre est *Management des connaissances en entreprise* (publié en 2004), écrit avec Imed Boughzala. Nous pensons que ce changement de noms n'est pas lié au hasard, sinon qu'il y a une explication profonde, que nous tenterons d'expliquer.

---

[15] Comme nous verrons plus bas pour Varela, « l'unité est la seule condition nécessaire à l'existence d'un domaine donné, quel qu'il soit » [Varela, 89]. Il parle de l'unité comme une condition déterminante. Cette détermination symbolise *la force*. Autrement dit, la force symbolise la structure (l'identité) et l'unité symbolise l'organisation.

### 2.1.1. Dualité *savoir/faire* de Ballay & dualité *tête/mains* de Bernard

Jean-François Ballay dans son livre, intitulé *Capitaliser et transmettre les savoir-faire de l'entreprise* (publié en 1997) a dit « nous n'avons besoin que de deux notions fondamentales, ce qui est déjà bien suffisant : le *savoir* (ou connaissance) et l'*action*. Vous noterez que cette définition nous évite l'habituel clivage entre "connaissance" et "savoir-faire" qui me semble peu opérant à cause de sa tendance à couper le monde en deux : celui des intellectuels et celui des manuels. A mes yeux le savoir-faire *est* du savoir (et réciproquement, tout savoir est susceptible de servir à un moment dans l'action ». Et puis, il a ajouté « nous sommes bien en peine de séparer action et pensée. Ecoutons ce que disait, au XIXème siècle, Claude Bernard, pionnier de la biologie moderne et promoteur de l'expérimentation scientifique : *il serait impossible de séparer ces deux choses : la tête et la main. Une main habile sans la tête qui la dirige est un instrument aveugle ; la tête sans la main qui réalise reste impuissante* » [Ballay, 97].

En conséquence, pour Jean-François Ballay ce qui compte est la dualité savoir/faire et non pas, d'après lui, la relation cartésienne savoir-faire[16]. Mais, également chez Claude Bernard avec la dualité tête/mains. En effet, pour lui, la tête et les mains habitent dans deux espaces fort différents, mais dans le sens de l'esprit l'un ne peut pas vivre (système vivant) sans l'autre, est dans le sens réel l'un ne peut pas être viable (système viable) sans l'autre. Voilà qu'un processus de causalité circulaire a de sens, seulement s'il est lié à une dualité cause/effet (tel que celle que nous avons définie plus haut).

Nous pensons aussi que le processus de causalité circulaire défini à travers une dualité cause/effet, enrichit davantage un autre propos de Claude Bernard, lorsqu'il a dit « les systèmes ne sont pas dans la nature, ils sont dans l'esprit des hommes ». C'est pour cette raison que nous avons choisi cette argumentation de Bernard comme citation du chapitre 5, afin de bien démarrer notre cas d'étude, c'est-à-dire notre champ d'expérimentation du modèle proposé dans un esprit sociotechnique.

### 2.1.2. Dualité *figure/fond* de Valéry & dualité *apparence/mécanisme* de Varela

Paul Valéry, dans son livre intitulé *De la simulation* (publié en 1927) a dit « il faut être deux pour inventer. L'un forme des combinaisons, l'autre choisit, reconnaît ce qu'il désire, et ce qui lui importe dans l'ensemble des produits premiers. Ce qu'on appelle le "génie" est bien moins l'acte de celui-ci – celui qui combine – que la promptitude du second à comprendre la valeur de ce qui vient de se produire et à saisir ce produit »[17].

---

[16] Dans la littérature on trouve *savoir-faire*, et donc pour Ballay le "-" a un effet cartésien (*des actions d'un couteau*, comme dit Varela pour illustrer ce paradigme analytique). Pour nous, dans cette thèse savoir-faire a la même signification que la dualité savoir/faire, c'est-à-dire, un processus de causalité circulaire, qui est valable aussi pour les autres savoirs, tels que : savoir-technique, savoir-être, savoir-vivre, etc. Néanmoins, nous avons préféré enlever le "-" du terme modèle socio-technique, pour montrer davantage, d'une part, le caractère soudé de "social" et de "technique", et d'autre part, pour insister davantage sur la dualité cause/effet, nous employons le terme : *modèle sociotechnique récursif*. Enfin, comme nous avons dit au chapitre 1, ce concept a été mis au point par Herbst dans son livre, intitulé *Socio-Technical Design* (publié en 1974).

A ce propos, Varela a dit (avant d'introduire le concept d'unité) « mon point de départ est cette citation de Valéry. Elle s'applique merveilleusement bien à ce que je cherche à dire. Et, même si Valéry faisait référence à la création et à l'imagination poétiques, je ne crois pas qu'il s'offusquerait de nous voir l'utiliser au cours d'un débat sur l'auto-organisation et sur l'apparition du nouveau dans le monde naturel ». Puis, il ajoute « la nécessité du nombre deux est la première idée importante de cette citation. Avant de pouvoir parler d'auto-organisation, il faut déjà supposer une certaine différence entre une unité (ou un système) et son milieu (ou son environnement, si vous préférez), un peu comme dans la relation figure/fond » [Varela, 89].

Cette dualité de Valéry est à la base du concept d'auto-organisation dans l'approche autopoïétique de Santiago. A ce propos, Varela a dit « en partant d'une intuition de Valéry, nous sommes arrivés à l'idée que l'auto-organisation suppose la distinction entre une unité et son fond, et que ces deux éléments sont en relation comme deux séries d'événements dotés d'un certain degré d'indépendance »[18].

Cette dualité de Varela permet d'introduire d'ores et déjà cinq concepts essentiels dans la modélisation autopoïétique, mais qu'est aussi vrai pour les autres modèles (modèle cybernétique et modèle systémique, mais dans un aucun cas pour le modèle analytique), à savoir : domaine, unité, distinction (opération de distinction), observateur, observé, qui à la fois donne naissance à trois dualités, à savoir : dualité domaine/unité, dualité unité/distinction, et dualité observateur/observé.

## 2.1.3. Dualité domaine/unité

Pour Maturana et Varela, « un domaine phénoménal est défini par les propriétés de l'unité ou des unités qui le constituent. Ainsi, dès qu'une unité est définie ou qu'est mise en place une classe ou des classes d'unités qui peuvent subir des transformations ou des interactions, un domaine phénoménal est défini » [Varela, 89].

Ainsi, un domaine a de sens si et seulement si, il est relatif à la description du phénomène observé : l'*unité*.

---

[17] Cette argumentation de Valery est la citation du chapitre IX (L'auto-organisation : de l'apparence au mécanisme) du livre de Varela, intitulé *Autonomie et connaissance* (publié en 1989).
[18] Nous pensons aussi au concept d'organisation (feed-back) de l'approche cybernétique (de deuxième ordre) de Heinz von Foerster dans la formulation du concept de l'auto-organisation dans l'approche autopoïétique de Santiago, car il faut ajouter deux autres ingrédients supplémentaires pour former l'autopoïèse comme nous argumentons plus bas.

Pour Maturana et Varela, « l'unité (le fait d'être distinguable de son environnement et donc des autres unités) est la seule condition nécessaire à l'existence d'un domaine donné, quel qu'il soit ». Pour eux « l'unité est intimement liée à l'organisation et au fonctionnement de l'unité, et elle a lieu dans l'espace physique où sont spécifiés cette organisation et ce fonctionnement … l'organisation de l'unité est orientée vers le maintien de son organisation »[19].

Ceci donne naissance à la *dualité domaine/unité*.

## 2.1.4. Dualité observateur/observé

Cette dualité est relative à la description du phénomène observé (l'unité). Cette description est formulée alors, d'une part, par rapport à un processus de causalité circulaire associé à la dualité cause/effet, et d'autre part, par rapport à une approche de la cognition attachée à l'observateur qui est définit par rapport à son projet téléologique et ses limites d'observation.

Ce qui pose la question : quel est le rôle de la cognition dans le processus de causalité circulaire ?[20]

Cela signifie que selon l'une des approches (cognitiviste, connexionniste, et enactiviste)[21] de la cognition choisie on aura une description du phénomène observé (l'unité).

Or, selon l'approche de la cognition choisie, la description de l'unité a lieu, simultanément et nécessairement, dans deux domaines. L'un est la *détermination interne*, l'autre est la *composition*. Autrement dit, pour l'observateur, l'unité existe (simultanément et nécessairement) dans le domaine de la *détermination interne* et dans le domaine de la *composition*. On a donc, deux existences parallèles (simultanément et nécessairement) d'une même unité, on parle alors de l'unité dans le domaine de la détermination interne, et simultanément et nécessairement, de l'unité dans le domaine de la composition.

---

[19] Ceci est vrai pour l'autopoïèse de Santiago. Nous n'avons aucun doute que ceci est vrai aussi pour l'autopoïèse du modèle CIBORGA, mais nous supposons que les autres modèles (OID, OIDC, AMS et MSV) ont été conçu dans ce même l'esprit, car dans la littérature nous trouvons aucune référence à l'unité de l'organisation, en plus le processus de structuration est relatif à un couplage des entrées et des sorties, et non pas par rapport à un couplage par clôture (couplage structurel), comme nous verrons plus loin.

[20] Dans le chapitre 3 nous abordons le concept de système complexe selon l'approche de l'enaction de Maturana et Varela. Néanmoins, si le processus de causalité circulaire est relatif à l'approche cognitiviste de la cognition, alors ce que l'on cherche à construire est une chaîne logique de cause à effet, tandis que si le processus est associé à l'approche enactiviste de la cognition, alors la relation ne se construit pas nécessairement à partir d'une représentation vraie (ou logique) de causes et d'effets, sinon qu'à partir de représentations de la connaissance qui maintiennent le système en vie et viable, puisque le problème de l'enaction n'apparaît pas lié à la représentation symbolique de la connaissance sur une réalité.

[21] Ceci pose la question aussi des approches hybrides de la cognition

---------------------------------------------------------------------------------------------------

Dans ce même contexte, Varela a dit « les processus qui déterminent et distinguent une unité spécifient sa nature et celle du domaine à l'intérieur duquel elle existe. Cela est vrai, qu'il s'agisse d'un processus conceptuel (où l'observateur, par une opération de distinction, définit une unité au sein de son domaine de discours et de description) ou d'un processus physique (où une unité, par la manifestation des propriétés mêmes qui la définissent, apparaît en se distinguant de son environnement) ». Puis il a ajouté « on redécouvre ici la dualité entre les descriptions organisationnelles et structurelles » [Varela, 89]. Ce qui fait apparaître la dualité organisation/structure, pour laquelle nous avons consacré une section, de même que pour la dualité organisation/système, afin de recueillir les points de vue d'autres auteurs.

Cette distinction de l'unité dans le domaine de la détermination interne (processus conceptuel), et de l'unité dans le domaine de la composition (processus physique) est possible grâce à une *distinction* relative à la dualité observateur/observé.

Ceci donne naissance à la *dualité observateur/observé*.

### 2.1.5. Dualité unité/distinction

Cette dualité est une conséquence de l'antérieur. En effet, pour Maturana et Varela, « une distinction est l'acte qui définit les composants d'une unité donnée » [Varela, 89], ce que nous appelons une *opération de distinction*, c'est-à-dire, le fait de distinguer l'unité de son environnement et donc d'autres unités.

Ceci donne naissance à la *dualité unité/distinction*.

D'ailleurs, la dualité organisation/structure nous permettra d'enrichir davantage cette définition. Ainsi nous dirons que l'opération de distinction permet à l'observateur de distinguer l'unité (simultanément et nécessairement) soit d'un point de vue de l'organisation, soit d'un point de vue de la structure.

Cette description est formulée alors, d'une part, par rapport à un processus de causalité circulaire associé à la dualité cause/effet, et d'autre part, par rapport à une approche de la cognition[22] associée à la dualité observateur/observé. D'après la classification de Varela (voir section 2.2.3), chaque approche scientifique de la cognition (cognitiviste, connexionniste, et enactiviste) peut être attachée à une opération de distinction, ce que nous appelons : l'*opération de distinction attachée à*

---

[22] Ceci pose la question aussi des approches hybrides de la cognition.

*l'approche cognitiviste de la cognition*, *l'opération de distinction attachée à l'approche connexionniste de la cognition*, et l'*opération de distinction attachée à l'approche enactiviste de la cognition*.

> **Opération de distinction attachée à l'approche cognitiviste de la cognition**

La description cherche à construire des symboles à partir de règles. Comme l'a souligné Varela « le cognitivisme utilise les symboles pour s'en doter d'un niveau sémantique ou représentationnel qui soit de nature physique » [Varela, 96].

Pour en dégager le signe et la signification des symboles, un processus de causalité circulaire associé à la dualité cause/effet (de type chaîne logique de causes à effets) est mis en place par l'observateur en fonction de la dualité observateur/observé.

> **Opération de distinction attachée à l'approche connexionniste de la cognition**

La description cherche à construire un réseau logique de causes et effets, relatif à la représentation connexionniste d'une réalité.

> **Opération de distinction attachée à l'approche enactiviste de la cognition**

La description ne cherche pas à construire des symboles, c'est-à-dire de représentation vraie ou logique des causes et des effets, puisque le problème de l'enaction n'apparaît pas lié à la représentation symbolique d'une réalité, sinon qu'au maintien du système en vie et viable.

Dans la section suivante, nous verrons à travers la réflexion de certains auteurs que les concepts d'*organisation*, de *système*, et de *structure* sont impliqués l'un par rapport à l'autre et vice versa, mais nous allons voir aussi qu'ils sont différenciés. Pour cela nous allons comparer une série de définitions que nous avons repérées de la littérature systémique, cybernétique, et autopoïétique. Notre idée est d'en faire ressortir des caractéristiques communes, en particulier dans la dualité organisation/système.

## 2.1.6. Dualité organisation/système

Pour montrer la dualité *organisation/système*, nous allons prendre l'approche système appliqué à la biologie chez Bertalanffy, comme point de départ de notre argumentation. Pour lui « le problème de la vie est celui de l'organisation »[23]. Dans ce contexte, nous constatons que dans le domaine biologique l'organisation se fait par rapport à un tout : *la vie*. Et donc, en général l'organisation se fait toujours par rapport à quelque chose, ce que l'on appelle la *contrainte*[24] (dans la dualité organisation/système la contrainte est le *système*). Par exemple, dans le domaine du management nous pouvons dire, et cela personne n'en doute, que l'organisation des relations entre composants se fait par rapport à la contrainte temps, coût, performance (selon le type d'objectif : stratégique, tactique et opérationnelle)[25]. Et donc, nous constatons une certaine dualité entre organisation et système, à savoir : l'unité (l'organisation) se fait par rapport à la totalité (le système), dans ce sens c'est la totalité qui donne l'identité à l'organisation.

Pour illustrer le fait que l'unité est liée à l'existence de ses contraintes nous avons emprunté l'exemple d'Ashby donné par Limone dans sa thèse « une "chaise" est un "objet" parce qu'elle a de la cohérence, parce qu'elle est une unité … le fait que la chaise soit une chaise, une unité, et non une collection de parties indépendantes est dû à la présence de contraintes » [Limone, 77].

Pour Bertalanffy (1) « un système est un complexe d'éléments en relation réciproque » [Bertalanffy, 76], et (2) « un système est un assemblage d'éléments formant un ensemble complexe »[26]. Nous constatons donc deux caractéristiques communes : les *échanges* d'après (1), et les *transformations* d'après (2). En d'autres termes, « un système est un ensemble organisé » [Mélèse, 72]. Nous compléterons ceci en ajoutant « d'échanges et transformations » pour faire quelque chose : *changements*[27]. Par exemple, dans le domaine biologique, les échanges sont entre le système et son environnement ou entre les relations des composants du système, tandis que les transformations sont relatives à l'énergie ou à la matière [Bertalanffy, 76], [Mélèse, 72]. A cet égard Mélèse dit « la description d'un organisme par l'approche système met l'accent sur les *changements* qui s'y produisent, donc sur les échanges entre le système et son environnement ainsi que sur les transformations qu'opère le système » [Mélèse, 72]. En généralisant cette caractéristique commune à tous les systèmes, il ajoute « un système est quelque chose qui opère une transformation entrées-sorties » » [Mélèse, 72]. Puis, il souligne « le système exprime simplement le fait que quelque chose entre et sort transformé » [Mélèse, 72]. Bref, dans l'approche système, la dualité organisation/système se trouve dans le fait que les relations entre composants sont organisées par rapport à la totalité (le système) pour faire quelque chose (entrées/transformation/sorties)[28]. Et donc, les relations sont limitées par des contraintes pour maintenir l'autonomie et l'identité du système, c'est-à-dire le fait d'appartenir ou non à une classe.

---

[23] Cité dans [Limone, 77].
[24] La contrainte, comme nous l'avons souligne au début du chapitre, est le résultat de la cybernétique (de premier ordre).
[25] L'organisation est indispensable à l'entreprise, et c'est peut être pour cette raison que dans le domaine du management parler d'entreprise ou d'organisation c'est parler de la même chose.
[26] Il s'agit plutôt d'une définition générale repérée par [Mélèse, 72] dans la littérature systémique.
[27] Limone dit en citant Mélèse « une façon rigoureuse de définir le changement est par le concept de transition. Une transition se définit comme le passage d'une chose d'un état à un autre et elle devient spécifiée par l'indication des deux états et l'indication de quoi change en quoi » [Limone, 77]. Ceci introduit le concept d'état que nous verrons dans l'approche cybernétique. Néanmoins dans l'approche systémique nous parlons des changements de l'organisation du système, tandis que dans l'approche cybernétique on est plus spécifique et l'on parle des états des entrées et des états des sorties.

Dans l'approche cybernétique (de deuxième ordre) appliquée au management, Mélèse dit « un système est un ensemble d'éléments pouvant chacun revêtir divers états, où l'état du système est défini, à un instant donné, par la liste des valeurs des variables, une ou plusieurs variables pouvant décrire chaque élément » [Mélèse, 72]. En d'autre termes, comme lui-même l'a souligné, « un système est un ensemble de variables pouvant prendre diverses valeurs » [Mélèse, 72]. De là, nous constatons que la caractéristique commune qui opère la transformation (ou processus de transformation) des entrées en sorties, est l'*information*[29]. Le contrôle des changements pour assurer l'organisation du système est déterminé par des *variables* qui doivent réguler les *états des entrées*, *états des transformations*, et les *états des sorties* selon certaines valeurs des variables essentielles (les objectifs). Comme l'a souligné Mélèse, « un état du système est lui-même défini par l'état de tous ses éléments, c'est-à-dire par l'ensemble des valeurs de toutes ses variables d'états » [Mélèse, 72], puis il ajoute « un système est déterminé si l'on sait dresser la liste des états des entrées (entrées extérieures et variables d'action) et la liste des états correspondants des sorties et des variables essentielles »[30]. Bref, dans l'approche cybernétique de deuxième ordre, la dualité organisation/système se trouve dans le fait que le contrôle des changements de l'organisation du système (du processus de transformation des entrées en sorties) est régulé par *l'information* (variables/valeurs).

Nous constatons que les approches système et cybernétique (de deuxième ordre) sont complémentaires, et donc que les concepts de système et d'organisation sont plus impliqués : (1) d'une part, par l'idée d'unité et totalité des composants[31], et d'autre part par l'idée des relations de ces composants entre elles, des relations entre les composants et la totalité, des relations entre les relations, etc. En effet, nous pensons que le fait que l'unité donne un caractère de totalité, doit nécessairement dire que la totalité détermine l'unité. Autrement dit, l'unité doit toujours respecter les contraintes imposées par la totalité, sinon l'on risque de ne plus différencier la totalité. Cela signifie aussi que la totalité pour être différenciée doit définir les interactions ou relations entre des composants pour construire l'unité. Comme l'a souligné Mélèse « un système est un *ensemble*

---

[28] Ceci correspond parfaitement à la description de l'organisation du système par le modèle OID de Le Moigne, niveaux 1 et 2. Dans le niveau 1 « le phénomène est identifiable », tandis que dans le niveau 2 « le phénomène est "actif " » [Le Moigne, 90].
[29] Ce même constat a été fait par Le Moige, dans la construction des troisième et quatrième niveaux de la description de l'organisation du système par le modèle OID, il dit « le phénomène est régulé » et « le phénomène s'informe sur son propre comportement », respectivement [Le Moigne, 77], [Le Moigne, 90].
[30] Les variables d'actions et essentielles sont définies plus bas. Ici, toutes les deux sont des variables d'états.
[31] Dans l'approche système on emploi aussi le terme éléments au lieu du terme composants, mais ce qui est important à retenir ici est que ces composants (ou éléments) peuvent être simples ou complexes, d'où le fait que l'on parle d'unité simple, d'unité complexe [Morin, 77], de totalité simple et de totalité complexe. Ceci nous renvoie à l'opération de distinction comme autre caractéristiques des systèmes. Nous la verrons plus loin.

*d'éléments* (de parties, de variables…) ; et un *ensemble de relations* appliquées à ces éléments, ce qui présuppose une *organisation* » [Mélèse, 79]. Ainsi, la différence entre organisation et système se trouve dans le fait que le système est la totalité (ou l'identité de classe), tandis que l'organisation est l'unité ; (2) par l'idée que pour le maintien de l'organisation il faut la régulation des variables d'actions dans le domaine physique (le système) à partir des variables essentielles (les objectifs que l'on doit contrôler et qui sont définies dans le domaine organisationnel). Ainsi, dans les grands lignes nous pensons que les arguments de Bertalanffy et Mélèse sont suffisants, pour montrer que l'organisation est nécessairement impliquée dans le système, et vice versa, soit par le fait que l'unité (l'organisation) détermine l'identité (la structure) par rapport à la totalité (l'autonomie), soit par le fait que la totalité détermine des contraintes qui la définissent comme unité.

Maintenant, nous allons développer la dualité organisation/système, sur la base de l'approche autopoïétique de Santiago et de Valparaiso, bien que plus loin nous argumentions l'aspect d'autonomie.

Dans l'approche autopoïétique de Santiago appliquée à la biologie, Maturana et Varela disent « l'organisation est la dynamique des interactions et relations, que comme configuration relationnelle entre des composants en étant conservé, elle sépare à un ensemble de composants d'autres, en permettant de distinguer un système » [Maturana et Varela, 72]. La caractéristique commune est donc donnée par le fait que la totalité (système) peut-être distingue par l'observateur en tant que manifestation de l'unité (l'organisation). La dualité organisation/système selon l'approche autopoïétique de Santiago suppose l'existence d'une *opération de distinction* (le fait de distinguer l'unité de son environnement et donc des autres unités) et permet de distinguer la frontière par le fonctionnement et comportement du système (en d'autres termes, les relations entre processus de production de composants qui forment l'unité dans une totalité).

Dans l'approche autopoïétique de Valparaiso appliquée au management, Limone dit « le concept d'organisation peut être défini comme l'ensemble des relations entre les composants d'un système, et les liens entre ces relations, qui définissent et déterminent le fonctionnement et le comportement du système comme unité dans le temps » [Limone, 77]. De là, la caractéristique commune est donnée par l'aspect temporel de changement de forme des relations entre processus de production de composants. La dualité organisation/système se trouve dans le fait d'une double existence de la phénoménologie comme unité (organisation) et comme identité (système).

### 2.1.7. Dualité organisation/structure

Pour montrer la dualité *organisation/structure*, nous allons mettre en évidence l'approche système d'Edgar Morin, sur ce qu'il appelle le *trinitaire système/interrelation/organisation*. Ce concept, nous l'avons choisi d'une part pour sa clarté et synthèse de la dualité *organisation/système* que nous venons d'argumenter, et d'autre part parce qu'il fait apparaître le concept de *structure* comme l'agencement de relations entre composants (ou d'interrelations selon le langage de Morin). A ce propos, il dit « l'idée d'organisation et l'idée de système sont encore, non seulement embryonnaires, mais dissociées. Je propose ici de les associer, puisque le système est le caractère phénoménal et global que prennent des interrelations dont l'agencement constitue l'organisation du système. Les deux concepts sont liés par des interrelations : toute interrelation dotée de quelque stabilité ou régularité prend un caractère organisationnel et produit un système. Il y a donc une réciprocité circulaire entre ces trois termes : interrelation, organisation, système ». Puis, il ajoute « l'organisation d'un système et le système lui-même sont constitués d'interrelations. La notion de système complète la notion d'organisation autant que la notion d'organisation complète la notion de système. L'organisation articule la notion de système laquelle phénoménalise la notion d'organisation, en la liant à des éléments matériaux et à un tout phénoménal. L'organisation est le visage intériorisé du système (interrelations, articulations, structure), le système est le visage extériorisé de l'organisation (forme, globalité, émergence) » [Morin, 77].

Nous constatons donc que la caractéristique commune est justement les *interrelations* ou *relations* du concept *trinitaire système/interrelation/organisation* chez Morin. Nous pensons que ce concept permet de spécifier d'une part, l'existence des interactions[32] entre relations des composants pour former (ou articuler selon Morin) l'unité, et d'autre part, le fait que dans l'organisation du système, il doit y avoir une volonté de construire l'unité et une volonté d'être unie. Comme l'a souligné Morin « l'organisation est l'agencement de relations entre composants ou individus qui produit une unité complexe ou système, dotée de qualités inconnus au niveau des composants ou individus. L'organisation lie de façon interrelationnelle des éléments ou événements ou individus divers qui dès lors deviennent les composants d'un tout. Elle assure solidarité et solidité relative à ces liaisons, donc assure au système une certaine possibilité de durée en dépit de perturbations aléatoires. L'organisation donc : *transforme*, *produit*, *relie*, *maintient* » [Morin, 77].

D'où l'autre caractéristique commune qui est l'existence de cette volonté de construire l'unité pour être uni dans une totalité mais aussi d'être différentiable d'autres totalités par le fait d'avoir une identité de classe. D'après l'approche système de Morin, nous pouvons dire que d'une volonté de solidarité (pour maintenir l'unité = l'organisation) et de solidité (pour maintenir l'identité = la structure) au cours du temps (la contrainte, l'autonomie), on voit apparaître l'idée de propriétés ou attributs des composants afin de construire l'organisation dans un domaine physique, capable d'être distingué comme une totalité (ou globalité selon Morin). En effet, cette volonté de "solidarité et solidité" de l'organisation comme l'on vient de dire doit être matérialisée sur un support physique afin que les relations puissent prendre une certaine forme pour maintenir l'unité, mais aussi comme l'a souligné Morin pour transformer, produire et relier l'unité. Et donc, l'on voit apparaître le concept de *structure*, lié aux concepts d'organisation (l'unité) et de système (la totalité ou globalité). Ainsi, la structure de l'organisation d'un système est constituée par l'articulation des relations, autrement dit par l'organisation même de la structure. En plus, nous pensons que cette *transformation* de l'unité implique en soi une sorte de changement de la forme des relations au cours du temps[33]. Ainsi, l'on voit apparaître (voir l'approche autopoïétique de Valparaiso plus haut) l'idée de *changement* de la structure pour maintenir l'organisation. Nous pensons que cette transformation peut être associée au concept de construction [Piaget, 74], ou au concept relations entre processus de production de composants [Maturana et Varela, 72].

---

[32] A ce propos dans l'approche autopoïétique de Santiago chez Maturana et Varela, comme nous avons vu plus haut on fait ressortir aussi l'idée d'interaction. En effet, pour eux « l'organisation est la dynamique des interactions et relations », voir plus haut le paragraphe complet.

En effet, comme l'a si bien dit Jean Piaget « il n'existe pas de structure sans une construction » [Piaget, 74]. A propos de cette affirmation, Limone dit aussi « il nous semble voir là une interrogation non formulée sur un "principe de structuration". Et bien, nous pensons que l'organisation pourrait être ce principe, puisqu'elle est à la fois l'agent qui relie la structure en tant qu'unité, en lui donnant sa cohérence et cohésion, et le "moteur" de sa dynamique en déterminant et régulant son fonctionnement » [Limone, 77]. De là, une autre caractéristique commune est l'existence de ce principe de *construction* [Piaget, 74], *structuration* [Limone, 77] ou *production* [Maturana et Varela, 72] interprété aussi selon d'autres termes, tels que : patron commun [Maturana et Varela, 72], invariant [Ashby, 72], [Limone, 77], principe d'organicité [Ashby, 72], principe organisateur [Morin, 77], etc.

A la lumière du temps, on voit apparaître une différence entre organisation et structure, que nous allons commenter par deux traits. Le premier relatif à la clôture de l'organisation de l'unité dans sa totalité par le fait que l'organisation de la structure est à la fois le produit et la productrice de l'organisation, selon une sorte de *principe de structuration* (l'argumentation de Limone ci-dessus), le deuxième est relatif à l'ouverture de l'organisation de l'unité dans sa totalité par le fait que la structure de l'organisation se transforme. En conséquence, la dualité organisation/structure se trouve dans la clôture organisationnelle du système et (simultanément et nécessairement) l'ouverture structurelle du système[34].

---

[33] Le fait du temps nous l'avons emprunté de l'approche autopoïétique de Valparaiso : « …le comportement du système comme unité dans le temps » [Limone, 77], voir plus haut le paragraphe complet.

Argumentons le trait de clôture organisationnelle du système. A cet égard, pour Morin « l'organisation apparaît comme une réalité quasi récursive, c'est-à-dire dont les produits finaux se bouclent sur les éléments initiaux ; d'où l'idée que l'organisation est toujours aussi, en même temps, organisation de l'organisation » [Morin, 77]. Nous constatons donc que selon l'approche système, la *récursivité* (l'organisation produit et elle est elle-même la productrice) est une caractéristique commune à la dualité organisation/structure. En effet, elle est suffisamment générique pour apparaître comme une caractéristique fondamentale de l'organisation de la structure. Elle apparaît matérialisée dans le concept *trinitaire système/interrelation/organisation*, puisque comme l'a souligné Morin « c'est une notion circulaire qui, tout en renvoyant au système, se renvoie à elle-même ; elle est constitutive des relations, formations, morphostases, invariances, etc., qui circulairement la constituent. L'organisation doit donc être conçue comme organisation de sa propre organisation, ce qui veut dire aussi qu'elle se referme sur elle-même en refermant le système par rapport à son propre environnement »[35] [Morin, 77].

Selon l'approche autopoïétique de Santiago, la récursivité (qui agit comme une boucle ou boucle récursive)[36] apparaît comme une *opération de distinction*[37], qui permet de distinguer les relations qui serviront pour former l'unité dans sa totalité et être distinguable d'autres totalités par le fait d'appartenir ou non à une *classe* : l'identité du système. A titre d'exemple ce concept de classe a la même signification, dans le domaine informatique, que le concept de classe du langage objet [Nanci *et al*, 96]. Dans ce sens, la classe spécifie : (1) la définition de sa structure (données) ; (2) les opérations de distinction (méthodes), autrement dit les types d'objets (membres de la classe) qu'elle peut abriter. Or, l'objet pour exister et être différenciable doit toujours appartenir à une classe. Et donc, nous pouvons dire que si la classe symbolise le système, alors l'objet symbolise l'organisation. Nous pensons que cette analogie avec le concept de classe en informatique permet de donner un sens plus pratique à ce que l'on entend par l'organisation de la structure dans un système.

Dans l'approche autopoïétique de Santiago, l'organisation de la structure est perçue comme « l'ensemble de relations qui doivent se produire entre les composants, pour être reconnu comme membre d'une classe spécifique » [Maturana et Varela, 72]. Egalement, sans l'approche autopoïétique de Valparaiso, l'organisation de la structure est perçue comme « l'ensemble des relations qui définit et détermine les conditions de constitution d'un système comme unité en spécifiant son identité de classe » [Limone et Bastias, 02a]. Cela signifie que la manifestation continuelle d'unité (d'organisation) de la structure confère à la classe une caractéristique d'identité (de totalité).

---

[34] Ceci est vrai, dans l'approche système où la distinction entre *système ouvert* et *système fermé* se fait en fonction de l'environnement, mais dans l'approche autopoïétique la distinction entre *système ouvert* et *système fermé* se fait en fonction de l'organisation de l'unité, le maintien de l'identité, et la contrainte de l'autonomie, ce qui donne naissance à un nouveau système : le système clos. Dans ce contexte on parle de clôture et non pas de fermeture.
[35] Cette argumentation chez Morin fait aussi ressortir l'idée de système ouvert et système fermé dont discuterons plus loin.
[36] Dans le domaine du management cette boucle peut prendre plusieurs traits, par exemple, un trait temporel (changement de forme et amplitude des relations), un trait en spirale (apprentissage organisationnel). Par exemple, chez Nonaka et Takeuchi, le trait en spirale se trouve dans la "boucle récursive" de transfert des connaissances.
[37] Nous pensons que cette opération de distinction laisse supposer l'existence d'une certaine loi à respecter lors de la construction (production) de la structure, comme nous l'avons dit plus haut.

En ce sens, l'organisation maintient (conserve selon Ashby) l'organisation de la structure par l'existence d'une certaine loi à respecter lors de la construction (production ou structuration) de la structure. Par exemple, si l'on fait une analogie avec la "production d'un système vivant" selon l'approche autopoïétique de Santiago « il y a un patron d'organisation commun qui peut être identifié dans tous les êtres vivants » [Maturana et Varela, 72], ou si l'on fait une analogie avec la "production d'un système viable" selon l'approche autopoïétique de Valparaiso, « il y a un patron d'organisation commun qui peut être identifié dans toutes les entreprises »[38] [Limone, 77], [Limone et Bastias, 02a]. Il semblerait alors, que la caractéristique commune soit donnée par la clôture organisationnelle du système qui existe dans un domaine conceptuel (espace social) qui comporte un *patron commun* des relations entre processus de production de composants qui doivent être respectées pour maintenir, tout au long du temps, l'organisation de l'unité et l'identité du système, dans un domaine physique (espace matériel).

Il nous faut maintenant argumenter le trait d'ouverture organisationnelle du système. En ce sens, l'organisation transforme (produit ou relie selon Morin) la structure de l'organisation par son fonctionnement, par son actionnement [Le Moigne, 90], [Mélèse, 72], par ce qu'elle sait faire [Limone et Bastias, 02a], par une volonté de "solidarité et solidité" [Morin, 86] ou de "cohérence et cohésion" [Limone, 77] comme l'on vient d'argumenter plus haut afin de construire [Piaget, 74], produire [Maturana et Varela, 72] ou structurer [Limone, 77] l'unité (l'organisation) et d'être unis dans sa totalité (système). Or, cette transformation demande une structure que matérialise l'organisation. En d'autres termes, selon l'approche autopoïétique de Santiago la structure de l'organisation est caractérisée par « les relations entre processus de production de composants qui doivent exister pour former une unité, comme produit d'une organisation » [Maturana et Varela, 72]. Egalement, pour l'approche autopoïétique de Valparaiso « à la manière concrète par laquelle est fait un système ou une unité dans l'espace dans lequel il peut exister » [Limone, 77]. Pourtant, la caractéristique commune est donnée par l'ouverture organisationnelle du système qui doit exister dans le domaine physique (espace matériel) de relations entre processus de production de composants qui matérialisent l'organisation (l'unité) dans un espace réel des propriétés et attributs de composants et leurs relations pour maintenir l'unité (le fait d'être distinguable de son environnement et donc des autres unités) et l'identité (le fait d'appartenir ou non à une classe) sous la contrainte de l'autonomie du système et le fait de l'identité.

---

[38] L'organisation est indispensable à l'entreprise, et c'est peut être pour cette raison que dans le domaine du management parler d'entreprise ou d'organisation, c'est parler de la même chose.

Au fond, la différence entre organisation et structure se trouve dans la dualité clôture et ouverture organisationnelle[39], puisque à la fois l'organisation maintient (conserve) et transforme (produit) la structure. Par exemple, si l'on suppose que l'université est un système (classe), alors il y a un patron de l'organisation commune qui garantit son identité comme université (classe). Chaque université (objet) doit alors respecter tout au long du temps, la charte édictée par le patron. Autrement dit, c'est le patron de l'organisation commune qui doit être l'héritier pour chaque université (objet) de la classe université afin qu'il soit le garant pour maintenir (conserver) la façon d'organiser la structure. Et donc, l'on voit bien apparaître le fait que l'organisation de la structure détermine le fait d'appartenir ou non à une classe (système = université). Ainsi, si nous parlons de deux universités, personne ne mettra en doute que la "solidarité et solidité" (en utilisant les termes de Morin) et la "cohérence et cohésion" (en utilisant les termes de Limone) entre ces deux entités sont différentes, et finalement, la structure de l'organisation de chaque université est distincte, mais toutes les deux appartiennent à la classe université, c'est-à-dire conservent la même organisation. Comme l'a dit Limone en citant Feibleman et Friend « c'est "cette" organisation spécifique qui détermine "ce" que la chose ou système est » [Limone, 77].

Bref, la dualité organisation/structure, selon les approches autopoïétiques de Santiago et de Valparaiso réside dans le fait que l'organisation de la structure est l'ensemble des relations entre processus de production de composants qui définissent un système comme une unité, tandis que la structure de l'organisation est l'ensemble des propriétés de ces composants et leurs relations qui matérialisent l'ensemble de relations définissant l'organisation de l'unité. Cela signifie que l'un ne peut pas vivre sans l'autre, et donc (simultanément et nécessairement) le système est décrit par son organisation (domaine conceptuel ou social : la topologie des relations) et par sa structure (domaine physique : les matériaux des composants et des relations). Néanmoins, dans la description organisationnelle et structurelle qu'on veut faire du système il faut tenir compte du commentaire de Varela, lorsqu'il dit « il semble que nous soyons incapables de caractériser une classe d'organisation si on ne sait pas mettre les relations qui la définissent en rapport avec une structure particulière. Inversement, aucune structure particulière ne peut être utilisée pour rendre compte de la phénoménologie qu'elle engendre si on n'a pas défini son appartenance à une classe d'organisation ». En effet, dans l'organisation de la structure du système, la structure est subordonnée à l'organisation qui définit une unité commune : le système. Par contre, dans la structure de l'organisation du système, l'organisation est réalisée dans l'espace matériel de ses composants et leurs relations, c'est-à-dire sa structure.

---

[39] Ceci nous renvoi aussi à l'idée de système ouvert et système fermé dont discuterons plus loin comme nous l'avons dit plus haute.

Enfin, selon l'approche cybernétique de Mélèse, nous pensons que la dualité organisation/structure se trouve dans le fait que l'organisation est liée à l'existence des variables essentielles (objectifs), tandis que la structure est liée aux valeurs de ces variables. Néanmoins, il est à noter que nous n'avons pas trouvé dans la littérature de Mélèse, la discussion sur la dualité organisation/structure (que nous avons trouvé ailleurs), pour justifier le double espace existentiel de l'entreprise.

Nous n'irons pas plus loin dans cette conceptualisation des caractéristiques communes et différenciation des dualités organisation/système et organisation/structure que nous avons retenues dans cette thèse. Une bonne synthèse et explication des autres concepts de l'approche système se trouvent dans le chapitre 2 « L'organisation (de l'objet au système) » du livre de Edgar Morin [Morin, 77].

## 2.1.8. Dualité gestion/management[40]

Si l'on regarde les sites Web de la *Revue française de gestion*[41] et du *Journal Information for Management Science*[42], on peut constater que les domaines de recherche de ces deux revues sont les mêmes. Nous sommes alors tentés de dire que le terme *gestion* correspond à la traduction du terme anglais *management*. Cependant, ces deux expressions n'ont pas la même signification comme nous l'avons constaté, lorsqu'on dit « la *Revue française de gestion* est la principale publication francophone de vulgarisation scientifique dans le domaine du management … le management concerne tous les secteurs de l'économie … dans les différents secteurs de la gestion (stratégie, marketing, finances, comptabilité, production…) … ce raisonnement gestionnaire s'applique aussi bien aux disciplines directement économiques qu'aux aspects humains de l'entreprise ». Cela prouve à notre avis que le management est plus que de la gestion, dans le sens qu'il est plus proche de l'individu.

Il nous parait intéressant d'argumenter davantage ce que l'on vient de dire à partir de deux points de vue : managérial et administratif.

---

[40] Non par hasard, il y a un changement de noms de deux ouvrages récents chez Jean-Louis Ermine : *La gestion des connaissances* (publié en 2003), et *Management des connaissances en entreprise* (publié en 2004), écrit avec Imed Boughzala.
[41] http://rfg.e-revues.com/acceuil.jsp
[42] http://www.jstor.org/journals/00251909.html

Du point de vue du managérial (ou business management), les businessmen des entreprises (les managers, les dirigeants, les administrateurs, les gestionnaires, les gérants, les cadres, etc.) dans le monde économique d'aujourd'hui en plus d'avoir des compétences techniques[43] pour gérer des tâches administratives des différentes départements (production, marketing, comptabilité, etc.) de l'entreprise, doivent avoir aussi des compétences personnelles et sociales[44] pour gérer les ressources humaines de l'entreprise. Dans le premier cas ce qui importe, est la connaissance des tâches (fonctions) administratives, tandis que dans le second cas, c'est la connaissance des relations humaines. Sans oublier aussi que les styles de management sont contraignants d'une part, par les modèles de gestion (ou management), par exemple, américain, français, japonais que l'on a choisi et d'autre part, par les aspects culturels de chaque pays où ces modèles sont appliqués ; d'où l'importance de l'approche managériale (ou business management) dans la globalisation des entreprises. Avant de passer à l'autre point de vue nous voulions signaler trois auteurs où la pensée système qui se fait sentir dans leurs réflexions, nous parlons de Henry Mintzberg, Peter Drucker et Michel Crozier, qui, pour nous, font partie de l'axe fondamental pour guider une approche managériale moderne. En ce sens, nous devons mentionner que d'après Henry Mintzberg la phénoménologie de l'entreprise est décrite par le couplage structurel de six composants d'un système organisé à savoir : (1) centre opérationnel ; (2) ligne hiérarchique ; (3) sommet stratégique ; (4) technostructure ; (5) fonctions de support logistique ; et (6) culture ou idéologie. Les relations qui lient ces composants pour former une unité, sont appelées *mécanismes de coordination*. Le couplage structurel de ces relations entre composant est le suivant : relation par supervision directe entre (3)/(1) ; relation par standardisation des procédés et des résultats entre (4)/(1) ; relation par standardisation de l'idéologie et des qualifications entre (6)/(1) ; relation par ajustement mutuel entre (1)/(1). Nous pensons que dans le langage de Mintzberg la frontière de l'organisation de l'unité dans sa totalité est la culture ou idéologie : les composants 2 et 5 ne participant pas dans le couplage structurel.

Du point de vue de la théorie administrative, nous voulions remarquer l'existence d'une volonté plutôt historique sur la popularité du terme management non seulement par rapport au terme gestion, mais aussi par rapport au terme administration, tel que le souligne Jacques-Marie Vaslin dans un article, intitulé : *Henri Fayol, théoricien du management*, lorsqu'il dit « aujourd'hui Taylor est considéré comme le père des théories du management … Fayol a tendance à sombrer dans l'oubli. Triste fin pour un inventeur recherchant la postérité »[45]. Il semblerait, d'après cet article, qu'aujourd'hui on parle plus de taylorisme que de fayolisme, et c'est peut être pour cette raison que le management a pris de la distance par rapport aux autres[46]. Dans cette thèse nous allons aussi l'utiliser, bien qu'à certains moments, nous préférions employer l'expression "domaine du management" pour impliquer beaucoup plus, le rôle des systèmes d'information dans le management.

---

[43] Dans la pensée administrative classique « administrer c'est prévoir, organiser, commander, coordonner et contrôler » disait Henri Fayol.
[44] L'Association pour l'emploi de cadres, donne 14 catégories génériques de compétences : décider, gérer, diriger, administrer, produire, organiser, communiquer, développer, chercher, former, contrôler, créer, négocier, conseiller. Où chacune des catégories est subdivisée, par exemple, l'aptitude d'innover se trouve classifier dans la catégorie créer. http://www.apec.fr/
[45] Apparu dans Le Monde du 30/01/04 http://www.lemonde.fr/web/article/0,1-0@2-3234,36-350463,0.html.

Ainsi, un projet de gestion des connaissances doit s'inscrire aussi bien dans une problématique de systèmes d'information ; que dans une problématique de business management. Nous avons préféré mettre le terme business dans le knowledge management pour impliquer davantage la créativité et l'innovation dans le management, puisque selon nous, l'expression *business management* implique en soi une approche pratique, une sorte de recherche de *quelque chose*, par exemple une idée capable de produire un succès économique. Cela signifie des fois, que le projet de gestion des connaissances peut être supporté technologiquement et organisationnellement selon une approche système d'information, il faut lui trouver un sens (une raison d'être) pour l'entreprise, c'est-à-dire il faut lui trouver une justification réelle d'un point de vue économique et une valeur ajoutée afin d'intégrer cette nouvelle technologie dans le parc informatique de l'entreprise.

## 2.2.    Les modèles de description de l'unité

Un modèle de description de l'unité peut être défini de façon générique comme une *description* d'un phénomène qu'on appel l'*unité*. Pour Varela, l'unité est « le fait d'être distinguable de son environnement et donc des autres unités » [Varela, 89]. Cette description du phénomène observé (l'unité), existe, simultanément et nécessairement, dans deux domaines. L'un est la *détermination*, l'autre est la *composition*. Autrement dit, pour l'observateur, l'unité existe (simultanément et nécessairement) dans le domaine de la *détermination* et dans le domaine la *composition*. On a donc, deux existences parallèle (simultanément et nécessairement) d'une même unité, on parle alors de l'unité dans le domaine de la détermination, et simultanément et nécessairement, de l'unité dans le domaine de la composition[47].

---

[46] Les racines de l'administration, de la gestion et du management se trouvent dans deux courants de la pensée administrative classique en France et aux Etats-Unis. Au début du siècle passé Henri Fayol parlait des 14 principes pour une bonne gestion des entreprises dans son livre *Administration industrielle et générale*, tandis qu'à la même époque Frederic Taylor parlait des principes du management scientifique des entreprises dans son ouvrage *Principles of Scientific Management*.
[47] Nous verrons que pour l'autopoïèse de Santiago, comme l'a dit Varela, « l'unité est la seule condition nécessaire à l'existence d'un domaine donné, quel qu'il soit » [Varela, 89]. Pour lui « l'unité est intimement liée à l'organisation et au fonctionnement de l'unité, et qu'elle a lieu dans l'espace physique où sont spécifiés cette organisation et ce fonctionnement … l'organisation de l'unité est orientée vers le maintien de son organisation ». Nous n'avons aucun doute que ceci est l'esprit du modèle CIBORGA, mais nous supposons que pour les autres modèles (OID, OIDC, AMS et MSV) vaut aussi.

Cette distinction de l'unité dans le domaine de la détermination, et de l'unité dans le domaine de la composition est possible grâce à une opération de distinction relative à la dualité observateur/observé, comme nous l'avons dit plus haut.

Dans cette section nous présentons les cinq modèles qui forment le cadre conceptuel de cette thèse, et qui sont à la base des approches organisationnelles, biologiques, managériales, et NTIC du KM de la gestion des connaissances, que nous avons développé au chapitre 1. Nous rappelons que dans tous ces modèles proposés (OID de Jean-Louis Le Moigne, OIDC de Jean-Louis Ermine, AMS de Jacques Mélèse, MSV de Stafford Beer, et CIBORGA d'Aquiles Limone et Luis Bastias), comme nous l'avons déjà dit au début du chapitre, la problématique centrale est la recherche « d'un patron d'organisation commune » qui permet d'expliquer l'organisation du système (ce patron d'organisation commune est mis en évidence, d'une part, à travers des relations entre composants pour les modèles OID, OIDC, AMS, et MSV, et d'autre part, à travers des relations entre processus de production de composants pour le modèle CIBORGA). Pour un souci de clarté et étant donné le pouvoir explicatif des modèles, nous avons décidé de les regrouper dans trois modèles génériques (système, cybernétique et autopoïétique) basés sous les approches système, cybernétique et autopoïétique qui les sustentent. Bien sûr nous n'allons pas reproduire ici l'explication des modèles OID, OIDC, AMS, MSV et CIBORGA avec les détails des auteurs concernés, mais plutôt mentionner sa source d'inspiration et utiliser certaines réflexions professionnelles (que nous avons faites au fil du temps comme maître de conférence en informatique de gestion dans l'Université Catholique du Maulé où nous utilisons principalement les modèles OID et OIDC pour expliquer l'organisation d'un système d'information et d'un système de connaissance).

## 2.2.1. Le modèle système

Le terme système vient du grec συνιστημι = *sunistànai* (permet de s'accorder, s'intégrer) [Le Moigne, 77]. Il prend ses racines dans la dualité observateur-observé[48], que nous symbolisons par deux modèles, à savoir : le *modèle système* et le *modèle analytique*.

Un modèle système (en opposition à un modèle analytique) permet de décrire la phénoménologie selon l'approche système (ou cybernétique d'ordre zéro) [Le Moigne, 90]. Cela signifie que dans un modèle système, on fait l'hypothèse, d'une part que la réalité physique peut être observée comme un système (entrées/transformation/sorties), et d'autre part que la réalité est dévoilée par l'observateur à partir d'une opération de distinction des relations entre composants, c'est-à-dire que l'observateur est impliqué ! En conséquence, un processus de causalité circulaire (la dualité cause/effet) entre l'observateur et l'observé va se mettre en place petite à petite dans l'explication du phénomène observé. Tandis, que dans un modèle analytique la réalité physique existe en dehors de l'observateur. La réalité est expliquée alors en termes de propriétés ou attributs des composants et des relations entre composants.

---

[48] L'observateur a en soi un projet téléologique (finalités) et ses limites d'observation. L'observé est la réalité. Par exemple, moi, je place mon point d'observation dans le génie industriel, et mes limites sont la gestion des connaissances imparfaites. Ma réalité est une entreprise du secteur industriel (réalité dans laquelle la connaissance existe simultanément et nécessairement dans un domaine : *flou* et *non flou*).

Bref, dans le modèle système l'observateur s'intéresse aux relations entre composants, tandis que dans le modèle analytique l'observateur s'intéresse aux propriétés des relations et des composants.

L'opération de distinction dans le modèle système permet à l'observateur de décrire l'unité dans sa totalité (l'organisation du système) à travers le *fonctionnement* des relations entre composants [Limone, 77], c'est-à-dire que la distinction est guidée par tout ce qui concerne la question fondamentale *qu'est-ce qui fait* l'unité ? Ce qui importe alors pour l'observateur qui a la charge de la construction du modèle, c'est la description des actions ou fonctions de l'unité. Comme l'a souligné Le Moigne « la modélisation systémique part de la question : "qu'est-ce que ça fait ?" : quelles sont les fonctions et les transformations, ou les opérations assurées ou à assurer pour cela » par l'unité dans sa totalité ? Nous pensons que cette distinction a un sens si elle est menée parallèlement ou en relation avec un modèle analytique. Rappelons que dans ce cas, ce qui importe pour l'observateur c'est de répondre à la question fondamentale soulignée par Le Moigne « la modélisation analytique : "de quoi c'est fait ?" : quels sont les éléments constitutifs, les objets ou les organes, dont la combinaison constitue ou peut constituer le phénomène (que l'on appelle plus souvent alors un objet) à modéliser » [Le Moigne, 90]. Et donc, l'opération de distinction dans la modélisation analytique permet à l'observateur de décrire l'unité dans sa totalité (l'organisation du système) à travers la *définition* des relations entre composants à partir des propriétés ou attributs [Limone, 77], tandis que dans la modélisation systémique, on décrit davantage le *fonctionnement* des relations entre composants.

Nous ne pensons pas que dans l'esprit de Le Moigne, il y ait une opposition de modèles, mais plutôt un souci méthodologique et pédagogique, afin de bien saisir la dualité de l'un par rapport à l'autre. Au fils du temps, la pratique a montré que les deux modèles, comme outils explicatifs pour décrire, représenter, expliquer, comprendre, bref pour modeler la phénoménologie (une réalité physique, quelque chose selon Le Moigne), sont finalement complémentaires [Jiménez *et al*, 99]. En effet, la description selon l'approche analytique (la structure du système : les attributs des relations entre composants) et l'approche système (l'organisation du système : les relations) forment un couple qui s'enrichit l'un par rapport à l'autre, mais l'usage dépend de l'observateur.

Dans ces modèles il y a une véritable stratégie de modélisation à suivre. En effet, l'un ou l'autre modèle a son utilité, selon le point de vue de l'observateur, autrement dit de la description qu'il veut faire de la phénoménologie étudiée. Comme l'a souligné Le Moigne « la représentation d'un phénomène perçu complexe par un système repose sur une hypothèse explicite d'opérationnalité irréversible, téléologique et récursive. Autrement dit, modéliser un système complexe, c'est modéliser d'abord un système d'actions. On ne cherche pas à représenter des choses, des objets, des éléments finis, des organes, comme on le ferait en modélisation analytique » [Le Moigne, 90]. Si le modèle permet de répondre à la question fondamentale *de quoi c'est fait* nous sommes dans la modélisation analytique, par contre si la question fondamentale est *qu'est-ce que ça fait* nous sommes dans la modélisation systémique. Autrement dit, dans la modélisation analytique, l'opération de distinction est la *définition* des propriétés ou attributs des composants et leurs relations, (afin de matérialiser la structure de l'organisation), tandis que dans la modélisation systémique, l'opération de distinction est le *fonctionnement* de l'organisation de sa structure qui permet de décrire la mise en action de la organisation du système. Comme l'a souligné Le Moigne « Le concept de base de la modélisation systémique n'est pas l'objet, ou la combinaison d'objets stables (la structure), mais l'action que l'on représentera donc systématiquement par la boîte noire ou le processeur symbolique qui rend compte de l'action » [Le Moigne,90].

Le Moigne lui-même a mis en évidence deux outils explicatifs pour mettre en valeur la complémentarité de la modélisation analytique et systémique. L'un est le triangle systémique, qui décrit la phénoménologie selon trois points de vue : (1) la structure, c'est-à-dire les attributs des relations entre composants selon l'approche analytique ; (2) l'organisation fonctionnelle, c'est-à-dire les relations entre composants selon l'approche système ; et (3) l'organisation comportementale, c'est-à-dire les actions des relations entre composants selon l'approche système. Nous constatons que dans la dualité organisation/structure de Le Moigne, l'organisation est attachée à l'approche système, tandis que la structure est attachée à l'approche analytique, se trouve la richesse du pouvoir explicatif triangle systémique de l'organisation du système.

Dans l'approche système la distinction entre *système ouvert* et *système fermé* se fait en fonction des rapports du système avec l'environnement. Si dans la transformation du système il n'y a pas de relations entre le système et son environnement, alors on parle de système fermé (ni entrées, ni sorties), dans le cas contraire on parle de système ouvert.

Ainsi, dans la dualité organisation/structure, peut-on savoir dans quel domaine on a une ouverture et une fermeture du système ?

D'ailleurs, cette question justifie davantage la nécessité d'une opération de distinction associée au point de vue cognitive chez l'observateur.

Dans les modèles système et analytique l'opération de distinction, est formulé par rapport à une dualité cause/effet sur la base de l'approche cognitiviste de la cognition.

Ainsi, cette opération de distinction permet indiquer que les causes et les effets sont distinguables dans des espaces fort différents. L'un est le domaine conceptuel (l'organisation) du processus organisationnel (c'est un processus conceptuel de description abstrait de relations entre composants de l'organisation). L'autre est le domaine physique (la structure) du processus structurel (c'est un processus physique de description matérielle de la structure à travers des propriétés des composants et leurs relations). Ainsi, le domaine conceptuel correspond à une modélisation systémique et le domaine physique correspond à une modélisation analytique. La figure 2.2 montre une synthèse du modèle système.

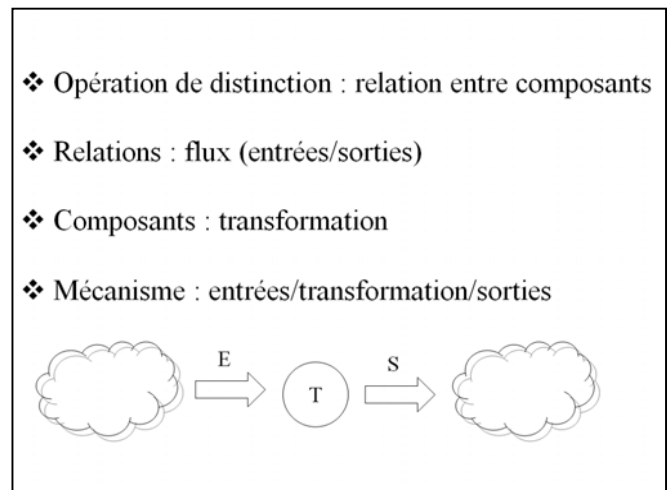

**Figure 2.2 :** Modèle système (source propre)

Ayant déjà clarifié le concept de triangle systémique chez Le Moigne, et le rôle de l'opération de distinction dans le modèle système et modèle analytique. Nous passons en revue, un autre outil conceptuel mis au point par Le Moigne, appelé *modèle d'un système organisé*, en abrégé modèle OID (Opération, Information, Décision) sur lequel s'appuie le modèle OIDC de Jean-Louis Ermine. Nous présentons ci-dessous les modèles OID et OIDC.

> ➢ **Le modèle OID**

Le modèle Opération, Information, Décision (OID) de Jean-Louis Le Moigne prend ses racines, d'une part dans le célèbre article, publié en 1956, de Keneth Boulding[49], intitulé : *General System Theory, the Skeleton of Sciences*, et d'autre part, sur la théorie générale des systèmes de Ludwig von Bertalanffy. Selon Le Moigne l'article de Boulding a ouvert pour la première fois une voie de recherche sur l'application de la modélisation systémique au management [Le Moigne, 90]. Mais c'est Bertalanffy qui a initié le débat et la réflexion sur le pouvoir explicatif de cette théorie dans d'autres domaines que la biologie, tel que l'observe Le Moigne à propos de la sortie en France du livre de Bertalanffy en 1968 « enfin rassemblés en un solide ouvrage, les textes, jusqu'ici introuvables du "père" de la "Théorie générale des systèmes" (les papiers les plus décisifs sont antérieurs à 1950). L'apport d'un grand biologiste à une réflexion qui affecte… toutes les disciplines de l'information comme celles de l'organisation. L'ouvrage vient d'être traduit en français : *Théorie général des systèmes*, Paris, Dunod, 1972 » [Le Moigne, 90]. Nous pensons que la richesse du pouvoir explicatif du modèle OID, mis en place par Le Moigne à partir des recherches de Bertalanffy et Boulding, se trouve dans deux traits, à savoir :

- l'organisation du système dans le modèle OID est décrit comme un complexe de composants en relation réciproque [Le Moigne, 90]. Ici, la complexité vient du fait que les composants sont eux-mêmes des processus ou transformations qui forment un réseau ou une chaîne de transformations, et comme nous le verrons dans le deuxième trait, au lieu de simplifier une réalité physique qu'on veut modéliser, le modèle OID est complexifié par neuf niveaux interrelationnés d'explication. C'est justement dans cette relation et l'émergence de cette relation, par une sorte de source "d'énergie"[50], entre des composants, entre des relations, avec le tout, les parties (composants), etc., que se trouve, selon nous, la richesse du modèle OID pour expliquer l'organisation du système général ;

- l'organisation du système dans le modèle OID est donnée par la distinction de neuf niveaux interrelationnés d'explication des composants et de leurs relations. La démarche de construction de ses neuf niveaux selon l'approche système (modélisation analytique et systémique) se trouve détaillée bien entendu dans [Le Moigne, 90] mais aussi dans [Durand, 98], [Lugan, 00] et [Nanci *et al*, 96]. Nous trouvons inutile de rentrer ici dans les détails de cette démarche car elle est connue de tous, mais ce que nous voulons faire est plutôt d'expliquer un peu plus la mécanique de l'organisation du système selon le modèle OID. Cette mécanique suppose que l'unité du modèle OID est définie par une opération de distinction.

---

[49] Les premiers travaux chez Boulding sont de 1935, c'est pour cette raison que nous plaçons la naissance de la théorie des systèmes à cette date-là.
[50] Ici nous pensons à l'approche biologiste mis en place par Joël de Rosnay pour expliquer une vision globale du monde à partir de ce qu'il appelle « le macroscope » [Rosnay, 75].

L'opération de distinction permet de dissocier trois composants dans ces neuf niveaux : le *système opérant*[51] (le fait de faire quelque chose, le business), le *système d'information*[52] (données précis et certaines du système opérant) et le *système de décision*[53] (opérationnel, tactique, stratégique), et leurs relations (Le Moigne emploie le terme "articulation" [Le Moigne, 77]), dans lequel le système opérant assure la transformation des entrées en sorties, selon le pilotage du système de décision décrit en termes de structure décisionnelle et de leurs processus de décision (l'organisation décisionnelle). Le système d'information mémorise d'une part les données en entrée et sortie des transformations du système opérant, et d'autre part le système d'information joue un rôle de système de support aux décisions stratégiques, tactiques, opérationnelles pour piloter le système opérant. Pour cela, le système de décision, est lui-même organisé à la fois en trois composants : le *système de finalisation*[54] (la fixation des buts, objectifs), le *système d'intelligence*[55] (la conception des buts) et le *système de coordination*[56] (la sélection des buts). Dans le modèle OID nous avons trois types de flux à la base des relations : les flux de données (entrées et sorties du système opérant), les flux d'informations (entrées et sorties du système d'information), et les flux de décisions (entrées et sorties du système décision).

Pour distinguer ces neuf niveaux, l'observateur doit appliquer deux opérations de distinction selon qu'il s'agit d'une modélisation analytique ou systémique comme nous l'avons vu plus haut. La première opération de distinction (définition de la structure) permet d'identifier de quoi est fait la structure de l'organisation du système opérant, d'information et de décision ; tandis que la deuxième opération de distinction (fonctionnement et comportement de l'organisation) permet d'identifier de quoi est faite l'organisation de la structure des systèmes opérant, d'information et de décision. Ainsi, le modèle OID permet de rendre compte de la dualité organisation/structure de ces trois composants dans un système général[57].

---

[51] Le niveau 1 « le phénomène est identifiable ». Le niveau 2 « phénomène est actif : il "fait" ».
[52] Le niveau 3 « le phénomène est régulé ». Le niveau 4 « le phénomène s'informe sur son propre comportement ». Le niveau 6 « le système mémorise ».
[53] Le niveau 5 « le système décide de son comportement ».
[54] Le niveau 9 « le système se finalise ».
[55] Le niveau 8 « le système imagine et conçoit de nouvelles décisions possibles ».
[56] Le niveau 7 « le système coordonne ses décisions d'action ».
[57] Le Moigne parle aussi des *systèmes complexes*, voir son livre *La modélisation des systèmes complexes* qui a popularisé le terme "complexité", largement appliqué aujourd'hui en management. Par exemple, dans le livre des deux cadres d'entreprises Alain Boyer et Guillaume Gozlan, intitulé *10 Repères essentiels pour une organisation en mouvement*, ils dissent « l'entreprise est un système complexe avec des interactions multiples dans leur typologie, dans leur fréquence et leur volume. Comme tout système "vivant", l'entreprise est en perpétuelle évolution avec alternance de périodes de pseudo-stabilités et de crise. Une bonne part de cette stabilité ou ces crises est due aux interactions complexes, parfois complémentaires entre elles, parfois antagonistes, avec l'environnement. L'entreprise puise une très grand partie de son énergie, de ses ressources, de ses moyens à l'extérieur » [Boyer et Gozlan, 00]. Il y a là, à notre avis, une définition

---------------------------------------------------------------------------------------------------------

D'après, Le Moigne, c'est justement le mélange (ou l'articulation), c'est-à-dire la complexification et non pas la simplification de tous ces composants et leurs relations dans les neuf niveaux qui permettra l'émergence de la description de l'organisation du "système". Il s'agit ici d'un système qui englobe les autres et que Le Moigne appelle *général* (pour indiquer un phénomène perçu complexe par un système, la réalité physique que l'on veut modéliser, l'unité dans sa totalité, l'organisation du système) [Le Moigne, 77]. Ce qui est le plus intéressant à noter dans le modèle OID est que chaque composant (système ou sous-système si l'on veut) du système général est décrit selon l'approche système, c'est-à-dire comme un flux d'entrées/transformation/sorties. Par exemple, si la réalité physique est une machine, le système opérant permet de décrire la machine comme un ensemble de flux de transformations pour transformer les entrées (les matières premières) en sorties (les produits). Par contre, le système de finalisation permet de décrire la machine comme une transformation d'objectifs de production. Le Moigne, décrit les entrées du système de finalisation comme les sorties du système d'intelligence, et inversement les sorties du système de finalisation sont les entrées du système d'intelligence.

Enfin, nous voudrions indiquer que cette démarche et canevas de pensée ont été largement utilisés en France et ailleurs au fils de temps pour modéliser des *systèmes complexes*, c'est-à-dire les systèmes où la contrainte (l'ordre) fait apparaître un processus de causalité circulaire entre les causes et les effets[58] propre pour maintenir l'unité (l'organisation), l'identité (la structure) et l'autonomie (la dynamique) du système. Or, selon notre expérience professionnelle (comme maître de conférence à l'Université Catholique du Maulé, en charge des cours de *systèmes d'information*[59] et de *systèmes de gestion*[60]), nous avons trouvé que l'avantage du modèle OID (Opération, Information, Décision) réside dans le fait de montrer que l'information qui compte, c'est-à-dire a un sens, est l'information qui permet de réguler une opération (activité) que l'on veut gérer par les données en entrée et sortie, et d'autre part que cette information a un sens seulement si elle contribue au processus de prise de décision. Nous avons utilisé cette constatation, d'une part dans la modélisation des systèmes d'information pour la gestion opérationnelle des flux d'entrées/transformation/sorties du système opérant à partir de la méthode Merise et de ses évolutions [Nanci *et al*, 96], et d'autre part dans la modélisation des systèmes de décision pour la gestion tactique et stratégique des flux d'entrées/transformation/sorties du système de décision sur la base d'une méthode popularisée au

---

opérationnelle de la complexité, bien que dans les discours chez Humberto Maturana, Francisco Varela et Aquiles Limone la complexité est toujours présente comme nous l'avons vu plus haut.

[58] Ce processus est attaché à la cognition, alors si le processus est relative à l'approche cognitiviste de la cognition, alors ce que l'on cherche à construire est une chaîne logique de cause à effet, tandis que si le processus est associé à l'approche enactiviste de la cognition, alors la relation ne se construit pas nécessairement d'une représentation vrai ou logique de causes et des effets, puisque le problème de l'enaction n'apparaît pas lié à la représentation symbolique d'une réalité, sinon qu'au maintien du système en vie et viable.

[59] C'est la Théorie générale des systèmes appliquée à la Théorie de l'information.

[60] C'est la Théorie générale des systèmes appliquée à la Gestion.

milieu des années 90 aux Etats-Unis par Robert Kaplan et David Norton, que l'on appelle *Tableau de bord prospectif*[61].

Pour Kaplan et Norton la *stratégie*, (c'est-à-dire la volonté et la nécessité de bien tracer un chemin stratégique afin d'avancer ou de continuer dans la direction envisagée dans la *vision* de l'entreprise pour maintenir l'unité dans sa totalité), est axée sur quatre plans : financier, clients, processus et apprentissage organisationnel. Nous tenons à remarquer que c'est dans le deuxième ouvrage de Robert Kaplan et David Norton, intitulé *Comment utiliser le tableau de bord prospectif*, que nous avons trouvé davantage d'informations sur l'utilisation du modèle OID dans la modélisation d'un système de gestion pour l'explication de la stratégie comme un système général, et ceci de façon implicite car ses auteurs ne font pas référence à la littérature de Le Moigne, et non plus aux autres auteurs systémiques qui l'utilisent pour expliquer les flux stratégiques que circulent au niveau opérationnel, tactique et stratégique dans l'entreprise.

L'inconvénient du modèle OID est qu'il ne permet pas le passage "automatique" à la construction d'une application informatique, comme support pour les systèmes d'information ou de décision de l'entreprise. En plus, dans la pluspart des projets informatiques, les données, générées et consommées par les transformations du système opérant, restent restreintes aux données précises et événements certains du domaine.

Après avoir présenté le modèle OID, nous allons aborder maintenant le modèle OIDC (Opération, Information, Décision, Connaissance) mis au point par Jean-Louis Ermine [Ermine, 96].

➢ **Le modèle OIDC**

Le modèle Opération, Information, Décision, Connaissance (OIDC) de Jean-Louis Ermine prend ses racines, comme nous l'avons déjà avancé, dans modèle OID de Jean-Louis Le Moigne. Il s'agit donc d'une extension du modèle OID pour la prise en compte d'un *système de connaissance* décrit au travers des *flux cognitifs* et *flux de compétences*, c'est-à-dire qu'en plus de décrire l'organisation du système général à travers des flux de données (entrées et sorties du système opérant), des flux d'informations (entrées et sorties du système d'information), des flux de décisions (entrées et sorties du système décision), il faut la décrire en termes des flux cognitifs et des flux de compétences entre le système de connaissance et les autres composants (système opérant, système d'information et système de décision) du système général. De là, on peut déduire que la Connaissance est un "ingrédient" supplémentaire du modèle OID : *données* (système Opérant), *informations* (système d'Information), décisions (système de Décision) qui introduit une complexité supplémentaire dans l'organisation du système par la prise en charge de la *connaissance* (système de connaissance). Ceci signifie que la connaissance dans le modèle OIDC de Jean-Louis Ermine est décrite à travers l'approche système (d'entrées/transformation/sorties), c'est-à-dire à travers la relation système-environnement.

---

[61] En anglais, *balance scorecard*.

La mécanique de l'organisation du système selon le modèle OIDC peut être schématisée de la façon suivante : (1) la génération de données pour le système opérant ; (2) la régulation et le contrôle qu'exerce sur lui le système de décision par le biais de la sélection des buts (système de coordination), la conception des buts (système d'intelligence) et la fixation des buts (système de finalisation) ; (3) la mémorisation dans le système d'information de données (système opérant) et décisions (système de décision); et enfin (4) la mémorisation dans le système de connaissances générées par le système opérant, système d'information et le système de décision. A cet égard, nous voudrions signaler que la démarche à suivre pour l'explication du modèle OIDC par la complexification de niveaux de distinction, tel que neuf niveaux donnés par Le Moigne pour le modèle OID, n'a pas été explicitée dans le livre de Ermine. Nous pensons que cette explication n'apparaît pas immédiatement au premier regard et c'est peut être pour cette raison qu'elle n'a pas été explicitée. Nous n'avons pas l'intention de le faire ici. Néanmoins, la figure 2.3 montre la dynamique des flux de ce modèle.

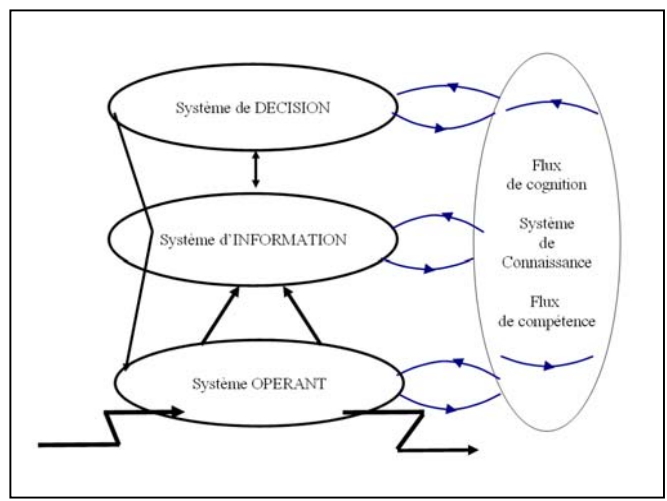

**Figure 2.3 :** Modèle OIDC (source [Ermine, 96])

La connaissance du modèle OIDC appartient au domaine du savoir-faire et du savoir-technique de l'expert dans son activité courante, c'est-à-dire de la connaissance métier qui prend forme à partir de l'action (le business exploité par les hommes et ses outils de production, soit le système opérant). Une application du modèle OIDC à la gestion des connaissances pour les projets de conception de produits innovants se trouve dans [Longueville *et al*, 01].

Enfin, il faut dire que le modèle OIDC a été la justification théorique pour la modélisation des *systèmes de connaissances* à partir des méthodes MASK/MKSM et ses évolutions. Il s'agit bien d'une représentation symbolique de la connaissance (selon l'approche cognitiviste de la connaissance) qui rejoint donc une partie du capital intellectuel, immatériel, intangible ou capital connaissances de l'entreprise.

Nous pensons que ces outils de l'intelligence artificielle (MASK/MKSM, CommonKADS/ KADS/, etc.) sont insuffisants pour la partie du capital connaissance qui ne rentre pas dans un modèle OIDC, et qui sont plus enracinés dans la nature humaine de l'individu, par exemple les compétences non technique requise pour établir une négociation avec un client, ou la prise de décision avec des données flous.

## 2.2.2. Le modèle cybernétique

La cybernétique voit le jour en 1940 au Massachusetts Institute of Technology (MIT) principalement par les recherches de Bigelow, Rosenblueth et Wiener sur les phénomènes de prédiction de comportement d'un système. Le cas particulier d'étude était : *la prédiction antiaérienne*[62]. Ils ont trouvé que *prédiction* et *contrainte* sont liées, comme l'a dit Limone, en citant Ashby « connaître les contraintes revient à connaître, en les identifiant, les transformations et les paramètres deviennent prévisibles ». Pour Ashby dans son livre, intitulé *Introduction à la cybernétique* (publié en 1958), cette étude sur la défense antiaérienne est à l'origine de la découverte du concept de la boucle (feed-back, rétroaction) négative qui a donné naissance au système de contrôle à circuit fermé[63] et à la mise en place de la cybernétique. Ce terme vient du grec κυβερνητικοσ = kubernétikos (pilotage). En 1945 Norbert Wiener la bâtit comme « la science du contrôle et la communication de la machine et l'animal ». Depuis-là, tel que le souligne Joël de Rosnay dans son livre *Le macroscope*, le mouvement cybernétique a été influencé par John von Newman (le paradigme de l'ordinateur) Warren McCulloch (modèles connexionistes), Gregory Bateson (le concept d'Ecology of Mind[64]), Jay Forrester (concept de dynamique industriel[65]), mais

---

[62] Eh oui! La machine de guerre depuis des siècles, nous a donnée nos plus belles découverts scientifiques. On dire que nous avons toujours besoin de détruire une partie de l'humanité pour créer et inventer. Le physicien américain Albert Einstein (que nous avons choisi comme citation du chapitre) est celui qui a payé le prix le plus fort pour le moment. Hélas pour lui !

[63] Sous l'angle du contrôle, un système est à circuit fermé quand il y a toujours réalimentation et circularité dans les opérations de contrôle, puisque les variables de sortie de la transformation agissent sur les variables d'entrée du système (transformation). En revanche, un système est à circuit ouvert quand il n'y a pas de réalimentation et circularité dans les opérations qui effectuent le contrôle, autrement dit, quand les variables de sortie de la transformation n'exercent aucune influence sur les variables d'entrée du système (transformation).

[64] Le livre de Bateson, intitulé *Steps to an Ecology of Mind* (publié en 1973) à été traduit en français comme *Vers une écologie de l'esprit* (publié en 1977).

[65] Chez Forrester l'activité industrielle est envisage par un réseau de boucles interconnectés. Ces boucles ont un symbolisme particulier pour décrire les différents dispositifs et flux d'activité d'une entreprise du secteur industriel, par exemple,

bien d'autres (où les noms n'apparaissent pas dans le schéma graphique de Rosnay dans son livre), ont contribué à apporter un regard nouveau dans le processus de contrôle et de régulation, tels que : Ashby (concept de variété[66]), Beer (concept de Brain of the Firm[67]), Mélèse (concept de variables essentielles[68]), etc.

Un modèle cybernétique enrichit alors le modèle système par le fait qu'il repère une autre caractéristique commune des systèmes la *contrainte*. En effet, c'est l'apparition de la contrainte qui régule l'*état des changements* de la structure de l'organisation du système (l'unité et les parties), autrement dit l'état des changements des *variables* (valeurs des attributs ou propriétés des composants et leurs relations) afin de maintenir l'unité dans sa totalité (pour maîtriser l'état des changements du système). Par exemple, si l'on pense au modèle OID, la prise en compte du changement du système se fait par les buts du système de décision et les données du système opérant contenues dans les variables et valeurs de ce système pour un environnement donné.

Comme nous l'avons dit, dans un modèle système, la description de l'organisation du système est caractérisée par : (1) les flux (échanges entrées/sorties) entre le système et leur environnement ; (2) les transformations du système ; et (3) par les opérations de distinction, d'une part dans la modélisation analytique (définition), et d'autre part dans la modélisation systémique (fonctionnement et comportement). Or, la nouveauté de la modélisation cybernétique réside dans la prise en compte des changements du système par de *variables d'état*. Dans ce contexte, nous pensons que la modélisation cybernétique a enrichi la modélisation systémique.

Cela signifie aussi qu'aux questions fondamentales des opérations de distinction : *qu'est-ce que ça fait ?* (modélisation systémique) pour décrire le fonctionnement et comportement de l'organisation (le faire du système) et *de quoi c'est fait ?* (modélisation analytique) pour décrire les composants et leurs relations de l'organisation (la structure du système), il convient d'ajouter la question fondamentale *dans quel état c'est ?* (modélisation cybernétique). Un système est déterminé, comme l'a souligné Mélèse, « si l'on sait dresser la liste des états des entrées (entrées extérieures et variables d'action) et la liste des états correspondants des sorties et des variables essentielles » [Mélèse, 72]. Ce qui revient a décrire l'unité dans sa totalité, selon nous, par le triangle systémique de

---

valvules, matières premières, temps d'attente, etc. Nous soulignons au passage que les modèles AMS de Jacques Mélèse et MSV de Stafford Beer ont été fortement influencés par le concept de dynamique industriel de Forrester.
[66] Ashby introduit le concept de variété en 1965. Il s'agit d'une mesure pour mesurer le nombre d'éléments différents que comporte un ensemble. Limone dit, en citant Mélèse, « on peut le définir encore comme le nombre d'états différents que peut revêtir le système, en sachant que dans un système il y aura un certain nombre de relations différentes entre ses éléments et, en conséquence, d'états différents de ses relations » [Limone, 77].
[67] Beer introduit le concept de management en gestion, comme liée au concept de contrôle en cybernétique, et donc pour lui, gérer c'est contrôler. Dans la section 2.2.2 nous présentons le modèle MSV (Modèle des Systèmes Viables).
[68] Mélèse introduit le concept de variables essentielles en gestion, comme une mesure de performance, à ce propos il dit « ce sont les variables essentielles, qu'on peut considérer comme des critères mesurant la réussite de la mission confiée à l'organisme » [Mélèse, 72]. Dans la même section nous présentons le modèle AMS (Analyse Modulaire des Systèmes).

---------------------------------------------------------------------------------------------------

Le Moigne. En effet, le point de vue de la modélisation analytique (la structure du système) décrit les composants et leurs relations au système, par contre le point de vue de la modélisation systémique (le faire du système) décrit le fonctionnement et le comportement du système, par contre le point de vue de la modélisation cybernétique (le devenir du système) décrit les changements du système.

Ainsi, la communauté cybernétique synthétise ses contributions dans trois grandes approches : la cybernétique d'ordre zéro qui correspond à la théorie des systèmes (environnement/système, entrées/transformation/sorties, unité/parties) de Keneth Boulding (1935), la cybernétique de premier ordre (la contrainte) de Norbert Wiener (1945), et la cybernétique de deuxième ordre (l'auto-organisation) de Heinz von Foerster (1962).

La distinction entre les approches de Wiener et Foerster se fait en fonction de la dualité observateur-observé. Lorsque l'observateur est une composante en plus de l'organisation du système, on parle de cybernétique de deuxième ordre. Dans ce cas, il y a une relation de causalité circulaire qui se créée entre l'unité, la contrainte (variables d'état), les parties et l'observateur (son projet téléologique et ses limites). En revanche, si l'observateur est extérieur à l'organisation du système, alors on parle de cybernétique de premier ordre. Enfin, si la description du système (transformation) se fait seulement par ses variables d'entrée et de sortie, alors on parle de cybernétique d'ordre zéro ou simplement théorie des systèmes ou approche système.

Dans l'approche cybernétique (de deuxième ordre) la distinction entre *système ouvert* et *système fermé* se fait en fonction de l'organisation de l'unité. Si dans la transformation du système (dualité système/environnement) il n'y a pas composants (dualité système/composants)[69] nouvelles qui arrivent de l'environnement ou qui partent vers l'environnement, alors on parle de système fermé. La structure du système ne change pas pour maintenir l'unité, car il n'y a pas de nouvelles contraintes. Dans le cas contraire on parle de système ouvert. La figure 2.4 montre une synthèse du modèle cybernétique.

---

[69] Nous préférons d'utiliser ici le terme *composants* (du système) au lieu d'utiliser les autres noms de l'approche systèmes, tels que : *parties* ou *éléments*, mais en gros signifient la même chose.

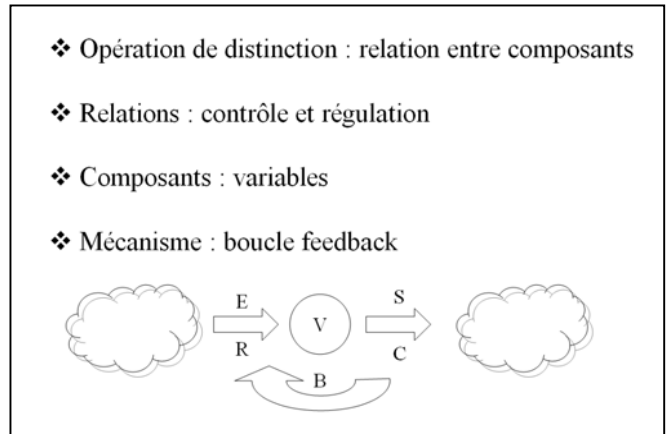

**Figure 2.4 :** Modèle cybernétique (source propre)

Les modèles cybernétiques (de deuxième ordre) que nous avons privilégiés sont les modèles de Jacques Mélèse et Stafford Beer[70], car ces auteurs se sont intéressés à observer les processus de contrôle et de régulation de l'entreprise d'un point de vue systémique et cybernétique. La base conceptuelle des différents modèles s'est faite à partir des travaux d'Ashby sur la cybernétique et de Bertalanffy à partir de la théorie générale des systèmes principalement. Spécifiquement leurs réflexions ont été centrées sur l'apparition, la détection et la maîtrise des divers problèmes d'efficacité et d'efficience, dont souffrent les entreprises dans leur production, coordination, planification, administration, direction, etc.[71]

Pour Mélèse et Beer penser l'entreprise comme un système vivant avait ses limites, il valait mieux la penser comme un système viable. En effet, comme nous avons dit un système vivant, d'une part est capable d'apprendre, d'évoluer, de s'adapter, etc., et d'autre part de survivre dans les conditions changeantes de son environnement. Par contre un système viable, en plus de vivant, est capable de se finaliser et de décider de son propre comportement pour survivre dans son environnement en changement dynamique permanent. Par exemple, les entreprises (ou les hommes) pour atteindre leurs objectifs doivent faire en permanence des choix pour survivre et leurs objectifs ne sont pas toujours opérationnels[72]. En d'autres termes, il faut observer l'entreprise par l'approche système et cybernétique agissant ensemble, c'est-à-dire comme un système (entrées/transformations/sorties) qui suivant les objectifs décisionnels s'organise, ainsi les changements du système sont contrôlés et régulés par les données décisionnelles et non pas par l'état de la machine. En effet, comme l'a souligné Mélèse « l'entreprise comme un système est plus qu'une machine, dans le sens que le système exprime simplement le fait que quelque chose entre et sort transformé » [Mélèse, 72], et où le contrôle (approche cybernétique) est possible par des mécanismes de régulation du système de pilotage.

---

[70] Ces auteurs ont contribué largement à montrer l'utilité de la modélisation cybernétique et systémique dans le domaine du management. Ceci principalement à partir des travaux d'Ashby sur la cybernétique et de Bertalanffy sur la théorie générale des systèmes. À propos de Beer, Le Moigne dit : « c'est un monument, qui est aussi le dossier le plus souvent cité à ce jour sur les applications de la cybernétique au management… » [Le Moigne, 90], et puis il ajoute à propos de la sortie du livre de Mélèse, intitulé : *La gestion par les systèmes,* « le principal mérite de cet ouvrage fut peut-être d'avoir été pratiquement le premier en langue française sur l'analyse systémique… » [Le Moigne, 90].
[71] L'esprit gestionnaire chez Mélèse et Beer se retrouve beaucoup moins dans la littérature de Le Moigne.
[72] Chez les hommes ceci nous renvoie à la hiérarchie des besoins selon Maslow.

Nous pensons que cette nécessité permanente de survie de l'entreprise est la justification même de l'existence d'une vision (sa raison d'être) pour construire sa mission (la stratégie pour atteindre les buts ou objectifs) qui sera matérialisée à travers un processus de décision (organisation) et leurs structures décisionnelles (structure), et ceci à tous les niveaux (stratégique, tactique, opérationnelle). La dualité organisation/structure montre bien que le processus décisionnel suppose nécessairement une structure concrète de support pour se réaliser [Limone, 77].

Pour observer l'entreprise comme un système viable, Jacques Mélèse a proposé un système organisé de *gestion*, en abrégé modèle AMS (Analyse Modulaire des Systèmes), tandis que Stafford Beer a proposé un système organisé de *management*, en abrégé modèle MSV (Modèle des Systèmes Viables). Nous avons laissé le terme gestion chez Mélèse et le terme management chez Beer pour respecter le côté historique des approches, l'une française et l'autre anglaise, mais comme nous l'avons dit plus haut, ces termes forment une dualité.

➤ **Le modèle AMS**

La structure de l'organisation du modèle AMS (Analyse Modulaire des Systèmes) de Mélèse est décrite par un ensemble de variables pouvant prendre diverses valeurs. Comme l'a souligné Mélèse « l'organisation est un "lieu de rencontre" » de variables et de valeurs [Mélèse, 72], [Limone, 77], [Mélèse, 79]. Le concept de variable considère alors de façon implicite, une certaine idée de mémorisation de la valeur de la variable à un instant donné de la transformation. On voit donc bien apparaître le rôle de l'information dans l'organisation de la structure du système (entrées/transformation/sorties). Le nombre de variables de ce modèle que le système de contrôle doit contrôler et réguler dans chaque niveau d'activité système de pilotage (stratégique, tactique, opérationnelle) a été regroupé en quatre, à savoir : les *variables d'entrée*, les *variables de sortie*, les *variables essentielles* et les *variables d'action*[73].

Variables d'entrées

Les variables d'entrée sont relatives aux entrées de la transformation (entrées/transformation/sorties), où leur valeur est une contrainte externe (l'environnement du système) et interne (l'opérateur qui opère la transformation du système). Autrement dit, le choix des variables d'entrée dépend de l'organisation du système (ce qu'elle doit faire), tandis que les valeurs des variables d'entrées sont fixées par la structure (ce qu'elle peut faire).

---

[73] Le variables d'entrée et de sortie son propres au système (entrées/transformation/sorties), tandis que les variables essentielles et d'action son propres au processus de contrôle et de régulation.

### Variables de sorties

Les variables de sortie sont relatives aux sorties de la transformation (entrées/transformation/sorties), où leur valeur est une contrainte externe (l'environnement du système) et interne (l'opérateur qui opère la transformation du système). Autrement dit, le choix des variables de sortie dépend de l'organisation du système (ce qu'elle doit faire), tandis que les valeurs des variables de sorties sont fixées par la structure (ce qu'elle peut faire).

### Variables essentielles

Les variables essentielles sont relatives à la mesure des objectifs (stratégiques, tactiques, opérationnels). Mélèse dit « ce sont les variables essentielles, qu'on peut considérer comme des critères mesurant la réussite de la mission confiée à l'organisme » [Mélèse, 72]. Le choit dépend de l'objectif qu'on veut mesurer, tandis que leurs valeurs renseignent sur la réalisation de la transformation (bon au mauvais), autrement dit sur la santé et l'efficacité du système [Limone, 77].

### Variables d'action

Les variables d'action sont relatives à la modification ou réglage des variables essentielles. Mélèse dit à propos des variables du système qu'elles sont à la disposition d'un opérateur pour modifier la transformation entrées-sorties [Mélèse, 72].

La mécanique générique du modèle AMS de Mélèse est la suivante : une fois qu'un opérateur stratégique du système de décision fixe la finalité du système, c'est-à-dire sa raison d'être, sa vision et sa mission, l'organisation de la structure du système est définie, et la construction de la structure de l'organisation du système peut commencer. Le système de contrôle fixe les objectifs des trois niveaux (stratégique, tactique, opérationnel) de décision et d'information de l'organisation. Ces objectifs seront matérialisés dans la structure de l'organisation du système par des variables essentielles. A son tour, le système de pilotage doit faire de même, mais cette fois avec des variables d'action. Le système est alors sous contrôle à tous les niveaux (stratégique, tactique, opérationnel), si et seulement si, le système est capable d'atteindre l'objectif, cela signifie que les valeurs des variables essentielles est alors atteint par les modifications et réglages des valeurs des variables d'action. La régulation doit ensuite modifier les valeurs des variables d'action pour réduire les écarts entre les valeurs visées et réalisées des variables essentielles dans un instant de temps. Autrement dit le système (entré/transformation/sorti) est sous contrôle si l'objectif est atteint. Ceci peut être décrit schématiquement : (transformation/contrôle) si et seulement si (transformation/objectif) à tous les niveaux (stratégique, tactique, opérationnel) de l'entreprise est atteint.

L'avantage du modèle AMS de Mélèse est de mettre en valeur le concept de variable dans le système de décision que nous ne voyons pas explicitement dans le modèle OID de Le Moigne et le modèle MSV de Beer. Le système de décision existe simultanément et nécessairement grâce au couplage du système de contrôle et du système de pilotage. En effet, le système de contrôle joue le rôle de contrôleur/contrôlé, afin de fixer des objectifs à tous les niveaux (stratégique, tactique, opérationnel) traduits par des valeurs des variables essentielles et simultanément des valeurs des variables d'action. Par contre, le système de pilotage a la charge de conduire les actions de contrôle des variables essentielles et simultanément de régulation des variables d'action, c'est-à-dire qu'il joue le rôle de régulateur/régulé des variables d'action afin de réduire les écarts des valeurs visées et réalisées des variables essentielles.

L'inconvénient du modèle AMS de Mélèse est la perte de la vision de l'organisation "totale" ou des "parties" du système selon une optique récursive *top-down* et *botton-up* que nous voyons dans le modèle OID de Le Moigne et le modèle MSV de Beer. En effet, le langage graphique du modèle AMS reste un peu lourd. On pourrait certes faciliter son utilisation comme outil d'analyse de l'organisation à partir de l'enchaînement de divers composants (connexions de modules, en utilisant le vocabulaire de l'AMS). Nous pensons que ceci doit rester un sujet de recherche, même si l'incorporation de nouveaux outils, par exemple une matrice avec des variables et valeurs, tel qu'un modèle BSP (Business Systems Planning) [Reix, 00] peut apparaître intéressant.

Un autre inconvénient du modèle AMS de Mélèse, mais aussi du modèle MSV de Beer et du modèle OID de Le Moigne est que dans la construction de la structure de l'organisation du système, l'aspect social n'est pas présent. A cet égard, Mélèse donne l'explication suivante « l'AMS n'introduit pas le "facteur humain" dans l'analyse, parce qu'elle introduit les hommes concernés en chair et en os ; le langage proposé devient un fait social dès lors qu'il est pratiqué et les aspects humains sont présents tout au long du déroulement d'une analyse » [Mélèse, 72].

C'est vrai qu'il n'y a pas de symbole dans l'analyse (que Mélèse appelle modulaire) pour représenter le facteur humain, tels que les symboles introduits pour représenter l'objectif, le pilotage et le contrôle, (comme dans les autres modèles MSV de Beer et OID de Le Moigne)[74]. Nous pensons que l'introduction d'un tel symbole au niveau de l'analyse afin de représenter les relations sociales (pouvoir, conflit, motivation, négociation, loyauté, etc.) n'offre aucun problème, mais la question est de savoir quelle en serait l'utilité. Par exemple dans le modèle de MSV de Beer, une ligne en zigzag permet de représenter les perturbations entre les composants du système opérant (que Beer appelle système d'implémentation), mais cela reste sur le plan conceptuel (autrement dit de l'analyse). Dans la conception de la structure de l'organisation du système cela peut se traduire, par le contrôle et la régulation de certaines variables d'action qui sur le domaine physique (par exemple une machine) sont possibles, alors que sur le domaine social (par exemple des émotions), le contrôle et la régulation cybernétique (de premier ordre) n'a pas beaucoup de sens. Le management des relations humaines pour l'atteinte d'un objectif d'organisation, échappe complètement à la métaphore de croire que l'entreprise est une machine mécaniciste.

En ce sens, nous sommes d'accord avec Mélèse : les aspects humains sont impliqués dans l'analyse de l'organisation pour diagnostiquer les disfonctionnement et les problèmes de santé dans l'entreprise, mais de là à introduire un symbole pour les représenter dans l'analyse, c'est inutile. Mais nous pensons que les stratégies de management du système opérant par le système de pilotage font partie de la culture, idéologie de chaque organisation, tout comme les relations sociales (pouvoir, conflit, motivation, négociation, etc.) qui ne peuvent pas être explicitées dans un modèle qui décrit la structure de l'organisation à travers des relations entre composants. Egalement, avec la pensée autopoïétique de Valparaiso (modèle CIBORGA) qui définit l'entreprise comme une machine autopoïétique, où la coopération est essentielle pour le maintien de l'unité, l'identité, et l'autonomie.

➢ **Le modèle MSV**

Le modèle MSV (Modèle des Systèmes Viables) de Stafford Beer est basé sur le même schéma conceptuel que les autres modèles (OID de Le Moigne et AMS de Mélèse) à savoir : la structure de l'organisation du système de décision[75] est cette fois formée par le couplage du système de contrôle et du système de pilotage. Le contrôle et la régulation existent alors simultanément et nécessairement grâce à la récursivité, d'une part comme "unité" dans une totalité, et d'autre part comme "parties" dans une totalité, entre le système de décision et le système opérant. Le système opérant est bien entendu comme nous l'avons dit, d'une part l'ensemble des transformations (entrées/transformation/sorties), et d'autre part l'ensemble des transformations (transformation/environnement) de chaque niveau de la structure de décision et d'information (stratégique, tactique, opérationnelle) de l'organisation. Le contrôle est fait alors de façon récursive et hiérarchique à tous les niveaux de la structure de l'organisation (stratégique, tactique, opérationnel), où chaque niveau agit comme régulateur par rapport à son immédiate hiérarchie supérieure et comme contrôle par rapport à son immédiate hiérarchie inférieure [Beer, 72].

---

[74] Beer fait dialoguer la cybernétique (de deuxième ordre) et le management dans son modèle. Le Moigne et Ermine se sont plongés davantage dans un domaine physique (de données, d'informations et de connaissances) plutôt que social (relations humaines). Ou, si l'on veut, ils ont été attirés davantage par un domaine cognitiviste qu'enactiviste de la cognition.
[75] C'est ici l'ensemble du processus de décision et d'information (stratégique, tactique, opérationnelle) qui permet de maîtriser et de guider les transformations du système opérant qui chez Beer correspond au système d'implémentation.

Le nombre de variables de ce modèle que le système de contrôle doit contrôler et réguler dans chaque niveau d'activité système de pilotage (stratégique, tactique, opérationnelle) a été regroupé en cinq, à savoir : les *variables politiques*, les *variables d'intelligence*, les *variables de contrôle* et les *variables de coordination*.

### Variables politiques

Les variables politiques forment le système politique, c'est-à-dire l'ensemble objectifs (stratégiques, tactiques, opérationnels), mission, raison d'être, de l'organisation, d'une part comme "unité" dans une totalité (par exemple, système = entreprise), et d'autre part comme "parties" dans une totalité (par exemple, système = département). Par exemple le système politique du niveau stratégique doit décider si la mission de l'entreprise a pour finalité (objectif) de produire ceci ou cela. La question fondamentale du système politique est la remise en question permanente de la direction de l'organisation, où l'on est, où l'on va, dans quel business nous sommes, dans quel business nous serons, et ainsi de suite, à chaque niveau de la structure de décision et d'information (stratégique, tactique, opérationnelle) de l'organisation.

### Variables d'intelligence

Les variables d'intelligence forment le système d'intelligence, c'est-à-dire l'ensemble objets et processus (business process & objets) nécessaires pour le maintien de l'organisation dans le temps. La structure de l'organisation du système d'intelligence est construite autour d'une cellule de business intelligence, c'est-à-dire de la recherche d'information, objets et processus (business process & objets) pour innover. Par exemple, le système d'intelligence du système opérationnel peut être organisé autour des nouvelles technologies.

### Variables de contrôle

Les variables de contrôle forment le système de contrôle, c'est-à-dire l'ensemble des variables qui sont nécessaires pour le maintien de l'équilibre et l'atteinte des objectifs par la régulation des variables contenues dans la structure de l'organisation.

Variables de coordination

Les variables de coordination forment le système de coordination, c'est-à-dire l'ensemble des décisions d'action entre les trois niveaux de décision et d'information (stratégique, tactique, opérationnelle) de la structure de l'organisation du système de pilotage, pour contrôler et réguler le système opérant, que Beer appelle système d'implémentation. La structure de l'organisation du système d'implémentation est formée par le couplage d'une part comme "unité" (l'organisation du système d'implémentation) dans une totalité (le système/l'environnement), et d'autre part comme "parties" (système politique, système d'intelligence, système de contrôle) dans une totalité (entrées/transformation/sorties).

La mécanique générique du modèle MSV de Beer fonctionne comme suit : le niveau 0 permet d'expliquer la "totalité" de l'organisation, tandis que les $n$ niveaux successifs (définis par le nombre de composants du système opérant que l'on veut distinguer), montrent les "parties" de l'organisation que l'on veut implanter. Ainsi, l'on voit apparaître dans la construction de la structure du système opérant des niveaux de complexité progressifs. Chaque niveau est formé par relation (contrôle et régulation) récursive d'une part comme "unité" dans une totalité, et d'autre part comme "parties" dans une totalité, entre deux composants : le système opérant ainsi que le système de décision (l'ensemble des transformations du système opérant), est contrôlé et régulé pour attendre les objectifs (stratégiques, tactiques, opérationnels) à travers le système de décision, qui est structuré alors par les variables politiques, les variables d'intelligence, les variables de contrôle, et les variables de coordination.

L'avantage, mais aussi l'inconvénient (du point de vue idéologique) du modèle MSV de Beer, est de mettre en valeur le fait que l'on peut planifier et administrer de façon cybernétique et systémique le cœur politique et financier de l'entreprise. En effet, une expérience concrète de l'application du modèle MSV de Beer, afin de planifier, contrôler et coordonner l'économie de tout un pays, a été menée pour la première fois au Chili, sous le nom de : projet SYSCO[76]. Ce projet a été initié en 1972. Néanmoins, il a été arrêté, comme d'ailleurs certains de ses acteurs, le 11/09/73. L'équipe de ce projet a été dispersée[77] et malheureusement la plupart des informations rassemblées ont été détruites. L'on imagine bien que dans une époque où la machine à écrire a été le seigneur, les traces des documents peuvent être effacées facilement. Ces documents parlaient du contrôle et du pilotage des processus productifs industriels de tout un pays sous régime marxiste-léniniste.

---

[76] 80 entreprises du secteur industriel chilien appartenant au groupe CORFO (Corporación de Fomento de la Producción) ont participé au projet SYSCO (c'est l'abréviation pour dire système de contrôle) [Espejo, 89], [Espejo *et al*, 96].
[77] Pour apercevoir un peu l'ambiance de l'exilé politique chilien, voici ce que dit Francisco Varela dans les remerciements de son livre *Invitation aux sciences cognitives* : « je voudrais aussi exprimer ici ma gratitude envers mes collègues de Paris qui m'ont accueilli en France, me permettant de poursuivre mes recherches alors qu'il devient impossible de le faire dans mon pays natal, le Chili, alors ravagé par une épidémie de fascisme » [Varela, 96].

Cependant, nous pensons qu'à cette époque, les technologies de l'information et de communication, en plus du mauvais esprit des gens pour préserver l'unité, ont été très loin de l'aspiration à former un système viable, bien que dans la tête des acteurs du projet SYSCO, tels que Fernando Flores, Raul Espejos, Aquiles Limone et bien d'autres, l'on puisse imaginer le contraire. A cet égard, Limone a dit « la tentative d'application du modèle de Beer à la réalité concrète, spécifiquement à tout un secteur de l'économie, réalisé au Chili en 1972, fournit des évidences empiriques quant à son applicabilité. Bien que l'expérience ait été interrompue par les événements politiques, et que cela nous prive de données concrètes sur son fonctionnement opérationnel, les résultats de la phase d'implantation sont hautement positifs à l'égard de sa viabilité » [Limone, 77].

En général, selon notre expérience pédagogique, les modèles OID, AMS et MSV sont assez similaires, puisqu'on a le même souci : la maîtrise du changement dans un système viable. La structure de l'organisation du système viable doit s'adapter à son environnement (en changement dynamique permanent) afin que l'organisation de la structure du système viable puisse survivre. En fait, ce que l'on cherche est le maintien de l'identité dans et par l'unité. Néanmoins, comme l'a dit Mélèse « la complexité, caractère fondamental, apparaît donc comme l'incapacité de décrire tout le système et de déduire son comportement á partir de la connaissance du comportement de ses parties » [Mélèse, 79]. Dans ce sens, le facteur clé pour cette maîtrise de la complexité est d'une part l'émergence de relations entre composants qui doivent véhiculer des informations de contrôle et de pilotage, et d'autre part, la modification des valeurs des variables d'actions du processus de décision et d'information (stratégique, tactique, opérationnelle) et des variable essentielles (objectifs stratégiques, objectifs tactiques, objectifs opérationnels), alors que simplement dans le modèle OID de Le Moigne l'on parle de régulation des entrées/sorties.

Ces trois modèles offrent un regard (systémique) pour observer (analyser et comprendre) l'entreprise comme un système organisé autour d'un objectif commun. Comme le dit Mélèse « l'AMS conduit finalement à établir *une maquette de l'entreprise* » [Mélèse, 72] sur laquelle il est possible de pousser l'analyse sur un sujet plus spécifique, par exemple dans la détection de flux et processus d'information, pour la construction d'un système d'information (par exemple, l'on peut appliquer Merise après AMS [Mélèse, 72]).

Le modèle MSV de Beer s'est plus répandu dans le monde que les autres dans les secteurs public et privé [Limone et Bastias, 02a], mais nous ne pensons pas que son champ d'utilisation ait été élargi à la réalité de plus d'une entreprise depuis son application au Chili bien que nous n'ayons pas de données concrètes. Néanmoins, nous pensons que les applications de ces modèles sur un plan industriel (par exemple : l'administration publique) ou pédagogique (par exemple : l'enseignement universitaire) doivent être toujours un sujet de recherche. En ce sens, dans le domaine du génie logiciel, le modèle OID de Le Moigne a été utilisé pour forger la méthode Merise et ses évolutions à la programmation orientée objet [Nanci *et al*, 96], et plus récemment dans [Kawalek, 02] on trouve une application du modèle MSV de Beer pour décrire l'organisation d'un processus de software. Le modèle AMS de Mélèse, de même que le modèle OID de Le Moigne, sont enseignés depuis leur apparition au public au milieu des années 70, dans toutes les Grandes Ecoles d'ingénieurs françaises. Au Chili[78], le modèle AMS de Mélèse a été introduit pour la première fois par Limone dans l'École de commerce de l'Université Catholique de Valparaiso. Il est actuellement appliqué dans le cours de contrôle de gestion de cette université [Limone et Cademártori, 98]. En revanche, nous l'avons introduit aussi dans les cours de systèmes d'information[79] et systèmes de gestion[80] pour nos élèves techniciens et ingénieurs. Dans l'état actuel de nos connaissances le nombre d'articles et ouvrages en espagnol consacrés au développement des travaux de Le Moigne et Mélèse sont malheureusement très rares.

Enfin, le modèle OIDC exprime la même préoccupation que les modèles cybernétiques et structuralistes des organisations des années 70, (chez Beer et Mélèse), en tenant compte que le contrôle et la coordination des activités de l'entreprise ne se fait pas seulement avec des informations internes ou externes du système qui le produit et le consomme ; mais qu'il convient d'ajouter un ingrédient supplémentaire : la *connaissance*, pour le maintien de l'identité dans et par l'unité au cours du temps. Néanmoins, la connaissance est présente dans le modèle MSV de Beer comme une variable d'intelligence dans le système de pilotage, et également dans le modèle OID de Le Moigne la connaissance fait partie du système de décision, comme un sous-système de d'intelligence (le niveau 8 « le système imagine et conçoit de nouvelles décisions possibles »). Le modèle OIDC ne fait que ressortir davantage cet aspect.

Ainsi, le modèle système et cybernétique dans le modèle OIDC nous permet de différencier d'ores et déjà, données, information, et connaissances :

- les données sont des variables du système opérant (O). Les valeurs de ces variables sont produites ou consommées par l'activité productive (le travail) du système opérant (O), et gérées par le système d'information (I). Le système opérant (O) permet la régulation des variables selon les objectifs du système de décision (D) ;

---

[78] Nous devons souligner que l'avance de l'enseignement de la systémique au Chili est dirigée (en dehors de Limone et Bastias) par Edmundo Leiva [Leiva *et al*, 99], Ricardo Acevedo [Acevedo, 99], Juan Bravo [Bravo, 03], [Bravo, 97], et Darío Rodriguéz [Rodríguez et Arnold, 91].
[79] Il s'agit de la théorie générale des systèmes appliquée aux systèmes d'information.
[80] Il s'agit de la théorie générale des systèmes appliquée aux systèmes de gestion (ou management).

- les informations sont des variables du système de décision (D). Les valeurs de ces variables sont produites ou consommées par l'activité décisionnelle (le processus de décision) du système de décision (D) et gérées par le système d'information (I). Le système de décision (D) permet le contrôle du système d'information (I) lequel doit réguler à son tour le système opérant (O) ;

- les connaissances sont des variables de deux natures différentes : cognitives et compétences du système de connaissance (C). Les valeurs de ces variables sont produites ou consommées par les activités cognitives (connaissance) et d'apprentissage (compétence) chez l'individu face à une situation de travail à tous les niveaux de l'entreprise (stratégique, tactique, opérationnel). Le système de connaissance (C) est alimentés à travers des flux de connaissances provenant des autres trois systèmes (O), (I) et (D), et en sens inverse les systèmes (O), (I) et (D) sont alimentée à travers des flux de compétences provenant de la capitalartion des connaissances du système de connaissance (C). Ainsi, ce système utilise les connaissances et les compétences pour contrôler et réguler les autres systèmes à tous les niveaux (stratégique, tactique, ou opérationnel) de l'entreprise.

### 2.2.3. Le modèle autopoïétique

L'autopoïèse voit le jour en 1969 à l'Université du Chili comme un élargissement de la cybernétique (de deuxième ordre) de Heinz von Foerster, où dans son livre, intitulé *Principles of Self-Organization* (publié en 1962), s'introduit le concept d'*auto-organisation*[81]. Ce concept sera enrichi par la suite par les concepts d'*auto-maintient* et d'*auto-gestion*, donnant naissance ainsi au concept d'*autopoïèse*. Puis en 1977 à l'Université Catholique de Valparaiso, l'autopoïèse est appliquée à la gestion. Ainsi, nous avons deux approches de l'autopoïèse : l'une est l'approche autopoïétique (ou théorie autopoïétique) de Maturana et Varela, connue sous le nom de l'*autopoïèse de Santiago*, l'autre est l'approche autopoïétique de Limone et Bastias, connue sous le nom de l'*autopoïèse de Valparaiso*[82].

La distinction entre modèle autopoïétique et modèle système se fait en fonction de la dualité observateur/observé. Dans un modèle autopoïétique l'observateur s'intéresse aux relations entre processus de production des composants pour définir l'unité (l'organisation), maintenir l'identité (la

---

[81] Bien que la formulation "poétique" du concept d'*auto-organisation* a été inspirée du livre de Paul Valéry, intitulé *De la simulation* (publié en 1927), comme nous avons argumenté plus haut.
[82] L'approche autopoïétique de Maturana et Varela est née à Santiago du Chili à partir de l'articlé *Autopoietic Systems* qui a été présenté dans un séminaire de recherche organisé par l'Université de Santiago en 1972, tandis que l'approche autopoïétique de Limone et Bastias a été popularisée à l'École de commerce de l'Université Catholique de Valparaiso, à partir de base de Aquiles Limone (publié en 1977) et le modèle CIBORGA (popularisé de 1998) avec la collaboration de Luis Bastias, Cardemártori et autres [Limone et Cademártori, 98]. D'autre part, le 16 novembre 2001, à l'occasion du congrès *Théorie des Systèmes*, organisé par l'Université Technique Federico Santa María à Valparaiso, je me suis aperçu pour la première fois de la contribution de Limone et Bastias à cette théorie. Et donc, depuis cette date-là, je partage avec mes amis et collègues de Valparaiso, la passion pour les systèmes autopoïétiques.

---

structure) et conserver l'autonomie (la dynamique) du système, tandis que dans un modèle système l'observateur s'intéresse aux relations entre composants à travers un processus d'interaction avec l'environnement.

La figure 2.5 montre les concepts d'auto-organisation, d'auto-maintien et d'auto-gestion du modèle autopoïétique. Dans ce modèle l'opération de distinction, est formulé par rapport à une dualité organisation/structure sur la base de l'approche enactiviste de la cognition. Cette opération de distinction permet indiquer que les causes et les effets sont distinguables dans des espaces fort différentes. L'un est le domaine conceptuel (l'organisation) du processus organisationnel (c'est un processus conceptuel de description abstrait de l'organisation). L'autre est le domaine physique (la structure) du processus structurel (c'est un processus physique de description matérielle de la structure, il s'agit bien ici de la description des propriétés des composants de la structure de l'organisation).

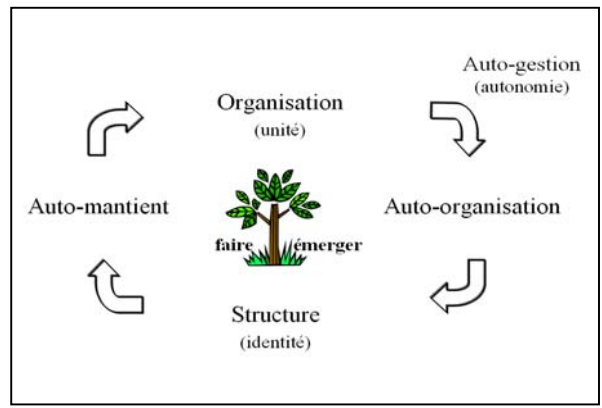

**Figure 2.5 :** Approche enactiviste de la cognition (source propre)

Dans l'approche autopoïétique de Santiago le domaine conceptuel (l'organisation) correspond à un domaine des relations autopoïétiques, et le domaine physique (la structure) correspond à un domaine matériel caractérisé par les propriétés des composants, comme la masse, la force, l'accélération, la distance, le champ, etc.

Dans l'approche autopoïétique de Valparaiso le domaine conceptuel (l'organisation) correspond à un domaine social défiinis à travers de relations humaines, et le domaine physique (la structure) correspond à un domaine matériel à travers des individus, matière, énergie, et symboles[83].

Avant de développer ces approches nous avons trouvé nécessaire ici d'ouvrir une parenthèse, afin d'expliquer les trois approches (plutôt scientifique que philosophique) de la cognition, puisqu'ils nous permettront, d'une part, de clarifier la question de l'intelligence chez l'homme et de l'intelligence l'artificielle chez la machine (où ces approches ont été largement utilisées), et d'autre part, pour comprendre de meilleur forme les outils autopoïétiques afin d'aborder la question de l'autopoïèse et la connaissance (dans le chapitre 3). Cette question est à la base du modèle proposé modèle.

---

[83] Les symboles ou signes sont des éléments d'un domaine linguistique, dans ce sens c'est l'observateur qui par une opération de distinction donne de sens aux objets qui les entourent.

➢ **Qu'est-ce que la cognition d'après Maturana et Varela ?**

La problématique d'étude de la pensée épistémologique, concerne la relation entre *connaissance* et *réalité physique*[84]. Cette relation a fait émerger la question épistémologique fondamentale depuis toujours à savoir : *comment l'on acquiert une connaissance de la réalité physique et combien cette connaissance peut être fiable et vraie*. La réponse à cette question qui reste d'ailleurs toujours ouverte a été abordée depuis des siècles par les approches philosophiques et scientifiques de la cognition.

Varela dans son livre, intitulé *Invitation aux sciences cognitives* (publié en 1989), a donné une classification scientifique de la cognition selon trois catégories, à savoir : les symboles (l'hypothèse cognitiviste : l'approche cognitiviste) ; l'émergence (une alternative à la manipulation de symboles : l'approche connexionniste) ; et l'enaction (une alternative à la représentation : l'approche de l'enaction de la cognition) [Varela, 96].

Nous synthétiserons l'approche cognitiviste et l'approche connexionniste de la cognition, puis nous donnerons la critique de Maturana et Varela sur le manque de nouveaux outils cognitifs pour la recherche biologique, ensuite nous présenterons la contribution de Maturana et Varela aux sciences cognitives : l'approche de l'enaction et l'autopoïèse de Santiago, pour en finir avec l'autopoïèse de Valparaiso.

L'approche cognitiviste

La connaissance dans l'approche cognitiviste est la description ou l'image d'une réalité physique. Comme l'a très bien dit Varela en citant Rorty « la connaissance est un miroir de la nature ». Cela signifie que la connaissance n'est connaissance que si elle reflète la réalité physique telle qu'elle est, alors seulement une réalité physique prédéfinie peut être représentée. A cet égard Glaserfeld dit « la connaissance serait alors le reflet ou l'image d'un monde qui est là, c'est-à-dire existe avant qu'une conscience ne le voie ou en fasse l'expérience d'une quelconque autre façon »[85]. Glaserfeld veut dire par là que dans un monde objectif, la réalité physique existe indépendamment de l'observateur. Mais si la connaissance est une description ou image d'une réalité physique en soi, nous avons alors besoin d'un critère qui nous permette de juger la véracité de cette représentation. Comme l'a dit Glaserfeld en citant Bateson « dès le début de la connaissance, se pose la question de la vérité. Son introduction fait du processus de la connaissance humaine un problème de savoir » [Glaserfeld, 95]. En d'autres termes, comment pouvons nous savoir si la description ou l'image de quelque chose transmise par nos sens correspond à la réalité physique ? C'est-à-dire cette description ou image est-elle correcte ou vraie ? Ou plus intéressant encore, dans quelle mesure cette description ou image déclanche-t-elle ou fait-elle émerger un comportement intelligent[86]? Pour illustrer le passé, Glaserfeld donne l'exemple de Sextus Empiricus sur la perception d'une pomme « une pomme apparaît à nos sens comme lisse, parfumée, sucrée et jaune, mais il ne va pas de soi du tout que la pomme possède effectivement ces propriétés, comme il n'est pas évident du tout qu'elle ne possède pas aussi d'autres propriétés que nos sens ne perçoivent simplement pas » [Glaserfeld, 95]. Comment le savoir sans faire l'expérience de goutter la pomme. Cette argumentation montre aussi que la connaissance n'est ni parfaite ni certaine. Enfin, Varela a dit « l'hypothèse cognitiviste prétend que la seule façon de rendre compte de l'intelligence et de l'intentionnalité est de postuler que la cognition consiste à agir sur la base de représentations qui ont une réalité physique sous forme de code symbolique dans un cerveau ou une machine » [Varela, 96].

---

[84] Glaserfeld utilise le terme "réalité ontologique objective" [Glaserfeld, 95].
[85] Cette argumentation fait aussi référence à l'expérience que nous verrons dans l'approche de la connaissance par l'épistémologie expérimentale.

Ainsi, un systèmes des connaissances selon l'approche cognitiviste correspond à la manipulation de symboles à partir de règles, normes, instructions, etc. Ce système fonctionne par n'importe quel dispositif pouvant représenter et manipuler des symboles (éléments physiques discontinus). Le système n'interagit qu'avec des codes symboliques (attributs physiques des symboles), et non leurs sens (signification). Un système de connaissance fonctionne de manière appropriée quand les symboles représentent de façon adéquate quelque aspect de la réalité, et que le traitement de l'information aboutit à une solution efficace du problème soumis au système.

### L'approche connexionniste

La connaissance dans l'approche connexionniste correspond à l'émergence d'états globaux dans un réseau de composants simples. La manipulation de symboles (de l'approche cognitiviste) est remplacée donc par des opérations numériques. L'émergence de schémas globaux, comme l'explique Varela en faisant une analogie au système nerveux « ici, chaque constituant fonctionne seulement dans son environnement *local* de sorte que le système ne peut être actionné par un agent extérieur qui en tournerait en quelque sort la manivelle. Mais grâce à la nature configurationnelle du système, une coopération *globale en émerge* spontanément lorsque les états de chaque neurone en cause atteignent un stade satisfaisant » [Varela, 96].

---

[86] L'aspect de l'intelligence sera vue avec la contribution de Jean Piaget.

L'émergence de nouveaux outils cognitifs de la théorie autopoïétique de Maturana et Varela a été possible grâce à une remise en questions des hypothèses de recherche sur l'étude des systèmes vivants à travers des systèmes cognitifs (parce que les outils cognitifs du moment ne permettaient pas l'étude du système nerveux). Pour synthétiser les savoirs en sciences cognitives courantes, Varela dit, en citant Kuffler et Nichols, « le cerveau est un assemblage actif de cellules qui reçoit constamment de l'information, qui la perçoit et la traite, et qui prend des décisions » [Varela, 96], puis il ajoute en citant Neisser « j'ai trouvé nécessaire de supposer que celui qui perçoit, possède certaines structures cognitives (nommées schémas), dont la fonction est de ramasser les informations offertes par l'environnement ». Nous constatons que les sciences cognitives du moment peuvent être généralisées sur deux axes : l'un relatif à la représentation de l'information (le quoi), et l'autre relatif au traitement de l'information (le faire) par le cerveau.

Ainsi, le système nerveux d'une part recueille des informations en provenance de son environnement qu'il traite, et d'autre part, leur traitement de l'information aboutit à une représentation de la réalité physique à l'intérieur du cerveau. Cette représentation de la réalité est vraie si le traitement de l'information fonctionne de manière appropriée [Varela, 89].

La critique fondamentale de Maturana et Varela à l'utilisation de l'approche symbolique pour la compréhension des systèmes vivants est que la connaissance est toujours la représentation adéquate d'une réalité physique prédéterminée, et donc le cerveau serait une machine qui produit une image exacte de cette réalité physique où la seule façon de rendre compte de l'intelligence et de l'intentionnalité est de postuler que la cognition consiste à agir sur la base de représentations qui ont une réalité physique sous forme de code symbolique dans ce cerveau[87]. A cet égard Varela dit « notre activité cognitive quotidienne révèle que cette image est incomplète. La plus importante faculté de toute cognition vivante est précisément, dans une large mesure, de *poser* les questions pertinentes qui surgissent à chaque moment de notre vie. Elles ne sont pas prédéfinies mais *enactées*, on les *fait émerger* sur un arrière plan, et les critères de pertinence sont dictés par notre sens commun » [Varela, 89]. Et Maturana ajoute « l'affirmation *a priori* selon laquelle la connaissance objective constitue une description de ce qui est connu … appelle les questions *qu'est-ce que savoir ?* et *comment savons-nous ?* » [Maturana et Varela, 73].

Comme nous pouvons le constater, la pensée de Maturana et Varela est très attachée aux grandes questions épistémologiques et cognitives fondamentales, que nous avons déjà anticipées (comment on acquiert une connaissance de la réalité physique et combien cette connaissance peut être fiable et vraie), que l'on peut résumer par trois voies de recherche à savoir : (1) comment acquiert-on une connaissance de la réalité et combien cette connaissance peut être fiable et vraie ; (2) dans quelle mesure l'image d'un objet transmise par nos sens correspond-elle à la réalité physique ; et (3) dans quelle mesure cette image déclanche-t-elle ou fait-elle émerger un comportement intelligent ?

---

[87] Ceci nous renvoie aussi à la classique analogie de l'intelligence artificielle qui compare le cerveau (soit de l'animal, soit de l'homme) avec l'ordinateur à partir de l'approche symbolique et de l'émergence.

Ces trois chemins réflexifs ont renvoyé les auteurs à se poser des questions sur le rôle de l'observateur qui perçoit cette réalité physique. Pour Maturana et Varela il y a deux chemins explicatifs possibles de la connaissance dans le domaine biologique : l'un est relatif à la *perception* et la *description*, l'autre à la *perception* et l'*action*. Le premier chemin explicatif, ou *ontologies transcendantes*, suppose que la réalité physique existe indépendamment de l'observateur. La connaissance est alors une *description* exacte (image fidèle) de la réalité physique. Cette connaissance est vraie si elle repose sur un ensemble de savoirs tenus pour acquis. Ici la connaissance est générée à partir des *symboles*. Le deuxième chemin explicatif, en *ontologies constitutives,* suppose au contraire que la réalité physique existe seulement pour l'observateur dans l'*action*, c'est-à-dire l'*acte de langage* (c'est-à-dire le langage et ses émotions) et la *culture* (c'est-à-dire la structure sociale). La connaissance est alors un *système d'actions* constitue par une *opération de distinction* qui est propre à l'observateur par ce qu'il fait, où parce qu'il est capable de distinguer. Ici la connaissance est générée à partir des *émotions*. Nous constatons que ces chemins explicatifs coïncident exactement avec la problématique de l'épistémologie traditionnelle et de l'épistémologie expérimentale que nous avons explicitées plus haut, mais les chemins d'explication sont fort différents.

Comme l'a si bien dit Varela « c'est notre réalisation sociale, par l'acte de langage, qui prête vie à notre monde. Il y a des actions linguistiques que nous effectuons constamment : des affirmations, des promesses, des requêtes et des déclarations. En fait, un tel réseau continu de gestes conversationnels, comportant leurs conditions de satisfaction, constitue non pas un outil de communication, mais la véritable trame sur laquelle se dessine notre identité » [Varela, 89]. Un exemple, de système d'actions est le *réseau de compromis social* de Fernando Flores[88]. Il s'agit d'un mode d'organisation du travail coopératif basé sur les modes d'être, les modes de faire pour achever l'objectif (le quoi faire) [Flores, 96a], [Flores, 96b].

Ces diverses critiques ont remis en cause le manque de nouveaux outils pour la compréhension des processus cognitifs des systèmes vivants, ce qui a donné naissance à une nouvelle approche de la cognition caractérisée, d'une part, par l'autonomie du système vivant, et d'autre part, par la création et l'évolution des connaissances à partir des émotions : l'*approche de l'enaction*. Maturana et Varela utilisent une analogie avec un "arbre" pour représenter ce concept [Maturana et Varela, 73].

---

[88] Ces idées sur les actes du langage ont été réfléchies par lui pendant son séjour de deux ans dans l'un des camps de concentration de la dictature militaire au Chili, à partir des visites que Humberto Maturana et Francisco Varela lui ont rendu. Néanmoins, son modèle autopoïétique a été développé plus tard dans le cadre de sa thèse sous la direction de Winograd à l'Université de Bercley aux Etats-Unis [Winograd et Flores, 89].

## L'approche de l'enaction

Dans l'épistémologie expérimentale, la connaissance est l'interprétation d'une réalité physique qui est constituée par notre expérience et non par la description ou l'image d'une réalité physique indépendante de toute expérience comme dans le cas antérieur. Ici, il y a la construction interprétative d'un modèle de la réalité physique grâce à notre expérience. Comme l'a dit Glaserfeld « le constructivisme radical … développe une théorie de la connaissance dans laquelle la connaissance ne reflète pas une réalité ontologique "objective", mais concerne exclusivement la mise en ordre et l'organisation d'un monde constitué par notre expérience » [Glaserfeld, 95]. Puis, il ajoute en faisant allusion à la pensée de Emmanuel Kant « étant donné les manières dont nous faisons l'expérience du réel, nous ne pouvons en aucun cas concevoir un monde indépendant de notre expérience » [Glaserfeld, 95]. Cela signifie que l'observateur est le seul responsable de sa pensée, de sa connaissance, et donc de ce qu'il veut et peut distinguer avec ce qu'il fait. Nous constatons alors que la véracité de la connaissance se trouve dans le domaine de l'expérience. Pour Le Moigne en citant Giambattista Vico « la seule manière de "connaître" une chose est de l'avoir faite, parce que alors seulement on sait quels sont ses composants et comment ils ont été assemblés » [Le Moigne, 99]. Pour Vico le passé a été résumé par lui dans un slogan : *verum ipsum factum* (le vrai est le même que le fait)[89]. Pour construire le domaine d'expérience Vico emploie le terme d'*opération*, et anticipe par là le concept d'*opération de distinction* (le fait d'être distinguable de son environnement et donc des autres unités) de Maturana et Varela[90]. Mais aussi d'après Glaserfeld, Vico anticipe les concepts d'*organisation* et de *structure* dans sa pensée, lorsqu'il a dit « l'utilisation explicite du terme *facere* par Vico, sa constante référence à la composition et au fait d'assembler » [Glaserfeld, 95].

Ainsi, l'idée centrale de l'approche de l'enaction de la cognition a été fondée sur trois constatations : (1) l'observateur occupe une place centrale dans toute connaissance ; (2) c'est l'observateur qui donne de sens à la réalité ; et (3) ce sont les observateurs qui constituent dans le langage et la culture notre identité.

L'approche de l'enaction considère que la connaissance est définie comme un système d'actions, qui prend en charge l'historique du *couplage structurel* qui enacte (fait émerger) un monde.

---

[89] *Factum* (fait) vient de *facere* (faire).
[90] Le terme *opération* nous le trouvons aussi chez d'autres constructivistes. Par exemple, Silvio Ceccato utilise le terme *consapevoleza operative* (conscience opérationnelle). Cité dans [Le Moigne, 99].

Ce système fonctionne comme un réseau d'éléments inter-connectés, capables de subir des changements structuraux au cours d'un historique non interrompu. Dans ce sens, un système de connaissance fonctionne de manière approprié quand il s'adjoint à un domaine de signification préexistant, en continuel développement (ou qu'il en forme un nouveau) [Varela, 89]. A cet égard Varela dit « c'est là une conception de la connaissance et de la réalité qui tient compte du fait que nous participons à leur élaboration, et nous pouvons voir qu'elles s'enracinent dans les formes cellulaires les plus élémentaires des processus cognitifs et informationnels » [Varela, 89].

Alors, selon l'approche de l'enaction : (1) le système nerveux est un système autonome de connaissance qui est défini par son organisation et qui fonctionne par clôture opérationnelle, c'est-à-dire les résultats des transformations du système sont les transformations du système. Le terme clôture se réfère au fait que le résultat de la transformation se situe à l'intérieur des frontières du système lui-même. Cela ne signifie pas pour autant que le système autonome soit fermé, c'est-à-dire qu'il n'y a pas d'interactions avec son environnement (sans entrées ni sorties) ; et (2) la connaissance est une conduite (mécanismes et moyens d'agir) qui permet à l'organisme de faire seulement des choses qui n'affectent pas sa survie. Par conséquent, la connaissance ne doit pas obligatoirement impliquer des représentations vraies de la réalité objective. Autrement dit, l'approche de l'enaction de la cognition correspond au fait de faire émerger un comportement intelligent. Comme si bien l'a dit Varela « l'intelligence ne se définit plus comme la faculté de résoudre un problème mais comme celle de pénétrer un monde partagé » [Varela, 89]. Et donc, l'enaction est un mécanisme "circulant" et "d'émergence de signification".

En conséquence, l'enaction (faire-émerger) de Maturana et Varela n'as pas le même sens que le concept d'enaction de Karl Weick, où le problème centrale la relation organisation-environnement comme un processus d'interaction, où chacun se construit elle-même par cette interaction. Cette idée de Weick se trouve dans son livre, intitulé *The Social Psychology of Organizing* (publié en 1979) [Weick, 79].

C'est pour cette raison que nous utilisions le terme *enactiviste* (faire-émerger) pour faire la différence avec l'approche de l'enaction (interaction organisation-environnement) de Weick. Mais aussi pour aligner ce nom avec les deux autres approches de la cognition. Nous reviendrons largement sur les concepts de *faire-émerger* et *interaction organisation-environnement* dans le chapitre 3.

Enfin, nous présentons une synthèse, que nous avons emprunté du livre chez Varela, intitulé *Invitation aux sciences cognitives* (publié en 1989) [Varela, 96].

L'approche cognitiviste

Question 1 :     Qu'est-ce que la cognition ?
Réponse :        Le traitement de l'information : la manipulation de symboles à partir de règles.
Question 2 :     Comment cela fonctionne-t-il ?
Réponse :        Par n'importe quel dispositif pouvant représenter et manipuler des éléments physiques discontinus : des symboles. Le système n'interagit qu'avec la forme des symboles (leurs attributs physiques), et non leur sens.
Question 3 :     Comment savoir qu'un système cognitif fonctionne de manière appropriée ?
Réponse :        Quand les symboles représentent adéquatement quelque aspect du monde réel, et que le traitement de l'information aboutit à une solution efficace du problème soumis au système.

L'approche connexionniste

Question 1 :     Qu'est-ce que la cognition ?
Réponse :        L'émergence d'états globaux dans un réseau de composantes simples.
Question 2 :     Comment cela fonctionne-t-il ?
Réponse :        Des règles locales gèrent les opérations individuelles et des règles de changement gèrent les liens entre les éléments.
Question 3 :     Comment savoir qu'un système cognitif fonctionne de manière appropriée ?
Réponse :        Quand les propriétés émergentes (et la structure résultante) sont identifiables à une faculté cognitive (une solution adéquate pour une tâche donnée).

L'approche enactiviste

Question 1 :     Qu'est-ce que la cognition ?
Réponse :        L'action productive : l'historique du couplage structurel qui enacte (fait-émerger) un monde.
Question 2 :     Comment cela fonctionne-t-il ?
Réponse :        Par l'entremise d'un réseau d'éléments inter-connectés, capables de subir des changements structuraux au cours d'un historique non interrompu.
Question 3 :     Comment savoir qu'un système cognitif fonctionne de manière appropriée ?
Réponse :        Quand il s'adjoint à un monde de signification préexistant, en continuel développement (comme c'est le cas des petits de toutes les espèces), ou qu'il en forme un nouveau (comme cela arrive dans l'histoire de l'évolution).

Nous fermons cette parenthèse, afin de passer à la section suivante.

➢ **Théorie autopoïétique de Santiago**

L'autopoïèse de Santiago, a été proposée aux débuts des années 70 par deux biologistes chiliens Humberto Maturana et Francisco Varela, comme un chemin explicatif de la phénoménologie[91] des systèmes vivants. Au travers de la vie au niveau cellulaire, d'une part dans un livre, intitulé : *De máquinas y seres vivos, Autopoiesis : La organización de lo vivo* [Maturana et Varela, 72][92] en 1972, et d'autre part dans un célèbre article, publié en 1974, intitulé : *Autopoiesis, The Organization of Living Systems, Its Characterization and a Model*. Pour Maturana et Varela, la question fondamentale de cette théorie dans le domaine biologique est la suivante « qu'y a-t-il de commun à tous les êtres vivants, qui permet de les qualifier de vivants ? En dehors de la force vitale, de quoi s'agit-il donc ? » [Maturana et Varela, 72], [Varela, 89]. Pour eux, ce qu'il y a de commun gravite autour de l'autonomie, l'identité et l'unité dans sa totalité, comme l'a dit Limone pour les êtres vivants « ils centrent leur analyse dans l'exploration de l'origine des manifestations d'autonomie, identité et unité, pour parvenir à formuler une explication axée sur la notion d'unité, dont la genèse se fonde sur l'existence et le fonctionnement d'une organisation particulière, commune à tous les êtres vivants » [Limone, 77]. En d'autres termes, c'est l'existence d'un modèle d'organisation qui peut expliquer la manifestation des propriétés caractéristiques des systèmes vivants au niveau biologique, c'est-à-dire d'un patron d'organisation commun des êtres vivants qui les caractérise au niveau cellulaire. Cette organisation (ou patron organisationnel) a été appelée : l'*autopoïèse*. Ce terme vient du grec αυτοσ = autos (soi) et ποιενιν = poiein (créer, produire) qui permet d'indiquer qu'une organisation autopoïétique est auto-créante (qui se crée elle-même) ou s'auto-produit (qui se produit elle-même) [Limone, 77], [Varela, 89]. Il est intéressant de remarquer que l'idée de l'autopoïèse a été possible grâce à une découverte fondamentale dans l'organisation et le fonctionnement du système nerveux à savoir : sa capacité d'auto-produire de façon permanente des cellules au niveau chimique afin de maintenir l'*autonomie*, l'*unité* et l'*identité* du système nerveux sans interaction avec son environnement. Dans ce contexte, l'autonomie est liée à la capacité (contrainte, ordre) d'auto-gestion du système nerveux pour auto-produire de façon permanente ses cellules au niveau chimique (espace matériel), tandis que l'unité est liée à l'organisation et au fonctionnement de l'unité (on parle donc d'auto-organisation), qui a lieu dans le domaine physique où cette unité est spécifiée (ceci nous renvoie à la dualité organisation/structure). Enfin, l'identité est liée à la structure qui s'auto-maintient dans et par l'organisation et le fonctionnement de l'unité. En plus, ils ont découvert l'existence d'une frontière qui est produite par la dynamique des cellules au niveau chimique qui simultanément produit les conditions nécessaires à l'existence de cette dynamique les relations qui caractérisent l'organisation du système dans le cas du système nerveux ce sont des relations de production des cellules au niveau chimique qui organise l'auto-conservation du système nerveux. L'organisation des cellules au niveau chimique demeure invariante dans le temps, ce qui change c'est sa structure (propriétés de composants).

---

[91] La phénoménologie est la science qui étudie les phénomènes. Dans le cas de la théorie autopoïétique ce sont des phénomènes biologiques qui sont à la base de la reproduction, l'ontogenèse, le comportement, l'évolution, l'autonomie, la diversité, la conservation, etc., des systèmes vivants [Limone, 77], [Varela, 89].
[92] En français « Sur des machines et des êtres vivants, Une théorie sur l'organisation biologique ». Le gros de la pensée autopoïétique a été présenté par ces auteurs dans de nombreux articles et en particulier dans les livres : *The Neurophysiology of Cognition* (1969), *Biology of Languague: The Epistemology of Reality* (1977), *Autopoiesis and Cognition* (1980), *Principles of Biological Autonomy* (1980), *El árbol del conocimiento. Las bases biológicas del entendimiento humano* (1984), *The Tree of Knowledge: The Biological Roots of Human Understanding* (1987), *Autonomie et Connaissance* (1989), et *l'Arbre de la Connaissance* (1994).

La question fondamentale sur le vivant, dans la théorie autopoïétique, reste toujours ouverte, et les liens très étroits avec la pensée constructiviste que nous discuterons[93] reste originale. Néanmoins, cette théorie a le mérite d'une part, d'avoir développé de nouveaux concepts tels que la *clôture opérationnelle*, le *couplage structurel*, et la *détermination structurelle*[94], et d'autre part, l'approche de l'enaction de la connaissance pour approcher des phénomènes biologiques des systèmes vivants comme une alternative à l'approche cognitiviste et à l'approche connexionniste de la cognition comme nous avons vu plus haut. Elle a permis également l'opportunité d'appliquer la théorie autopoïétique à d'autres domaines pour lesquels on cherche à comprendre les problèmes d'autonomie, d'unité et d'identité[95]. A cet égard, nous voudrions citer les chercheurs "autopoïétiques" qui nous ont aidé dans la rédaction de cette thèse et dans notre vie professionnelle aussi : Aquiles Limone et Luis Bastias dans le domaine du management [Limone et Bastias, 02a], Fernando Flores dans le domaine linguistique (les actes conversationnels dans l'entreprise) [Flores, 96a], [Flores, 96b], Darío Rodríguez dans le domaine des processus décisionnels [Rodríguez, 91], Edmundo Leiva dans le domaine des processus de coordination [Leiva *et al*, 99], et El-Sayed Abou-Zeid dans le domaine des systèmes d'information [Abou-Zeid, 02].

Cette remarque faite, nous voudrions revenir au paragraphe précédent pour dire que les découvertes de Maturana et Varela sur l'organisation et le fonctionnement du système nerveux ont été utilisées pour formuler l'hypothèse de base de la théorie autopoïétique, à savoir : l'autopoïèse (la capacité qu'a l'organisation pour s'auto-produire de façon permanente) est nécessaire et suffisante pour définir l'organisation des systèmes vivants dans le domaine biologique. En conséquence, l'autopoïèse serait le modèle d'organisation des systèmes vivants. En généralisant cette constatation l'autopoïèse est le modèle d'organisation des systèmes autopoïétiques. De ce fait pour nous l'autopoïèse peut être reçue comme le modèle d'organisation des systèmes, et donc applicable à l'entreprise.

---

[93] Néanmoins, il faut souligner que le seul moyen que nous avons trouvé pour le faire a été l'utilisation massive de références aux auteurs concernés, parce que ces concepts sont nés dans un domaine qui est loin de notre champ de recherche : le domaine biologique, bien que l'application de ces concepts au management comme nous allons le voir dans l'approche autopoïétique de Limone et Bastias reste plus compréhensible. Notre source d'information privilégiée a été l'article de Ernst Von Glasersfeld, intitulé : *Introduction à un constructivisme radical*, apparu dans le livre de Paul Watzlawick, intitulé : *L'invention de la réalité. Contributions au constructivisme* et les livres de Francisco Varela, intitulés : *Invitations aux sciences cognitives* et *Autonomie et Connaissance. Essai sur le Vivant* qui sont une compilation (en français) de recherches avec Humberto Maturana.

[94] Ces concepts seront nos outils pour formuler le modèle autopoïétique de la gestion des connaissances imparfaites que nous ferons au chapitre 3.

[95] Bien que l'utilisation de l'*autopoïèse* en dehors du domaine biologique ait été critiquée par ses auteurs comme nous verrons à la fin de cette approche, avant de présenter l'approche autopoïétique de Limone et Bastias.

Pour Maturana et Varela un système autopoïétique « est organisé comme un réseau de processus de production de composants qui (a) régénèrent continuellement par leurs transformations et leurs interactions le réseau qui les a produits, et qui (b) constituent le système en tant qu'unité concrète dans l'espace où il existe, en spécifiant le domaine topologique où il se réalise comme réseau » [Varela, 89]. En d'autres termes, un système autopoïétique est caractérisé par une organisation et une structure autopoïétique. Les composants de ce système sont des processus (transformations) qui s'auto-créent (qui se créent eux-mêmes) ou s'auto-produisent (qui se produisent eux-mêmes) de telle sorte que :

- l'organisation autopoïétique est déterminée en termes de relations entre processus de production des composants dans un domaine de détermination interne (domaine conceptuel), c'est-à-dire de relations autopoïétiques qui dépend récursivement les uns des autres participant à la réalisation des processus de production des composants et à la génération d'une frontière dans le domaine physique qui délimite le système en tant qu'unité par son fonctionnement ;

- la structure autopoïétique est la spécification de ces relations autopoïétiques dans un domaine de composition (domaine physique). Et donc, il y a une description structurelle de propriétés et d'attributs dans un espace matériel des processus de production des composants pour maintenir l'identité de l'unité et l'autonomie. L'espace matériel est caractérisé par les propriétés des composants, comme la masse, la force, l'accélération, la distance, le champ, etc.

Dans l'organisation autopoïétique il y a trois types de relations autopoïétiques, c'est-à-dire des relations de production de composants qui définissent un type particulier de composant qui doit se matérialiser dans le domaine physique (la structure autopoïétique), à savoir : les *relations constitutives*, de *spécification*, et d'*ordre*.

Relations constitutives[96]

Les relations constitutives sont nécessaires à la production des composants ayant pour but la définition de la forme spatiale (topologique) du réseau de processus de production de composants, à partir des contraintes telles que la proximité, le voisinage, la séparation, etc., entre composants qui doivent être respectés au moment de la réalisation de la structure dans le domaine physique. Ces composants particuliers fixent la topologie du réseau, et donc la frontière physique de l'organisation autopoïétique est fixée par eux. Cette spécification topologique implique la définition de l'unité (l'organisation) du réseau des processus de production de composants qui opère sous clôture opérationnelle, couplage structurel, et détermination structurelle.

---

[96] On parle aussi de *processus de constitution*. Il a pour but la production de relations constitutives.

Les relations constitutives permettent de répondre à la question : *de quoi est faite la solidarité de l'organisation autopoïétique ?*

Les relations constitutives sont à la charge de l'auto-organisation du système autopoïétique.

Relations de spécification[97]

Les relations de spécification sont nécessaires à la production de composants ayant pour but la définition des propriétés des composants dans le domaine physique. Cette spécification des propriétés implique la définition de l'identité (la structure) du réseau des processus de production de composants qui opère sous clôture opérationnelle, couplage structurel, et détermination structurelle.

Les relations de spécification permettent de répondre à la question : *de quoi est faite la solidité de l'organisation autopoïétique ?*

Les relations de spécification sont à la charge de l'auto-maintient du système autopoïétique.

Relations d'ordre[98]

Les relations d'ordre sont nécessaires à la production des composants ayant pour but le contrôle, la régulation et la coordination de la dynamique du réseau de processus de production de composants (y compris la production des relations constitutives, de spécification et d'ordre). Cette spécification des propriétés implique la définition de l'autonomie (la contrainte) par la gestion du réseau des processus de production de composants qui opère sous clôture opérationnelle, couplage structurel, et détermination structurelle.

Les relations d'ordre permettent de répondre à la question : *de quoi est faite la coopération de l'organisation autopoïétique ?*

Les relations d'ordre sont à la charge de l'auto-gestion du système autopoïétique.

Ainsi, l'organisation de la structure autopoïétique est décrite (constituée) par ces trois types de relations autopoïétiques qui existent et fonctionnent simultanément comme nous l'avons vu dans deux domaines : le domaine conceptuel et le domaine physique. Le domaine conceptuel est l'espace théorique ou abstrait de l'organisation autopoïétique. Le domaine physique est l'espace matériel de la structure autopoïétique, c'est-à-dire la structure des processus d'auto-production des composants.

---

[97] On parle aussi de *processus de spécification*. Il a pour but la production de relations de spécification.
[98] On parle aussi de *processus d'ordre*. Il a pour but la production de relations d'ordre.

Pour cela les relations autopoïétiques doivent respecter simultanément et nécessairement trois conditions, à savoir : la *clôture opérationnelle*, le *couplage structurel*, et la *détermination structurelle*.

## Clôture opérationnelle

La clôture opérationnelle se réfère au fait que les sorties des processus d'auto-production des composants (qui par l'autopoïése sont des processus de production des composants eux-mêmes), se situent à l'intérieur des frontières du système autopoïétique lui-même. Bien sûr le résultat d'un système autopoïétique est le produit de ce qui se trouve déjà en lui-même (c'est-à-dire, son organisation). En d'autres termes la clôture opérationnelle est relative à l'organisation (le réseau de relations entre processus d'auto-production des composants). Cela signifie qu'un système autopoïétique est un système clos au niveau de son organisation, c'est-à-dire qu'il n'y a pas d'interactions avec son environnement (sans entrées ni sorties) pour la génération continue du réseau de processus d'auto-production de composants.

## Couplage structurel

Le couplage structurel signifie que l'espace matériel (la structure) de la réalisation des processus d'auto-production des composants par son fonctionnement, peut subir une série de perturbations qui en général peuvent être classées dans deux types : composition ou concaténation. Les perturbations de composition sont relatives au changement des propriétés de composants, tandis que les perturbations de concaténation sont relatives au disfonctionnement dans l'auto-production des composants. Ces perturbations, causées par des événements de son environnement, doivent être contrôlées et régulées afin d'éviter la désintégration du réseau de relations autopoïétiques (entre processus d'auto-production des composants), c'est-à-dire pour éviter la perte de l'organisation de l'unité, et donc la disparition du système. Pour cela le couplage structurel est un concept de la théorie autopoïétique pour indiquer que la structure de l'organisation peut subir des transformations structurelles, afin de compenser ces perturbations au niveau de la composition ou la concaténation entre processus d'auto-production des composants du réseau qui opèrent sous clôture opérationnelle pour éviter la désintégration du système autopoïétique. En d'autres termes, un système autopoïétique pour maintenir l'unité de l'organisation et afin de ne pas perdre son identité, doit modifier sa structure, donc en adaptation permanente avec son environnement. Un système autopoïétique est un système ouvert au niveau de sa structure (l'espace des propriétés des processus d'auto-production des composants), mais autonome car les changements de la structure sont subordonnés au maintien de l'identité de l'organisation, sans perdre son unité (distinguable comme une classe).

Détermination structurelle

La détermination structurelle, signifie que la frontière d'un système autopoïétique est spécifiée par son fonctionnement. Autrement dit, les processus d'auto-production des composants du réseau qui opère sous clôture opérationnelle, assurent une transformation définie par son organisation, spécifiée ou déterminée dans sa structure. Ainsi, le concept de détermination structurelle est utilisé pour indiquer que le système autopoïétique a la propriété de fixer ses propres limites. Voici l'idée dans une analogie superficielle :

Le pont de la place Saint Pierre

Imaginons un pont où la structure était faite pour résister un poids maximum de 2 tonnes était traversé pour un camion de 3 tonnes. Or, si pendant la traversée le pont se collapse alors ce n'est pas la faute du pont qui n'a pas résisté au poids du camion, mais plutôt c'est le fait que sa structure n'était pas prévue pour supporter un poids de 3 tonnes.

La figure 2.6 montre une synthèse du modèle autopoïétique.

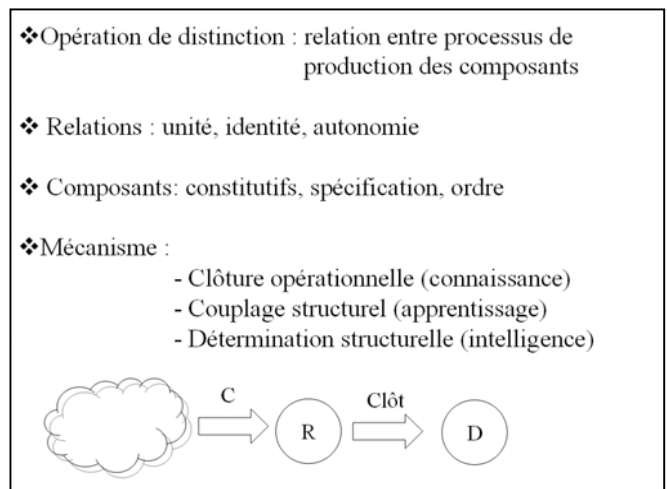

**Figure 2.6 :** Modèle autopoïétique (source propre)

Dans ce modèle l'intelligence correspond au maintien de la structure de l'organisation par un mécanisme de détermination structurelle (D). L'apprentissage correspond au maintien du système en vie, vis-à-vis des perturbations de l'environnement, par un mécanisme de couplage structurel (C). La connaissance correspond au maintien du système, viable (avec en sens) par un mécanisme de clôture opérationnelle (Clôt), vis-à-vis des relations (R) : unité, identité, et autonomie.

En conséquence, dans la conception de la connaissance selon l'approche de l'enaction il y a une non séparabilité du *vivant* et du *cognitif*. Nous pensons que cela est une raison suffisante pour appliquer l'autopoïèse en dehors du domaine biologique, par exemple au niveau social, bien que ceci ait été critiqué par Maturana et Varela comme nous l'avons montré plus haut. En effet, l'autopoïèse se réfère à l'auto-production (chimique) des relations entre processus de production de composants dans le champ (biologique), et cette capacité constitue la propriété centrale des systèmes vivants au niveau cellulaire mais non pas au niveau social. En fait, nous pensons que la question fondamentale de la théorie autopoïétique (qu'y a-t-il de commun à tous les êtres vivants, qui permet de les qualifier de vivants ?) se trouve autour de l'autonomie, l'identité et l'unité des relations entre processus de production de composants de l'organisation. Cette théorie explique la manifestation des propriétés caractéristiques des systèmes vivants (un patron commun, une classe) dans le domaine conceptuel et dans le domaine physique, et peut être généralisée en termes *d'autonomie* et de *connaissance* (ce n'est pas un hasard si le livre de Varela, s'intitule en français : *Autonomie et Connaissance. Essai sur le Vivant* qui correspond à la traduction du livre *Principles of Biological Autonomy* qui est basé sur l'ouvrage de Maturana et Varela, intitulé *Autopoiesis and Cognition)*, pour appliquer la mise en œuvre de l'*autopoïèse* dans le champ social.

Nous présentons ensuite la contribution de Limone qui a ouvert le débat sur : l'entreprise matérialise-t-elle une organisation autopoïétique ?

Pour cela nous prenons comme point de départ, l'autopoïèse de Valparaiso pour concevoir l'entreprise comme une organisation autopoïétique. Cette hypothèse (qui peut être énoncée comme suit « Y-a-t-il un patron d'organisation commun qui peut être identifié dans toutes les entreprises ? » [Limone 77] a été mise au point par notre ami et collègue Aquiles Limone, afin d'introduire pour la première fois la théorie autopoïétique de Santiago comme un sujet de recherche en management, à l'occasion de sa thèse intitulée *L'autopoïèse dans les organisations*, sous la direction de Jacques Mélèse et publiée en 1977 par l'Université Paris IX Dauphine.

La thèse de Limone a posé un regard nouveau sur les questions fondamentales des organisations sociales, en particulier l'entreprise, en se rapprochant d'un point de vue autopoïétique, de la phénoménologie des entreprises. Et ceci a été fait à partir des conditions de sa thèse, comme il l'a souligné lui-même « nous allons centrer nos préoccupations dans l'étude d'une théorie où nos recherches dans le domaine de la biologie, un peu par "hasard" et aussi un peu par "nécessité" du sujet, nous ont amenées à : la théorie autopoïétique » [Limone 77]. Ainsi, l'application de la théorie autopoïétique de Santiago à l'entreprise lui a permis d'élaborer plusieurs hypothèses de recherche, qui ont bâti plus tard avec la collaboration de Luis Bastias, la théorie autopoïétique de Valparaiso et le modèle CIBORGA[99] [Limone et Bastias, 02a], [Limone et Cademártori, 98], [Bastias, 00] que nous présentons ensuite.

> ➢ **Théorie autopoïétique de Valparaiso & modèle CIBORGA**

La théorie autopoïétique de Valparaiso, connue aussi sous le nom de *théorie de l'autopoïèse de l'entreprise*, peut être formulée sous trois hypothèses. La première hypothèse est que l'entreprise s'organise parce qu'elle a des objectifs, des buts à atteindre qui garantissent sa survie, et donc elle peut être considérée comme un système viable. La deuxième hypothèse est que l'entreprise est un système vivant parce que sa force motrice, ce sont des hommes. La troisième hypothèse est que l'entreprise est un système clos à l'intérieur d'un système ouvert parce qu'elle est ouverte par rapport à son fonctionnement pour le maintien de sa structure mais clos par rapport à son organisation de l'unité, de telle sorte que l'organisation est matérialisée par un réseau de processus de production de composants qui opère sous clôture opérationnelle, couplage structurel, et détermination structurelle pour le maintien de l'identité dans et par l'unité, c'est-à-dire qui produit par sa transformation les composants eux-mêmes, et ce résultat reste à l'intérieur de sa frontière. Comme l'a dit Limone « c'est précisément dans cette coexistence d'un système clos au sein d'un système ouvert, que semble se trouver la clé pour la compréhension du phénomène que nous appelons organisation humaine ». Puis il ajoute « l'existence d'un invariant qui semblerait coexister avec les changements continus ». En d'autres termes, d'après Limone l'autopoïèse est le modèle d'organisation que les relations entre processus de production de composants de l'entreprise doivent respecter pour maintenir, tout au long du temps, l'autonomie, l'identité et l'unité de l'entreprise, bien entendu quelle que soit l'entreprise. Nous n'irons pas plus loin sur cette théorie et leur démarche et canevas de pensée mise en place pour la développer (qui peut être consultée dans [Limone, 77]), car nous verrons davantage le résultat de l'application de cette théorie dans l'entreprise : le *modèle CIBORGA*.

Le modèle CIBORGA (Cybernétique de l'Organisation et de l'Apprentissage) existe simultanément et nécessairement, d'une part, dans le domaine conceptuel (l'organisation). Il s'agit d'un domaine social défini à travers de relations humaines, et d'autre part, dans un domaine physique (la structure). Il s'agit d'un domaine matériel à travers des individus, matière, énergie, et symboles[100] [Limone et Bastias, 02a], [Limone et Cademártori, 98], [Bastias, 00].

Dans le modèle CIBORGA, l'organisation de la structure est composée par trois types de relations qui opèrent sous clôture opérationnelle, couplage structurel, et détermination structurelle pour le maintien de l'identité dans et par l'unité, à savoir : les *relations primaires*, *structurelles* et *décisionnelles*.

---

[99] Modèle Cybernétique de l'Organisation et de l'Apprentissage.
[100] Les symboles ou signes sont des éléments d'un domaine linguistique, dans ce sens c'est l'observateur qui par une opération de distinction donne de sens aux objets qui les entourent.

Dans un modèle autopoïétique l'opération de distinction, est formulé par rapport à une dualité organisation/structure sur la base de l'approche enactiviste de la cognition. Cette opération de distinction permet indiquer que les causes et les effets sont distinguables dans des espaces fort différentes. L'un est le domaine conceptuel (l'organisation) du processus organisationnel (c'est un processus conceptuel de description abstrait de l'organisation). L'autre est le domaine physique (la structure) du processus structurel (c'est un processus physique de description matérielle de la structure, il s'agit bien ici de la description des propriétés des composants de la structure de l'organisation).

Relations primaires

Les relations primaires sont nécessaires à l'auto-production de composants primaires tels que : financiers, productifs, ressources humaines et commerciales. Les processus primaires ont pour but la régénération des fonds consommés dans le maintien de sa structure et des fonds utilisés dans ces processus primaires. Les relations primaires correspondent aux relations constitutives de la théorie autopoïétique de Santiago.

Relations structurelles

Les relations structurelles sont nécessaires à l'auto-production de composants structurels ayant pour but la définition des propriétés des composants des processus primaires (financiers, productifs, humains et commerciaux) dans le domaine physique. La structuration est une tâche de différenciation et d'intégration des composants dans la structure par le choix des propriétés physiques telles que, l'expérience (pour des ressources humaines), le type de machine (pour des ressources productives), etc. Les relations structurelles correspondent aux relations de spécification de la théorie autopoïétique de Santiago.

Relations décisionnelles

Les relations décisionnelles sont nécessaires à l'auto-production de composants décisionnels ayant pour but le contrôle, la régulation et la coordination des processus primaires, structurels et décisionnels par la *décision*. La décision est un acte conversationnel de compromis pour faire quelque chose, soit au niveau technologique, travail, ou économique. Les relations décisionnelles correspondent aux relations d'ordre de la théorie autopoïétique de Santiago.

Or, comme dans la théorie autopoïétique de Santiago, ces trois relations doivent respecter trois conditions pour exister dans le domaine social et physique de la dualité organisation/structure afin d'auto-produire (de façon permanente) les composants qui sont eux-mêmes des processus ou transformations qui forment un réseau ou une chaîne de transformations, à savoir : la *clôture opérationnelle*, le *couplage structurel*, et la *détermination structurelle*.

## Clôture opérationnelle

La clôture opérationnelle se réfère au fait que l'entreprise est un système clos au niveau de son organisation (le résultat de la transformation se situe à l'intérieur des frontières du système lui-même), mais ouvert au niveau de sa structure (le résultat de la transformation se situe à l'extérieur des frontières du système lui-même). Essayons de voir ceci par un exemple, si nous pensons à une relation décisionnelle de l'entreprise, par exemple la négociation, cette relation est toujours close, c'est-à-dire engendrée dans un réseau fermé de compromis, engendrée dans un domaine clos de transformations de négociation, mais le produit de ce processus de transformation (la négociation) aura une influence sur l'activité, et donc la négociation sera matérialisée soit au niveau de la structure de l'organisation, soit au niveau des processus de transformation.

## Couplage structurel

Le couplage structurel signifie que l'entreprise pour maintenir son organisation doit modifier sa structure. Par exemple, les relations structurelles d'apprentissage ou d'intelligence économique sont nécessaires pour maintenir la structure en capacité permanente avec son environnement technologique. Dans cet exemple, les attributs des relations entre processus de production de composants sont des expertises qui forment une structure de connaissance.

## Détermination structurelle

La détermination structurelle veut dire que l'entreprise assure une transformation définie par l'organisation de ses relations entre processus de production de composants. Par exemple, les relations primaires (dans la pensée fayoliste) sont contraignantes pour la détermination structurelle. C'est le milieu naturel de certains technocrates de la gestion, qui essayent de tout planifier, organiser, diriger et contrôler pour bien gérer. Avec cela, nous ne voudrions pas dire que pour gérer une affaire il ne faut pas tenir compte de ces quatre fonctions clé des relations primaires de la pensé fayoliste, mais que le modèle du business dans un sens large, doit tenir compte aussi des aléas de la réalité quotidienne de ces relations primaires.

Par conséquent, selon l'approche autopoïétique de Valparaiso : (1) l'entreprise est un système clos qui fonctionne avec *clôture opérationnelle* (le résultat de la transformation du système clos se situe nécessairement à l'intérieur du même système) ; et (2) l'entreprise existe simultanément et nécessairement sur deux domaines [Limone, 77], [Limone et Bastias, 02a]. L'un est le *domaine social* constitué par un espace clos de relations humaines, l'autre est le *domaine physique* constitué par un espace ouvert de ressources (humaines, matérielles et énergétiques), ainsi que de symboles[101]. L'identité de l'entreprise comme structure, est définie dans le domaine physique, ceci conditionne la spécification des relations entre processus de production de composants dans l'espace matériel de ses propriétés. Le résultat du fonctionnement de l'entreprise d'une part dans le domaine social est le maintien de l'organisation de l'unité, et d'autre part dans le domaine physique ce sont des relations entre processus de production de composants qui définissent l'entreprise comme unité. Le fonctionnement de l'organisation est donc clos, tandis que la structure de l'organisation du système est ouverte, ce que nous renvoie à la dualité organisation/structure.

L'application de la théorie autopoïétique de Valparaiso à l'entreprise a permis d'élaborer le modèle CIBORGA qui a été conçu comme un modèle explicatif de l'entreprise à partir de relations autopoïétiques (constitutives, spécification, et ordre) qui opèrent sous clôture opérationnelle, couplage structurel, et détermination structurelle. Ce modèle nous a ouvert un champ de réflexion sur l'interprétation des modèles traditionnels (principalement OID et OIDC) sous l'angle de l'autopoïèse. Il a également permis d'élargir le champ d'explication des systèmes à base de connaissance proposé par Jean-Louis Ermine. Nous devons préciser néanmoins que l'utilisation de l'autopoïèse à cette fin a déjà été mise en oeuvre par Limone. En effet, pour observer l'entreprise, les modèles traditionnels (OID, OIDC, AMS et MSV), comme nous le verrons par la suite, ont supposé que l'entreprise est un système, et que ceci reste vrai, quelle que soit l'approche (système et/ou cybernétique). Cela nous amène à observer l'entreprise, d'une part, en termes de son organisation comme un système clos, et d'autre part, en termes de sa structure comme un système ouvert, ce qui pose la question de l'unité, l'identité et l'autonomie, d'une part, sur la frontière entre les entrées/transformation/sorties de la structure et leur environnement selon l'approche système, et d'autre part, l'approche cybernétique (de deuxième ordre) introduit les concepts de variables, de contrôle et de régulation. Malgré cela, aucune de ces approches ne sont suffisantes pour expliquer complètement l'organisation de l'unité dans sa totalité imparfaite.

**Conclusion du chapitre**

Nous avons présenté dans ce chapitre :

---

[101] Les symboles ou signes sont des éléments d'un domaine linguistique, dans ce sens c'est l'observateur qui par une opération de distinction donne de sens aux objets qui les entourent.

- premièrement, les dualités qui ont servi pour expliquer les modèles de description de l'unité, telles que : domaine/unité, observateur/observé, unité/distinction, organisation/système, organisation/structure, et gestion/management. Ces dualités ont été considérées comme un processus de causalité circulaire propre à la dualité cause/effet, et non pas comme une dualité cartésienne. Néanmoins, d'une façon général, ce processus est attaché à la cognition. Ainsi, (a) si le processus est relatif à l'approche cognitiviste de la cognition, alors ce que l'on cherche à construire est une chaîne logique de cause à effet, et (b) si le processus est associé à l'approche enactiviste de la cognition, alors la relation ne se construit pas nécessairement à partir d'une représentation vraie ou logique de causes et d'effets, puisque le problème de l'enaction n'apparaît pas lié à la représentation symbolique d'une réalité. La figure 2.7 permet de situer ces approches et ses éléments associés.

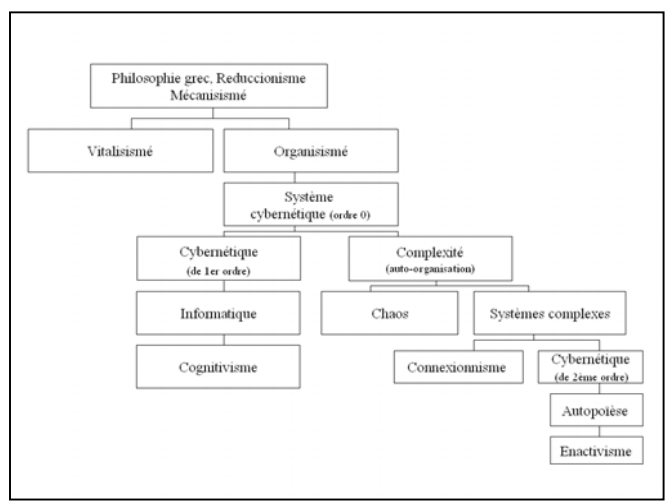

**Figure 2.7 :** Synthèse des approches cognitivistes et ses éléments associés (source [Bastias, 00])[102]

- deuxièmement, on a présentés les modèles, que nous avons retenu pour la de description de l'unité, tels que : OID (Opération, Information, Décision), OIDC (Opération, Information, Décision, Connaissance), AMS (Analyse Modulaire des Systèmes), MSV (Modèle des Systèmes Viables), et CIBORGA (Modèle Cybernétique de l'Organisation et de l'Apprentissage). Pour chacun de ces modèles nous avons fait ressortir une caractéristique générique, à savoir un concept d'activité pour le modèle OID, un concept de flux cognitif et de flux de compétence pour le modèle OIDC, un concept de contrôle et de régulation des variables pour le modèle AMS, le concept d'intelligence et de pouvoir pour le modèle MSV, et enfin un concept d'organisation et d'apprentissage pour le modèle CIBORGA.

---

[102] Au XVIIIème siècle Lawrence Henderson introduit le concept de "collectivité" comme la base de l'organisme (organicisme). En revanche, Hans Driesch parle de "force vitale" pour définir l'organisme (vitalisme).

A partir de ces concepts associés, ces modèles ont été considérés comme des modèles génériques de la gestion des connaissances. Ainsi, les racines de l'arbre de la gestion des connaissances correspondent aux trois sources complémentaires, à savoir :

- le modèle système a permis d'observer la connaissance dans l'entreprise comme un flux d'entrées/transformation/sorties. L'approche système introduit la relation entre système et environnement ;

- le modèle cybernétique (de deuxième ordre) a permis d'observer la connaissance dans l'entreprise comme une organisation de relations entre composants. L'approche cybernétique (de deuxième ordre) introduit le concept d'organisation de l'unité à travers les mécanismes de contrôle et de régulation des variables qui correspondent aux différents flux du système ;

- le modèle autopoïétique de Santiago et de Valparaiso a permis d'observer la connaissance dans le système (ou l'entreprise) comme une organisation de relations entre processus de production de composants, à savoir : constitutives (ou primaires), de spécification (ou structurelles), et d'ordre (ou décisionnelles). Ces relations (qui correspondent aux mécanismes du modèle autopoïétique) opèrent sous clôture opérationnelle, couplage structurel, et détermination structurelle. L'approche autopoïétique introduit le concept d'opération de distinction. Le tableau 2.1 montre une comparaison de ces trois modèles.

|  | Description de l'unité | Type de système | Exemples |
|---|---|---|---|
| Modèle système | environnement | système ouvert (entrées, sorties) système fermé (ni entrées, ni sorties) | OID, OIDC |
| Modèle cybernétique | organisation | système viable (sens) | AMS, MSV |
| Modèle autopoïétique | autonomie | système clos (sens, organisation, structure) | CIBORGA |

**Tableau 2.1 :** Description de l'unité (système, cybernétique, autopoïétique)

La conclusion générale de ce chapitre est que cette organisation du système nous permet de faire le clivage entre deux sorts de connaissances :

- la connaissance d'un point de vue système (OID, OIDC) et cybernétique (AMS, MSV). La connaissance est un système ouvert d'actions. Le contrôle et la régulation doivent garantir l'équilibre homéostatique du système. La création de nouvelles connaissances et d'apprentissage organisationnel, doivent garantir d'une part, l'innovation continue (dans le sens du modèle de Nonaka et Takeuchi), et d'autre part, la capacité du système de connaissances pour "faire-évoluer" les connaissances et les compétences (selon l'approche de l'enaction de Karl Weick) dans un système d'organisation du travail coopératif.

- la connaissance d'un point de vue autopoïétique est un système d'actions clos, où l'action existe, pour l'acteur, comme un acte de langage (c'est-à-dire le langage et ses émotions) et de culture (c'est-à-dire la structure sociale). Ici, la connaissance n'est pas considérée comme la "dernière" couche comme dans le modèle OIDC, ni comme une dualité cartésienne cognition/action, mais plutôt comme le fait que la connaissance génère les informations et les données nécessaires pour maintenir le système en vie (avec une structure et une organisation) et viable (avec un sens). Le contrôle et la régulation doivent garantir l'autonomie, l'identité et l'unité du système autopoïétique. La création de nouvelles connaissances et d'apprentissage organisationnel, doivent garantir d'une part, l'organisation, et d'autre part, la structure du système de connaissance pour "faire-émerger" les connaissances et les compétences (selon l'approche de l'enaction de Maturana et Varela) dans un système d'organisation du travail dans un paradigme d'avantage collectif.

Pour nous, le modèle autopoïétique reste plus en accord avec la réalité imparfaite, car le langage est un moyen flou d'action qui permet de reproduire le savoir-faire d'un expert dans son activité, par exemple en racontant à ses collègues ce qu'il faut faire dans le cas où certains événements redoutés se produisent dans la production ou dans une négociation avec un client.

Finalement, ce chapitre ne représente en fait qu'une amorce pour pouvoir finalement établir un véritable *framework* autour de la gestion des connaissances.

> *Autopoïèse ... c'est là une conception de la connaissance et de la réalité qui tient compte du fait que nous participons à leur élaboration, et nous pouvons voir qu'elle s'enracine dans les formes cellulaires les plus élémentaires des processus cognitifs et informationnels.*
> *Francisco Varela, biologiste et informaticien chilien*

Nous soulignons à ce stade que la problématique scientifique de cette thèse comporte trois volets. Premièrement, peut-on envisager la question des mécanismes génériques de la gestion des connaissances ? Si la réponse est oui, comment peut-on la formuler ? Deuxièmement, sur la base de ces mécanismes, comment peut-on envisager un modèle pour la gestion des connaissances ? Troisièmement, comment peut-on envisager le retour d'expérience du modèle proposé (en partant sur un besoin industriel) ? C'est ainsi que :

- Le chapitre 1 a permis de dégager les mécanismes génériques de la gestion des connaissances à partir des mécanismes de création des connaissances nouvelles et d'apprentissage organisationnel d'un point de vue social et technique. Ce que nous avons fait à partir de quatre approches, à savoir : (1) l'approche organisationnelle de Nonaka et Takeuchi fondée sur le concept de *knowledge creating-company* ; (2) l'approche biologique de Maturana et Varela fondée sur le concept de l'*enaction* ou l'*arbre des connaissances* ; (3) l'approche managériale de Jean-Louis Ermine fondée sur le concept de la *marguerite* ; et (4) l'approche de NTIC du KM fondée sur les nouvelles technologies de l'information et de la communication du Knowledge Management (Internet, Intranet, et Extranet). Cette démarche, nous a permis de formuler les mécanismes génériques de la gestion des connaissances que nous avons identifiée par des verbes à l'infinitif ;

- Le chapitre 2 a permis de dégager les modèles génériques de la gestion des connaissances à partir des approches systémique, cybernétique, et autopoïétique. Nous avons repéré cinq modèles, à savoir : le *modèle des systèmes organisés* (en abrégé modèle OID[1]) de Jean-Louis Le Moigne, le *modèle des systèmes de connaissances* (en abrégé modèle OIDC[2]) de Jean-Louis Ermine, le *modèle des systèmes de gestion* (en abrégé modèle AMS[3]) de Jacques Mélèse, le *modèle des systèmes de management* (en abrégé modèle MSV[4]) de Stafford Beer, et enfin le *modèle des organisations autopoïétiques* (en abrégé modèle CIBORGA[5]) de Aquiles Limone et Luis Bastias. Dans le sens, que les modèles OID et OIDC sont associés à l'approche système, les modèles AMS et MSV sont associés à l'approche cybernétique, tandis que le modèle CIBORGA est associé à l'approche autopoïétique de Valparaiso (qui prend ses racines dans l'approche autopoïétique de Santiago).

---

[1] Opération, Information, Décision.
[2] Opération, Information, Décision, Connaissance.
[3] Analyse Modulaire des Systèmes.
[4] Modèle des Systèmes Viables.
[5] Modèle Cybernétique de l'Organisation et de l'Apprentissage.

Ces modèles dans ce chapitre 3, vont nous servir comme un véritable *framework* autour de la gestion des connaissances, c'est-à-dire que ces modèles vont guider la démarche et les canevas de pensée de notre démarche, afin de construire un modèle Les fondements théoriques du modèle proposé se trouvent dans l'approche de "l'enaction" de Maturana et Varela et le modèle autopoïétique (de Santiago et de Valparaiso), ainsi que dans les mécanismes génériques de la gestion des connaissances (présentés dans la section 1.4 du chapitre 1). Dans ce contexte, l'enaction est un mécanisme "circulant" et "d'émergence de signification", et non pas de "sélection d'information", selon l'approche de l'enaction de Karl Weick.

L'objectif de ce chapitre est donc de proposer un modèle pour la gestion des connaissances.

Ce chapitre est organisé en quatre parties. La première partie appelée « évolution de la connaissance dans le modèle de gestion des connaissances de Ermine » permet d'explorer les hypothèses de base de l'évolution de la connaissance, à partir la relation entre connaissance et environnement, l'organisation et le fonctionnement du modèle de gestion des connaissances, l'opération de distinction de ce modèle, et la relation entre environnement et organisation, et environnement et patrimoine des connaissances. La deuxième partie appelée « évolution de la connaissance dans le modèle autopoïétique de la gestion des connaissances » permet d'explorer les hypothèses de base de l'évolution de la connaissance, à partir la relation entre autopoïèse et connaissance, création de sens dans un système clos (clôture opérationnelle), et apprentissage et évolution du système clos (couplage structurel). La troisième partie appelée « cadre de référence du modèle proposé par rapport à l'aspect social et technique de la gestion des connaissances » permet de présenter les quatre hypothèses du modèle autopoïétique de la gestion des connaissances imparfaites, à savoir : l'hypothèse de l'enaction, l'hypothèse de spontanéité des relations, l'hypothèse du noyau invariant, et l'hypothèse de la connaissance imparfaite, ensuite nous pressentons le modèle proposé, premièrement, par rapport à l'aspect social de la gestion des connaissances que nous faisons à travers l'aspect dynamique, l'aspect action, l'aspect culturel, et l'aspect chaos de la gestion des connaissances, et deuxièmement selon l'aspect technique de la gestion des connaissances à travers les SBC issus de l'approche cognitiviste, les SBC issus de l'approche connexioniste, et les SBC issus de l'approche enactiviste. La quatrième partie appelée « 4 Repères essentiels pour la gestion des connaissances dans l'entreprise » permet de présenter nos quatre repères essentiels pour positionner une problématique de gestion des connaissances dans l'entreprise. Le premier repère est que la gestion des connaissances est un système qui peut être décrit à travers un système sociotechnique récursif. Le deuxième repère est que la gestion des connaissances est un objet de gestion par rapport à la dualité organisation/structure. Le troisième repère est relative au fait que la gestion des connaissances peut être observé, selon trois approches : l'approche organisationnelle de Nonaka et Takeuchi, l'approche biologique de Maturana et Varela, et l'approche managériale de Jean-Louis Ermine. Le quatrième repère est que la gestion des connaissances existe dans un plan moral et éthique plus que sociale ou technique

## 3.1. Evolution de la connaissance dans le modèle de gestion des connaissances de Ermine[6]

Dans cette section nous passons en revue les arguments chez Tounkara[7] relatifs à l'évolution de la connaissance dans le modèle de gestion des connaissances. Nous avons repéré Ces arguments dans sa thèse, chapitre 4 (proposition d'un modèle d'interaction en réponse à la problématique d'évolution du patrimoine de connaissances), section 4.1 (environnement et connaissances : les hypothèses de travail). Ces arguments-là seront utilisés pour montrer que l'approche de l'enaction de Karl Weick, est le fondement théorique du modèle de gestion des connaissances de Jean-Louis Ermine.

L'approche de l'enaction de Karl Weick a été proposée dans son livre, intitulé *The Social Psychology of Organizing* (publié en 1979) [Weick, 79]. En gros, cette approche décrit la relation organisation-environnement comme un processus d'interaction.

### 3.1.1. Connaissance et environnement

L'hypothèse de base du modèle de gestion des connaissances de Ermine peut être expliquée à partir de la relation connaissance-environnement. A ce propos Tounkara a dit « lorsque l'on cherche à positionner une organisation dans un contexte socié-économique, la vision qui s'impose est celle de l'entreprise qui est incluse dans son environnement. Cependant il ne s'agit que d'une vision spatiale, qu'on aurait tort d'utiliser comme référence. Quand une entreprise doit se positionner dans son environnement, il est bon de remettre en cause ce point de vue purement topologique, et de considérer que l'entreprise et son environnement sont deux sous-systèmes distincts, mais en interaction forte, et que l'un des deux systèmes ne doit pas "dicter sa loi" à l'autre, ne serait ce que par une relation d'inclusion unilatérale. La pondération équivalente des interactions de l'environnement vers l'organisation et vice versa est un facteur fondamental de stabilité » [Tounkara, 02].

---

[6] Dans le chapitre 1 (section 1.2.3) nous avons introduit ce modèle par le biais du modèle de la marguerite, notre propos était de repérer les mécanismes génériques de création de connaissance et d'apprentissage organisationnel. Ici, nous sommes intéressés au processus d'organisation et de fonctionnement du système de connaissance de ce modèle.

[7] Nous rappelons que la thèse de Tounkara porte sur la problématique d'évolution du patrimoine de connaissances du modèle de la marguerite chez Jean-Louis Ermine. C'est pour cette raison que nous situons les hypothèses du travail de thèse chez Tounkara au cœur de ce modèle. En plus, Ermine lui-même est le Codirecteur de cette thèse.

Dans cette relation connaissance-environnement, nous observons deux sortes de phénomènes, que nous appelons : *processus d'interaction par inclusion* et *processus d'interaction par parallélisme*[8] :

- le processus d'interaction par inclusion, représente le fait que la connaissance est incluse dans l'environnement. L'inclusion signifie que l'environnement est un seul domaine de connaissance, alors l'identification et la solution des problèmes doivent avoir lieu dans ce domaine de connaissance et pas dans un autre. En conséquence, le processus d'interaction par inclusion ne permet pas l'évolution de connaissances, mais la création de connaissances dans le même domaine. Et donc, il n'y a pas création d'autres domaines de connaissances ;

- le processus d'interaction par parallélisme représente le fait que la connaissance n'est pas incluse dans l'environnement, autrement dit la connaissance à un domaine propre qui n'appartient pas à l'environnement. Dans ce contexte, le parallélisme signifie que l'identification d'un problème est lié à son domaine de connaissance, mais sa solution peut se trouver dans l'environnement. La seule façon d'ouvrir un contact avec l'environnement se trouve à travers un "dialogue". Ce dialogue correspond à un processus d'interaction par parallélisme. Dans ce contexte, nous avons à la fois création et évolution aussi, bien pour le domaine de connaissance en question, que pour l'environnement.

Le processus d'interaction par parallélisme correspond au processus d'interaction (confrontation) EP chez Tounkara. Dans ce contexte il a dit « le processus de confrontation[9] de l'entreprise à son environnement peut être modélisé en trois phases : la projection, le renseignement et la création de connaissances » [Tounkara, 02].

Nous constatons donc que le processus d'interaction EP est le fondement du processus de création et évolution de connaissances du modèle de gestion des connaissances de Ermine.

A partir de cet argument-là, nous faisons l'hypothèse que le support théorique du modèle de la gestion des connaissances chez Ermine, est l'approche de l'enaction de Karl Weick.

Passons maintenant à une analyse plus approfondie d'un tel effet, à l'aide de trois aspects du modèle de gestion des connaissances de Ermine : (1) organisation et fonctionnement ; (2) opération de distinction ; et (3) environnement-organisation et environnement-patrimoine de connaissances, afin de décrire le modèle de Ermine, comme un système composé de deux sous-systèmes. L'un est le système de connaissances, et l'autre est le système opérationnel.

---

[8] Nous verrons que le processus d'interaction par parallélisme est à la base du concept de couplage ponctuel, chez Maturana et Varela.
[9] Dans le chapitre 1 nous utilisons le terme "dialogue" plutôt que "confrontation" pour décrire la marguerite (voir chapitre 1, section 1.2.2).

### 3.1.2. Organisation et fonctionnement

> **L'organisation du modèle**

Si l'on fait l'hypothèse que l'approche de l'enaction de Karl Weick est la base théorique du modèle de la gestion des connaissances chez Ermine, reste à savoir s'il y a une autre approche de la cognition en jeu dans ce modèle ?

Pour répondre à cette question-là, regardons un peu plus en détails le système de connaissances du modèle de Ermine.

D'après Tounkara, l'organisation du modèle peut être décrite comme « l'ensemble des connaissances de l'organisation repose sur une "mémoire collective", un capital intangible que nous avons appelé le "patrimoine de connaissances". Ce patrimoine n'est pas réductible à des systèmes déjà existante comme le système d'information, de documentation, de ressources humaines (formation, compétences) etc. Le patrimoine de connaissances est un système spécifique, au même titre que le système cognitif chez l'être humain. Il a ses objectifs propres assignés par l'entreprise (Repérer, Préserver, Valoriser, Evoluer), son organisation et sa structure propre »[10].

Nous constatons donc deux traits caractéristiques du système de connaissances. L'un est que le système de connaissances est au cœur de l'entreprise. L'autre est que le système de connaissances fonctionne comme le système cognitif des individus. Cela signifie, que les mécanismes de création de connaissances et d'apprentissage individuel (mais aussi collectif) sont caractérisés par quatre fonctions, que nous symbolisons par les verbes : Repérer, Préserver, Valoriser, et Evoluer.

> **Le fonctionnement du modèle**

Le fonctionnement de la structure du modèle est formé par l'interaction circulaire de trois processus[11] autour du système de connaissances (ou patrimoine de connaissances), en suivant la

---

[10] Dans la thèse de Tounkara les concepts d'organisation et de structure ne sont pas spécifiés. Dans cette thèse, organisation = unité (classe) et structure = identité (un élément de la classe). Nous utilisons aussi les termes "solidarité" et "solidité" pour désigner l'organisation et la structure, respectivement. L'autonomie correspond à la dualité unité/identité, que parfois nous symbolisons par le terme "dans et par".

[11] Dans le chapitre 1 nous avons introduit ce modèle par le biais de cinq processus, car notre propos était de repérer les mécanismes de création de connaissance et d'apprentissage organisationnel, à savoir : (1) le processus de capitalisation et de partage des connaissances ; (2) le processus d'interaction avec l'environnement ; (3) le processus de sélection par l'environnement ; (4) le processus d'apprentissage et de création de connaissances ; (5) le processus d'évaluation du

démarche de la marguerite, à savoir : (1) le processus de projection. L'objectif de ce processus (symbolisé par le verbe *repérer*) est la formulation des requêtes vers l'environnement externe (sélection) afin de détecter un besoin, les outils privilégiés sont la veille ; (2) le processus de renseignement. L'objectif de ce processus (symbolisé par le verbe *préserver*) est l'élaboration de corpus d'information, les outils privilégiés sont la capitalisation et le partage de connaissances ; et (3) le processus d'apprentissage et de création de connaissances. L'objectif de ce processus (symbolisé par les verbes *valoriser* et *évoluer*) est double, d'une part, la valorisation de connaissances acquises avec l'environnement interne (interaction), et d'autre part, l'évolution du patrimoine de connaissances, les outils privilégiés sont le retour d'expérience et la créativité.

Cette démarche peut encore être affinée, ainsi nous avons six étapes, à savoir : la projection, la distorsion, l'identification, le rétro-ajustement, la représentation, et la construction de sens.

Tounkara a illustré dans le cadre de sa thèse ces six étapes, à travers la construction d'un système de connaissances dans le domaine botanique, comme lui-même il a souligné « le but de celle-ci, en se basant sur son système de connaissances, est de découvrir de nouvelles connaissances dans le domaine de la botanique ».

Nous constatons donc deux choses. D'abord le système de connaissances du modèle de gestion des connaissances chez Ermine, est construit par rapport à un domaine spécifique de connaissances, et donc il doit toujours être attiré par *le besoin* (c'est la *connaissance métier*, comme nous avons dit dans le chapitre 1).

En conséquence, dans l'aspect organisation et fonctionnement du modèle de la gestion des connaissances chez Ermine, la démarche de construction du système de connaissances, en plus d'être guidé, par l'approche de l'enaction de Karl Weick, est guidée par l'approche cognitiviste de la cognition, c'est-à-dire que les six étapes de la démarche, en particulier les étapes de représentation de connaissances, et de construction de sens sont liées aux explications symboliques[12].

---

patrimoine de connaissances (ce processus ne se trouve pas détaillé dans le modèle de la marguerite, mais nous l'avons trouvé ailleurs d'une façon implicite. Ainsi, les complexités de la gestion des connaissances chez Wendi Bukowitz, Ruth Williams [Bukowitz et Williams, 00], Karl Sveiby et Leif Edvinsson [Sveiby, 00], que nous avons identifié dans le même chapitre sont relatives au processus (5). Ici, nous sommes intéressés au processus d'organisation et de fonctionnement du système de connaissance de ce modèle, et donc nous somme passé à trois processus.
[12] Comme Maturan et Varela l'ont dit, par rapport à l'approche cognitiviste « une conception de la connaissance comme image et vision du monde … un monde donné d'instructions et de représentations » [Varela, 89].

### 3.1.3. Opération de distinction[13]

Au niveau des opérations de distinction, nous pouvons dire que le processus de renseignement de la démarche de la marguerite compte avec des outils de capitalisation et de partage de connaissances, tels que MKSM et MASK. Néanmoins, ces outils sont basés sur une approche cognitiviste de la connaissance. Pour les deux autres processus, il n'y a pas un discours très pragmatique, comme l'a fait sentir Tounkara, lorsqu'il a dit :

- pour le processus de projection : « la phase de projection implique le patrimoine de connaissances car c'est nécessairement la vision interne de l'entreprise qu'on met en correspondance avec l'extérieur … elle sera ainsi d'autant plus efficace que ce patrimoine sera mieux connu et mieux exploité. On rejoint ainsi la règle de bon sens du "connais-toi toi-même" : pour mieux connaître ce qui nous entoure, il est nécessaire de bien se connaître » [Tounkara, 02] ;

- pour le processus d'apprentissage et de création de connaissances : « la phase de création de connaissances est directement rattachée au patrimoine de connaissances. C'est pourtant encore la phase la plus mystérieuse : comment se créent de nouvelles connaissances utiles à l'entreprise lorsque celle-ci observe son environnement ? Comment l'entreprise se représente-t-elle ces connaissances et comment les intègre-t-elle dans son patrimoine ? » [Tounkara, 02].

Ces questions[14], nous renvoient toujours aux sources épistémologiques (approche philosophique, scientifique ou autre) de la cognition, comme l'ont souligné Aquiles Limone et Luis Bastias, dans un article[15] récent, intitulé Autopoïèse et connaissance dans l'organisation. Fondement conceptuel pour une authentique gestion des connaissances, ils ont dit « l'épistémologie … a comme objectif central : qu'est-ce que la connaissance ? Comment se partage-t-elle ? Comment se conserve-t-elle ? Comment évolue-t-elle ? » [Limone et Bastias, 02b][16].

Nous pensons que le discours de la connaissance, réactivé de nos jours comme discours de la gestion des connaissances, a remis en valeur la problématique de création des connaissances nouvelles et d'apprentissage organisationnel. Cette réactivité a besoin d'une opération de distinction[17], c'est-à-dire d'un lieu pour observer, ou pour mieux dire, d'un lieu pour expliquer la construction de sens et sa représentation.

---

[13] Le fait de distinguer l'unité de son environnement et donc d'autres unités. Autrement dit, l'opération de distinction permet à l'observateur de distinguer l'unité soit d'un point de vue de l'organisation, soit d'un point de vue de la structuration.
[14] Qui (heureusement !) restent ouverte et loin d'un prix marchand, comme nous l'avons fait sentir au chapitre 1, section 1.5 (les origines de la connaissance industrielle).
[15] L'article original est : *Autopoiésis y Conocimiento en la Organización. Fundamento Conceptual para una Auténtica Gestión del Conocimiento*, « la epistemología … ha tenido como objeto de estudio central la interrogatión acerca de qué es el conocimiento, cómo se difunde el conocimiemnto, cómo se conserva y cómo se impugna » [Limone et Bastias, 02b].
[16] Pour nous ces quatre sources épistémologiques ont été une véritable démarche d'explorateur, afin de retrouver un repère de réflexion, c'est ainsi que notre fil conducteur était conçu à travers trois directions de recherche, comme nous l'avons vu au chapitre 1, à savoir : (1) l'origine de la connaissance ; (2) la nature de la connaissance ; et (3) la validité de la connaissance, qui ont formé en gros le corpus de ce chapitre 1, de même que pour les modèles associés que nous avons regroupés au chapitre 2.

A ce sujet, Tounkara a dit « il s'agit, en effectuant les capacités cognitives individuelles ou collectives, de mettre en marche un processus d'interprétation et de création de connaissances dites "actionnables" au sens de Argyris (utiles dans les processus de production de la valeur de l'entreprise, à partir de la représentation de l'environnement résultant de la boucle projection/rétroaction). C'est un processus très mal connu au cœur du problème de l'organisation "cognitive", sorte de "sémiose collective" sur laquelle presque tout reste à dire » [Tounkara, 02].

Nous constatons donc, que le concept d'*actionable knowledge* chez Argyris est utilisé comme opération de distinction, afin d'observer le problème de la construction de sens et sa représentation dans le système de connaissances du modèle de la gestion des connaissances de Ermine.

Nous proposons, d'utiliser le concept de l'*autopoïèse de Santiago* et *de Valparaiso*, comme opération de distinction afin d'étudier ce problème-là.

### 3.1.4. Environnement-organisation et environnement-patrimoine de connaissances

Si nous partons de l'hypothèse que le processus d'interaction EP est le fondement du processus de création et d'évolution de connaissances du modèle de gestion des connaissances de Ermine, reste à définir le concept de système utilisé dans le modèle, s'agit-il d'un système ouvert ou d'un système fermé ?

Pour répondre à cette question-là, regardons un peu plus en détails l'hypothèse de base du modèle de la gestion des connaissances de Ermine. A ce propos, Tounkara a dit « la relation environnement-organisation et la relation environnement-patrimoine de connaissances … nous postulons que la confrontation entre l'entreprise et son environnement se fait via le patrimoine de connaissances et qu'il existe donc un processus d'interaction entre le patrimoine de connaissances et l'environnement de l'entreprise … à travers les acteurs détenteurs du savoir de l'entreprise le patrimoine de connaissances n'est pas un système isolé de l'environnement. Il est au contraire fortement relié à l'environnement et agit sur lui en créant de nouvelles connaissances qui vont modifier ainsi quelque peu l'environnement »[18].

---

[17] Ici ce concept prend l'importance d'un "macroscope" (vers une vision globale), en utilisant le terme chez Joël de Rosnay.

Par conséquent, le système de connaissances du modèle de la gestion des connaissances de Ermine, est un système ouvert : entrées/transformation/sorties. Ainsi, les entrées (appelées flux de compétences), la transformation, et les sorties (appelées flux cognitifs) sont définis par une approche système : l'*approche OIDC*[19].

Néanmoins, cette définition est un peu limitée, parce qu'elle ne permet pas de définir avec exactitude, la *frontière* entre l'environnement et le système OIDC, autrement dit, les limites entre ce qui est à l'intérieur (ce qui appartient au système) et, simultanément et nécessairement, à l'extérieur (ce qui appartient à l'environnement). A ce propos, par rapport aux relations entre le système de connaissances (C) et le système opérant (O), le système d'information (I), et le système de décision (D), Ermine a dit « le sous-système de connaissances est vu comme un sous-système actif du système. Ce processus se traduit classiquement par des flux qui créent des interrelations actives avec les autres sous-systèmes du systèmes … les flux de compétence, représentent justement la compétence (au sens classique et intuitif) en terme de connaissance, des différents sous-systèmes du système, c'est en quoi ils enrichissent le patrimoine de connaissances … les flux de cognition correspondent à l'appropriation implicite (le plus souvent) ou explicite de ce patrimoine en vue de l'utiliser dans le processus de transformation propre au système ». Nous constatons donc que les relations entre composants (O,I,D,C) du système OIDC sont les porteuses de compétences et de connaissances.

D'autre part, pour Tounkara les relations environnement-organisation et environnement-patrimoine de connaissances, peuvent être organisées en trois perspectives « (1) l'entreprise et son environnement sont deux sous-systèmes distincts, en interaction forte ; (2) l'intelligence de l'entreprise repose sur une mémoire collective que l'on appelle le "patrimoine de connaissances" ; et (3) il existe un processus d'interaction entre le patrimoine de connaissances de l'entreprise et son environnement » [Tounkara, 02].

Dans cette argumentation, reste à savoir quelle est la signification du concept d'*interaction forte* dans un processus d'interaction par parallélisme ?

A partir de ces trois aspects : (1) organisation et fonctionnement ; (2) opération de distinction ; et (3) environnement-organisation et environnement-patrimoine de connaissances, le modèle de gestion des connaissances de Ermine peut être expliqué en termes d'organisation et de structure des connaissances à travers un système de gestion des connaissances, composé de deux sous-systèmes, à savoir :

- le système de connaissances. Ce système est conçu d'une part, dans le domaine de l'organisation, par la description de l'unité à travers le modèle de la marguerite[20], et d'autre part, dans le domaine de la structure, par les composants de l'unité conçus par l'approche de l'enaction de la cognition de Karl Weick.

- le système opérationnel. Ce système est conçu, d'une part, dans le domaine de l'organisation, par la description de l'unité à travers le modèle OIDC, et d'autre part, dans le domaine de la structure, par les composants de l'unité conçus par l'approche cognitiviste de la cognition.

---

[18] Dans cette argumentation, nous supposons qu'un *système isolé* correspond à un système fermé.
[19] Cette approche est détaillée au chapitre 2.

Nous postulons que ce système de gestion des connaissances doit être décrit en termes d'organisation et de structure des connaissances, autour d'un paradigme d'unité, d'identité, et d'autonomie, c'est-à-dire comme un *système autopoïétique*.

Autrement dit, nous proposons de décrire le système de gestion des connaissances, non pas par une relation connaissance-environnement, mais par une relation connaissance-autopoïèse, à travers l'unité, l'identité et l'autonomie du système de connaissances et du système opérationnel.

Une partie de cette proposition est validée du fait que le paradigme d'unité, d'identité, et d'autonomie se trouve déjà de forme implicite dans le modèle OIDC, comme le souligne Ermine « les flux de cognition, qui transitent du système de connaissance vers les autres sous-systèmes peuvent exister de manière diffuse, opérant pas osmose, interpénétrant le système de connaissances et les autres sous-systèmes, selon l'hypothèse que cette pénétration est "profitable" (contribue au projet) globalement à tout le processus du système » [Ermine, 96]. Puis, il a ajouté « les flux de cognition du système de connaissance vers le système d'information sont d'une nature particulière … puisqu'ils sont le fruit de l'activité qui consiste à identifier, caractériser et expliciter les connaissances, afin que le système d'information puisse les stocker et les mettre à disposition suivant sa mission première » [Ermine, 96].

Le système autopoïétique ne fait que ressortir davantage le caractère vivant et viable du système de gestion des connaissances, vis à vis de son organisation et de sa structure, mais non pas par rapport à la relation entre composants, mais par rapport à la relation entre processus de production des composants, selon l'approche de l'enaction de Maturana et Varela.

---

[20] Dans le modèle de la marguerite on préfère utiliser le terme "patrimoine de connaissances" au lieu de système des connaissances, car dans le modèle OIDC, le composant (C) du modèle est un système de connaissances. Nous avons bien à l'esprit que ces deux systèmes sont différentes, donc pour ne pas introduire un terme supplémentaire, nous utilisons de façon générique système de connaissance.

### 3.2. Evolution de la connaissance dans le modèle autopoïétique de la gestion des connaissances

Un système fermé correspond à une transformation qui n'a pas entrées, ni sorties. Mais, qu'est-ce que cela peut signifier ? En effet, si dans la transformation du système il n'y a pas d'interactions (de relations) entre le système et son environnement, alors on parle de système fermé (ni entrées, ni sorties), dans le cas contraire on parle de système ouvert.

D'autre part, un système peut avoir une connotation scientifique (un processus de causalité circulaire sur la base d'une dualité cause/effet), une connotation sociale et politique (l'union fait la force), ou autre.

En conséquence, un système est un outil composé d'une dynamique circulaire pour : développer la compréhension de phénomènes, produire de la connaissance, et produire de l'apprentissage. Dans ce contexte, le système ne peut pas être détaché de la dualité observateur/observé, c'est-à-dire, de la conception (conscient ou inconscient) de la *cognition*[21] chez l'observateur[22]. Autrement dit, l'observateur doit choisir un champ. Soit, pour lui « notre monde et nos actions sont inséparables », ce qui correspond à l'approche cognitiviste ou connexionniste[23] de la connaissance, ou bien « un monde émergent de signification » ce qui correspond à la conception d'un système autopoïétique selon l'approche de l'enaction de Maturana et Varela[24].

Dans cette section nous allons étudier l'approche de l'enaction de Maturana et Varela, d'une part, par la description de la représentation et la création de sens dans un système clos, et d'autre part, par l'apprentissage et l'évolution du système clos. Ceci est fait à partir d'une analogie superficielle avec le système nerveux de l'être humain ou l'animal, selon le modèle autopoïétique. Cette description nous permet de comprendre, d'une part, le rôle de la clôture opérationnelle dans la création de sens, et d'autre part, le rôle de la détermination structurelle dans l'apprentissage et l'évolution du système clos. Ceci nous ferons avec les outils du modèle autopoïétique (relations

---

[21] Les trois approches scientifiques de la cognition ont été abordées dans le chapitre 2.
[22] Il nous semble que dans cette dynamique, un *système de connaissances* amplifie davantage la création, on peut donc parler de la connaissance de la connaissance (on monte d'une échelle). Néanmoins, cet effet amplificateur n'existe pas lorsque nous parlons de *système de bases de données* ou *système d'information*, car la problématique est la gestion de données et non pas de gérer des connaissances. D'ailleurs, le terme *système de bases de connaissances* est un héritage du terme *bases de données*. Dans cette thèse un *système de bases de connaissances* de type *base de règles* ou *base de cas*, correspond plutôt au *système opérationnel* et non pas au *système de connaissances* du modèle proposé. D'autre part, il faut souligner aussi qu'à certains moments nous préférons utiliser ce terme sans "s", afin d'emphatiser davantage l'unité, l'identité et l'autonomie du système.
[23] On parle plutôt d'un réseau cognitiviste.
[24] A ce sujet Varela a fait un commentaire entre le paradigme feed-back de Wiener, le paradigme de l'ordinateur chez Neumann et le paradigme autopoïétique « chez Wiener et von Neumann, les conceptions sur la connaissance et le vivant véhiculent aussi une signification politique et sociale. C'est notre ferme conviction que l'orientation épistémologique et scientifique proposée ici pourrait être un élément positif pour lutter contre les diverses formes de dogmatisme qui enserrent partout notre monde et qui peuvent nous mener à la destruction mutuelle » [Varela, 89].

---

entre processus de production des composants) qui a été présenté dans le chapitre 2 afin de l'aligner avec les autres modèles (relations entre composants).

### 3.2.1. Autopoïèse et connaissance

Dans le champ managérial, Aquiles Limone et Luis Bastias, dans un article récent (dont nous avons déjà fait référence dans l'introduction du chapitre 1), intitulé : *Autopoiésis y Conocimento en la Organización. Fundamento Conceptual para una Auténtica Gestión del Conocimiento*, et que nous traduisons comme « Autopoïèse et connaissance dans l'organisation. Fondement conceptuel pour une authentique gestion des connaissances » [Limone et Bastias, 02b], ont situé, à notre avis, pour la première fois, la justification de la relation *autopoïèse* et *connaissance*. Nous pensons que cette avancée intellectuelle, est utile pour justifier davantage le modèle proposé.

Comme nous l'avons vu dans le chapitre 2, le système autopoïétique permet (1) de décrire l'unité dans un domaine de détermination interne, c'est-à-dire, à travers une description organisationnelle de l'unité, à partir de trois processus génériques (de production de ces mêmes composants génériques) : constitutifs, de spécification, et d'ordre ; et simultanément et nécessairement, (2) de décrire le fonctionnement de l'unité dans un domaine de composition, c'est-à-dire, une description structurelle de l'unité à travers l'espace matériel des propriétés des composants. Ce fonctionnement opère sous trois contraintes génériques, à savoir : clôture opérationnelle, couplage structurel, et détermination structurelle.

Dans le système autopoïétique, la distinction entre *système ouvert* et *système fermé* se fait en fonction de l'organisation de l'unité, le maintien de l'identité, et la contrainte de l'autonomie, et non pas en fonction de la relation avec l'environnement, ce qui donne naissance au *système clos* et à la *clôture opérationnelle* qui correspond au fonctionnement du *système clos*. Dans ce contexte, on parle de clôture et non pas de fermeture du système.

### 3.2.2. Création de sens dans un système clos : clôture opérationnelle

Un système clos a-t-il la capacité de créer de sens ? Un système clos est une explication symbolique de l'unité (organisation) et du fonctionnement de l'unité, afin de maintenir l'identité (structure) du système clos sous la contrainte d'autonomie entre l'unité et l'identité. L'autonomie traduit alors la capacité du système clos pour gérer la dualité organisation/structure.

Pour expliquer un système clos, on fait une analogie avec un système ouvert, qui est décrit comme : entrées/transformation/sorties. Autrement dit le résultat de la transformation se situe à l'extérieur des frontières du système lui-même.

En revanche, les entrées d'un système clos sont des perturbations et non pas des inputs, tandis que les sorties sont le résultat d'une opération de transformation. Ce résultat se situe à l'intérieur de la frontière du système clos, c'est-à-dire que les sorties d'un système clos ne font pas partie de l'environnement.

Il y a deux cas particuliers :

- Si le résultat de l'opération de transformation est l'opération de transformation alors on parle de *système autopoïétique*[25]. En fait, il s'agit d'1/3 de *système autopoïétique*, car cette condition est seulement relative à la clôture opérationnelle. L'autre 2/3 est le couplage structurel et la détermination structurelle.

- Si le résultat de l'opération de transformation est autre chose que l'opération elle-même, mais qu'il reste à l'intérieur de la frontière du système clos, alors on parle de *système allopoïétique*[26].

Un exemple de système clos est le système autonome. D'où, la relation autonomie et connaissance.

Pour Maturana et Varela « l'idée d'autonomie fait référence à un système à forte détermination interne, ou auto-affirmation … l'identité du système s'affirme dans et par son fonctionnement » [Varela, 89]. Et donc, nous pouvons dire qu'un système autonome possède une unité (le fait d'être distinguable de son environnement et donc des autres unités) qui lui permet d'être distinguable comme un système autonome, et une structure qui lui permet de fonctionner pour maintenir son identité et son autonomie. Pour gérer l'autonomie « la structure d'un système autonome spécifie un domaine d'interaction possible avec son environnement. Les seules interactions permises sont celles qui sont spécifiées par les composants du système et qui sont compatibles avec le maintien de sa clôture. Ce domaine d'interactions, nous l'appelons *domaine cognitif* »[27].

---

[25] A la différence du terme *allopoïétique* le terme *autopoïétique* veut dire auto-création (qui se crée elle-même) ou auto-produit (qui se produit lui-même) [Limone, 77], [Varela, 89].
[26] Le terme *allopoïétique* veut dire autre-création (allo = autre, et poiein = création). [Limone, 77], [Varela, 89].
[27] Ceci justifie en peu ce que nous avons dit au chapitre 2, relatif à l'auto-organisation de l'unité (l'organisation), d'auto-maintient de l'identité (la structure), et l'auto-gestion de l'autonomie (la dynamique) d'un système autopoïétique.

Ce qui pose la question de savoir : qu'est-ce qu'un domaine cognitif pour un système de connaissance ?

Nous explorons maintenant cette question avec les arguments chez Maturana et Varela, pour la description du système nerveux. Dans chaque paragraphe nous faisons une analogie avec un système de connaissance.

Pour Maturana et Varela, « la caractérisation classique d'un système repose sur l'existence d'un flux : une entrée, une transformation, une sortie » [Varela, 89].

Ce qui est le cas pour le système de connaissances du modèle OIDC. Ce système est décrit à travers des flux cognitifs et des flux de compétences.

Selon Maturana et Varela, le système nerveux ne peut pas être décrit par un modèle système. En effet, ils ont dit par rapport à ce modèle « cette caractérisation est pertinente dans de nombreux cas, comme les ordinateurs et les organes de commande ; mais elle n'est pas appropriée pour d'autres comme le système nerveux … l'idée que nous emmagasinons des représentations de l'environnement ou que nous emmagasinons des informations au sujet de l'environnement ne correspond en rien au fonctionnement du système nerveux … de notions comme celles de mémoire ou de souvenir, ce sont des descriptions qui relèvent du domaine de l'observateur et non du domaine d'opération du système nerveux » [Varela, 89]. Puis, ils ont ajouté « ces systèmes peuvent être utilement décrits, non en termes d'input/output et de transformations, mais ayant une forme spécifique de clôture opérationnelle : les résultats des opérations du système sont les opérations du système, dans une situation autoréférentielle » [Varela, 89].

Dans ce contexte, l'hypothèse de la clôture opérationnelle, est formulée de la façon suivante : *tout système autonome est opérationnellement clos*. Autrement dit, un *système clos* est un système autonome qui fonctionne par clôture opérationnelle. Maturana et Varela insistent sur le fait que la clôture opérationnelle n'a rien à voir avec la fermeture ou l'isolement du système nerveux. A ce propos, ils disent « il faut noter que la clôture n'est pas une fermeture : le terme de clôture se réfère au fait que le résultat d'une opération se situe à l'intérieur des frontières du système lui-même ; il ne présuppose pas que le système n'a pas d'interaction avec l'extérieur, ce qui serait la fermeture. Notre étude n'a pas pour objet les systèmes isolés ».

Nous constatons donc, qu'un système clos fonctionne sans inputs ni outputs, et d'autre part, que le résultat de la transformation est la transformation elle-même. En effet, ils ont dit « dans le cas du système nerveux, le résultat de l'activité neuronale est l'activité neuronale … ainsi les activités des neurones se définissent mutuellement, et leur activité coopérative produit un état global autodéterminé ou comportement propre »[28].

Dans ce cas, si le système de connaissance est un système clos, alors les entrées ne sont pas d'inputs mais des perturbations. Mais qu'est-ce qu'une perturbation ?

D'autre part, un système clos, ne peut pas être décrit comme un système fermé (ou isolé, ni entrées, ni sorties), on ne peut pas dire non plus que la distinction entre un système clos et un système ouvert se fait par rapport à l'environnement (comme c'est le cas pour un système fermé), ou par rapport à la connaissance imparfaite de l'un par rapport à l'autre, etc., mais par rapport à son autonomie.

Cela signifie que c'est dans l'autonomie du système clos que réside la connaissance. A ce propos Maturana et Varela ont dit « la clôture opérationnelle du système nerveux … les conclusions auxquelles nous arrivons à ce sujet auront un impact plus grand sur notre compréhension de ce qu'est la connaissance … il nous faut concevoir le système nerveux comme un système autonome ; et il faut accepter les conséquences qui en découlent au niveau de notre conception de la communication et de la représentation … afin que vous ne pensiez pas qu'il est complètement fou de croire que le système nerveux fonctionne sans inputs ni outputs » [Varela, 89].

Pour eux, le système nerveux est « un réseau d'interactions, déterminant récursivement un domaine cognitif macromoléculaire de l'organisme » [Varela, 89].

Dans ce contexte, la création de sens dans le système nerveux est un phénomène propre de la clôture opérationnelle car il n'y a pas ni inputs ni outputs.

Or, cette capacité de clôture du système est la justification théorique de l'approche de l'enaction de Maturana et Varela, à ce propos ils ont dit « en postulant que la clôture d'un système est l'aspect majeur à considérer, nous avons abandonné les notions d'entrée et de sortie ; et la direction du flux d'information a perdu toute signification … par exemple, notre monde visuel est plein d'objets : chaises, voitures, personnes, etc. L'attitude représentationnelle suppose que nous parvenons à les connaître en construisant une image interne de ce qu'ils sont. Mais, si nous renonçons à regarder le cerveau comme ayant une entrée bien définie d'informations, nous ne pouvons plus postuler une représentation … comment un système existe dans un monde (ou dans son monde), s'il n'en a pas construit une représentation interne. La réponse consiste à s'apercevoir que la clôture d'un système peut *faire émerger un monde*, et créer son propre monde de signification » [Varela, 89].

---

[28] Par rapport, à d'autres exemples de clôture opérationnelle (systèmes qui ont une activité coopérative et un comportement propre), ils ont dit « le système nerveux n'est qu'un exemple d'une telle classe de systèmes. Le système immunitaire et les animaux multicellulaires appartiennent aussi à cette classe » [Varela, 89].

Cela signifie que la détermination interne de l'unité d'un système clos est un domaine cognitif[29]. Et donc, pour un système de connaissance ce doit être vrai aussi.

Dans cette argumentation, notre analogie entre système clos et système de connaissance d'une entreprise, peut être formulée de la façon suivante : Le système de connaissance est-il un lieu de rencontre ? C'est-à-dire, le système de connaissance est-il un domaine d'interactions cognitives ? Ce lieu cognitif, crée-t-il en sens entre ses interactions ? Si la réponse est oui. Reste à savoir d'où viennent-elles ces interactions, qui les génèrent,…

Nous pensons que le problème de la validité de cette analogie se pose au niveau des sources de ces interactions porteuses de sens, mais elle se pose au niveau des sources de ces interactions, car ces interactions sont à la base de création de connaissances nouvelles.

Au niveau de la description d'un système de connaissance par rapport à un système clos, l'imagination peut donner un point de réflexion et d'explication. Par exemple, pour le système nerveux, Maturana et Varela ont dit « le système nerveux doit être conçu comme une unité autonome, c'est-à-dire comme un réseau d'interactions cellulaires qui à chaque instant, détermine sa propre identité. Ce qu'il importe de comprendre, c'est comment cette identité et maintenue et réalisée » [Varela, 89].

Mais, au niveau de la source de ces interactions du système clos, ils ont dit « la clôture opérationnelle engendre une unité, qui à son tour spécifie un domaine phénoménal ». Ainsi, « la vie a émergé à travers la constitution d'unités autopoïétiques, capables d'engendrer leurs propres frontières » [Varela, 89].

Alors, on peut dire que la création de sens dans un système de connaissance se fait à travers la constitution d'unités autopoïétiques dans le système de connaissance. Mais, qu'est-ce que cela peut signifier ?

Néanmoins, le modèle autopoïétique offre un regard nouveau, car avant l'apparition de ce modèle la description du système nerveux se faisait par la modélisation analytique, système ou cybernétique. A ce propos Maturana et Varela ont dit « le système nerveux était expliqué par la distinction entre le soi et le non-soi. A l'inverse, nous partons de l'autonomie du système nerveux, de son identité qui n'est qu'un processus d'interactions coopératives. Cette idée nous conduit tout naturellement à la notion d'un réseau, puis à sa spécification génétique, à son domaine cognitif, à son histoire récursive » [Varela, 89].

---

[29] Dans le chapitre 2 on a dit que la description de l'unité a lieu, simultanément et nécessairement, dans deux domaines. L'un est la *détermination interne* (l'organisation), l'autre est *composition* (la structure).

D'où l'idée de formuler une explication du modèle OIDC, non pas à partir d'un modèle système (relations entre composants) comme c'est souvent le cas, mais plutôt à partir du modèle autopoïétique (relations entre processus de production des composants). Cela implique de faire la description d'un système de connaissance non pas à travers des entrées et des sorties, mais bien à travers l'autonomie et l'identité du système, puis au niveau de la clôture opérationnelle.

Dans une telle situation, l'organisation de l'unité de ces quatre composants (O,I,D,C), doit être pensée au niveau de l'organisation de l'unité et du fonctionnement de la structure de l'unité, ce qui pose une description à travers l'autonomie et l'identité de ces composants, et non pas à travers les entrées et sorties de chaque composant.

### 3.2.3. Apprentissage et évolution du système clos : couplage structurel

Un système clos a-t-il la capacité d'apprendre et d'évoluer ? Comme dans le paragraphe antérieur, nous explorons cette question avec les arguments chez Maturana et Varela, relatifs à la description du système nerveux. Dans chaque paragraphe nous faisons une analogie avec un système de connaissance.

A propos du système nerveux Maturana et Varela ont dit « on n'insistera jamais assez sur cette dépendance du système immunologique à l'égard de son histoire récursive. Ne pas s'y intéresser équivaudrait à étudier le système nerveux sans tenir compte de sa capacité à apprendre, ou à étudier la différenciation d'une cellule sans tenir compte de son milieu environnant … l'idée que nous proposons (l'autopoïèse) ici, vient de certaines considérations biologiques et cybernétiques. Parler d'information génétique ou moléculaire n'a de sens que dans le contexte d'un système d'interactions coopératives … et il semble justifié de supposer que les organismes héritent de leurs ancêtres le savoir accumulé par l'évolution phylogénétique » [Varela, 89]. Puis, ils ont ajouté par rapport au système immunitaire « dans cette description du système immunitaire, j'ai insisté sur l'aspect coopératif de phénomènes typiques de cellules lymphoïdes ».

Pour eux, « l'apprentissage est un phénomène de transformation du système nerveux lié à des transformations du comportement … il est le résultat du couplage structurel continu de la

---

phénoménologie du système nerveux et de la phénoménologie de l'environnement » [Varela, 89]. Mais, c'est aussi un phénomène coopératif.

Nous constatons donc, que l'apprentissage et l'évolution du système clos sont associés au couplage structurel du système clos.

Pour expliquer le couplage structurel du système clos, comme dans le cas antérieur, on fait une analogie avec un système ouvert, décrit comme : entrées/transformation/sorties.

Dans un système ouvert le processus d'interaction entre le système et l'environnement correspond bien à la dualité système/environnement, et donc il y a un processus de causalité circulaire (voir chapitre 2, section 2.1.5) qui créée une différence entre le système et l'environnement, ou entre l'unité et le milieu, comme Varela a dit « il faut déjà supposer une certaine différence entre une unité (ou un système) et son milieu (ou son environnement, si vous préférez), en peu comme dans la relation figure/fond … l'unité et le milieu sont couplés en certains points. Il existe une surface de couplage, où se croisent les influences mutuelles, mais cette surface de couplage n'est pas toute l'unité, elle ne constitue qu'une ou que quelques-unes de ces dimensions. C'est ce que je nommerai dorénavant le *couplage ponctuel* » [Varela, 89]. Cette différence est possible grâce à un *processus d'interaction* entre les deux composants, qu'on appelle : *couplage ponctuel*.

Pour Maturana et Varela, il y a deux sortes de couplage ponctuel entre le système et l'environnement : le *couplage par inputs* et le *couplage par clôture*[30].

➤ **Couplage par input**

Ce type de couplage entre le système et l'environnement a lieu seulement si le résultat de la transformation se situe à l'extérieur des frontières du système lui-même. Il s'agit donc du couplage des systèmes ouverts (entrées/transformation/sorties). La transformation est définie par les entrées du système. Autrement dit, c'est le couplage de l'approche système et cybernétique. A ce propos, Maturana et Varela ont dit « la théorie des systèmes nous fournit un paradigme du couplage ponctuel : un *input* transforme la dynamique des états d'un système … un input fait partie intégrante de la définition d'une unité … un input spécifie la seule façon dont une transformation d'état donnée peut avoir lieu … le couplage par input consiste á considérer que le système est essentiellement défini par des inputs »[31]. Ainsi, les modèles systèmes (OID, OIDC) sont un exemple de couplage par input.

---

[30] Pour Varela « longtemps, on a négligé le couplage par clôture, en faveur du couplage par input, parce que ce dernier prédominait en physique et dans les sciences de l'ingénieur » [Varela, 89]. Mais ce n'est pas le cas pour le génie industriel, où l'homme et la machine sont en dialogue permanent.

Autrement dit, ce couplage est caractérisé par le fait que chaque composant se construit lui-même par ce couplage. Le couplage par input est donc une sorte de *processus d'interaction par parallélisme* (voir section 3.1.1). Autrement dit, le couplage par input correspond à l'approche de l'enaction de Karl Weick. Dans le sens que l'unité et le milieu progressent parallèlement, l'un en face de l'autre, et à un moment donné s'établit un "dialogue" entre les deux, et donc il y a ce phénomène d'interaction (parallèle), source d'apprentissage et de connaissance pour l'un et l'autre, s'il existe bien entendu la capacité de conversions de problèmes et de solutions. Le couplage entre le système de connaissances du modèle OIDC et l'environnement est un couplage par input, et donc l'apprentissage se trouve dans un flux de compétences.

➢ **Couplage par clôture**

Ce type de couplage entre le système et l'environnement a lieu seulement si le résultat de la transformation se situe à l'intérieur des frontières du système lui-même. Il s'agit donc du couplage des systèmes clos. Le couplage par clôture introduit le concept de *perturbation*, pour définir l'entrée du système clos. D'après Maturana et Varela « une perturbation ne spécifie pas l'agent, elle ne prend en compte que son effet sur la structure de l'unité » [Varela, 89]. Autrement dit, la perturbation n'introduit pas une donnée nécessaire à la transformation du système clos, sinon qu'il s'agit d'un événement qui peut modifier la structure, ce qui correspond à une perte de l'identité du système clos. Puis, ils ont ajouté « le couplage par clôture consiste à penser que le système est défini essentiellement par ses divers modes de cohérence interne, lesquels découlent de son interconnectivité » [Varela, 89].

L'apprentissage se trouve dans la manipulation de la perturbation pour maintenir la structure. Mais, qu'est-ce que cela peut signifier pour un système de connaissance ?

Bref, le couplage par inputs est relatif à la transformation : entrées/transformation/sorties de l'approche système, afin de représenter un système ouvert, tandis que le couplage structurel est relatif aux entrées du système clos : les perturbations. Ce qui entraîne un effet sur la structure du système clos.

---

[31] En plus, la cybernétique de premier ordre, introduit le concept de feed-back (par rapport à l'output) afin de réguler l'état du système et la dynamique interne, ce que revient en définitif à la même situation, qu'on peut l'appeler *couplage par outputs*.

A partir de ces deux aspects : (1) la création de sens dans un système clos (clôture opérationnelle) ; et (2) l'apprentissage et l'évolution du système clos (couplage structurel), le modèle proposé peut être expliqué en termes d'organisation et de structure des connaissances à travers un système de gestion des connaissances, composé de deux sous-systèmes, à savoir :

- le système de connaissance. Ce système est conçu d'une part, dans le domaine de l'organisation, par la description de l'unité à travers le modèle autopoïétique de la marguerite, et d'autre part, dans le domaine de la structure, par les composants de l'unité conçus par l'approche de l'enaction de Maturna et Varela ;

- le système opérationnel. Ce système est conçu, d'une part, dans le domaine de l'organisation, par la description de l'unité à travers le modèle autopoïétique OIDC, et d'autre part, dans le domaine de la structure, par les composants de l'unité conçus par l'approche cognitiviste selon une représentation imparfaite de la connaissance.

Le fonctionnement du modèle proposé au niveau du système de connaissance est expliqué à partir le modèle autopoïétique de la marguerite :

<u>Modèle autopoïétique de la marguerite</u>

Dans le modèle autopoïétique de la marguerite, la marguerite est formée d'un cœur et autour de celui-ci se trouvent quatre pétales. Le cœur de la marguerite est appelé *patrimoine des connaissances* (ou *systèmes des connaissances*). Les quatre pétales ont pour noms : *repérer*, *préserver*, *valoriser* et *évoluer*. Le processus de projection permet la production de relations de type *repérer*. Une opération de distinction de type projection (sélection) est créée pour faire les requêtes vers l'environnement externe. Le processus de renseignement permet la production de relations de type *préserver*. Une opération de distinction de type capitalisation et partage doit être créée pour faire l'élaboration de corpus d'information. Le processus d'apprentissage et de création de connaissances permet la production de relations de type *valoriser* et *évoluer*. Une opération de distinction de type retour d'expérience et de créativité doit être créée, d'une part, pour faire la valorisation du corpus d'information avec l'environnement interne (interaction), et d'autre part, pour faire l'évolution du patrimoine de connaissances.

Le fonctionnement du modèle proposé au niveau du système opérationnel est expliqué à partir le modèle autopoïétique OIDC :

Modèle autopoïétique OIDC

Le modèle autopoïétique OIDC est formé par un système clos au sein d'un système ouvert. Le système d'information (I), le système de décision (D), et le système de connaissance (C) opèrent sous clôture opérationnelle, ce sont des systèmes clos au sein d'un système ouvert qui est le système opérant (O). Les relations constitutives (ou primaires), de spécification (ou structurelles), et d'ordre (ou décisionnelles) sont nécessaires pour avoir un couplage structurel, et une détermination structurelle entre les composants pour le maintien de l'identité dans et par l'unité.

Nous pensons que cette interprétation de l'organisation du modèle OIDC est beaucoup plus riche que si elle était constituée uniquement de flux de données (entrées et sorties du système opérant), de flux d'informations (entrées et sorties du système d'information), de flux de décisions (entrées et sorties du système de décision), de flux de compétences (entrées du système de connaissance), et de flux cognitifs (sorties du système de connaissance).

Ainsi, un modèle OIDC en termes d'un modèle autopoïétique, est perçu comme un système clos au sein d'un système ouvert qui opère sous clôture opérationnelle, couplage structurel, et détermination structurelle. Nous voulons dire par-là, que le modèle autopoïétique de la gestion des connaissances doit décrire l'unité dans une totalité formée par un système clos à l'intérieur d'un système ouvert, à l'aide des relations constitutives (ou primaires), de spécification (ou structurelles), et d'ordre (ou décisionnelles). Ainsi, le modèle prend en charge l'unité, c'est-à-dire la totalité dans le temps, qui comme Maturana et Varela l'ont si bien formulé « l'organisation … qui est à chaque instant l'unité dans sa totalité » au quotidien, mais aussi à court et long terme, qui existe simultanément et nécessairement par une opération de distinction sur deux domaines. L'un est le domaine conceptuel ou social (l'organisation), l'autre est le domaine physique (la structure). La dualité organisation/structure est décrite à travers des relations autopoïétiques (constitutives ou primaires, de spécification ou structurelles, et d'ordre ou décisionnelles), que nous observons d'un point de vue autopoïétique, Pour nous cette vision des choses, est le seul moyen de garantir à tout moment l'entreprise (le business, l'activité) comme unité dans le temps, l'unité dans et par la totalité.

### 3.3. Cadre de référence du modèle proposé par rapport à l'aspect social et technique de la gestion des connaissances

Comme nous avons vu dans le chapitre 2 (section 2.2.3) le modèle autopoïétique a débuté en 1969 à l'Université du Chili, pour une application dans le domaine biologique ce qui a donné naissance à l'*autopoïèse de Santiago* de Maturana et Varela. Ensuite, ce modèle a été appliqué en 1977 au domaine de la gestion, ce qui a donné naissance à l'*autopoïèse de Valparaiso* ou théorie *de l'autopoïèse de l'entreprise* de Limone. Comme résultat de cette expérience le modèle CIBORGA[32] est apparu, puis enrichi et popularisé depuis 1998 avec la collaboration de Bastias, Cardemártori et autres collègues de l'École de commerce de l'Université Catholique de Valparaiso.

Nous sommes bien d'accord que l'autopoïèse de Santiago a été conçue, dans un champ de phénoménologie biologique. Ce modèle permet d'observer un système vivant comme un système autopoïétique, c'est-à-dire une organisation de relations entre processus de production de composants qui s'auto-créent (ou s'auto-produisent) eux-mêmes de façon permanente dans le temps dans et par une unité, une identité, et une autonomie. Ainsi :

- L'organisation autopoïétique définit son organisation dans le temps dans et par une unité qui s'auto-organise elle-même afin de ne pas perdre son identité (d'où le phénomène d'auto-organisation du système autopoïétique) ;

- L'organisation autopoïétique maintient sa strucure par la fixation des propriétés de ses composantes et leurs relations, afin de ne pas perdre son identité (d'où le phénomène d'auto-maintient du système autopoïétique) ;

- L'organisation autopoïétique gère sa dynamique propre d'ordre pour le respect de la contrainte, afin de ne pas perdre son autonomie (d'où le phénomène d'auto-contrainte du système autopoïétique).

Nous sommes bien conscients que l'utilisation de l'autopoïèse de Santiago, en dehors du domaine des phénomènes biologiques a été critiquée par ses auteurs, en particulier par Varela qui a dit « les relations qui caractérisent l'autopoïèse sont des relations de production des composants, et que l'idée de production des composants se réfère fondamentalement à la production chimique. Étant donné cette notion de production des composants, les cas d'autopoïèse tel que les systèmes vivants ont comme critère de distinction une frontière topologique ; et les processus qui le définissent ont lieu dans un espace semblable à l'espace matériel, espace réel ou simulé par l'ordinateur. Ainsi, l'idée d'autopoïèse se trouve par définition restreinte à des relations de production d'un genre ou d'un autre et renvoie à des frontières topologiques » [Varela, 89]. Puis, il ajoute et c'est ici que surgit sa critique « les autres systèmes autonomes ne satisfont pas manifestement à ces deux conditions. Prenons par exemple une société animale, les frontières de l'unité ne sont pas topologiques et il semble saugrenu de décrire les interactions sociales en termes de "production" de composants. Ce n'est sûrement pas ce genre de notion qu'utilisent les ontomologistes, par exemple, lorsqu'ils étudient les sociétés d'insectes. On a suggéré que certains systèmes, comme les institutions, pourraient être compris comme des systèmes autopoïétiques (Beer 1975 ; Zeleny et Pierre 1976 ; Zeleny 1977). Je pense que ces caractérisations reposent sur des erreurs de catégories. Elles confondent l'autopoïèse et l'autonomie » [Varela, 89].

---

[32] Modèle Cybernétique de l'Organisation et de l'Apprentissage.

A notre avis, le doute qui s'installe chez Varela (mais également chez Maturana), vient du fait que dans les systèmes sociaux, par exemple l'entreprise, nous aurons des relations entre composants (nous le constatons tous les jours), mais pas des relations de production des composants. En plus, selon Maturana et Varela les relations entre processus de production de composants se produisent de façon "spontanée"[33]. Par conséquent, il n'y a pas la preuve scientifique qu'un tel phénomène se produise réellement dans un labo de recherche, bien que Maturana et Varela argumentent le contraire[34].

Nous pensons finalement que l'utilisation de l'approche autopoïétique de Santiago, en dehors du domaine des phénomènes biologiques, peut être justifiée comme une sorte d'analogie superficielle et non pas comme une analogie profonde, c'est-à-dire dans un monde plutôt réfléchi que réel. Et donc, pour nous et bien d'autres, l'autopoïèse est une ontologie, un chemin explicatif, tout simplement, que nous utilisons dans un domaine industriel pour expliquer la gestion des connaissances imparfaites.

Mais, si d'autres hommes de sciences, ont rêvé avec l'application de l'*autopoïèse de Santiago* (dans d'autres domaines que celui de la biologie), pourquoi pas nous. En toute modestie nous développons dans ce chapitre-ci un chemin pour intégrer l'*autopoïèse* dans une problématique de gestion des connaissances imparfaites.

En effet, pour étudier des systèmes complexes, c'est-à-dire les systèmes où la contrainte (l'ordre) fait apparaître une causalité circulaire (une chaîne logique de cause à effet) propre pour maintenir l'unité (l'organisation), l'identité (la structure) et l'autonomie (la dynamique) du système[35]. Un certain nombre d'auteurs ont utilisé l'autopoïèse de Santiago comme le fondement théorique de leurs modèles. Par exemple, le modèle autopoïétique de communications de Niklas Luhmann (dans le domaine social [Luhmann, 91]), le modèle autopoïétique des processus décisionnels de Darío Rodríguez[36] (dans le domaine organisationnel [Rodríguez, 91]), le modèle autopoïétique des actes conversationnels dans l'entreprise de Fernando Flores (dans le domaine du management [Flores, 96a], [Flores, 96b]), le modèle CIBORGA de Aquiles Limone et Luis Bastias (dans le domaine de la gestion [Limone et Cademártori, 98]), et le modèle autopoïétique des systèmes d'information de El-Sayed Abou-Zeid (dans le domaine informatique).

---

[33] Ce terme est né d'une analogie avec le système nerveux. En effet, comme le souligne Varela « ici, chaque constituant fonctionne seulement dans son environnement *local* de sorte que le système ne peut être actionné par un agent extérieur qui en tournerait en quelque sort la manivelle. Mais grâce à la nature configurationnelle du système, une coopération *globale en émerge* spontanément lorsque les états de chaque neurone en cause atteignent un stade satisfaisant ». C'est ce que nous appelons l'*hypothèse de spontanéité des relations autopoïétiques*.

[34] C'est peut-être, pour cette raison, qu'il n'a pas reçu le prix Nobel de sciences, car il était nommé par l'Université Libre de Bruxelles, une raison de plus pour nous d'être fiers de lui et de cette théorie, qui a été au cour du temps enrichi par Francisco Varela, comme lui même a dit « *c'est d'Humberto Maturana que j'ai appris à concevoir le système nerveux comme un système opérationnellement clos. Il proposa explicitement cette idée dès 1969. Je considère aujourd'hui que cette intuition de Maturana est fondamentale. Il a établi un rapport essentiel entre les processus matériels et systématiques qui ont lieu à l'intérieur du système nerveux et propose une conception profonde et riche de la connaissance chez l'homme* » [Varela, 89]

[35] La complexité n'a rien avoir avec la taille, le nombre de relations, ou les phénomènes compliqués du système.

### 3.3.1. Le modèle autopoïétique de la gestion des connaissances imparfaites

La phénoménologie qu'il nous intéresse d'observer à l'aide d'un modèle autopoïétique est la gestion des connaissances imparfaites dans une entreprise. Néanmoins, nous attirons l'attention sur le fait qu'il ne s'agit pas ici de proposer un modèle "généralisé" de gestion des connaissances imparfaites, mais plutôt toute une réflexion sur l'utilité et la faisabilité d'une telle démarche.

Notre canevas de pensée dans le modèle autopoïétique de la gestion des connaissances imparfaites sont guidés par :

- l'hypothèse de l'enaction ;
- l'hypothèse de spontanéité des relations ;
- l'hypothèse du noyau invariant ;
- l'hypothèse de la connaissance imparfaite.

➢ **L'hypothèse de l'enaction dans le modèle autopoïétique de la gestion des connaissances imparfaites**

Cette hypothèse prend ses racines dans l'hypothèse de la clôture opérationnelle de l'approche de l'enaction de Maturana et Varela.

---

[36] Nous soulignons au passage que Darío Rodríguez était le thésard de Luhmann en Allemagne, et donc trés influencé par lui dans son discours. Un autre exemple d'influence est le cas de Aquiles Limone qui était le thésard de Jacques Mélèse en France. Dans ce sens, Francisco Varela était l'ancien élève d'Humberto Maturana à l'Université du Chili, de même Luis Bastias était l'ancien élève de Aquiles Limone à l'Université Catholique de Valparaiso. Enfin, Fernando Flores a été initié à l'autopoïèse par Maturana et Varela, et lui à son tour a initié Stafford Beer (lorsque Beer était de passage au Chile pour le projet SYSCO). Voilà, un bel exemple du mécanisme "circulant" et "d'émergence de signification" de l'arbre de la gestion des connaissances, que nous résumons avec un texte de Varela, que nous empruntons aussi pour remercier ici mon ami et maître : Germain Lacoste « *Como ha ocurrido a menudo en la historia de la ciencia, la dinámica creativa entre Maturana y yo fue una resonancia en espiral ascendente, en la que participaba un interlocutor ya maduro que aportaba su bagaje de experiencia y pensamiento previo, y un joven científico que aportaba ideas y perspectivas frescas* ».

Notre hypothèse de base[37] est que l'approche de l'enaction (faire-émerger) de Maturana et Varela est le fondement théorique du *système de gestion des connaissances*, et non pas l'approche de l'enaction (selection) de Karl Weick.

Pour montrer que la connaissance dans le modèle autopoïétique de la gestion des connaissances imparfaites, est conçu par l'approche de l'enaction (faire-émerger), c'est-à-dire, par un mécanisme "circulant" et "d'émergence de signification", nous faisons trois sous-hypothèses, à savoir :

- la première sous-hypothèse : le système de gestion des connaissances imparfaites est un *système clos* ;
- la deuxième sous-hypothèse : le système de gestion des connaissances imparfaites est un *système vivant* selon l'approche autopoïétique de Santiago ;
- la troisième sous-hypothèse : le système de gestion des connaissances imparfaites est un *système viable* selon l'approche autopoïétique de Valparaiso.

<u>Première sous-hypothèse : le système de gestion des connaissances imparfaites est un système clos</u>

La première sous-hypothèse suppose que le système de gestion des connaissances imparfaites doit avoir une organisation et une structure du type système clos.

Autrement dit, le *système de connaissance* et le *système opérationnel* du modèle proposé sont des *systèmes clos*, par rapport à son organisation et structure, c'est-à-dire que l'organisation est close (le résultat de la transformation se situe à l'intérieur des frontières du système lui-même), et simultanément et nécessairement la structure est ouverte (le résultat de la transformation se situe à l'extérieur des frontières du système lui-même). En conséquence, la dualité organisation/structure se trouve dans la clôture organisationnelle du système et (simultanément et nécessairement) l'ouverture structurelle du système.

Dans le modèle proposé la structure est de type entrées/transformation/sorties, mais non pas l'organisation.

---

[37] Ceci est vrai pour le système de connaissance. Néanmoins, pour le système opérationnel du modèle proposé reste confiné dans l'approche cognitiviste de la cognition.

<u>Deuxième sous-hypothèse : le système de gestion des connaissances imparfaites est un système</u>
<u>vivant</u>

Nous utilisons la même analogie antérieure afin de supposer que le système de gestion des connaissances imparfaites est un système vivant. En effet, d'après l'autopoïèse de Valparaiso l'entreprise en plus d'être un système, est un système vivant.

Un système vivant dans la pensée autopoïétique de Santiago est un système défini comme une concaténation (de réseaux) des processus de production des composants qui constituent le système (en tant qu'unité dans l'espace matériel) et qui engendre en retour les processus qui les ont produit. Cela signifie d'une part que les systèmes vivants sont organisés par un réseau de relations de "production des composants" qui sont matérialisées dans un domaine physique (sa structure), et le résultat du fonctionnement du système est le système lui-même. La dynamique de fonctionnement est donc récursive avec clôture opérationnelle.

Nous pensons que dans cette argumentation, il y a trois types de description possibles pour l'observateur, à savoir : la modélisation analytique, la modélisation systémique et la modélisation autopoïétique, car la perception de la cognition chez Maturana et Varela est plus enactiviste (faire-émerger) et connexionniste (réseaux) que cognitiviste (instructions), ou enaction (sélection) chez Weick.

En effet, l'enaction (faire-émerger) et non pas l'enaction (sélection) est une approche de la connaissance qui nous interroge sur la construction du sens. Cette approche nous invite à travailler (ou construire un monde partage) sans avoir de représentations vrai de la réalité. Et donc, l'enaction remet en question le cognitiviste de la connaissance et non pas le connexionnisme de la cognition (tout au moins dans le sens du partage), car il faut toujours être deux pour construire.

Néanmoins, sous l'hypothèse autopoïétique, l'observateur n'a qu'un seul outil pour expliquer l'organisation. Cet outil est en termes de relations entre processus de production de composants, ce que nous appelons la modélisation autopoïétique.

Nous postulons que dans le modèle proposé, l'observateur a trois outils qu'il doit faire dialoguer pour expliquer l'organisation : (1) en termes de propriétés ou attributs des composants (modélisation analytique) ; (2) en termes de relations entre composants (modélisation systémique) ; (3) en termes de relations entre processus de production de composants (modélisation autopoïétique).

D'autre part, la dynamique du modèle proposé doit être récursive, avec clôture opérationnelle (le résultat de la transformation se situe nécessairement à l'intérieur du même système). Et donc, cela implique l'existence d'un ou plusieurs systèmes clos dans le modèle proposé.

<u>Troisième sous-hypothèse : le système de gestion des connaissances imparfaites est un système viable</u>

Nous utilisons la même analogie antérieure afin de supposer que la gestion des connaissances imparfaites est un système viable. En effet, d'après l'autopoïèse de Valparaiso l'entreprise en plus d'être un système vivant, est un système viable.

Un système viable dans la pensée autopoïétique de Valparaiso est un système dynamique, complexe et ouvert, manifestant toujours, d'une part, l'unité de l'organisation afin de maintenir l'identité et préserver, à chaque instant, l'autonomie du système, et, d'autre part, la phénoménologie (fonctionnement et comportement) de l'unité, existant simultanément et nécessairement sur le domaine social (l'organisation) défini à travers de relations humaines, et le domaine physique (la structure) défini à travers des individus, matière, énergie, et symboles. Ainsi, dans une entreprise nous avons des individus qui s'organisent pour faire quelque chose. Cette organisation seulement est possible par des relations humaines et les actes du langage (compromis, promesses, etc.). Le résultat de cette organisation est l'organisation elle-même, le domaine social est donc un système clos. Néanmoins, la structure de l'organisation est de type : entrées/transformation/sorties, sur la base de trois ressources : matière (matières premières, fournitures, etc.), énergie (électricité, eau, etc.), symboles (documents, systèmes d'information, bases de données, etc.).

Supposons que la matière, l'énergie, et les symboles appartiennent à la classe objets, et les individus appartiennent à la classe individus. Nous avons donc une relation entre ces classes. Pour simplifier, soit la relation individus-objets. De cette relation on peut concevoir trois systèmes complexes, c'est-à-dire un processus de causalité circulaire défini par la dualité causes/effets. Nous avons alors trois types de complexité, à savoir : (1) les relations entre individus ; (2) les relations entre objets ; et (3) les relations entre individus et objets (et vice versa). Ainsi, la complexité fait preuve des relations de causalité circulaire entre l'aspect social (les individus) et l'aspect technique (les objets), et donc, la complexité n'a rien à voir avec la taille du système, le nombre de relations, ou les phénomènes compliqués du système.

Nous constatons aussi que ces trois complexités coïncident avec le modèle de Nonaka et Takeuchi, ce que justifie davantage cette hypothèse. En effet, si l'on change le terme "objet" de l'approche autopoïétique de Valparaiso, par le terme "concept", nous avons crée, d'une part, une relation entre "connaissance explicite" et "concept", et d'autre part, entre "connaissance tacite" et "individu".

Enfin, le système de gestion des connaissances imparfaites existe simultanément et nécessairement dans le domaine social (relations humaines) et physique (individus, matière, énergie, et symboles), et il fonctionne avec une clôture opérationnelle.

Reste à savoir la signification de ces éléments dans la dualité organisation/structure du modèle proposé. Cette partie nous la développerons dans le chapitre 5.

> **L'hypothèse de spontanéité des relations du modèle autopoïétique de la gestion des connaissances imparfaites**

Cette hypothèse prend ses racines dans hypothèse de la clôture opérationnelle du modèle autopoïétique, que dit : *tout système autonome est opérationnellement clos*. Comme Maturana et Varela soulignent « la clôture opérationnelle engendre une unité, qui à son tour spécifie un domaine phénoménal » [Varela, 89]. Autrement dit, une fois défini l'organisation, il faut trouver la structure que maintient l'identité dans et par l'organisation de l'unité. Comme nous l'avons dit plus haut, ce processus de production des composants est fait de façon "spontanée".

Nous venons d'affirmer plus haut que sans connaissance collective il n'y a pas d'apprentissage organisationnel, et sans apprentissage collectif il n'y a pas de connaissances nouvelles.

Nous pensons que c'est dans la dualité création/apprentissage du système clos qu'il faut trouver une explication de cette relation "spontanée", car nous venons d'affirmer plus haut que sans connaissance collective il n'y a pas d'apprentissage organisationnel, et sans apprentissage collectif il n'y a pas de connaissances nouvelles. Voici l'idée dans une analogie superficielle :

Il faut toujours être deux pour construire

Imaginons l'affirmation :
« *Aujourd'hui j'ai appris quelque chose* »

Nous sommes d'accord que la « *chose* » est bien la connaissance, et que cette connaissance, créée est, premièrement le produit de l'expérience de l'autre, et deuxièmement découle de la volonté de l'autre de partager avec nous cette connaissance.

En définitive, nous pouvons dire que:

« *Le bien être de l'autre passe par le bien être de nous-mêmes, et vice versa* »

Bref, de cette relation « spontanée » est née la connaissance[38].

Selon l'approche de l'enaction, cette connaissance ne doit pas être nécessairement le reflet d'une réalité absolue (qui serait le cas de l'approche symbolique de la cognition), ni calculée (mathématiquement) de façon locale afin de faire émerger un comportement nouveau (ce qui serait le cas de l'approche connexionniste de la cognition), mais simplement enancte dans un monde qui pour nous a de la signification. C'est justement dans le retour de la signification avec l'autre, et l'enaction de cette signification dans son esprit[39], qu'apparaît d'après nous : la connaissance. Cette connaissance, peut très bien prendre la forme d'une analogie, d'un concept, par exemple dans un concept nouveau. C'est dans ce contexte-là, que nous disons que l'enaction est un mécanisme "circulant" et "d'émergence de signification".

Dans le chapitre 5 l'étude de cas permettra de vérifier, au moins de donner une explication empirique du phénomène de la "spontanéité" du processus de production des composants du modèle autopoïétique, que nous ferons par l'acte de langage ou flux conversationnel chez les individus.

> **L'hypothèse du noyau invariant du modèle autopoïétique de la gestion des connaissances imparfaites**

Cette hypothèse prend ses racines dans le modèle autopoïétique. L'élément fédérateur du modèle autopoïétique c'est le "patron d'organisation commun" de l'autopoïèse de Santiago chez Maturana et Varela, c'est le "noyau invariant" de l'autopoïèse de Valparaiso chez Limone et Bastias, c'est-à-dire, c'est ce "quelque chose" qui ne change pas, afin de maintenir un système viable en vie[40].

Nous postulons dans le modèle proposé, ce "patron d'organisation commun" ou "noyau invariant" soit le *kernel* du *système de connaissance* du modèle proposé. En effet, la connaissance est au centre de la vie, car s'il n'y a pas d'expérience accumulée il n'y aura pas d'évolution possible[41]. Un

---

[38] Nous partons du principe que rien ne nous oblige à rester et partager avec l'autre. En effet, dans un monde enacté, c'est-à-dire dans un monde plutôt réfléchi que réel, c'est le monde de l'autopoïèse (un système d'unité, d'identité et d'autonomie).

[39] Nous utilisons le terme *esprit*, du fait que le livre de Bateson, intitulé *Steps to an Ecology of Mind* (publié en 1973) a été traduit en français comme *Vers une écologie de l'esprit* (publié en 1977), puis l'esprit est quelque chose qui existe dans le plan abstrait.

[40] La justification de cette invariant se trouve dans l'un des mécanismes de création des connaissances nouvelles et d'apprentissage organisationnel, que nous avons identifié par le verbe "stabiliser". Voir chapitre 1, section 1.4 (les mécanismes de généralisation de la gestion des connaissances selon l'aspect social et l'aspect technique).

[41] Néanmoins, pour rester fidèle au discours autopoïétique de Maturana et Varela, l'avenir n'a pas d'existence dans un sens de représentation symbolique des objets.

---

exemple de "patron d'organisation commun" ou de « noyau invariant » est donné par le champ technologique[42], en effet il suffit de regarder les objets de notre maison ou les outils de production de notre travail pour constater qu'ils ne sont que des améliorations du passé[43].

Dans cet esprit, la séparation entre connaissance et action ne se produit jamais, car au cœur de l'entreprise se trouvent les relations entre individus. Sans ces relations il n'y a pas d'activité possible dans un domaine conversationnel. C'est justement cet espace de travail (linguistique) qui constitue, pour nous, le moyen privilégié de résolution de problèmes complexes, c'est-à-dire de problèmes qui ont besoin de l'interaction avec l'autre pour sa mise en contexte, son analyse et l'identification en définitif du problème à résoudre[44]. En conséquence, c'est de ce processus conversationnel que vont dériver les problèmes compliqués pour lesquels la machine peut jouer un rôle intéressant. Un exemple de cela est l'ordinateur : pour lui 2 plus 2 fait toujours 4, mais pour l'homme habitué à la pensée systémique, 5 est une valeur que l'on souhaiterait obtenir.

Dans le modèle proposé nous déplaçons la connaissance. En effet, le centre de la connaissance collective n'est plus l'individu (comme nous l'avons argumenté dans le chapitre 1), mais la connaissance collective se trouve dans la relation entre processus (individus) de production de composants (nous reviendrons sur ce point plus loin).

Dans la même perspective notre argumentation, en plus d'être justifiée dans le champ biologique par l'approche enactiviste (faire-émerger) de Maturana et Varela, se trouve justifiée davantage, dans le champ social, par Edgar Morin[45], lorsqu'il parle de « la connaissance de la connaissance », c'est d'ailleurs le titre de son livre *La méthode 3. La connaissance de la connaissance* [Morin, 86].

Dans le chapitre 5, l'étude de cas permettra de vérifier l'existence d'un "noyau invariant" ou "patron d'organisation commun" sur lequel doit graviter le système de connaissance du modèle proposé.

---

[42] D'après Gousty, la technologie, en citant à Nollet *et al*., est « un ensemble de méthodes, de procédures, d'équipements et même d'approches utilisés pour fournir un service ou produire un bien » [Gousty, 98].
[43] Nous avons justifié davantage, dans la section 1.5 du chapitre 1 (les origines de la connaissance industrielle), la production de connaissance dans l'industrie, d'abord comme un levier de productivité, puis comme un levier d'avantage concurrentiel ou compétitif, ensuite comme un levier d'avantage coopératif.
[44] C'est justement ce canevas de pensée (contexte, analyse et identification) que nous avons choisi comme démarche pour présenter notre expérience sur le terrain.
[45] Dans ce contexte, nous voudrions souligner aussi que la thèse de Limone se trouve référencée dans *La méthode 2* [Morin, 80].

> **L'hypothèse de la connaissance imparfaite dans le modèle autopoïétique de la gestion des connaissances**

Cette hypothèse prend ses racines dans l'enaction (faire-émerger). Comme nous l'avons dit plus haut, l'enaction est une approche de la connaissance qui nous interroge sur la construction du sens. Cette approche nous invite à travailler (ou construire un monde partage) sans avoir de représentation vraie de la réalité. Et donc, l'enaction remet en question le cognitiviste de la connaissance et non pas le connexionnisme de la cognition (tout au moins dans le sens du partage), car il faut toujours être deux pour construire.

Ainsi, l'approche enactiviste de la cognition implique que dans le processus de causalité circulaire entre les causes et les effets, la relation ne se construit pas nécessairement d'une représentation vraie ou logique de causes et d'effets, car le problème de l'enaction n'apparaît pas lié à la représentation symbolique d'une réalité, mais au maintien du système en vie et viable. Mais ceci implique la mise en place d'un support technique pour l'enaction. Ce support est la *connaissance imparfaite*. Voici l'idée dans une analogie superficielle :

La boite de confiture de cerises

Pour maintenir un système en vie, une boite de confiture de cerises ne doit pas forcément contenir des cerises, mais peut très bien contenir seulement de la "chimie" de cerises. Ici (l'approche de l'enaction) le problème du vrai ou du faux de la représentation symbolique (approche cognitiviste) ne se pose pas, car le vrai problème est la survie et la viabilité d'un système vivant et viable.

Cet exemple permet d'établir une relation entre *enaction* et *connaissance imparfaite*, comme moyen d'explication technique de l'enaction. En effet, si l'on revient à l'exemple de la boite de cerises, la connaissance parfaite serait le cas d'avoir des cerises ou de la "chimie" de cerises, l'un ou l'autre (vrai ou faux), mais non pas une partie de chacun à la fois avec un certain degré d'appartenance à l'un et l'autre (vrai et faux), qui serait le cas de la connaissance imparfaite[46].

Nous avons décidé donc de prendre la connaissance imparfaite comme un support technique de l'explication de l'approche de l'enaction de Maturana et Varela, ainsi que le modèle autopoïétique pour le fondement théorique du modèle proposé[47], appelé *modèle autopoïétique de la gestion des connaissances imparfaites*.

---

[46] Dans notre boite de cerises on peut lire « préparée avec 50g de fruits pour 100g », puis dans la même boite on peut lire « satisfait ou remboursé deux fois ». Nous pesons qu'il n'y pas un esprit corporatiste du business dans cet exemple (dans l'introduction générale nous avons cité Donald Trump pour introduire le concept de *business*, que mieux lui pour le faire). Pour nous, la preuve est que dans le sac en plastique on peut lire « petit casino, mon épicier est un type formidable ».

D'autre part nous avons vu que le modèle OID[48] de Jean-Louis Le Moigne a été élargi pour la prise en charge de la connaissance par le biais d'un système de connaissance, connecté par des flux cognitifs et flux de compétences au système opérant, au système d'information et au système de décision du modèle OID, ce qui a donné naissance au modèle Opération, Information, Décision, Connaissance (OIDC) de Jean-Louis Ermine, et donc, ce modèle pourrait être élargi davantage pour abriter la connaissance imparfaite. En fait, s'il y a de la place pour la connaissance parfaite pourquoi n'y aurait-il pas de la place pour la connaissance imparfaite. Reste à définir alors, les nouveaux composants du modèle et la mécanique d'intégration afin de représenter les connaissances d'une réalité (imparfaite), et la façon de les gérer.

Dans ce contexte, nous avons étendu l'aspect social et technique de la gestion des connaissances aux données imprécises et incertaines dans une dualité organisation/structure. Ainsi :

- l'imperfection de la connaissance dans l'organisation, se réfère à la pense et le raisonnement en termes flou chez l'individu ;

- l'imperfection de la connaissance dans la structure, se réfère à la prise en charge du flou sous l'angle des bases de données relationnelles floues, par le biais de la représentation des objets flous (traitement de données imprécises), et l'interrogation des événements flous (traitement de requêtes floues). La gestion des connaissances imparfaites signifie pour nous l'extraction des connaissances de données étendue aux données floues (objets imprécis et événements incertains) d'une base de données relationnelles floues.

En conséquence, la structure du système de connaissance existe, simultanément et nécessairement, dans un double milieu à savoir : *flou* et *non flou*. Le milieu non flou est l'espace plus structuraliste de l'information où le business est plongé, c'est le milieu naturel des modèles (OID, OIDC, AMS et MSV) où se sont justifiés les systèmes d'information et de connaissance traditionnels, mais contrairement, le milieu flou du business, doit tenir compte de la réalité imparfaite l'organisation du système de connaissance.

---

[47] Le modèle proposé se fonde aussi dans les mécanismes génériques de la gestion des connaissances (présentés dans la section 1.4 du chapitre 1).
[48] Opération, Information, Décision.

Les aspects théoriques de l'imprécis et de l'incertain l'imperfection de la connaissance (point de vue social et technique) seront étudiés dans le chapitre 4.

Dans le chapitre 5, l'étude de cas permettra de vérifier l'existence d'un contexte industriel dans lequel la phénoménologie peut être expliquée à partir de l'imprécis et de l'incertain.

### 3.3.2. Le modèle proposé selon un point de vue social

Dans cette section nous présentons quatre sous-aspects qui témoignent les enjeux de la complexité de la gestion des connaissances par rapport à son aspect social, et que le modèle proposé doit tenir compte, à savoir : l'aspect dynamique de la gestion des connaissances, l'aspect action de la gestion des connaissances, l'aspect culturel de la gestion des connaissances, et l'aspect chaos de la gestion des connaissances[49].

➤ **Aspect dynamique de la gestion des connaissances**

Comme nous l'avons dit au chapitre 1, la génération des mécanismes de création des connaissances nouvelles et d'apprentissage organisationnel passe par la capacité de l'entreprise et ses ressources humaines[50]. Il y a donc pour nous, dans la capacité de l'entreprise un premier aspect qui témoigne de la complexité de la gestion des connaissances, à savoir un aspect dynamique de la gestion des connaissances que l'on doit intégrer (combiner, coordonner) de façon "circulante" d'une part, dans la *gestion des connaissances* (capitaliser, partager, créer), et d'autre part, dans la *gestion de l'innovation* (créer, produire, offrir) afin de faire-émerger de la connaissance nouvelle et de nouveaux produits ou services pour l'entreprise. L'intégration "circulante" signifie pour nous que dans cette dynamique de gestion l'un doit alimenter l'autre et vice versa, en plus de la circularité interne de la *gestion des connaissances* et de la *gestion de l'innovation*[51].

---

[49] Le premier aspect prend se source dans notre recherche. Le deuxième aspect prend se source dans le modèle de connaissance et action de Edgar Morin que d'après lui l'action est la connaissance de la connaissance, mais également dans le modèle de connaissance et action de Maturana et Varela que d'après eux savoir est faire et faire est savoir. Le troisième aspect prend se source dans le modèle de gestion des connaissances de Prusak et Davenport sur la base d'un nouveau style de management et de comportement organisationnel de l'entreprise. Le quatrième aspect prend se source dans les problématiques autour de la gestion de l'information dans les entreprises des années 70. Jacques Mélèse et Jean-Louis Le Moigne ont bien argumenté les contraintes humaines face à l'inconnu : perte d'emploi, de pouvoir, etc., à cause de l'intégration de systèmes d'information dans l'entreprise.

[50] Dans ce même sens Emmanuel Caillaud parle plutôt de "réactivité" des systèmes industriels de production, lorsqu'il dit « le principal levier de réactivité … est constitué par les ressources humaines. En effet, la réactivité repose sur la capacité des acteurs à exploiter leurs connaissances pour répondre aux événements perturbateurs. Ceci implique généralement une coopération au sein d'une activité ou entre plusieurs activités et créée de nouvelles connaissances » [Caillaud, 00]. Et donc, pour lui, et pour nous aussi d'ailleurs « la réactivité dépend principalement des hommes de l'entreprise ». Plus loin, il ajoute « nous souhaitons mettre en valeur les conditions nécessaires à la réactivité (coopération, connaissances, compétences), aider cette réactivité et contribuer à son évaluation (performance, risques) » [Caillaud, 00].

[51] Nous justifierons plus loin que dans cette dynamique de gestion il est nécessaire d'ajouter la *gestion des compétences*.

En ce qui concerne ce premier aspect de la complexité de la *gestion des connaissances*, nous avons utilisé, d'une part, les termes de Jean-Louis Ermine (c'est-à-dire les verbes à l'infinitif : capitaliser, partager et créer) qu'il emploie pour caractériser la *gestion des connaissances*, lorsqu'il dit « la gestion des connaissances ("Knowledge Management") s'inscrit dans la réalité de l'entreprise : la connaissance est un enjeu économique majeur de demain. Créer, capitaliser et partager son capital de connaissances est une préoccupation de toute organisation performante », et d'autre part, les termes de Jean-Claude Tarondeau (c'est-à-dire les verbes à l'infinitif : créer, produire et offrir), qu'il utilise pour définir la capacité d'une entreprise lorsqu'il dit « les capacités sont définies comme des routines de mise en œuvre d'actifs pour créer, produire et/ou offrir des produits ou services sur un marché » [Tarondeau, 98]. En plus, nous avons préféré, à l'inverse de Tisseyre, laisser "produire" comme une tâche de gestion, car la gestion est liée à la capacité de l'entreprise d'après Tarondeau.

> **Aspect action de la gestion des connaissances**

Un deuxième aspect qui caractérise la complexité de la *gestion des connaissances*, réside dans l'impossibilité de dissocier la connaissance et l'application de cette connaissance dans l'action. En effet, la connaissance qui compte pour la capacité de l'entreprise est la connaissance qui est transformée en action pour l'acteur (individu, groupe, entreprise)[52]. C'est l'aspect action de la gestion des connaissances, tel que l'ont souligné d'une part, Barthelme-Trapp dans sa thèse, lorsqu'elle dit « au premier rang des difficultés, limites et risques d'un management des connaissances au sein d'une entreprise, on trouve l'impossibilité de dissocier la connaissance de son application et la nécessité d'intégrer la dynamique dans cette gestion » [Barthelme-Trapp, 03], et d'autre part, Tounkara en citant Argyris dans sa thèse, « pour Argyris l'apprentissage est indissociable de l'action. En effet, il considère qu'il y aura toujours un écart entre la connaissance que nous avons stockée et la connaissance qu'il faut pour agir avec efficacité dans des circonstances données. Et c'est lorsque nous détectons et corrigeons cet écart que nous apprenons. En résumé, il faut mettre en mouvement les connaissances, les confronter au monde de pratique pour apprendre » [Tounkara, 02]. Alors, le fait de "mettre en mouvement les connaissances" est lié à la dynamique circuler (interne et externe) de *gestion des connaissances* (capitaliser, partager, créer), la gestion de l'innovation (créer, produire, offrir), et la gestion des compétences (apprentissage, action). Et donc, les connaissances, les compétences, et l'innovation sont au cœur de l'entreprise[53]. Il y a une dynamique interne, d'une part, entre capitaliser, partager, et créer dans la *gestion des connaissances*, ainsi qu'entre créer, produire, et offrir dans la gestion de l'innovation, et entre apprentissage et action dans la gestion des compétences, et d'autre part, une dynamique externe entre *gestion des connaissances*, gestion de l'innovation et la gestion des compétences, dans le sens que l'un alimente l'autre. Enfin, l'aspect action de la connaissance est enraciné dans le concept *Actionable Knowledge* de Argyris, qu'il développe dans son livre *Savoir pour agir. Surmonter les obstacles à l'apprentissage organisationnel* [Argyris, 95].

---

[52] Nous avons privilégié cette définition de l'acteur, car elle correspond à peu près à la dimension ontologique du modèle de création des connaissances nouvelles et d'apprentissage organisationnel de Nonaka et Takeuchi. En effet, la dimension ontologique correspond à l'individu, le groupe, l'organisation et l'inter-organisation, cela signifie que pour eux l'aspect humain : l'individu, le groupe, l'entreprise est à la base de la création des connaissances nouvelles et d'apprentissage organisationnel pour l'entreprise. Dans ce modèle la connaissance est approchée par un système social.
[53] Plus loin nous parlons plutôt de système d'organisation du travail coopératif (l'entreprise, le *business*).

> **Aspect culturel de la gestion des connaissances**

Un troisième aspect qui caractérise la complexité de la *gestion des connaissances* (et qui est lié à la capacité de l'entreprise), concerne le trait culturel de chaque entreprise et de ses membres. Nous voulons dire par là que si une démarche pour une certaine approche a bien marché ici, rien ne garantit qu'il va bien marcher ailleurs. En effet, les relations entre ses membres pour capitaliser, partager et créer des connaissances sont propres à chaque pays ou région, ainsi qu'à la structure de l'entreprise pour créer, produire et offrir de nouveaux produits ou services. Ce troisième aspect est relié à l'aspect culturel de la gestion des connaissances.

Selon l'aspect culturel, la généralisation des approches et démarches de la *gestion des connaissances* reste un sujet de recherche sur le plan académique, des livres et articles sur le sujet en témoignent. Dans un article de juin 2001 intitulé *Analyse comparée de méthodes de gestion des connaissances pour une approche managériale*, Barthelme-Trapp dit « la gestion des connaissances est un domaine encore peu formalisé dans les sciences de gestion. Développée au travers de démarches pragmatiques, elle recouvre aujourd'hui un vaste ensemble de méthodes, d'outils et de pratiques organisationnelles » [Barthelme-Trapp, 02]. Et plus récemment dans un article d'octobre 2003 intitulé *De l'extraction des connaissances au Knowledge Management*, Dominique Crié dit « la notion de KM se décompose en un ensemble de processus organisationnels et culturels qui cherche à organiser une combinaison synergique entre données puis informations et la capacité créative et innovatrice des individus à l'aide de supports technologiques. Bien qu'il existe une multitude de définitions différentes du KM, le concept peut être approché dans (1) ses aspects organisationnels, humains, culturels et technologiques ; et (2) ses enjeux de capitalisation des connaissances, de création de valeur innovation, et d'avantage stratégique » [Crié, 03]. Ce qu'elle appelle « le cadre général du management des connaissances » [Crié, 03].

En revanche, sur le plan économique, pour un chef d'entreprise la gestion des connaissances est une problématique de gestion et un enjeu économique plein de difficultés, limites et risques, comme nous avons pu le constater avec les cinq problématiques de gestion des connaissances analysées par Cap Gemini Ernst & Young en France, les difficultés se trouvent détaillées par Tisseyre dans son livre, en plus sur la confidentialité de ces études de cas cependant il dit « il n'est pas possible de citer les noms de ces entreprises pour la raison simple que le Knowledge Management faisant partie des nouvelles démarches stratégiques, elles ne souhaitent pas dévoiler leurs secrets. Par contre il a été possible d'utiliser les résultats de la mise en place de la démarche de Knowledge Management à des fins pédagogiques » [Tisseyre, 99]. Voici les projets cités par Tisseyre (1) transformer un centre de documentation en centre stratégique d'information ; (2) mettre en place un serveur de connaissances pour partager un savoir-faire rare ; (3) changer la culture d'un groupe mondial par une gestion des ressources humaines transversale ; (4) créer une synergie entre les activités d'avant-vente et de production dans le cadre d'une activité de conseil ; et enfin (5) mettre en place un référentiel mondial des connaissances. Nous attirons ici l'attention que le projet (3) montre bien la relation directe entre gestion des connaissances et gestion des ressources humaines (GRH), et plus particulièrement la gestion des compétences, tel que nous l'avons déjà avancé dans notre deuxième aspect qui caractérise la complexité de la gestion des connaissances.

➢ **Aspect chaos de la gestion des connaissances**

C'est un aspect qui caractérise la complexité de la gestion des connaissances, qui a été par Robert Reix, lorsqu'il dit que c'est « un domaine à la mode, mais encore mal défini, une littérature abondante (et pas toujours dénuée de préoccupations commerciales) qui tend a confondre, sous ce titre, des réalisations fort différentes en jouant sur l'ambiguïté des termes informations, connaissances, voir compétences » [Reix, 00]. C'est dans ce chaos face aux pressions externes et internes qui peuvent influencer défavorablement la capacité de l'entreprise (la mode, l'intérêt commercial) que réside l'ambiguïté des termes liés à la gestion des connaissances.

En conclusion, la diversité de ces cinq cas d'entreprises que l'on vient de citer (en plus René-Charles Tisseyre membre de Capgemini, et comme nous verrons plus bas, Jean-Yves Prax est PDG de la société CorEdge[54]), prouve que les approches pédagogiques utilisées (afin de dégager une démarche générique) peuvent être influencées par la façon de faire de chaque société de conseil. Nous constatons aussi que pour Tisseyre et Prax, ainsi que chez d'autres auteurs qui se sont intéressés à la *gestion des connaissances*, ce terme correspond à la traduction littérale du terme *Knowledge Management*, qui reste un mot "made in USA", avec un sentiment de mode ou de rejet social, tel comme nous le voyons dans la vie de tous le jours avec Coca Cola, ou McDonalds.

Avant de continuer avec le résumé de ces quatre aspects, nous voudrions justifier davantage ce quatrième aspect qui concerne le chaos ou la confusion que Reix annonce autour de la *gestion des* connaissances, et faire une analogie avec l'intelligence artificielle à partir de la définition de Patrick Winston. D'après lui « l'intelligence artificielle est l'étude des idées qui permettent aux ordinateurs d'être intelligents » [Winston, 94]. Or, au moment de l'apparition de l'intelligence artificielle, beaucoup ont cru voir dans cette nouvelle science la fin de l'homme, et l'affleurement d'un nouvel être : une machine (l'ordinateur) avec la capacité d'un cerveau humain. Cependant le film *2001 : A Space Odyddey* de Stanley Kubrick sorti en 1968, et récemment le film *A.I. Artificial Intelligence* de Steven Spielberg sorti en 2001, ont montré avec un décalage technologique assez important que l'intelligence artificielle est encore une science mal comprise par le grand public. Le fait de voir dans l'ordinateur la capacité d'aller au-delà de l'intelligence de l'être humain le prouve.

---

[54] Il s'agit d'un autre cabinet de conseil comme Capgemini http://www.coredge.fr/

La critique de Francisco Varela au paradigme simonien, nous paraît aussi illustrative de cette confusion autour de l'intelligence artificielle, lorsqu'il dit « la métaphore populaire désignant le cerveau comme une machine de traiter de l'information n'est pas seulement ambiguë, elle est totalement fausse ». Plus tard dans un entretien réalisé par Jeanne Mallet à l'occasion de la sortie de son livre *L'inscription corporelle de l'esprit* (publié en 1993), Varela dit « il s'avère que ce type de machine, de robot typiquement incarné dans l'image de l'ordinateur "Hall" dans le film "2001, Odyssée de l'espace", c'est sympathique mais après 30 ans d'expérience, on se rend compte que dans cette logique on atteint rapidement les limites. Pourquoi ? Parce qu'en fait la connaissance, la capacité de faire quelque chose, de bouger, de se risquer à bouger, d'aller d'ici et là, est tellement contextuelle que l'on ne peut jamais tout faire contenir dans un programme qui prévoirait toutes les éventualités »[55].

Pour nous la confusion autour de l'intelligence artificielle est de même nature que la confusion autour de la gestion des connaissances, ici et là la problématique est attachée aux concepts de cognition et d'apprentissage, d'une part chez l'individu (l'intelligence humaine) et dans la machine (l'intelligence artificielle) dans une relation existentielle, et d'autre part chez l'individu (système cognitif géré par un processus d'apprentissage et d'adaptation à leur environnement) et dans l'organisation (système de connaissances géré par un processus d'apprentissage organisationnel et d'évolution avec leur environnement) dans une relation de travail. En effet, pour Alliot et Schiex, en citant Rapaport, disent « la science cognitive en général cherche à comprendre les fonctions cognitives humaines en termes d'états mentaux et de processus c'est-à-dire, en termes d'algorithmes qui réalisent la transformation des données d'entrée en données de sortie » [Alliot et Schiex, 93]. Cette définition classique de la cognition selon l'approche de l'intelligence artificielle, remet en question le concept d'apprentissage et ses mécanismes d'évolution chez l'être humain. La confusion est constatée encore avec Patrick Winston, lorsqu'il dit « apprendre aux ordinateurs à être plus intelligentes permettra sans doute d'apprendre à l'homme à être plus intelligent » [Winston, 94]. Par contre, l'apprentissage dans la gestion des connaissances est relatif à l'apprentissage organisationnel qui « exprime le fait que l'organisation doit adapter, modifier sans cesse ses comportements et ses compétences » [Ballay, 97] ou comme l'a dit Tarondeau en citant Koenig « l'apprentissage organisationnel est un phénomène collectif d'acquisition et d'élaboration de compétences qui, plus ou moins profondément, plus ou moins durablement, modifie la gestion des situations et les situations elles-mêmes » [Tarondeau, 98]. En conséquence, l'apprentissage organisationnel n'a rien à voir avec les techniques d'apprentissage en intelligence artificielle, par exemple en utilisant des réseaux neuronaux, bien que l'on puisse envisager ces techniques pour représenter la connaissance.

---

[55] L'entretien complet se trouve dans le site
http://lambesc.educaix.com/enseignants/mallet/dossier_texte/textesmallet/ouvrages/developpementpersonneetorg/txt_dpdo/dpdo08.htm

Et donc, pour nous, la cognition et l'apprentissage doivent suivre des chemins séparés pour approcher l'évolution de l'être humain, sinon l'on risque d'alimenter de faux progrès. Par exemple penser qu'aujourd'hui qu'il est possible d'encoder dans un programme d'ordinateur le côté affectif d'un être humain, reste encore dans le domaine de la science fiction, comme l'ont si bien montré Stanley Kubrick et Steven Spielberg dans leurs films.

En réalité, rien n'empêche que cela reste un pari pour les nouvelles générations des informaticiens, tel qu'en 1970, le défi pour eux a été d'écrire un programme pour simuler le comportement d'un expert face à une tâche à résoudre dans le contexte d'une activité professionnelle. En général, le formalisme utilisé pour reproduire l'expertise sur un savoir spécifique de l'expert a été les *règles de production*, le programme qui en découle a été appelé *systèmes experts*, et la démarche générique de construction a été connu sur le nom *d'ingénierie des connaissances* ou l'*ingénierie cognitive*. Cependant, étant donné la « complexité du raisonnement humain » ces outils informatiques ont montré leurs limites, et pour certaines entreprises, un mauvais investissement. Néanmoins, il semble que les *systèmes experts* ont le mérite d'avoir ouvert le chemin pour la suite, comme en témoigne Jean-Louis Ermine dans une entretien chez Neteconomie, lorsqu'il dit « le CEA … a notamment beaucoup investi dans les technologies de l'intelligence artificielle, qui, si elles ont abouti plutôt à un constat d'échec, ont permis leur réorientation vers le domaine de la gestion des connaissances »[56].

Ainsi, plus récemment, en 1990, des informaticiens, des cogniticiens et des professionnels des sciences humaines ont pris le défi des systèmes experts, mais sur une autre approche que l'on appelle *systèmes à base de connaissances*, où l'objectif comme l'a dit Michel Grundstein[57] est « de modéliser la connaissance d'un ou plusieurs experts sans qu'elle perde sa spécificité et sa structure cognitive. Les bases de connaissances recensent et organisent l'ensemble des connaissances pour les valoriser, les diffuser et les réutiliser » [Grundstein, 96]. Et donc, en suivant une telle démarche l'on voit apparaître une implication organisationnelle que les systèmes experts n'ont pas considéré pour leur développement dans l'entreprise.

[56] Le CEA est le Commissariat à l'Energie Atomique. L'entretien complet se trouve dans le site http://www.neteconomie.fr/les_entretiens/entretienGES.php3?id=80
[57] Depuis 1996 il est vice-président de l'Institut international pour l'intelligence artificielle (IIIA).

Cette évolution apparaît de nos jours, grâce à l'émergence des nouvelles technologies de l'information et de la communication (NTIC)[58], sous les noms de *systèmes de connaissances* ou *systèmes de gestion des connaissances* qui ont pour objectif justement la *gestion des connaissances*, c'est-à-dire *capitaliser*, *partager* et *créer* des connaissances nouvelles dans l'entreprise, comme l'a souligne Jean-Louis Ermine « j'appellerai d'une manière générale système de gestion des connaissances tout système (au sens le plus général) qui permet de capitaliser, partager ou créer des connaissances » [Ermine, 03], et puis dans un sens plus spécifique « les connaissances dans une organisation s'organisent en système propre (au sens de la théorie des systèmes) au même titre que le système cognitif des êtres humains, que nous appelons le "patrimoine de connaissances" de l'organisation (ou "système de connaissances"). Ce patrimoine n'est pas réductible à des systèmes déjà existants, comme le système d'information, de documentation, de ressources humaines (formation, compétences), etc. Il a ses fonctions propres assignées par l'organisation (créer, capitaliser, partager les connaissances), son organisation et sa structure propre. Bien sûr il est en interrelation constante et puissante avec tous les autres systèmes de l'organisation, notamment le système d'information, le système de coopération et le ou les systèmes d'interaction avec l'environnement (systèmes de veille ...) etc. » [Ermine, 03]. Cette démarche implique aussi la mise au point au fur et à mesure de méthodes et outils de l'*ingénierie et capitalisation des connaissances* beaucoup plus performant chaque jour[59].

D'ailleurs chacun de ces trois systèmes a des objectifs fort différents. Le système expert permet la capitalisation des connaissances d'un expert ou d'un groupe d'expert pour un domaine spécifique. Le système à base de connaissances permet la capitalisation et le partage des connaissances d'un expert ou d'un groupe d'experts sur un domaine ou plusieurs domaines. Néanmoins, les techniques de l'intelligence artificielle, en particulier (1) de l'ingénierie des connaissances (systèmes experts) ; (2) l'ingénierie cognitive (systèmes à base de connaissances) ; et (3) l'ingénierie et capitalisation des connaissances (systèmes de connaissances) sont utilisées pour modéliser, justement, les connaissances. Par exemple en utilisant des réseaux neuronaux, lors de la construction de systèmes experts, de systèmes à base de connaissances ou de systèmes de connaissances.

---

[58] Selon Prax « les NTIC n'ont plus pour finalité de "traiter" l'information (au sens data computing) ; elles se doivent de proposer un lieu "virtuel" facilitant les échanges, la capitalisation et la diffusion de la connaissance » [Prax, 00].
[59] Soulignons au passage le livre de Jean-Yves Prax Le manuel du Knowledge Management : Une approche de 2e génération (publié en 2003), ainsi que La gestion des connaissances de Jean-Louis Ermine (publié en 2003) et *Management des connaissances en entreprise* d'Imed Boughzala et Jean-Louis Ermine (publié en 2004). En fin pour une vision très technique de méthodes et outils pour la gestion des connaissances, nous avons les trois ouvrages de Rose Dieng et son équipe de recherche ACACIA à l'INRIA Sophia Antipolis : *Méthodes et outils pour la gestion des connaissances* (publié en 2000), *Méthodes et outils pour la gestion des connaissances : une approche pluridisciplinaire du Knowledge Management* (publié en 2001), et *Knowledge Management & Organizational Memories* (publié en 2004).

Nous fermons cette parenthèse, afin de présenter le résumé de ces quatre aspects qui témoignent la complexité de la gestion des connaissances afin de l'intégrer dans la capacité de l'entreprise, à tous ses niveaux : stratégique, tactique, et opérationnel mais dans un sens de réseau des savoirs qui est plutôt fondée sur la pyramide hiérarchique, (1) l'aspect dynamique de la gestion permet de combiner, d'intégrer et de coordonner de façon à faire "circuler" la gestion des connaissances (capitaliser, partager, créer), la gestion de l'innovation (créer, produire, offrir), et la gestion des compétences (apprentissage, action) ; (2) l'aspect action de la connaissance permet d'intégrer de façon à faire "circuler" la connaissance à l'application de cette connaissance dans l'action (gestion des compétences) ; (3) l'aspect culturel de la gestion des connaissances permet d'intégrer la gestion des connaissances au sein de l'entreprise et auprès de ses membres avec succès ; (4) l'aspect chaos de la gestion des connaissances permet d'attirer l'attention sur les pressions externes et internes face à un projet de gestion des connaissances dans l'entreprise.

Il nous paraît opportun de signaler ici une première définition de la gestion des connaissances pour le modèle proposé : **la gestion des connaissances (capitaliser, partager et créer des connaissances) pour la rendre utile (gestion des compétences) doit être liée à l'évolution de l'acteur (individu, groupe, entreprise) dans son espace de travail, c'est-à-dire la connaissance métier donnée par l'environnement : le travail et ses outils de production afin de créer, produire et offrir de nouveaux produits ou services (gestion de l'innovation).** Pour nous cette définition doit être indépendante des outils de NTIC du KM.

En conséquence, c'est sur cette synergie entre la gestion des connaissances, la gestion des compétences, et la gestion de l'innovation nous pouvons dire que se joue de nos jours la performance de l'entreprise dans une économie globalisée (l'outsourcing du travail).

Nous postulons que cette synergie dans le modèle proposé est possible seulement à travers l'individu et l'organisation dans une relation de travail, que nous matérialisons dans un *système d'organisation du travail* dans un paradigme d'*avantage collectif*. Ainsi, ce système du travail est caractérisé par deux systèmes, l'un est le *système cognitif* (l'individu), et l'autre est le *système de connaissances* (l'organisation). Le système cognitif explique notre capacité de faire émerger nos sens, nos chemins d'apprentissage, notre capacité à créer de métaphores, etc., en gros la gestion des connaissances chez l'individu, tandis que le système de connaissances explique principalement la création de connaissance collective, l'apprentissage organisationnel, ainsi que la gestion des connaissances. Au sujet, Prax a souligné « pour favoriser la création de connaissance collective, un concept créée par un individu doit être partagé avec les autres alors que ceux-ci n'en ressentent pas l'utilité immédiate et ne sont donc pas prêts à l'intégrer dans leurs propres systèmes de représentation » [Prax, 00]. Finalement, le système d'organisation du travail par avantage collectif nous permet (1) de trouver la signification du mot connaissance dans le travail ; (2) de souligner l'importance de la connaissance comme un facteur clé dans l'organisation du travail dans un paradigme d'avantage collectif ; et (3) d'expliquer la capacité de l'entreprise pour être organisée dans le temps comme un tout.

Deux exemples, donnés par Prax nous permettent d'illustrer l'impact des NTIC sur le système d'organisation par avantage collectif. L'un est relative à « la réorganisation des services déconcentrés de l'Etat et de la définition même du territoire » [Prax, 00]. L'idée ici est que par le biais des NTIC l'on peut créer un espace territorial virtuel organisé intelligemment en pôles de compétences. L'autre est relative à l'entreprise élargie qui « montre comment entreprises substituent aujourd'hui des avantages coopératifs aux anciens avantages compétitifs » [Prax, 00]. L'idée ici est que par le biais des NTIC l'on peut créer une entreprise virtuelle entre clients et fournisseurs.

Néanmoins, dans les systèmes de connaissances la création des connaissances nouvelles reste toujours en défi humain, comme l'argumente très bien Prax « *stricto sensu*, la création de connaissance nouvelle est uniquement le fait d'individus. Une organisation ne peut pas créer elle-même de la connaissance, sans individus » [Prax, 00]. Puis il ajoute « l'organisation appuie son processus et lui procure un contexte spécifique ». Ce qu'implique une relation directe entre l'acteur (individu, groupe, entreprise) et son espace de travail (l'environnement : le travail et ses outils de production), ce que nous la caractérisons, simultanément et nécessairement, dans une dualité organisation/structure (voir chapitre 2, section 2.1.7) à travers un processus d'action, en que la "matière grise" maintient une relation de couplage structurelle dans le système cognitif de l'individu lequel reste la base de la connaissance collective et du savoir collectif couplé aussi structurellement dans le système de connaissance de l'entreprise afin de créer un avantage concurrentiel, compétitif, coopératif durable ou autre, quel que soit l'horizon stratégique de l'entreprise, leur mission et leur vision du futur.

D'ailleurs, le système d'organisation du travail par avantage collectif nous le percevons aussi comme un modèle en couches qui s'est enrichi par notre histoire industrielle à travers de méthodes de travail et de modes de management. Au centre du modèle se trouve toujours l'individu, d'abord en termes de force de travail, que nous caractérisons par les paradigmes tayloriste et fordiste, principalement, puis en termes de sa capacité limitée de traiter de l'information, cette époque nous la caractérisons par le paradigme simonien, ensuite en termes de sa capacité, cette fois-ci, illimitée de créer de la connaissance, dans un passé récent pour développer un avantage concurrentiel ou compétitif durable pour l'entreprise, de nous jours pour développer un avantage coopératif durable, ces époques nous les caractérisons (1) par le modèle de création des connaissances nouvelles et d'apprentissage organisationnel de Nonaka et Takeuchi ; et (2) par l'émergence de l'entreprise élargie[60] et les NTIC[61].

Des ces quatre aspects qui témoignent la complexité de la gestion des connaissances, nous proposons une deuxième définition pour le modèle proposé : **la gestion des connaissances signifie "mettre en mouvement les connaissances" d'un système d'organisation du travail par avantage collectif à travers une dynamique circulante (interne et externe) de gestion des connaissances (capitaliser, partager, créer), gestion de l'innovation (créer, produire, offrir), et gestion des compétences (apprentissage, action) à partir de méthodes de travail et mode de management pas fondées sur une pyramide (séparation des fonctions, des postes, de tâches,…) mais plutôt sur un réseau des savoirs[62] (qui fait quoi, qui sait quoi,…) et un réseau de compromis social[63] (l'union fait la force, le langage, la culture de partage, la culture de diffusion,…) en couplage structurel avec l'environnement pour garantir l'ontogenèse et l'évolution[64] du système comme un tout organisé**.

Malheureusement, nous n'avons pas développé un langage graphique pour représenter ce modèle, où au centre se situerait un système d'organisation du travail par avantage collectif, autour duquel nous pouvons faire graviter les complexités antérieures liées à notre définition de la gestion des connaissances, mais si nous faisons une analogie avec le modèle de la marguerite[65] de Jean-Louis

---

[60] Selon Prax « l'idée de "l'entreprise élargie" est d'améliorer le fonctionnement d'un réseau professionnel par une transparence totale d'information, de savoirs et de savoir-faire entre les différents acteurs de la chaîne de conception-production … même s'ils sont concurrents » [Prax, 00].

[61] Selon Prax « les NTIC n'ont plus pour finalité de "traiter" de l'information (au sens data computing) ; elles se doivent de proposer un lieu "virtuel" facilitant les échanges, la capitalisation et la diffusion de la connaissance » [Prax, 00].

[62] Le terme *réseau des savoirs* nous l'avons adapté du terme *le mode de travail en réseau* de René-Charles Tysseyre, lorsqu'il a dit « le mode de travail en réseau : tout acteur est capable de parler à tout autre acteur sans aucun frein ; il n'y a plus de hiérarchie formelle à respecter mais au contraire une diffusion de manière transverse de l'information » [Tisseyre, 99].

[63] Le terme *réseaux de compromis social* nous l'avons emprunté à Fernando Flores, en gros, cela signifie un mode d'organisation du travail coopératif basé sur les modes d'être, le modes de faire pour achever l'objectif (le quoi faire). Il s'agit donc d'un outil de flux conversationnel d'une entreprise. Ceci élargie en peu l'approche de l'*enaction* de Maturana et Varela qui considère la connaissance comme un système d'actions, où l'action existe, pour l'acteur, comme un acte de langage (c'est-à-dire le langage et ses émotions) et la culture (c'est-à-dire la structure sociale). Comme l'a dit Varela « c'est notre réalisation sociale, par l'acte de langage, qui prête vie à notre monde. Il y a des actions linguistiques que nous effectuons constamment : des affirmations, des promesses, des requêtes et des déclarations. En faite, un tel réseau continu de gestes conversationnels, comportant leurs conditions de satisfaction, constitue non pas un outil de communication, mais la véritable trame sur laquelle se dessine notre identité » [Varela, 96].

[64] Pour Maturana et Varela « l'ontogenèse est l'histoire de la transformation structurelle d'une unité » [Varela, 86], tandis que l'évolution est « l'histoire de ses changements » [Varela, 86], et donc, le concept l'évolution est applicable à l'espèce, par contre ontogenèse est applicable à l'individu. Dans un tel sens, ces deux concepts sont applicables à la phénoménologie biologique, mais Aquiles Limone l'applique à l'organisation, lorsqu'il a dit « en ce qui concerne l'entreprise, il devient extrêmement difficile de préciser où finit l'ontogenèse et où commence l'évolution » [Limone, 77].

[65] Dans le chapitre1 nous montrons que le modèle de la marguerite de Jean-Louis Ermine le patrimoine de connaissances (système de connaissances) est gravité par cinq processus (1) le processus de capitalisation et de partage des connaissances ; (2) le processus d'interaction avec l'environnement ; (3) le processus de sélection par l'environnement ; (4) le processus d'apprentissage et de création de connaissances ; (5) le processus d'évaluation du patrimoine de connaissances (ce processus

---

Ermine, pour nous, au cœur du système d'organisation du travail par avantage collectif (l'entreprise, le *business*) se trouve l'individu, et bien entendu le travail aussi, et donc la connaissance comme un objet de gestion doit habiter chez l'individu, mais aussi chez le travail, autrement dit dans le travail réalisé par l'individu, et donc nous parlons plutôt de la *connaissance métier* que du savoir de l'individu. En effet, du fait que « les compétences sont des habilitées, des savoir-faire, susceptibles d'être mobilisés par l'acteur pour l'intervention » [Prax, 00], les compétences sont propres à la relation entre l'individu et le travail, tandis que les connaissances sont propres à la relation entre le travail et l'individu. Cela signifie que la gestion des connaissances « se concentre sur les savoirs et savoir-faire sous-jacents à une activité et non aux effectifs qui la réalisent ... car il s'agit ici de gérer les connaissances et non pas les compétences individuelles ou organisationnelles de l'entreprise » [Prax, 00]. En conséquence, la gestion des connaissances est relative à la gestion du savoir, savoir-faire, ou autre savoir (individuel ou organisationnelle) relative à un processus ou à un réseau de processus de l'entreprise. En revanche, la gestion des compétences est relative à la gestion du savoir ou savoir-faire ou autre savoir chez l'individu qui doit être mobilisé pour accomplir un objectif dans son espace de travail, c'est-à-dire pour créer, produire et offrir de nouveaux produits ou services (gestion de l'innovation).

Ainsi, l'individu est au centre de la gestion des connaissances et au centre de la gestion des compétences, car l'individu est au centre de l'organisation, et donc il est au centre du système d'organisation du travail par avantage collectif. Néanmoins, dans la mise au point du système d'organisation du travail coopératif, comme l'a dit Prax « il faut que le management des compétences se situe comme un prolongement naturel du management des connaissances » [Prax, 00]. En effet, il faut une certaine dynamique pour faire circuler les connaissances dans l'entreprise par le biais de la création des connaissances nouvelles et l'apprentissage organisationnel. A cette égard, Tarondeau dit « le savoir et l'application du savoir dans l'action constituent les fondements des capacités et des compétences ... c'est l'accumulation de savoirs individuels et collectifs et l'apprentissage obtenu dans leur mise en action qui génèrent les aptitudes, les capacités et les compétences » [Tarondeau, 98].

Et donc, la connaissance et la compétence deviennent une capacité pour l'entreprise lorsqu'elle veut, peut et sait les gérer dans le contexte d'un activité productive (le travail). Le verbe "gérer" signifie pour nous (1) la création des connaissances et des compétences nouvelles par domaines, thèmes ou processus au niveau stratégique, tactique, et opérationnel de l'entreprise ; (2) la combinaison des connaissances et des compétences par niveaux de l'entreprise (stratégique, tactique, opérationnel) de façon verticale et transversale ; (3) l'intégration des connaissances et des

---

ne se trouve pas détaillé dans le modèle de la marguerite, mais nous l'avons trouvé ailleurs d'une façon implicite. Ainsi, les complexités de la gestion des connaissances chez Wendi Bukowitz, Ruth Williams, Karl Sveiby et Leif Edvinsson, que nous avons identifié plus haut sont relatives au processus (5).

compétences dans des produits et services nouveaux ; et (4) la coordination des connaissances et des compétences au niveau de l'entreprise élargie.

### 3.3.3. Le modèle proposé selon un point de vue technique

Dans cette section nous présentons une classification du système opérationnel associé au système de connaissance, à partir des trois approches cognitives que nous avons vu au chapitre 1, à savoir : les *SBC issus de l'approche cognitiviste*, les *SBC issus de l'approche connexioniste*, et les *SBC issus de l'approche enactiviste*.

➢ **Les SBC issus de l'approche cognitiviste**

Un système de connaissance de type cognitiviste fonctionne selon l'approche de la cognition (le traitement symbolique de la connaissance), d'où le nom de cogniticiens donné dans les années 80 aux informaticiens chargés de recueillir l'expertise chez l'expert afin de constituer un cahier des charges pour la construction d'un système expert ou d'un système à base de connaissance. Ces systèmes de connaissance sont bâtis sur la dualité savoir-faire/connaissance.

Ainsi, un système de connaissance prend en charge une réalité physique sous forme de code symbolique. Comme l'a dit Ermine, « tout phénomène perceptible (signe) s'observe selon trois niveaux indissociables : le référent ou signe (la manifestation), le signifié (la désignation), le signifiant (le sens) ou encore se perçoit selon trois dimensions : syntaxique, sémantique, pragmatique. Cette conjonction de points de vue est inséparable ». Puis il a dit, en citant Le Moigne « on ne peut pas … manipuler un symbole en faisant comme s'il n'était qu'un signe dénué a priori de signification et de configurabilité » [Ermine, 96].

En conséquence, toute l'informatique (*computing* en anglais, *computación* en espagnol) de nos jours, se fait par le biais du traitement symbolique de l'information (base de données, réseaux, etc.), et du traitement symbolique de la connaissance (l'émulation du savoir-faire des experts par des systèmes experts, les SBC de type base de règles ou base de cas, etc.), mais également la bureautique (le traitement symbolique de documents), la télématique (le traitement symbolique du son, de l'image et de la parole dans un même support), etc., s'appuient aussi sur l'approche cognitiviste de la cognition. De plus, ce n'est pas seulement le *software* qui suit une ligne symbolique, mais aussi le *hardware* de la machine [Narvarte, 90]. En effet, la représentation binaire est partout dans la structure de la machine.

Nous attirons l'attention aussi, que le fait d'avoir cette contrainte symbolique pour la représentation de la connaissance, a limité notre travail de thèse à l'extraction des connaissances de données étendue aux données floues (objets imprécis et événements incertains) d'une base de données relationnelles floues de type FSQL. Cela signifie, que la seule façon de faire le flou dans une machine logique (0 ou 1), c'est à travers la simulation d'une couche logique. Cette couche sert de pont entre la couche conceptuel et la couche physique. D'ailleurs, cela implique la construction de *tables de conversion*[66] (dans la couche logique) pour faire le passage entre les autres deux couches, cela implique aussi un gaspillage de ressources énormes en termes de temps CPU[67].

Il semblerait que le contexte scientifique de cette thèse (où l'objectif est la gestion de connaissances imparfaites) soulève la question se savoir si dans un futur (proche pour l'humanité) aurons-nous une machine flou, c'est-à-dire une machine où les propriétés des composants dans l'espace matériel. Nous pesons, que la direction de ces recherches à la *bioinformatique*, car nous vivons tous les jours avec le flou (et le non flou aussi d'ailleurs).

C'est pour cette raison, que dans la citation du chapitre 1, nous avons mis à Francisco Varela, car lui est un *biologiste et informaticien*. En plus, lorsqu'il dit « le monde du vivant, la logique de l'autoréférence et tout l'histoire naturelle de la circularité devrait nous dire que la tolérance et le pluralisme sont le véritable fondement de la connaissance. Ici, les actes valent mieux que les mots ».

Pour nous ce paragraphe est clé dans la démarche et le canevas de pensée chez Varela, car c'est une véritable invitation à explorer d'autres chemins pour représenter la connaissance. Et, comme l'a si bien dit Gaston Bachelard dans son livre, intitulé *Le nouvel esprit scientifique* (publié en 1934) : *Pourquoi pas ?* Eh oui, pourquoi ne serait-il pas opportun d'imaginer l'informatique en dehors de l'approche cognitiviste. Cela signifie qu'il faut laisser derrière le paradigme de l'ordinateur (informatique ou computationnel) chez John von Newman. Pas si longtemps Warren McCulloch l'a fait avec le modèle connexioniste, fondé sur l'approche connexionniste (que nous verrons par la suite). L'avenir pour nous se trouve, comme nous l'avons dit plus haut, dans la *bioinformatique*, et donc l'approche de l'enaction (que nous verrons ici aussi), pourrait devenir un arbre si solide pour construire une nouvelle approche de l'informatique, et de nouveaux systèmes de connaissance.

---

[66] Dans le chapitre 5 nous développons une stratégie pour la construction des couches. La couche conceptuelle que permet la représentation de données imprécises, et la couche logique que permet la représentation de tables de conversion, pour le passage de la couche conceptuel à la couche physique (que dans cette expérience est Oracle 8). Les aspects théoriques de l'imprécis et de l'incertain sont donné dans le chapitre 4. Principalement, du livre de Dubois et Prade, intitulé *Théorie des Possibilités - Applications à la représentation des connaissances en informatique* (publié pour la première fois en 1985, puis en 1988).
[67] Dans l'annexe 2 nous avons incorporé en exemple pour le traitement flou de 19 tuples (registres) à l'aide du système FSQL sous Oracle 8. Néanmoins, il faut soulgine que les temps d'exécution pour une interrogation flou (avec un ou deux filtres flous) sont loin d'être fier, si l'on sait que l'entreprise gère en batch, entre 5000 et 10000 registres pour un simple procédure de gestion de stocks.

> **Les SBC issus de l'approche connexioniste**

Un système de connaissance de type connexioniste fonctionne selon une approche de la connexion (le traitement en réseau de la connaissance). Ces systèmes de connaissance sont bâtis sur la dualité environnement/connaissance, dans lesquels un phénomène de veille participe dans le recueillement d'informations dans l'environnement interne et externe (les travaux de Jean-Louis Ermine vont dans ce sens). Dans cette logique le modèle de la marguerite qui ne fait qu'enrichir davantage les modèles MASK et MKSM sont issus d'un milieu cognitiviste comme le montrent ses deux ouvrages fondamentaux de l'époque. L'un est *Systèmes experts, théorie et pratique* (publié en 1989) [Ermine, 89], l'autre est *Génie logiciel et génie cognitif pour les systèmes à base de connaissances* (publié en 1993) [Ermine, 93]. Ce virage à droite, dans le changement de pensée, peut s'observer aussi chez Michel Grundstein, en effet lui est passé du cognitivisme au connexionnisme, comme nous l'avons constaté dans son livre *Les systèmes à base de connaissances, systèmes experts pour l'entreprise* (publié en 1988) [Grundstein *et al*, 88], et dans son article *La capitalisation des connaissances de l'entreprise, une problématique de management* (publié en 1996) [Grundstein, 96][68].

> **Les SBC issus de l'approche enactiviste**

L'approche de l'enaction de Maturana et Varela considère la connaissance comme un système d'actions [Maturana et Varela, 73], où l'action existe, pour l'acteur, comme un acte de langage (c'est-à-dire le langage et ses émotions) et la culture (c'est-à-dire la structure sociale). Et donc, dans l'approche enactiviste la connaissance est le fruit du flux conversationnel d'une entreprise [Flores, 96a], [Flores, 96b].

Ainsi, notre hypothèse de recherche pour la construction d'un modèle autopoïétique de la gestion des connaissances imparfaites se base sur le fait que la connaissance ne peut pas être considérée comme un système de connaissance qui fonctionne selon une approche cognitiviste (le traitement symbolique de la connaissance, bâti sur la dualité savoir-faire/connaissance), ni selon une approche connexionniste (le traitement en réseau de la connaissance, bâti sur la dualité environnement/connaissance), mais comme un système de connaissance qui fonctionne selon une approche enactiviste (le traitement du flux conversationnel de la connaissance, bâti sur la dualité *autopoïèse*/connaissance).

---

[68] Nous soulignons au passage que les travaux de Alexandre Pachulski (2001), Thierno Tounkara (2002) et Barthelme-Trapp (2003) ont été très influencés par Ermine et Grundstein. Néanmoins, aucun des trois n'a fait appel à la théorie autopoïétique dans la construction de ses approches pour la gestion des connaissances, bien que dans la section référence bibliographiques de leurs thèses respectives, nous trouvons la citation aux ouvrages de Maturana (Biology of cognition) chez Tounkara, et de Maturana et Varela (L'arbre des connaissances) chez Barthelme-Trapp, en plus de quatre ouvrages chez Varela (Autonomie et connaissances, Connaître les sciences cognitives, The Embodied Mind, et Quel savoir pour l'éthique). Enfin, Pachulski fait référence à une seule livre de Varela.

Nous partons du principe qu'un modèle de gestion des connaissances dans et par l'entreprise est imparfaite, autrement dit, la connaissance existe simultanément et nécessairement dans un domaine : *flou* et *non flou*, ce que nous simplifions en disant : *gestion des connaissances imparfaites*.

Le modèle autopoïétique de la gestion des connaissances que nous proposons prend en charge la connaissance issue d'un domaine : *flou* et *non flou*.

L'originalité de cette approche est que nous regardons la connaissance à partir de l'approche enactiviste, et non pas congnitiviste ou connexionniste. A notre avis, cette approche doit rester un canevas de pensée pour la construction du système de connaissance afin de gérer les connaissances dans et par l'entreprise, bien que la construction du système opérationnel associé, soit moins évidente, en fait nous sommes dans une phase de réflexion sur l'utilité et la faisabilité d'un tel projet.

## 3.4.  4 Repères essentiels pour la gestion des connaissances dans l'entreprise

Dans cette section nous présentons sous une forme de "repère"[69], quelque point de réflexion lorsqu'on est confronté à un projet de gestion des connaissances dans l'entreprise.

### ➢ Repère 1

Le premier repère est que la gestion des connaissances est un système qui peut être décrit à travers un système sociotechnique récursif. Où (1) la composante sociale est relative au système d'organisation du travail par avantage collectif décrit à la fois par deux systèmes. L'un est le système cognitif (l'individu) qui est géré par un processus d'apprentissage et d'adaptation à leur espace de travail, et l'autre est le système de connaissances (l'organisation) qui est géré par un processus d'apprentissage organisationnel et d'évolution avec leur environnement au travers de méthodes de travail et de modes de management ; et (2) la composante technique est relative au système opérationnel qui peut être informatisé ou pas à travers les NTIC du KM. En conséquence, la récursivité dans ce système implique que la connaissance est perçue par la composante sociale comme un système d'organisation du travail par avantage collectif, organisé à travers un système de connaissances, et structurer au travers la gestion des connaissances. Par contre, la connaissance est perçue par la composante technique, comme un objet de gestion, organisé à travers l'information, et structurer à travers de données.

---

[69] L'idée est venue à partir du livre de Boyer et Gozlan, intitulé *10 Repères essentiels pour une organisation en mouvement* (publié en 2000) [Boyer et Gozlan, 00].

Dans ce système sociotechnique récursif, nous supposons que les composants existent simultanément et nécessairement dans une dualité organisation/structure (voir chapitre 2, section 2.1.7), dans le sens que la composante sociale s'appuie sur la composante technique, et la composante technique matérialise la composante sociale, ainsi nous avons un couplage structurel du système. Autrement dit, la *gestion des connaissances* est formée par (1) le couplage entre la composante sociale (l'acteur en rapport avec son espace de travail) et le système d'organisation du travail par avantage collectif, c'est-à-dire le couplage entre le système cognitif et le système de connaissances ; et (2) le couplage entre la composante technique (l'acteur en rapport avec les NTIC du KM) et le système d'organisation du travail par avantage collectif, c'est-à-dire le couplage entre le système opérationnel et le système de connaissances. Au plus simplement, d'une part, nous avons l'organisation de la connaissance sur le plan de méthodes de travail et modes de management, et d'autre part, nous avons la structure de la connaissance sur le plan technologique. Ainsi, la récursivité du système sociotechnique nous la trouvons dans la gestion des connaissances par une double existence, d'abord dans la dualité organisation/structure, puis dans la dualité sociale/technique. Autrement dit, dans l'aspect social la gestion des connaissances existe dans la dualité organisation/structure, tandis que dans l'aspect technique la gestion des connaissances existence aussi dans la dualité organisation/structure.

> **Repère 2**

Le deuxième repère est que la gestion des connaissances est un objet de gestion par rapport à dualité organisation/structure, et donc l'aspect social et technique de l'organisation/structure peut être décrit au travers de verbes à l'infinitif. Par exemple gérer l'entreprise dans le sens classique de l'administration des organisations mise en évidence par Henri Fayol, signifie planifier, organiser, diriger et contrôler. Dans cette démarche qui est encore valable aujourd'hui pour gérer l'entreprise à tous ses niveaux (stratégique, tactique, opérationnel), l'on voit bien que la gestion est composée par des verbes qui agissent sur une « chose », ainsi par exemple, le résultat d'organiser la « chose » est son organisation, ce résultat se justifie encore par les travaux d'Edgar Morin dans son livre *La méthode 3. La connaissance de la connaissance* [Morin, 86]. Ainsi, dans la gestion des connaissances, la « chose » est bien la connaissance qu'il faut gérer.

➤ **Repère 3**

Le troisième repère est relative au fait que la gestion des connaissances peut être observé, d'après nous, selon trois approches (1) l'approche organisationnelle de Nonaka et Takeuchi fondée sur le concept *knowledge creating-company* ; (2) l'approche biologique de Maturana et Varela fondée sur le concept de l'*arbre de connaissance* ; et (3) l'approche managériale de Jean-Louis Ermine fondée sur le concept de la *marguerite*. Nous supposons que dans ces trois domaines (organisationnel, biologique et managérial) qu'il est possible (1) d'établir des mécanismes de la *gestion des connaissances* pour chaque approche en particulier, et (2) d'établir un "pont" entre eux, afin d'en dégager une généralisation de la *gestion des connaissances* selon les points de vue sociale et technique. Ces mécanismes nous le trouvons en termes d'un processus évolutif (Barthelme-Trapp), d'un processus dynamique (Ermine) ou d'un processus circulante (Nonaka et Takeuchi, Maturana et Varela) de *gestion des connaissances* que nous caractérisons par des verbes à l'infinitif, tandis que le point en commun nous le trouvons, d'une part, en termes de la cognition et de l'apprentissage au niveau du système d'organisation du travail par avantage collectif, c'est-à-dire au niveau de l'individu (système cognitif) et de l'organisation (système de connaissances), et d'autre part, en termes de la dualité organisation/structure. Cela signifie, que selon l'aspect social, la connaissance dans l'entreprise est perçue, d'une part comme un *système*, organisé à travers un *système de connaissances*, et d'autre part comme un *objet de gestion*, structuré à travers la *gestion des connaissances* (capitaliser, partager, créer), la gestion de l'innovation (créer, produire, offrir), et la gestion des compétences (apprentissage, action). Par contre, selon l'aspect technique, la connaissance dans l'entreprise est perçue aussi, d'une part, comme un *système*, mais organisé à travers un système opérationnel[70] (informatisé ou pas) des NTIC du KM, et d'autre part aussi comme un *objet de gestion*, mais structuré à travers (1) la gestion de la communication, de la coordination et de la coopération entre les acteurs (individu, groupe, entreprise) et les relations du système d'organisation du travail par avantage collectif qui sont nécessaires pour créer des connaissances nouvelles ; et (2) la gestion de l'information et de données pour en extraire la connaissance et créer aussi des connaissances nouvelles pour l'entreprise.

Dans cette hypothèse, le fait de supposer que la connaissance est un système[71], signifie, pour nous, que la connaissance peut être décrit selon le triangle systémique de Le Moigne « la structure, (comment se structurent les objets et les concepts de la connaissance ?) la fonction (dans quelle fonction, quelle activité s'inscrit la connaissance ?) et l'évolution (quel est l'historique de la connaissance ?) ». Et donc, des outils de l'intelligence artificielle en particulier de l'ingénierie des connaissances, de l'ingénierie cognitive, de l'ingénierie et capitalisation des connaissances peuvent être utilisées pour modéliser, justement, la connaissance.

---

[70] Le bons sens indique que la solution informatique fait le trie entre les tâches qui seront automatisés et les tâches que seront manuelles. Pour éviter le chaos d'une usine à gaz. Ermine parle de « la conception d'un système opérationnel de gestion des connaissances » plutôt que de système informatisé ou système informatique.

[71] Selon la méthode MKSM de Jean-Louis Ermine « d'abord la connaissance se perçoit comme un signe, qui contient de l'information (quelle est la forme codée ou perçue du signe que je reçois ?), du sens (quelle représentation l'information engendre-t-elle dans mon esprit ?), et du contexte (quel environnement conditionne le sens que je mets sur l'information reçue ?). Ensuite la connaissance se perçoit comme un système, avec toujours trois points de vue : la structure, (comment se structurent les objets et les concepts de la connaissance ?) la fonction (dans quelle fonction, quelle activité s'inscrit la connaissance ?) et l'évolution (quel est l'historique de la connaissance ?) ».

Par contre, le fait de supposer que la connaissance est un objet de gestion, implique pour nous, que l'information est un objet de connaissance, et alors les données sont un objet d'information[72], et aussi l'existence des outils des NTIC du KM, d'une part, pour la matérialisation du système de connaissances à travers de la communication, de la coordination et de la coopération des acteurs du système d'organisation du travail par avantage collectif impliqués dans la création des connaissances nouvelles, par exemple avec des outils du groupware, et d'autre part, pour l'extraction des connaissances à partir des données. Par exemple, avec (1) des outils de bases de données, tels que le Web mining, le text mining et le data mining qui permettent l'extraction de connaissances à partir du web, du texte, et de données non structurées, respectivement ; (2) des outils de veille qui permettent l'extraction de connaissances à partir de données du marché, c'est le cas notamment des outils du CRM qui permettent l'extraction de connaissances à partir de données clients, des outils de l'ERP qui permettent l'extraction de connaissances à partir de données internes de la chaîne de valeur de l'entreprise, etc.

> **Repère 4**

Le quatrième repère (*last but not least*) est que la *gestion des connaissances* existe dans un plan moral et éthique plus que sociale ou technique. Pour nous les termes *knowledge worker* (ou *knowledge society*), *learning organization* (ou *systems thinking*), *actionable knowledge*, *knowledge-creating company*, *information ecology* (ou *information age*), *knowledge based economy*, *corporate knowledge*, *corporate longitude*, *knowledge-based assets*, etc., forgés par Drucker, Senge, Argyris, Nonaka et Takeuchi, Davenport, Prusak, Prax, Edvinsson, Sveiby respectivement, font parti du bien collective de l'humanité, et non pas des *corporation* du fait que l'on parle de « la création de connaissance collective ». Par exemple, aux Etats-Unis parler de *knowledge worker* des nos jours n'a aucun sens lorsque le coût du salaire d'un *job overseas* se fait par rapport au coût d'un plat de riz capable de maintenir un homme ou une femme dans son poste de travail pendant les 7 jours de la semaine. En effet, l'Amérique 2004 est loin de l'économie d'après guerre (période 50-60) et des *baby-*

---

[72] Pour le Club informatique des grandes entreprises françaises (Cigref http://www.cigref.fr) où l'objectif est de promouvoir l'usage des systèmes d'information comme facteur de création de valeur pour l'entreprise, l'information est objet de connaissance, dans les sens que pour eux l'information « un ensemble de données non structurées et organisées pour donner forme à un message résultant d'un contexte donné ».

*booms* des années 65 et 75, dans lesquels la classe ouvrière américaine n'avait par la peur de voir ses *jobs overseas* comme le fait aujourd'hui. Pour nous la *corporation* comme mode d'organisation du travail et mode de management prend en charge seulement les solutions organisationnelles "gagnant-gagnant" pour eux-mêmes, c'est-à-dire pour les gros patrons d'entreprises et le gouvernement que les appuient, en négligeant les *knowledges workers* et la *knowledge society* de la répartition juste du salaire (le prix du travail). Dans un tel scénario parler de *gestion des connaissances* ou d'innovation organisationnelle, sur un plan moral et éthique, n'aucun sens. A cet égard, Prax a dit « le Knowledge Management correspond ainsi au passage de l'individu à l'organisation pour la recherche d'un bénéfice collectif ».

Pour justifier davantage ce quatrième repère, mais cette fois-ci liée à la création des connaissances, nous avons repéré dans un article récent de la revue *Le journal du management* (d'avril 2004), le concept de *think tank*, ou "groupe d'experts" ou "réservoirs à pensées". Selon l'auteur de l'article Christian Harbulot « les premiers think tanks ont vu le jour aux Etats-Unis au début du siècle dernier. Puis, à la fin des années 40, est née la *Rand Corporation*, le think tank américain le plus connu et le plus important. Créé en pleine guerre froide, il est spécialisé en stratégie militaire … Ils ont alors compris l'importance de la production de connaissances, bien avant la société de l'information »[73].

Les membres du club sont de grands patrons de corporations, des universitaires et chercheurs renommés, unis par le principe "l'union fait la force" afin de (1) réfléchir ensemble sur un point précis dans le but d'atteindre un objectif ; et (2) faire parvenir cette réflexion au gouvernement, tel comme l'a souligné Harbulot « la réflexion débouche sur l'action et doit mener à des résultats, par exemple un projet de loi ». Ce qui reste à savoir c'est le pouvoir d'influence des *think tanks* près des responsables politiques pour faire passer cette loi, et les gagnants de cette loi sur un plan moral et éthique.

**Conclusion du chapitre**

Nous avons présenté dans ce chapitre :

- premièrement, les arguments sur l'évolution de la connaissance dans le modèle de la gestion des connaissances de Ermine, à partir de la relation entre connaissance et environnement, l'organisation et le fonctionnement du modèle a été particulièrement regardé : l'opération de distinction de ce modèle, et la relation entre environnement et organisation, et environnement et patrimoine des connaissances. Ceci a permis de montrer que l'approche de l'enaction de Karl Weick est le fondement théorique du modèle de la marguerite et du modèle OIDC, c'est-à-dire que le modèle de gestion des connaissances de Ermine est basé sur un paradigme de "sélection d'information" dans un système ouvert.

---

[73] L'article complet se trouve dans le site http://management.journaldunet.com/dossiers/040435thinktanks/think_tanks.shtml

- deuxièmement, la démarche antérieure a permis aussi de présenter les arguments sur l'évolution de la connaissance dans le modèle proposé, mais à partir de la relation entre autopoïèse et connaissance, création de sens dans un système clos (clôture opérationnelle), et apprentissage et évolution du système clos (couplage structurel). Ceci a permis de formuler le modèle autopoïétique de la gestion des connaissances imparfaites. Ce modèle a été bâti, d'une part, sur un paradigme "d'émergence de signification" selon l'approche de l'enaction de Maturana et Varela, et d'autre part, sur un élargissement du modèle OIDC, afin d'abriter une réalité de gestion des connaissances imparfaites.

- troisièmement, nous avons présenté les hypothèses de base de l'évolution de la connaissance dans le modèle proposé de gestion des connaissances, à partir de l'hypothèse de l'enaction, l'hypothèse de spontanéité des relations, l'hypothèse du noyau invariant, et l'hypothèse de la connaissance imparfaite. L'hypothèse de l'enaction a permis de supposer que le système de connaissance du modèle proposé est un système clos (selon la dualité organisation/structure), un système vivant (selon l'approche autopoïétique de Santiago), et un système viable (selon l'approche autopoïétique de Valparaiso). Cette hypothèse a permis de supposer que le système de gestion des connaissances imparfaites existe simultanément et nécessairement dans le domaine social (relations humaines) et physique (individus, matière, énergie, et symboles), et il fonctionne avec une clôture opérationnelle. L'hypothèse de spontanéité des relations a permis de supposer que dans le domaine social (relations humaines) il existe une sorte de "spontanéité" (par l'acte de langage ou flux conversationnel chez les individus) pour créer des relations entre processus de production des composants. L'hypothèse du noyau invariant a permis de définir un "noyau invariant" ou "patron d'organisation commun" sur lequel doit graviter le système de connaissance du modèle proposé. L'hypothèse de la connaissance imparfaite a permis d'interpréter l'hypothèse de l'enaction à partir de la théorie de l'imprécis et de l'incertain dans le système opérationnel du modèle proposé.

Dans ce contexte, le modèle proposé a été encadré :

- par rapport à l'aspect social de la gestion des connaissances. Ceci a été fait à travers quatre sous-aspects, afin de décrire les enjeux de la complexité de la gestion des connaissances, à savoir l'aspect dynamique, l'aspect action, l'aspect culturel, et l'aspect chaos de la gestion des connaissances. L'aspect dynamique de la gestion des connaissances a permis d'intégrer dans une dynamique circulante (implication mutuelle) la gestion des connaissances, la gestion de l'innovation, ainsi que la gestion des compétences de l'entreprise. Autrement dit, la capacité de l'entreprise de capitaliser, partager, créer… la connaissance (gestion des connaissances), de créer, produire, offrir… de nouveaux produits ou services (gestion de l'innovation), d'apprentissage, d'action… de ressources humaines (gestion des compétences) sont un même enjeu pour l'entreprise. L'aspect action de la gestion des connaissances a permis d'établir une relation entre la connaissance et son application dans l'action (gestion des compétences) de l'entreprise. L'aspect culturel de la gestion des connaissances a permis de mettre en valeur le rapport à la structure sociale comme le véritable responsable du succès de l'intégration de la gestion des connaissances au sein de l'entreprise. L'aspect chaos de la gestion des connaissances a permis de mettre l'accent sur les contraintes externes et internes face à un projet de gestion des connaissances dans l'entreprise. En effet, les problématiques de gestion telles que CRM[74] et SCM (au niveau de contraintes externes) et ERP (au niveau de contraintes internes) en sont la preuve. Nous voulons dire par là, que la gestion des connaissances peut être aussi associée à un discours marchand, et beaucoup de sociétés de conseils l'ont assez bien compris depuis fort-long temps[75].

Nous pensons que ces différents aspects sont de véritables paramètres génériques pour analyser et positionner une problématique de gestion des connaissances dans l'entreprise.

- par rapport à l'aspect technique de la gestion des connaissances que nous avons fait à travers quatre systèmes opérationnels de la gestion des connaissances, à savoir : les SBC issus de l'approche cognitiviste, les SBC issus de l'approche connexioniste, et les SBC issus de l'approche enactiviste. Cette classification des outils a permis de montrer le retard technologique par rapport à la gestion des connaissances dans un paradigme "d'émergence de signification" et non pas de "sélection d'information". Par conséquent, le défi est lancé pour le développement de nouvelles théories de l'imprécis et de l'incertain afin de décrier le paradigme de "l'émergence de signification". En effet, les outils d'aujourd'hui pour maîtriser l'imprécis et l'incertain de la connaissance, comme nous le verrons dans le chapitre 4, ont été développés pour les systèmes à base de connaissance en suivant une approche cognitiviste[76], et non pas enactiviste comme l'exige le modèle autopoïétique de la gestion des connaissances.

---

[74] CRM *Customer Relationship Management* (système de gestion de la relation client de l'entreprise). SCM *Supply Chain Management* (système de gestion de la chaîne logistique de l'entreprise). ERP *Enterprise Ressource Planning* (système de gestion de la chaîne valeur de l'entreprise).

[75] Notre intention, n'est pas de faire de la polémique dans cette thèse, mais nous devons souligner que nous avons voulu retenir ces discours marchands, tout au début du chapitre 1 (Qu'est-ce que la gestion des connaissances ?), pour bien sentir cet enjeu commercial. Cet enjeu a été matérialisé davantage grâce à l'approche de NTIC du KM. Les auteurs qui font partie de sociétés de conseils sont : Wendi Bukowitz et Ruth Williams de la société PricewaterhouseCoopers, Karl Sveiby et Leif Edvinsson de la société Sveiby, René-Charles Tisseyre de la société Capgemini, Jean-Yves Prax de la société CorEdge, et Jean-Louis Ermine du Club Gestion des Connaissances.

- enfin quatrièmement, nous avons positionné une problématique de gestion des connaissances dans l'entreprise à partir de quatre repères essentiels. Le premier repère a mis en évidence le fait que la gestion des connaissances est un système capable d'être décrit à travers un système sociotechnique récursif. Le deuxième repère a mis en évidence le fait que la gestion des connaissances est un objet de gestion par rapport à la dualité organisation/structure. Le troisième a mis en évidence le fait que la gestion des connaissances peut être observée, selon trois approches : l'approche organisationnelle de Nonaka et Takeuchi, l'approche biologique de Maturana et Varela, et l'approche managériale de Jean-Louis Ermine. Le quatrième repère a mis en évidence le fait que la gestion des connaissances existe dans un plan moral et éthique plus que social ou technique.

La conclusion générale de ce chapitre est que le modèle autopoïétique de la gestion des connaissances imparfaites est fondé sur un principe fondamental : la réalisation simultanément et nécessairement de trois dualités dans et par le système de connaissance, à savoir :

- la première est la dualité organisation/système ; cette dualité qui prend ses racines dans l'approche cybernétique (de deuxième ordre), permet l'identification de l'unité dans et par une totalité. Ainsi, le système de connaissance est un système vivant (caractérisé par l'unité et l'identité)[77] et viable (caractérisé par l'autonomie) ;

- la deuxième est la dualité organisation/structure ; cette dualité qui prend ses racines dans l'approche autopoïétique, permet l'application d'une opération de distinction. Ainsi, le système de connaissance existe sous un domaine : social (processus d'organisation) et physique (processus de structuration). En d'autres termes, le système de connaissance est décrit (1) selon l'organisation sociale de la structure des relations entre processus de production des composants du système ; (2) selon la structure physique de l'organisation du système, c'est-à-dire par la matérialisation dans l'espace physique de propriétés ou attributs de composants et leurs relations ;

---

[76] La preuve de cette argumentation se trouve, à mon avis, dans deux thèses et un livre. La thèse de Dubois, intitulée *Modèles mathématiques de l'imprécis et de l'incertain en vue d'application aux techniques d'aide à la décision* (publiée en 1983), et la thèse de Prade, intitulée *Modèles mathématiques de l'imprécis et de l'incertain en vue d'applications au raisonnement naturel* (publié en 1982), ainsi que dans l'ouvrage de ces deux auteurs, intitulé *Théorie des Possibilités - Applications à la représentation des connaissances en informatique* (publié pour la première fois en 1985, puis en 1988).

[77] Autrement dit, un système est "intelligent" s'il est capable, d'une part, d'être reconnu comme unité, et d'autre part, de maintenir l'identité de son unité.

- la troisième est la dualité organisation/environnement ; cette dualité qui prend ses racines dans l'approche système, permet le maintient de l'enaction du système, autrement dit, la création des connaissances nouvelles et d'apprentissage organisationnel relatives à un même processus (noyau invariant), qui a pour objectif la mise sur le marché d'un nouveau produit ou service, qui existe simultanément et nécessairement pour la satisfaction d'un besoin réel et le maintient de la totalité comme une unité.

| Chapitre 4 | Le système (connaissance et opérationnel) dans un environnement imprécis et incertain |
|---|---|

*Le langage de l'homme comme la pensée est flou et/ou logique*
*A. Kaufmann, mathématicien belge*

Dans le chapitre 3 nous avons fait l'hypothèse qu'une relation existe entre autopoïèse et connaissance, ce que nous a permit d'intégrer le modèle de gestion des connaissances de Ermine (système de connaissance : modèle de la marguerite et système opérationnel : modèle OIDC) avec l'approche de l'enaction de Maturana et Varela dans une dualité organisation/structure.

L'*hypothèse de la connaissance imparfaite dans le modèle autopoïétique de la gestion des connaissances*, nous permet d'interpreter l'approche de l'enaction (faire-émerger) à travers la connaissance imparfaite, et de proposer un modèle appelé *modèle autopoïétique de la gestion des connaissances imparfaites*.

L'objectif de ce chapitre 4 est d'une part, donner un cadre social de l'imprécis et de l'incertain pour le système de connaissance, et d'autre part, donner un cadre technique de l'imprécis et de l'incertain pour le système opérationnle, ceci est fait par la prise en charge du flou sous l'angle des bases de données relationnelles floues, par le biais de la représentation des objets flous (traitement de données imprécises), et l'interrogation des événements flous (traitement de requêtes floues). Donc, la gestion des connaissances est étendue aux données imprécises et incertaines, dans l'espoir de faire une extraction des connaissances à partir de ces données.

Ce chapitre 4 est organisé en cinq parties. La première partie appelée « Le système de connaissance : Aspect social de la gestion des connaissances » traite du problème de la représentation de connaissances imparfaites dans un système de connaissance à travers des modèles, en tant que moyen d'appréhender le monde réel (flou et non flou), de le représenter et d'en utiliser son image comme un outil de communication et de raisonnement dans l'entreprise. Pour cela, nous observons l'activité sous l'angle des modèles, tout d'abord à partir d'une image fidèle d'une certaine réalité (et donc d'une recherche pour maintenir une précision dans la représentation des objets et des événements), afin que la consultation de cette image soit certaine. Ensuite nous abordons la problématique de l'imperfection de cette connaissance sous l'angle de bases de données relationnelles floues (BDRF), en introduisant des objets flous, des événements flous, des requêtes floues. La deuxième partie appelée « La théorie de l'imprécis et de l'incertain » traite cette théorie sous trois aspects: (1) la théorie des sous-ensembles flous, (2) la théorie des possibilités, et (3) le filtrage flou. L'originalité de notre approche est que nous expliquons la modélisation mathématique des mesures de possibilité et de nécessité sur un même exemple, que nous utilisons pour expliquer la notion de filtrage flou. Nous soulignons que ce n'est pas l'objectif de ce chapitre de parcourir les divers chemins théoriques et applicatifs de cette théorie, mais plutôt de mettre en évidence sa contribution applicative dans le domaine de l'ingénierie des systèmes d'information. La troisième partie appelée « Les applications de la théorie de l'imprécis et de l'incertain dans l'industrie » essaye de clarifier les concepts de l'imprécis et de l'incertain (autrement dit de l'imperfection de l'information), à travers l'usage de cette théorie dans un certain nombre d'applications, principalement dans l'industrie manufacturière (retenu dans le cadre de cette thèse). Nous nous référons au système F-MRP (Fuzzy-Manufacturing Ressource Planning) [Reynoso, 04]; le système SIMCAIR (Système d'Interrogation Multi-Critères Avec Importances Relatives) [Andres, 89]; le système TOULMED [Buisson, 87]; le système FLORAN (Filtrage flou et Objets pour Raisonner par Analogie) [Salotti, 92]; le système DOLMEN (DéfectOLogie et Mémoire d'Entreprise) [Simon, 97] ; et le système FSQL (Fuzzy Structured Query Language) [Galindo, 99]. Ce dernier système a été choisi, dans le cadre de cette thèse, pour la représentation et l'interrogation des données dans une base de données relationnelles floues (BDRF) afin d'en extraire des connaissances sur un terrain industriel du domaine de la manufacture du carton. La quatrième partie appelée « Représentation et interrogation des données floues dans FSQL » présente les outils FSQL pour la modélisation de données imprécises et incertaines dans une base de données relationnelles floues (BDRF). La cinquième partie appelée « Le système opérationnel : Aspect technique de la gestion des connaissances étendues au flou » présente l'aspect technique de la gestion des connaissances, en particulier le système opérationnel auquel nous nous sommes intéressés, ainsi que l'argumentation de problématiques industrielles liées à la manufacture du carton, domaine où l'imprécis et l'incertain est présent dans la dynamique du processus.

## 4.1    Le système de connaissance: Aspect social

➢    **Les connaissances imparfaites comme un modèle**

L'être humain perçoit, pense et raisonne à travers des modèles. En effet, comme le dit Paul Valéry « nous ne raisonnons que sur des modèles », cela signifie que le monde réel est perçu, pensé et raisonné sur la base d'images de la réalité, et que ces images sont l'outil de tout processus d'information, de cognition et de décision. Or, d'après Hebert Simon « tout, dans la nature, peut être organisé par niveaux » [Simon, 91], cela signifie qu'en général ces modèles sont empilés les uns sur les autres, selon des lois de modélisation des systèmes complexes qui régulent la complexité de chaque niveau de représentation de la réalité [Le Moigne, 90]. Par exemple, dans un contexte de conception de systèmes d'information pour l'industrie, un modèle est un langage de communication tout d'abord entre les acteurs humains de l'entreprise (ce que l'on veut faire au niveau du *software* de l'ordinateur), et puis entre les composants de la machine (ce que l'on peut faire au niveau du *hardware* de l'ordinateur). En guise d'exemple, citons les quatre niveaux d'abstraction de la méthode MERISE pour la conception de systèmes d'information : niveau conceptuel, niveau organisationnel, niveau logique et niveau physique [Nanci *et al*, 96], et les sept couches du modèle OSI : application, présentation, session, transport, réseau, liaison et physique [Mourier, 96]. Dans le premier cas, on voit que le modèle joue un rôle de moyen de communication entre les différents acteurs du processus de développement d'un logiciel aux niveaux du modèle conceptuel, modèle organisationnel, modèle logique et modèle physique, tandis que dans le deuxième cas le modèle joue un rôle de moyen de communication entre les différents composants d'un protocole standard (c'est-à-dire l'interface) de communication entre ordinateurs.

➢ **Penser et raisonner en termes flous**

Le paradigme simonien fait une analogie entre l'être humain et l'ordinateur, dans le sens que l'ordinateur fonctionne comme un système de symboles [Simon, 91], c'est-à-dire que l'ordinateur est une machine pour stocker, traiter et manipuler en entrée et sortie des symboles qui représentent une certaine réalité humaine. Nous constatons que le vrai problème se pose ici, d'une part, au niveau de l'*artificialité* qui intervient dans les systèmes complexes et de son interrelation avec son environnement pour accomplir un projet téléologique (ce qui nous renvoie au domaine de l'intelligence artificielle), et d'autre part au niveau de la capacité de l'être humain à penser et raisonner en termes *flous* (ce qui nous renvoie à la théorie de sous-ensembles flous). En effet, si l'on me en œuvre le triangle systémique de Le Moigne, nous pouvons dire que l'être humain, et en général tout être vivant, doit fonctionner en accommodant ses structures au changement lié à l'évolution de son environnement à partir d'une volonté d'action, c'est-à-dire de faire quelque chose, tandis que dans le cas de l'ordinateur, sa structure, c'est-à-dire le hardware, est matérialisée par la logique binaire, son fonctionnement, alors que le software est matérialisé par une programmation séquentielle, orientée (objet ou autre), et donc l'évolution est guidée par l'innovation technologique du hardware et/ou du software.

En résumé, le comportement de l'ordinateur est géré par un raisonnement booléien soit au niveau du hardware, soit au niveau du software, tandis que le comportement de l'être humain est géré par un raisonnement plus ou moins certain et précis. Comme l'a dit Kaufmann dans son ouvrage *Introduction à la théorie des sous-ensembles flous* « le langage de l'homme comme sa pensée est flou et/ou logique » [Kaufmann, 77], et puis il rajoute « nos modèles sont flous, notre pensée formée de modèles plus ou moins indépendants est floue, nous sommes tellement différents d'un ordinateur! Un ordinateur est une machine non-floue par définition. Mais l'homme possède, en plus de la faculté de prise en compte et de traitement logique, la prise en compte globale ou parallèle, comme tous les êtres vivants. Cette prise en compte globale ou parallèle, à l'encontre de la prise en compte logique, est floue et elle doit être floue. L'être vivant doué d'une possibilité d'initiative, perçoit et traite une information plus ou moins floue et s'adapte ». De plus, il s'interroge sur le fait « sera-t-il possible de traiter des problèmes flous à l'aide d'ordinateurs qui sont des machines séquentielles à logique binaire ? ». Dans ce qui est de notre point de vue, Nous avons constaté qu'un grand nombre de recherches sur la logique floue ont abouti à une réponse positive à cette question :

- au niveau du software, il est possible de simuler le flou dans le modèle conceptuel et le modèle logique (par exemple la gestion de la date d'arrivée d'une commande d'un client autour de la fin du mois) ;

- au niveau du modèle physique (par exemple la surveillance d'une panne avec le système opératif Windows) bien que dans l'industrie on peut trouver des circuits logiques flous sur des machines à laver et des appareils de climatisation, et en fait, ce sont les Japonais qui en premiers ont utilisé la "fuzzy technology" avec un réel succès.

En pratique le problème de la conception d'un système opérationnel qui permet de stocker, de traiter et de consulter des informations incomplètes, imprécises et incertaines a été résolu, au niveau physique, par une extension du langage SQL. Cela signifie qu'au niveau physique il y a une voie de recherche chez les constructeurs des bases de données relationnelles pour appliquer la logique floue.

➢ **Les images des objets et des événements**

L'entreprise perçoit et raisonne la réalité non floue et floue (incomplète, imprécise et incertaine) à travers des systèmes d'informations. Pour donner une idée générale sur la signification du terme "système d'information", l'organisme américain de codification des systèmes de bases de données CODASYL donne la définition (plutôt la recommandation) suivante « un système d'information traite des objets et des événements du monde réel intéressant les utilisateurs de ce système d'information. Ces objets réels et événements, appelés entités, sont représentés dans le système par des données. L'information relative à une entité particulière se présente sous la forme de valeurs qui décrivent quantitativement et/ou qualitativement un ensemble d'attributs qui ont une signification dans le système » [Golvers, 90]. Dans ce contexte, l'objet est "ce qui existe entre deux événements", par exemple dans une réalité de gestion, l'événement "livraison" clôture l'objet "commandes en cours" (qui cesse d'être en cours dès qu'elle est arrivée) et génère l'objet de "livraison en cours". Cela signifie que les systèmes d'information représentent une réalité de gestion sous forme d'objets de gestion et d'événements de gestion, mais aussi sous forme de règles de gestion (qui vont gérer la création et la clôture des objets par des événements) dont le but est de maintenir une image d'une réalité de gestion qui peut être consultée à tout moment pour pouvoir agir, prévoir ou savoir. En conséquence, l'entreprise perçoit et raisonne à travers des systèmes d'information qui sont l'image fidèle d'une certaine réalité de gestion.

Afin d'illustrer la précision et la certitude des systèmes d'information, supposons qu'on veuille consulter un système de gestion de bases de données relationnelles pour savoir avant de faire la livraison, les produits et les quantités commandées par un client à une usine de l'entreprise à un moment donné de la semaine. Dans un scénario certain et précis, la consultation de la base de données peut nous donner par exemple la réponse suivante "le jeudi 13/02/03 à 15h30, M. Dupont a passé une commande de 350 tonnes d'un produit X". Maintenant si nous nous plaçons dans une optique de prévision, grâce à l'information passée du client donné par une série chronologique [Jiménez, 95] nous pouvons avoir la réponse suivante "autour de la mi-février M. Dupont espère passer une commande entre 250 et 400 tonnes d'un produit X". Nous constatons que dans le premier cas le résultat de la consultation à la base de données est donné par un chiffre (une valeur crisp) tandis que dans le deuxième cas, la réponse est donnée, d'une part par une étiquette linguistique "autour de", et d'autre part par une tranche de valeurs possibles. Ce dernier scénario, est un scénario incertain et imprécis. Nous pouvons avoir aussi un scénario incomplet, dans ce cas la réponse est incomplète, par exemple la base d'information a stocké la quantité d'un produit commandé par M. Dupont, mais on n'y a pas enregistré la date et l'heure de la commande. Nous reviendrons au paragraphe 4.4 sur la représentation et l'interrogation du traitement de l'information incertaine, imprécise voir incomplète dans les bases de données relationnelles. Enfin nous signalons que la problématique non floue de traitement et de manipulation de l'information a été résolue de façon commerciale par les constructeurs des systèmes de gestion de base de données depuis longtemps, par exemple le SGBD Oracle est largement utilisé dans l'industrie, mais le contexte flou d'une problématique de gestion a été abordé seulement d'une manière expérimentale. Une possible explication de ce fait peut se trouver dans l'évolution de l'informatique de gestion dans l'entreprise.

➢ **Les systèmes d'information de gestion**

Problématique organisationnelle : les relations humaines

Dans les années 60 l'informatique de gestion, c'est-à-dire "les systèmes d'information pour la gestion", a pénétré l'industrie pour gérer tout d'abord les bases de données de l'entreprise, c'est-à-dire les objets et les événements de gestion autour des tâches routinières et administratives de son activité de production. Ensuite pendant les années 70 l'informatique de gestion a géré les bases d'informations de l'entreprise autour des clients, des fournisseurs, etc. Depuis les années 80 l'informatique de gestion a géré les bases de connaissances de l'entreprise, tout d'abord avec les systèmes experts pour automatiser les tâches cognitives de l'expert, et plus tard les systèmes à base de connaissances pour gérer un domaine de connaissance du savoir et du savoir-faire produit (savoir-faire nouveau) et consommé (savoir-faire acquis) pendant un cycle de production d'un bien ou d'un service de l'entreprise en interaction avec son environnement. Par exemple, à travers un système à base de cas, on peut ajouter un cas nouveau mais il n'y a aucune garantie de trouver une solution innovante dans ce cas. Or, cette évolution de la gestion de l'entreprise par l'outil informatique, c'est-à-dire l'ordinateur, a été marquée au cours du temps par plusieurs paradigmes au niveau humain, organisationnel et technologique de l'entreprise. Au niveau technologique, on a par exemple, l'informatique centralisée ou distribuée, l'informatique client-serveur classique ou world wide web. Au niveau organisationnel, on trouve par exemple les phénomènes de veille stratégique, et au niveau humain on peut citer le travail coopératif assisté par ordinateur. Au carrefour de ces trois niveaux, on trouve le système d'information (I) qui informe au système de décision (D) les activités en entrée et en sortie du système opérant (O) de l'entreprise afin de prendre une action, telle qu'elle a été mise en évidence par Jean-Louis Le Moigne dans la modélisation des systèmes complexes (modèle OID). Dans la modélisation des systèmes d'information, l'analyse conceptuelle essaye, d'une part de délimiter la partie du monde réel à laquelle s'intéressent les futurs utilisateurs du système d'information, et d'autre part de préciser les objets de gestion constitutifs de cette partie du monde réel, ainsi que les événements qui les modifient suivant des règles de gestion [Golvers, 90]. Cette délimitation et précision du monde réel a été recherchée à partir de la théorie des ensembles et le raisonnement booléen. En conséquence, il faut une évolution de l'informatique de gestion vers la gestion de l'imperfection de l'information dans l'entreprise, c'est-à-dire dans des *bases de données*. En effet, d'après Gardarin « les bases de données ont pris aujourd'hui une place essentielle dans l'informatique, plus particulièrement en gestion » [Gardarin, 99].

Problématique technologique : bases de données relationnelles

Gardarin a une définition passe-partout du type « une base de données est un ensemble qui modélise les objets d'une partie du monde réel et sert de support à une application informatique » [Gardarin, 99]. Une autre définition générale est donnée par Bosc « les bases d'informations (base de données, bases de connaissances …) sont des systèmes utilisés pour représenter les croyances et connaissances que l'on peut avoir sur le monde réel ou tout au moins, sur une partie du monde réel. Cette représentation doit être le plus fidèle possible afin que la seule interrogation de ce système permette à son utilisateur de se faire une idée la plus exacte possible et de prendre des décisions adéquates » [Bosc *et al*, 02]. En termes pratiques, les bases de données relationnelles sont des ensembles de *tables* (dans le sens de la théorie des ensembles), une table étant un ensemble de *tuples* (lignes d'une table) de valeurs, chacune des ces valeurs appartenant à des domaines de valeurs bien définies et éventuellement restreintes par des contraintes d'intégrité, que l'on appelle *attributs* (colonnes d'une table). Or, les 8 opérations de l'algèbre relationnelle (projection, sélection, jointure, division, union, intersection, différence, produit) [Codd, 70] ont tous un opérateur spécifique du SQL (Structured Query Language), sauf la division relationnelle où il faut l'imbrication de plusieurs SELECT. Un SELECT est une *requête* dans le langage SQL. Une *requête* SQL a la syntaxe suivante :

```
SELECT liste des colonnes (attributs)
FROM liste des tables (relations)
WHERE liste des lignes (conditions de la tuple)
```

**Figure 4.1 :** Requête SQL

Dans le cadre des bases de données floues et afin d'alléger cette thèse, nous allons prendre la description suivante :

Bases de données floues = Bases de données + Théorie des sous-ensembles flous

Pour le système opérationnel nous proposons de faire l'analyse conceptuelle d'une base de données floues à partir de l'outil FuzzyCase [Urrutia, 03] puisqu'il permet la construction d'un modèle conceptuel EER (Enhanced Entity Relationship) flou, c'est-à-dire la notion de flou a été considérée sous trois aspects: (a) dans la spécification des attributs; (b) dans la spécification des valeurs de ces attributs; et (c) dans la spécification des contraintes de ces attributs, pour représenter les objets de gestion (données, informations et connaissances) et les événements de gestion du monde réel. Ceci nous l'avons étudié dans [Urrutia *et al*, 01a] et [Urrutia *et al* ,02].

> **L'imperfection de l'information sous l'angle des bases de données relationnelles floues**

L'imperfection de l'information est prise en compte par la théorie de représentation de l'imprécision et de l'incertitude (ou théorie de l'imprécis et de l'incertain) composée de la théorie des sous-ensembles flous, la théorie des possibilités (développés par Zadeh à partir des années 60 et 70 respectivement), les logiques floues et possibilistes, … La théorie de l'imprécis et de l'incertain est devenu ces dernières années un domaine de recherche reconnu. La preuve est qu'un nombre important de travaux scientifiques ont été discutés dans des livres, articles, colloques, conférences et un nombre croissant de thèses ont été consacrées au sujet. Le domaine d'application classique de cette théorie est vaste. A titre illustratif nous citons les langages flous, les bases de données flous, les automates flous, les algorithmes flous, le contrôle flou, les problèmes de décision dans un univers flou, la reconnaissance de formes, les problèmes de classement et de sélection, la recherche documentaire, … Parmi les applications nouvelles on trouve par exemple la prise en compte de l'incertitude dans les projets innovants [Bougaret, 02] l'incertitude de mesure pour la compréhension des résultats mesurés [Mauris, 00], [Ioannou *et al*, 02] la prise en compte de l'incertitude dans certains outils d'ordonnancement classiques [Letouzey, 01], [Grabot *et al*, 02], [Reynoso, 04].

Dubois et Prade ont étudié l'imperfection de l'information à travers des modèles logico-mathématiques de l'imprécision et de l'incertitude [Dubois et Prade, 85, 88], en se basant sur la théorie des sous-ensembles flous et sur la de théorie des possibilités (développés par Zadeh à partir des années 60 et 70 respectivement). L'imprécis a été modélisé sous la forme d'ensembles flous et l'incertain a été modélisé sous la forme d'une mesure de possibilité. Dans ce contexte, « l'imprécis et l'incertain peuvent être considérés comme deux points de vue antagonistes sur une même réalité qu'est l'imperfection de l'information » [Dubois et Prade, 85, 88]. Pour ces auteurs l'imperfection de l'information peut être produite selon trois points de vue équivalents: structure, contenu et événement. En effet, « on a trois façon équivalentes d'envisager un ensemble d'informations, selon que l'on met l'accent sur la structure (le point de vue de la logique), le contenu (le point de vue ensembliste) de ces informations, ou leur relation à des faits réels (le point de vue événementiel) ». L'imperfection de l'information est représentée, d'une part, sous la forme d'une proposition logique comportant des prédicats (et éventuellement des quantificateurs) pour indiquer l'imprécision de l'information, et d'autre part sous la forme des événements relatifs à la validité de ces prédicats pour indiquer l'incertitude de l'information. Dans un article récent [Bosc *et al*, 02] ces auteurs, au lieu de parler de l'imperfection de l'information, ont préféré utiliser (dans des bases de données) le terme "information incomplète" comme terme de référence pour les informations imprécises, floues, incertaines, vagues, nulles et incomplètes. Nous pouvons appliquer le triangle systémique de Jean-Louis Le Moigne à l'imperfection de l'information en termes de proposition, prédicat et événement pour fixer la place de l'information incomplète vis à vis l'information incertaine et imprécise. Nous constatons à ce moment là que :

- du point de vue de la structure (information imprécise) l'imperfection de l'information est représenté sous la forme d'une proposition logique, ses prédicats et ses quantificateurs. Ensuite, du point de vue de la fonction (information incertaine) l'imperfection de l'information est représentée à travers la validité des événements relatifs aux prédicats de la proposition logique ;

- du point de vue de l'évolution (information incomplète) l'imperfection de l'information est représentée par des états de la base de données en termes de complétude par rapport au temps, et donc l'imperfection de l'information correspond à l'imperfection de sa représentation structurelle, fonctionnelle ou évolutive, ou bien par le mélange de ces trois caractéristiques au niveau du contenu et la consultation de la base de données.

Dubois et Prade proposent de représenter l'imperfection de l'information dans une base de données relationnelles comme un quadruplet [Dubois et Prade, 85, 88] :

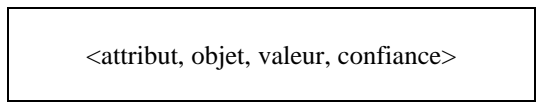

<attribut, objet, valeur, confiance>

**Figure 4.2 :** Modèle de Dubois et Prade pour l'imperfection de l'information

Dans ce modèle l'objet correspond à une ligne (ou tuple) de la relation (ou table), l'attribut correspond à une colonne de la relation, la valeur est à l'intersection ligne et colonne, et la confiance (autre colonne de la relation) est un degré (une valeur réelle de l'intervalle [0,1]) pour indiquer la fiabilité d'appartenance à un ensemble flou. Par exemple, supposons qu'on ait la proposition suivante: "la température de la machine est normale", avec un prédicat "135ºC", et un événement "X = savoir si une température de 135ºC est plus ou moins normale". Alors, cette situation peut être modélisée par le quadruplet <température,machine,normale,confiance(X=135 sachant que la température est normale)=0.7>. Nous constatons donc que la valeur de la température est imprécise. En effet, dans cet exemple, on a 70% de possibilité qu'une température de 135ºC appartient à l'ensemble flou "normal". Autrement dit, il y a 30% d'incertitude sur la vérité de cette proposition. Nous reviendrons au paragraphe 4.2.2 sur les mesures de possibilité et de nécessité de la théorie de l'incertain et la définition de l'étiquette linguistique "normale". Dans un contexte général de la figure 4.2 d'après Dubois et Prade « l'imprécis concerne le contenu de l'information (la composante "valeur" du quadruplet) tandis que l'incertain est relatif à sa vérité, entendue au sens de sa conformité à une réalité (la composante "confiance" du quadruplet ». En conséquence, nous constatons tout d'abord que cette représentation est plus riche que la représentation classique de l'information dans une base de données relationnelle (triplet <attribut, objet, valeur> [Gardarin, 99]). Si l'on revient à l'exemple précédent, la triplette <température,machine,135> permet seulement de savoir avec précision (température = 135ºC) si oui ou non la machine fonctionne ou si elle est en panne, et donc la confiance dans ce cas peut être représentée par oui ou par non (ou par vrai ou par faux, ou par 1 ou par 0), c'est-à-dire on a la quadruplet <température,machine,135,confiance(X=135)=1 ou 0>. Cela signifie que dans cette représentation, il n'y a pas de subjectivité, par exemple qu'il est impossible de répondre par vrai ou faux à la question "la température est-elle normale ?", et nous ne pouvons pas représenter, par exemple la notion de température normale et de température anormale sur un référentiel ordonné dans laquelle la machine ne sera plus en état d'assurer correctement sa fonction. En général, cette différence peut être associée à une différence entre une représentation ensembliste classique et une représentation ensembliste floue, et peut être justifiée par ce qu'a dit Zadeh au sujet de la logique du raisonnement humain dans la préface de l'ouvrage de Kaufmann « l'aptitude du cerveau humain (une aptitude que les actuels ordinateurs ne possèdent pas) à penser et à raisonner en termes imprécis, non-qualitatifs, "flous" » [Kaufmann, 77]. Ensuite nous constatons que la représentation de la figure 4.2 est commode car elle permet de visualiser l'imperfection de l'information comme une *ontologie*, c'est-à-dire comme un référentiel (ou structure) unique du domaine où l'on peut mesurer la distance et le recouvrement entre les concepts [Gardarin, 99] et donc, mettre en relation quelques qualificatifs de l'incertain et l'imprécis. En effet, d'une part la notion de mesure de distance peut être associée à la notion de mesure de l'incertitude (qui pour [Fabiani, 96] a été étudiée sous les aspects de l'ignorance, de la confiance, voire même de l'imprécis), et d'autre part la notion de recouvrement peut être associée à la notion des opérations ensemblistes ($\cap, \cup, \subseteq$, etc.) entre des ensembles flous dans la spécification des attributs et de leurs valeurs, et donc finalement l'imprécis peut être qualifié de vague ou flou, avec la notion de limites indéterminées ou d'absence de contour bien délimité entre des ensembles flous [Dubois et Prade, 85, 88].

Une autre forme d'imprécision est l'ambigüité qui correspond à une forme d'imprécision liée au langage [Dubois et Prade, 85, 88], [Mottro, 94]. Par exemple, une température de 135° peut être en °C ou °F, et donc l'ambigüité vient bien du fait que le référentiel n'est pas precisé.

## 4.2 La théorie de l'imprécis et de l'incertain

On vient de voir que Dubois et Prade (voir figure 4.2), définissent l'imperfection de l'information comme un quadruplet <attribut, objet, valeur, confiance>, où l'imprécis renvoie à la valeur de l'information, tandis que l'incertain se rapporté à la confiance. Autrement dit, l'imprécision est relative à la valeur de l'attribut de l'objet de la base de données, tandis que l'incertitude est applicable aux événements ou occurrences qui agissent sur l'objet. On parle alors d'une situation incertaine ou flou lorsque les paramètres (attributs) qui la définissent sont inconnus ou mal connus.

Zadeh a étudié le flou dans le cadre d'un modèle de l'incertain [Zadeh, 78]. Dans ce contexte, l'incertain (ou son synonyme l'incertitude), est perçu comme l'occurrence de deux événements, l'un est plus ou moins *possible* et l'autre est plus ou moins *certain*. Pour chaque événement et son contraire, on calcule finalement un degré de possibilité. Dans ce cas, le complément à *1* du degré de possibilité de l'occurrence d'un événement est interprété comme un degré de nécessité (ou certitude) de l'événement contraire.

La théorie des sous-ensembles flous et des possibilités a été imaginée par des mathématiciens (Zadeh, Kaufmann, Dubois, Prade,…) et utilisée par des ingénieurs au fils des années. Pendant les années 90 le Japon a introduit la "fuzzy technology", pour exprimer d'une façon non probabiliste, les perceptions subjectives du réel, c'est-à-dire, l'incertain. En fait, l'incertain a été étudié toujours selon une approche probabiliste, car l'incertain est lié au caractère aléatoire d'un résultat[1], où face

---

[1] Pour Bonnal « l'incertain est lié au caractère aléatoire d'un résultat, et à ce titre revêt une dimension probabiliste ; l'imprécis est lié au caractère imparfaitement défini d'un résultat, et il est de nature déterministe ». Dans le cadre de la planification d'un projet industriel, il rajout « on a fait aussi trop souvent l'amalgame entre deux concepts qui sont vraiment distincts : l'imprécis et l'incertain. Le premier revêt une dimension purement déterministe pour laquelle les acteurs disposent de moyens d'actions. Le second renvoie à une dimension aléatoire, donc effectivement probabiliste, pour laquelle les acteurs n'ont aucun moyen d'action ». Puis il ajoute « y a-t-il unanimité pour considérer qu'entre incertitude,

aux événéments on a aucune information (incertitude totale) ou une information partielle (incertitude partielle) pour agir.

Dans cette section, la théorie de l'imprécis et de l'incertain est étudiée sous trois aspects: la théorie des sous-ensembles flous, la théorie des possibilités et le filtrage flou.

## 4.2.1    Théorie des sous-ensembles flous

La théorie des sous-ensembles flous a été proposée par Zadeh dès 1965 à partir du concept d'ensemble (c'est-à-dire une collection d'objets[2]). Cette théorie remet en cause la notion d'ensemble, dans le sens que dans la théorie classique des ensembles, la notion d'appartenance est liée à une valeur binaire 1 ou 0 (un objet est, ou bien dans l'ensemble, ou bien en-dehors). L'idée centrale de la réflexion de Zadeh porte sur le fait qu'il existe des ensembles qui se définissent au travers d'attributs subjectifs, dans lesquels il n'est pas clairement déterminé si un élément appartient ou non à l'ensemble. En effet, comme il dit dans la préface de l'ouvrage de Kaufmann « nous avons mis longtemps à réaliser que beaucoup, ou la presque totalité du savoir et de l'action réciproque des humains avec le monde extérieur, implique des constructions abstraites qui ne sont pas des "ensembles" dans le sens classique du mot, mais plutôt des "ensembles flous" (ou "sous-ensembles flous"), c'est-à-dire des classes aux limites indéterminées, dans lesquelles la transition d'appartenance à non-appartenance est plutôt graduelle que brusque » [Kaufmann, 77]. Afin d'illustrer ceci, nous allons commenter l'exemple donné dans [Villiger, 01]. L'ensemble des personnes grandes est un ensemble flou, parce que la valeur du seuil (une taille limite) à partir duquel une personne est grande ou ne l'est pas n'est pas claire. Donc, la valeur du seuil est floue et, par conséquent, l'ensemble qu'il délimite le sera aussi. Si la taille limite est 1m75, alors dans la théorie des ensembles, l'ensemble des personnes grandes est à partir de 1m75, et donc une personne de 1m85 fait partie de l'ensemble des personnes grandes mais une personne de 1m60 non. Par contre, dans la théorie des sous-ensembles flous, une personne de 1m60 est grande, par exemple, d'un degré 0.3, et une personne de 1m85 est grande, par exemple, d'un degré 0.9. Cela signifie que la théorie des sous-ensembles flous permet une quantification de l'imprécision de la perception et du raisonnement de l'être humain en termes d'attributs subjectifs (ou étiquettes linguistiques) sur un référentiel. Si l'on revient à l'exemple il peut y avoir autour de l'étiquette linguistique "grande" les étiquettes: "plutôt grande", "assez grande", "très grande", etc. En conséquence, pour les ensembles

---

notion foncièrement probabiliste, et imprécision, notion déterministe, c'est plutôt cette dernière qui s'applique au contexte de la planification de projet ».
[2] Ici la signification d'objet est différente de la figure 4.2.

flous la notion d'appartenance est liée à une fonction pouvant prendre toutes les valeurs réelles entre 0 et 1 (...,0.1,...,0.2,...,0.3,...,0.9,...) indiquant de degrés ou le taux d'appartenance[3].

Plus formellement, un sous-ensemble flou $A$ est défini sur un ensemble $U$, appelé référentiel (ou domaine, ou univers du discours), par une fonction, appelée *fonction d'appartenance*, notée $\mu_A$, de la manière suivante :

$$A = \{\mu_A(x)/x : x \in U, \mu_A(x) \in [0,1]\}$$

Oú pour tout objet $x$ de $U$, $\mu_A(x)$ est le *degré d'appartenance* de l'objet $x$ au sous-ensemble flou $A$. Un degré égal à 0 est interprété comme une non-appartenance absolue au sous-ensemble flou $A$ et un degré égal à 1 comme une appartenance absolue ou totale au sous-ensemble flou $A$. Le référentiel $U$ est un ensemble quelconque, discret ou continu, fini ou infini. La théorie des sous-ensembles flous généralise la théorie des ensembles classiques, car la fonction d'appartenance $\mu(x)$ renvoie les objets $x \in U$ non seulement sur l'ensemble $\{0,1\}$, mais sur tout l'intervalle réel $[0,1]$. Une étude détaillée sur les aspects théoriques de base que nous avons utilisé dans cette thèse, comme les propriétés et les opérations ensemblistes floues sont discutées, par exemple dans [Kaufmann, 77]. Dans la littérature, par exemple [Vincke, 73] on trouve plutôt le terme "ensembles flous" que "sous-ensembles flous", mais pour Kaufmann et autres il y a une différence. En effet, d'après Kaufmann « un autre point de détail à commenter, pourquoi employer les termes "sous-ensembles flous" et non pas "ensembles flous". Parce que, un ensemble flou ne serait pas le concept qui convient à la présente théorie, le référentiel étant toujours un ensemble vulgaire » [Kaufmann, 77].

Dans cette thèse nous n'avons pas fait de différence, car notre problématique de recherche est industrielle plutôt que mathématique. Mais nous avons employé, d'une part le terme domaine pour désigner le référentiel $U$ qui est plutôt le nom utilisé dans les bases de données, et d'autre part le terme élément pour désigner l'objet $x \in U$ pour ne pas être ambigu vis à vis de la figure 4.2.

Dans [Galindo, 99] nous avons trouvé cinq significations pour le degré d'appartenance que nous allons illustrer ceci avec l'ensemble flou ou étiquette linguistique "grande".

---

[3] Le concept d'appartenance peut être illustré avec la norme international : S (small), M (medium), X (large), XL (extra large), XXL (extra extra large) que l'industrie du textile a adoptée pour classifier ses produits (selon la taille).

➤ **Le degré d'appartenance comme un degré de préférence**

Dans ce cas la sémantique associée au degré d'appartenance est dans le sens d'une valeur. Par exemple, nous recherchons quelqu'un de 1m75, mais une personne de 1m73 est préférable à une personne de 1m71. Cela signifie que l'importance est donnée par une relation de préférence de la forme:

$$\mu_{grande}(1m71)=0.1 < \mu_{grande}(1m73)=0.5 < \mu_{grande}(1m75)=1$$

➤ **Le degré d'appartenance comme un degré d'accomplissement**

Dans ce cas la sémantique associée au degré d'appartenance est dans le sens d'une satisfaction d'une propriété (un ensemble flou). Si la propriété est accomplie totalement, alors on associe un degré 1, tandis que si la propriété n'est pas accomplie absolument alors on associe un degré 0. Par exemple, supposons que la propriété à satisfaire soit que la taille de gens soit comprise absolument entre 1m75 et 1m80. Donc dans ce cas on a :

$$\mu_{grande}(1m76)=1 \text{ ou } \mu_{grande}(1m73)=0 \text{ ou } 0< \mu_{grande}(1m73)<1$$

➤ **Le degré d'appartenance comme un degré de nécessité**

Dans ce cas, la sémantique associée au degré d'appartenance est dans le sens d'un jugement incertain. Par exemple si on a $\mu_{grande}(1m73)=0.5$, cela signifie qu'il est plus ou moins certain que la taille soit 1m73 à un degré 0.5.

➤ **Le degré d'appartenance comme un degré de possibilité**

Dans ce cas, la sémantique associée au degré d'appartenance est dans le sens d'un jugement possible. Par exemple si on a $\mu_{grande}(1m73)=0.5$, cela signifie que il est plus ou moins possible (mais pas certain) que la taille soit 1m73 à un degré 0.5.

➢ **Le degré d'appartenance comme un degré de similitude**

Dans ce cas la sémantique associée au degré d'appartenance est dans le sens d'une comparaison entre deux propriétés (deux ensembles flous). Pour cela on utilise le degré de possibilité et le degré de nécessité.

## 4.2.2 Théorie des posibilités

La théorie des possibilités est l'un des outils mathématiques pour représenter, gérer et évaluer des incertitudes à l'aide des ensembles flous définis sur un domaine ordonné. Elle a été introduite et formalisée aussi par Zadeh au travers de la *mesure de possibilité* et de la *mesure de nécessité*, puis développée par Prade et Dubois.

Mesure de possibilité

La mesure de possibilité, notée $\Pi$, est une mesure non probabiliste de l'incertitude d'un événement plus ou moins possible sur un domaine ordonné. Supposons que $A$ est un événement défini sur un domaine $U$, et alors la mesure de possibilité de l'événement $A$, notée $\Pi(A)$, est une fonction de $U$ dans l'intervalle réel [0,1]. Si $\Pi(A)=0$ alors l'événement $A$ est complètement impossible, si $\Pi(A)=1$ alors l'événement $A$ est complètement possible, et si $0<\Pi(A)<1$ alors l'événement $A$ est plus ou moins possible. Maintenait, si la probabilité de l'événement $A$ est égale à 1, cela signifie que l'événement $A$ est complètement certain (ou sûr) et la probabilité de l'événement contraire de $A$ dans $U$ est égale à 0, mais $\Pi(A)=1$ ne signifie pas que l'événement $A$ soit complètement certain. Par ailleurs soit $\bar{A}$ l'événement contraire de $A$ avec $\Pi(A$ , $\bar{A})=max(\Pi(A),\Pi(\bar{A}))^4=1$ cela signifie que l'événement $A$ peut être complètement possible et l'événement contraire de $A$ aussi (cas où $\Pi(A)=\Pi(\bar{A})=1$). Dans ce contexte, l'incertitude d'un événement est décrite à la fois par une mesure de possibilité de cet événement et par la mesure de possibilité de l'événement contraire dans un domaine ordonné. Il faut donc introduire une nouvelle mesure pour exprimer le fait qu'un événement est complètement certain.

---

[4] max(X,Y) = maximum de X et Y.

<u>Mesure de nécessité</u>

La mesure de nécessité, notée $N$, est définie comme $N(A)=1-\Pi(\bar{A})$ qu'il faut lire comme la nécessité d'un événement est égale à 1 moins la possibilité de l'événement contraire. Cette définition contient le fait que d'une manière générale, la mesure de nécessité, ou certitude, de l'événement $A$ sur un domaine $U$, notée $N(A)$, est une fonction de $U$ dans l'intervalle réel [0,1]. Si $N(A)=0$ alors l'événement $A$ est complètement incertain, si $\Pi(A)=1$ alors l'événement $A$ est complètement certain, et si $0<N(A)<1$ alors l'événement $A$ est plus ou moins certain. Soit $N(A, \bar{A})=min(N(A),N(\bar{A}))^5=0$ cela signifie que l'événement $A$ peut être complètement incertain et son contraire aussi (cas où $N(A)=N(\bar{A})=0)$ ou au moins un événement n'est pas certain. Il en résulte que la possibilité d'un événement est plus grande que sa certitude, puisque $\Pi(A)\geq N(A)$. Dans [Dubois et Prade, 85,88] sont détaillées d'autres relations et propriétés entre les mesures de possibilité et de nécessité.

D'un point de vue pratique, le domaine $U$ est interprété comme le domaine de valeurs d'un paramètre (c'est-à-dire une variable), et donc l'événement $A$ de ce domaine $U$ est une partie de $U$ que l'on interprète comme le fait que la valeur du paramètre appartient à $A$. Il en résulte que $\Pi(A)$ mesure la possibilité de l'occurrence de l'événement $A$, que l'on peut interpréter comme la possibilité que la valeur du paramètre soit dans $A$, tandis que $N(A)$ mesure la nécessité, ou certitude, de l'occurrence de l'événement $A$, que l'on peut interpréter comme la certitude que la valeur du paramètre soit dans $A$. Dans le cadre d'un système à base de cas, voir par exemple [Salotti, 92] on considère que ces événements sont flous, car l'objectif est de pouvoir manipuler des informations imparfaites (ou mal définies) dans la base de cas. Un événement flou (ou mal défini) peut être représenté par un ensemble flou sur un domaine ordonné que l'on interprète comme le fait, imprécis, que la valeur d'un paramètre donné appartient à cet ensemble flou. Dans [Dubois et Prade, 85, 88] et [Salotti, 92] se trouve la justification mathématique qui permet d'étendre les mesures de possibilité et de nécessité à des ensembles flous, et donc dans cette thèse nous avons gardé la même sémantique pour ces mesures. Or, dans le contexte d'un système à base de règles, voir par exemple [Buisson, 87] on préfère le terme proposition floue au lieu d'événement flou, plus adéquat quand on emploie la logique possibiliste pour prendre en compte l'imprécision et l'incertitude des faits flous et des règles floues. La logique possibiliste[6] est une logique classique (booléenne) qui permet d'associer à des propositions floues décrivant des faits flous et des règles floues des mesures de possibilité et de nécessité indiquant dans quelle mesure il est possible et il est certain que des faits flous soient vrais et des règles flous valides. En conséquence, dans cette thèse nous avons utilisé le terme proposition floue dans le cas d'un système à base de règles et le terme événement flou est utilisé pour un système à base de cas, mais cela n'est plus qu'un choix sémantique. En effet, Dubois et Prade disent « une proposition peut être vue aussi comme une affirmation relative à l'occurrence d'un événement » [Dubois et Prade, 85, 88] alors il n'y a pas de différence entre les deux et non plus entre proposition floue et événement flou, puisque les propositions et les événements sont représentés par des ensembles et les propositions floues et les événements flous sont représentés par des ensembles flous.

---

[5] $min(X,Y)$ = minimum de X et Y.

> **Modélisation mathématique des mesures de possibilité et de nécessité**

Nous allons maintenant commenter avec un exemple de [Dubois et Prade, 85, 88] l'approche intuitive de modélisation mathématique de la mesure de l'incertain à travers des mesures de possibilité et de nécessité, pour évaluer des incertitudes affectant un jugement incertain d'une grandeur imprécise d'une mesure, l'un des aspects les plus courants de la vie de tous le jours d'un expert et des gens en général, c'est-à-dire les incertitudes de mesure [Mauris, 00] :

*« si on ignore la valeur exacte d'un paramètre, on connaît précisément les limites de son domaine de variation. On remarquera qu'étant donnée une mesure imprécise M de la grandeur X, les propositions du type "X appartient à l'intervalle I" seront naturellement qualifiées à l'aide des modalités du possible et du nécessaire car*

> *i)    Si $M \cap I$ est non vide alors "$X \in I$" est possiblement vrai*
>
> *ii)   Si $M \subseteq I$ alors "$X \in I$" est nécessairement vrai.*

*...Le possible est évalué à partir d'une intersection ensembliste entre le contenus M et I des deux propositions "$X \in M$" et "$X \in I$"; le nécessaire est évalué à partir de l'inclusion ensembliste »[7].*

---

[6] La logique floue est la discipline qui regroupe tous les théories de l'incertain comme les ensembles flous, la théorie des possibilités, la théorie des fonctions de croyance, etc. Or, la logique possibiliste est une partie de la théorie des possibilités.
[7] Ici le mot "proposition" peut être changé à tout moment par "événement". La condition *i)* correspond à un degré d'intersection de deux sous-ensembles flous. Le système FSQL traduit ce degré par sept filtres flous de possibilité (voir tableau 4.2), tandis que la condition *ii)* correspond à un degré d'inclusion de deux sous-ensembles flous. Le système FSQL traduit aussi ce degré par sept filtres flous de nécessité (voir tableau 4.3).

Dans ce contexte, $X$ est une variable dont la valeur est imprécise, c'est-à-dire qu'elle est connue seulement par l'ensemble flou $M$, et $I$ est un ensemble flou donnée. Afin de clarifier la situation, soit par exemple $X = température$ et $M = normale$ alors la proposition floue $"X \in M"$ se traduit par "*température* est *normale*", de même soit $I = chaud$ alors la proposition floue $"X \in I"$ se traduit par "*température* est *chaud*". Or, pour évaluer la possibilité et la nécessité que la proposition floue $"X \in I"$ (ou $"X$ est $I"$) soit vraie sachant que $"X \in M"$ (ou $"X$ est $M"$) on utilise deux mesures d'incertitude (ou de confiance selon la terminologie de [Dubois et Prade, 85, 88]) pour exprimer l'incertitude de la mesure $I$. La mesure de possibilité permet d'évaluer l'incertitude, tandis que la mesure de nécessité permet d'évaluer la certitude. Cela signifie que dans la condition $M \cap I$ *est non vide* on évalue le fait imprécis que les valeurs de $M$ possiblement appartiennent à $I$, tandis que dans la condition $M \subseteq I$ on évalue le fait imprécis que les valeurs de $M$ nécessairement appartiennent à $I$. Il en résulte que ces deux mesures non probabilistes de l'incertitude sont deux concepts différents.

➤ **Lien entre possibilités et ensembles flous**

Le lien entre possibilités et ensembles flous se fait par les concepts de *degré de possibilité* et de *distribution de possibilités* que nous allons illustrer à l'aide de l'exemple antérieur. Supposons que $I$ est un ensemble flou et $X$ est un paramètre (aussi dans la littérature on emploie le nom variable, donnée ou attribut), tout les deux définies sur un domaine $U$ (l'ensemble de toutes les valeurs possibles que peuvent prendre $X$ et $I$). Or, pour évaluer la possibilité que la proposition floue $"X$ est $I"$ (voir paragraphe au dessus) soit vraie, c'est-à-dire que chaque élément $x$ sur un domaine $U$ du paramètre $X$ appartient à l'ensemble flou $I$, on utilise le degré d'appartenance de l'élément $x$ à l'ensemble flou $I$. En effet, supposons que $\mu_I(x)=0$ pour tout $x$ sur un domaine $U$, alors il est complètement impossible que l'élément $x$ appartienne au paramètre $X$, autrement dit, il est tout à fait impossible que l'élément $x$ soit la valeur du paramètre $X$. Maintenant supposons que $\mu_I(x)=1$ pour tout $x$ sur un domaine $U$, alors il est complètement possible que l'élément $x$ appartienne au paramètre $X$, autrement dit il est tout à fait possible que l'élément $x$ soit la valeur du paramètre $X$. Dans le cas intermédiaire $0<\mu_I(x)<1$, plus $\mu_I(x)$ est proche de 1 et plus la possibilité que $x$ soit la valeur du paramètre $X$ est grande, tandis qu'au contraire plus $\mu_I(x)$ est proche de 0 et plus la possibilité que $x$ soit la valeur du paramètre $X$ est petite. En conséquence, pour tout élément $x$ sur un domaine $U$, le *degré de possibilité* de la proposition floue $"X \in I"$, c'est-à-dire que l'élément $x$ appartient au paramètre $X$ ($x \in X$), est défini, par le degré d'appartenance de l'élément $x$ à l'ensemble flou $I$.

En généralisant ce résultat, une *distribution de possibilités* sur un domaine $U$ attachée au paramètre $X$, notée $\pi_X(x)$, pour tout élément $x$ sur un domaine $U$, est définie par la fonction d'appartenance $\mu_I(x)$, telle que:

$$\forall x \in U, \ \forall X, I \subseteq U : \pi_X(x) = \mu_I(x)$$

Cela signifie que si $I$ est un ensemble flou et $X$ est un paramètre, avec $X$ et $I$ définies sur un domaine $U$, alors une distribution de possibilités sur un domaine $U$ attachée au paramètre $X$, est définie par un ensemble flou des éléments possibles du paramètre $X$. Il en résulte qu'une distribution de possibilités peut s'interpréter comme une fonction d'appartenance.

En général une distribution de possibilités sur un domaine $U$ attachée au paramètre $X$, est une fonction $\pi$ de $U$ dans l'intervalle réel [0,1] qui est normalisée, c'est-à-dire vérifiant pour tout élément $x$ sur le domaine $U$ que $sup(\pi_X(x))=1$, permet de construire une mesure de possibilité pour $X$ de la forme $\Pi(X)=sup\pi_X(x)$. Où, $\pi_X(x)$ est le degré de possibilité que la valeur du paramètre $X$ soit $x$, et $sup$ est la borne supérieure de $\pi_X(x)$. Par dualité, la mesure de nécessité pour $X$ est définie par $N(X)=1-\Pi(\ \bar{X})=inf(1-\pi_{\bar{X}}(x))$. Où, $inf$ est la borne inférieure de $1-\pi_{\bar{X}}(x)$. En conséquence, $1-\pi_{\bar{X}}(x)$ est le degré de nécessité (certitude) que la valeur du paramètre $X$ soit $x$.

Revenons maintenant au même exemple pour vérifier cela. En effet, supposons que $\pi_M(x)=1$ alors cela ne signifie pas que l'élément $x$ doit y être nécessairement en $M$, puisque rien ne garantit la condition $M \subseteq I$. Autrement dit, il n'y a pas de certitude que $x$ soit la valeur de $M$. Tandis que si $\pi_M(x)=0$ alors cela signifie avec certitude que l'élément $x$ n'est pas la valeur de $M$. Il s'ensuit que cette constatation nous permet de voir intuitivement l'existence, d'une part, d'un degré de possibilité auquel il est possible que la proposition floue ″$X$ est $I$″ soit possiblement vraie et d'attacher une distribution de possibilités sur un domaine $U$ au paramètre $X$, et d'autre part, d'un degré de nécessité auquel il est certain que la proposition floue ″$X$ est $I$″ soit nécessairement vraie et d'associer une distribution de nécessités sur un domaine $U$ au paramètre $X$.

En conséquence, étant donné la distribution de possibilités $\pi_X(x)$, pour tout élément $x$ sur un domaine $U$ et pour tout ensemble flou $X, M$ et $I \subseteq U$, qui traduit, d'une part, la restriction de la proposition floue "$X$ est $M$" par la fonction d'appartenance $\mu_M(x)$, et d'autre part, la restriction de la proposition floue "$X$ est $I$" par la fonction d'appartenance $\mu_I(x)$, nous pouvons maintenant évaluer la mesure de possibilité et la mesure de nécessité de la proposition floue (ou événement flou) "$X$ est $I$" sachant que "$X$ est $M$", par les équations de [Zadeh, 78] :

$$\Pi(I,X) = sup_{x \in U} min(\pi_X(x), \mu_I(x)) = sup_{x \in U} min(\mu_M(x), \mu_I(x))^8 \in [0,1] \qquad (4.1)$$

$$N(I,X) = 1 - \Pi(\bar{I},X) = 1 - sup_{x \in U} min(\mu_M(x), 1 - \mu_I(x)) = inf_{x \in U} max(1 - \mu_M(x), \mu_I(x))^9 \in [0,1] \qquad (4.2)$$

On constate donc que dans l'équation (4.1), $\Pi(I,X)$ est la mesure de possibilité qui évalue la possibilité que la valeur de $X$ soit $I$ sachant que la valeur de $X$ est $M$. Une autre façon de la noter d'après [Gacôgne, 90] :

possibilité ("$X$ est $I$" sachant que "$X$ est $M$") ou ps($X$ est $I$) ou ps($X / I$)

Que l'on lit "la possibilité d'avoir $I$ sachant que $X$ est $M$". Cette mesure correspond à une borne supérieure parmi l'ensemble des fonctions d'appartenance pour $M$ et $I$ dont le degré d'appartenance est le plus petit pour tout $x$ sur un domaine $U$. Tandis que dans l'équation (4.2), la mesure de nécessité $N(I,X)$ est calculée par dualité par rapport à $\Pi(I,X)$, et donc elle évalue la nécessité (certitude ou obligation) que la valeur de $X$ soit $I$ sachant que la valeur de $X$ est $M$. Une autre façon de la noter d'après [Gacôgne, 90] :

nécessité ("$X$ est $I$" sachant que "$X$ est $M$") ou nc($X$ est $I$) ou nc($X / I$)

Que l'on lit "la nécessité d'avoir $I$ sachant que $X$ est $M$". Cette mesure de nécessité (ou certitude) correspond à une borne inférieure parmi l'ensemble des fonctions d'appartenance pour $M$ et $I$ dont le degré d'appartenance est le plus grand pour tout $x$ sur un domaine $U$.

Une étude plus détaillée sur les mesures de possibilité et de nécessité sur base de la théorie des possibilités peut être consultée par exemple dans [Dubois et Prade, 85, 88]. Néanmoins, il faut noter que le contenu (le point de vue ensembliste) de ces propositions floues permet de donner une interprétation graphique des mesures de possibilité et de nécessité. En effet, la mesure de possibilité $\Pi(I,M) = \Pi(M,I)$ peut être interprétée comme un degré d'intersection entre $M$ et $I$, c'est-à-dire il s'agit d'un indice maximal de chevauchement [Gacôgne, 90] tandis que la mesure de nécessité $N(I, M) \neq N(I, M)$ peut être interprétée comme un degré d'inclusion de $M$ dans $I$, c'est-à-dire il s'agit d'un indice minimal d'inclusion [Gaçôgne, 90]. Un point de détail à commenter est que ceci justifie la conclusion donnée dans l'exemple qui a permis d'illustrer les mesures de possibilité et de nécessité. En conséquence, ces deux mesures de confiance sont les outils essentiels du filtrage flou (pour évaluer à quel point $M$ et $I$ sont compatibles ou similaires, c'est-à-dire $M \subseteq I$ ou $I \subseteq M$) que nous présentons au paragraphe 4.2.3. Mentionnons que Jean Piaget lui aussi a été intéressé par l'étude des modalités du possible et du nécessaire dans les mécanismes des constructions cognitives de la pensée formelle et le savoir raisonné chez l'enfant (entre 7 et 10 ans environ). Ces études ont été inspirées à partir des recherches du philosophe français Léon Brunschvicg vis à vis la proposition "*est*". En effet « dans une thèse publiée en 1897, Brunschvicg en était arrivé à s'interroger sur le possible, le réel et le nécessaire lorsque, cherchant à déterminer la nature des l'activité intellectuelle, il avait identifié celle-ci à l'acte de jugement, et plus particulièrement à l'acte de jugement théorique par lequel la pensée humaine vise à affirmer l'être. Analysant la modalité du couple "*est*" dans les jugements de perception, dans le jugements mathématiques, et dans les jugements de la physique, il commençait par montrer que la réalité affirmée par ce jugement manque à la fois de certitude (quant à ce qui est affirmé) et d'intelligibilité. Examinant ensuite la nécessité associée à l'être des jugements mathématiques il montre que, en dépit de l'intériorité qui leur est propre et de la compréhension intellectuelle relativement complète qu'ils permettent de leur contenu, ces jugements ne peuvent satisfaire la visée primordiale de l'acte de jugement théorique. Enfin, examinant le jugement (de la science) physique, il y découvrit que l'être affirmé par celui-ci, s'il participe à la fois de l'intelligibilité du jugement mathématique sur lequel il s'appuie et de la réalité du jugement de perception, n'a pas la certitude spontanément recherchée par la pensée. La modalité du jugement de réalité le plus satisfaisant qui soit pour l'intelligence humaine n'est, en conséquence, autre que la simple possibilité » [Piaget *et al*, 83].

---

8 min(X,Y) = minimum de X et Y; sup(X,Y) = borne supérieure de X et Y.

Or, nous constatons que dans la théorie des possibilités de Zadeh, Prade et Dubois, la réflexion porte aussi sur les modalités du possible, du nécessaire et du réel. En effet, la mesure de l'incertain est en fait imprécise car il s'agit d'un jugement incertain du réel. Et donc il y a toujours l'existence d'une subjectivité floue de l'être humain pour manipuler les concepts d'imprécision et d'incertitude autour d'une proposition floue (ou événement flou) de la réalité. Les outils logico-mathématiques proposés par Zadeh permettent de gérer cette subjectivité de l'individu autour d'une proposition floue (c'est-à-dire possible et/ou nécessaire) du monde réel, ou bien pour quantifier l'incertitude d'une affirmation relative à l'occurrence d'un événement flou, à l'aide des mesures de possibilités et de nécessités dans le cadre de la théorie des possibilités. Une discussion profonde sur le sujet se trouve en [Dubois et Prade, 85, 88]. Nous nous contentons maintenant de faire le lien entre possibilités, ensemble flous et filtrage flou.

---

[9] max(X,Y) = maximum de X et Y; inf(X,Y) = borne inférieure de X et Y.

### 4.2.3    Filtrage flou

D'après Andres « au sens le plus général, le filtrage est l'opération qui consiste à rechercher dans une collection d'informations, une information dont le contenu correspond le mieux à une certaine description (le filtre) » [Andres, 89]. Il en résulte que si cette description est floue alors on parle plutôt de *filtrage floue*. Le filtrage flou est une technique de comparaison entre deux ensembles flous (ou distributions de possibilités) convexes[10] et normalisés[11], ou bien entre la valeur d'une variable et un ensemble flou. La comparaison se réalise à partir des mesures de possibilités et de nécessités de la théorie des possibilités [Zadeh, 78].

Or, revenons à l'exemple du possible et du nécessaire pour évaluer la compatibilité (ou similarité) entre les ensembles flous *"M et I"* et *"I et M"*, c'est-à-dire l'intersection et l'inclusion entre les valeurs des températures *"normale et chaud"* et *"chaud et normale"*. En effet, dans cet exemple on fait la supposition que $X = température$ et $M = normale$ alors la proposition floue *"X ∈ M"* est traduit par "*température* est *normale*", de même nous avons mis $I = chaud$ alors la proposition floue *"X ∈ I"* est traduit par "*température* est *chaud*". Or, en utilisant les mesures de possibilité et de nécessité définies par [Zadeh, 78] dans l'intervalle réel [0,1] que nous avons donné par les équations 4.1 et 4.2. On a donc trois situations possibles pour tout élément $x$ sur un domaine $U$ :

<u>Première situation</u>

Pour évaluer à quel point $M$ et $I$ sont compatibles du point de vue de la mesure de possibilité nous devons calculer la ps($M$ est $I$) = ps(*normale* est *chaud*)[12]. Or d'après l'équation (4.1) on a :

---

[10] Un ensemble flou $A$ est dit convexe si sa fonction d'appartenance vérifie la relation : $\forall a,b \in U,\ x \in [a,b] \Rightarrow \mu_A(x) \geq min(\mu_A(a),\mu_A(b))$.

[11] Un ensemble flou $A$ est dit normalisé si il existe au moins un élément du domaine pour lequel la fonction d'appartenance $\mu_A(x)$ est égale à 1.

[12] Ps et nc est une autre notation pour indiquer $\Pi$ et $N$, respectivement.

---

$$\Pi(I,M) = sup_{x \in U} min(\mu_M(x)), \mu_I(x))^{13} \in [0,1] \qquad (4.3)$$

Où, $\mu_M(x)$ est la fonction d'appartenance de l'ensemble flou $M$ et $\mu_I(x)$ est la fonction d'appartenance de l'ensemble flou $I$. Or, dans cette expression il faut noter que $\Pi(I, M) = \Pi(M,I)$, c'est-à-dire, on a ps(*normale* est *chaud*) = ps(*chaud* est *normale*).

Deuxième situation

Pour évaluer à quel point $M$ et $I$ sont compatibles du point de vue de la mesure de nécessité nous devons calculer la nc($M$ est $I$) = nc(*normale* est *chaud*). Or d'après l'équation (4.2) on a :

$$N(I,M) = inf_{x \in U} max(1 - \mu_M(x)), \mu_I(x))^{14} \in [0,1] \qquad (4.4)$$

Où, $\mu_M(x)$ est la fonction d'appartenance de l'ensemble flou $M$ et $\mu_I(x)$ est la fonction d'appartenance de l'ensemble flou $I$. D'autre part, cette expression correspond bien à l'énoncé de l'exemple de départ. En effet, $M$ est la donnée observée et $I$ la donnée de référence.

Troisième situation

Pour évaluer à quel point $I$ et $M$ sont compatibles du point de vue de la mesure de nécessité nous devons calculer la nc($I$ est $M$) = nc(*chaud* est *normale*). Or d'après l'équation (4.2) on a :

$$N(M,I) = inf_{x \in U} max(\mu_M(x)), 1 - \mu_I(x))^{15} \in [0,1] \qquad (4.5)$$

Où, $\mu_M(x)$ est la fonction d'appartenance de l'ensemble flou $M$ et $\mu_I(x)$ est la fonction d'appartenance de l'ensemble flou $I$. On constate donc que cette expression ne correspond pas à l'énoncé de l'exemple.

Pour le calcul de la mesure de nécessité, le système FSQL (voir section 4.3.6) emploi plutôt la deuxième situation pour la construction d'un filtre flou. Nous reviendrons au paragraphe 4.5 sur la justification de ce choix.

---

[13] min(X,Y) = minimum de X et Y; sup(X,Y) = borne supérieure de X et Y.
[14] max(X,Y) = maximum de X et Y; inf(X,Y) = borne inférieure de X et Y.

A cet effet, les filtres flous ont été utilisés dans la plupart des systèmes à base de connaissances confrontés aux problèmes de comparer une donnée à une référence afin d'apprécier à quel degré la donnée observée vérifie la donnée de référence (le filtre). Dans le système opérationnel, nous nous intéresserons à l'application du filtrage flou dans deux types de systèmes. Les systèmes à base de connaissances de type base de cas et de type base de règles.

➢ **Filtrage flou dans un système à bases de connaissances de type base de cas**

Dans le langage de requêtes à une base de données relationnelles floues le terme de filtrage flou est utilisé pour désigner une consultation ou une interrogation. Le filtre flou est la condition graduelle (voir figure 4.3) de la requête ainsi formulée à la base de données relationnelles floues. Revenons à l'exemple de départ et supposons par exemple la requête "trouver les personnes de taille plutôt supérieure à 1m71", alors le filtre flou ("plutôt supérieur à 1m71") est comparé à l'ensemble des distributions des possibilités, associées à chaque donnée imprécise (attribut "taille") de la base de données, de manière à avoir en réponse, les tuples correspondant le mieux à ce filtre flou. En général, le filtrage est l'opération qui consiste à rechercher dans une base de données, les données qui correspondent le mieux au filtre [Andres, 89].

Dans ce travail les divers cas d'exemples ont été stockés dans une base de données relationnelle Oracle 8 étendue aux données floues par le système FSQL [Galindo, 99] alors le filtrage flou est la consultation à cette base de données, c'est ce que l'on appelle une *requête floue* ou *interrogation flexible* [Connann, 99]. Une requête floue FSQL a la syntaxe suivante :

SELECT liste des colonnes (attributs flous et/ou crisp)
FROM liste des tables (relations)
WHERE liste des filtres flous (conditions graduelles)

**Figure 4.3 :** Requête floue FSQL

Dans la suite nous spécifirons le rôle du filtrage flou dans une base de données relationnelles floues Pour cela nous avons retenu dans le cadre de cette thèse les systèmes suivantes :

---

[15] max(X,Y) = maximum de X et Y; inf(X,Y) = borne inférieure de X et Y.

Le système FLORAN

Pour Salotti « un filtre flou portant sur un attribut élémentaire est défini par un ensemble flou sur le domaine de l'attribut. Les valeurs dont le dégré d'appartenance au filtre est égal à 1 sont complètement admises, celles dont le degré est égal à 0 sont complètement refusées, celles dont le degré est compris strictement entre 0 y 1 sont des valeurs "limites". Plus précisément, le filtre flou étant défini par la fonction caractéristique $\mu_F$ sur le domaine U, pour tout élément u de U, $\mu_F(u)$ représente le degré de compatibilité de u avec le filtre » [Salotti, 92].

Nous constatons que dans le système FLORAN le filtrage flou est interprété comme un degré d'accomplissement entre la valeur d'une variable et un ensemble flou qui joue le rôle de filtre flou. Ceci est calculé par une mesure de compatibilité. Par exemple, si l'on revient à l'exemple de départ sur la taille des gens. Supposons que le domaine $U$ =[1m50,2m50] et que l'attribut élémentaire "grande" est définie par rapport à l'intervalle flou [1m75,1m85]. Alors, la technique de filtrage consiste à comparer un $x \in U$ avec le filtre "grande". Soit $x=1m73$, alors la possibilité que $X$ soit $x$ est donnée par:

$$\text{possibilité } (X = 1m73 \ / \ X \in grade \ ) = \mu_{grande}(1m73)=0.5$$

Le système DOLMEN

Pour Simon « une mesure de similarité partielle est le résultat d'une comparaison entre la courbe critique d'un paramètre critique et la valeur de son homologue dans la fabrication d'acier. Cette mesure doit permettre d'évaluer à quel point la valeur de fabrication est influente sur le défaut indépendamment des autres paramètres. La technique choisie pour opérer cette comparaison est dérivée du filtrage flou ... elle permet de juger de l'adéquation d'un ensemble de données à un ensemble de filtres, les filtres sont les courbes critiques el les données, les valeurs des paramètres dans la fabrication » [Simon, 97].

Nous constatons que dans le système DOLMEN le filtrage flou est interprété cette fois comme un degré de similitude entre les filtres qui sont des courbes critiques et les données qui sont les valeurs des paramètres dans la fabrication. Dans ce système et autres [Ruet, 99] les filtres flous sont modélisés à partir des ensembles flous et les données imprécises par des distributions de possibilités. Le calcul se fait donc par une mesure de similarité.

➤ **Filtrage flou dans un système à bases de connaissances de type base de règles**

Dans une optique système expert, par exemple dans le système TOULMED [Buisson, 87] le filtrage flou est utilisé comme un outil de mesure de compatibilité entre propositions (les règles et la base de faits). Dans [Jiménez, 02a], [Jiménez, 02b] nous avons détaillé la prise en compte de l'imprécis et de l'incertain dans CommonKADS [Schreiber *et al*, 00].

Dans le but d'étudier la représentation des connaissances floues dans les systèmes à bases de connaissances de type base de cas ou de type base de règles en utilisant le filtrage flou, nous allons clarifier dans la suite l'usage des concepts incomplet, imprécis et incertains vis à vis du traitement de l'imperfection de l'information sous l'angle des bases de données relationnelles floues. Pour cela nous avons retenu dans le cadre de cette thèse un certain nombre de systèmes.

## 4.3 Les applications de la théorie de l'imprécis et de l'incertain dans l'industrie

### 4.3.1 Le système F-MRP

Le système F-MRP (Fuzzy-Manufacturing Ressource Planning) [Reynoso, 04] propose une application de la théorie des ensembles flous et des possibilités pour traiter l'imprécision et l'incertitude relatives à des commandes dans une problématique de gestion de la production MRP2. Dans cette méthode, une commande imprécise est définie comme une commande dont la quantité commandée n'est pas connue avec exactitude, par exemple *M. Dupont va passer une commande d'environ 400 pièces*, ou bien *M. Dupont va passer une commande entre 500 et 600 pièces*. Dans le premier exemple, l'imprécision se trouve au niveau du langage vu que "environ" a une signification associée à la réalité des acteurs, tandis que dans le second exemple, l'imprécision provient du fait d'avoir une information partielle de la réalité. Ici comme ailleurs, l'imprécision est représentée sous forme d'ensembles flous. D'autre part, une commande incertaine est définie comme l'occurrence d'un événement incertain, autrement dit l'occurrence de l'événement n'est pas assurée où sa vérité est incertaine par rapport à une réalité que dans ce cas est la gestion de la production. Le calcul de l'occurrence de l'événement se fait à l'aide de la mesure de possibilité. Afin de simplifier les calculs, la méthode F-MRP utilise une fonction trapézoïdale dans la modélisation de l'incertain, par exemple si on prend la réalité de gestion antérieure, l'incertitude est représentée par une mesure de possibilité d'occurrence de la commande où la valeur 1 (le dessus du trapèze) représente la certitude d'occurrence de la commande (soit 400 pièces dans le première exemple ou soit entre 500 et 600 pièces du deuxième exemple antérieur), la valeur 0 (le dessous du trapèze) représente la possibilité d'occurrence que la commande soit annulée (0 pièces commandé), tandis qu'une valeur entre 0 et 1(hauteur du trapèze) représente le degré de possibilité d'occurrence de la commande. L'objectif de la méthode F-MRP est de pouvoir établir un plan de charge flou, par exemple si on a 20 commandes d'un produit sur 5 jours, parmi ces commandes, on peut avoir 60% de commandes précises et certaines, 20% de commandes imprécises et certaines, 10% de commandes imprécises et incertaines, 10% de commandes précises et incertaines.

### 4.3.2    Le système SIMCAIR

Le système SIMCAIR (Système d'Interrogation Multi-Critères Avec Importances Relatives) [Andres, 89] propose une application de la théorie des ensembles flous et des possibilités au problème de la représentation de connaissances imparfaites, soit sur le contenu de l'information, soit sur la véracité de l'information ; dans le premier cas on parle d'imprécision et dans le second cas on parle d'incertitude. Ceci nous renvoie aux mêmes idées développées par Dubois et Prade au sujet de l'imperfection de l'information et dont nous avons discuté au paragraphe antérieur (voir figure 4.2).

### 4.3.3    Le système TOULMED

Le système TOULMED [Buisssson, 87] propose une application de la théorie des ensembles flous et des possibilités au problème de la représentation de connaissances imparfaites dans un système expert médical dans le domaine de la diabétologie. L'information représentée par ce système est formalisée sous forme des propositions logiques comportant des prédicats (qu'ici on appelle des variables) et des valeurs relatives aux objets médicaux (faits, conditions et conclusions de règles). L'imprécis est une proposition où la valeur de la variable est désignée par un sous-ensemble ordinaire de valeurs. Par exemple: "l'âge du malade est strictement entre 20 et 40 ans", ou "le malade a 20 ans ou 25 ans". Le flou est une proposition où la valeur de la variable forme un ensemble flou, c'est-à-dire les frontières de la variable ne sont pas précisément délimitées, par exemple: "l'âge du malade est d'environ 20 ans" ou "le malade est plutôt âgé". L'incertain est une proposition où il existe un doute quant à sa vérité, par exemple: "l'âge du malade est probablement de 20 ans" ou "il est très possible que le malade soit âge". Dans TOULMED on prévoit un certain mélange dans les propositions. Par exemple, une proposition imprécise et incertaine peut être "l'âge du malade est très certainement inférieur à 20 ans", tandis que "le malade est probablement âgé" représente une proposition floue et incertaine. Nous constatons que dans ces définitions le flou est porté plus sur l'imprécis que sur l'incertain, car le contenu de la proposition est flou, donc on a une équivalence avec la notion d'imprécis de Dubois et Prade (voir figure 4.2).

### 4.3.4    Le système FLORAN

Le système FLORAN (Filtrage flou et Objets pour Raisonner par Analogie) [Salotti, 92] propose un modèle général basé sur le raisonnement analogique pour sélectionner un cas à partir d'une base de cas. Dans ce système les cas de références de la base de cas sont décrits de façon incomplète et imprécise. Le langage utilisé pour représenter le cas à été SHIRKA [Rechenmann, 85], langage de représentation centrée sur l'objet développé à l'INRIA. Dans ce langage, l'unité de la représentation est le schéma, qui permet de décrire une classe ou une instance. Un schéma est décrit par un ensemble d'attributs. Un attribut peut représenter une propriété, une caractéristique ou un lien vers d'autres classes ou instances. Dans un schéma de classe, les attributs sont décrits par un ensemble de facettes, permettant de définir le type de l'attribut, son domaine, etc. Dans un schéma d'instance, les attributs sont associés à une valeur ou une liste de valeurs, qui doivent vérifier les restrictions exprimées par les facettes. Finalement, une méthode de filtrage flou fondée sur la théorie des ensembles flous et des possibilités a été mise au point pour aller rechercher les cas similaires à la base de cas. Au début, ce système était destiné à l'aide à la décision en zone minière en sélectionnant dans une base de connaissances comprenant des descriptions de gisements aurifères (déjà été prospectés), les gisements les plus similaires à une zone d'étude décrite de façon incomplète et imprécise. L système devait proposer les cas sélectionnés à l'expert du Bureau de Recherches Géologiques et Minières d'Orléans pour l'aider dans sa prise de décision dans le diagnostic sur la présence ou non de minerais aurifère et les caractéristiques de cette éventuelle minéralisation sur la zone d'étude. Malheureusement, ce système n'a jamais été achevé car la base de connaissances présentait une complexité non résolue au niveau de la description des cas de références. Néanmoins, ce système a été appliqué à deux exemples pour sélectionner d'une part des cas concernant, d'une part des étudiants, et d'autre part, des voitures d'occasion, cas similaire au cas étudié, dans le but de déterminer un avis de poursuite d'études et un avis d'achat d'une voiture d'occasion. Dans la suite, nous allons discuter cette notion d'imprécision manipulée dans le système à partir de l'exemple de la voiture donnée dans la thèse de Salotti. L'imprécision se situe au niveau de la valeur d'un attribut d'un objet lorsqu'il n'est pas précisément connue mais que l'on peut délimiter une partie du domaine de l'attribut, dans laquelle la valeur se trouve certainement. Par exemple, le kilométrage d'une voiture d'occasion se situe entre 75000 et 85000 kms. La valeur de l'attribut "kilométrage" de l'objet "voiture" se situe certainement dans l'intervalle [75000, 85000].

Étant donné que le domaine de l'attribut est contenu, on peut définir un ensemble flou à partir du domaine de l'attribut vers l'intervalle réel [0,1].

### 4.3.5    Le système DOLMEN

Le système DOLMEN (DéfectOLogie et Mémoire d'Entreprise) [Simon, 97] calcule le risque d'apparition d'un défaut métallurgique par une mesure de similarité variant dans l'intervalle réel [0,1]. Cette mesure est exprimée en termes d'un degré de possibilité et un degré de nécessité entre un défaut source et un défaut de référence représentés par des distributions de possibilité sur l'intervalle réel [0,1]. Le système propose un modèle basé sur le raisonnement analogique pour sélectionner un cas à partir d'une mémoire d'entreprise (en défectologie) représentée sous la forme d'une base de cas de défauts métallurgiques lors des fabrications d'acier passées. La détection de défauts consiste à rechercher la fabrication d'acier défectueuse la plus proche de la fabrication testée. Le calcul de la mesure de similarité est fondé sur des techniques de filtrage flou. Le résultat fournit par ce système est la liste des défauts détectés. Le travail montre comme exemple la modélisation du défaut "gros grains". Les descriptions de défauts sont souvent incomplètes (lorsqu'il manque par exemple les valeurs sensibles de certains paramètres) et imprécises (lorsque l'on a des tendances). Dans ce cas, un ensemble flou permet de représenter l'ensemble des valeurs du paramètre qui sont influentes sur le défaut. Le degré d'appartenance d'une valeur d'un paramètre est interprété comme un degré d'influence de ce paramètre sur le défaut. Une valeur 0 signifie que le paramètre est neutre vis à vis du défaut modélisé, tandis qu'une valeur 1 du paramètre signifie que ce paramètre est le plus influent possible sur le défaut. Une valeur entre 0 et 1 représente le degré d'influence du paramètre dans le défaut. Ici on a une sémantique associée au paramètre pour indiquer son rôle dans la formation du défaut (déclencheur, aggravant, inhibiteur...).
Finalement, ce modèle peut être adapté au problème des caractéristiques mécaniques de l'acier afin de construire un système capable de prédire les caractéristiques mécaniques d'un acier à partir d'une description de fabrication.

Le processus de détection (le calcul d'apparition) d'un défaut est le suivant. Un défaut de référence est caractérisé en termes des paramètres qui sont représentés par des ensembles flous et une formule d'agrégation de ces paramètres, par exemple le défaut A = (P1 OR P2) AND P3 AND P4. Ensuite on teste la similarité partielle entre les valeurs des paramètres en cours de fabrication et les valeurs des paramètres modélisés, en calculant pour chaque cas un degré de possibilité et un degré de nécessité sur l'apparition du défaut pour ces valeurs des paramètres testés. Le degré d'apparition du défaut est donné par le calcul de la similarité globale dans la formule d'agrégation. Dans ce système, chaque paramètre est représenté par une seule courbe critique.

### 4.3.6    Le système FSQL

Le système FSQL (Fuzzy Structured Query Language) [Galindo, 99] permet la représentation et l'interrogation des données dans une base de données relationnelles floues (BDRF) de type GEFRED (GEneralized model for Fuzzy RElational Databases) [Medina, 94], [Medina *et al*, 94]. C'est justement ce type de scénario que nous avons choisi pour justifier d'avantage l'aspect technique de la gestion des connaissances étendue aux données imprécises et incertaines [Urrutia *et al*, 01a], c'est-à-dire, d'une part, la valeur de l'attribut de l'objet représenté dans la BDRF est imprécise, et d'autre part, les événements qui agissent sur les objets sont incertains. Pour la représentation de données dans la BDRF nous parlons de données imprécises, en revanche pour l'interrogation de la BDRF nous parlons de données incertaines. Dans la section suivante nous présentons le cadre conceptuel de la représentation et l'interrogation  des données dans une base de données relationnelles floues de type FSQL.

### 4.4 Représentation et interrogation des données floues dans FSQL

Les bases de données relationnelles floues ont fait l'objet de nombreuses recherches afin de construire des systèmes d'information permettant la représentation et l'interrogation de données floues complètes, incomplètes et nulles. Dans la plupart des cas, la théorie des sous-ensembles flous et la théorie des possibilités a été largement utilisée pour la modélisation de valeurs d'attributs connues, mal connues et nulles[16]. En général dans ces modèles, les informations sur les données peuvent être complètes, incomplètes ou nulles. Le système FSQL propose l'utilisation du modèle GEFRED (GEneralized model for Fuzzy RElational Databases) [Medina, 94] pour la modélisation de données (floues) car son implémentation est possible dans une base de données relationnelles Oracle [Jiménez *et al*, 03]. Ce modèle permet (1) le stockage d'objets contenant des valeurs d'attributs nulles, inconnues, inapplicables, étiquettes linguistiques, intervalles, valeurs triangulaires, et valeurs trapézoïdales. Ces valeurs sont décrites par des distributions de possibilité ; et (2) l'interrogation floue (flexible) basée sur une extension du langage SQL (Structured Query Language) développé sous le nom de FSQL (Fuzzy Structured Query Language) par [Galindo, 99].

---

[16] Par exemple dans l'étude de [Testemale, 84] on a défini un modèle relationnel étendu dans lequel sont représenté des valeurs d'attributs précises, imprécises, complètement inconnues, et partiellement connues ou floues.

Cette section est organisée en trois parties. La première partie appelée « représentation des attributs flous » montre le protocole de représentation d'attributs du modèle GEFRED. La deuxième partie appelée « représentation des données flous » montre le protocole de représentation de données du modèle GEFRED. La troisième partie appelée « interrogation de données flous » montre les sept filtres flous de possibilité et les sept filtres flous de nécessité, ainsi que la structure d'une requête floue du système FSQL.

### 4.4.1    Représentation des attributs flous

Le modèle GEFRED permet la representation de trois types d'attributs flous :

Attributs flous de Type 1 : ces attributs sont des "données précises" (classiques, sans imprécision) qui peuvent avoir des étiquettes linguistiques définis sur eux. Les attributs flous de Type 1 reçoivent une représentation égale aux données précises, mais ceux-ci peuvent être utilisés dans des conditions floues[17], c'est-à-dire, en utilisant des comparateurs flous, constants flous, des seuils d'accomplissement, etc. Par exemple, la température de l'eau est tiède.

Attributs flous de Type 2 : ces attributs sont des "données précises" données imprécises sur référentiel ordonné. Ces attributs admettent des données classiques comme floues, sous forme de distributions de possibilité sur un domaine sous-jacent ordonné. Par exemple, l'attribut "température" peut avoir les étiquettes linguistiques {"froid", "tiède", "chaud"}, définies sur un ensemble entre 5ºC et 45ºC.

Attributs flous de Type 3 : ces attributs sont des "données imprécises sur référentiel non ordonné régularisés". Ces attributs sont définis sur un domaine sous-jacent non ordonné, par exemple, l'attribut "couleur du papier" peut avoir les étiquettes linguistiques {"blanc", "café", "jaune"}.

---

[17] Ces attributs reçoivent le nom de *crisp* pour les différencier des attributs classiques.

### 4.4.2    Représentation des données floues

➢ **Les données complètes**

Le système d'information a une connaissance précise de la valeur de l'attribut pour un objet donné. En général, nous avons trois conditions à vérifier, d'abord que la valeur existe dans le domaine de l'attribut, que le domaine de l'attribut soit applicable à l'objet et enfin que la valeur soit connue (donc elle peut être stockée dans une relation de la base de données). Ces trois conditions sont à la base de la modélisation de données classiques que le modèle GEFRED a élargie aux données floues.

<u>La valeur de l'attribut est précise</u>

Les données complètes sont représentées dans le modèle GEFRED comme attributs flous de Type 1 sur un référentiel ordonné. La figure 4.4 montre la syntaxe pour une valeur précise qui admet traitement flou.

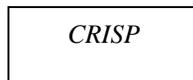

**Figure 4.4 :** Valeur précise

La distribution de possibilité sur un intervalle réel [0,1] dans un domaine ordonné $U$ associé à une *valeur précise* est définie la manière suivante :

$$CRISP = \{\mu_{CRISP}(x)/x :\ x \in U\}$$

Où pour tout élément $x$ du domaine ordonné $U$, $\mu_{CRISP}(x) = 1$.

➢ **Les données incomplètes**

Le système d'information a une connaissance imparfaite de la valeur de l'attribut pour un objet donné, dans le sens où les valeurs de l'attribut sont mal connues. Pour un objet donné le modèle GEFRED considère les situations suivantes :

La valeur de l'attribut est inconnue

La valeur existe et le domaine de l'attribut est applicable à l'objet mais la valeur est complètement inconnue. La figure 4.5 montre la syntaxe pour une valeur inconnue.

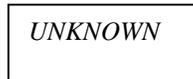

**Figure 4.5 :** Valeur inconnue

La distribution de possibilité sur un intervalle réel [0,1] dans un domaine ordonné $U$ associé à une *valeur inconnue* est définie la manière suivante :

$$UNKNOWN = \{\mu_{UNKNOWN}(x)/x : x \in U\}$$

Où pour tout élément $x$ du domaine ordonné $U$, $\mu_{UNKNOWN}(x) = 1$. Alors, $UNKNOWN = \{1/x : x \in U\}$.

Cela signifie que la valeur inconnue pour un attribut flou de Type 1 ou Type 2 peut prendre n'importe quelle valeur dans un domaine ordonné $U$. Et donc, UNKNOWN est utilisé quand il y a une ignorance totale par rapport à la valeur qui prend un attribut flou de Type 1 ou Type 2.

La valeur de l'attribut est inapplicable

La valeur existe et elle est connue, mais le domaine de l'attribut est inapplicable à l'objet. La figure 4.6 montre la syntaxe pour une valeur inapplicable.

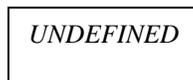

**Figure 4.6 :** Valeur inapplicable

La distribution de possibilité sur un intervalle réel [0,1] dans un domaine ordonné $U$ associé à une *valeur inapplicable* est définie la manière suivante :

$$UNDEFINED = \{\mu_{UNDEFINED}(x) / x : \ x \in U\}$$

Où pour tout élément $x$ du domaine ordonné $U, \mu_{UNDEFINED}(x) = 0$. Alors, $UNDEFINED = \{0/x : \ x \in U\}$.

Cela signifie que la valeur inapplicable pour un attribut flou de Type 1 ou Type 2 ne peut pas prendre aucune valeur dans un domaine ordonné $U$. Et donc, UNDEFINED est utilisé quand aucune valeur n'est possible pour un attribut flou de Type 1 ou Type 2.

<u>La valeur de l'attribut est partiellement connue</u>

La valeur existe et elle applicable à l'objet, mais elle est incomplètement connue, c'est-à-dire la donnée est imprécise, dans le sens où l'on connaît seulement un sous-ensemble du domaine de l'attribut auquel elle appartient.

Dans le modèle GEFRED nous avons quatre situations possibles pour représenter les valeurs *imprécises*. La figure 4.7 montre la syntaxe pour une valeur trapézoïdale. La figure 4.8 montre la syntaxe pour une étiquette linguistique. La figure 4.9 montre la syntaxe pour une valeur triangulaire, tandis que la figure 4.10 montre la syntaxe pour un intervalle.

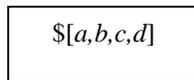

**Figure 4.7 :** Valeur trapézoïdale

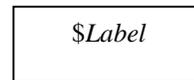

**Figure 4.8 :** Étiquette linguistique

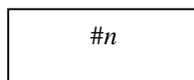

**Figure 4.9 :** Valeur triangulaire

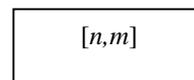

**Figure 4.10 :** Intervalle

❖ <u>Valeur trapézoïdale</u>

Soit [*a,b,c,d*] un intervalle de valeurs précises sur un domaine ordonné *U* associé à un attribut flou Type 2 (c'est-à-dire, avec un pré ordre *a*<*b*<*c*<*d*), alors une valeur trapézoïdale permet de représenter des données imprécises par une distribution de possibilité sur un intervalle réel [0,1] dans un domaine ordonné *U* de la manière suivante :

$$\$[a,b,c,d] = \{\mu_{\$[a,b,c,d]}(x)/x : \ x \in U, \ \mu_{\$[a,b,c,d]}(x) \in [0,1]\}$$

Où pour tout élément *x* du domaine ordonné *U*, $\mu_{\$[a,b,c,d]}(x)$ est la fonction d'appartenance, telle que :

$$\mu_{\$[a,b,c,d]}(x) = \begin{cases} [0,1] & \text{si } a \le x \le b \\ 1 & \text{si } b \le x \le c \\ [0,1] & \text{si } c \le x \le d \end{cases}$$

Alors,

$$\$[a,b,c,d] \ = \{[0,1]/ \ a \le x \le b, \ 1/ \ b \le x \le c, \ [0,1]/ \ c \le x \le d\}$$

❖ <u>Étiquette linguistique</u>

Soit *Label* une variable (ou descripteur) linguistique, par exemple, « grand », « très grand ». Or, L'étiquette linguistique permet de symboliser une valeur imprécise d'un attribut flou de Type 2 par une distribution de possibilités sur un domaine ordonné *U*, où la valeur est représentée par une valeur trapézoïdale.

❖ <u>Valeur triangulaire</u>

Soit *n* une valeur précise sur un domaine ordonné *U* associé à un attribut flou de Type 2, alors une valeur triangulaire (autour-*n* ou environ-*n*) permet de représenter des données imprécises par une distribution de possibilité sur un intervalle réel [0,1] dans un domaine ordonné *U* de la manière suivante :

$$\#n = \{\mu_{\#n}(x)/x : \ x \in U = [n-\gamma, \ n+\gamma], \ \mu_{\#n}(x) \in [0,1]\}$$

Où $\gamma$ est une marge pour tout élément *x* du domaine ordonné *U*, et $\mu_{\#n}(x)$ est la fonction d'appartenance, telle que :

$$\mu_{\#n}(x) = \begin{cases} [0,1] & \text{si } n\text{-}\gamma \leq x \\ 1 & \text{si } x = n \\ [0,1] & \text{si } x \leq n+\gamma \end{cases}$$

Alors,

$$\#n = \{[0,1]/\ n\text{-}\gamma \leq x,\ 1/n=x,\ [0,1]/\ n+\gamma \geq x\}$$

❖ Intervalle

Soient $n$ et $m$ deux valeurs précises (par exemple 25 et 45) sur un domaine ordonné $U$ associé à un attribut flou de Type 1, alors l'intervalle réel $[n,m]$ permet de représenter des données imprécises par une distribution de possibilité sur un intervalle réel $[0,1]$ dans un domaine ordonné $U$ de la manière suivante :

$$[n,m] = \{\mu_{[n,m]}(x)/x :\ x \in U,\ \mu_{[n,m]}(x) \in [0,1]\}$$

Où pour tout élément $x$ du domaine ordonné $U$, $\mu_{[n,m]}(x)$ est la fonction d'appartenance, telle que :

$$\mu_{[n,m]}(x) = \begin{cases} 0 & \text{si } m < x \\ 1 & \text{si } n \leq x \leq m \\ 0 & \text{si } x < n \end{cases}$$

Alors,

$$[n,m] = \{0/\ x < n,\ 1/n \leq x \leq m,\ 0/\ m < x \}$$

➢ **Les données nulles**

La valeur de l'attribut est nulle

La valeur peut être existante ou inexistante, ou bien elle peut être connue ou inconnue, c'est-à-dire qu'il n'y a pas de connaissance de la valeur de l'attribut flou de Type 1 ou Type 2. La figure 4.11 montre la syntaxe pour une valeur inapplicable.

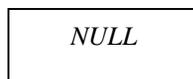

**Figure 4.11 :** Valeur nulle

---

La distribution de possibilité sur un intervalle réel [0,1] dans un domaine ordonné $U$ associé à une valeur *nulle* est définie la manière suivante :

$$NULL = \{\mu_{NULL}(x)/x : x \in UNKNOWN \text{ ou } x \in UNDEFINED\}$$

Où pour tout élément $x$ du domaine ordonné $U$, d'un attribut flou de Type 1 ou Type 2, $\mu_{UNDEFINED}(x) = 1$.

Alors,

$$NULL = \{1/x : x \in UNKNOWN \text{ ou } x \in UNDEFINED\}$$

Et donc, NULL est utilisé quand il y a une ignorance totale de la valeur d'un attribut flou de Type 1 ou Type 2. Le tableau 4.1 résume les quatre situations possibles.

| Valeur | Connue | Inconnue |
|---|---|---|
| Existante | *NULL* | *NULL* |
| Inexistante | *NULL* | *NULL* |

**Tableau 4.1 :** Valeur nulle

### 4.4.3    Interrogation des données floues

➢ **Filtrage flou dans FSQL pour attributs flous de Type 1 ou Type 2**

Le filtrage flou dans FSQL pour des attributs flous de Type 1 ou Type 2 est possible grâce aux filtres (ou comparateurs) flous afin de comparer des attributs flous entre eux ou avec une constante (floue ou crisp) [Galindo, 99]. Ces filtres flous sont une extension de filtres traditionnels tels que ≥, ≤, >, <, et =, utilisés dans l'algèbre relationnelle des bases de données.

Il s'agit donc de 14 comparateurs flous utilisés pour comparer deux données floues entre eux ou avec une constante[18]. Ces filtres flous ont été divisés en deux groupes. Le tableau 4.2 montre

---

[18] Dans cette thèse nous avons testé ces 14 comparateurs flous (mis au point par [Galindo, 99] dans FSQL pour des attributs flous de Type 1 ou de Type 2) sur un exemple, que nous présentons dans l'annexe II, à partir de la représentation mathématique de chaque opérateurs, ce qui nous a permis d'ailleurs de vérifier l'algorithme en langage PL/SQL de chaque opérateur. Nous devons signaler aussi que le nombre de ligne du code est énorme et son style de programmation est plus analytique que systémique, peut être pour cette raisons nous avons détectés de nombreuse résultats erronés par rapport à la définition de l'opérateur, ce qui a permis de débuguer le code du programme. Dans la première version de FSQL [Medina, 94] seulement ont été implémentés les filtres flous FEQ et NFEQ qui correspondent aux équations 4.1 et 4.2 (voir ci-dessus) respectivement. Les autres filtres ont été proposés plus tard dans la deuxième version de FSQL.

---

les 7 filtres flous de possibilité, et le tableau 3.3 montre les 7 filtres flous de nécessité avec son interprétation respective.

| Filtres Flous | Interprétation |
|---|---|
| FEQ | Fuzzy EQual, possiblement égal |
| FGT | Fuzzy Greater Than, possiblement plus grand que |
| FGEQ | Fuzzy Greater or EQual, possiblement plus grand ou égal |
| FLT | Fuzzy Less Than, possiblement plus petit que |
| FLEQ | Fuzzy Less or EQual, possiblement plus petit ou égal |
| MGT | Much Greater Than, possiblement beaucoup plus grand que |
| MLT | Much Less Than, possiblement beaucoup plus petit que |

**Tableau 4.2 : Filtres flous de possibilité**

| Filtres Flous | Interprétation |
|---|---|
| NFEQ | Necessarily Fuzzy EQual, nécessairement égale |
| NFGT | Necessarily Fuzzy Greater Than, nécessairement plus grande que |
| NFGEQ | Necessarily Fuzzy Greater or EQual, nécessairement plus grande ou égale |
| NFLT | Necessarily Fuzzy Less Than, nécessairement plus petite que |
| NFLEQ | Necessarily Fuzzy Less or EQual, nécessairement plus petite ou égale |
| NMGT | Necessarily Much Greater Than, nécessairement beaucoup plus grande que |
| NMLT | Necessarily Much Less Than, nécessairement beaucoup plus petite que |

**Tableau 4.3 : Filtres flous de nécessité**

➢ **Mesure de possibilité et de nécessité dans FSQL**

Le calcul de la similarité est donné à partir de la fonction CDEG (Compatibility DEGree) de FSQL [Galindo, 99] qui calcule pour chaque tuple d'une table dans une base de données relationnelle FSQL sous Oracle, le degré d'accomplissement d'un attribut flou Type 1 ou Type 2 (qui correspond à une colonne de la table) avec la condition d'une requête floue, qui est spécifiée dans l'expression WHERE de la sentence SELECT par des filtres flous. La comparaison se fait par rapport à une autre colonne de la table (autre attribut flou du même type) ou une constante. Autrement dit, la fonction CDEG exprime le degré d'accomplissement d'une comparaison floue. Le résultat de cette comparaison dépend du filtre (ou comparateur) flou utilisé (voir tableaux 4.2 et 4.3). La figure 4.12 montre la syntaxe d'une requête floue FSQL :

| | |
|---|---|
| SELECT *nom des colonnes* (*A*,...) | SELECT *nom des colonnes* (*A*,...) |
| CDEG(*A*) | CDEG(*A*) |
| FROM *nom de la table* | FROM *nom de la table* |
| WHERE *A filtre flou B* | WHERE *A filtre flou B* |
| THOLD γ | THOLD γ |

**Figure 4.12 :** Requêtes floues FSQL

Où THOLD γ (threshold γ) est une constante qui permet de fixer un seuil $\gamma \in [0,1]$ dans la requête floue FSQL.

## 4.5 Le système opérationnel : Aspect technique de la gestion des connaissances étendu au flou

Dans cette section nous nous intéresserons à l'aspect technique de la gestion de connaissances, c'est-à-dire le système opérationnel. Pour explorer cette voie nous avons deux alternatives possibles. La première relative aux NTIC du KM pour la gestion de la communication, de la coordination et de la coopération (afin de définir un réseau de travail), la deuxième relative aux outils de gestion de l'information et de données (pour en extraire la connaissance dans l'entreprise), dans le sens ou le système opérationnel est couplé structurellement au système de connaissances.

Nous, avons choisi la deuxième alternative, c'est-à-dire la gestion de l'information et de données pour en extraire la connaissance. Dominique Crié dans l'article intitulé *De l'extraction des connaissances au Knowledge Management* (publié en 2003), fait d'ailleurs une classification de ces outils par rapport à l'aspect technique[19] (*text mining*, *web mining*, et *data mining*…). Nous avons constaté que dans la plupart des cas les outils utilisés pour le *data mining*, et en général les outils d'exploitation de bases de données relationnelles, ne traitent que de données précises et certaines. En cela, les bases de données relationnelles floues sont une alternative au problème de l'imprécision et de l'incertitude de données sur la base de (1) la théorie des sous-ensembles flous ; (2) la théorie des possibilités ; et (3) le filtrage flou, c'est-à-dire sur la base de méthodes et outils de la logique floue et de la théorie de l'incertain.

Dans ce contexte, nous avons étendu l'aspect technique de la gestion des connaissances aux données imprécises et incertaines, dans l'espoir de faire une extraction des connaissances à partir de ces données. Néanmoins nous n'avons pas eu l'ambition de travailler à partir de plusieurs sources de données pour en faire un *fuzzy data mining*, notre ambition est restée modeste et nous avons préféré opter pour un seul format de données, afin de faciliter le traitement de données floues.

---

[19] Une classification de l'aspect social de la gestion des connaissances au travers de méthodes et outils se trouve chez Barthelme-Trapp dans le texte de sa thèse *Une approche constructiviste des connaissances : contribution à la gestion dynamique des connaissances* (publié en 2003), et dans son article *Analyse comparée de méthodes de gestion des connaissances pour une approche managériale* (publié en 2001), ainsi que dans les deux ouvrages de Rose Dieng et de son équipe de recherche ACACIA à l'INRIA Sophia Antiopolis, l'un est *Méthodes et outils pour la gestion des connaissances* (publié en 2000), l'autre est *Méthodes et outils pour la gestion des connaissances : une approche pluridisciplinaire du Knowledge Management* (publié en 2001)).

Notre intérêt pour l'imperfection de la connaissance remonte à l'article, intitulé *Proposal for the extension of MER for knowledge management and fuzzy consultations* [Urrutia *et al*, 01b], dans cette réflexion avec Angélica Urrutia et Mario Piattini nous avons mise en place un cadre conceptuel pour la prise en charge du flou à partir d'un modèle MER (dans une problématique de gestion des connaissances, liées à la localisation de compétences organisationnels par rapport à une tâche donnée). Notre démarche divisé une problématique de gestion des connaissances en trois étapes. La première étape, appelée *analyse des connaissances* avait comme objectif l'identification des connaissances critiques pour l'entreprise. La deuxième étape, appelée *structuration des connaissances* avait comme objectif l'identification des variables en termes d'*entités* et de *relations* pour un scénario donné. C'est justement dans cette étape qu'un modèle MER peut jouer un rôle important pour la représentation de ces variables identifiés. La troisième étape, appelée *exploitation des connaissances* avait comme objectif l'identification des requêtes à une base de données relationnelle (BDR). Dans cette étape nous avons perçu que la connaissance imparfaite du domaine étudié (les compétences des ressources humaines) exigé l'incorporation des requêtes floues (aussi connu par les spécialistes de bases de ; données comme flexibles) pour l'exploitation du modèle MER. Ceci pose la question de la représentation et de l'interrogation de données floues dans une base de donnés relationnelle (BDR).

La représentation de données floues, a été étudié dans trois articles : (1) *Representación de Información Imprecisa en el Modelo Conceptual EER Difuso* [Urrutia *et al*, 01a] ; (2) *Extensión del Modelo Conceptual EER para Representar Tipos de Datos Difuso* [Urrutia *et al*, 02]; et (3) *Extensión del Conocimiento del Dominio de CommonKADS con Lógica Difusa* [Jiménez et Urrutia, 02a].

L'interrogation de données floues a été possible grâce à un système client-serveur FSQL (Fuzzy SQL)[20] de bases de données relationnelles sous plate-forme Oracle 8, développé et mis en service, à des fins de recherche, par José Galindo dans le contexte de sa thèse intitulé *Tratamiento de la imprecisión en bases de datos relacionales: Extensión del modelo y adaptación de los SGBD actuales*, publié en 1999 par l'Université de Granada[21]. Ceci permet l'utilisation d'un moteur d'interrogation floue à travers des attributs flous, et des comparateurs flous sur un modèle conceptuel pour des bases de données relationnelles floues[22]. Ce modèle conceptuel a été défini par Angélica Urrutia dans le cadre de sa thèse intitulé *Defininición de un Modelo Conceptual para Bases de Datos Difusas* publié en 2003 par l'Université de Castilla-La Mancha[23]. Ainsi, on a les trois couches d'un système de bases de données (couche conceptuelle, couche logique et couche physique) sur un même système (le système client-serveur FSQL sous Oracle 8).

---

[20] Dans cette thèse nous parlons indifféremment de système FSQL, langage FSQL, modèle FSQL, système client-serveur FSQL, Progiciel FSQL, etc. Donc, dans un sens générique "FSQL" est simplement une abréviation.
[21] Cette thèse [Galindo, 99] ainsi que d'autres travaux relatifs aux bases de données relationnelles floues se trouvent dans le site Web http://www.lcc.uma.es/~ppgg/galindo.html

Ainsi, nous proposons une démarche en trois étapes, afin de gérer les connaissances d'un système opérationnel. La première étape consiste à la détection du besoin utilisateur. La deuxième étape consiste à la traduction de ce besoin dans un langage Fuzzy Structured Query Language (FSQL). La troisième étape est l'extraction de connaissance d'une base de données relationnelles floues (BDRF).

## Conclusion du chapitre 4

Nous avons présenté dans ce chapitre :

- premièrement, le problème de la modélisation dans un système de connaissances imparfaites selon l'aspect social de la gestion des connaissances. Ceci a été envisagé à travers cinq enjeux, à savoir : la connaissance imparfaite comme un modèle (en tant que moyen de l'être humain pour appréhender le monde réel : flou et non flou), penser et raisonner en termes flous (en tant que capacité de l'être humain pour gérer le flou et le non flou dans ses activités ordinaires), les images des objets et des événements (en tant que capacité du système d'information pour gérer les objets et les événements sur lesquels il faut agir), les systèmes d'information de gestion (en tant qu'outil pour gérer d'abord une problématique organisationnelle : les relations humaines, et ensuite une problématique technologique : les bases de données relationnelles), et l'imperfection de l'information sous l'angle des bases de données relationnelles floues (en tant qu'outil "élargi" pour faire le flou sur le non flou, autrement dit construire une BDRF sur une BDR). Cette démarche nous a permis de mettre en évidence qu'une capacité naturelle de l'être humain (la gestion de données floues) est impossible à faire sur une machine floue, sans passer pour une conversion de données (objets flous, événements flous, requêtes floues), ce qui est une contrainte technologique incontournable de nos jours.

---

[22] Dans l'annexe I nous avons donnée la définition de ces opérateurs, tandis que dans l'annexe II nous avons donnée un exemple avec l'utilisation de ces opérateurs, que nous les utilisons comme de filtres flous dans une base de données relationnelles floues de type FSQL.

- deuxièmement, nous avons présenté l'aspect théorique de la connaissance imparfaite sous l'angle de la théorie de l'imprécis et de l'incertain (à travers la théorie des sous-ensembles flous, la théorie des possibilités, et le filtrage flou). La théorie des sous-ensembles flous nous a permis d'introduire le concept de degré d'appartenance, à partir cinq points de vues, à savoir : comme un degré de préférence, un degré d'accomplissement, un degré de nécessité, un degré de possibilité, et un degré de similitude. La théorie des possibilités d'une part, nous a permis de faire la modélisation mathématique des mesures de possibilité et de nécessité, et d'autre part, d'établir le lien entre ces mesures et le concept d'ensemble flous. Le filtrage flou nous a permis d'illustrer la procédure d'interrogation de données imprécises dans une base de données relationnelles floues à partir de requêtes floues. Deux types de systèmes ont été envisagés pour l'application du filtrage flou. L'un a été le système à base de connaissances de type base de cas, et l'autre est le système à base de connaissances de type base de règles. Cette démarche nous a permis, en particulier, de comprendre la logique de programmation des mesures de possibilité et de nécessité qui ont été intégrées dans les opérateurs flous du progiciel FSQL, et en général, de mettre en évidence la contribution applicative de la théorie de l'imprécis et de l'incertain dans le domaine de l'ingénierie des systèmes d'information.

- troisièmement, nous avons passé un revue un certain nombre de systèmes qui ont intégré la théorie de l'imprécis et de l'incertain, tels que : le système F-MRP (Fuzzy-Manufacturing Ressource Planning) [Reynoso, 04], le système SIMCAIR (Système d'Interrogation Multi-Critères Avec Importances Relatives) [Andres, 89], le système TOULMED [Buisson, 87], le système FLORAN (Filtrage flou et Objets pour Raisonner par Analogie) [Salotti, 92], le système DOLMEN (DéfectOLogie et Mémoire d'Entreprise) [Simon, 97], et le système FSQL (Fuzzy Structured Query Language) [Galindo, 99]. Ceci nous a permis d'illustrer (avec des applications industrielles), les outils mis en place de la théorie de l'imprécis et de l'incertain.

- quatrièmement, nous avons positionné notre problématique de gestion des connaissances dans un système opérationnel à partir du système FSQL. La modélisation de données imprécises et incertaines dans une base de données relationnelles floues (BDRF) de type FSQL a été faite à partir de la représentation des attributs flous, la représentation des données floues, et l'interrogation des données floues. Ceci a été possible par la construction d'une BDRF à partir de trois couches : la couche conceptuelle, la couche logique et la couche physique. La couche conceptuelle a été définie

---

[23] Cette thèse [Urrutia, 03] ainsi que d'autres travaux relatifs à la modélisation conceptuelle de données floues se trouvent

par le modèle conceptuel FuzzyEER [Urrutia, 03] et le progiciel FuzzyCase [Urrutia, 03] associé pour la représentation des entités et relations floues. La couche logique a été définie par le modèle GEFRED [Medina, 94], [Medina *et al*, 94] et le langage FSQL [Galindo, 99] pour la représentation des données floues. La couche physique a été définie par le système Oracle 8 et l'interface FIRST (Fuzzy Interface for Relational SysTems) qui sert de protocole de communication entre la couche logique et la couche physique (oracle 8).

- cinquièmement, nous avons formulé une démarche composée de trois étapes, afin de gérer les connaissances d'un système opérationnel. La première étape consiste en la détection du besoin utilisateur. La deuxième étape consiste en la traduction de ce besoin dans un langage Fuzzy Structured Query Language (FSQL). La troisième étape est l'extraction de connaissance d'une base de données relationnelles floues (BDRF). Cette démarche sera mise en pratique dans une problématique industrielle liée à la manufacture du carton, dans le domaine de l'imprécis et de l'incertain, que nous présenterons dans le chapitre 5.

La conclusion générale de ce chapitre est que le système (connaissance et opérationnel) dans un environnement imprécis et incertain est enfermé dans une démarche d'ingénierie des systèmes d'information, où l'on suppose que le monde réel, dont on veut garder une image fidèle (certaine et précise, autrement dit non floue) afin d'agir sur cette réalité ou une partie de cette réalité, est formé d'objets et d'événements, c'est-à-dire de déclencheurs d'actions sur les objets. Il nous faut donc des méthodes et outils pour coder le flou, et aussi des supports pour les stocker et les manipuler, d'une autre façon qu'au travers de protocoles de conversion pour simuler le flou vis-à-vis du non flou.

*Les systèmes ne sont pas dans la nature, ils sont dans l'esprit des hommes.*
*Claude Bernard, biologiste français*

Le chapitre 3 a permis de proposer le modèle autopoïétique de la gestion des connaissances imparfaites, sur la base de quatre hypothèses, à savoir :

- l'hypothèse de l'enaction ;
- l'hypothèse de spontanéité des relations ;
- l'hypothèse du noyau invariant du système de connaissance ;
- l'hypothèse de la connaissance imparfaite.

Le chapitre 4 a permis d'étudier davantage l'hypothèse de la connaissance imparfaite dans le système opérationnel du modèle proposé. Ainsi, pour le système opérationnel nous avons mis au point une démarche en trois étapes pour gérer les connaissances. La première étape consiste à la détection du besoin utilisateur. La deuxième étape consiste en la traduction de ce besoin dans un langage Fuzzy Structured Query Language (FSQL). La troisième étape est l'extraction de connaissance d'une base de données relationnelles floues (BDRF).

Le chapitre 5 a pour objectif de valider ces quatre hypothèses, autrement dit le *modèle autopoïétique de la gestion des connaissances imparfaites*. Cette validation est faite sur la base d'une problématique industrielle. Ainsi :

- pour valider l'hypothèse de l'enaction, il nous faut supposer l'existence d'une dualité organisation/structure, dans laquelle le domaine social (relations humaines) et physique (individus, matière, énergie, et symboles) sont enactés, c'est-à-dire la dynamique de la dualité organisation/structure correspond à une clôture opérationnelle. L'étude de cas nous permet d'interpréter le domaine social et physique sur un champ expérimental ;

- pour valider l'hypothèse de spontanéité des relations, il nous faut supposer l'existence d'un système d'actions sur lequel la clôture opérationnelle se manifeste. L'étude de cas nous permet d'interpréter cette spontanéité des relations sur un champ expérimental ;

- pour valider l'hypothèse du noyau invariant, il nous faut supposer l'existence d'un système de connaissance autour d'un patron commun sur lequel les activités sont organisées. L'étude de cas nous permet d'identifier ce noyau invariant sur un champ expérimental ;

- pour valider l'hypothèse de la connaissance imparfaite, il nous faut supposer, d'une part, l'existence de données imprécises et incertaines, et d'autre part qu'il est possible d'extraire des connaissances à partir de ces données. L'étude de cas nous permet de mettre en place la démarche conçue dans le chapitre 4.

L'objectif de ce chapitre 5 est la validation du modèle autopoïétique de la gestion des connaissances imparfaites, à partir d'un cas d'étude. Ce cas a eu lieu dans deux sites industriels du domaine de la manufacture du papier-carton. L'un est situé à Santiago du Chili (manufacture du papier), et l'autre dans la région du Maulé au sud du pays (manufacture du carton). C'est le deuxième site de l'entreprise CMPC que nous avons choisi davantage pour mettre sur "papier" cette expérience industrielle.

Ce chapitre 5 est organisé en trois parties. La première partie appelée « Mise en œuvre du système de connaissance » permet de valider les hypothèses du modèle proposé dans le chapitre 3 dans un contexte industriel. La deuxième partie appelée « Mise en œuvre du système opérationnel » permet d'exécuter la stratégie développée au chapitre 4 pour la représentation de données imprécises et l'interrogation d'une base de cas dans une base de données relationnelles floues de type FSQL à l'aide d'un filtre flou (c'est-à-dire un degré de possibilité). La troisième partie appelée « Analyse de la problématique industrielle » traite de la description du contexte de la gestion des connaissances et des aspects des connaissances imparfaites que les acteurs ont de la réalité (le *business*) qu'ils doivent gérer, afin d'identifier une problématique industrielle capable d'être prise en charge par un système opérationnel. Nous avons choisi un scénario relatif au *processus de traçabilité des défauts du carton*. Ce scénario nous a permis de rassembler des informations pendant notre stage sur le sujet.

## 5.1. Mise en œuvre du système de connaissance

### 5.1.1. Hypothèse du noyau invariant du système de connaissance

Le business de l'entreprise CMPC Maulé est la manufacture et la vente de carton. Pour cela l'entreprise compte avec une machine à papier[1] dont la laize (largeur) permet de fabriquer une feuille de carton de 4,80 m de large à 450 m à la minute, cela signifie une production de 160.000 tonnes par an répartie dans une gamme d'en général 29 types de cartons avec un rang standard de grammage[2] qui varie entre 200 g/m2 et 450 g/m2. Sa production est destinée aux marchés national et international des matériaux d'emballage pour l'industrie alimentaire (pour des produits surgelés et secs) la pharmacie, la filière beauté hygiène parfumerie, les industries des équipements du foyer, les industries de biens d'équipements, les industries de déménagement, les industries du tabac, etc. d'une part, et d'autre part des matériaux de papeteries pour l'industrie graphique (livres, cahiers, magazines, catalogues, brochures, cartes de visites, cartes postales, affiches, etc.).

Dans ce contexte industriel, nous avons repéré quatre systèmes de connaissances qui ont permis de vérifier l'existence d'un noyau invariant, c'est-à-dire une connaissance qui ne change pas, afin de définir un patron d'organisation commun autour de cette connaissance[3]. Parmi ces systèmes, nous avons le système de connaissance autour des *recettes*, le système de connaissance autour des *conflits*, le système de connaissance autour des *réclamations*, le système de connaissance autour des *défauts*.

➢ **Système de connaissance autour des recettes**

Dans ce système, le patron commun d'organisation, c'est la connaissance[4] sur la fabrication du carton. Nous avons constaté que pour les gens de l'unité de planification et contrôle des opérations, ce qui importe le plus ce sont les connaissances autour de la formation de la structure des fibres de la cellulose. En somme, comme nous le verrons dans la section 5.3.3, le carton est un agglomérat de fibres de cellulose qui sont reparties en général entre quatre et sept couches (voir tableau 5.14), et dont la connaissance réside dans la façon dont ces fibres sont organisées pour former la totalité. En fait, les mécanismes de prise de décision au niveau des procédés et paramètres rattachés à la conversion de la matière pour former ces fibres sont appelés en fonction de la gamme de carton à fabriquer (voir tableaux 5.1 et 5.12). Par conséquent, ces gens là doivent bien maîtriser les procédés de manufacture, en d'autres termes la structuration des fibres et sa répartition dans les couches du carton et le mélange des pâtes. A cet égard, ce que nous avons constaté ce sont deux traits de la connaissance qui nous semblent extrêment importants pour comprendre la complexité de la traçabilité des défauts. L'un de ces traits concerne la capitalisation de connaissances sous la forme d'expériences du *pourquoi* et du *comment* impliqués dans les procédés et paramètres sous-jacents. Par exemple, pourquoi nous avons décidé de choisir telle matière première et non pas une autre pour la formation de la pâte ; comment nous avons fait le mélange des pâtes, avec quel additif, quelle quantité, à quel moment. L'autre trait important réside dans la réutilisation de cette expérience à partir d'une *recette*. Il s'agit d'une mémoire d'entreprise qui prend la forme d'un document qui conserve la connaissance qui a été explicitée et qui peut être exploitée. Cette recette constitue, bien entendu, un secret industriel qui est gardé précieusement par l'entreprise, mais nous pensons qu'elle reste pour l'entreprise un résultat idéal final[5], car elle est toujours en train de l'améliorer à partir de l'expérience accumulée, des nouveaux besoins du marché et de l'innovation technologique de la chaîne produits-procédés-processus.

---

[1] Le carton est un cas particulier du papier, c'est pourquoi la machine a conservé son nom d'origine.
[2] Le grammage correspond au poids en grammes d'un mètre carré de papier.
[3] Bien entendu au niveau de la structure les attributs des composants vont varier pour maintenir l'unité de l'organisation.
[4] Nous voulons dire par là, l'ensemble des savoirs, savoir-faire, savoir-technique autour des produits, procédés, et processus du système de fabrication.

Nous constatons que ces deux traits du système de fabrication (produits-procédés-processus), permettent de verifier l'existence d'un noyau invariant. C'est la gestion de la recette le noyau invariant du système de connaissance.

➢ **Système de connaissance autour des conflits**

Dans ce système le patron commun d'organisation, c'est la connaissance[6] sur la commercialisation du carton, nous avons constaté que pour les gens de l'unité commerciale ce qui importe ce sont les connaissances du réseau de compromis entre les clients, les agents externes et internes, et l'unité de planification et contrôle des opérations. Le système de planification permet de gérer les conflits. Les mécanismes de prise de décision de ce système ont la charge de résoudre, par exemple, les conflits suivants : Est-ce que la commande du client sera fabriquée pour la date de nécessité[7] ? Y aura-t-il un retard dans la prise en compte de sa fabrication ? Est-ce que ce retard peut entraîner le changement de la date de compromis[8] de la commande ?

---

[5] Par exemple, au sens de la théorie TRIZ.
[6] Nous voulons dire par là, l'ensemble des savoirs, savoir-faire, savoir-technique autour des produits et clients.
[7] C'est la date où le produit doit être dans l'entrepôt.
[8] C'est la date où le produit doit être chez le client.

Nous constatons que dans le système commercial (produits), c'est la gestion des conflits le noyau invariant.

> ➢ **Système de connaissance autour des réclamations**

Dans ce système le patron commun d'organisation, c'est la connaissance[9] sur la relation client. Cette relation est relative à la prise en charge de l'insatisfaction du client par rapport au produit acheté, c'est-à-dire cette relation est relative au concept CRM (Customer Relationship Management). A cet égard, nous avons constaté que l'outil informatique mis à la disposition de l'unité d'attention au client pour l'enregistrement de la réclamation est une sorte de fiche technique sur un tableur Excel avec l'historique des réclamations pour l'année en cours, qu'on appelle base des réclamations clients. Donc, dans le système commercial (produits), c'est la gestion des réclamations le noyau invariant.

> ➢ **Système de connaissance autour des défauts**

Dans ce système le patron commun d'organisation, c'est la connaissance[10] des défauts, nous avons constaté que ce qui intéresse les gens de l'unité de planification et contrôle des opérations, ce sont les connaissances autour des mécanismes[11] de formation des défauts. Ici, nous avons constaté qu'il y a trois niveaux de connaissance autour de la chaîne produit-procédé-processus : (1) au niveau produit, ce sont les connaissances sur les défauts détectés chez le client, par exemple au cours de la production d'emballages, de la production des applications graphiques, ou bien dans les techniques d'impression utilisées; (2) au niveau procédé, ce sont les connaissances sur les défauts survenus lors de la conversion de la matière, par exemple si le procédé est la formation de la pâte, alors ce sont les connaissances relatives au fait de pouvoir détecter des risques de formation de défauts dans la recette; et (3) au niveau processus, ce sont les connaissances sur les défauts survenus lors de la conversion du produit, par exemple dans le transfert du produit vers les entrepôts, la codification du produit, le marquage du produit, etc.

Ici, nous constatons que le noyau invariant, c'est la gestion des défauts. Cette gestion est décalée à trois niveaux. Le premier niveau permet de détecter des risques de formation de défauts dans l'utilisation du carton chez le client, le deuxième niveau permet de détecter des risques de formation de défauts dans la recette, et le troisième niveau permet de détecter des risques de formation de défauts dans les opérations et l'enchaînement des opérations dans les ateliers discontinus.

---

[9] Nous voulons dire par là, l'ensemble des savoirs, savoir-faire, savoir-technique autour de la description des défauts, sans oublier le côte humaine : savoir-être.
[10] Nous voulons dire par là, l'ensemble des savoirs, savoir-faire, savoir-technique autour des produits, procédés, et processus de fabrication.

En conclusion, nous avons présenté quatre champs des connaissances (recettes, conflits, réclamations, et défauts). Ces connaissances sont un noyau invariant de quatre systèmes (le système de connaissance autour des *recettes*, le système de connaissance autour des *conflits*, le système de connaissance autour des *réclamations*, le système de connaissance autour des *défauts*).

Nous pensons que ces quatre systèmes de connaissance offrent un cadre conceptuel intéressant pour faire une réflexion sur la gestion des connaissances imparfaites dans un système opérationnel, et ceci pour deux raisons. La première raison, parce qu'il y a dans ces quatre systèmes (le système de connaissance autour des *recettes*, le système de connaissance autour des *conflits*, le système de connaissance autour des *réclamations*, le système de connaissance autour des *défauts*) l'existence d'un environnement imprécis et incertain, pas pris en compte par l'entreprise dans sa gestion, et à notre avis, c'est le « maillon[12] faible » de la synergie avec les autres systèmes d'information de l'entreprise, tels que : le système de gestion de la logistique, le système de gestion des entrepôts, le système de gestion de la codification des produits, le système de gestion de marquage des produits, et le système de gestion de la production. En fait, tous ces systèmes, nous les avons imaginés dans notre réflexion, d'abord comme des systèmes opérants (données précises et imprécises) du modèle OID de Le Moigne, et puis comme des systèmes d'information (événements certains et incertains). La deuxième raison est que nous avons préféré aller rechercher une problématique commune dans ce maillon faible afin de faire réfléchir l'entreprise à la gestion des connaissances imparfaites. En effet, comme nous le verrons dans le cas 1 (Carton 301B) de la section 5.3.3 que nous avons appelé *Rôle du grammage dans le risque d'apparition d'un défaut*, cette problématique est relative à la traçabilité des défauts du carton par la prise en compte de l'imprécision dans la définition du profil du grammage du client à partir des informations du grammage des produits semi-élaborés ou élaborés (y compris les commandes qui ont été annulées).

En fait, comme nous pouvons le voir dans le tableau 5.2, le grammage a la signification d'un poids et rien de plus. Cependant, ce poids a été au cœur de notre réflexion sur la traçabilité des défaus détectés chez le client comme nous le verrons dans la section 5.3.3. Ici, le noyau invariant du système de connaissance réside dans la *traçabilité*, c'est-à-dire dans le suivi des flux d'information et non pas dans les défauts eux-mêmes ni dans leurs mécanismes de formation, car les structures et processus de structuration de composants du carton vont évoluer avec les nouvelles recettes.

---

[11] Principalement les principes physico-chimiques mis en jeu.
[12] Ce terme englobe le transport des produits élaborés, le stockage dans des entrepôts et la production.

Enfin, ces quatre systèmes de connaissance nous les avons imaginé, principalement, à partir de l'étude de quatre systèmes de l'entreprise en question, à savoir : le système de fabrication, le système de planification, le système de gestion des réclamations, et le système de gestion des défauts. Dans le paragraphe suivant nous allons essayer de les expliquer afin de vérifier les autres hypothèses du modèle proposé.

## 5.1.2. Hypothèse de l'enaction du système de connaissance

Cette hypothèse, est relative à la définition des relations de production des composants que nous avons formulé à partir de trois sous-hypothèses, à savoir : système clos, système vivant, et système viable. C'est la troisième sous-hypothèse qui nous intéresse ici. Pour la vérifier nous devons vérifier l'existence d'un domaine social à travers des relations humaines et d'un domaine physique composé des individus, matière, énergie, et symboles.

Le système de fabrication a permis de vérifier cette hypothèse.

L'approche système OID, permet de décrire le système de fabrication, en premier lieu, comme un système opérant (ce que l'on cherche, ce sont les activités, le "quoi"), et puis la complexité de ces activités est expliquée autour des informations et décisions qu'ils produisent et génèrent.

➢ **Le système de fabrication**

Ce système gère les informations autour de la manufacture (produits-procédés-processus) du carton. D'abord dans la définition de la gamme et sub-gamme de fabrication, puis dans l'enchaînement des opérations de la ligne de production de la machine à papier (procédés-carton continus et discontinus) ensuite dans l'enchaînement des opérations dans les ateliers discontinus (processus-carton) et enfin dans la traçabilité des produits par un suivi des flux des informations dans toute la chaîne produits-procédés-processus de manufacture du carton. Néanmoins, il faut souligner que nous n'avons pas l'intention ici d'entrer dans les détails, par exemple en génie des procédés papetiers, ou bien dans les problèmes de gestion de la production et de l'ordonnancement de tâches dans les ateliers discontinus, mais plutôt de donner une vision générale des connaissances[13] liée à l'organisation des fibres, aux défauts et leurs mécanismes de formation, ce qui nous aidera à comprendre la gestion de la traçabilité des défauts et les améliorations qu'on peut envisager dans ce système.

Produits-carton

La famille de cartons de l'entreprise CMPC Maulé est organisée autour de quatre gammes de cartons repartis dans une sub-gamme d'en général 29 produits, tels que le montre le tableau 5.1.

| Gammes des cartons | Code | Grammage (g/m2) | N° | Recommandations d'usage | Méthodes d'impression |
|---|---|---|---|---|---|
| Carton Maulé Reverso Café (maulé RC) | 1 | 200 - 450 | 13 | *Données confidentielles* | Offset ou huecograbado. |
| Carton Maulé Reverso Blanco (maulé RB) | 5 | 210 - 360 | 7 | *Données confidentielles* | Offset ou huecograbado. |
| Carton Maulé Reverso Manila (maulé GC2) | 7 | 230 - 340 | 7 | *Données confidentielles* | Offset ou huecograbado. |
| Carton Maulé Reverso Estucado (maulé Graphics) | 4 | 220 - 275 | 2 | *Données confidentielles* | Offset ou huecograbado. |

**Tableau 5.1 :** Famille des cartons de l'entreprise CMPC Maulé

La colonne code correspond à l'identificateur de la gamme ; la colonne grammage représente la variation de grammage de la gamme, et dont le grammage le plus faible fait 200 g/m2 et le plus lourd fait 450 g/m2 ; la colonne N° représente la sub-gamme, c'est-à-dire le nombre de produits de la gamme (voir tableau 5.12), il faut noter ici que chaque carton de la sub-gamme a sa propre spécification technique de manufacture, car les valeurs des paramètres, relatives aux caractéristiques physiques (voir tableau 5.2) de la sub-gamme vont être différentes ; la colonne recommandations d'usage suggère l'utilisation adéquate de la gamme chez le client ; enfin la colonne méthodes d'impression définit le type d'impression recommandé. D'ailleurs les recommandations d'usage et méthodes d'impression sont un point clé dans la formation des défauts chez le client, puisque les propriétés du carton (voir tableau 5.2) sont définies par rapport à ceux-ci.

---

[13] Nous voulons dire par là le flux des données et le flux des informations qui ont une signification pour la traçabilité des produits, en particulier des produits avec défauts.

Bref, il s'agit d'un système de production complexe, car comme le montre le tableau 5.1 en plus d'avoir une gestion de la production selon quatre gammes, nous avons une sub-gamme de produits à gérer, où chaque produit de la gamme a des spécifications techniques de manufacture différentes par rapport aux autres produits de cette même gamme. Cela signifie que chaque sub-gamme de cartons possède une recette propre et donc une séquence d'élaboration spécifique de la machine à papier, que l'on appelle bloc. En fait, pour le système de fabrication, un bloc est une tâche, qui est gérée par le système de planification, comme nous le verrons dans le paragraphe suivant, d'une part, sur une Carte Gantt à travers une série de dates, au plutôt ou au plus tard, de début ou fin des opérations, et d'autre part, est géré l'état de la tâche en question : finie, programmée, active.

Procédés-carton

Les procédés-carton correspondent aux procédés continus et discontinus de la machine à papier, où la gamme d'usinage de la ligne de production à enchaîner, c'est-à-dire que nous avons regroupé la séquence des opérations de la machine à papier dans trois étapes génériques : (1) les procédés continus de préparation de la pâte ; (2) les procédés continus de formation de la feuille ; et (3) les procédés discontinus de conversion de la feuille.

Dans un premier temps, les procédés continus de préparation de la pâte suivent ce cheminement : désintégration (la pâte est ramenée dans un état liquide), raffinage (séparation des fibres de la pâte), épuration (élimination des impuretés de la pâte).

Dans un deuxième temps, les procédés continus de formation de la feuille se font dans la machine à papier par un procédé que l'on appelle partie humide et qui correspond à l'élimination de l'eau par pression de la feuille dans une toile de presse, et ensuite on a un procédé qu'on appelle la partie sèche et qui correspond à l'élimination du reste de l'eau par séchage de la feuille dans une sécherie.

Ces procédés ont pour mission de définir les caractéristiques du carton selon un certain nombre de paramètres physiques. Le tableau 5.2 montre les principaux paramètres physiques du papier ou carton et leur signification.

| Paramètres | Signification |
|---|---|
| Gramaje | Es el peso en gramos de un metro cuadrado de papel. |
| Calibre | Es el espesor de un papel o distancia entre una cara y otra, el chequeo de esta propiedad puede evitar numerosos problemas para el impresor ya que una diferencia de espesor en la misma hoja provoca descalce, falta de impresión y arrugas. |
| Blancura | Se define como una alta y uniforme reflectancia del papel en todas las longitudes de onda del espectro visible. |
| Humedad | Con este ensayo se determina el contenido de agua del papel, un papel seco presenta bajas en las propiedades mecánicas (Resistencia a los dobleces, elongación, etc.). |
| Rigidez | Consiste en la facultad que presentan los papeles o cartulina para soportar una fuerza que tiende a curvarlo para darle una forma distinta a la anterior. Esta rigidez debe ir acompañada de una buena dosis de flexibilidad para evitar que se produzca la ruptura en el doblado, tanto el espesor como el gramaje tiene una influencia directa sobre la rigidez. |
| Resistencia Superficial | Para la impresión ofsset se necesita un papel de alta resistencia superficial, dada la viscosidad de las tintas empleadas, a fin de superar las fuerzas adhesivas del tack de las tintas. |

**Tableau 5.2 :** Paramètres physiques du papier ou carton

Dans un troisième temps, nous avons les procédés discontinus de conversion de la feuille qui se font à la sortie de la machine à papier et qui correspondent d'une part, à l'enroulement de la feuille, sur une bobine ou rouleau que l'on appelle bobine mère[14], et d'autre part on a la coupe de la feuille selon un format spécifique à chaque bloc. Pour cela la bobine mère est déroulée puis coupée et ensuite enroulée sur des bobines plus petites que l'on appelle bobines filles. Ces bobines filles sont stockées dans un entrepôt soit pour être vendues directement au client, soit, comme nous le verrons dans le paragraphe suivant, pour être transformées dans deux produits, qu'on appelle *pallets* ou *rolls* (voir figure 5.5). L'ensemble de ces produits (bobines filles, pallets ou rolls) élaborés à partir d'une commande sont appelles produits élaborés.

Processus-carton

Les processus-carton correspondent à la phase d'enchaînement des opérations dans les ateliers discontinus des produits élaborés, où nous avons deux processus discontinus de conversion de la feuille qui suit le cheminement suivant : coupage et emballage, selon le type de produit fini que l'on veut fabriquer.

---

[14] Une bobine mère a un poids maximum de 25 tonnes.

Dans la manufacture des rolls, l'atelier est composé de deux machines : la machine réenrouleuse des bobines et la machine emballeuse des rolls, où l'enchaînement des opérations est le suivant : la machine réenrouleuse des bobines est alimentée avec une bobine fille que l'on a récupérée d'un entrepôt des bobines filles pour être déroulée et coupée puis enroulée sur des bobines plus petites que l'on appelle rolls. Ensuite, la machine emballeuse des rolls est alimentée avec un roll pour être emballée et puis être conduite à l'entrepôt des produits élaborés.

Dans la manufacture des pallets, l'atelier est composé de deux machines : la machine coupeuse des bobines et la machine emballeuse des pallets, où l'enchaînement des opérations est le suivant : la machine coupeuse de bobines est alimentée avec une bobine fille que l'on a récupérée dans l'entrepôt des bobines filles pour être coupée en forme de carrés puis empilés dans un palet en bois. Ensuite, la machine emballeuse des pallets est alimentée avec un pallet pour être emballée et puis être conduite à l'entrepôt des produits élaborés.

Avant de vérifier l'hypothèse de l'enaction du système de connaissance, nous pouvons dire que dans toutes ces activités que nous avons exposées succinctement dans ce paragraphe-ci, la traçabilité des produits se fait par un suivi des flux des informations dans toute la chaîne produits-procédés-processus de manufacture du carton. Nous allons voir dans la section 5.3.2 le protocole de codage (voir tableau 5.13) qui est utilisé par l'entreprise, mais nous pouvons avancer que cette recherche a abouti à la résolution d'un conflit, et à la mise en question du système de codage. En fait, il est sujet à réflexion et à de possibles améliorations.

En conclusion, l'approche systémique OID, nous a permis de repérer quatre activités du système opérant qui donnent vie au système de fabrication, à savoir : produits-carton, procédés-carton, et processus-carton.

Si l'on suppose qu'une activité dans la modélisation systémique corresponde à une dualité organisation/structure dans la modélisation autopoïétique. Donc, nous avons la dualité carton/produits, la dualité carton/procédés, et la dualité carton/processus. Cela signifie que dans le domaine social, c'est le *carton* qui définit l'organisation, par contre ce sont les *produits*, les *procédés*, et le *processus* qui sont les composants de la structure dans le domaine physique. Par conséquent, l'unité (le carton) maintient son identité à partir de sa structure, laquelle est définie par les *produits*, les *procédés*, et les *processus*.

Nous proposons donc de définir le domaine physique de la dualité organisation/structure du modèle proposé[15] à partir d'un domaine industriel, composé par *procédés*, *processus*, et *produits*. Autrement dit, la *matière* dans le modèle proposé correspond aux *procédés*, l'*énergie* se traduit par des *processus*, et les *symboles* sont des *produits*. Les individus du domaine physique sont les individus du domaine industriel, de même que les relations humaines du domaine social sont les relations humaines du domaine social du carton.

En conséquence, l'hypothèse de l'enaction du système de connaissance du modèle proposé est vérifié, car il existe bien une dualité organisation/structure sur laquelle on peut bâtir le système de connaissance.

### 5.1.3. Hypothèse de la connaissance imparfaite du système de connaissance

Cette hypothèse est relative à la vérification de l'existence d'un contexte industriel dans lequel la phénoménologie peut être expliquée à partir de l'imprécis et de l'incertain.

Les systèmes de planification et de gestion des réclamations ont permis de vérifier cette hypothèse.

> ➢ **Le système de planification**

Ce système gère les informations autour de la coordination et régulation des activités de manufacture (produits-procédés-processus). Ainsi, le système de planification permet de composer les blocs de production, c'est-à-dire de programmer les tâches dans un cycle de production de la machine à papier à partir des commandes clients. Dans ce sens, l'unité de planification et contrôle des opérations est la responsable de cette programmation et du contrôle de la production de la machine à papier, et des machines à couper et à emballer.

La programmation de ces cycles de production, consiste d'une part, dans l'ordonnancement des tâches dans le temps (dates de début ou fin des opérations au plutôt et/ou au plus tard, etc.) c'est-à-dire dans la définition d'une Carte Gantt avec l'état de la tâche en question : finie, programmée, active pour les machines (à papier, à couper et à emballer) et d'autre part, dans le contrôle de la performance de chaque machine selon des critères de production associés à chaque bloc.

---

[15] Nous rappelons que le modèle proposé est défini par rapport à un domaine social et technique, cela signifie que la dualité organisation/structure existe (simultanément et nécessairement), dans ces deux domaines. Nous décrivons ce phénomène par le terme récursif.

La taille et la séquence des blocs de production sont définies par rapport : (1) à la gamme de fabrication (Carton Maulé Reverso Café, Carton Maulé Reverso Blanco, Carton Maulé Reverso Manila, Carton Maulé Reverso Estucado) ; (2) au grammage standard associé à chaque sub-gamme, et comme on l'a déjà dit, les 29 cartons ou sub-gammes sont élaborés par une seule machine à papier pour un rang standard de grammage qui va de 200 g/m2 à 450 g/m2 ; et (3) au nombre des commandes, mais aussi il faut respecter un certain nombre de contraintes telles que la variation de grammage d'un carton à l'autre pour une même gamme, les tailles minimum des lotes de production, les périodes de maintenance et de nettoyage de la machine, etc. A cet effet, des techniques de programmation par contrainte sont utilisées.

L'entrée du système de planification est la commande client, celle-ci s'obtient à partir de deux agents qui se trouvent dans l'unité commerciale. Les premières ce sont des agents externes qui sont placés dans chaque pays avec laquelle l'entreprise maintient une affaire commerciale. Ces agents externes ont une côte (en tonnes par mois) qui représente le volume de vente pour l'année en cours, la côte est définie par rapport à l'historique de ventes pour le pays en question. Les secondes ce sont des agents internes qui ont la charge de valider et passer la commande dans le système de planification. Les informations principales qui sont à valider par les agents internes ce sont les informations commerciales du client et la côte de vente du mois en cours de l'agent externe. Si la côte est dépassée alors l'agent interne doit reporter la commande pour le mois suivant ou renégocier avec un autre agent externe pour lequel la côte de vente n'a pas été dépassée pour ce mois-ci.

Or, étant donné la planification des blocs par cycle de production de la machine à papier, les agents internes de l'unité commerciale peuvent introduire dans le système planification les commandes par leurs clients. Ensuite, le système les informe sur les blocs qui ont été programmés[16] pour le cycle (période de temps) en question, mais pas sur les commandes qui ont été déjà assignées au bloc de production. A notre avis, ceci est fait pour éviter les conflits entre les agents internes de l'unité commerciale.

---

[16] Ici, l'état de la tâche de la machine à papier sur la Carte Gantt du système de planification est programmé.

La prise de la commande par le système de planification est définie à travers une série de dates. Nous en avons retenu six : (1) la date de commande qui correspond à la date courante de la mise en place de la commande par les agents internes de l'unité commerciale ; (2) la date de programmation qui correspond au moment où l'état de la tâche est « programmée » dans un bloc de production ; (3) la date de fabrication qui correspond au moment où l'état de la tâche est « active » pour un bloc en question ; (4) la date de nécessité qui correspond à la date où les produits (bobines filles, pallets ou rolls) doivent être dans l'entrepôt de stockage pour être délivrés au client; (5) la date de compromis qui correspond à la date où le produit doit être chez le client ; et (6) la date de re compromis qui correspond à une nouvelle date de compromis.

Les autres dates correspondent aux différentes dates que nous appelons dans un sens large de "compromis" par rapport au réseau de compromis maillé par les relations entre les ateliers discontinus où se trouvent les diverses machines à couper et à emballer, mais aussi avec les responsables des entrepôts de stockage.

En général, nous avons constaté qu'il y a une série de règles par rapport à toutes ces dates qui permettent de gérer les conflits dans l'enchaînement des opérations dans les ateliers continus et discontinus de l'entreprise, par exemple si la date de nécessité est proche de la date de commande alors la priorité dans le bloc de production est augmentée, un autre exemple est que la date de manufacture de la commande est définie comme le maximum entre la date de nécessité et la date de programmation de la commande client dans le bloc de production plus cinq jours. Néanmoins, le fait que la commande soit prise ou pas dans un bloc de production est défini par l'unité de planification et contrôle des opérations, de même que la date de programmation de la tâche dans le système de planification et l'ordre de lancement de la tâche dans la machine à papier. Il faut souligner ici que les agents internes n'ont pas la possibilité d'aller modifier cette date dans le système. Ceci nous montre que le système opérant régule sa propre production d'activité par rapport aux informations du système de fabrication. Sinon cela serait une source de conflits entre l'unité commerciale et l'unité de planification et contrôle des opérations. En effet, les agents internes ont tendance à définir les dates de nécessité plutôt que les réelles dates de nécessité du client afin de ne pas avoir de retard avec la commande.

Or, une fois que la commande a été programmée dans un bloc, on rentre dans un espace d'incertitude (événement vrai ou faux) sur la possibilité d'accomplir ou non à temps celle-ci selon les dates de nécessité et de compromis[17] avec le client.

Afin de gérer cette situation tout au moins une fois que la tâche est active, c'est-à-dire que la fabrication d'un bloc a été lancée, on utilise la date de recompromis qui permet d'avertir le client du retard. Mais, en général on a ce qu'on appelle des produits semi-élaborés, que sont des bobines filles, pallets ou rolls élaborés à partir des produits les plus vendus selon l'horizon de prévision, ces données sont générées par l'étude de séries chronologiques sur les ventes passées, ou bien ce sont des produits où la commande a été annulée. Néanmoins, cela implique une dégradation des produits stockés pour les effets de l'entropie du carton et la possibilité de confusion avec autres produits stockés car la recherche se fait de façon manuelle dans les entrepôts.

La sortie du système de planification est la commande du client satisfait, autrement dit le carton sous forme de bobines filles, pallets ou rolls tout prêt pour être délivré au client. En effet, sur la base des dates que l'on vient de voir, prises en compte dans le système, et aussi sur la base d'autres informations mises dans le système de planification, par exemple la sub-gamme, la taille, les dimensions de la coupe, etc., l'unité de planification et contrôle des opérations doit transformer la commande du client dans des ordres spécifiques de manufacture, cela implique deux aspects. L'un de ces aspects est la modularisation du bloc, c'est-à-dire le nombre des bobines mères du bloc, d'une part, et d'autre part la mise en forme des bobines filles, autrement dit la fixation des dimensions de la coupe dans l'enrouleuse de la machine à papier. Ainsi, une fois que la bobine mère est coupée en bobines filles par l'enrouleuse de la machine à papier, elles sont acheminées vers l'entrepôt de stockage des bobines filles dont elles se tiennent près, soit pour la coupe en pallets ou rolls selon le format spécifié par le client, soit pour la délivrance directement chez le client. L'autre aspect est un chronogramme sous forme d'une Carte Gantt qui indique dans le temps le cheminement des bobines filles sur l'une ou l'autre des machines à couper et/ou à emballer selon les spécifications de la commande. En effet, une fois que la bobine mère est coupée en bobines filles par l'enrouleuse de la machine à papier, elles sont acheminées vers l'entrepôt de stockage de bobines filles où elles se tiennent près, soit pour la coupe en pallets ou rolls selon le format spécifié par le client, soit pour la délivrance directement chez le client.

---

[17] Dans notre approche ces dates doivent être traitées comme des données imprécises.

Or, comme nous l'avons un peu avancé auparavant, des techniques de programmation par contrainte sont utilisées : (1) pour maximiser la surface de la feuille dans le processus de coupe du bloc ou des bobines filles ; et (2) pour faire l'ordonnancement des tâches dans les ateliers discontinus, par exemple faire le choix de la machine à couper la plus propre selon les spécifications de la commande. Mais aussi, la programmation par contrainte et des autres outils de gestion, sont utilisés pour assurer la performance du système de planification et la survie du système dans sa totalité, c'est-à-dire le *business* de l'entreprise CMPC Maulé.

Néanmoins, le système de planification ne prévoit pas la prise en compte de l'imprécis et de l'incertain dans sa programmation. En effet, par exemple si l'on tient compte de l'ordonnancement des tâches dans les ateliers discontinus, les dates de nécessité et de compromis que nous avons exposées plus haut sont définies comme des données précises, tandis que dans la réalité du traitement de la commande d'un client, comme nous avons pu le constater, ces dates sont des données imprécises. Nous pensons que la mise au point d'un degré de possibilité permettra la prise en compte de la connaissance imparfaite que les acteurs ont sur la réalité (le *business*) qu'ils doivent gérer. Autrement dit, ceci permettrait d'améliorer les rapports sociaux entre l'unité commerciale et l'unité de planification et de contrôle des opérations. En effet, on aurait un espace de travail avec moins d'appels téléphoniques et/ou emails, cela implique un gain de temps. Néanmoins, comme tout changement visant les processus dans une structure organisationnelle cela oblige la remise en question des responsabilités des acteurs dans la définition d'un nouveau réseau de compromis, c'est-à-dire dans l'ordonnancement des tâches générées, grâce à la connaissance imparfaite qu'ont ces acteurs sur les différentes dates du système de planification.

En conséquence, l'hypothèse de la connaissance imparfaite du système de connaissance du modèle proposé est vérifiée, car il existe bien une phénoménologie dans le système de planification qui peut être expliquée à partir de l'imprécis et de l'incertain.

➢ **Le système de gestion de réclamations**

Ce système gère les informations autour de la relation client au niveau de la gestion des réclamations. Pour cela, l'unité d'attention au client possède une Base des Réclamations Clients qui correspond à une table Excel où l'on enregistre en ligne les réclamations. Les données de la base, ce sont le code de réclamation, la date de réclamation, le code client, le code produit, la description du défaut et enfin on a une colonne observations. Dans ce sens, la description de la réclamation se fait en terme du « nom du défaut » que plus au moins les agents de l'unité d'attention au client essaient d'approcher à partir de l'observation du client. Nous avons constaté que cette description doit être précise, néanmoins une imprécision, c'est-à-dire une certaine flexibilité dans la description du défaut peut être décrite dans la colonne observation.

L'entrée des données dans la base de réclamation se fait le plus souvent par appels téléphoniques, emails ou fax. Cette base de réclamation est le point d'entrée du système de gestion des défauts que nous allons décrire dans la suite.

En conséquence, l'hypothèse de la connaissance imparfaite du système de connaissance du modèle proposé est vérifiée, car il existe bien une phénoménologie dans le système de gestion des réclamations qui peut être expliquée à partir de l'imprécis et de l'incertain.

### 5.1.4. Hypothèse de spontanéité des relations du système de connaissance

Cette hypothèse, est relative à la vérification du phénomène de la "spontanéité" du processus de production des composants du modèle autopoïétique.

Le système de gestion des défauts a permis de vérifier cette hypothèse.

Nous avons choisi un scénario relatif au *processus de traçabilité des défauts du carton*, pour vérifier cette hypothèse. Dans ce sens, le terme traçabilité englobe les mécanismes explicatifs de "quelque chose", c'est-à-dire une fois que la "chose" est distinguable et identifiable comme unité, alors on essaie d'expliquer son fonctionnement, par la mise en place d'une chaîne logique de faits. Autrement dit, la traçabilité est un suivi des flux d'informations autour d'une "chose". Pour cela nous avons fait deux sous-hypothèses, à savoir :

- la première sous-hypothèse est qu'un défaut peut être considéré comme un système, et donc peut être étudié comme un système complexe[18] ;

---

[18] Le système où s'établit un processus de causalité circulaire propre à la dualité cause/effet. Ce processus est nécessaire afin de maintenir l'unité (l'organisation), l'identité (la structure) et l'autonomie (la dynamique) du système. La dualité cause/effet se construit ici entre défauts, entre défauts et individus, et entre individus.

---

- la deuxième sous-hypothèse est que la manufacture du carton, comme d'ailleurs de tout produit manufacturier, gravite autour de trois relations, tout d'abord le produit, c'est-à-dire le « tout » qui nous regardons dans une optique de besoin, de veille, d'innovation, etc., puis les procédés de manufacture, c'est-à-dire les "parties" qui correspondent à la conversion des pâtes pour fabriquer les quatre ou sept couches du carton, et finalement les processus de manufacture, c'est-à-dire les relations entre composants et entre relations qui correspondent à l'organisation et l'ordonnancement des tâches. Où l'ordre « produit, procédés, processus » dépend de la problématique de gestion que nous sommes en train d'analyser et de comprendre.

> **Le système de gestion des défauts**

Ce système gère les informations autour des défauts et ses mécanismes de formation. Le concept de « défaut », nous l'avons utilisé jusqu'ici dans un sens large, mais d'ores et déjà il faut le clarifier, tout au mois nous l'espérons, avant de parler de « traçabilité des défauts ». A cet égard nous avons trouvé le terme « défaut » dans un document appelé « contrôle de la qualité ». Mais, dans ce document, nous n'avons pas trouvé une définition explicite sur ce que l'on entend par un « défaut », ce que nous avons trouvé a été plutôt une sorte de connaissance opératoire[19] sur les défauts, au niveau de la manufacture (procédés-processus-produit) du carton, attachés à chaque sub-gamme de fabrication (voir tableaux 5.1 et 5.12).

En ce sens, chaque connaissance opératoire sur le défaut établit une sorte de schéma, sur les actions à accomplir si l'on est confronté à une telle situation ou une autre, selon le savoir-faire et le savoir-technique d'une tâche.

Ces informations, en termes de connaissances sur les défauts, nous pouvons les représenter sous la forme de deux relations : l'un pour les procédés ou processus, et l'autre pour les paramètres associés.

Ainsi, la première relation prend la forme :

---

[19] Ici, nous avons très bien pu mettre le mot « techniques », mais pour bien rester fidèles au modèle OID ce que nous observons ce sont les informations relatives aux connaissances sur les défauts survenant lors de la manufacture du carton, ou bien chez le client, par exemple, pendant les phases de formation de l'emballage.

```
                    « procédé-défaut-action »
                              ou
                    « processus-défaut-action »
```

**Figure 5.1 :** Relation procédés ou processus

Et la deuxième relation prend la forme :

```
                    « paramètre-défaut-action »
```

**Figure 5.2 :** Relation paramètre

Alors, dans les figures 5.1 et 5.2 on voit bien que le problème qui est à l'origine un défaut peut se trouver dans un procédé, dans un processus ou dans un paramètre. Or, ce que l'on veut exprimer par ces relations est que les défauts de fabrication sont relatifs au couplage (procédé, paramètres), ou processus, paramètres, tandis que l'action pour résoudre un défaut dépend du raisonnement de l'expert selon le savoir-faire et savoir-technique autour d'une tâche, cela signifie que la solution d'un problème est toujours associée à une action. Ainsi, on a une structure générique:

```
                    « problème-raisonnement-solution »
```

**Figure 5.3 :** Structure générique

Si nous faisons une analogie avec une base de cas, la relation de la figure 5.3 peut être représentée et exploitée par une approche objet dans une base de cas, qui a la structure générique :

```
                    « objet, paramètre, valeurs »
```

**Figure 5.4 :** Structure générique de la base de cas

Les objets qui nous intéressent ici, ce sont la sub-gamme, le paramètre et le grammage, et les valeurs sont le rang standard de grammage associé à chaque sub-gamme (voir tableau 5.12).

Revenons maintenant au début de ce paragraphe, pour mieux situer le contexte d'un défaut, comme nous l'avons dit, ce concept est rattaché au contrôle de qualité. Dans les documents que nous avons analysés, la qualité est définie par rapport à la norme ISO 8402, qui dit « totalidad de las caracteristicas de una entidad que le confieren la aptitud para satisfacer necesidades entablecidas e implicitas », dans cette définition, à notre avis, l'un des points clé par rapport au concept de défaut est la satisfaction d'une nécessité dans le sens qu'en général tous les produits, et mieux encore tous les couples (paramètres, valeurs) pour une sub-gamme (c'est-à-dire, les objets selon la structure de la figure 5.4) sont définis comme nous l'avons avancé dans le tableau 5.1 à partir des recommandations d'usage et méthodes d'impression du carton chez le client, et donc c'est la totalité (le système) qui confère au défaut son sens de pouvoir destructif. Autrement dit, le domaine d'utilisation du carton, qui est caractérisé par les recommandations d'usage et les méthodes d'impression, doit définir la structure du carton. A cet égard le tableau 5.3 montre une liste des défauts qui affectent la méthode d'impression offset chez le client.

| Défauts | Signification |
|---|---|
| Contaminación Superficial | Material no fibroso (generalmente carachas) débilmente adherido a la superficie del papel que puede desprenderse en la impresión. |
| Englobado | Deslaminación que se presenta en zonas de forma redonda o irregular, pero no como franjas longitudinales. |
| Deslaminación | Las diversas capas que constituyen la cartulina se separan entre sí o están débilmente unidas y pueden separarse en los procesos. |
| Humedad Alta o Baja | La humedad del papel es mayor o menor que lo especificado. |
| Espesor fuera de norma | El grosor o calibre del papel es superior o inferior a lo especificado. |
| Pintas | Contaminación o suciedad en el papel de tamaño pequeño y firmemente adherido a la superficie. Este defecto no se desprenderá en el proceso de impresión. |
| Gramaje fuera de norma | El peso base del papel es inferior o superior a lo especificado. |
| Resistencia superficial baja | La superficie del papel o del estuco no soporta adecuadamente la tracción perpendicular, por lo que puede haber desprendimiento de partículas durante la impresión. |
| Rayas | Líneas finas en dirección longitudinal, en bajorrelieve. |
| Tono Variado | Diferencias de tono (blancura) del papel entre rollos distintos o pliegos de una misma resma. |
| Curvatura | Perdida de planitud de la cartulina por levantamiento del plano superior o inferior. Puede ocurrir por diferencias de humedad de las capas u otras causas. |

**Tableau 5.3 :** Un exemple des défauts pour la méthode d'impression offset

En conséquence, avec un carton on peut faire ce qui était prévu dans sa structure, et donc si on veut faire autre chose qui n'était pas prévu dans sa structure alors il peut y avoir un risque d'apparition d'un défaut. Dans ce point, nous sommes d'accord avec Simon[20] dans sa large étude consacrée aux défauts métallurgiques pour la construction d'un modèle de défauts, elle dit « nous avons constaté qu'un défaut métallurgique est une appellation qui recouvre en fait diverses formes de dégradations des aciers produits ». A cet égard, d'après l'analyse des documents relatifs au contrôle de la qualité du carton, nous avons trouvé aussi le mot « dégradation », mais davantage le mot « déformation ».

---

[20] Simon G., thèse intitulée *Modèles et méthodes pour la conception des mémoires d'entreprise. Le système DOLMEN : une application en métallurgie.*

Revenons maintenant sur les deux sous-hypothèses de l'hypothèse de spontanéité des relations du système de connaissance. La première sous-hypothèse est relative au fait qu'un défaut peut être imaginé comme un système complexe, tandis que la deuxième hypothèse est relative au fait que dans la manufacture du carton, il y a toujours trois relations : produit-procédés-processus, où l'ordre dépend de la problématique de gestion que l'on doit gérer. Alors, si l'on réfléchit sur les défauts en termes de systèmes complexes, la dégradation ou déformation peut avoir lieu au niveau des composants et aussi au niveau des relations entre ces composants, cela signifie que la structure du système est dégradée, dans un première temps, et puis suite à cette dégradation de la structure, le processus de structuration sera lui aussi dégradé, et donc l'organisation de la structure du système sera déformée par la suite de cette dégradation. Il y a là, deux aspects qui nous semblent extrêment importants pour valider ces deux hypothèses de travail.

Le premier aspect est relatif à l'évolution du défaut. En ce sens, Simon dit, au sujet des défauts métallurgiques « la notion de défaut évolue dans le temps. En effet, à une certaine période, un phénomène peut être considéré comme normal puis, plus tard, considéré comme un défaut qu'il faudra supprimer ». Nous constatons ici que cette évolution du défaut métallurgique est prise dans le sens du processus de fabrication, mais non pas dans sa recommandation d'usage de l'acier, car cela signifierait que l'acier est utilisé pour autre chose que ce qu a été prévu dans sa structure, et donc la survenance du défaut.

Chez nous, cela serait simplement un non respect des recommandations d'usage du produit mais pas un défaut en soi que l'on va mettre dans la base des réclamations clients de l'unité d'attention au client. En plus, le sens évolutif du défaut doit être pris dans le sens du patrimoine des connaissances sur les défauts, puisqu'une fois que le défaut est maîtrisé, ce n'est plus un défaut. Ou bien, le défaut peut apparaître ailleurs.

En effet dans la recherche d'explication d'un défaut, un grand nombre d'auteurs s'appuient sur une chaîne logique de cause-effet, que graphiquement on peut mettre en place, par exemple, à l'aide d'un diagramme d'Isikawa afin d'établir la relation avec les autres possibles défauts (effets) et ceci à plusieurs niveaux dans la schématisation des défauts. A titre d'exemple, supposons qu'une des causes nous montre que le défaut agit sur un paramètre quelconque, et donc pour ce paramètre nous allons modifier sa valeur, alors cette variation de valeur va faire émerger une nouvelle relation avec les autres paramètres, et nous sommes bien d'accord que cela peut être la cause d'un autre défaut qui avant n'existait pas, car la structure a changé[21].

Le deuxième aspect est le fait que nous pouvons très bien imaginer l'utilisation du concept des « cônes de définition »[22] largement utilisé dans la pensée systémique dans la modélisation de la complexité, pour modeler un défaut, à partir de l'hypothèse que la manufacture du carton gravite autour des trois niveaux (produit-procédés-processus) ou comme nous l'avons dit l'ordre dépend de la problématique de gestion que l'on doit gérer.

Néanmoins, l'utilisation du concept des « cônes de définition » chez Stafford Beer pour la modélisation des défauts implique que dans la construction de ce modèle, le niveau inférieur est une complexité plus riche que dans les niveaux supérieurs, cela signifie que l'on doit définir les niveaux de complexité.

En ce sens, afin de définir les niveaux de complexité nous avons établi une analogie avec le modèle de défaut chez Simon, pour cela il nous faut changer le mot « cônes » du modèle de Stafford Beer par « niveaux » utilisé chez Simon comme nous le verrons plus bas.

Dans le système DOLMEN, dont le but d'après Simon a été « la détection de défauts à partir de descriptions de fabrications d'aciers », le modèle de défaut de ce système permet de structurer les connaissances sur les défauts à partir de la description des causes d'apparition du défaut selon l'expert[23] défectologique de la direction métallurgique de l'entreprise. Simon dit à propos du rôle de la direction « l'activité principale de ce service consiste à contrôler, maintenir et améliorer la qualité des aciers produits. Cela le conduit entre autres à surveiller et répertorier constamment l'ensemble des défauts détectés soit au cours de la production, soit, une fois la production terminée, chez le client ». A cet égard, elle dit « il est arrivé que des sites de production différents identifient le même défaut mais sous un nom différent ». Et puis elle ajoute « la mise en évidence d'un défaut n'est pas quelque chose de simple et l'explication de ses causes de formation l'est encore moins ».

---

[21] Ceci nous renvoie au concept de détermination structurelle, que nous avons illustré à partir de l'analogie suivante : un pont où la structure était faite pour résister à un poids maximum de 2 tonnes était traversé pour un camion de 3 tonnes. Or, si pendant la traversée le pont se collapse alors ce n'est pas la faute du pont qui n'a pas résisté au poids du camion, mais plutôt c'est le fait que sa structure n'était pas prévue pour supporter un poids de 3 tonnes.
[22] Nous avons emprunté ce concept à Stafford Beer, mais on le trouve chez autres systémiques comme Le Moigne et Mélèse.
[23] En général, Simon parle des experts défectologiques dans son étude.

Ceci nous montre, qu'il y a plusieurs regards possibles sur les défauts. En général nous constatons que pour l'identification d'un défaut, nous devons tenir compte de deux critères, et nous supposons que ceci est vrai dans n'importe quel domaine de connaissance.

Le premier critère concerne la position du défaut (procédé, processus, produit). Par exemple, ce qu'il nous intéresse d'identifier, ce sont les défauts au cours de la transformation de la matière dans ses différentes états, ou bien au cours de l'enchaînement d'une tâche à une autre dans les ateliers discontinus selon l'ordonnancement des tâches, ou bien dans l'utilisation du produit chez le client, ce que l'on appelle couramment un « défaut de fabrication ».

Le deuxième critère concerne le lieu du défaut. En effet, il s'agit de savoir si les connaissances sur les défauts proviennent d'un même site de production, ou plutôt si ce sont des sites de production différents. Par exemple, dans le modèle de défaut du système DOLMEN le lieu d'observation d'un défaut est décrit sur ce qu'on appelle « contexte du défaut », c'est-à-dire les informations relatives au domaine du défaut, afin de bien préciser que le défaut a un sens, tandis que la position du défaut est décrite sur ce qu'on appelle « mécanismes d'explication d'un défaut ».

A cet égard, Simon dit qu'il s'agit de « trois points de vue différents d'explication des défauts : l'apparence physique, les mécanismes de formation physico-chimiques et l'ensemble des paramètres métallurgiques sous-tendant ces mécanismes ». Et elle ajoute « Le premier niveau décrit le défaut en termes de constatations physiques sur la forme qu'il prend sur l'acier. On peut à ce niveau savoir si ce défaut est une tâche, une surépaisseur, s'il prend une couleur particulière, s'il apparaît à un endroit précis de la bande d'acier, etc. Le second niveau consiste à décrire des mécanismes de types physique ou chimique pouvant expliquer l'apparence du défaut, décrite dans le premier niveau. Enfin le troisième niveau est le niveau d'explication le plus fin. Il permet de relier les mécanismes physico-chimiques du niveau précédent aux paramètres métallurgiques qui les sous-tendent et sur lesquels il est donc éventuellement possible d'agir afin de remédier au défaut ».

A partir de ces trois points de vue chez Simon, il est possible de construire un modèle de défaut qui peut être utilisé pour la compréhension des mécanismes de formation du défaut. Ce qui est intéressant à noter dans ces trois niveaux est que leur statut explicatif n'est pas le même. A cet égard, Simon dit « du point de vue du système que nous concevons, le niveau le plus intéressant est le dernier car il permet vraiment de savoir comment se produit le défaut ».

Nous pensons que cette démarche explicative à travers ces trois niveaux d'explication, ou trois points de vue de distinction, peuvent être une source d'explication empirique au phénomène de la "spontanéité" du processus de production des composants du modèle autopoïétique.

Au niveau produit qui correspond à la réalité (c'est un niveau tout à fait humain). Le seul outil de distinction est la perception de l'opérateur, par exemple, la vision, le toucher, ici sont fondamentaux pour apprécier la présentation des bobines mères ou des produits élaborés (bobines filles, pallets ou rolls). A ce niveau de distinction, ce que nous pensons est que cette perception humaine doit être guidée par le domaine d'utilisation du carton qui correspond comme nous l'avons dit, aux « recommandations d'usage » et « méthodes d'impression » du carton. En effet, ce que nous pouvons dire est que le risque d'apparition d'un défaut est dû au fait du non respect du domaine d'utilisation du carton.

Au niveau procédé, nous avons constaté qu'avec la même recette, les équipes de travail obtiennent de qualités différentes, pour être plus précis cela se passe plutôt au site de Santiago de la compagnie, tandis que dans la région du Maulé cela l'est beaucoup moins.

## 5.2. Mise en œuvre du système opérationnel

Nous avons proposé dans le chapitre 4 trois étapes pour gérer les connaissances d'un système opérationnel. La première étape consiste dans la détection du besoin utilisateur. La deuxième étape consiste dans la traduction de ce besoin dans un langage Fuzzy Structured Query Language (FSQL). La troisième étape est l'extraction de connaissance d'une base de données relationnelles floues (BDRF).

### 5.2.1. Détection du besoin de l'utilisateur

La phase d'analyse du besoin est une phase de chantier, c'est-à-dire qu'à partir de la rencontre avec les gens sur le lieu du travail, la problématique industrielle doit émerger. C'est ce que nous avons mis tout au début du chapitre, le contexte de la problématique industrielle, l'analyse de la problématique industrielle et l'identification de la problématique industrielle.

Gestion d'une commande floue dans un entrepôt de stockage

La figure 5.5 montre les deux types de produits (pallets ou rolls) qui sont stockés dans l'entrepôt pour être délivrés au client.

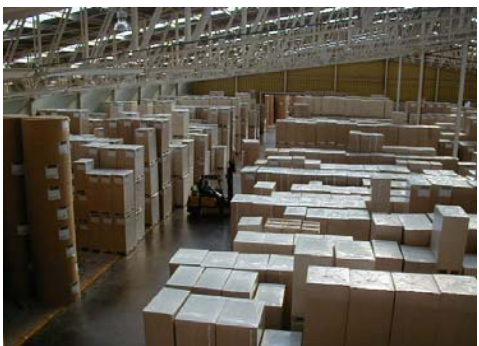 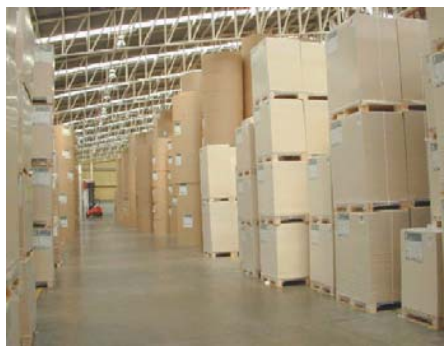

**Figure 5.5 :** Entrepôt des produits stockés (pallets ou rolls)

Pour nous cet entrepôt, bien entendu, est géré par un système d'information qui permet, d'une part, de représenter les objets (pallets ou rolls) stockés, et d'autre part, manipuler ces objets selon les événements associés. Pour nous ces événements sont incertains (en plus d'etre certains) c'est-à-dire relatifs aux commandes floues, tandis que les objets sont décrits d'une façon imprécise, en plus de précise.

Pour illustrer des événements incertains, voici deux exemples qui expriment le besoin utilisateur en langage naturel, exprimés pendant notre stage en entreprise :

- « *Obtenga los rollos de cartulina almacenados en la bodega que estén sucios o húmedos* ». En français, cela veut dire « *rechercher les rolls qui sont sales ou humides* ». Dans cette phrase nous constatons le caractère vague ou ambigu[24] de l'expression du besoin de l'utilisateur, car ce qui est sale (ou humide) pour un opérateur peut ne pas l'être pour un autre.

- « *Obtenga todos los datos de las cartulinas estucadas que su tono de cara sea posiblemente igual a blanco y que el tono reverso también sea posiblemente igual a blanco* ». En français, cela veut dire « *rechercher les données du carton RC dans lequel la tonalité[25] de la couche*

---

[24] Pour Bonnal en citant Dubois et Prade « il y a unanimité pour considérer le vague comme quelque chose qui a une absence de contour précis, pour lequel on éprouve de la difficulté à fixer des limites. L'ambigü peut etre considéré sous une forme subjective ; il est alors lié à l'imprécision du langage » [Bonnal, 02].
[25] Nous avons quatre types de tonalités du carton {blanco, amarillo, café, manila}.

*"cara" (voir tableau 5.14) et de la couche "reverso" (voir tableau 5.14) soient possiblement égales à blanche »*. Le caractère vague ou ambigü de cette requête, se trouve pour nous, dans le terme "possiblement égale".

Au niveau de la représentation précise et imprécise des objets (pallets ou rolls) stockés dans l'entrepôt, le tableau 5.4 montre quelques attributs propres du carton.

| Atributs des pallets | Atributs des rolls |
|---|---|
| Pliegos parejos y planos. | Bobinado parejo. |
| Pliegos planos. | Limpio. |
| Tarima seca y con dimensiones correctas. | Empalmes correctos (bien pegados e identificados). |
| Cantidad exacta. | Corte limpio (no pelusiento). |
| Formato especificado (ancho, largo, altura). | Formato especificado (diámetro, altura). |
| Identificación correcta. | Identificación correcta. |

**Tableau 5.4 :** Attributs précis et imprécis du carton (pallets ou rolls)

Les attributs du tableau 5.4 représentent les caractéristiques ou qualités qui doivent être présentes dans les produits finis, soit sous forme de pallet ou roll. Certaines de ces caractéristiques peuvent être considérées comme attributs imprécis ou flous, du fait que la mesure de la valeur de l'attribut n'est pas réalisée par un instrument physique mais plutôt par la perception humaine, principalement le tact et la vue des opérateurs. Ces caractéristiques du carton font partie du savoir-faire de l'entreprise et pourtant la maîtrise de ces paramètres joue un rôle important pour assurer la qualité d'utilisation du produit chez les clients de l'entreprise, par l'identification préventive de certains défauts. Pour les pallets nous avons : {golpeado, mojado, orilla picada, englobado, sucio, picaduras, rayas en la superficia}, et pour les rolls nous avons : {englobado, deslaminado, húmedo, sucio, rayas, curvas, empalme defectuoso, orilla crespa, disparejo}.

### 5.2.2.    Traduction du besoin utilisateur dans un langage FSQL

Comme nous l'avons avancé dans le chapitre 4 (voir sections 4.3.6 et 4.4), le système FSQL nous permet de construire un système opérationnel sur la base de deux modèles. Au niveau conceptuel, nous avons le modèle conceptuel FuzzyEER [Urrutia, 03], [Galindo *et al*, 04a] et leur progiciel de design associé FuzzyCase [Urrutia *et al*, 03]. Le modèle FuzzyEER permet la construction d'un modèle conceptuel EER (Enhanced Entity Relationship) flou, c'est-à-dire la notion de floue a été considérée sous trois aspects : (a) dans la spécification des attributs [Urrutia *et* *al* ,02]; (b) dans la spécification des valeurs de ces attributs [Urrutia *et al*, 01a]; et (c) dans la spécification des contraintes de ces attributs [Galindo *et al*, 04b], pour représenter les objets de gestion (données, informations ou connaissances) et les événements de gestion du business model. Au niveau logique, nous avons d'une part, le modèle GEFRED (GEneralized model for Fuzzy RElational Databases) [Medina, 94], [Medina *et al*, 94] qui permet le stockage des objets du modèle FuzzyEER dans la BDRF de FSQL [Galindo, 99], [Carrasco *et al*, 02], et d'autre part, l'interface FIRST (Fuzzy Interface for Relational SysTems) qui correspond au protocole de communication entre le modèle logique et le modèle physique de la BDR.

➢ **Modèle conceptuel FuzzyEER**

La figure 5.6 montre le schéma conceptuel FuzzyEER du business model retenu à l'aide du progiciel FuzzyCase. Dans ce schéma nous distinguons la gamme complète des produits que nous avons identifiés dans les tableaux 5.1 et 5.12, ainsi que la structure en couches du carton (dans le tableau 5.14 nous avons identifié les cinq couches du carton maulé RC).

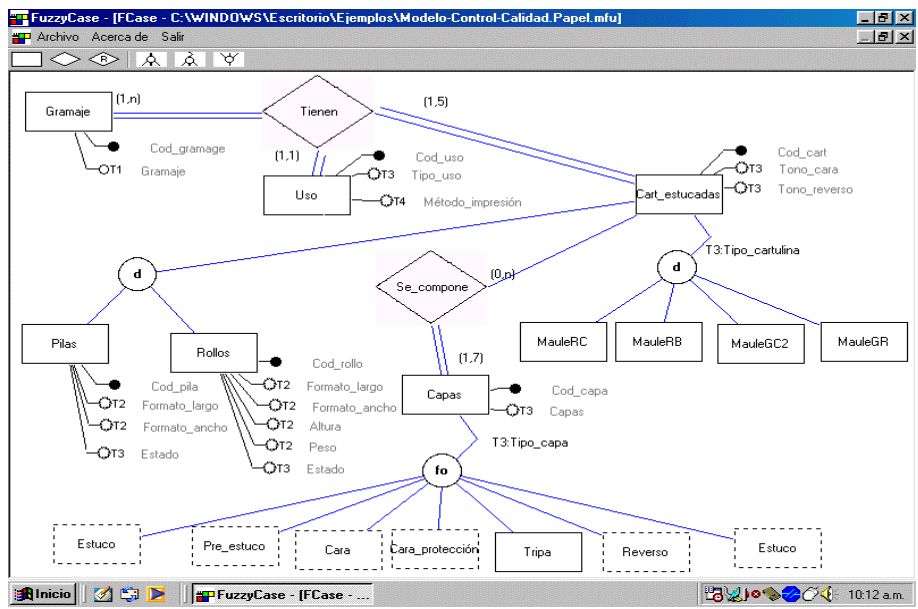

**Figure 5.6 :** Schéma conceptuel FuzzyEER de la gamme de fabrication

La partie du modèle qui nous intéresse est relative à deux classes d'objets : pallets ou rolls (pilas ou rollos en espagnol). Cette partie du schéma conceptuel FuzzyEER est reprise dans la figure 5.7. Néanmoins nous avons ajouté davantage d'attributs pour enrichir le modèle.

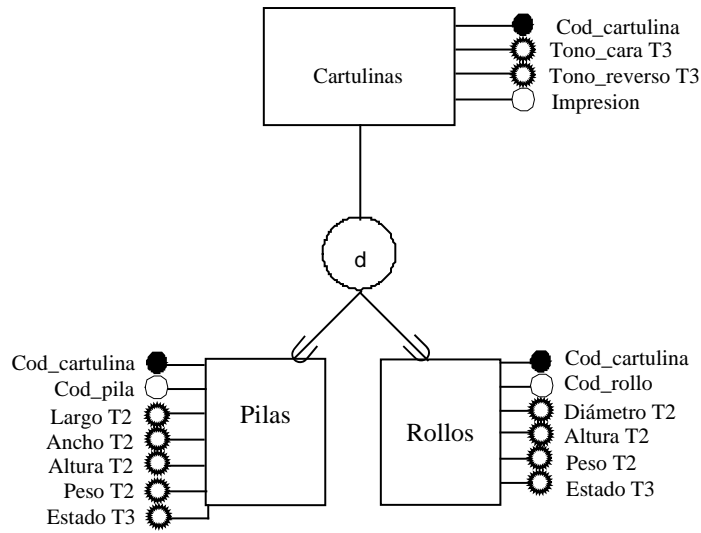

**Figure 5.7 :** Schéma conceptuel FuzzyEER pour pallets ou rolls (pilas ou rollos)

Le modèle FuzzyEER de la figure 5.7 nous permet de décrire deux sub-classes disjointes de l'entité carton (cartulinas)[26], du fait qu'un roll ne peut pas être à la fois un pallet, et vice-versa.

Pour la description des entités, nous utilisons, d'une part, les attributs flous de Type 1, de Type 2 et de Type 3 (voir chapitre 4, section 4.4.1)[27], et d'autre part, les données floues associées (voir chapitre 4, section 4.4.3) :

Entité cartulinas (cartons)

Cette entité est décrite par le schéma {*Cod_cartulina*, *Tono_Cara*, *Tono_Reverso*, *Impresion*}. L'attribut *Cod_cartulina* est un attribut précis, car il permet d'identifier le code du carton par une donnée précise (voir tableau 5.12). Les attributs *Tono_Cara* et *Tono_Reverso* sont d'attributs flous de Type 3, car le domaine sous-jacent aux attributs est non ordonné. Ces attributs sont définis par les étiquettes linguistiques {blanco, amarillo, café, manila}. Ceci permet la représentation de données imprécises sur un domaine non ordonné régularisé, afin de représenter la couleur du carton pour chaque côté (cara ou reverso, voir tableau 5.14). L'attribut *Impresion* est un

---

[26] La disjonction de classe est indiquée par la lettre "d" dans la figure 5.7.
[27] Les attributs flous de Type 1 sont des données précises représentés par des étiquettes linguistiques, les attributs flous de Type 2 sont des données imprécises sur un domaine ordonné, et les attributs flous de Type 3 sont des données imprécises sur un domaine non ordonné régularisé.

---

attribut flou de Type 1, c'est-à-dire un attribut précis défini par deux types de caractères[28] : huecograbado ou offset, d'après le tableau 5.1.

### Entité pilas (pallet)

Cette entité est décrite par le schéma {*Cod_cartulina*, *Cod_pila*, *Largo*, *Ancho*, *Altura*, *Peso*, *Estado*}. Les attributs *Cod_cartulina* et *Cod_pila* sont précis, car ils permetent d'identifier le code du carton et du pallet par une donnée précise. Les attributs *Largo*, *Ancho*, *Altura*, et *Peso* sont des attributs flous de Type 2, que nous considérons dans le cadre de notre modèle comme des données imprécises définies sur un domaine ordonné. Or, ces attributs sont définis par les étiquettes linguistiques {corto, largo, muy largo}; {angosto, ancho, muy ancho}; {alta, muy alta, baja, muy baja}; et {bajo, muy bajo, sobre, muy sobre} respectivement. Ceci permet la représentation de données imprécises sur un domaine ordonné régularisé, afin de représenter la longueur (*Largo*), largeur (*Ancho*), la taille (*Altura*) et le poids (*Peso*) d'un pallet (pila). L'attribut *Estado* est un attribut flou de Type 3, car le domaine sous-jacent de l'attribut est non ordonné, défini par les étiquettes linguistiques {golpeado, mojado, orilla picada, englobado, sucio, picaduras, rayas en la superficie}. Ceci permet la représentation de données imprécises sur un domaine non ordonné régularisé, afin de représenter les états des défauts des pallets (voir tableau 5.4).

### Entité rollos (rolls)

Cette entité est décrite par le schéma {*Cod_cartulina*, *Cod_*rollo, *Diametro*, *Altura*, *Peso*, *Estado*}. Les attributs *Cod_cartulina* et *Cod_rollo* sont précis, car ils permetent d'identifier le code du carton et du roll par une donnée précise. Les attributs *Diametro*, *Altura*, et *Peso* sont des attributs flous de Type 2, que nous considérons dans le cadre de notre modèle comme des données imprécises définies sur un domaine ordonné. Or, ces attributs sont définis par les étiquettes linguistiques {rango mínimo, normal, rango máximo}; {baja, mediana, alta}; et {bajo, optimo, sobre} respectivement. Ceci permet la représentation de données imprécises sur un domaine ordonné régularisé, afin de représenter le diamètre (*Diametro*), la taille (*Altura*) et le poids (*Peso*) d'un roll (rollo). L'attribut *Estado* est un attribut flou de Type 3, car le domaine sous-jacent de l'attribut est non ordonné, défini par les étiquettes linguistiques {englobado, deslaminado, húmedo, sucio, rayas, curvas, empalme defectuoso, orilla crespa, disparejo}. Ceci permet la représentation de données imprécises sur un domaine non ordonné régularisé, afin de représenter les états des défauts des rolls (voir tableau 5.4).

---

[28] Il ne faut pas les confondre avec de possibles étiquettes linguistiques {huecograbado, offset}.

Or, nous avons considéré les attributs *Largo*, *Ancho*, *Altura*, et *Peso* des pallets et les attributs *Diametro*, *Altura*, et *Peso* comme flous, seulement pour illustrer la problématique floue, mais dans la réalité (formulaire de commande) ils ne le sont pas. A titre d'exemple supposons que la norme fixe la valeur de l'attribut *Diametro* (diamètre) à 400 cm qui est la valeur dictée, supposons maintenant que dans l'entrepôt de stockage nous avons des rolls qui font 387 cm (au-dessous de la norme) ou bien 412 cm (au-dessus de la norme) alors la problématique est évidente : comment peut-on classifier ces rolls dans le système d'information ? C'est justement dans ce type de situation que le BDRF devient intéressant car nous pouvons utiliser des étiquettes linguistiques, par exemple : {dessous-de-la-norme, dans-la-norme, dessus-de-la-norme} afin de calculer un degré d'appartenance (voir chapitre 4, section 4.2.1) pour chaque donnée précise du domaine. Ainsi, nous pouvons avoir, pour une même valeur précise, par exemple 402, deux degrés d'appartenances différentes, par exemple $\mu_{dessous-de-la-norme}$ *(402)* = 0,4 et $\mu_{dans-la-norme}$ *(402)* = 0,6.

➢ **Modèle logique GEFRED**

Le modèle GEFRED (GEneralized model for Fuzzy RElational Databases) permet le stockage des objets (entités cartulinas, pilas et rollos) du modèle FuzzyEER (voir figure 5.7) dans la BDRF de FSQL.

Ce modèle permet la représentation de trois types d'attributs flous : Type 1, Type 2 et Type 3 dans une BDR.

Représentation d'attributs flous de Type 1

La représentation de ces attributs flous est réalisée soit directement dans la BDR, ou bien à l'aide des étiquettes linguistiques. Pour cela on utilise la représentation LABEL du tableau 5.5.

Représentation d'attributs flous de Type 2

La représentation de ces attributs flous dans la BDR est réalisée par un protocole de conversion que nous présentons dans le tableau 5.5.

| Attributs flous de Type 2 | Protocole de conversion | | | | |
|---|---|---|---|---|---|
| | Id | V1 | V2 | V3 | V4 |
| UNKNOWN | 0 | NULL | NULL | NULL | NULL |
| UNDEFINED | 1 | NULL | NULL | NULL | NULL |
| NULL | 2 | NULL | NULL | NULL | NULL |
| CRISP | 3 | d | NULL | NULL | NULL |
| LABEL | 4 | FUZZY_ID | NULL | NULL | NULL |
| INTERVALO (n,m) | 5 | n | NULL | NULL | m |
| APROXIMADAMENTE (d) | 6 | d | d-margen | d+margen | margen |
| TRAPECIO | 7 | $\alpha$ | $\beta$-$\alpha$ | $\gamma$-$\delta$ | $\delta$ |

**Tableau 5.5 :** Protocole de conversion pour attributs flous de Type 2

La colonne *Attributs flous Type 2* montre les 8 descripteurs qui sont associés aux attributs flous de Type 2 pour les identifier. Dans le chapitre 4 nous avons passé en revue la représentation des attributs flous (voir section 4.4.1). Ainsi :

- UNKNOWN (voir figure 4.5) est utilisé lorsque la valeur de l'attribut est inconnue.

- UNDEFINED (voir figure 4.6) est utilisé lorsque la valeur de l'attribut est inapplicable.

- NULL (voir figure 4.11) est utilisé lorsque la valeur de l'attribut est nulle[29].

- CRISP (voir figure 4.4) est utilisé lorsque la valeur de l'attribut est précise.

- LABEL (voir figure 4.8) est utilisé lorsque la valeur de l'attribut est partiellement connue et la donnée imprécise est représentée par une étiquette linguistique.

- INTERVALO (n,m) (voir figure 4.10) est utilisé lorsque la valeur de l'attribut est partiellement connue et la donnée imprécise est représentée par un intervalle réel [*n,m*].

- APROXIMADAMENTE (d) (voir figure 4.9) est utilisé lorsque la valeur de l'attribut est partiellement connue et la donnée imprécise est représentée par un valeur triangulaire #d.

- TRAPECIO (voir figure 4.7) est utilisé lorsque la valeur de l'attribut est partiellement connue et la donnée imprécise est représentée par un valeur trapézoïdale $[*a,b,c,d*].

La colonne *Protocole de conversion* montre le protocole du modèle GEFRED. Cette colonne est divisée en 5 autres pour montrer, d'une part, l'identificateur de l'attribut flou de Type 2 considéré, et d'autre part, les 4 valeurs de la variable pour définir la donnée en question. Par exemple, pour l'attribut flou de Type 2 APROXIMADAMENTE (d), il nous faut définir V1, V2, V3 et V4. Ainsi, V1 permet de stocker la valeur précise de la donnée, V2 stocke sa limite gauche, V3 stocke sa limite droite, et V4 la valeur de variation (margen). En revanche, pour l'attribut flou de Type 2 CRISP, on a besoin seulement de V1 pour stocker la valeur.

---

[29] Les attributs flous Type 2 UNKNOWN, UNDEFINED et NULL sont stockés comme une valeur nulle dans une BDR.

Représentation d'attributs flous de Type 3

La représentation de ces attributs flous dans la BDR est réalisée par un protocole de conversion que nous présentons dans le tableau 5.6.

| Attributs flous de Type 3 | Protocole de conversion | | | | | |
|---|---|---|---|---|---|---|
| | FT | FP1 | F1 | ... | $FP_n$ | $F_n$ |
| UNKNOWN | 0 | NULL | NULL | ... | NULL | NULL |
| INDEFINED | 1 | NULL | NULL | ... | NULL | NULL |
| NULL | 2 | NULL | NULL | ... | NULL | NULL |
| SIMPLE | 3 | p | d | ... | NULL | NULL |
| DISTRIBUCION POSIBILIDAD | 4 | $p_1$ | $d_1$ | ... | $p_n$ | $d_n$ |

**Tableau 5.6 :** Protocole de conversion pour attributs flous de Type 3

La colonne *Attributs flous Type 3* montre les 5 descripteurs qui sont associés aux attributs flous de Type 3 pour les identifier. Les trois premiers attributs sont identiques aux attributs flous de Type 2 (voir tableau 5.5). En revanche l'attribut flou de Type 3 SIMPLE est défini par rapport à deux valeurs. L'un est le degré de possibilité (stocké dans la colonne FP1), l'autre est l'étiquette linguistique (stocké dans la colonne F1) associée à cette valeur. Enfin l'attribut flou de Type 3 DISTRIBUCION POSIBILIDAD permet de stocker plus d'une étiquette linguistique dans une relation (FP1, F1) ... (FPn, Fn).

➢ **Modèle physique FIRST**

L'interface FIRST (Fuzzy Interface for Relational SysTems) permet la formulation des scripts en langage PL/SQL pour la création, le chargement, et la consultation des tables dans la BDR Oracle[30]. Il s'agit d'un protocole de communication entre le modèle logique et le modèle physique de la BDR.

D'après notre modèle FuzzyEER (voir figure 5.7) nous avons trois relations à créer qui correspondent aux entités : *cartulinas*, *pilas* et *rollos*.

La figure 5.8 montre le script en langage PL/SQL pour la création de la table *cartulinas* selon le schéma {*Cod_cartulina*, *Tono_Cara*, *Tono_Reverso*, *Impresion*} du modèle FuzzyEER. Dans ce schéma l'attribut *Cod_cartulina* est un attribut précis, tandis que les attributs *Tono_Cara* et

---

[30] Cette partie du travail a été amené à bien à l'aide de Yolande Valdés et de Jorge Vistoso [Valdés et Vistoso, 03].

*Tono_Reverso* sont des attributs flous de Type 3 définis par les étiquettes linguistiques {blanco, amarillo, café, manila}, enfin l'attribut *Impresion* est un attribut flou de Type 1 (crisp).

```
CREATE TABLE cartulinas (cod_cartulina number(3) not null, impresion varchar2(15) not null, tono_carat number(1) not null,
check (tono_carat between 0 and 4), tono_carap1 number(3,2), tono_cara1 number(3), tono_carap2 number(3,2), tono_cara2
number(3), tono_carap3 number(3,2), tono_cara3 number(3), tono_carap4 number(3,2), tono_cara4 number(3), tono_reversot
number(1) not null, check (tono_reversot between 0 and 4), tono_reversop1 number(3,2), tono_reverso1 number(3),
tono_reversop2 number(3,2), tono_reverso2 number(3),tono_reversop3 number(3,2), tono_reverso3 number(3),
tono_reversop4 number(3,2), tono_reverso4 number(3))
```

**Figure 5.8 :** Script pour la création de l'entité *cartulinas* du schéma conceptuel FuzzyEER

La figure 5.9 montre le script en langage PL/SQL pour la création de la table *pilas* selon le schéma {*Cod_cartulina*, *Cod_pila*, *Largo*, *Ancho*, *Altura*, *Peso*, *Estado*} du modèle FuzzyEER. Dans ce schéma les attributs *Cod_cartulina* et *Cod_pila* sont précis, tandis que les attributs *Largo*, *Ancho*, *Altura*, et *Peso* sont des attributs flous de Type 2 définis par les étiquettes linguistiques {corto, largo, muy largo}; {angosto, ancho, muy ancho}; {alta, muy alta, baja, muy baja}; et {bajo, muy bajo, sobre, muy sobre} respectivement, enfin l'attribut *Estado* est un attribut flou de Type 3 défini par les étiquettes linguistiques {golpeado, mojado, orilla picada, englobado, sucio, picaduras, rayas en la superficie}.

```
CREATE TABLE pilas (cod_cart number(3) not null, cod_pila number(2) not null, largot number(1) not null, largo1
number(3),largo2 number(3),largo3 number(3), largo4 number(3), anchot number(1) not null, ancho1 number(3), ancho2
number(3), ancho3 number(3), ancho4 number(3), alturat number(1) not null, altura1 number(3), altura2 number(3), altura3
number(3), altura4 number(3), estadot number(1) not null, estadop1 number(3,2), estado1 number(3), estadop2 number(3,2),
estado2 number(3), estadop3 number(3,2), estado3 number(3),estadop4 number(3,2), estado4 number(3), estadop5
number(3,2), estado5 number(3), estadop6 number(3,2), estado6 number(3), estadop7 number(3,2), estado7 number(3),
estadop8 number(3,2), estado8 number(3), estadop9 number(3,2), estado9 number(3))
```

**Figure 5.9 :** Script pour la création de l'entité *pilas* du schéma conceptuel FuzzyEER

La figure 5.10 montre le script en langage PL/SQL pour la création de la table *rollos* selon le schéma {*Cod_cartulina*, *Cod_*rollo, *Diametro*, *Altura*, *Peso*, *Estado*} du modèle FuzzyEER. Dans ce schéma les attributs *Cod_cartulina* et *Cod_rollo* sont précis, tandis que les attributs *Diametro*, *Altura*, et *Peso* sont des attributs flous de Type 2 définis par les étiquettes linguistiques {rango mínimo, normal, rango máximo}; {baja, mediana, alta}; et {bajo, optimo, sobre} respectivement, enfin l'attribut *Estado* est un attribut flou de Type 3 défini par les étiquettes linguistiques {englobado, deslaminado, húmedo, sucio, rayas, curvas, empalme defectuoso, orilla crespa, disparejo}.

```
CREATE TABLE rollos (cod_cart number(3) not null, cod_rollo number(2) not null, tipo varchar2(15) not null, largot
number(1) not null, largo1 number(3), largo2 number(3), largo3 number(3), largo4 number(3), anchot number(1) not null,
ancho1 number(3), ancho2 number(3), ancho3 number(3), ancho4 number(3), alturat number(1) not null, altura1 number(3),
altura2 number(3), altura3 number(3), altura4 number(3), pesot number(1) not null,peso1 number(3), peso2 number(3), peso3
number(3), peso4 number(3), estadot number(1) not null, estadop1 number(3,2), estado1 number(3), estadop2 number(3,2),
```

estado2 number(3), estadop3 number(3,2), estado3 number(3), estadop4 number(3,2), estado4 number(3), estadop5 number(3,2), estado5 number(3), estadop6 number(3,2), estado6 number(3), estadop7 number(3,2), estado7 number(3), estadop8 number(3,2), estado8 number(3),estadop9 number(3,2), estado9 number(3)

**Figure 5.10 :** Script pour la création de l'éntité *rollos* du shéma conceptuel FuzzyEER

De façon graphique nous pouvons expliquer ces scripts par une série de links des tables. La figure 5.11 montre la configuration schématique des tables *pilas* et *rollos* nécessaire pour stocker ses attributs flous.

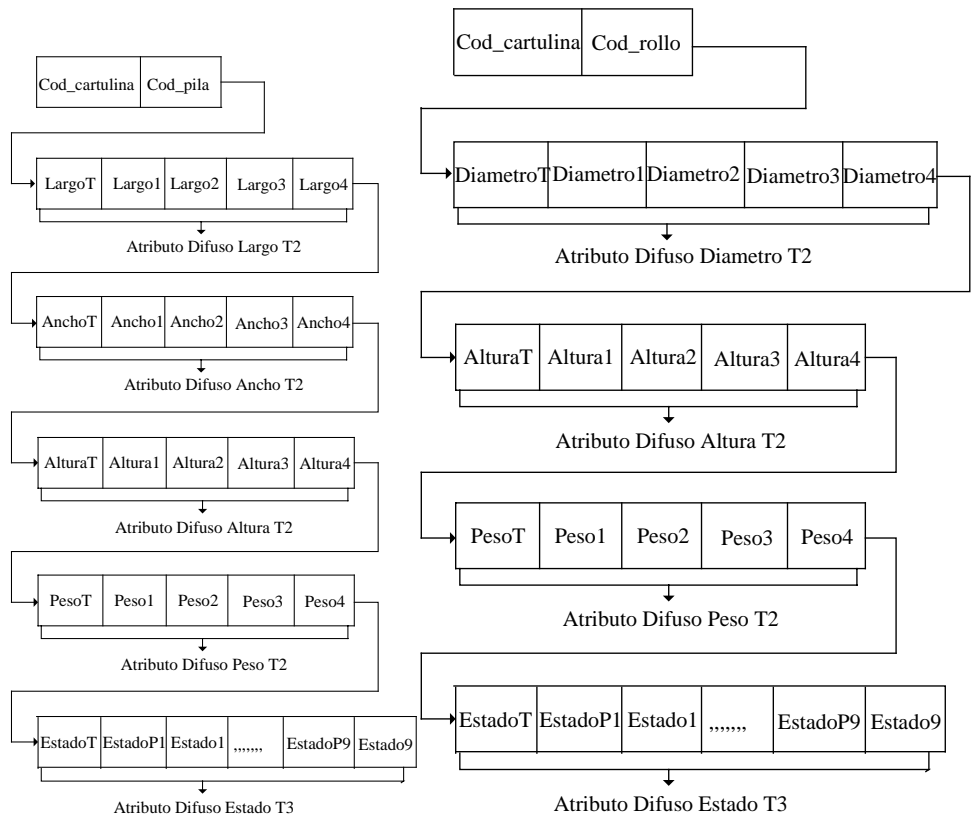

**Figure 5.11 :** Schema logique pour la création des tables (*pilas* et *rollos*)

Une fois les tables créées et les attributs flous chargés vient la phase de chargement de données floues, ceci est expliqué dans la paragraphe suivant.

> ➢ **Données floues**

Le chargement de données floues se fait aussi à l'aide d'une série de tables. L'interface FIRST est composé des tables : FUZZY_COL_LIST (FCL), FUZZY_OBJECT_LIST (FOL), FUZZY_LABEL_DEF (FLD), et FUZZY_NEARNESS_DEF (FND), entre autres.

---

Table FUZZY_COL_LIST (FCL)

Cette table contient la description de tous les attributs flous du modèle FuzzyEER. Ainsi, nous avons, par exemple, pour la table *pilas*, d'une part, le schéma {*Largo*, *Ancho*, *Altura*, *Peso*} relatif aux attributs flous de Type 2, et d'autre part, le schéma {*Estado*} relatif aux attributs flous de Type 3 définis par les étiquettes linguistiques {golpeado, mojado, orilla picada, englobado, sucio, picaduras, rayas en la superficie}. En revanche, la table *rollos* est composée d'une part par le schéma {*Diametro*, *Altura*, *Peso*} relatif aux attributs flous de Type 2, et d'autre part, le schéma {*Estado*} relatif aux attributs flous de Type 3 définis par les étiquettes linguistiques {englobado, deslaminado, húmedo, sucio, rayas, curvas, empalme defectuoso, orilla crespa, disparejo}.

Le tableau 5.7 montre la table FCL pour *pilas* et *rollos*. Les colonnes OBJ# et COL# sont utilisées pour charger le nom de l'entité et ses attributs associés. La colonne F_Type est utilisée pour charger le type d'attribut flou considéré[31]. La colonne LEN est utilisée seulement pour les attributs flous de Type 3 afin d'indiquer le nombre de relations de la distribution de possibilité considérée[32].

| OBJ# | COL# | F_TYPE | LEN |
|------|------|--------|-----|
| Pilas | Largo | 2 | 1 |
| Pilas | Ancho | 2 | 1 |
| Pilas | Altura | 2 | 1 |
| Pilas | Peso | 2 | 1 |
| Pilas | Estado | 3 | 9 |
| Rollos | Diametro | 2 | 1 |
| Rollos | Altura | 2 | 1 |
| Rollos | Peso | 2 | 1 |
| Rollos | Estado | 3 | 9 |

**Tableau 5.7 :** Table FCL de la FIRST du système FSQL

La figure 5.12 montre le script associé pour le chargement des données floues dans la table FCL, pour quelques tuples.

---

[31] Le nombre 1 est utilisé pour les attributs flous Type 1, le nombre 2 est utilisé pour les attributs flous Type 2, et le nombre 3 est utilisé pour les attributs flous Type 3.
[32] Pour les attributs flous Type 1 et Type 2, la colonne LEN montre un 1 par défaut.

```
INSERT into FCL values (t_PILAS,c_PLARGO,2,1,USER||'.Pilas.largo');
INSERT into FCL values (t_PILAS,c_PANCHO,2,1,USER||'.Pilas.ancho');
INSERT into FCL values (t_PILAS,c_PPESO,2,1,USER||'.Pilas.Peso');
INSERT into FCL values (t_PILAS,c_PESTADO,3,9,USER||'.Pilas.Estado');
```

**Figure 5.12 :** Script pour le le chargement des données floues dans la table FCL

Table FUZZY_OBJECT_LIST (FOL)

Cette table contient la description de toutes les étiquettes linguistiques du modèle FuzzyEER. Par exemple, pour l'objet (OBJ#) *rollos* de la table FCL, nous avons quatre attributs flous qui sont représentés dans la colonne COL# de la table FCL. Ainsi, la table FOL permet de représenter les étiquettes linguistiques {rango mínimo, normal, rango máximo}; {baja, mediana, alta}; {bajo, optimo, sobre}; et {englobado, deslaminado, húmedo, sucio, rayas, curvas, empalme defectuoso, orilla crespa, disparejo} des attributs *Diametro*, *Altura*, *Peso* et *Estado* respectivement.

Le tableau 5.8 montre la table FOL pour *rollos*. Les colonnes OBJ# et COL# sont les *foreing key* de la table FCL, utilisées pour charger le nom de l'entité et ses attributs associés. La colonne Fuzzy_ID est le descripteur de l'attribut flou considéré[33]. La colonne Fuzzy_NAME est utilisée pour indiquer la liste des étiquettes linguistiques considérées. La colonne Fuzzy_TYPE est utilisée pour indiquer le type de donnée floue considérée[34].

| OBJ# | COL# | FUZZY_ID | FUZZY_NAME | FUZZY_TYPE |
|------|------|----------|------------|------------|
| Rollos | Diametro | 0 | 'Rango_min' | 7 |
| Rollos | Diametro | 1 | 'Normal' | 7 |
| Rollos | Diametro | 2 | 'Rango_max' | 7 |
| Rollos | Altura | 0 | 'Baja' | 7 |
| Rollos | Altura | 1 | 'Mediana' | 7 |
| Rollos | Altura | 2 | 'Alta' | 7 |
| Rollos | Peso | 0 | 'Bajo' | 7 |
| Rollos | Peso | 1 | 'Optimo' | 7 |
| Rollos | Peso | 2 | 'Sobre' | 7 |
| Rollos | Estado | 0 | 'Englobado' | 4 |
| Rollos | Estado | 1 | 'Deslaminado' | 4 |
| Rollos | Estado | 2 | 'Humedo' | 4 |
| Rollos | Estado | 3 | 'Sucio' | 4 |
| Rollos | Estado | 4 | 'Rayas' | 4 |
| Rollos | Estado | 5 | 'Curvas' | 4 |
| Rollos | Estado | 6 | 'Empalme_defectuoso | 4 |

---

[33] Un nombre 0, 1, 2, …

[34] Le tableau 5.5 (voir colonne Id) montre les identificateurs pour les attributs flous Type 2, tandis que tableau 5.6 (voir colonne FT) montre les identificateurs pour les attributs flous Type 3. Dans notre cas, nous avons 7 pour TRAPECIO et 4 pour DISTRIBUCION POSIBILIDAD.

| | | | | |
|---|---|---|---|---|
| Rollos | Estado | 7 | 'Orilla_crespa' | 4 |
| Rollos | Estado | 8 | 'Disparejo' | 4 |

**Tableau 5.8 :** Table FOL de la FIRST du système FSQL

La figure 5.13 montre le script associé pour le chargement des données floues dans la table FOL, pour quelques tuples.

```
INSERT into FOL values(t_ROLLOS,c_RDIAMETRO,0,'RANGO_MIN',7);
INSERT into FOL values(t_ROLLOS,c_RDIAMETRO,1,'NORMA',7);
INSERT into FOL values(t_ROLLOS,c_RDIAMETRO,2,'RANGO_MAX',7);
INSERT into FOL values(t_ROLLOS,c_RESTADO,7,'ORILLA_CRESPA',4);
INSERT into FOL values(t_ROLLOS,c_RESTADO,8,'DISPAREJO',4);
```

**Figure 5.13 :** Script pour le le chargement des données floues dans la table FOL

### Table FUZZY_LABEL_DEF (FLD)

Cette table contient les données de toutes les étiquettes linguistiques pour les attributs flous de Type 2 du modèle FuzzyEER. Par exemple, d'après les tables FCL et FOL pour l'objet (OBJ#) *rollos*, nous avons trois attributs flous de Type 2 qui sont représentés dans la colonne COL# de la table FOL, avec son identificateur respectif donné dans la colonne FUZZY_ID de la même table. Ainsi, la table FLD permet de représenter les donnés (voir tableau 5.8) pour un FUZZY_TYPE 7, qui selon le tableau 5.5 correspond à TRAPECIO. Les données sont stockées respectivement dans les colonnes ALFA, BETA, GAMMA, et DELTA de la table FLD comme le montre le tableau 5.9.

| OBJ# | COL# | FUZZY_ID | ALFA | BETA | GAMMA | DELTA |
|---|---|---|---|---|---|---|
| Rollos | Diametro | 0 | 50 | 70 | 100 | 130 |
| Rollos | Diametro | 1 | 100 | 150 | 170 | 220 |
| Rollos | Diametro | 2 | 190 | 220 | 250 | 300 |
| Rollos | Altura | 0 | 3 | 4 | 5 | 7 |
| Rollos | Altura | 1 | 5 | 8 | 10 | 11 |
| Rollos | Altura | 2 | 10 | 12 | 15 | 17 |
| Rollos | Altura | 3 | 14 | 16 | 17 | 19 |
| Rollos | Peso | 0 | 15 | 20 | 35 | 40 |
| Rollos | Peso | 1 | 30 | 45 | 65 | 75 |
| Rollos | Peso | 2 | 70 | 85 | 95 | 100 |

**Tableau 5.9 :** Table FLD de la FIRST du système FSQL

La figure 5.14 montre le script associé pour le chargement des données floues dans la table FOL, pour quelques tuples.

**Figure 5.14 :** Script pour le le chargement des données floues dans la table FLD

### Table FUZZY_NEARNESS_DEF (FND)

Cette table contient les données de toutes les étiquettes linguistiques pour les attributs flous de Type 3 du modèle FuzzyEER. Par exemple, d'après les tables FCL et FOL pour l'objet (OBJ#) *rollos*, nous avons un attribut flou de Type 3 qui est représenté dans la colonne COL# de la table FOL, avec son identificateur respectif donné dans la colonne FUZZY_ID de la même table. Ainsi, la table FND permet de représenter les données (voir tableau 5.8) pour un FUZZY_TYPE 4, qui selon le tableau 5.6 correspond à DISTRIBUCION POSIBILIDAD. Les données sont stockées respectivement dans les colonnes FUZZY_ID1, FUZZY_ID2 et DEGREE de la table FND comme le montre le tableau 5.10. Ici on utilise les mêmes identificateurs FUZZY_ID de la table FOL pour FUZZY_ID1 et FUZZY_ID2. Ainsi la relation (0,2,0) signifie que FUZZY_ID1 = 0, FUZZY_ID2 = 2 et DEGREE = 0. Cela veut dire que 'Englobado' et 'Humedo' non aucune relation de similitude.

| OBJ# | COL# | FUZZY_ID1 | FUZZY_ID2 | DEGREE |
|------|------|-----------|-----------|--------|
| Rollos | Estado | 0 | 2 | 0 |
| Rollos | Estado | 0 | 3 | 0 |
| Rollos | Estado | 0 | 4 | 0 |
| Rollos | Estado | 0 | 5 | 0.3 |
| Rollos | Estado | 0 | 6 | 0.5 |
| Rollos | Estado | 0 | 7 | 0.6 |
| Rollos | Estado | 0 | 8 | 0 |
| Rollos | Estado | 1 | 2 | 0 |
| Rollos | Estado | 1 | 3 | 0 |
| Rollos | Estado | 1 | 4 | 0 |
| Rollos | Estado | 1 | 5 | 0 |
| Rollos | Estado | 1 | 6 | 0.8 |
| Rollos | Estado | 1 | 7 | 0 |
| Rollos | Estado | 1 | 8 | 0.1 |

**Tableau 5.10 :** Table FND de la FIRST du système FSQL

La figure 5.15 montre le script associé pour le chargement des données floues dans la table FND, pour quelques tuples.

INSERT into FND values(t_ROLLOS,c_RESTADO,0,2,0);
INSERT into FND values(t_ROLLOS,c_RESTADO,0,3,0);

```
INSERT into FND values(t_ROLLOS,c_RESTADO,0,4,0);
INSERT into FND values(t_ROLLOS,c_RESTADO,0,5,.3);
INSERT into FND values(t_ROLLOS,c_RESTADO,0,6,.5);
```

**Figure 5.15 :** Script pour le le chargement des données floues dans la table FND

### 5.2.3. Extraction de connaissance du système FSQL

- « *Obtenga todos los datos de las cartulinas estucadas que su tono de cara sea posiblemente igual a blanco y que el tono reverso también sea posiblemente igual a blanco* ». En français, cela veut dire « *rechercher les données du carton RC dans lequel la tonalité[35] de la couche "cara" et de la couche "reverso" soient possiblement égales à blanche* ».

Ce besoin utilisateur se traduit par :

```
SELECT cartulina.%
FROM cartulina
WHERE tono_cara FEQ $blanco THOLD 0.5 and tono_reverso FEQ $blanco THOLD 0.5;
```

La figure 5.16 montre le resultat de l'interrogation au système FSQL

**Figure 5.16 :** Interface FSQL

### 5.3. Analyse de la problématique industrielle

Dans le paragraphe antérieur nous avons présenté (certainement de façon très grossière mais en même temps très utile pour vérifier les hypothèses), d'une part, la mise en oeuvre du système de connaissance autour des recettes, des conflits, des réclamations, des défauts, et d'autre part, la stratégie de la mise en ouvre du système opérationnel dans un système FSQL. Il nous reste maintenant à identifier un noyau invariant (recettes, conflits, réclamations, défauts) capable d'être modélisé sous conditions d'imprécision et d'incertitude afin de définir un système de connaissance imparfaite.

---

[35] Nous avons quatre types de tonalités du carton {blanco, amarillo, café, manila}.

### 5.3.1. Contexte de la problématique industrielle

Nous avons défini un certain nombre de scénarios pour préciser le contexte de la problématique industrielle autour du système de fabrication, système de planification, système de gestion de réclamations, et système de gestion de défauts. Ces scénarios ont été classés dans deux groupes : contexte-objet et contexte-événement.

➢ **Contexte-objet : système de fabrication**

La manufacture du carton contient un ensemble de procédés, processus et paramètres qui intervient dans l'élaboration d'une sub-gamme de produit donné. Or, pour capturer les données relatives à la manufacture, le système de fabrication compte comme une série de systèmes d'information en temps réel capables de maîtriser la gestion de la production. Dans ces systèmes, la pluspart des informations sont relatives aux procédé et processus en cours de manufacture et ses paramètres associés. Par exemple, on a des informations relatives à la production en ligne de la machine à papier, les bobines filles en cours de chemin vers l'entrepôt, les produits (pallets ou rolls) stockés dans leur entrepôt, les bobines filles consumées par les machines à couper, les matières premières utilisées, etc.

Cela signifie que ces systèmes d'information sont destinés à mesurer la performance du système de production globale, hommes et machines confondus, et à contrôler la qualité de la production.

➢ **Contexte-objet : système de planification**

La programmation des tâches de la machine à papier est faite sur la base, d'une part des cycles de production, qui normalement existent en trois cycles : l'année, le trimestre et le mois, et d'autre part d'un certain nombre de critères. On a en général deux critères de programmation.

L'un des critères de programmation est l'horizon de ventes de chaque sub-gamme de produits pendant les 12, 6, ou 3 mois passés. Pour chaque cycle de production, les tâches sont organisées par blocs de production, cela signifie que chaque bloc de production correspond à un ensemble de bobines mères où le poids maximum comme nous l'avons dit est de 25 tonnes et en général un bloc de production pèse entre cent et mille tonnes.

L'autre critère de programmation est la gamme du produit et le grammage associé. En effet, l'ordonnancement de tâches se fait d'abord à partir de la gamme et puis on suit un ordre qui varie vers le plus, par rapport au grammage de chaque sub-gamme.

Pour cela, le système de planification est composé de plusieurs systèmes d'information capables d'optimiser les temps de machines, les dimensions de la coupe, et en général de maîtriser les coûts et les temps de permanence des produits dans les différents ateliers et entrepôts. Cependant, pour le moment, le système de planification ne compte pas avec un système d'information qui lui permet d'aller rechercher de façon automatique dans l'entrepôt de stockage les commandes (bobines filles, pallets ou rolls) qui ont été annulées afin de les accommoder, pour satisfaire une commande en cours, voire même dans l'ordre de lancement d'un bloc de production. Cela signifie que cette procédure se fait de façon manuelle.

➢ **Contexte-objet : système de gestion de réclamations**

L'enregistrement des réclamations dans la base des réclamations clients par les gens de l'unité d'attention au client ne prévoit pas la prise en compte de l'incertain dans la saisie des informations sur les défauts du produit avec des outils informatiques.

➢ **Contexte-objet : système de gestion de défauts**

Le processus de traçabilité des défauts du carton est à la charge de trois unités : l'unité d'attention au client, l'unité commerciale et l'unité de planification et de contrôle des opérations. En effet, l'unité d'attention au client doit gérer la base des réclamations clients, tandis que l'unité commerciale doit gérer la commande du client, et l'unité de planification et de contrôle des opérations doit gérer la manufacture du carton selon la gamme d'usinage : préparation de la pâte, formation de la feuille, et conversion. Néanmoins, nous avons constaté aussi un manque d'outils informatiques relatifs à l'aide de la décision innovante dans ce processus.

Ces quatre contextes-processus nous confirment que les données en ligne, générées par le système de fabrication ne sont pas bien exploitées par les trois autres systèmes, principalement par le système de planification. En effet, ce système est à la charge de la gestion d'une série d'événements entre l'unité commerciale et l'unité de planification et de contrôle des opérations principalement.

D'après l'analyse des documents correspondants, nous avons retenu trois de ces contextes-événements pour identifier la problématique industrielle.

➤ **Contexte-événement : Annulation de la commande**

Les raisons d'annulation d'une commande par l'unité commerciale sont multiples, par exemple une grève dans le service de transport maritime. Or, pour l'unité de planification et contrôle des opérations, la cause de cette annulation n'est pas quelque chose d'important mais ce qui les intéresse est plutôt le moment où cet événement est généré par l'unité commerciale. En effet, la magnitude de l'impact dans le système de planification dépend du moment de la survenance du fait redouté, selon l'état de la tâche : finie, programmée, active.

Tâche finie

Cela signifie que la commande a été déjà fabriquée, soit comme bobine fille, pallet ou roll. Dans ce cas l'unité de planification et de contrôle des opérations le prend comme une perte d'espace de stockage dont il faut se débarrasser car la détérioration du carton, comme nous l'avons dit plus haut, due aux effets climatiques (humidité, chaleur, etc.) est une chose courante, comme l'est aussi le danger de confusion avec d'autres produits élaborés ou semi-élaborés, selon l'entrepôt de stockage. Pour le système de planification, il s'agit d'un excès de production dont il faut tenir compte dans la programmation des tâches de la machine à papier. Cela signifie qu'avant d'initier la gamme d'usinage pour un produit quelconque, il faut aller regarder d'abord si ce produit, on ne l'a pas dans l'entrepôt de stockage comme produit (élaboré ou semi-élaboré) annulé.

Tâche programmée

Si la commande est annulée tout près de la date de programmation, alors il y a une certaine perturbation dans la programmation du bloc de la machine à papier car le bloc doit être remodularisé ce qui entraîne un changement de dates des autres commandes programmées dans le même bloc, cela signifie un retard dans la production des bobines filles et par conséquent une prolongation de la date de compromis.

<u>Tâche active</u>

Si la commande est annulée en cours de manufacture, alors la totalité de la commande est gérée comme un excès de production qui est destiné à l'entrepôt de stockage des produits élaborés.

➢ **Contexte-événement : Changement des besoins du client**

Cet événement est perçu par le système de planification comme : (1) un changement des quantités commandées par le client ; (2) une modification au plus tôt de sa date de compromis ; et (3) un changement dans le statut de la commande par prioritaire. Dans ces trois cas, cela entraîne une perturbation dans le système de planification, c'est-à-dire une reprogrammation de la machine à papier et des machines à couper. En ce sens, l'unité de planification et contrôle des opérations doit informer l'unité commerciale si cela est possible ou non, et faire la mise à jour des dates si c'est le cas. Par exemple, si c'est possible d'augmenter la quantité d'un produit demandé dans un bloc alors cela entraîne la reprogrammation des autres blocs. En effet, le retard dans la production d'un bloc implique le retard de tous les blocs programmés. De manière similaire, le fait d'introduire dans le système de planification une date plus tôt que la date de compromis ou bien un changement dans le statut de la commande par prioritaire, cela se traduit par la prise en charge d'une commande supplémentaire dans le bloc postérieur programmé pour la même gamme de fabrication.

➢ **Contexte-événement : Apparition de défauts**.

<u>Procédés-carton</u>

Supposons que dans les procédés-carton, l'un des tests en ligne ou en laboratoire remette en question la qualité de la pâte pour une certaine sub-gamme de fabrication, cela peut signifier dans certains cas que la bobine mère qui a été déjà assignée pour la manufacture d'une commande ne soit pas élaborée par la machine à papier, ou bien que celle-ci soit fabriquée mais pas dans sa totalité pour satisfaire la commande du client. Une fois que les paramètres du procédé ont été réglés et le défaut réparé, l'unité de planification et de contrôle des opérations doit générer un nouvel ordre de manufacture, cela signifie la reprogrammation d'un bloc de production. Ici, nous avons donc les mêmes problèmes que dans le changement du besoin du client qui a été présenté précédemment.

Processus-carton

Supposons qu'une bobine fille ait un défaut, alors, cette bobine fille doit rester dans l'entrepôt de stockage des bobines filles jusqu'à ce qu'une solution ait été prise, selon la gravité du défaut, généralement la date de passage pour la machine à couper est reprogrammée afin de retirer la partie avec défaut et d'utiliser le reste de la bobine pour satisfaire la commande ; bien entendu l'unité de planification et de contrôle des opérations doit prendre en charge la quantité manquante de la commande Mais si le défaut est trop important, la bobine fille est envoyée à l'origine du processus de manufacture, c'est-à-dire à la désintégration de la bobine dans le pulper. Dans ce cas, l'unité de planification et de contrôle des opérations a deux alternatives à suivre : (1) produire une bobine de remplacement. Cela signifie une perturbation des dates dans le système de planification ; ou (2) aller chercher dans la zone de stockage un produit semi-élaboré pour satisfaire la commande.

Or, afin de satisfaire les commandes et de pouvoir répondre aux aléas, d'une part de changement des besoins du client, et d'autre part d'apparition de défauts, le système de planification fait appel aux produits semi-élaborés, ou bien aux produits élaborés pour lesquels la commande a été annulée. Tandis que, pour l'annulation de la commande, dans le cas où la tâche est active, le système de planification les gère comme un excès de production qui est destiné à l'entrepôt de stockage des produits élaborés.

Néanmoins, avec ces produits qui ont déjà été fabriqués à l'avance ou stockés dans les entrepôts, nous avons constaté qu'il y a chez le client l'apparition de défauts. Nous pensons que cela est dû au fait que l'unité de planification et de contrôle des opérations ne prend pas en compte l'imprécision du grammage défini dans les spécifications techniques des cartons afin d'établir une relation avec les informations du grammage des produits semi-élaborés ou élaborés (y compris les commandes qui ont été annulées) qui sont envoyés chez le client pour satisfaire la commande.

## 5.3.2. Identification de la problématique industrielle

Nous allons étudier le rôle du grammage dans le risque d'apparition d'un défaut. Nous ferons ceci à travers deux cas. Mais avant de faire cela, nous essaierons maintenant d'expliquer la procédure de codification et de marquage. Les deux paragraphes suivants ont été consacrés à cette explication.

> **Procédure de codification**

Le processus de traçabilité d'un produit élaboré ou semi-élaboré est garanti par une procédure de codification dont le but est de retrouver l'ensemble de procédés et paramètres qui interviennent dans sa manufacture. L'entrée du processus, ce sont les étiquettes qui sont collées dans les emballages de ces produits. Une étiquette est pourtant l'identification de la tâche qui a été programmée dans la machine à papier pour fabriquer une sub-gamme particulière de carton. Or, comme nous pouvons le voir dans l'entrepôt de stockage des pallets et rolls (voir figure 5.5), au moins deux étiquettes égales sont collées dans ces produits, et ceci est fait à la fin du processus d'emballage juste avant de le rentrer à l'entrepôt des produits élaborés pour être délivrés au client.

Les données qui sont précisées dans les étiquettes correspondent d'une part aux informations de la commande client qui doit gérer l'unité commerciale, par exemple le code client, le code commande, le code produit, le code bobine/roll/pallet, etc., et d'autre part, aux caractéristiques techniques du produit qui correspondent aux paramètres de manufacture que doit gérer l'unité de planification et de contrôle des opérations. Par exemple, dans le cas d'un pallet, les caractéristiques techniques sont le grammage, le poids, le format spécifié par la largeur x longueur (width x leght) et le nombre de feuilles, etc. Dans le cas d'un roll ou d'une bobine fille, le format est décrit par la largeur x diamètre x mandrin (width x diameter x core).

Ainsi, la procédure de codification au niveau des étiquettes est caractérisée par deux descripteurs principalement.

Le premier est le descripteur du code produit qui est défini par un modèle composé de 4 champs : ZZXM. Le tableau 5.11 montre la décomposition du descripteur et la signification de chaque champ.

| Champ | Signification du champ |
|-------|------------------------|
| ZZ | Ces deux champs là correspondent à la valeur que peut prendre le grammage dans son intervalle de valeurs possibles défini pour la gamme en question (voir colonne grammage du tableau 5.1). Il s'agit donc d'une approximation au grammage standard de la gamme. |
| X | Ce champ indique le code de la gamme (voir colonne code du tableau 5.1). |

| M | Ce champ est utilisé pour identifier la plante de fabrication. La lettre A correspond à l'identification de la plante Valdivia, tandis que la lettre B correspond à l'identification de la plante Maulé. |
|---|---|

**Tableau 5.11 :** Modèle de codification du descripteur code produit

Or, comme nous l'avons dit plus haut (voir tableau 5.1) les types des cartons dans une même gamme sont considérés comme des valeurs approchées par rapport au grammage, c'est-à-dire, on passe de la catégorie de grammages faibles à la catégorie de grammages lourds. Ainsi, ce descripteur permet de codifier la sub-gamme de fabrication, le tableau 5.12 montre le code produit et le grammage total standard associé à chacun des produits de la gamme.

| Carton Maulé RC | |
|---|---|
| **Code produit** | **Grammage** |
| 201B | 200 |
| 231B | 230 |
| 251B | 250 |
| 261B | 265 |
| 281B | 280 |
| 291B | 295 |
| 301B | 305 |
| 321B | 325 |
| 341B | 345 |
| 361B | 365 |
| 391B | 395 |
| 421B | 425 |
| 451B | 450 |

| Carton Maulé RB | |
|---|---|
| **Code produit** | **Grammage** |
| 215B | 210 |
| 235B | 230 |
| 265B | 260 |
| 285B | 280 |
| 315B | 310 |
| 335B | 335 |
| 365B | 360 |

| Carton Maulé GC2 | |
|---|---|
| **Code produit** | **Grammage** |
| 234B | 230 |
| 254B | 250 |
| 264B | 265 |
| 284B | 285 |
| 304B | 305 |
| 324B | 320 |
| 344B | 340 |

| Carton Maulé Graphics | |
|---|---|
| **Code produit** | **Grammage** |
| 227B | 220 |
| 277B | 275 |

**Tableau 5.12 :** Code produit et grammage de la gamme de fabrication

Nous attirons l'attention sur le fait que d'avoir un même grammage pour des catégories différentes ne signifie pas qu'on soit en face d'un même produit. Par exemple, les produits 231B et 235B selon le tableau 5.12 ont un grammage de 230 g/m2 mais ces produits (les 29 types de carton fabriqués pour l'entreprise CMPC dans son site de production de Maulé) sont différents sur deux plans principalement. L'un est par rapport aux paramètres physiques (voir tableau 5.2) et l'autre plan de distinction, que nous présentons plus tard, est le nombre de couches et la composition des pâtes de chaque couche.

Le deuxième est le descripteur pour le code bobine/roll/pallet. Ce descripteur est caractérisé par un modèle composé de 14 champs : M-A-FFF-JJJ-BB-RR-PP. Le tableau 5.13 montre la décomposition du descripteur et la signification de chaque champs du code.

| Champ | Signification du champ |
|-------|------------------------|
| M | Ce champs est utilisé pour identifier la plante de fabrication (C'est le même champs M du descripteur code produit, voir tableau 5.1) |
| A | Ce champs est utilisé pour identifier l'année d'élaboration du bloc. Les numéros 0...9 sont utilisés pour coder l'année en cours[36]. |
| FFF | Ces champs sont utilisés pour identifier les blocs de production de la machine à papier. Il s'agit d'un nombre entre 001 et 999. |
| JJJ | Ces champs sont utilisés pour identifier les bobines filles fabriquées dans un bloc. Il s'agit d'un nombre entre 001 et 999. |
| BB | Ces champs sont utilisés pour identifier les bobines filles dans une bobine mère. Le premier caractère indique le nombre de l'arrêt de la bobine mère, tandis que le deuxième caractère indique sa position à l'intérieur de l'arrêt. |
| RR | Ces champs sont utilisés pour identifier les rolls dans une bobine fille. Le premier caractère indique le nombre de l'arrêt de la bobine fille, tandis que le deuxième caractère indique sa position à l'intérieur de l'arrêt. |
| PP | Ces champs sont utilisés pour identifier les palets dans une bobine fille. Le premier caractère indique le nombre de l'arrêt de la bobine fille, tandis que le deuxième caractère indique sa position à l'intérieur de l'arrêt. |

**Tableau 5.13 :** Modèle de codification du descripteur produits élaborés

➢ **Procédure de marquage**

La deuxième procédure qui joue un rôle important dans le processus de traçabilité d'un produit élaboré ou semi-élaboré est la procédure de marquage dont le but est d'enregistrer les informations sur la relation : commande/produit, entre l'unité commerciale et l'unité de planification et contrôle des opérations à chaque étape du processus-carton.

Nous avons vu que le processus-carton a pour but la conversion du carton, c'est-à-dire la mise au format des bobines filles selon les ordres spécifiques de manufacture de l'unité de planification et de contrôle des opérations, ces ordres sont saisis en gros du bon de commande du client qui gère l'unité commerciale. Ainsi, l'entrée du processus de conversion, ce sont les bobines filles stockées dans l'entrepôt, et la sortie du processus, ce sont les produits élaborés qui seront stockés dans l'entrepôt de produits élaborés, ces produits ne peuvent avoir que deux types de formes : pallet ou roll.

---

[36] L'horizon de traçabilité est de 10 ans.

Or, l'ensemble des connaissances relatives aux processus et leurs paramètres de réglage dans la chaîne de transformation d'un pallet ou roll, nous les appelons *domaines*.

A ce stade de l'investigation, nous avons identifié deux processus relatifs à la traçabilité des produits élaborés, que nous allons présenter selon le chaînage des activités du processus de conversion d'une bobine fille, soit dans un ensemble de pallets ou soit dans un ensemble de rolls.

Dans un premier temps, nous avons le processus « coupage ». Or, si le domaine du processus est un roll, cela signifie que la bobine fille doit être coupée par la machine réenrouleuse des bobines, autrement dit les bobines filles sont déroulées pour être ensuite enroulées sur des bobines plus petites que l'on appelle rolls.

Par contre, si le domaine du processus est un pallet, cela signifie que la bobine fille doit être coupée par la machine coupeuse des bobines, autrement dit les bobines filles sont transformées en feuilles de carton et ensuite empilées dans un pallet en bois.

Ensuite, nous avons le processus « emballage ». Or, si le domaine du processus est un roll cela signifie que la bobine fille doit être emballée par la machine emballeuse des bobines filles. Dans le cas où le domaine du processus est un pallet, alors il doit être emballé par la machine emballeuse des pallets.

Ces deux processus doivent être caractérisés par des mécanismes d'actions et par des mécanismes de régulation. Les mécanismes d'actions du processus « coupage » sur le domaine « pallet » correspondent, par exemple à l'encollage d'une étiquette à la sortie du pallet de la machine, tandis que les mécanismes de régulation correspondent, par exemple au réglage de la machine selon les dimensions de la coupe, définies par l'unité de planification et de contrôle des opérations.

### 5.3.3. Rôle du grammage dans le risque d'apparition d'un défaut

Dans le paragraphe antérieur, nous venons de voir, grosso modo, la procédure de codification et de marquage. Il nous reste alors à aller chercher les informations, en terme de connaissances, qui nous intéressent pour dégager le rôle du grammage dans le risque d'apparition d'un défaut.

Ceci, nous le ferons à travers deux cas. Ces cas ont été construits par rapport à un modèle que nous appelons *modèle carton* et qui nous a permis de réflechir sur la traçabilité des défauts et en particulier sur le risque d'apparition d'un défaut chez le client.

Maintenant, parler de modélisation (au sens informatique) aujourd'hui, c'est parler de l'approche orientée objet et cela est vrai pour un grand nombre d'informaticiens qui l'utilisent afin de comprendre des phénomènes complexes à modéliser dans sa totalité, c'est pour cela que nous avons choisi une structure objet pour représenter les informations relatives aux procédures de codification et de marquage, en terme de connaissances, qui nous intéressent pour étudier le rôle du grammage dans la recette.

Il y a dans ce modèle deux traits qu'il nous intéresse de représenter. L'un de ces traits est bien entendu le grammage qui a été représenté dans les tableaux 5.1 et 5.12. Pour cela, les familles de cartons du tableau 5.1, nous pouvons les considérer comme des classes d'objets auxquelles nous pouvons associer des attributs et des valeurs dans le sens descriptif d'un modèle orienté objet.

Ainsi les informations du tableau 5.13 seront traitées comme des objets. Cela signifie que la structure de connaissance requise doit avoir une représentation de la forme « objet, attribut, valeur » selon un domaine de connaissances qui dans ce cas-là est implicite au produit (carton).

A titre d'exemple, prenons la gamme 301B, c'est-à-dire, un objet qui appartient à la classe 1 (Maulé RC), et qui est caractérisé par trois attributs, le premier est le grammage, le deuxième est le code de la classe, et le troisième est le code du site de production, schématiquement nous pouvons représenter cette connaissance dans une structure de type objet. La figure 5.17 montre la représentation de la classe d'objet produit (carton)[37].

| | | |
|---|---|---|
| Objet | : | 301B |
| Attributs | : | grammage, code-classe, code-site |
| Valeurs | : | $300^{38}$, 1, B |

**Figure 5.17 :** Classe produit (carton)

---

[37] Pour un souci de simplicité nous avons pris le même nom de la sub-gamme pour identifier l'objet.

L'autre trait est la composition des pâtes de la couche. En effet, le carton est fait d'un mélange des pâtes, composé de différentes matières premières et celles-ci sont réparties dans un certain nombre de couches qui forment la structure du carton. Le nombre caractéristique de couches varie entre quatre et sept selon la gamme.

A titre d'exemple le tableau 5.14 montre la structure en cinq couches[39] du carton maulé RC et les matières premières qui font partie de la couche, autrement dit le type de pâtes qu'il faut utiliser pour sa manufacture.

| Nom de la couche | Matières premières |
|---|---|
| Estuco | Fibra estucada en dos capas |
| Cara | Fibra de celulosa blanca (Fibra Corta) |
| Proteccion | Fibra de pulpa mecánica refinada, Fibra de recorte propio |
| Tripa | Fibra de pulpa mecánica refinada, Fibra de recorte tratado |
| Reverso | Fibra de celulosa cruda (Fibra Larga) |

**Tableau 5.14 :** Structure en cinq couches du carton maulé RC

La question se pose maintenant, au niveau de la représentation du grammage par couche (voir tableau 5.12) et la répartition de ses matières premières (voir tableau 5.14), car la valeur 300 de la figure 5.17 est une valeur approximative du grammage total pour cette sub-gamme de produits. Pour le faire, nous proposons une structure pour représenter la classe d'objet grammage et la classe d'objet fibre, telle qu'elle est montrée dans les figures 5.18 et 5.19.

| | | |
|---|---|---|
| Objet | : | GRAMMAGE |
| Attributs | : | Gramaje Estuco, Gramaje Cara, Gramaje Protección, Gramaje Tripa, Gramaje Reverso |
| Valeurs | : | 20, 50, 40, 165, 30 |

**Figure 5.18 :** Classe grammage

| | | |
|---|---|---|
| Objet | : | FIBRE |
| Attributs | : | Fibra estucada en dos capas, Fibra de celulosa blanca (fibra corta), Fibra de pulpa mecánica refinada, Fibra de recorte propio, Fibra de pulpa mecánica refinada, Fibra de recorte tratado, Fibra de celulosa cruda (fibra larga) |
| Valeurs | : | Spécifie dans la recette |

**Figure 5.19 :** Classe fibre

Dans la figure 5.19 nous ne pouvons pas divulguer les valeurs de la recette car il s'agit bien entendu d'un secret industriel (dont nous ne connaissons rien). Cependant, cela n'a aucune importance parce que nous ne nous sommes pas intéressés dans cette thèse par la problématique de capitalisation des connaissances de fabrication mais plutôt par la problématique de gestion des connaissances autour de la traçabilité des défauts détectés chez le client. En d'autres termes, les informations que nous nous sommes fixées de rechercher dans l'analyse de documents ont été relatives à la construction d'un modèle qui nous permet de faire la traçabilité, et dont les valeurs de la recette pour les objets impliqués ne seront pas spécifiées.

---

[38] Nous ajoutons toujours un 0 pour représenter le grammage approximatif total: 300 g/m2.

Nous pensons que la mise en évidence de cette structure (voir figure 5.17) nous permet de mieux saisir la complexité de la formation de la pâte et de comprendre le rôle du grammage dans la recette de fabrication.

<u>Cas 1 : Carton 301B</u>

D'après le tableau 5.12, le grammage total du carton 301B doit être 305 g/m2. Or, comme nous avons cinq couches (voir tableau 5.14) pour cette sub-gamme de carton, il est clair que ce poids-là doit être réparti parmi ces couches, la question se pose maintenant au niveau de cette répartition, s'agira-t-il d'une réparation équitable ou au contraire y aura-t-il des couches qui prendront plus de poids que les autres. La réponse à cette question est donnée dans la fiche des spécifications techniques de manufacture de la structure du carton maulé RC. En effet, la figure 5.15 montre bien que le grammage total du carton 301B est reparti de façon différenciée entre ces couches.

En ce sens, la couche estuco a 20 g/m2 ($\approx 7\%$), la couche cara a 50 g/m2 ($\approx 16\%$), la couche proteccion a 40 g/m2 ($\approx 13\%$), la couche tripa a 165 g/m2 ($\approx 54\%$), la couche reverso a 30 g/m2 ($\approx 10\%$). Or, si nous faisons la somme de tous ces grammages partiels nous devons bien avoir 305 g/m2, ce qui correspond au grammage total du carton 301B.

Pourquoi avons-nous mis tous ces détails de calculs, parce que c'est précisément dans cette analyse que nous trouvons deux traits qui nous semblent extrêment importants pour comprendre la complexité de la traçabilité des défauts, mais aussi pour mettre en évidence la complexité, d'une part, de la détection de l'apparition d'un défaut au niveau de sa structure de cinq couches (voir tableau 5.14) et, d'autre part, l'explication de ses mécanismes (cause-effet) de formation.

---

[39] Nous avons laissé les noms en espagnol car il s'agit d'un jargon propre de l'entreprise, mais la traduction serait à peu près : estuco, face, protection, trippe, revers.

L'un de ces traits est le fait que la recette, comme nous venons de le dire, et nous l'avons évoqué dans la section 5.1.1, est le noyau invariant du système de connaissance, cela signifie que pour chaque sub-gamme (voir tableau 5.12) de manufacture, il y a une recette qui permet de gérer les procédés et ses paramètres associés de manufacture. En d'autres termes, le savoir-faire de l'entreprise CMPC Maulé réside alors dans la composition de la recette, c'est-à-dire dans le pourcentage de chaque composante de la couche pour faire le mélange des pâtes.

L'autre trait important est la frontière floue entre les cartons d'une même classe. En effet, ce constat, nous l'avons fait en consultant les spécifications techniques de chaque produit. Ainsi, dans le tableau 5.15, nous pouvons voir la composition par couche au niveau du grammage des deux produits proches du carton 301B. En ce sens, au-dessus, nous avons le carton 321B avec un grammage total de 325 g/m2 et au dessous nous avons le carton 291B avec un grammage total de 295 g/m2. Nous constatons aussi que la différence de la recette au niveau du poids se fait sentir au niveau de la couche tripa, et cela est vrai aussi pour les autres produits de la gamme de carton maulé RC (voir tableau 5.12). Alors, si l'entreprise souhaite la mise en marche d'un nouveau produit pour une gamme donnée, ce produit sera alors un raffinement des autres. En effet, en ce qui concerne le grammage (ou poids) nous pouvons dire qu'avec un grammage (ou poids) supérieur, il y a une plus grande consistance des fibres de la pâte et cela permet d'avoir une amélioration dans les caractéristiques des paramètres (voir tableau 5.2), par exemple : la rigidité.

| Carton Maulé RC | 291B | 301B | 321B |
|---|---|---|---|
| Grammage total (g/m2) | 295 | 305 | 325 |
| Grammage estuco (g/m2) | 20 | 20 | 20 |
| Grammage cara (g/m2) | 50 | 50 | 50 |
| Grammage proteccion (g/m2) | 40 | 40 | 40 |
| Grammage tripa (g/m2) | 155 | 165 | 180 |
| Grammage reverso (g/m2) | 30 | 30 | 35 |

**Tableau 5.15 :** Un exemple de la spécification du grammage par couche

Nous allons nous intéresser maintenant à l'exploration des limites floues autour du carton 301B. D'après l'analyse des documents nous avons constaté que l'entreprise utilise les concepts de « valeurs acceptées » et « valeurs critiques ». Le premier concept est utilisé pour définir un intervalle de valeurs possibles au niveau du grammage total pour le produit en question. Par contre le deuxième concept est utilisé pour définir la frontière des valeurs critiques sur lesquelles le produit ne rentre pas dans cette classification. Le tableau 5.16 nous permet de mieux comprendre la mécanique d'utilisation de ces deux concepts appliqués au carton 301B, cependant ce choix-là ne reste pas de généralité à notre approche.

| Carton Maulé RC | grammage total (g/m2) | valeurs acceptés (g/m2) | | valeurs critiques (g/m2) | |
|---|---|---|---|---|---|
| 291B | 295 | 286 | 304 | 280 | 310 |
| 301B | 305 | 296 | 314 | 290 | 320 |
| 321B | 325 | 315 | 335 | 309 | 341 |

**Tableau 5.16 :** Un exemple de la norme du grammage

Ainsi, supposons que le carton qu'on doit fabriquer soit le carton 301B, cela signifie que le grammage objectif est de 305 g/m2. Or, d'après le tableau 5.16 la valeur du grammage peut varier dans l'intervalle [296,314], autrement dit ce paramètre est autorisé à grossir ou bien à maigrir 9 g/m2, si cela est le cas, et nous n'en avons aucun doute, le grammage restera dans son poids, pour ainsi dire, c'est-à-dire qu'on est toujours dans la sub-gamme 301B.

Or, l'histoire n'est pas finie, car le paramètre peut encore continuer à grossir ou à maigrir et cela jusqu'à 6 g/m2 de plus. Mais, au delà de ce supplément-là ce paramètre n'appartient plus à la sub-gamme 301B, à moins qu'il ait vraiment changé de taille, ou bien il est maintenant de la sub-gamme 291B ; dans le cas où ce paramètre a maigri de plus de 15 g/m2 par rapport au poids idéal (305 g/m2), ou dans le cas contraire il a grossi de plus de 15 g/m2 par rapport au même poids, et dans ce cas le grammage de la sub-gamme est 321B.

C'est pourquoi il y a là, à notre avis, l'une des causes de risque d'apparition d'un défaut chez le client, parce que si dans le bon de commande d'un client l'unité commerciale a introduit dans le système de planification spécifié que le produit est 301B, alors l'unité de planification et de contrôle des opérations doit appliquer la politique des « valeurs acceptées » et « valeurs critiques ». La question ici ne se pose pas au niveau de la politique des valeurs critiques car la procédure de contrôle de qualité rejette automatiquement les valeurs des paramètres qui sont en dehors de l'intervalle critique, mais au niveau de l'intervalle des valeurs acceptées et valeurs critiques.

En effet, nous avons 6 g/m2 pour lesquels la valeur du grammage peut encore varier, mais cela signifie que le produit tend à se transformer petit à petit dans un autre produit, si l'on revient à notre cas, si l'on perd du poids, on va vers un 291B, ou bien dans le cas contraire si l'on gagne du poids, on va vers un 321B.

Autrement dit, il y a un changement dans la structure du produit, et cela présente pour l'entreprise un risque de formation de défauts, autrement dit ce changement de structure peut être un réel danger au niveau de l'apparition d'un défaut. En effet, d'après le tableau 5.11 nous savons que chaque produit a un ensemble de caractéristiques qui sont gérées par l'ensemble des paramètres qui la conforment, autrement dit les valeurs de ces paramètres sont à la base de l'émergence de ces caractéristiques dans sa totalité.

Or, l'usage du produit introduit une entropie constante dans sa structure, donc si l'entreprise veut maîtriser la qualité de ces produits, elle doit définir d'abord, ce que d'ailleurs elle fait, les types d'usage que la structure du produit est capable de supporter et ses méthodes d'impression (voir tableau 5.1). En conséquence, dans l'exemple du tableau 5.16, la différence des cartons 291B, 301B et 321B est fixée chez le client par son usage, et en général les plus courants défauts se trouvent dans la méthode d'impression.

Cas 2 : Analogie avec l'industrie du déménagement

Or, afin d'illustrer le rôle du grammage dans le risque d'apparition d'un défaut dans un autre domaine, supposons que le paramètre critique soit la « résistance superficielle » (voir tableau 5.2). A cet effet, supposons que notre client soit une entreprise qui fabrique des boîtes de cartons pour les industries de déménagement. Le client, passe toujours une commande pour le produit 301B, dont le poids des trois dernières commandes a été : 310, 312, et 317 g/m2, nous voyons bien dans le tableau 5.16 que ces trois valeurs du grammage pour le carton 301B sont bien admises dans la norme.

Supposons maintenant que les entreprises de déménagement pour lesquelles notre client fabrique des boîtes avec le carton 301B n'ont pas enregistré de problèmes de défauts au cours de son activité, cela peut signifier si l'on regarde le tableau 5.15, que le carton 321B pour fabriquer ces boîtes peut lui convenir aussi, mais cela à un prix majeur, et donc qu'il n'est pas nécessaire de le faire.

En plus, supposons que dans la commande en cours, le grammage enregistre un poids de 290 g/m2, et donc un poids acceptable pour le produit 301B, et supposons aussi que pendant l'usage propre de l'activité de déménagement dans certaines boîtes on ait détecté l'apparition d'un défaut qui a été aussitôt rapporté par la suite à l'unité d'attention au client. Cette unité enregistre la réclamation dans la base des réclamations et puis l'unité de planification et de contrôle des opérations prend en charge la réclamation pour son suivi.

Nous avons un peu exagéré notre supposition d'une prise en charge rapide de la réclamation, car nous savons par expérience que ce que font les entreprises de déménagement est plutôt de changer de fabricant de boîtes si le problème de défauts persiste.

Néanmoins, cette supposition peu réaliste nous permet quand même de continuer notre réflexion sur la traçabilité des défauts. Ainsi, essayons maintenant de rechercher la cause du défaut á partir du grammage et de la réclamation, qui a été enregistrée dans la base de réclamations pour le produit 301B, que comme nous l'avons déjà dit, le grammage total était de 290 g/m2.

Pour cela, revenons au tableau 5.16 pour constater que ce grammage appartient aussi au rang des valeurs acceptées pour le produit 291B. En effet, 290 appartient à l'intervalle [296,314], et donc cela peut signifier que la cause du défaut détecté par l'entreprise de déménagement peut être due au fait que la structure du produit 291B ne lui convient pas du tout pour les activités courantes d'une entreprise de déménagement. C'est ce que nous avons dit plus haut au sujet de l'analogie avec le pont.

Bref, nous voilà confrontés á un problème impossible à gérer pour l'entreprise aujourd'hui car les données du tableau 5.16 ne sont intégrées dans aucun système d'information de l'entreprise.

En plus, une fois que les produits ont franchi le processus de « passage » l'unité de planification et de contrôle des opérations doit prendre la décision de les rejeter car ils ne sont pas dans le poids, ou bien le faire avancer dans la gamme d'usinage, mais quelle que soit la décision, ce qu'il faut retenir, est que la donnée du grammage est enregistrée automatiquement par le système de contrôle comme une donnée opératoire, c'est-à-dire que sa valeur historique n'intéresse personne.

En effet, aujourd'hui si un client passe une commande pour un produit, par exemple notre 301B, alors l'unité de planification et de contrôle des opérations vérifie s'il y a un produit 301B qui a déjà été fabriqué comme produit semi-élaboré ou si la commande a déjà été annulée, si c'est le cas, alors si l'affaire est réglée. La question qui se pose maintenant est sur le risque d'apparition d'un défaut.

Il y a là, pourtant, un point sur lequel nous pouvons faire une première justification sur l'utilisation d'un système à base de cas où les cas de références nous permettront de stocker des informations floues relatives aux grammages des produits commandés pour le client.

En conséquence, ce que nous proposons est de garder dans un cas de référence l'intervalle flou du grammage par produit et par client, puis d'aller rechercher dans la base de cas de commandes annulées, le cas (le produit) le plus adapté. Pour ce faire, dans la section 5.2 nous avons présenté la représentation et l'interrogation dans un système FSQL afin de faire la gestion d'une commande floue dans l'entrepôt de stockage de produits finis.

**Conclusion du chapitre**

Nous avons présenté dans ce chapitre :

- premièrement, la mise en œuvre du système de connaissance dans un contexte industriel. Ceci a permis de valider les quatre hypothèses du modèle proposé, à savoir : l'hypothèse du noyau invariant du système de connaissance, l'hypothèse de l'enaction du système de connaissance, l'hypothèse de la connaissance imparfaite du système de connaissance, et l'hypothèse de spontanéité des relations du système de connaissance. L'hypothèse du noyau invariant du système de connaissance a permis de définir quatre noyaux autour desquels le système de connaissance doit être construit, à savoir : le noyau recettes, le noyau conflits, le noyau réclamations, le noyau défauts. Ces noyaux font partie du savoir-faire de trois unités, à savoir : l'unité de planification et de contrôle des opérations (recettes et défauts), l'unité commerciale (conflits), l'unité d'attention au client (réclamations). L'hypothèse de l'enaction du système de connaissance a permis de définir la dualité organisation/structure du système de connaissance. Le domaine social a été défini par les relations humaines autour du *carton* qui définit l'organisation, par contre le domaine physique a été défini à travers les *produits*, les *procédés*, et le *processus*. Ainsi, l'unité (le carton) maintient son identité à partir de sa structure, laquelle a été définie par les *produits*, les *procédés*, et les *processus*. L'hypothèse de la connaissance imparfaite du système de connaissance a permis de vérifier dans le système de planification et dans le système de gestion des réclamations, l'existence de deux problématiques autour de l'imprécis et de l'incertain. L'hypothèse de spontanéité des relations du système de connaissance a été étudiée dans le système de gestion des défauts à travers le processus de traçabilité des défauts du carton. Pour vérifier cette hypothèse, les relations « produit, procédés, processus » ont formé trois niveaux d'explication de relations autopoïétiques.

- deuxièmement, nous avons présenté la mise en œuvre du système opérationnel pour la représentation de données imprécises et l'interrogation d'une base de données relationnelles floues (BDRF) de type FSQL, à travers trois étapes, à savoir : la détection du besoin de l'utilisateur, la traduction du besoin utilisateur dans un langage FSQL, l'extraction de connaissance du système FSQL. La détection du besoin de l'utilisateur a été envisagée à partir de la gestion d'une commande floue dans un entrepôt de stockage. La traduction du besoin utilisateur dans un langage FSQL a été faite à travers (1) le modèle conceptuel FuzzyEER et son progiciel FuzzyCase associé qui nous a permis de representer trois entités : entité cartulinas (cartons), entité pilas (pallet), et entité rollos (rolls), et (2) le modèle GEFRED qui a permis le stockage des objets (entités cartulinas, pilas et rollos) du modèle FuzzyEER dans la BDRF de FSQL. L'extraction de connaissance du système FSQL a été possible grâce à l'interface FSQL.

- troisièmement, nous avons identifié dans le *processus de traçabilité des défauts du carton*, un certaine nombre de scénarios (qui ont été classés dans deux groupes : contexte-objet et contexte-événement), autour des systèmes de fabrication, de planification, de gestion de réclamations, et de gestion des défauts, capables d'être pris en charge par le système opérationnel qui a été expliqué dans le point antérieur.

La conclusion générale de ce chapitre est que dans le site de Maulé nous avons constaté deux sortes de connaissances qu'il faut gérer, à savoir :

- la connaissance (tacite ou explicite) du business[40] réside : (1) dans les savoir, savoir-faire et savoir-technique du produit-procédé-processus de manufacture ; (2) dans le processus de commercialisation du produit et (3) dans le processus d'attention au client. Les acteurs qui participent à la gestion de cette connaissance (manufacture et vente du carton) gravitent autour de trois unités principalement : l'unité de planification et contrôle des opérations, l'unité commerciale et l'unité d'attention au client ;

- la connaissance émotionnelle[41] qui réside dans le savoir-être des acteurs de ces unités et qui participe au réseau de compromis[42] dans la gestion du business (manufacture et vente du carton).

---

[40] Ici, nous utilisons le mot « business » et non pas « entreprise » afin d'attirer l'attention sur le fait que dans cette étude de cas, et d'ailleurs dans toute la thèse, nous suivons une démarche de *gestion*. Autrement dit, ce qui nous intéresse ce sont les mécanismes de prise de décision sur un domaine (manufacture, vente,...) plutôt que la structuration des connaissances de l'entreprise, par exemple dans une mémoire d'entreprise.
[41] Ici, nous avons préféré utiliser le terme « connaissance » et non pas « intelligence », afin de parler de capital émotionnel et capital intellectuel dans ce paragraphe, mais il s'agit plutôt d'un choix de rédaction, puisque derrière la connaissance émotionnelle se trouve la théorie de l'intelligence émotionnelle mise en évidence par Daniel Goleman dans son livre « L'intelligence émotionnelle 2 » [Goleman, 99].

Cela n'est pas une nouveauté, car dans la cartographie générique de gestion, les trois leviers stratégiques de l'entreprise, sont composés de : (1) la gestion du capital intellectuel qui prend forme à travers la gestion des flux de données, d'informations et de connaissances ; (2) la gestion des ressources telles que l'environnement, la pollution, et les coûts ; et (3) la gestion du capital émotionnel de l'entreprise. En ce sens, ces trois aspects posent des pressions quotidiennes à l'entreprise pour maîtriser sa rentabilité, autrement dit pour garantir sa survie.

Donc, la connaissance émotionnelle (et davantage la connaissance tacite) est impliquée dans ce processus. Ceci peut être une explication au fait que nous avons constaté qu'avec la même recette, les équipes de travail obtiennent des qualités différentes, pour être plus précis cela se passe plutôt au site de Santiago de la compagnie, tandis que dans la région du Maulé cela l'est beaucoup moins.

---

[42] Nous avons emprunté le terme « réseaux de compromis » à Fernando Flores [Flores, 96a], en gros, cela signifie un mode d'organisation du travail coopératif basé sur les modes d'être, les modes de faire pour achever l'objectif (le quoi faire).

# Conclusion Générale

Nous avons présenté dans cette thèse cinq chapitres :

Le chapitre 1 a permis de montrer qu'une approche cartésienne de la gestion des connaissances est insuffisante pour rendre compte de la complexité du phénomène, mieux vaut une approche sociotechnique. Cette approche a permis de visualiser les enjeux de la gestion des connaissances, à partir de la mise en évidence de deux concepts : "faire-évoluer" les connaissances (qui traduit le fait de créer de nouvelles connaissances là où il n'y a pas de savoir) et "faire-émerger" la connaissance, (qui traduit le fait de créer des connaissances nouvelles à partir d'une représentation non symbolique de la réalité). Egalement, l'enjeu de la gestion des connaissances se trouve dans la capacité de l'entreprise pour créer, apprendre et mobiliser son savoir (ou connaissances) à travers un système d'organisation du travail collectif qui permet à la fois la gestion des compétences, des connaissances et l'innovation durable. Pour nous, ces trois conditions sont essentielles pour bâtir une vraie culture d'entreprise autour d'un avantage collectif, en laissant derrière les avantages productifs, concurrentiels, compétitifs, et coopératifs.

Le chapitre 2 a permis de montrer que la racine du concept "faire-évoluer" les connaissances est associée au paradigme de "sélection d'information" d'après l'approche de l'enaction de Karl Weick, tandis que la racine du concept "faire-émerger" les connaissances, est associée au paradigme "d'émergence de signification" d'après l'approche de l'enaction de Humberto Maturana et Francisco Varela. La connaissance imparfaite a été justifiée justement à partir de cette dernière approche. Dans ce contexte, nous postulons une collaboration de trois approches pour expliquer l'organisation d'un système de connaissance, à savoir : l'approche système (entrées/transformation/sorties sur la base d'une dualité système/environnement) ; l'approche cybernétique de deuxième ordre (organisation de l'unité à travers les mécanismes de contrôle et de régulation des variables qui correspondent aux différents flux du système de connaissance) qui a permis d'observer la connaissance dans l'entreprise comme une organisation de relations entre composants, l'approche autopoïétique de Santiago et de Valparaiso (définition de l'unité, maintien de l'identité et l'autonomie du système clos).

Le chapitre 3 a permis de formuler le modèle autopoïétique de la gestion des connaissances imparfaites, à travers le dialogue de trois dualités, à savoir : dualité organisation/système (approche cybernétique de deuxième ordre), dualité organisation/structure (approche autopoïétique de Santiago et de Valparaiso), et dualité organisation/environnement (approche système).

Le chapitre 4 a permis de formuler l'approche de l'enaction de Humberto Maturana et Francisco Varela en terme de connaissance « imparfaite » sous l'angle de la théorie de l'imprécis et de l'incertain. Ceci a permis la définition d'un système connaissance et la matérialisation d'un système opérationnel à travers un système FSQL.

Le chapitre 5 enfin a permis de formuler la validation du modèle proposé dans un secteur industriel : l'industrie du carton en Amérique Latine. Ce qui a été possible par la validation de quatre hypothèses, à savoir :

-   l'hypothèse de l'enaction ;
-   l'hypothèse de spontanéité des relations ;
-   l'hypothèse du noyau invariant du système de connaissance ;
-   l'hypothèse de la connaissance imparfaite.

Pour ce qui est des perspectives d'évolution de ce travail elles peuvent être apparaître au nombre de trois :

- la mise en évidence et la quantification du concept retour d'expérience à partir de bases de cas et bases de règles. En particulier, comment ce concept enrichie et modifie le système de connaissance sous une lecture d'autopoiese ;

- l'expérimentation à d'autres secteurs sociotechniques mettant en évidence plus de connaissances tacites en vérifiant les noyaux invariants d'un tel système,

- l'organisation de concepts d'action-réaction tels qu'on peut les rencontrer dans les plans actions-corrections dans la constitution de plans qualité (afin de faire correspondre des situations de comportement (réaction) à la lecture d'informations en apparence indépendante).

Dans ces trois perspectives, la connaissance explicite est plus proche d'être modélisée par un système de connaissance de type base de règle, base de cas, alors que la connaissance tacite (où se trouve la vraie connaissance pour faire l'innovation) est difficile à intégrer dans un système d'information. En plus, le modèle autopoietique a permis de vérifier l'existence d'une autre source de connaissance appelée connaissance émotionnelle, liée aux rapports sociaux. Nous pouvons ainsi schématiser cette mécanique par une dynamique gestion des connaissances, gestion de l'innovation (connaissance tacite), gestion de compétences par les savoir-être, savoir-faire, savoir-vivre ensemble, savoir communiquer, savoir partager, ...

La gestion des connaissances se retrouve davantage dans un réseau de collectivité, que dans un système de compétition entre individus. Elle se situera demain dans la création et la consommation au sein d'un système d'organisation de travail collectif.

Voilà un changement d'horizon que nous avons voulu explorer dans le cadre de cette thèse, et que nous aimerions poursuivre.

Je voudrais pour finir ce long travail, démarré en septembre 2000, conclure sur les mêmes inquiétudes que mon ami et collègue de travail Aquiles Limone [Limone, 77] « Nous attendons maintenant, d'une part la critique et la discussion autour des idées que nous avons exposées et, d'autre part, avec le temps de décantation nécessaire à l'obtention des fruits de ce travail, la maturation créatrice permettant de guider la continuation de cette recherche et de faire ressortir les répercutions et projections qu'il n'est pas possible de voir immédiatement. »

# Bibliographie

# ANNEXE 1

## Degré de possibilité et de nécessité dans FSQL sous Oracle 8

### I.1 Degré de possibilité du comparateur flou FEQ (Fuzzy EQual)

Le degré de possibilité du comparateur flou FEQ est défini de la façon suivante :

$$\text{CDEG}(R \textbf{ FEQ } S) = \textit{sup min}\ (\mu_R(x),\ \pi_S(x)) \quad \text{pour tout } x \in U \qquad (I.1)$$

Ce degré est utilisé pour obtenir une mesure de compatibilité des valeurs $x$ dans lequel la fonction d'appartenance $\mu_R(x)$ est *possiblement égale* dans la distribution de possibilité $\pi_S(x)$ pour tout $x \in U$. Dans le filtrage flou cela signifie que la requête FSQL retournera les tuples de la Banque de Données (BD) avec un degré 1 si $C_R \geq D_S$; avec un degré entre 0 et 1 si $C_R < D_S$ et $D_R > C_S$; et avec un degré 0 dans les autres cas.

La figure A.1.1 permet de représenter les critères de comparaison de cette définition.

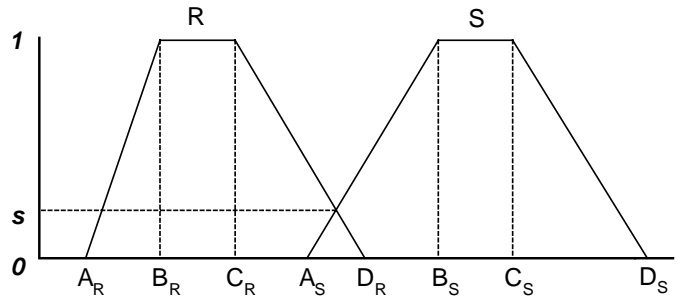

**Figure A.1.1 :** Mesure de compatibilité FEQ entre R et S

### I.2 Degré de nécessité du comparateur flou NFEQ (Necessarily Fuzzy EQual)

Le degré de nécessité du comparateur flou NFEQ est défini de la façon suivante :

$$\text{CDEG}(R \textbf{ NFEQ } S) = \textit{inf max}\ (1\text{-}\mu_R(x),\ \pi_S(x)) \quad \text{pour tout } x \in U \qquad (I.2)$$

Ce degré est utilisé pour obtenir une mesure de compatibilité des valeurs $x$ dans lequel la fonction d'appartenance $\mu_R(x)$ est *nécessairement égale* dans la distribution de possibilité $\pi_S(x)$ pour tout $x \in U$. Dans le filtrage flou cela signifie que la requête FSQL retournera les tuples de la BD avec un degré 1 si $C_R \geq D_S$; avec un degré entre 0 et 1 si $C_R < D_S$ et $D_R > C_S$; et avec un degré 0 dans les autres cas.

La figure A.1.2 permet de représenter les critères de comparaison de cette définition.

---

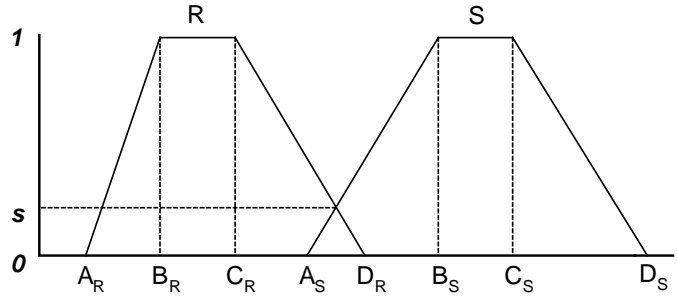

**Figure A.1.2 :** Mesure de compatibilité NFEQ entre R et S

## I.3 Degré de possibilité du comparateur flou FGT (Fuzzy Greater Than)

Le degré de possibilité du comparateur flou FGT est défini de la façon suivante :

$$
\text{CDEG } (R \textbf{ FGT } S) \quad = \quad
\begin{cases}
1 & \text{pour } C_R \geq D_S \\[2ex]
\dfrac{D_R - C_S}{(D_S - C_S) - (C_R - D_R)} & \text{pour } C_R < D_S \text{ et } D_R > C_S \\[2ex]
0 & \text{pour } D_R \leq C_S
\end{cases}
$$

Ce degré est utilisé pour obtenir une mesure de compatibilité des valeurs $x$ dans lequel la fonction d'appartenance $\mu_R(x)$ est *possiblement plus grand* dans la distribution de possibilité $\pi_S(x)$ pour tout $x \in U$. Dans le filtrage flou cela signifie que la requête FSQL retournera les tuples de la BD avec un degré 1 si $C_R \geq D_S$; avec un degré entre 0 et 1 si $C_R < D_S$ et $D_R > C_S$; et avec un degré 0 dans les autres cas.

La figure A.1.3 permet de représenter les critères de comparaison de cette définition.

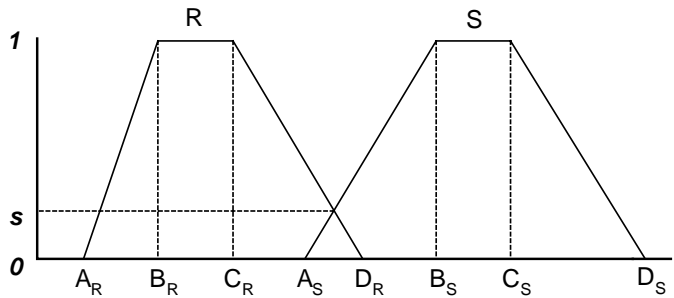

**Figure A.1.3 :** Mesure de compatibilité FGT entre R et S

## I.4 Degré de nécessité du comparateur flou NFGT (Necessarily Fuzzy Greater Than)

---

Le degré de nécessité du comparateur flou NFGT est défini de la façon suivante :

$$
\text{CDEG (R NFGT S)} = \begin{cases}
1 & \text{pour } A_R \geq D_S \\[2em]
\dfrac{B_R - C_S}{(D_S - C_S) - (A_R - B_R)} & \text{pour } A_R < D_S \text{ et } B_R > C_S \\[2em]
0 & \text{pour } B_R \leq C_S
\end{cases}
$$

Ce degré est utilisé pour obtenir une mesure de compatibilité des valeurs $x$ dans lequel la fonction d'appartenance $\mu_R(x)$ est *nécessairement plus grand* dans la distribution de possibilité $\pi_S(x)$ pour tout $x \in U$. Dans le filtrage flou cela signifie que la requête FSQL retournera les tuples de la BD avec un degré 1 si $A_R \geq D_S$; avec un degré entre 0 et 1 si $A_R < D_S$ et $B_R > C_S$; et avec un degré 0 dans les autres cas.

La figure A.1.4 permet de représenter les critères de comparaison de cette définition.

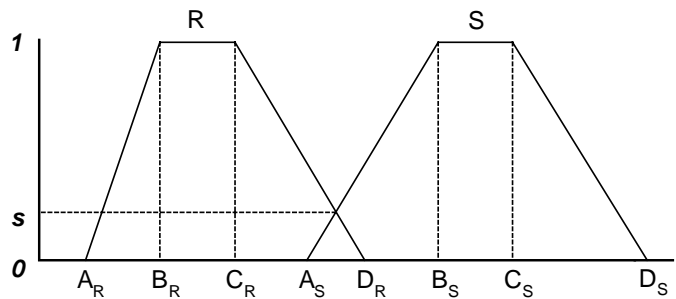

**Figure A.1.4 :** Mesure de compatibilité NFGT entre R et S

## I.5 Degré de possibilité du comparateur flou FGEQ (Fuzzy Greater EQual)

Le degré de possibilité du comparateur flou FGEQ est défini de la façon suivante :

$$
\text{CDEG (R FGEQ S)} = \begin{cases}
1 & \text{pour } C_R \geq B_S \\[2em]
\dfrac{D_R - A_S}{(B_S - A_S) - (C_R - D_R)} & \text{pour } C_R < B_S \text{ et } D_R > A_S \\[2em]
0 & \text{pour } D_R \leq A_S
\end{cases}
$$

Ce degré est utilisé pour obtenir une mesure de compatibilité des valeurs $x$ dans lequel la fonction d'appartenance $\mu_R(x)$ est *possiblement plus grand au égale* dans la distribution de possibilité $\pi_S(x)$ pour tout $x$

---------------------------------------------------------------------------------------------------

$\in U$. Dans le filtrage flou cela signifie que la requête FSQL retournera les tuples de la BD avec un degré 1 si $C_R \geq B_S$; avec un degré entre 0 et 1 si $C_R < B_S$ et $D_R > A_S$; et avec un degré 0 dans les autres cas.

La figure A.1.5 permet de représenter les critères de comparaison de cette définition.

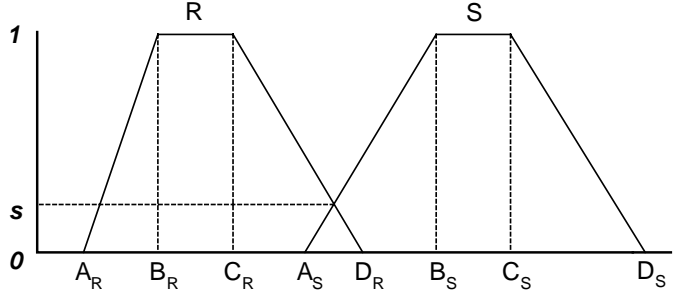

**Figure A.1.5 :** Mesure de compatibilité FGEQ entre R et S

## I.6 Degré de nécessité du comparateur flou NFGEQ (Necessarily Fuzzy Greater EQual)

Le degré de nécessité du comparateur flou NFGEQ est défini de la façon suivante :

$$\text{CDEG (R NFGEQ S)} = \begin{cases} 1 & \text{pour } A_R \geq B_S \\[2em] \dfrac{B_R - A_S}{(B_S - A_S) - (A_R - B_R)} & \text{pour } A_R < B_S \text{ et } B_R > A_S \\[2em] 0 & \text{pour } B_R \leq A_S \end{cases}$$

Ce degré est utilisé pour obtenir une mesure de compatibilité des valeurs $x$ dans lequel la fonction d'appartenance $\mu_R(x)$ est *nécessairement plus grand au égale* dans la distribution de possibilité $\pi_S(x)$ pour tout $x \in U$. Dans le filtrage flou cela signifie que la requête FSQL retournera les tuples de la BD avec un degré 1 si $A_R \geq B_S$; avec un degré entre 0 et 1 si $A_R < B_S$ et $B_R > A_S$; et avec un degré 0 dans les autres cas.

La figure A.1.6 permet de représenter les critères de comparaison de cette définition.

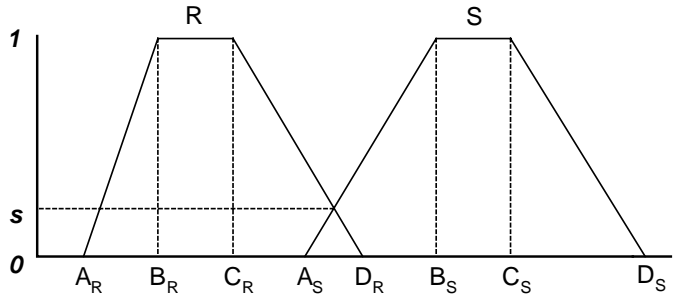

**Figure A.1.6 :** Mesure de compatibilité NFGEQ entre R et S

## I.7 Degré de possibilité du comparateur flou FLT (Fuzzy Less Than)

Le degré de possibilité du comparateur flou FLT est défini de la façon suivante :

$$CDEG\ (R\ FLT\ S) = \begin{cases} 1 & \text{pour } B_R \leq A_S \\[2em] \dfrac{A_R - B_S}{(A_S - B_S) - (B_R - A_R)} & \text{pour } B_R > A_S \text{ et } A_R < B_S \\[2em] 0 & \text{pour } A_R \geq B_S \end{cases}$$

Ce degré est utilisé pour obtenir une mesure de compatibilité des valeurs $x$ dans lequel la fonction d'appartenance $\mu_R(x)$ est *possiblement plus petite* que la distribution de possibilité $\pi_S(x)$ pour tout $x \in U$. Dans le filtrage flou cela signifie que la requête FSQL retournera les tuples de la BD avec un degré 1 si $B_R \leq A_S$; avec un degré entre 0 et 1 si $B_R > A_S$ et $A_R < B_S$; et avec un degré 0 dans les autres cas.

La figure A.1.7 permet de représenter les critères de comparaison de cette définition.

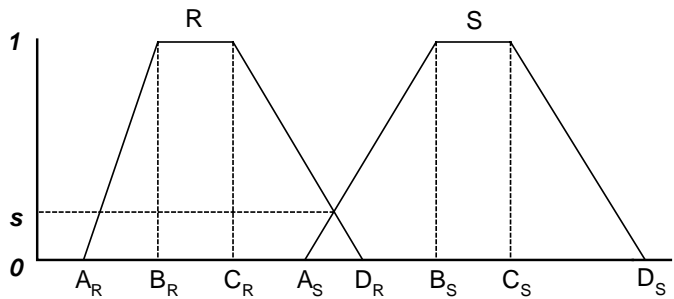

**Figure A.1.7 :** Mesure de compatibilité FLT entre R et S

---

### I.8 Degré de nécessité du comparateur flou NFLT (Necessarily Fuzzy Less Than)

Le degré de nécessité du comparateur flou NFLT est définie de la façon suivante :

$$
CDEG\ (R\ NFLT\ S)\quad =\quad
\begin{cases}
1 & \text{pour } D_R \leq A_S \\[2ex]
\dfrac{C_R - B_S}{(A_S - B_S) - (D_R - C_R)} & \text{pour } D_R > A_S \text{ et } C_R < B_S \\[2ex]
0 & \text{pour } C_R \geq B_S
\end{cases}
$$

Ce degré est utilisé pour obtenir une mesure de compatibilité des valeurs $x$ dans lequel la fonction d'appartenance $\mu_R(x)$ est *nécessairement plus petite* dans la distribution de possibilité $\pi_S(x)$ pour tout $x \in U$. Dans le filtrage flou cela signifie que la requête FSQL retouena les tuples de la BD avec un degré 1 si $D_R \leq A_S$; avec un degré entre 0 et 1 si $D_R > A_S$ et $C_R < B_S$; et avec un degré 0 dans les autres cas.

La figure A.1.8 permet de représenter les critères de comparaison de cette définition.

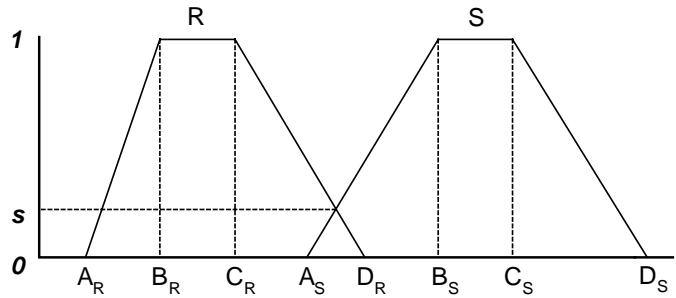

**Figure A.1.8 :** Mesure de compatibilité NFLT entre R et S

### I.9 Degré de possibilité du comparateur flou FLEQ (Fuzzy Less EQual)

Le degré de possibilité du comparateur flou FLEQ est défini de la façon suivante :

$$
CDEG\ (R\ FLEQ\ S)\quad =\quad
\begin{cases}
1 & \text{pour } B_R \leq C_S \\[2ex]
\dfrac{D_R - A_S}{(B_S - A_S) - (C_R - D_R)} & \text{pour } B_R > C_S \text{ et } A_R < D_S \\[2ex]
0 & \text{pour } A_R \geq D_S
\end{cases}
$$

Ce degré est utilisé pour obtenir une mesure de compatibilité des valeurs $x$ dans lequel la fonction d'appartenance $\mu_R(x)$ est *possiblement plus petite au égale* dans la distribution de possibilité $\pi_S(x)$ pour tout $x \in U$. Dans le filtrage flou cela signifie que la requête FSQL retouena les tuples de la BD avec un degré 1 si $B_R \leq C_S$; avec un degré entre 0 et 1 si $B_R > C_S$ et $A_R < D_S$; et avec un degré 0 dans les autres cas.

La figure A.1.9 permet de représenter les critères de comparaison de cette définition.

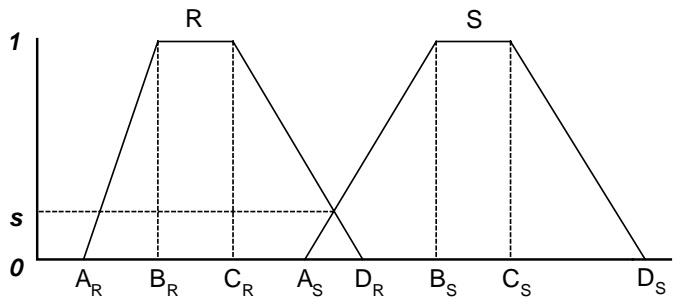

**Figure A.1.9 :** Mesure de compatibilité FLEQ entre R et S

## I.10 Degré de nécessité du comparateur flou NFLEQ (Necessarily Fuzzy Less EQual)

Le degré de nécessité du comparateur flou NFLEQ est défini de la façon suivante :

$$
\text{CDEG (R NFLEQ S)} = \begin{cases} 1 & \text{pour } D_R \leq C_S \\[2mm] \dfrac{C_R - D_S}{(C_S - D_S) - (D_R - C_R)} & \text{pour } D_R > C_S \text{ et } C_R < D_S \\[2mm] 0 & \text{pour } C_R \geq D_S \end{cases}
$$

Ce degré est utilisé pour obtenir une mesure de compatibilité des valeurs $x$ dans lequel la fonction d'appartenance $\mu_R(x)$ est *nécessairement plus petite au égale* dans la distribution de possibilité $\pi_S(x)$ pour tout $x \in U$. Dans le filtrage flou cela signifie que la requête FSQL retournera les tuples de la BD avec un degré 1 si $D_R \leq C_S$; avec un degré entre 0 et 1 si $D_R > C_S$ et $C_R < D_S$; et avec un degré 0 dans les autres cas.

La figure A.1.10 permet de représenter les critères de comparaison de cette définition.

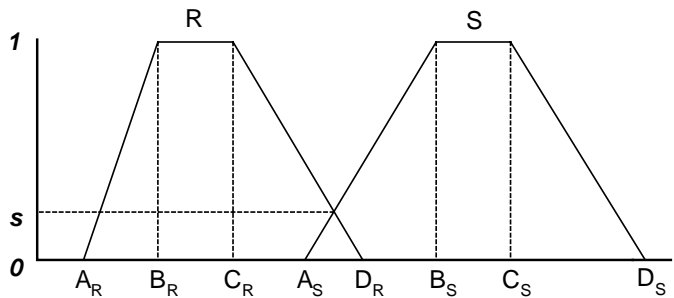

**Figure A.1.10 :** Mesure de compatibilité NFLEQ entre R et S

---

### I.11  Degré de possibilité du comparateur flou MGT (Much Greater Than)

Le degré de possibilité du comparateur flou MGT est défini de la façon suivante :

$$
\text{CDEG (R MGT S)} = \begin{cases} 1 & \text{pour } C_R \geq D_S + M \\[2mm] \dfrac{C_S + M - D_S}{(C_R - D_R) - (D_S - C_S)} & \text{pour } C_R < D_S + M \text{ et } D_R > C_S + M \\[2mm] 0 & \text{pour } D_R \leq C_S + M \end{cases}
$$

Ce degré est utilisé pour obtenir une mesure de compatibilité des valeurs $x$ dans lequel la fonction d'appartenance $\mu_R(x)$ est *possiblement beaucoup plus grand* dans la distribution de possibilité $\pi_S(x)$ pour tout $x \in U$. Dans le filtrage flou cela signifie que la requête FSQL retournera les tuples de la BD avec un degré 1 si $C_R \geq D_S + M$; avec un degré entre 0 et 1 si $C_R < D_S + M$ et $D_R > C_S + M$; et avec un degré 0 dans les autres cas.

La figure A.1.11 permet de représenter les critères de comparaison de cette définition.

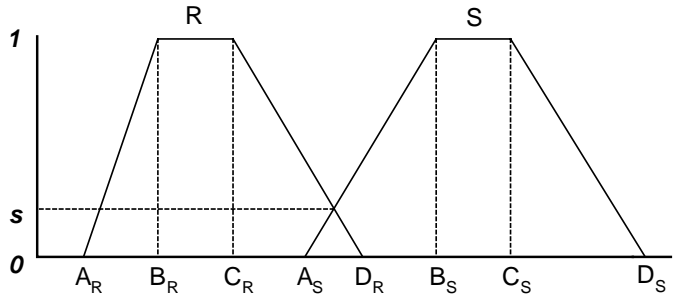

**Figure A.1.11 :** Mesure de compatibilité MGT entre R et S

### I.12  Degré de nécessité du comparateur flou NMGT (Necessarily Much Greater Than)

Le degré de nécessité du comparateur flou NMGT est défini de la façon suivante :

$$
\text{CDEG (R NMGT S)} = \begin{cases} 1 & \text{pour } A_R \geq D_S + M \\[2mm] \dfrac{C_S + M - B_R}{(A_R - B_R) - (D_S - C_S)} & \text{pour } A_R < D_S + M \text{ et } B_R > C_S + M \\[2mm] 0 & \text{pour } B_R \leq C_S + M \end{cases}
$$

---

Ce degré est utilisé pour obtenir une mesure de compatibilité des valeurs $x$ dans lequel la fonction d'appartenance $\mu_R(x)$ est *nécessairement beaucoup plus grand* dans la distribution de possibilité $\pi_S(x)$ pour tout $x \in U$. Dans le filtrage flou cela signifie que la requête FSQL retournera les tuples de la BD avec un degré 1 si $A_R \geq D_S + M$; avec un degré entre 0 et 1 si $A_R < D_S + M$ et $B_R > C_S + M$; et avec un degré 0 dans les autres cas.

La figure A.1.12 permet de représenter les critères de comparaison de cette définition.

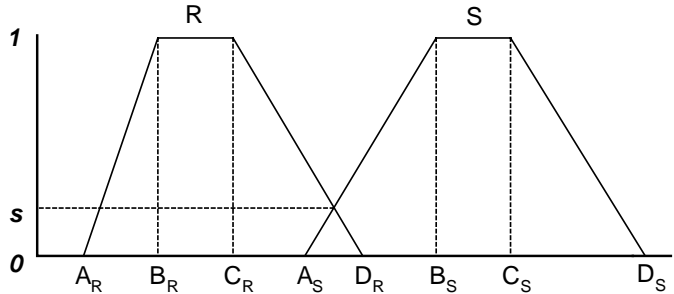

**Figure A.1.12 :** Mesure de compatibilité NMGT entre R et S

## I.13  Degré de possibilité du comparateur flou MLT (Much Less Than)

Le degré de possibilité du comparateur flou MLT est défini de la façon suivante :

$$
\text{CDEG (R MLT S)} = \begin{cases} 1 & \text{pour } B_R \leq A_S - M \\[2ex] \dfrac{B_S - M - A_R}{(B_R - A_R) - (A_S - B_S)} & \text{pour } B_R > A_S - M \text{ et } A_R < B_S - M \\[2ex] 0 & \text{pour } A_R \geq B_S - M \end{cases}
$$

Ce degré est utilisé pour obtenir une mesure de compatibilité des valeurs $x$ dans lequel la fonction d'appartenance $\mu_R(x)$ est *possiblement beaucoup plus petit* que la distribution de possibilité $\pi_S(x)$ pour tout $x \in U$. Dans le filtrage flou cela signifie que la requête FSQL retournera les tuples de la BD avec un degré 1 si $B_R \leq A_S - M$; avec un degré entre 0 et 1 si $B_R > A_S - M$ et $A_R < B_S - M$; et avec un degré 0 dans les autres cas.

La figure A.1.13 permet de représenter les critères de comparaison de cette définition.

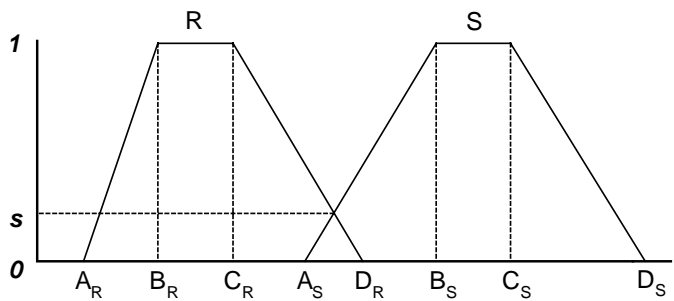

**Figure A.1.13 :** Mesure de compatibilité MLT entre R et S

---

### I.14  Degré de nécessité du comparateur flou NMLT (Necessarily Much Much Less Than)

Le degré de nécessité du comparateur flou NMLT est défini de la façon suivante :

$$
CDEG\ (R\ NMLT\ S) \quad = \quad
\begin{cases}
1 & \text{pour } D_R \leq A_S - M \\[2ex]
\dfrac{B_S - M - C_R}{(D_R - C_R) - (A_S - B_S)} & \text{pour } D_R > A_S - M \text{ et } C_R < B_S - M \\[2ex]
0 & \text{pour } C_R \geq B_S - M
\end{cases}
$$

Ce degré est utilisé pour obtenir une mesure de compatibilité des valeurs $x$ dans lequel la fonction d'appartenance $\mu_R(x)$ est *nécessairement beaucoup plus petite* dans la distribution de possibilité $\pi_S(x)$ pour tout $x \in U$. Dans le filtrage flou cela signifie que la requête FSQL retournera les tuples de la BD avec un degré 1 si $D_R \leq A_S - M$; avec un degré entre 0 et 1 si $D_R > A_S - M$ et $C_R < B_S - M$; et avec un degré 0 dans les autres cas.

La figure A.1.14 permet de représenter les critères de comparaison de cette définition.

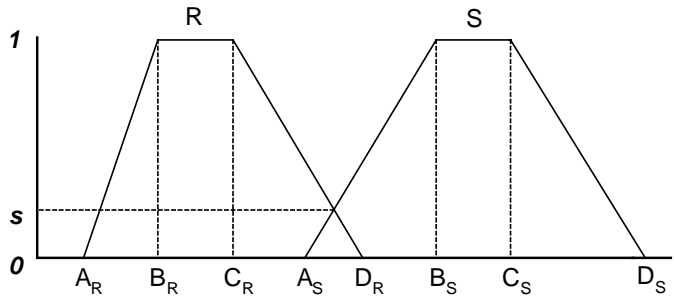

**Figure A.1.14 :** Mesure de compatibilité NMLT entre R et S

# ANNEXE 2

## Scénario de représentation et consultation des informations incomplètes, imprécises et incertaines dans FSQL pour attributs flous de Type 1 ou 2

### II.1    Représentation des données

Considérons l'exemple qui a été donné dans [GALINDO, 99] relatif au traitement des compétences techniques de joueurs de basket-ball (voir tableau II.1). Cet exemple nous paraît intéressant car il permet de penser dans un modèle de gestion de connaissances autour d'une tâche dans un environnement relationnel [URRUTIA et al, 01b]. En plus, il est similaire à l'exemple étudié dans [TESTEMALE, 84], [DUBOIS et PRADE, 85,88] dans le cadre de la manipulation et du traitement d'informations précises, imprécises, incomplètes, incertaines, nulles, inconnus et inapplicables dans une base de données relationnelle, qui permet de stocker et de traiter des questions vagues relatives à un groupe d'étudiants par rapport à leur niveau, à leur âge et à leurs sentiments mutuels.

**Tableau II.1** : Interface FSQL table joueurs

La relation du tableau II.1 est composée par quatre attributs. L'attribut « jugador » représente un indice pour identifier la tuple. L'attribut « equipo » représente le nom de l'équipe de basket-ball. L'attribut « altura » représente la taille d'un joueur où le domaine attaché est l'intervalle [160,240]. Les valeurs pour cet attribut sont des valeurs précises (192, 198, 170 et 177), d'étiquettes linguistiques (Bas, Normal, Grand, Très Grand) définies selon le tableau II.2, et de valeurs inconnus. Il s'agit donc d'un attribut flou de Type 2 sur un domaine continu car il admet des données précises et des données imprécises sur un domaine ordonné et continu. L'attribut « calidad » représente la qualité du joueur déterminée selon le nombre de points par match. Le domaine attaché à cet attribut est l'intervalle [0,50]. Les valeurs pour cet attribut sont des valeurs autour de $n$ (#6, #8, #10, #15, #25, #35), d'étiquettes linguistiques (malo, regular, bueno, muy bueno), de valeurs trapézoïdales ($[2,7,10,15], $[8,12,15,25], $[30,38,40,45], $[31,34,35,38]) et une valeur inconnu. Il s'agit donc aussi d'un attribut flou de Type 2 sur un domaine continu car il admet des données imprécises sur un domaine ordonné et continu.

| Taille | $A_R$ | $B_R$ | $C_R$ | $D_R$ |
|--------|-------|-------|-------|-------|
| **Bas** | **0** | **0** | **175** | **180** |
| **Normal** | **175** | **180** | **185** | **195** |
| **Grand** | **185** | **195** | **200** | **210** |
| **Très-Grand** | **200** | **210** | **1000** | **1000** |
| **170** | **170** | **170** | **170** | **170** |
| **177** | **177** | **177** | **177** | **177** |
| **192** | **192** | **192** | **192** | **192** |
| **198** | **198** | **198** | **198** | **198** |

**Tableau II.2 :** Étiquettes linguistiques et valeurs Crisp FSQL

## II.2    Consultation des données

Nous allons illustrer les 14 filtres flous de FSQL [GALINDO, 99] à l'aide d'une interrogation flexible à la relation « joueurs » (voir tableau II.1). Afin de simplifier les modèles nous allons prendre seulement en compte l'attribut « taille », avec également un seul comparateur flou dans la condition WHERE du SELECT. Par ailleurs, la comparaison floue se fait entre cet attribut et une constante, une valeur trapézoïdal FSQL, une étiquette linguistique FSQL, un intervalle FSQL, et une valeur autour-de-$n$ FSQL. Ses valeurs sont montrées dans le tableau II.3. Le seuil $\gamma$ vaut 0.05 dans toutes les requêtes floues FSQL.

| Constante FSQL | Symbole FSQL | $A_S$ | $B_S$ | $C_S$ | $D_S$ |
|---|---|---|---|---|---|
| **Valeur trapézoïdale** | **$[181, 186, 191, 220]** | **181** | **186** | **191** | **220** |
| **Étiquette linguistique** | **$GRAND** | **185** | **195** | **200** | **210** |
| **Intervalle** | **[180, 220]** | **180** | **180** | **220** | **220** |
| **Valeur autour-de-$n$** | **#197** | **197 – 6** | **197** | **197** | **197 + 6** |

**Tableau II.3 :** Constantes FSQL

## II.3    Modèle comportant une valeur trapézoïdale FSQL comme constante

Soit la valeur trapézoïdal $[181, 186,191, 220] comme constante donc d'après le tableau II.3, on a : $A_S = 181$, $B_S = 186$, $C_S = 191$ et $D_S = 220$.

### II.3.1    Filtres flous FEQ/NFEQ : Fuzzy/Necessarily EQual, possiblement/nécessairement égal

Soit la requête FEQ/NFEQ « Sélectionner toutes les informations des joueurs qui sont possiblement/nécessairement égale au valeur trapézoïdal $[181, 186, 191,220] ». La figure II.1 montre la syntaxe FSQL associé au modèle FEQ/NFEQ, tandis que les tableaux II.4 et II.5 montrent les résultats de ces requêtes floues FSQL.

| | |
|---|---|
| SELECT JOUEURS. %, CDEG (TAILLE) FROM JOUEURS<br>WHERE TAILLE **FEQ** $[181, 186, 191,220] THOLD 0.05; | SELECT JOUEURS. %, CDEG (TAILLE) FROM JOUEURS<br>WHERE TAILLE **NFEQ** $[181, 186, 191,220] THOLD 0.05; |

**Figure II.1** : Requêtes avec filtres flous TAILLE **FEQ**/**NFEQ** $[181, 186, 191,220]

| Nº Fila | JOUEUR | ÉQUIPE | TAILLE | QUALITÉ | CDEG(TAILLE) |
|---|---|---|---|---|---|
| 1 | P2 | Córdoba | TRÈS_GRAND | [2,7,10,15] | 0,51 |
| 2 | P3 | Granada | NORMAL | RÉGULIER | 0,93 |
| 3 | P4 | Granada | 192 | RÉGULIER | 0,97 |
| 4 | P5 | Granada | GRAND | 10±10 | 0,9 |
| 5 | P6 | Málaga | 198 | MAUVAIS | 0,76 |
| 6 | P7 | Málaga | TRÈS_GRAND | 35±10 | 0,51 |
| 7 | P10 | Sevilla | NORMAL | BON | 0,93 |
| 8 | P11 | Cádiz | TRÈS_GRAND | 25±10 | 0,51 |
| 9 | P13 | Almería | GRAND | TRÈS_BON | 0,9 |
| 10 | P14 | Almería | TRÈS_GRAND | 8±10 | 0,51 |
| 11 | P16 | Huelva | GRAND | TRÈS_BON | 0,9 |
| 12 | P17 | Huelva | UNKNOWN | UNKNOWN | 1 |
| 13 | P18 | Jaén | UNKNOWN | [8,12,15,25] | 1 |
| 14 | P19 | Jaén | NORMAL | 25±10 | 0,93 |

**Tableau II.4** : Résultat de la requête avec filtre flou TAILLE **FEQ** $[181, 186, 191,220]

Le tableau II.4 montre 14 joueurs vérifiants *possiblement égale* à $[181, 186, 191,220]. Le filtre flou TAILLE FGT $[181, 186, 191,220] cherche les tuples à partir de l'équation I.1 (voir paragraphe I.1). Nous utilisons alors les règles de cette équation dans la condition WHERE de la requête de la figure II.1. La règle $C_R \geq D_S$ avec $D_S = 220$ cherche les tuples avec une mesure de possibilité égale à 1. Nous avons alors, d'après le tableau II.2, la valeur $C_R$ qui satisfait à cette règle et qui vaut 1000. Ceci correspond à l'étiquette linguistique TRÈS_GRAND. Ensuite, la règle $C_R < D_S$ avec $D_S = 220$ et $D_R > C_S$ avec $C_S = 191$ cherche les tuples avec une mesure de possibilité entre 0 et 1. Or, d'après le tableau II.2, les valeurs C$_R$ et D$_R$ qui satisfont cette condition vallent 185 et 195 pour l'étiquette linguistique NORMAL, et 200 et 210 pour l'étiquette linguistique GRAND. La valeur crisp 198 satisfait aussi à cette règle.

La mesure de possibilité donne une valeur 0 dans les autres cas, c'est-à-dire oú D$_R$ ≤ C$_S$ avec C$_S$ = 191. Il s'agit des valeurs crisp 170, 177 et 192, et de l'étiquette linguistique BAS comme le montre le tableau II.2. Le comparateur FEQ assigne par définition la valeur 1 à la mesure de possibilité de UNKNOWN.

| Nº Fila | JOUEUR | ÉQUIPE | TAILLE | QUALITÉ | CDEG(TAILLE) |
|---|---|---|---|---|---|
| 1 | P4 | Granada | 192 | RÉGULIER | 0,97 |
| 2 | P5 | Granada | GRAND | 10±10 | 0,51 |
| 3 | P6 | Málaga | 198 | MAUVAIS | 0,76 |
| 4 | P13 | Almería | GRAND | TRÈS_BON | 0,51 |
| 5 | P16 | Huelva | GRAND | TRÈS_BON | 0,51 |

**Tableau II.5 :** Résultat de la requête avec filtre flou TAILLE **NFEQ** $[181,186,191,220]

Le tableau II.5 montre 5 joueurs qui sont *nécessairement égale* à $[181, 186, 191, 220]. Le filtre flou TAILLE NFEQ $[181, 186, 191, 220] cherche les tuples à partir de l'équation I.2 (voir paragraphe I.2). Nous utilisons alors les règles de cette équation dans la condition WHERE de la requête. La règle A$_R$ ≥ D$_S$ avec D$_S$ = 220 cherche les tuples avec une mesure de nécessité égale à 1. Nous observons, d'après le tableau II.2, qu'il n'y a aucune valeur A$_R$ qui satisfait cette règle. Ensuite, la règle A$_R$ < D$_S$ avec D$_S$ = 220 et B$_R$ > C$_S$ avec C$_S$ = 191 cherche les tuples avec une mesure de nécessité entre 0 et 1. Or, d'après le tableau II.2 les valeurs A$_R$ et B$_R$ qui satisfont cette condition vallent 185 et 195 pour l'étiquette linguistique GRAND, et 200 et 210 pour l'étiquette linguistique TRÈS_GRAND. Les valeurs crisp 192 et 198 satisfont aussi cette règle.

La mesure de nécessité donne 0 dans les autres cas, c'est-à-dire oú B$_R$ ≤ C$_S$ avec C$_S$ = 191. Il s'agit des valeurs crisp 170 et 177, et des étiquettes linguistiques BAS et NORMAL selon le montre le tableau II.2. Le comparateur NFEQ assigne par définition la valeur 0 à la mesure de possibilité de UNKNOWN.

## II.3.2  Filtres flous FGT (FGEQ)/NFGT (NFGEQ) : Fuzzy/Necessarily Greater (Equal) Than, possiblement/nécessairement plus grand (égal) que

Soient les requêtes FGT/NFGT et FGEQ/NFGEQ « Sélectionner toutes les informations des joueurs qui sont possiblement/nécessairement plus grand (égal) à la valeur trapézoïdale $[181, 186, 191, 220] ». Les figures II.2 et II.3 montrent la syntaxe FSQL associée aux modèles FGT/NFGT et FGEQ/NFGEQ respectivement, tandis que les tableaux II.6, II.7, II.8 et II.9 montrent les résultats de ces requêtes floues FSQL.

| SELECT JOUEURS. %, CDEG (TAILLE) FROM JOUEURS<br>WHERE TAILLE **FGT** $[181, 186, 191, 220] THOLD 0.05; | SELECT JOUEURS. %, CDEG (TAILLE) FROM JOUEURS<br>WHERE TAILLE **NFGT** $[181, 186, 191, 220] THOLD 0.05; |
|---|---|

**Figure II.2 :** Requêtes avec filtres flous TAILLE **FGT/NFGT** $[181, 186, 191, 220]

| SELECT JOUEURS. %, CDEG (TAILLE) FROM JOUEURS<br>WHERE TAILLE **FGEQ** $[181, 186, 191, 220] THOLD 0.05; | SELECT JOUEURS. %, CDEG (TAILLE) FROM JOUEURS<br>WHERE TAILLE **NFGEQ** $[181, 186, 191, 220] THOLD 0.05; |
|---|---|

**Figure II.3 :** Requêtes avec filtres flous TAILLE **FGEQ/NFGEQ** $[181, 186, 191, 220]

| Nº Fila | JOUEUR | ÉQUIPE | TAILLE | QUALITÉ | CDEG(TAILLE) |
|---|---|---|---|---|---|
| 1 | P2 | Córdoba | TRÈS_GRAND | [2,7,10,15] | 1 |
| 2 | P3 | Granada | NORMAL | RÉGULIER | 0,1 |
| 3 | P5 | Granada | GRAND | 10±10 | 0,49 |
| 4 | P6 | Málaga | 198 | MAUVAIS | 0,24 |
| 5 | P7 | Málaga | TRÈS_GRAND | 35±10 | 1 |
| 6 | P10 | Sevilla | NORMAL | BON | 0,1 |
| 7 | P11 | Cádiz | TRÈS_GRAND | 25±10 | 1 |
| 8 | P13 | Almería | GRAND | TRÈS_BON | 0,49 |
| 9 | P14 | Almería | TRÈS_GRAND | 8±10 | 1 |
| 10 | P16 | Huelva | GRAND | TRÈS_BON | 0,49 |
| 11 | P17 | Huelva | UNKNOWN | UNKNOWN | 1 |
| 12 | P18 | Jaén | UNKNOWN | [8,12,15,25] | 1 |
| 13 | P19 | Jaén | NORMAL | 25±10 | 0,1 |

**Tableau II.6 :** Résultat de la requête avec filtre flou TAILLE **FGT** $[181, 186, 191, 220]

Le tableau II.6 montre 13 joueurs qui sont possiblement plus grand que 2m20. Le filtre flou TAILLE FGT $[181, 186,191, 220] cherche les tuples à partir de l'équation I.3 (voir paragraphe I.3). Nous utilisons alors les règles de cette équation dans la condition WHERE de la requête. La règle $C_R \geq D_S$ avec $D_S = 220$ cherche les tuples avec une mesure de possibilité égale à 1. Nous observons, d'après le tableau II.2, que la valeur $C_R$ qui satisfait cette règle vaut 1000, ceci correspond à l'étiquette linguistique TRÈS_GRAND. Ensuite, la règle $C_R < D_S$ avec $D_S = 220$ et $D_R > C_S$ avec $C_S = 191$ cherche les tuples avec une mesure de possibilité entre 0 et 1. Or, d'après le tableau II.2 les valeurs $C_R$ et $D_R$ qui satisfont cette condition valent 185 et 195 pour l'étiquette linguistique NORMAL, et 200 et 210 pour l'étiquette linguistique GRAND. La valeur crisp 198 satisfait aussi cette règle. Nous avons résumé les calculs de la mesure de possibilité dans le tableau suivant :

| TAILLE | $p = D_R - C_S$ | $r = D_S - C_S$ | $s = C_R - D_R$ | $q = r - s$ | CDEG (TAILLE) $p/q$ |
|---|---|---|---|---|---|
| TRES_GRAND | - | - | - | - | 1 |
| NORMAL | 195 – 191 | 220 – 191 | 185 – 195 | 29 + 10 | 0.10 |
| GRAND | 210 – 191 | 220 – 191 | 200 – 210 | 29 + 10 | 0.49 |
| 198 | 198 – 191 | 220 – 191 | 198 – 198 | 29 | 0.24 |
| UNKNOWN | - | - | - | - | 1 |

La mesure de possibilité donne 0 dans les autres cas, c'est-à-dire oú $D_R \leq C_S$ avec $C_S = 191$. Il s'agit des valeurs crisp 170, 177 et 192, et de l'étiquette linguistique BAS comme le montre le tableau II.2. Le comparateur FGT assigne par définition la valeur 1 à la mesure de possibilité de UNKNOWN.

| Nº Fila | JOUEUR | ÉQUIPE | TAILLE | QUALITÉ | CDEG(TAILLE) |
|---|---|---|---|---|---|
| 1 | P2 | Córdoba | TRÈS_GRAND | [2,7,10,15] | 0,49 |
| 2 | P5 | Granada | GRAND | 10±10 | 0,1 |
| 3 | P6 | Málaga | 198 | MAUVAIS | 0,24 |
| 4 | P7 | Málaga | TRÈS_GRAND | 35±10 | 0,49 |
| 5 | P11 | Cádiz | TRÈS_GRAND | 25±10 | 0,49 |
| 6 | P13 | Almería | GRAND | TRÈS_BON | 0,1 |
| 7 | P14 | Almería | TRÈS_GRAND | 8±10 | 0,49 |
| 8 | P16 | Huelva | GRAND | TRÈS_BON | 0,1 |

**Tableau II.7 :** Résultat de la requête avec filtre flou TAILLE **NFGT** $[181, 186,191, 220]

Le tableau II.7 montre 8 joueurs qui sont nécessairement plus grand que 2m20. Le filtre flou TAILLE NFGT $[181, 186, 191, 220] cherche les tuples à partir de l'équation I.4 (voir paragraphe I.4). Nous utilisons alors les règles de cette équation dans la condition WHERE de la requête. La règle $A_R \geq D_S$ avec $D_S = 220$ cherche les tuples avec une mesure de nécessité égale à 1. Nous avons alors, d'après le tableau II.2, qu'il n'y a aucune valeur $A_R$ qui satisfait cette règle. Ensuite, la règle $A_R < D_S$ avec $D_S = 220$ et $B_R > C_S$ avec $C_S = 191$ cherche les tuples avec une mesure de nécessité entre 0 et 1. Or, d'après le tableau II.2 les valeurs $A_R$ et $B_R$ qui satisfont cette condition valent 185 et 195 pour l'étiquette linguistique GRAND, et 200 et 210 pour l'étiquette linguistique TRÈS_GRAND. Les valeurs crisp 192 et 198 satisfont aussi cette règle. Nous avons résumé les calculs de la mesure de nécessité dans le tableau suivant :

| TAILLE | $p = B_R - C_S$ | $r = D_S - C_S$ | $s = A_R - B_R$ | $q = r - s$ | CDEG (TAILLE) $p/q$ |
|---|---|---|---|---|---|
| TRÈS_GRAND | 210 – 191 | 220 – 191 | 200 – 210 | 29 + 10 | 0.49 |

| GRAND | 195 – 191 | 220 – 191 | 185 – 195 | 29 + 10 | 0.10 |
|---|---|---|---|---|---|
| 198 | 198 – 191 | 220 – 191 | 198 – 198 | 29 | 0.24 |

La mesure de nécessité vaut 0 dans les autres cas, c'est-à-dire oú $B_R \leq C_S$ avec $C_S = 191$. Il s'agit des valeurs crisp 170 et 177, et des étiquettes linguistiques BAS et NORMAL selon le montre le tableau II.2. Il faut souligner que par souci de simplification le comparateur NFGT assigne la valeur 0 à la mesure de nécessité CDEG(192) et par définition la mesure de nécessité de UNKNOWN vaut 0.

| Nº Fila | JOUEUR | ÉQUIPE | TAILLE | QUALITÉ | CDEG(TAILLE) |
|---|---|---|---|---|---|
| 1 | P2 | Córdoba | TRÈS_GRAND | [2,7,10,15] | 1 |
| 2 | P3 | Granada | NORMAL | RÉGULIER | 0,93 |
| 3 | P4 | Granada | 192 | RÉGULIER | 1 |
| 4 | P5 | Granada | GRAND | 10±10 | 1 |
| 5 | P6 | Málaga | 198 | MAUVAIS | 1 |
| 6 | P7 | Málaga | TRÈS_GRAND | 35±10 | 1 |
| 7 | P10 | Sevilla | NORMAL | BON | 0,93 |
| 8 | P11 | Cádiz | TRÈS_GRAND | 25±10 | 1 |
| 9 | P13 | Almería | GRAND | TRÈS_BON | 1 |
| 10 | P14 | Almería | TRÈS_GRAND | 8±10 | 1 |
| 11 | P16 | Huelva | GRAND | TRÈS_BON | 1 |
| 12 | P17 | Huelva | UNKNOWN | UNKNOWN | 1 |
| 13 | P18 | Jaén | UNKNOWN | [8,12,15,25] | 1 |
| 14 | P19 | Jaén | NORMAL | 25±10 | 0,93 |

**Tableau II.8 :** Résultat de la requête avec filtre flou TAILLE **FGEQ** \$[181,186,191,220]

Le tableau II.8 montre 14 joueurs qui sont possiblement plus grand ou égal à 2m20. Le filtre flou TAILLE FGEQ \$[181, 186, 191, 220] cherche les tuples à partir de l'équation I.5 (voir paragraphe I.5). Nous utilisons alors les règles de cette équation dans la condition WHERE de la requête. La règle $C_R \geq B_S$ avec $B_S = 186$ cherche les tuples avec une mesure de possibilité égale à 1. Il s'agit des valeurs crisp 192 et 198, et des étiquettes linguistiques GRAND et TRÈS_GRAND comme le montre le tableau II.2. Ensuite, la règle $C_R < B_S$ avec $B_S = 186$ et $D_R > A_S$ avec $A_S = 181$ cherche les tuples avec une mesure de possibilité entre 0 et 1. D'après le tableau II.2 les valeurs $C_R$ et $D_R$ qui satisfont cette condition vallent 185 et 195 pour l'étiquette linguistique NORMAL. La mesure de possibilité est donnée dans le tableau suivant :

| TAILLE | $p = D_R – A_S$ | $r = B_S – A_S$ | $s = C_R – D_R$ | $q = r – s$ | CDEG (TAILLE) p/q |
|---|---|---|---|---|---|
| TRES_GRAND | - | - | - | - | 1 |
| NORMAL | 195 – 181 | 186 – 181 | 185 – 195 | 5 + 10 | 0,93 |
| 192 | - | - | - | - | 1 |
| GRAND | - | - | - | - | 1 |
| 198 | - | - | - | - | 1 |
| UNKNOWN | - | - | - | - | 1 |

La mesure de possibilité donne 0 dans les autres cas, c'est-à-dire $D_R \leq A_S$ avec $A_S = 181$. Il s'agit des valeurs crisp 170 et 177, et de l'étiquette linguistique BAS comme le montre le tableau II.2. Le comparateur FGEQ assigne par définition la valeur 1 à la mesure de possibilité pour UNKNOWN.

| Nº Fila | JOUEUR | ÉQUIPE | TAILLE | QUALITÉ | CDEG(TAILLE) |
|---|---|---|---|---|---|
| 1 | P2 | Córdoba | TRÈS_GRAND | [2,7,10,15] | 1 |
| 2 | P4 | Granada | 192 | RÉGULIER | 1 |
| 3 | P5 | Granada | GRAND | 10±10 | 0,93 |
| 4 | P6 | Málaga | 198 | MAUVAIS | 1 |
| 5 | P7 | Málaga | TRÈS_GRAND | 35±10 | 1 |
| 6 | P11 | Cádiz | TRÈS_GRAND | 25±10 | 1 |
| 7 | P13 | Almería | GRAND | TRÈS_BON | 0,93 |
| 8 | P14 | Almería | TRÈS_GRAND | 8±10 | 1 |
| 9 | P16 | Huelva | GRAND | TRÈS_BON | 0,93 |

**Tableau II.9 :** Résultat de la requête avec filtre flou TAILLE **NFGEQ** \$[181, 186, 191, 220]

Le tableau II.9 montre 9 joueurs qui sont nécessairement plus grand ou égale au 2m20. Le filtre flou TAILLE NFGEQ \$[181, 186, 191, 220] cherche les tuples à partir de l'équation I.6 (voir paragraphe I.6). Nous utilisons alors les règles de cette équation dans la condition WHERE de la requête. La règle $A_R \geq B_S$ avec $B_S = 186$ cherche les tuples avec une mesure de nécessité égale à 1. Il s'agit des valeurs crisp 192 et 198, et de l'étiquette linguistique

---

TRÈS_GRAND comme le montre le tableau II.2. Ensuite, la règle $A_R < B_S$ avec $B_S = 186$ et $B_R > A_S$ avec $A_S = 181$ cherche les tuples avec une mesure de nécessité entre 0 et 1. D'après le tableau II.2 les valeurs $A_R$ et $B_R$ qui satisfont cette condition vallent 185 et 195 pour l'étiquette linguistique GRAND. Les valeurs crisp 192 et 198 satisfont aussi cette règle. La mesure de nécessité est donnée dans le tableau suivant :

| TAILLE | $p = B_R - A_S$ | $r = B_S - A_S$ | $s = A_R - B_R$ | $q = r - s$ | CDEG (TAILLE) p/q |
|---|---|---|---|---|---|
| TRES_GRAND | - | - | - | - | 1 |
| 192 | - | - | - | - | 1 |
| GRAND | 195 – 181 | 186 – 181 | 185 – 195 | 5 + 10 | 0.93 |
| 198 | - | - | - | - | 1 |

La mesure de nécessité donne 0 dans les autres cas, c'est-à-dire oú $B_R \leq A_S$ avec $A_S = 181$. Il s'agit des valeurs crisp 170 et 177, et des étiquettes linguistiques BAS et NORMAL comme le montre le tableau II.2. Le comparateur NFGEQ assigne par définition la valeur 0 à la mesure de nécessité du paramètre UNKNOWN.

## II.3.3 Filtres flous FLT (FLEQ)/NFLT(NFLEQ) : Fuzzy/Necessarily Less (Equal) Than, Possiblement/nécessairement plus petit (égal) que

Soient les requêtes FLT/NFLT et FLEQ/NFLEQ « Sélectionner toutes les informations des joueurs qui sont possiblement/nécessairement plus petite (égale) au valeur trapézoïdal $[181, 186, 191, 220]$ ». Les figures II.4 et II.5 montrent la syntaxe FSQL associé aux modèles FGT/NFGT et FGEQ/NFGEQ respectivement, tandis que les tableaux II.10, II.11, II.12 et II.13 montrent les résultats de ces requêtes floues FSQL.

| | |
|---|---|
| SELECT JOUEURS. %, CDEG (TAILLE) FROM JOUEURS WHERE TAILLE FLT $[181, 186, 191, 220]$ THOLD 0.05; | SELECT JOUEURS. %, CDEG (TAILLE) FROM JOUEURS WHERE TAILLE NFLT $[181, 186, 191, 220]$ THOLD 0.05; |

**Figure II.4 :** Requêtes avec filtres flous FLT/NFLT et valeur trapézoïdal

| | |
|---|---|
| SELECT JOUEURS. %, CDEG (TAILLE) FROM JOUEURS WHERE TAILLE FLEQ $[181, 186, 191, 220]$ THOLD 0.05; | SELECT JOUEURS. %, CDEG (TAILLE) FROM JOUEURS WHERE TAILLE NFLEQ $[181, 186, 191, 220]$ THOLD 0.05; |

**Figure II.5 :** Requêtes avec filtres flous FLEQ/NFLEQ et valeur trapézoïdale

| Nº Fila | JOUEUR | ÉQUIPE | TAILLE | QUALITÉ | CDEG(TAILLE) |
|---|---|---|---|---|---|
| 1 | P1 | Córdoba | BAS | [30,38,40,45] | 1 |
| 2 | P3 | Granada | NORMAL | RÉGULIER | 1 |
| 3 | P5 | Granada | GRAND | 10±10 | 0,07 |
| 4 | P8 | Málaga | 170 | [31,34,35,38] | 1 |
| 5 | P9 | Sevilla | BAS | 15±10 | 1 |
| 6 | P10 | Sevilla | NORMAL | BON | 1 |
| 7 | P12 | Cádiz | BAS | TRÈS_BON | 1 |
| 8 | P13 | Almería | GRAND | TRÈS_BON | 0,07 |
| 9 | P15 | Almería | 177 | 6±10 | 1 |
| 10 | P16 | Huelva | GRAND | TRÈS_BON | 0,07 |
| 11 | P17 | Huelva | UNKNOWN | UNKNOWN | 1 |
| 12 | P18 | Jaén | UNKNOWN | [8,12,15,25] | 1 |
| 13 | P19 | Jaén | NORMAL | 25±10 | 1 |

**Tableau II.10 :** Résultat de la requête avec filtre flou TAILLE **FLT** $[181, 186, 191, 220]$

Le tableau II.10 montre 13 joueurs qui sont possiblement plus petite que 1m81. Le filtre flou TAILLE FLT $[181, 186, 191, 220]$ cherche les tuples à partir de l'équation I.7 (voir paragraphe I.7). Nous utilisons alors les règles de cette équation dans la condition WHERE de la requête. La règle $B_R \leq A_S$ avec $A_S = 181$ cherche les tuples avec une mesure de possibilité égale à 1. Il s'agit des valeurs crisp 170 et 177, et des étiquettes linguistiques BAS et NORMAL selon le montre le tableau II.2. Ensuite, la règle $B_R > A_S$ avec $A_S = 181$ et $A_R < B_S$ avec $B_S = 186$ cherche les tuples avec une mesure de possibilité entre 0 et 1. D'après le tableau II.2 les valeurs $A_R$ et $B_R$ qui satisfont cette condition vallent 185 et 195 pour l'étiquette linguistique GRAND. La mesure de possibilité est donnée dans le tableau suivant :

| TAILLE | p = A$_R$ – B$_S$ | r = A$_S$ – B$_S$ | s = B$_R$ – A$_R$ | q = r – s | CDEG (TAILLE) p/q |
|--------|----------|----------|----------|----------|----------|
| GRAND | 185 – 186 | 181 – 186 | 195 – 185 | -5 – 10 | 0.07 |

La mesure de possibilité donne 0 dans les autres cas, c'est-à-dire oú A$_R$ ≥ B$_S$ avec B$_S$ = 186. Il s'agit des valeurs crisp 192 et 198, et de l'étiquette linguistique TRÈS_GRAND comme le montre le tableau II.2. Le comparateur FLT assigne par définition la valeur 1 à la mesure de possibilité du paramètre UNKNOWN.

| Nº Fila | JOUEUR | ÉQUIPE | TAILLE | QUALITÉ | CDEG(TAILLE) |
|---------|--------|--------|--------|---------|--------------|
| 1 | P1 | Córdoba | BAS | [30,38,40,45] | 1 |
| 2 | P3 | Granada | NORMAL | RÉGULIER | 0,07 |
| 3 | P8 | Málaga | 170 | [31,34,35,38] | 1 |
| 4 | P9 | Sevilla | BAS | 15±10 | 1 |
| 5 | P10 | Sevilla | NORMAL | BON | 0,07 |
| 6 | P12 | Cádiz | BAS | TRÈS_BON | 1 |
| 7 | P15 | Almería | 177 | 6±10 | 1 |
| 8 | P19 | Jaén | NORMAL | 25±10 | 0,07 |

**Tableau II.11 :** Résultat de la requête avec filtre flou TAILLE **NFLT** $[181, 186, 191, 220]

Le tableau II.11 montre 8 joueurs qui sont nécessairement plus petit que 1m81. Le filtre flou TAILLE NFLT $[181, 186, 191, 220] cherche les tuples à partir de l'équation I.8 (voir paragraphe I.8). Nous utilisons alors les règles de cette équation dans la condition WHERE de la requête. La règle D$_R$ ≤ A$_S$ avec A$_S$ = 181 cherche les tuples avec une mesure de nécessité égale à 1. Il s'agit des valeurs crisp 170 et 177, et de l'étiquette linguistique BAS comme le montre le tableau II.2. Ensuite, la règle D$_R$ > A$_S$ avec A$_S$ = 181 et C$_R$ < B$_S$ avec B$_S$ = 186 cherche les tuples avec une mesure de nécessité entre 0 et 1. Or, d'après le tableau II.2 les valeurs C$_R$ et D$_R$ qui satisfont cette condition vallent 185 et 195 pour l'étiquette linguistique NORMAL. La mesure de nécessité est donnée dans le tableau suivant :

| TAILLE | p = C$_R$ – B$_S$ | r = A$_S$ – B$_S$ | s = D$_R$ – C$_R$ | q = r – s | CDEG (TAILLE) p/q |
|--------|----------|----------|----------|----------|----------|
| NORMAL | 185 – 186 | 181 – 186 | 195 – 185 | -5 – 10 | 0.07 |

La mesure de nécessité donne 0 dans les autres cas, c'est-à-dire oú C$_R$ ≥ B$_S$ avec B$_S$ = 186. Il s'agit des valeurs crisp 192 et 198, et des étiquettes linguistiques GRAND et TRÈS_GRAND comme le montre le tableau II.2. Le comparateur NFLT assigne par définition la valeur 0 à la mesure de nécessité de UNKNOWN.

| Nº Fila | JOUEUR | ÉQUIPE | TAILLE | QUALITÉ | CDEG(TAILLE) |
|---------|--------|--------|--------|---------|--------------|
| 1 | P1 | Córdoba | BAS | [30,38,40,45] | 1 |
| 2 | P2 | Córdoba | TRÈS_GRAND | [2,7,10,15] | 0,51 |
| 3 | P3 | Granada | NORMAL | RÉGULIER | 1 |
| 4 | P4 | Granada | 192 | RÉGULIER | 0,97 |
| 5 | P5 | Granada | GRAND | 10±10 | 0,9 |
| 6 | P6 | Málaga | 198 | MAUVAIS | 0,76 |
| 7 | P7 | Málaga | TRÈS_GRAND | 35±10 | 0,51 |
| 8 | P8 | Málaga | 170 | [31,34,35,38] | 1 |
| 9 | P9 | Sevilla | BAS | 15±10 | 1 |
| 10 | P10 | Sevilla | NORMAL | BON | 1 |
| 11 | P11 | Cádiz | TRÈS_GRAND | 25±10 | 0,51 |
| 12 | P12 | Cádiz | BAS | TRÈS_BON | 1 |
| 13 | P13 | Almería | GRAND | TRÈS_BON | 0,9 |
| 14 | P14 | Almería | TRÈS_GRAND | 6±10 | 0,51 |
| 15 | P15 | Almería | 177 | 6±10 | 1 |
| 16 | P16 | Huelva | GRAND | TRÈS_BON | 0,9 |
| 17 | P17 | Huelva | UNKNOWN | UNKNOWN | 1 |
| 18 | P18 | Jaén | UNKNOWN | [8,12,15,25] | 1 |
| 19 | P19 | Jaén | NORMAL | 25±10 | 1 |

**Tableau II.12 :** Résultat de la requête avec filtre flou TAILLE **FLEQ** $[181, 186, 191, 220]

Le tableau II.12 montre 19 joueurs qui sont possiblement plus petite ou égale à 1m91. Le filtre flou TAILLE FLEQ $[181, 186, 191, 220] cherche les tuples à partir de l'équation I.9 (voir paragraphe I.9). Nous utilisons alors les règles de cette équation dans la condition WHERE de la requête. La règle B$_R$ ≤ C$_S$ avec C$_S$ = 191 cherche les tuples avec une mesure de possibilité égale à 1. Il s'agit des valeurs crisp 170 et 177, et des étiquettes linguistiques BAS et

NORMAL comme le montre le tableau II.2. Ensuite, la règle $B_R > C_S$ avec $C_S = 191$ et $A_R < D_S$ avec $D_S = 220$ cherche les tuples avec une mesure de possibilité entre 0 et 1. D'après le tableau II.2 les valeurs $A_R$ et $B_R$ qui satisfont cette condition vallent 185 et 195 pour l'étiquette linguistique GRAND, et 200 et 210 pour l'étiquette linguistique TRÈS_GRAND. Les valeurs crisp 192 et 198 satisfont aussi cette règle. Nous avons résumé les calculs de la mesure de possibilité dans le tableau suivant :

| TAILLE | $p = D_S - A_R$ | $r = B_R - A_R$ | $s = C_S - D_S$ | $q = r - s$ | CDEG(TAILLE) p/q |
|---|---|---|---|---|---|
| GRAND | 220 – 185 | 195 – 185 | 191 – 220 | 10 + 29 | 0.9 |
| TRÈS_GRAND | 220 – 200 | 210 – 200 | 191 – 220 | 10 + 29 | 0.51 |
| 192 | 220 – 192 | 192 – 192 | 191 – 220 | 29 | 0.97 |
| 198 | 220 – 198 | 198 – 198 | 191 – 220 | 29 | 0.76 |

Le comparateur FLEQ assigne par définition la valeur 1 à la mesure de possibilité deUNKNOWN.

**Tableau II.13 :** Résultat de la requête avec filtre flou TAILLE **NFLEQ** $[181, 186, 191, 220]

Le tableau II.13 montre 13 joueurs qui sont nécessairement plus petite ou égale à 1m91. Le filtre flou TAILLE NFLEQ $[181, 186, 191, 220] cherche les tuples à partir de l'équation I.10 (voir paragraphe I.10). Nous utilisons alors les règles de cette équation dans la condition WHERE de la requête. La règle $D_R \leq C_S$ avec $C_S = 191$ cherche les tuples avec une mesure de nécessité égale à 1. Il s'agit des valeurs crisp 170 et 177, et de l'étiquette linguistique BAS comme le montre le tableau II.2. Ensuite, la règle $D_R > C_S$ avec $C_S = 191$ et $C_R < D_S$ avec $D_S = 220$ cherche les tuples avec une mesure de nécessité entre 0 et 1. Or, d'après le tableau II.2 les valeurs $C_R$ et $D_R$ qui satisfont cette condition vallent 185 et 195 pour l'étiquette linguistique NORMAL, et 200 et 210 pour l'étiquette linguistique GRAND. Les valeurs crisp 192 et 198 satisfont aussi cette règle. Nous avons résumé les calculs de la mesure de possibilité dans le tableau suivant :

| TAILLE | $p = C_R - D_S$ | $r = C_S - D_S$ | $s = D_R - C_R$ | $q = r - s$ | CDEG(TAILLE) p/q |
|---|---|---|---|---|---|
| NORMAL | 185 – 220 | 191 – 220 | 195 – 185 | -29 – 10 | 0.9 |
| GRAND | 200 – 220 | 191 – 220 | 210 – 200 | -29 – 10 | 0.51 |
| 192 | 192 – 220 | 191 – 220 | 192 – 192 | -29 | 0.97 |
| 198 | 198 – 220 | 191 – 220 | 198 – 198 | -29 | 0.76 |

La mesure de nécessité donnne 0 dans les autres cas, c'est-à-dire oú $C_R \geq D_S$ avec $D_S = 220$. Il s'agit de l'étiquette linguistique TRÈS_GRAND comme le montre le tableau II.2. Le comparateur NFLEQ assigne par définition la valeur 0 à la mesure de nécessité de UNKNOWN.

## II.3.4 Filtres flous MGT (MLT)/NMGT(NMLT) : Much Greater (Less) Than, Possiblement /nécessairement beaucoup plus grand (petit) que

Soient les requêtes MGT/NMGT et MLT/NMLT « Sélectionner toutes les informations des joueurs qui sont possiblement/nécessairement beaucoup plus grande (petite) qu'au valeur trapézoïdal $[181, 186, 191, 220] ». Les

---

figures II.6 et II.7 montrent la syntaxe FSQL associée aux modèles MGT/NMGT et MLT/NMLT respectivement, tandis que les tableaux II.14, II.15, II.16 et II.17 montrent les résultats de ces requêtes floues FSQL.

| SELECT JOUEURS. %, CDEG (TAILLE) FROM JOUEURS WHERE TAILLE **MGT** $[181, 186, 191, 220] THOLD 0.05; | SELECT JOUEURS. %, CDEG (TAILLE) FROM JOUEURS WHERE TAILLE **NMGT** $[181, 186, 191, 220] THOLD 0.05; |
|---|---|

**Figure II.6 :** Requêtes avec filtres flous **MGT/NMGT** et valeur trapézoïdale

| SELECT JOUEURS. %, CDEG (TAILLE) FROM JOUEURS WHERE TAILLE **MLT** $[181, 186,191, 220] THOLD 0.05; | SELECT JOUEURS. %, CDEG (TAILLE) FROM JOUEURS WHERE TAILLE **NMLT** $[181, 186,191, 220] THOLD 0.05; |
|---|---|

**Figure II.7 :** Requêtes avec filtres flous **MLT/NMLT** et valeur trapézoïdale

| Nº Fila | JOUEUR | ÉQUIPE | TAILLE | QUALITÉ | CDEG(TAILLE) |
|---|---|---|---|---|---|
| 1 | P2 | Córdoba | TRÈS_GRAND | [2,7,10,15] | 1 |
| 2 | P5 | Granada | GRAND | 10±10 | 0,18 |
| 3 | P7 | Málaga | TRÈS_GRAND | 35±10 | 1 |
| 4 | P11 | Cádiz | TRÈS_GRAND | 25±10 | 1 |
| 5 | P13 | Almería | GRAND | TRÈS_BON | 0,18 |
| 6 | P14 | Almería | TRÈS_GRAND | 8±10 | 1 |
| 7 | P16 | Huelva | GRAND | TRÈS_BON | 0,18 |
| 8 | P17 | Huelva | UNKNOWN | UNKNOWN | 1 |
| 9 | P18 | Jaén | UNKNOWN | [8,12,15,25] | 1 |

**Tableau II.14 :** Résultat de la requête avec filtre flou TAILLE **MGT** $[181, 186,191, 220]

Le tableau II.14 montre 9 joueurs qui sont possiblement beaucoup plus grand que 2m20. Le filtre flou TAILLE MGT $[181, 186, 191, 220] cherche les tuples à partir de l'équation I.11 (voir paragraphe I.11). Nous utilisons alors les règles de cette équation dans la condition WHERE de la requête. La règle $C_R \geq D_S + M$ avec $D_S = 220$ et $M = 12$ cherche les tuples avec une mesure de possibilité égale à 1. Ceci correspond à l'étiquette linguistique TRÈS_GRAND comme le montre le tableau II.2. Ensuite, la règle $C_R < D_S + M$ avec $D_S = 220$ et $D_R > C_S + M$ avec $C_S = 191$ cherche les tuples avec une mesure de possibilité entre 0 et 1. Or, d'après le tableau II.2 les valeurs $C_R$ et $D_R$ qui satisfont cette condition vaux 200 et 210 pour l'étiquette linguistique GRAND. La mesure de possibilité est donnée dans le tableau suivant :

| TAILLE | $p = C_S + M - D_R$ | $r = C_R - D_R$ | $s = D_S - C_S$ | $q = r - s$ | CDEG(TAILLE) p/q |
|---|---|---|---|---|---|
| **GRAND** | **191+ 12 – 210** | **200 – 210** | **220 – 191** | **-10 – 29** | **0.18** |

La mesure de possibilité donne 0 dans les autres cas, c'est-à-dire oú $D_R \leq C_S + M$ avec $C_S = 186$ et $M = 12$. Il s'agit des valeurs crisp 170, 177, 192 et 198, et des étiquettes linguistiques BAS et NORMAL selon le montre le tableau II.2. Le comparateur MGT assigne par définition la valeur 1 à la mesure de possibilité de UNKNOWN.

| Nº Fila | JOUEUR | ÉQUIPE | TAILLE | QUALITÉ | CDEG(TAILLE) |
|---|---|---|---|---|---|
| 1 | P2 | Córdoba | TRÈS_GRAND | [2,7,10,15] | 0,18 |
| 2 | P7 | Málaga | TRÈS_GRAND | 35±10 | 0,18 |
| 3 | P11 | Cádiz | TRÈS_GRAND | 25±10 | 0,18 |
| 4 | P14 | Almería | TRÈS_GRAND | 8±10 | 0,18 |

**Tableau II.15 :** Résultat de la requête avec filtre flou TAILLE **NMGT** $[181,186,191,220]

Le tableau II.15 montre 4 joueurs qui sont nécessairement beaucoup plus grand que 2m20. Le filtre flou TAILLE NMGT $[181,186, 191, 220] cherche les tuples à partir de l'équation I.12 (voir paragraphe I.12). Nous utilisons alors les règles de cette équation dans la condition WHERE de la requête. La règle $A_R \geq D_S + M$ avec $D_S = 220$ et $M = 12$ cherche les tuples avec une mesure de nécessité égale à 1. Nous avons alors, d'après le tableau II.2, qu'il n'y a aucune valeur $A_R$ qui satisfait cette règle. Ensuite, la règle $A_R < D_S + M$ avec $D_S = 220$ et $B_R > C_S + M$ avec $C_S = 191$ cherche les tuples avec une mesure de nécessité entre 0 et 1. Or, d'après le tableau II.2 les valeurs $A_R$ et $B_R$ qui satisfont cette condition vaux 200 et 210 pour l'étiquette linguistique TRÈS_GRAND. La mesure de nécessité est donnée dans le tableau suivant :

| TAILLE | $p = C_S + M - B_R$ | $r = A_R - B_R$ | $s = D_S - C_S$ | $q = r - s$ | CDEG (TAILLE) p/q |
|---|---|---|---|---|---|
| TRÈS_GRAND | 191+ 12 – 210 | 200 – 210 | 220 – 191 | -10 – 29 | 0.18 |

La mesure de nécessité donne 0 dans les autres cas, c'est-à-dire oú $B_R \leq C_S + M$ avec $C_S = 191$ et $M = 12$. Il s'agit des valeurs crisp 170, 177, 192 et 198, et des étiquettes linguistiques BAS, NORMAL et GRAND comme le montre le tableau II.2. Le comparateur NMGT assigne par définition la valeur 0 à la mesure de nécessité de UNKNOWN.

| Nº Fila | JOUEUR | ÉQUIPE | TAILLE | QUALITÉ | CDEG[TAILLE] |
|---|---|---|---|---|---|
| 1 | P1 | Córdoba | BAS | [30,38,40,45] | 1 |
| 2 | P8 | Málaga | 170 | [31,34,35,38] | 0,91 |
| 3 | P9 | Sevilla | BAS | 15±10 | 1 |
| 4 | P12 | Cádiz | BAS | TRÈS_BON | 1 |
| 5 | P15 | Almería | 177 | 6±10 | 0,27 |
| 6 | P17 | Huelva | UNKNOWN | UNKNOWN | 1 |
| 7 | P18 | Jaén | UNKNOWN | [8,12,15,25] | 1 |

**Tableau II.16 :** Résultat de la requête avec filtre flou TAILLE **MLT** $[181, 186, 191, 220]

Le tableau II.16 montre 7 joueurs qui sont possiblement beaucoup plus petite que 1m81. Le filtre flou TAILLE MLT $[181, 186, 191, 220] cherche les tuples à partir de l'équation I.13 (voir paragraphe I.13). Nous utilisons alors les règles de cette équation dans la condition WHERE de la requête. La règle $B_R \leq A_S - M$ avec $A_S = 181$ et $M = 12$ cherche les tuples avec une mesure de possibilité égale à 1. Ceci correspond à l'étiquette linguistique BAS comme le montre le tableau II.2. Ensuite, la règle $B_R > A_S - M$ avec $A_S = 181$ et $A_R < B_S - M$ avec $B_S = 186$ cherche les tuples avec une mesure de possibilité entre 0 et 1. Or, d'après le tableau II.2 les valeurs $A_R$ et $B_R$ qui satisfont cette règle correspondent aux valeurs crisp 170 et 177. La mesure de possibilité est donnée dans le tableau II.2 suivant :

| TAILLE | $p = B_S - M - A_R$ | $r = B_R - A_R$ | $s = A_S - B_S$ | $q = r - s$ | CDEG (TAILLE) p/q |
|---|---|---|---|---|---|
| 170 | 186 – 12 – 164 | 170 – 164 | 181 – 186 | 6 + 5 | 0.91 |
| 177 | 186 – 12 - 171 | 177 - 171 | 181 - 186 | 6 + 5 | 0.27 |

La mesure de possibilité vaut 0 dans les autres cas, c'est-à-dire oú $A_R \geq B_S - M$ avec $B_S = 186$ et $M = 12$. Il s'agit des valeurs crisp **177**, 192 et 198, et des étiquettes linguistiques NORMAL, GRAND et TRÈS_GRAND comme le montre le tableau II.2. Le comparateur MLT assigne par définition la valeur 1 à la mesure de possibilité de UNKNOWN.

| Nº Fila | JOUEUR | ÉQUIPE | TAILLE | QUALITÉ | CDEG[TAILLE] |
|---|---|---|---|---|---|
| 1 | P8 | Málaga | 170 | [31,34,35,38] | 0,36 |
| 2 | P17 | Huelva | UNKNOWN | UNKNOWN | 1 |
| 3 | P18 | Jaén | UNKNOWN | [8,12,15,25] | 1 |

**Tableau II.17 :** Résultat de la requête avec filtre flou TAILLE **NMLT** $[181, 186, 191, 220]

Le tableau II.17 montre 3 joueurs qui sont nécessairement beaucoup plus petit que 1m81. Le filtre flou TAILLE NMLT $[181, 186, 191, 220] cherche les tuples à partir de l'équation I.14 (voir paragraphe I.14). Nous utilisons alors les règles de cette équation dans la condition WHERE de la requête. La règle $D_R \leq A_S - M$ avec $A_S = 181$ et $M = 12$ cherche les tuples avec une mesure de nécessité égale à 1. Alors, d'après le tableau II.2 il n'y a aucune valeur $D_R$ qui satisfait cette règle. Ensuite, la règle $D_R > A_S - M$ avec $A_S = 181$ et $C_R < B_S - M$ avec $B_S = 186$ cherche les tuples avec une mesure de nécessité entre 0 et 1. Or, d'après le tableau II.2 les valeurs $C_R$ et $D_R$ qui satisfont cette règle correspondent à la valeur crisp 170. La mesure de nécessité est donnée dans le tableau suivant :

| TAILLE | $p = B_S - M - C_R$ | $r = D_R - C_R$ | $s = A_S - B_S$ | $q = r - s$ | CDEG (TAILLE) p/q |
|---|---|---|---|---|---|
| 170 | 186 – 12 – 170 | 176 – 170 | 181 – 186 | 6 + 5 | 0.36 |

La mesure de nécessité vaut 0 dans les autres cas, c'est-à-dire oú $C_R \geq B_S - M$ avec $B_S = 186$ et $M = 12$. Il s'agit des valeurs crisp 177, 192 et 198, et des étiquettes linguistiques BAS, NORMAL, GRAND et TRÈS_GRAND comme le montre le tableau II.2. Le comparateur NMLT assigne par définition la valeur **1** à la mesure de nécessité de UNKNOWN.

## II.4 Modèle comportant une étiquette linguistique FSQL comme constante

Soit la valeur grand comme constante donc d'après le tableau II.4, on a : $A_S = 185$, $B_S = 195$, $C_S = 200$ et $D_S = 210$.

### II.4.1 Filtres flous FEQ/NFEQ : Fuzzy/Necessarily EQual, possiblement/nécessairement égal à

Soit la requête FEQ/NFEQ « Sélectionner toutes les informations des joueurs qui sont possiblement/nécessairement égale à une grande taille ». La figure II.8 montre la syntaxe FSQL associé au modèle FEQ/NFEQ, tandis que les tableaux II.18 et II.19 montrent les résultats de ces requêtes floues FSQL.

| | |
|---|---|
| SELECT JOUEURS. %, CDEG (TAILLE) FROM JOUEURS WHERE TAILLE **FEQ** $GRAND THOLD 0.05; | SELECT JOUEURS. %, CDEG (TAILLE) FROM JOUEURS WHERE TAILLE **NFEQ** $GRAND THOLD 0.05; |

**Figure II.8 :** Requêtes avec filtres flous **FEQ**/**NFEQ** et étiquette linguistique

**Tableau II.18** : Résultat de la requête avec filtre flou **FEQ** et étiquette linguistique

**Tableau II.19** : Résultat de la requête avec filtre flou **NFEQ** et étiquette linguistique

### II.4.2 Filtres flous FGT (FGEQ)/NFGT(NFGEQ) : Fuzzy/Necessarily Greater (Equal) Than, possiblement/nécessairement plus grand (égal) que

Soient les requêtes FGT/NFGT et FGEQ/NFGEQ « Sélectionner toutes les informations des joueurs qui sont possiblement/nécessairement plus grand (égale) qu'une grand taille ». Les figures II.9 et II.10 montrent la syntaxe FSQL associé aux modèles FGT/NFGT et FGEQ/NFGEQ respectivement, tandis que les tableaux II.20, II.21, II.22 et II.23 montrent les résultats de ces requêtes floues FSQL.

| | |
|---|---|
| SELECT JOUEURS. %, CDEG (TAILLE) FROM JOUEURS WHERE TAILLE **FGT** $GRAND THOLD 0.05; | SELECT JOUEURS. %, CDEG (TAILLE) FROM JOUEURS WHERE TAILLE **NFGT** $GRAND THOLD 0.05; |

**Figure II.9 :** Requêtes avec filtres flous **FGT/NFGT** et étiquette linguistique

| SELECT JOUEURS. %, CDEG (TAILLE) FROM JOUEURS WHERE TAILLE **FGEQ** $GRAND THOLD 0.05; | SELECT JOUEURS. %, CDEG (TAILLE) FROM JOUEURS WHERE TAILLE **NFGEQ** $GRAND THOLD 0.05; |
|---|---|

**Figure II.10 :** Requêtes avec filtres flous **FGEQ/NFGEQ** et étiquette linguistique

| N° Fila | JOUEUR | ÉQUIPE | TAILLE | QUALITÉ | CDEG(TAILLE) |
|---|---|---|---|---|---|
| 1 | P2 | Córdoba | TRÈS_GRAND | [2,7,10,15] | 1 |
| 2 | P5 | Granada | GRAND | 10±10 | 0,5 |
| 3 | P7 | Málaga | TRÈS_GRAND | 35±10 | 1 |
| 4 | P11 | Cádiz | TRÈS_GRAND | 25±10 | 1 |
| 5 | P13 | Almería | GRAND | TRÈS_BON | 0,5 |
| 6 | P14 | Almería | TRÈS_GRAND | 8±10 | 1 |
| 7 | P16 | Huelva | GRAND | TRÈS_BON | 0,5 |
| 8 | P17 | Huelva | UNKNOWN | UNKNOWN | 1 |
| 9 | P18 | Jaén | UNKNOWN | [8,12,15,25] | 1 |

**Tableau II.20 :** Résultat de la requête avec filtre flou TAILLE **FGT** $GRAND

Le tableau II.20 montre 9 joueurs qui sont possiblement plus grand que 2m10. Le filtre flou TAILLE FGT $GRAND cherche les tuples à partir de l'équation I.3 (voir paragraphe I.3). Nous utilisons alors les règles de cette équation dans la condition WHERE de la requête. La règle $C_R \geq D_S$ avec $D_S = 210$ cherche les tuples avec une mesure de possibilité égale à 1. Nous avons alors, d'après le tableau II.2, que la valeur $C_R$ qui satisfait cette règle vaut 1000, ceci correspond à l'étiquette linguistique TRÈS_GRAND. Ensuite, la règle $C_R < D_S$ avec $D_S = 210$ et $D_R > C_S$ avec $C_S = 200$ cherche les tuples avec une mesure de possibilité entre 0 et 1. Or, d'après le tableau II.2 les valeurs $C_R$ et $D_R$ qui satisfont cette condition valent 200 et 210 pour l'étiquette linguistique GRAND. La mesure de possibilité est donnée dans le tableau suivant :

| TAILLE | $p = D_R - C_S$ | $r = D_S - C_S$ | $s = C_R - D_R$ | $q = r - s$ | CDEG(TAILLE) $p/q$ |
|---|---|---|---|---|---|
| **GRAND** | **210 – 200** | **210 – 200** | **200 – 210** | **10 + 10** | **0.5** |

La mesure de possibilité donne 0 dans les autres cas, c'est-à-dire oú $D_R \leq C_S$ avec $C_S = 200$. Il s'agit des valeurs crisp 170, 177, 192 et 198, et des étiquettes linguistiques BAS et NORMAL selon le montre le tableau II.2. Le comparateur FGT assigne par définition la valeur 1 à la mesure de possibilité de UNKNOWN.

| N° Fila | JOUEUR | ÉQUIPE | TAILLE | QUALITÉ | CDEG(TAILLE) |
|---|---|---|---|---|---|
| 1 | P2 | Córdoba | TRÈS_GRAND | [2,7,10,15] | 0,5 |
| 2 | P7 | Málaga | TRÈS_GRAND | 35±10 | 0,5 |
| 3 | P11 | Cádiz | TRÈS_GRAND | 25±10 | 0,5 |
| 4 | P14 | Almería | TRÈS_GRAND | 8±10 | 0,5 |

**Tableau II.21 :** Résultat de la requête avec filtre flou TAILLE **NFGT** $GRAND

Le tableau II.21 montre 4 joueurs qui sont nécessairement plus grand que 2m10. Le filtre flou TAILLE NFGT $GRAND cherche les tuples à partir de l'équation I.4 (voir paragraphe I.4). Nous utilisons alors les règles de cette équation dans la condition WHERE de la requête. La règle $A_R \geq D_S$ avec $D_S = 210$ cherche les tuples avec une mesure de nécessité égale à 1. Nous avons alors, d'après le tableau II.2, qu'il n'y a aucune valeur $A_R$ qui satisfait cette règle. Ensuite, la règle $A_R < D_S$ avec $D_S = 210$ et $B_R > C_S$ avec $C_S = 200$ cherche les tuples avec une mesure de nécessité entre 0 et 1. Or, d'après le tableau II.2 les valeurs $A_R$ et $B_R$ qui satisfont cette condition vaux 185 et 195 pour l'étiquette linguistique GRAND, et 200 et 210 pour l'étiquette linguistique TRÈS_GRAND. La mesure de nécessité est donnée dans le tableau suivant :

| TAILLE | $p = B_R - C_S$ | $r = D_S - C_S$ | $s = A_R - B_R$ | $q = r - s$ | CDEG (TAILLE) $p/q$ |
|---|---|---|---|---|---|
| **TRÈS_GRAND** | **210 – 200** | **210 – 200** | **200 – 210** | **10 + 10** | **0.5** |

La mesure de nécessité vaut 0 dans les autres cas, c'est-à-dire oú $B_R \le C_S$ avec $C_S = 200$. Il s'agit des valeurs crisp 170, 177, 192 et 198, et des étiquettes linguistiques BAS, NORMAL et GRAND comme le montre le tableau II.2. Le comparateur NFGT assigne par définition la valeur 0 à la mesure de nécessité de UNKNOWN.

| Nº Fila | JOUEUR | ÉQUIPE | TAILLE | QUALITÉ | CDEG(TAILLE)] |
|---|---|---|---|---|---|
| 1 | P2 | Córdoba | TRÈS_GRAND | [2,7,10,15] | 1 |
| 2 | P3 | Granada | NORMAL | RÉGULIER | 0.5 |
| 3 | P4 | Granada | 192 | RÉGULIER | 0.7 |
| 4 | P5 | Granada | GRAND | 10±10 | 1 |
| 5 | P6 | Málaga | 198 | MAUVAIS | 1 |
| 6 | P7 | Málaga | TRÈS_GRAND | 35±10 | 1 |
| 7 | P10 | Sevilla | NORMAL | BON | 0.5 |
| 8 | P11 | Cádiz | TRÈS_GRAND | 25±10 | 1 |
| 9 | P13 | Almería | GRAND | TRÈS_BON | 1 |
| 10 | P14 | Almería | TRÈS_GRAND | 8±10 | 1 |
| 11 | P16 | Huelva | GRAND | TRÈS_BON | 1 |
| 12 | P17 | Huelva | UNKNOWN | UNKNOWN | 1 |
| 13 | P18 | Jaén | UNKNOWN | [8,12,15,25] | 1 |
| 14 | P19 | Jaén | NORMAL | 25±10 | 0.5 |

**Tableau II.22 :** Résultat de la requête avec filtre flou TAILLE **FGEQ** $GRAND

Le tableau II.22 montre 14 joueurs qui sont possiblement plus grand ou égale à 1m95. Le filtre flou TAILLE FGEQ $GRAND cherche les tuples à partir de l'équation I.5 (voir paragraphe I.5). Nous utilisons alors les règles de cette équation dans la condition WHERE de la requête. La règle $C_R \ge B_S$ avec $B_S = 195$ cherche les tuples avec une mesure de possibilité égale à 1. Il s'agit de crisp 198, et des étiquettes linguistiques GRAND et TRÈS_GRAND selon le montre le tableau II.2. Ensuite, la règle $C_R < B_S$ avec $B_S = 195$ et $D_R > A_S$ avec $A_S = 185$ cherche les tuples avec une mesure de possibilité entre 0 et 1. Or, d'après le tableau II.2 les valeurs $C_R$ et $D_R$ qui satisfont cette règle correspondent à crisp 192 et aux valeurs 185 et 195 pour l'étiquette linguistique NORMAL. Nous avons résumé les calculs de la mesure de possibilité dans le tableau suivant :

| TAILLE | $p = D_R - A_S$ | $r = B_S - A_S$ | $s = C_R - D_R$ | $q = r - s$ | CDEG (TAILLE) $p/q$ |
|---|---|---|---|---|---|
| NORMAL | 195 – 185 | 195 – 185 | 185 – 195 | 10 + 10 | 0.5 |
| 192 | 192 – 185 | 195 – 185 | 192 – 192 | 10 | 0.7 |

La mesure de possibilité vaut 0 dans les autres cas, c'est-à-dire oú $D_R \le A_S$ avec $A_S = 185$. Il s'agit des valeurs crisp 170 et 177, et de l'étiquette linguistique BAS comme le montre le tableau II.2. Le comparateur FGEQ assigne par définition la valeur 1 à la mesure de possibilité de UNKNOWN.

| Nº Fila | JOUEUR | ÉQUIPE | TAILLE | QUALITÉ | CDEG(TAILLE)] |
|---|---|---|---|---|---|
| 1 | P2 | Córdoba | TRÈS_GRAND | [2,7,10,15] | 1 |
| 2 | P4 | Granada | 192 | RÉGULIER | 0.7 |
| 3 | P5 | Granada | GRAND | 10±10 | 0.5 |
| 4 | P6 | Málaga | 198 | MAUVAIS | 1 |
| 5 | P7 | Málaga | TRÈS_GRAND | 35±10 | 1 |
| 6 | P11 | Cádiz | TRÈS_GRAND | 25±10 | 1 |
| 7 | P13 | Almería | GRAND | TRÈS_BON | 0.5 |
| 8 | P14 | Almería | TRÈS_GRAND | 8±10 | 1 |
| 9 | P16 | Huelva | GRAND | TRÈS_BON | 0.5 |

**Tableau II.23 :** Résultat de la requête avec filtre flou TAILLE **NFGEQ** $GRAND

Le tableau II.23 montre 9 joueurs qui sont nécessairement plus grand ou égale au 1m95. Le filtre flou TAILLE NFGEQ $GRAND cherche les tuples à partir de l'équation I.6 (voir paragraphe I.6). Nous utilisons alors les règles de cette équation dans la condition WHERE de la requête. La règle $A_R \ge B_S$ avec $B_S = 195$ cherche les tuples avec une mesure de nécessité égale à 1. Il s'agit de crisp 198, et de l'étiquette linguistique TRÈS_GRAND comme le montre le tableau II.2. Ensuite, la règle $A_R < B_S$ avec $B_S = 195$ et $B_R > A_S$ avec $A_S = 185$ cherche les tuples avec une mesure de nécessité entre 0 et 1. Or, d'après le tableau II.2 les valeurs $A_R$ et $B_R$ qui satisfont cette règle correspondent à crisp 192 et aux valeurs 185 et 195 pour l'étiquette linguistique GRAND. Nous avons résumé les calculs de la mesure de nécessité dans le tableau suivant :

| TAILLE | $p = B_R - A_S$ | $r = B_S - A_S$ | $s = A_R - B_R$ | $q = r - s$ | CDEG (TAILLE) p/q |
|---|---|---|---|---|---|
| GRAND | 195 – 185 | 195 – 185 | 185 – 195 | 10 + 10 | 0.5 |
| 192 | 192 – 185 | 195 – 185 | 192 – 192 | 10 | 0.7 |

La mesure de nécessité donne 0 dans les autres cas, c'est-à-dire oú $B_R \leq A_S$ avec $A_S = 185$. Il s'agit des valeurs crisp 170 et 177, et des étiquettes linguistiques BAS et NORMAL selon le montre le tableau II.2. Le comparateur NFGEQ assigne par définition la valeur 0 à la mesure de nécessité de UNKNOWN.

## II.4.3 Filtres flous FLT (FLEQ)/NFLT(NFLEQ) : Fuzzy/Necessarily Less (Equal) Than, Possiblement/nécessairement plus petit (égal) que

Soient les requêtes FLT/NFLT et FLEQ/NFLEQ « Sélectionner toutes les informations des joueurs qui sont possiblement/nécessairement plus petit (égal) d'une grand taille ». Les figures II.11 et II.12 montrent la syntaxe FSQL associé aux modèles FGT/NFGT et FGEQ/NFGEQ respectivement, tandis que les tableaux II.24, II.25, II.26 et II.27 montrent les résultats de ces requêtes floues FSQL.

| | |
|---|---|
| SELECT JOUEURS. %, CDEG (TAILLE) FROM JOUEURS WHERE TAILLE **FLT** $GRAND THOLD 0.05; | SELECT JOUEURS. %, CDEG (TAILLE) FROM JOUEURS WHERE TAILLE **NFLT** $GRAND THOLD 0.05; |

**Figure II.11 :** Requêtes avec filtres flous **FLT/NFLT** et étiquette linguistique

| | |
|---|---|
| SELECT JOUEURS. %, CDEG (TAILLE) FROM JOUEURS WHERE TAILLE **FLEQ** $GRAND THOLD 0.05; | SELECT JOUEURS. %, CDEG (TAILLE) FROM JOUEURS WHERE TAILLE **NFLEQ** $GRAND THOLD 0.05; |

**Figure II.12 :** Requêtes avec filtres flous **FLEQ/NFLEQ** et étiquette linguistique

| Nº Fila | JOUEUR | ÉQUIPE | TAILLE | QUALITÉ | CDEG[TAILLE] |
|---|---|---|---|---|---|
| 1 | P1 | Córdoba | BAS | [30,38,40,45] | 1 |
| 2 | P3 | Granada | NORMAL | RÉGULIER | 1 |
| 3 | P4 | Granada | 192 | RÉGULIER | 0,3 |
| 4 | P5 | Granada | GRAND | 10±10 | 0,5 |
| 5 | P8 | Málaga | 170 | [31,34,35,38] | 1 |
| 6 | P9 | Sevilla | BAS | 15±10 | 1 |
| 7 | P10 | Sevilla | NORMAL | BON | 1 |
| 8 | P12 | Cádiz | BAS | TRÈS_BON | 1 |
| 9 | P13 | Almería | GRAND | TRÈS_BON | 0,5 |
| 10 | P15 | Almería | 177 | | 6±10 | 1 |
| 11 | P16 | Huelva | GRAND | TRÈS_BON | 0,5 |
| 12 | P17 | Huelva | UNKNOWN | UNKNOWN | 1 |
| 13 | P18 | Jaén | UNKNOWN | [8,12,15,25] | 1 |
| 14 | P19 | Jaén | NORMAL | 25±10 | 1 |

**Tableau II.24 :** Résultat de la requête avec filtre flou TAILLE **FLT** $GRAND

Le tableau II.24 montre 14 joueurs qui sont possiblement plus petits que 1m85. Le filtre flou TAILLE FLT $GRAND cherche les tuples à partir de l'équation I.7 (voir paragraphe I.7). Nous utilisons alors les règles de cette équation dans la condition WHERE de la requête. La règle $B_R \leq A_S$ avec $A_S = 185$ cherche les tuples avec une mesure de possibilité égale à 1. Il s'agit des valeurs crisp 170 et 177, et des étiquettes linguistiques BAS et NORMAL selon le montre le tableau II.2. Ensuite, la règle $B_R > A_S$ avec $A_S = 185$ et $A_R < B_S$ avec $B_S = 195$ cherche les tuples avec une mesure de possibilité entre 0 et 1. Or, d'après le tableau II.2 les valeurs $A_R$ et $B_R$ qui satisfont cette règle correspondent à la valeur crisp 192 et aux valeurs 185 et 195 pour l'étiquette linguistique GRAND. Nous avons résumé les calculs de la mesure de nécessité dans le tableau suivant :

| TAILLE | $p = A_R - B_S$ | $r = A_S - B_S$ | $s = B_R - A_R$ | $q = r - s$ | CDEG (TAILLE) p/q |
|---|---|---|---|---|---|
| GRAND | 185 – 195 | 185 – 195 | 195 – 185 | -10 – 10 | 0.5 |

| 192 | 192 – 195 | 185 – 195 | 192 – 192 | -10 | 0.3 |

La mesure de possibilité donne 0 dans les autres cas, c'est-à-dire où $A_R \geq B_S$ avec $B_S = 195$. Il s'agit de crisp 198, et de l'étiquette linguistique TRÈS_GRAND selon le montre le tableau II.2. Le comparateur FLT assigne par définition la valeur 1 à la mesure de possibilité de UNKNOWN.

| Nº Fila | JOUEUR | ÉQUIPE | TAILLE | QUALITÉ | CDEG(TAILLE) |
|---|---|---|---|---|---|
| 1 | P1 | Córdoba | BAS | [30,38,40,45] | 1 |
| 2 | P3 | Granada | NORMAL | RÉGULIER | 0,5 |
| 3 | P4 | Granada | 192 | RÉGULIER | 0,3 |
| 4 | P8 | Málaga | 170 | [31,34,35,38] | 1 |
| 5 | P9 | Sevilla | BAS | 15±10 | 1 |
| 6 | P10 | Sevilla | NORMAL | BON | 0,5 |
| 7 | P12 | Cádiz | BAS | TRÈS_BON | 1 |
| 8 | P15 | Almería | 177 | 6±10 | 1 |
| 9 | P19 | Jaén | NORMAL | 25±10 | 0,5 |

**Tableau II.25 :** Résultat de la requête avec filtre flou TAILLE **NFLT** $GRAND

Le tableau II.25 montre 9 joueurs qui sont nécessairement plus petits que 1m81. Le filtre flou TAILLE NFLT $GRAND cherche les tuples à partir de l'équation I.8 (voir paragraphe I.8). Nous utilisons alors les règles de cette équation dans la condition WHERE de la requête. La règle $D_R \leq A_S$ avec $A_S = 185$ cherche les tuples avec une mesure de nécessité égale à 1. Il s'agit des valeurs crisp 170 et 177, et de l'étiquette linguistique BAS comme le montre le tableau II.2. Ensuite, la règle $D_R > A_S$ avec $A_S = 185$ et $C_R < B_S$ avec $B_S = 195$ cherche les tuples avec une mesure de nécessité entre 0 et 1. Or, d'après le tableau II.2 les valeurs $C_R$ et $D_R$ qui satisfont cette règle correspondent à la valeur crisp 192 et aux valeurs 185 et 195 pour l'étiquette linguistique NORMAL. Nous avons résumé les calculs de la mesure de nécessité dans le tableau suivant :

| TAILLE | $p = C_R - B_S$ | $r = A_S - B_S$ | $s = D_R - C_R$ | $q = r - s$ | CDEG (TAILLE) P/q |
|---|---|---|---|---|---|
| NORMAL | 185 – 195 | 185 – 195 | 195 – 185 | -10 – 10 | 0.5 |
| 192 | 192 – 195 | 185 – 195 | 192 – 192 | -10 | 0.3 |

La mesure de nécessité donne 0 dans les autres cas, c'est-à-dire où $C_R \geq B_S$ avec $B_S = 195$. Il s'agit de crisp 198 et des étiquettes linguistiques GRAND et TRÈS_GRAND selon le montre le tableau II.2. Le comparateur NFLT assigne par définition la valeur 0 à la mesure de nécessité de UNKNOWN.

| Nº Fila | JOUEUR | ÉQUIPE | TAILLE | QUALITÉ | CDEG(TAILLE) |
|---|---|---|---|---|---|
| 1 | P1 | Córdoba | BAS | [30,38,40,45] | 1 |
| 2 | P2 | Córdoba | TRÈS_GRAND | [2,7,10,15] | 0,5 |
| 3 | P3 | Granada | NORMAL | RÉGULIER | 1 |
| 4 | P4 | Granada | 192 | RÉGULIER | 1 |
| 5 | P5 | Granada | GRAND | 10±10 | 1 |
| 6 | P6 | Málaga | 198 | MAUVAIS | 1 |
| 7 | P7 | Málaga | TRÈS_GRAND | 35±10 | 0,5 |
| 8 | P8 | Málaga | 170 | [31,34,35,38] | 1 |
| 9 | P9 | Sevilla | BAS | 15±10 | 1 |
| 10 | P10 | Sevilla | NORMAL | BON | 1 |
| 11 | P11 | Cádiz | TRÈS_GRAND | 25±10 | 0,5 |
| 12 | P12 | Cádiz | BAS | TRÈS_BON | 1 |
| 13 | P13 | Almería | GRAND | TRÈS_BON | 1 |
| 14 | P14 | Almería | TRÈS_GRAND | 8±10 | 0,5 |
| 15 | P15 | Almería | 177 | 6±10 | 1 |
| 16 | P16 | Huelva | GRAND | TRÈS_BON | 1 |
| 17 | P17 | Huelva | UNKNOWN | UNKNOWN | 1 |
| 18 | P18 | Jaén | UNKNOWN | [8,12,15,25] | 1 |
| 19 | P19 | Jaén | NORMAL | 25±10 | 1 |

**Tableau II.26 :** Résultat de la requête avec filtre flou TAILLE **FLEQ** $GRAND

Le tableau II.26 montre 19 joueurs qui sont possiblement plus petits ou égale à 2m00. Le filtre flou TAILLE FLEQ $GRAND cherche les tuples à partir de l'équation I.9 (voir paragraphe I.9). Nous utilisons alors les règles de cette équation dans la condition WHERE de la requête. La règle $B_R \leq C_S$ avec $C_S = 200$ cherche les tuples avec une mesure de possibilité égale à 1. Il s'agit des valeurs crisp 170, 177, 192 et 198, et des étiquettes linguistiques BAS, NORMAL et GRAND comme le montre le tableau II.2. Ensuite, la règle $B_R > C_S$ avec $C_S = 200$ et $A_R < D_S$ avec $D_S = 210$ cherche les tuples avec une mesure de possibilité entre 0 et 1. Or, d'après le tableau II.2 les valeurs $A_R$ et $B_R$ qui satisfont cette règle vaux 200 et 210 pour l'étiquette linguistique TRÈS_GRAND. La mesure de nécessité est donnée dans le tableau suivant :

---

| TAILLE | $p = D_S - A_R$ | $r = B_R - A_R$ | $s = C_S - D_S$ | $q = r - s$ | CDEG (TAILLE) $p/q$ |
|---|---|---|---|---|---|
| TRÈS_GRAND | 210 – 200 | 210 – 200 | 200 – 210 | 10 + 10 | 0.5 |

Le comparateur FLEQ assigne par définition la valeur 1 à la mesure de possibilité de UNKNOWN.

| Nº Fila | JOUEUR | ÉQUIPE | TAILLE | QUALITÉ | CDEG(TAILLE) |
|---|---|---|---|---|---|
| 1 | P1 | Córdoba | BAS | [30,36,40,45] | 1 |
| 2 | P3 | Granada | NORMAL | RÉGULIER | 1 |
| 3 | P4 | Granada | 192 | RÉGULIER | 1 |
| 4 | P5 | Granada | GRAND | 10±10 | 0.5 |
| 5 | P6 | Málaga | 198 | MAUVAIS | 1 |
| 6 | P8 | Málaga | 170 | [31,34,35,38] | 1 |
| 7 | P9 | Sevilla | BAS | 15±10 | 1 |
| 8 | P10 | Sevilla | NORMAL | BON | 1 |
| 9 | P12 | Cádiz | BAS | TRÈS_BON | 1 |
| 10 | P13 | Almería | GRAND | TRÈS_BON | 0.5 |
| 11 | P15 | Almería | 177 | 6±10 | 1 |
| 12 | P16 | Huelva | GRAND | TRÈS_BON | 0.5 |
| 13 | P19 | Jaén | NORMAL | 25±10 | 1 |

**Tableau II.27 :** Résultat de la requête avec filtre flou TAILLE **NFLEQ** $GRAND

Le tableau II.27 montre 13 joueurs qui sont nécessairement plus petits ou égale à 2m00. Le filtre flou TAILLE NFLEQ $GRAND cherche les tuples à partir de l'équation I.10 (voir paragraphe I.10). Nous utilisons alors les règles de cette équation dans la condition WHERE de la requête. La règle $D_R \leq C_S$ avec $C_S = 200$ cherche les tuples avec une mesure de nécessité égale à 1. Il s'agit des valeurs crisp 170, 177, 192 et 198, et des étiquettes linguistiques BAS et NORMAL selon le montre le tableau II.2. Ensuite, la règle $D_R > C_S$ avec $C_S = 200$ et $C_R < D_S$ avec $D_S = 210$ cherche les tuples avec une mesure de nécessité entre 0 et 1. Or, d'après le tableau II.2 les valeurs $C_R$ et $D_R$ qui satisfont cette condition vaux 200 et 210 pour l'étiquette linguistique GRAND. La mesure de nécessité est donnée dans le tableau suivant :

| TAILLE | $p = C_R - D_S$ | $r = C_S - D_S$ | $s = D_R - C_R$ | $q = r - s$ | CDEG (TAILLE) $p/q$ |
|---|---|---|---|---|---|
| GRAND | 200 – 210 | 200 – 210 | 210 – 200 | -10 – 10 | 0.5 |

La mesure de nécessité donne 0 dans les autres cas, c'est-à-dire oú $C_R \geq D_S$ avec $D_S = 210$. Il s'agit de l'étiquette linguistique TRÈS_GRAND comme le montre le tableau II.2. Le comparateur NFLEQ assigne par définition la valeur 0 à la mesure de nécessité de UNKNOWN.

## II.4.4 Filtres flous MGT (MLT)/NMGT(NMLT) : Much Greater (Less) Than, Possiblement/nécessairement beaucoup plus grand (petit) que

Soient les requêtes MGT/NMGT et MLT/NMLT « Sélectionner toutes les informations des joueurs qui sont possiblement/nécessairement beaucoup plus grande (petite) qu'une grand taille». Les figures II.13 et II.14 montrent la syntaxe FSQL associé aux modèles MGT/NMGT et MLT/NMLT respectivement, tandis que les tableaux II.28, II.29, II.30 et II.31 montrent les résultats de ces requêtes floues FSQL.

| SELECT JOUEURS. %, CDEG (TAILLE) FROM JOUEURS WHERE TAILLE **MGT** $GRAND THOLD 0.05; | SELECT JOUEURS. %, CDEG (TAILLE) FROM JOUEURS WHERE TAILLE **NMGT** $GRAND THOLD 0.05; |
|---|---|

**Figure II.13 :** Requêtes avec filtres flous **MGT**/**NMGT** et étiquette linguistique

| SELECT JOUEURS. %, CDEG (TAILLE) FROM JOUEURS WHERE TAILLE **MLT** $GRAND THOLD 0.05; | SELECT JOUEURS. %, CDEG (TAILLE) FROM JOUEURS WHERE TAILLE **NMLT** $GRAND THOLD 0.05; |
|---|---|

**Figure II.14 :** Requêtes avec filtres flous **MLT/NMLT** et étiquette linguistique

| Nº Fila | JOUEUR | ÉQUIPE | TAILLE | QUALITÉ | CDEG(TAILLE) |
|---------|--------|--------|--------|---------|--------------|
| 1 | P2 | Córdoba | TRÈS_GRAND | [2,7,10,15] | 1 |
| 2 | P7 | Málaga | TRÈS_GRAND | 35±10 | 1 |
| 3 | P11 | Cádiz | TRÈS_GRAND | 25±10 | 1 |
| 4 | P14 | Almería | TRÈS_GRAND | 8±10 | 1 |
| 5 | P17 | Huelva | UNKNOWN | UNKNOWN | 1 |
| 6 | P18 | Jaén | UNKNOWN | [8,12,15,25] | 1 |

**Tableau II.28 :** Résultat de la requête avec filtre flou TAILLE **MGT** $GRAND

Le tableau II.28 montre 6 joueurs qui sont possiblement beaucoup plus grands que 2m10. Le filtre flou TAILLE MGT $GRAND cherche les tuples à partir de l'équation I.11 (voir paragraphe I.11). Nous utilisons alors les règles de cette équation dans la condition WHERE de la requête. La règle $C_R \geq D_S + M$ avec $D_S = 210$ et $M = 12$ cherche les tuples avec une mesure de possibilité égale à 1. Ceci correspond à l'étiquette linguistique TRÈS_GRAND comme le montre le tableau II.2. Ensuite, la règle $C_R < D_S + M$ avec $D_S = 210$ et $D_R > C_S + M$ avec $C_S = 200$ cherche les tuples avec une mesure de possibilité entre 0 et 1. Or, d'après le tableau II.2 il n'y a aucune valeur pour $C_R$ et $D_R$ qui satisfont cette règle. La mesure de possibilité vaut 0 dans les autres cas, c'est-à-dire oú $D_R \leq C_S + M$ avec $C_S = 200$ et $M = 12$. Il s'agit des valeurs crisp 170, 177, 192 et 198, et des étiquettes linguistiques BAS, NORMAL et GRAND comme le montre le tableau II.2. Le comparateur MGT assigne par définition la valeur 1 à la mesure de possibilité de UNKNOWN.

| Nº Fila | JOUEUR | ÉQUIPE | TAILLE | QUALITÉ | CDEG(TAILLE) |
|---------|--------|--------|--------|---------|--------------|
|  |  |  |  |  |  |

**Tableau II.29 :** Résultat de la requête avec filtre flou TAILLE **NMGT** $GRAND

Le tableau II.29 nous montre une requête vide, cela signifie qu'il n' y a aucun joueur nécessairement beaucoup plus grand que 2m10. En effet, le filtre flou TAILLE NMGT $GRAND cherche les tuples à partir de l'équation I.12 (voir paragraphe I.12). Nous utilisons alors les règles de cette équation dans la condition WHERE de la requête. La règle $A_R \geq D_S + M$ avec $D_S = 210$ et $M = 12$ cherche les tuples avec une mesure de nécessité égale à 1. Nous avons alors, d'après le tableau II.2, qu'il n'y a aucune valeur $A_R$ qui satisfait cette règle. Ensuite, la règle $A_R < D_S + M$ avec $D_S = 210$ et $B_R > C_S + M$ avec $C_S = 200$ cherche les tuples avec une mesure de nécessité entre 0 et 1. Or, d'après le tableau II.2 il n'y a aucune valeur $A_R$ et $B_R$ qui satisfont cette règle. En revenche, la règle $B_R \leq C_S + M$ avec $C_S = 200$ et $M = 12$ est vérifie dans tous les cas, c'est-à-dire pour les valeurs crisp 170, 177, 192 et 198, et les étiquettes linguistiques BAS, NORMAL, GRAND et TRÈS_GRAND, cela signifie que la mesure de nécessité vaut 0. Enfin, le comparateur NMGT assigne par définition la valeur 0 à la mesure de nécessité de UNKNOWN.

| Nº Fila | JOUEUR | ÉQUIPE | TAILLE | QUALITÉ | CDEG(TAILLE) |
|---------|--------|--------|--------|---------|--------------|
| 1 | P1 | Córdoba | BAS | [30,38,40,45] | 1 |
| 2 | P3 | Granada | NORMAL | RÉGULIER | 0,53 |
| 3 | P8 | Málaga | 170 | [31,34,35,38] | 1 |
| 4 | P9 | Sevilla | BAS | 15±10 | 1 |
| 5 | P10 | Sevilla | NORMAL | BON | 0,53 |
| 6 | P12 | Cádiz | BAS | TRÈS_BON | 1 |
| 7 | P15 | Almería | 177 | 6±10 | 0,75 |
| 8 | P17 | Huelva | UNKNOWN | UNKNOWN | 1 |
| 9 | P18 | Jaén | UNKNOWN | [8,12,15,25] | 1 |
| 10 | P19 | Jaén | NORMAL | 25±10 | 0,53 |

**Tableau II.30 :** Résultat de la requête avec filtre flou TAILLE **MLT** $GRAND

Le tableau II.30 montre 10 joueurs qui sont possiblement beaucoup plus petits que 1m85. Le filtre flou TAILLE MLT $GRAND cherche les tuples à partir de l'équation I.13 (voir paragraphe I.13). Nous utilisons alors les règles de cette équation dans la condition WHERE de la requête. La règle $B_R \leq A_S - M$ avec $A_S = 185$ et $M = 12$ cherche les tuples avec une mesure de possibilité égale à 1. Ceci correspond à la valeur crisp 170 et à l'étiquette linguistique BAS comme le montre le tableau II.2. Ensuite, la règle $B_R > A_S - M$ avec $A_S = 185$ et $A_R < B_S - M$ avec $B_S = 195$ cherche les tuples avec une mesure de possibilité entre 0 et 1. Or, d'après le tableau II.2 les valeurs $A_R$ et $B_R$ qui satisfont cette règle correspondent à la valeur crisp 177 et aux valeurs 175 et 180 pour l'étiquette linguistique NORMAL. Nous avons résumé les calculs de la mesure de nécessité dans le tableau suivant :

| TAILLE | $p = B_S - M - A_R$ | $r = B_R - A_R$ | $s = A_S - B_S$ | $q = r - s$ | CDEG (TAILLE) $p/q$ |
|--------|------|------|------|------|------|
| NORMAL | 195 – 12 – 175 | 180 – 175 | 185 – 195 | 5 + 10 | 0.53 |
| 177 | 195 – 12 – 171 | 177 – 171 | 185 – 195 | 6 + 10 | 0.75 |

La mesure de possibilité vaut 0 dans les autres cas, c'est-à-dire oú $A_R \geq B_S - M$ avec $B_S = 195$ et M = 12. Il s'agit des valeurs crisp **192** et 198, et des étiquettes linguistiques **GRAND** et TRÈS_GRAND comme le montre le tableau II.2. Le comparateur MLT assigne par définition la valeur 1 à la mesure de possibilité de UNKNOWN.

| Nº Fila | JOUEUR | ÉQUIPE | TAILLE | QUALITÉ | CDEG(TAILLE)] |
|---|--------|--------|--------|---------|--------------|
| 1 | P1 | Córdoba | BAS | [30,38,40,45] | 0,53 |
| 2 | P8 | Málaga | 170 | [31,34,35,38] | 0,81 |
| 3 | P9 | Sevilla | BAS | 15±10 | 0,53 |
| 4 | P12 | Cádiz | BAS | TRÈS_BON | 0,53 |
| 5 | P15 | Almería | 177 | 6±10 | 0,38 |
| 6 | P17 | Huelva | UNKNOWN | UNKNOWN | 1 |
| 7 | P18 | Jaén | UNKNOWN | [8,12,15,25] | 1 |

**Tableau II.31 :** Résultat de la requête avec filtre flou TAILLE **NMLT** $GRAND

Le tableau II.31 montre 7 joueurs qui sont nécessairement beaucoup plus petits que 1m85. Le filtre flou TAILLE NMLT $GRAND cherche les tuples à partir de l'équation I.14 (voir paragraphe I.14). Nous utilisons alors les règles de cette équation dans la condition WHERE de la requête. La règle $D_R \leq A_S - M$ avec $A_S = 185$ et M = 12 cherche les tuples avec une mesure de nécessité égale à 1. Ceci correspond au valeur crisp 170 comme le montre le tableau II.2. Ensuite, la règle $D_R > A_S - M$ avec $A_S = 185$ et $C_R < B_S - M$ avec $B_S = 195$ cherche les tuples avec une mesure de nécessité entre 0 et 1. Or, d'après le tableau II.2 les valeurs $C_R$ et $D_R$ qui satisfont cette règle correspondent à la valeur crisp 177 et aux valeurs 175 et 180 pour l'étiquette linguistique BAS. Nous avons résumé les calculs de la mesure de nécessité dans le tableau suivant :

| TAILLE | $p = B_S - M - C_R$ | $r = D_R - C_R$ | $s = A_S - B_S$ | $q = r - s$ | CDEG (TAILLE) $p/q$ |
|--------|------|------|------|------|------|
| BAS | 195 – 12 – 175 | 180 – 175 | 185 – 195 | 5 + 10 | 0.53 |
| 177 | 195 – 12 – 177 | 183 – 177 | 185 – 195 | 6 + 10 | 0.38 |
| 170 | 195 – 12 – 170 | 176 –170 | 185 – 195 | 6 + 10 | 0.81 |

La mesure de nécessité donne 0 dans les autres cas, c'est-à-dire oú $C_R \geq B_S - M$ avec $B_S = 195$ et M = 12. Il s'agit des valeurs crisp 192 et 198, et des étiquettes linguistiques GRAND et TRÈS_GRAND comme le montre le tableau II.2. Le comparateur NMLT assigne par définition la valeur **1** à la mesure de nécessité de UNKNOWN.

## II.5 Modèle comportant un intervalle FSQL comme constante

Soit la valeur [180,220] comme intervalle donc d'après le tableau II.4, on a : $A_S = B_S = 180$ et $C_S = D_S = 220$.

## II.5.1 Filtres flous FEQ/NFEQ : Fuzzy/Necessarily EQual, possiblement/nécessairement égal

Soit la requête « Sélectionner les tuples de la BD dont la taille des joueurs est possiblement/nécessairement dans l'intervalle [180,220] ». La figure II.15 montre la syntaxe FSQL associé au modèle FEQ/NFEQ, tandis que les tableau II.32 et II.33 montrent les résultats de ces requêtes floues FSQL.

| SELECT JOUEURS. %, CDEG (TAILLE) FROM JOUEURS WHERE TAILLE **FEQ** [180,220] THOLD 0.05; | SELECT JOUEURS. %, CDEG (TAILLE) FROM JOUEURS WHERE TAILLE **NFEQ** [180,220] THOLD 0.05; |
|---|---|

**Figure II.15 :** Requêtes avec filtres flous **FEQ**/**NFEQ** et l'intervalle

| Nº Fila | JOUEUR | ÉQUIPE | TAILLE | QUALITÉ | CDEG(TAILLE) |
|---|---|---|---|---|---|
| 1 | P2 | Córdoba | TRÈS_GRAND | [2.7,10,15] | 1 |
| 2 | P3 | Granada | NORMAL | RÉGULIER | 1 |
| 3 | P4 | Granada | 192 | RÉGULIER | 1 |
| 4 | P5 | Granada | GRAND | 10±10 | 1 |
| 5 | P6 | Málaga | 198 | MAUVAIS | 1 |
| 6 | P7 | Málaga | TRÈS_GRAND | 35±10 | 1 |
| 7 | P10 | Sevilla | NORMAL | BON | 1 |
| 8 | P11 | Cádiz | TRÈS_GRAND | 25±10 | 1 |
| 9 | P13 | Almería | GRAND | TRÈS_BON | 1 |
| 10 | P14 | Almería | TRÈS_GRAND | 8±10 | 1 |
| 11 | P16 | Huelva | GRAND | TRÈS_BON | 1 |
| 12 | P17 | Huelva | UNKNOWN | UNKNOWN | 1 |
| 13 | P18 | Jaén | UNKNOWN | [8,12,15,25] | 1 |
| 14 | P19 | Jaén | NORMAL | 25±10 | 1 |

**Tableau II.32 :** Résultat de la requête avec filtre flou **FEQ** et l'intervalle

| Nº Fila | JOUEUR | ÉQUIPE | TAILLE | QUALITÉ | CDEG(TAILLE) |
|---|---|---|---|---|---|
| 1 | P4 | Granada | 192 | RÉGULIER | 1 |
| 2 | P5 | Granada | GRAND | 10±10 | 1 |
| 3 | P6 | Málaga | 198 | MAUVAIS | 1 |
| 4 | P13 | Almería | GRAND | TRÈS_BON | 1 |
| 5 | P16 | Huelva | GRAND | TRÈS_BON | 1 |

**Tableau II.33 :** Résultat de la requête avec filtre flou **NFEQ** et l'intervalle

## II.5.2 Filtres flous FGT(FGEQ)/NFGT(NFGEQ) :Fuzzy/Necessarily Greater (Equal) Than, possiblement/nécessairement plus grand (égal) que

Soient les requêtes FGT/NFGT et FGEQ/NFGEQ « Sélectionner toutes les informations des joueurs qui sont possiblement/nécessairement plus grand (égale) à l'intervalle [180,220] ». Les figures II.20 et II.21 montrent la syntaxe FSQL associé aux modèles FGT/NFGT et FGEQ/NFGEQ respectivement, tandis que les tableaux II.34, II.35, II.36 et II.37 montrent les résultats de ces requêtes floues FSQL. 1

| SELECT JOUEURS. %, CDEG (TAILLE) FROM JOUEURS WHERE TAILLE **FGT** [180,220] THOLD 0.05; | SELECT JOUEURS. %, CDEG (TAILLE) FROM JOUEURS WHERE TAILLE **NFGT** [180,220] THOLD 0.05; |
|---|---|

**Figure 4.20** : Requêtes avec filtres flous **FGT/NFGT** et intervalle

| SELECT JOUEURS. %, CDEG (TAILLE) FROM JOUEURS WHERE TAILLE **FGEQ** [180,220] THOLD 0.05; | SELECT JOUEURS. %, CDEG (TAILLE) FROM JOUEURS WHERE TAILLE **NFGEQ** [180,220] THOLD 0.05; |
|---|---|

**Figure 4.21 :** Requêtes avec filtres flous **FGEQ/NFGEQ** et intervalle

| Nº Fila | JOUEUR | ÉQUIPE | TAILLE | QUALITÉ | CDEG(TAILLE) |
|---|---|---|---|---|---|
| 1 | P2 | Córdoba | TRÈS_GRAND | [2.7,10,15] | 1 |
| 2 | P7 | Málaga | TRÈS_GRAND | 35±10 | 1 |
| 3 | P11 | Cádiz | TRÈS_GRAND | 25±10 | 1 |
| 4 | P14 | Almería | TRÈS_GRAND | 8±10 | 1 |
| 5 | P17 | Huelva | UNKNOWN | UNKNOWN | 1 |
| 6 | P18 | Jaén | UNKNOWN | [8,12,15,25] | 1 |

**Tableau II.34 :** Résultat de la requête avec filtre flou TAILLE **FGT** [180,220]

Le tableau II.34 montre 6 joueurs qui sont possiblement plus grand que 2m20. Le filtre flou TAILLE FGT [180,220] cherche les tuples à partir de l'équation I.3 (voir paragraphe I.3). Nous utilisons alors les règles de cette équation dans la condition WHERE de la requête. La règle $C_R \geq D_S$ avec $D_S = 220$ cherche les tuples avec une mesure de possibilité égale à 1. Nous avons alors, d'après le tableau II.2, que la valeur $C_R$ qui satisfait cette règle vaut 1000, ceci correspond à l'étiquette linguistique TRÈS_GRAND. Ensuite, la règle $C_R < D_S$ avec $D_S = 220$ et $D_R > C_S$ avec $C_S = 220$ cherche les tuples avec une mesure de possibilité entre 0 et 1. Nous avons alors, d'après le tableau II.2, qu'il n'y a aucun valeur $C_R$ et $D_R$ qui satisfont cette règle. La mesure de possibilité vaut 0 dans les autres cas, c'est-à-dire oú $D_R \leq C_S$ avec $C_S = 220$. Il s'agit des valeurs crisp 170, 177, 192 et 198, et des étiquettes linguistiques BAS, NORMAL et GRAND comme le montre le tableau II.2. Le comparateur FGT assigne par définition la valeur 1 à la mesure de possibilité de UNKNOWN.

| Nº Fila | JOUEUR | ÉQUIPE | TAILLE | QUALITÉ | CDEG(TAILLE) |
|---|---|---|---|---|---|
|  |  |  |  |  |  |

**Tableau II.35 :** Résultat de la requête avec filtre flou TAILLE **NFGT** [180,220]

Le tableau II.35 nous montre une requête vide, cela signifie qu'il n' y a aucun joueur nécessairement plus grand que 2 m 20. En effet, le filtre flou TAILLE NFGT [180,220] cherche les tuples à partir de l'équation I.4 (voir paragraphe I.4). Nous utilisons alors les règles de cette équation dans la condition WHERE de la requête. La règle $A_R \geq D_S$ avec $D_S = 220$ cherche les tuples avec une mesure de nécessité égale à 1. Nous avons alors, d'après le tableau II.2, qu'il n'y a aucune valeur $A_R$ qui satisfait cette règle. Ensuite, la règle $A_R < D_S$ avec $D_S = 220$ et $B_R > C_S$ avec $C_S = 220$ cherche les tuples avec une mesure de nécessité entre 0 et 1. Or, d'après le tableau II.2 il n'y a aucune valeur $A_R$ et $B_R$ qui satisfont cette règle.

La mesure de nécessité donne 0 dans les autres cas, c'est-à-dire où $B_R \leq C_S$ avec $C_S = 220$. Il s'agit des valeurs crisp 170, 177, 192 et 198, et des étiquettes linguistiques BAS, NORMAL, GRAND et TRÈS_GRAND comme le montre le tableau II.2. Le comparateur NFGT assigne par définition la valeur 0 à la mesure de nécessité de UNKNOWN.

| Nº Fila | JOUEUR | ÉQUIPE | TAILLE | QUALITÉ | CDEG(TAILLE) |
|---|---|---|---|---|---|
| 1 | P2 | Córdoba | TRÈS_GRAND | [2,7,10,15] | 1 |
| 2 | P3 | Granada | NORMAL | RÉGULIER | 1 |
| 3 | P4 | Granada | 192 | RÉGULIER | 1 |
| 4 | P5 | Granada | GRAND | 10±10 | 1 |
| 5 | P6 | Málaga | 198 | MAUVAIS | 1 |
| 6 | P7 | Málaga | TRÈS_GRAND | 35±10 | 1 |
| 7 | P10 | Sevilla | NORMAL | BON | 1 |
| 8 | P11 | Cádiz | TRÈS_GRAND | 25±10 | 1 |
| 9 | P13 | Almería | GRAND | TRÈS_BON | 1 |
| 10 | P14 | Almería | TRÈS_GRAND | 8±10 | 1 |
| 11 | P16 | Huelva | GRAND | TRÈS_BON | 1 |
| 12 | P17 | Huelva | UNKNOWN | UNKNOWN | 1 |
| 13 | P18 | Jaén | UNKNOWN | [8,12,15,25] | 1 |
| 14 | P19 | Jaén | NORMAL | 25±10 | 1 |

**Tableau II.36 :** Résultat de la requête avec filtre flou TAILLE **FGEQ** [180,220]

Le tableau II.36 montre 14 joueurs qui sont possiblement plus grand ou égal à 1m80. Le filtre flou TAILLE FGEQ [180,220] cherche les tuples à partir de l'équation I.5 (voir paragraphe I.5). Nous utilisons alors les règles de cette équation dans la condition WHERE de la requête. La règle $C_R \geq B_S$ avec $B_S = 180$ cherche les tuples avec une mesure de possibilité égale à 1. Il s'agit des valeurs crisp 192 et 198, et des étiquettes linguistiques NORMAL, GRAND et TRÈS_GRAND comme le montre le tableau II.2. Ensuite, la règle $C_R < B_S$ avec $B_S = 180$ et $D_R > A_S$ avec $A_S = 180$ cherche les tuples avec une mesure de possibilité entre 0 et 1. Or, d'après le tableau II.2 il n'y a aucune valeur $A_R$ et $B_R$ qui satisfont cette règle. La mesure de possibilité donne 0 dans les autres cas, c'est-à-dire où $D_R \leq A_S$ avec $A_S = 180$. Il s'agit des valeurs crisp 170 et 177, et de l'étiquette linguistique BAS comme le montre le tableau II.2. Le comparateur FGEQ assigne par définition la valeur 1 à la mesure de possibilité de UNKNOWN.

| Nº Fila | JOUEUR | ÉQUIPE | TAILLE | QUALITÉ | CDEG(TAILLE) |
|---|---|---|---|---|---|
| 1 | P2 | Córdoba | TRÈS_GRAND | [2,7,10,15] | 1 |
| 2 | P4 | Granada | 192 | RÉGULIER | 1 |
| 3 | P5 | Granada | GRAND | 10±10 | 1 |
| 4 | P6 | Málaga | 198 | MAUVAIS | 1 |
| 5 | P7 | Málaga | TRÈS_GRAND | 35±10 | 1 |
| 6 | P11 | Cádiz | TRÈS_GRAND | 25±10 | 1 |
| 7 | P13 | Almería | GRAND | TRÈS_BON | 1 |
| 8 | P14 | Almería | TRÈS_GRAND | 8±10 | 1 |
| 9 | P16 | Huelva | GRAND | TRÈS_BON | 1 |

**Tableau II.37 :** Résultat de la requête avec filtre flou TAILLE NFGEQ [180,220]

Le tableau II.37 montre 9 joueurs qui sont nécessairement plus grand ou égale au 1m80. Le filtre flou TAILLE NFGEQ [180,220] cherche les tuples à partir de l'équation I.6 (voir paragraphe I.6). Nous utilisons alors les règles de cette équation dans la condition WHERE de la requête. La règle $A_R \geq B_S$ avec $B_S = 180$ cherche les tuples avec une mesure de nécessité égale à 1. Il s'agit des valeurs crisp 192 et 198, et des étiquettes linguistiques GRAND et TRÈS_GRAND comme le montre le tableau II.2. Ensuite, la règle $A_R < B_S$ avec $B_S = 180$ et $B_R > A_S$ avec $A_S = 180$ cherche les tuples avec une mesure de nécessité entre 0 et 1. Or, d'après le tableau II.2 il n'y a aucun valeur $A_R$ et $B_R$ qui satisfont cette règle. La mesure de nécessité vaut 0 dans les autres cas, c'est-à-dire où $B_R \leq A_S$ avec $A_S = 180$. Il s'agit des valeurs crisp 170 et 177, et des étiquettes linguistiques BAS et NORMAL comme le montre le tableau II.2. Le comparateur NFGEQ assigne par définition la valeur 0 à la mesure de nécessité de UNKNOWN.

## II.5.3 Filtres flous FLT (FLEQ)/NFLT(NFLEQ) : Fuzzy/Necessarily Less (Equal) Than, Possiblement/nécessairement plus petit (égal) que

Soient les requêtes FLT/NFLT et FLEQ/NFLEQ « Sélectionner toutes les informations des joueurs qui sont possiblement/nécessairement plus petit (égal) à l'intervalle [180,220] ». Les figures II.16 et II.17 montrent la syntaxe FSQL associé aux modèles FGT/NFGT et FGEQ/NFGEQ respectivement, tandis que les figures II.38, II.39, II.40 et II.41 montrent les résultats de ces requêtes floues FSQL.

| SELECT JOUEURS. %, CDEG (TAILLE) FROM JOUEURS<br>WHERE TAILLE **FLT** [180,220] THOLD 0.05; | SELECT JOUEURS. %, CDEG (TAILLE) FROM JOUEURS<br>WHERE TAILLE **NFLT** [180,220] THOLD 0.05; |
|---|---|

**Figure II.16 :** Requêtes avec filtres flous **FLT**/**NFLT** et intervalle

| SELECT JOUEURS. %, CDEG (TAILLE) FROM JOUEURS<br>WHERE TAILLE **FLEQ** [180,220] THOLD 0.05; | SELECT JOUEURS. %, CDEG (TAILLE) FROM JOUEURS<br>WHERE TAILLE **NFLEQ** [180,220] THOLD 0.05; |
|---|---|

**Figure II.17 :** Requêtes avec filtres flous **FLEQ**/**NFLEQ** et intervalle

| Nº Fila | JOUEUR | ÉQUIPE | TAILLE | QUALITÉ | CDEG(TAILLE) |
|---|---|---|---|---|---|
| 1 | P1 | Córdoba | BAS | [30,38,40,45] | 1 |
| 2 | P3 | Granada | NORMAL | RÉGULIER | 1 |
| 3 | P8 | Málaga | 170 | [31,34,35,38] | 1 |
| 4 | P9 | Sevilla | BAS | 15±10 | 1 |
| 5 | P10 | Sevilla | NORMAL | BON | 1 |
| 6 | P12 | Cádiz | BAS | TRÈS_BON | 1 |
| 7 | P15 | Almería | 177 | 6±10 | 1 |
| 8 | P17 | Huelva | UNKNOWN | UNKNOWN | 1 |
| 9 | P18 | Jaén | UNKNOWN | [8,12,15,25] | 1 |
| 10 | P19 | Jaén | NORMAL | 25±10 | 1 |

**Tableau II.38 :** Résultat de la requête avec filtre flou TAILLE FLT [180,220]

Le tableau II.38 montre 10 joueurs qui sont possiblement plus petits que 1m80. Le filtre flou TAILLE FLT [180,220] cherche les tuples à partir de l'équation I.7 (voir paragraphe I.7). Nous utilisons alors les règles de cette équation dans la condition WHERE de la requête. La règle $B_R \leq A_S$ avec $A_S = 180$ cherche les tuples avec une mesure de possibilité égale à 1. Il s'agit des valeurs crisp 170 et 177, et des étiquettes linguistiques BAS et NORMAL selon le montre le tableau II.2. Ensuite, la règle $B_R > A_S$ avec $A_S = 180$ et $A_R < B_S$ avec $B_S = 180$ cherche les tuples avec une mesure de possibilité entre 0 et 1. Or, d'après le tableau II.2 il n'y a aucune valeur $A_R$ et $B_R$ qui satisfont cette règle. La mesure de possibilité donne 0 dans les autres cas, c'est-à-dire où $A_R \geq B_S$ avec $B_S = 180$. Il s'agit des valeurs crisp 192 et 198, et des étiquettes linguistiques GRAND et TRÈS_GRAND comme le montre le tableau II.2. Le comparateur FLT assigne par définition la valeur 1 à la mesure de possibilité de UNKNOWN.

| Nº Fila | JOUEUR | ÉQUIPE | TAILLE | QUALITÉ | CDEG(TAILLE) |
|---|---|---|---|---|---|
| 1 | P1 | Córdoba | BAS | [30,38,40,45] | 1 |
| 2 | P8 | Málaga | 170 | [31,34,35,38] | 1 |
| 3 | P9 | Sevilla | BAS | 15±10 | 1 |
| 4 | P12 | Cádiz | BAS | TRÈS_BON | 1 |
| 5 | P15 | Almería | 177 | 6±10 | 1 |

**Tableau II.39 :** Résultat de la requête avec filtre flou TAILLE NFLT [180,220]

Le tableau II.39 montre 5 joueurs qui sont nécessairement plus petit que 1m80. Le filtre flou TAILLE NFLT [180,220] cherche les tuples à partir de l'équation I.8 (voir paragraphe I.8). Nous utilisons alors les règles de cette équation dans la condition WHERE de la requête. La règle $D_R \leq A_S$ avec $A_S = 180$ cherche les tuples avec une mesure de nécessité égale à 1. Il s'agit des valeurs crisp 170 et 177, et de l'étiquette linguistique BAS comme le montre le tableau II.2. Ensuite, la règle $D_R > A_S$ avec $A_S = 180$ et $C_R < B_S$ avec $B_S = 180$ cherche les tuples avec une mesure de nécessité entre 0 et 1. Or, d'après le tableau II.2 il n'y a aucune valeur $C_R$ et $D_R$ qui satisfont cette règle.

La mesure de nécessité donne 0 dans les autres cas, c'est-à-dire où $C_R \geq B_S$ avec $B_S = 180$. Il s'agit des valeurs crisp 192 et 198, et des étiquettes linguistiques NORMAL, GRAND et TRÈS_GRAND comme le montre le tableau II.2. Le comparateur NFLT assigne par définition la valeur 0 à la mesure de nécessité de UNKNOWN.

| Nº Fila | JOUEUR | ÉQUIPE | TAILLE | QUALITÉ | CDEG(TAILLE) |
|---|---|---|---|---|---|
| 1 | P1 | Córdoba | BAS | [30,38,40,45] | 1 |
| 2 | P2 | Córdoba | TRÈS_GRAND | [2,7,10,15] | 1 |
| 3 | P3 | Granada | NORMAL | RÉGULIER | 1 |
| 4 | P4 | Granada | 192 | RÉGULIER | 1 |
| 5 | P5 | Granada | GRAND | 10±10 | 1 |
| 6 | P6 | Málaga | 198 | MAUVAIS | 1 |
| 7 | P7 | Málaga | TRÈS_GRAND | 35±10 | 1 |
| 8 | P8 | Málaga | 170 | [31,34,35,38] | 1 |
| 9 | P9 | Sevilla | BAS | 15±10 | 1 |
| 10 | P10 | Sevilla | NORMAL | BON | 1 |
| 11 | P11 | Cádiz | TRÈS_GRAND | 25±10 | 1 |
| 12 | P12 | Cádiz | BAS | TRÈS_BON | 1 |
| 13 | P13 | Almería | GRAND | TRÈS_BON | 1 |
| 14 | P14 | Almería | TRÈS_GRAND | 8±10 | 1 |
| 15 | P15 | Almería | 177 | 6±10 | 1 |
| 16 | P16 | Huelva | GRAND | TRÈS_BON | 1 |
| 17 | P17 | Huelva | UNKNOWN | UNKNOWN | 1 |
| 18 | P18 | Jaén | UNKNOWN | [8,12,15,25] | 1 |
| 19 | P19 | Jaén | NORMAL | 25±10 | 1 |

**Tableau II.40** : Résultat de la requête avec filtre flou TAILLE **FLEQ** [180,220]

Le tableau II.40 montre 19 joueurs qui sont possiblement plus petite ou égal au 2m20. Le filtre flou TAILLE FLEQ [180,220] cherche les tuples à partir de l'équation I.9 (voir paragraphe I.9). Nous utilisons alors les règles de cette équation dans la condition WHERE de la requête. La règle $B_R \leq C_S$ avec $C_S = 220$ cherche les tuples avec une mesure de possibilité égale à 1. Il s'agit des valeurs crisp 170, 177, 192 et 198, et des étiquettes linguistiques BAS, NORMAL, GRAND et TRÈS_GRAND comme le montre le tableau II.2. Ensuite, la règle $B_R > C_S$ avec $C_S = 220$ et $A_R < D_S$ avec $D_S = 220$ cherche les tuples avec une mesure de possibilité entre 0 et 1. Or, d'après le tableau II.2 il n'y a aucun valeur $A_R$ et $B_R$ qui satisfont cette règle. Le comparateur FLEQ assigne par définition la valeur 1 à la mesure de possibilité de UNKNOWN.

| Nº Fila | JOUEUR | ÉQUIPE | TAILLE | QUALITÉ | CDEG(TAILLE) |
|---|---|---|---|---|---|
| 1 | P1 | Córdoba | BAS | [30,38,40,45] | 1 |
| 2 | P3 | Granada | NORMAL | RÉGULIER | 1 |
| 3 | P4 | Granada | 192 | RÉGULIER | 1 |
| 4 | P5 | Granada | GRAND | 10±10 | 1 |
| 5 | P6 | Málaga | 198 | MAUVAIS | 1 |
| 6 | P8 | Málaga | 170 | [31,34,35,38] | 1 |
| 7 | P9 | Sevilla | BAS | 15±10 | 1 |
| 8 | P10 | Sevilla | NORMAL | BON | 1 |
| 9 | P12 | Cádiz | BAS | TRÈS_BON | 1 |
| 10 | P13 | Almería | GRAND | TRÈS_BON | 1 |
| 11 | P15 | Almería | 177 | 6±10 | 1 |
| 12 | P16 | Huelva | GRAND | TRÈS_BON | 1 |
| 13 | P19 | Jaén | NORMAL | 25±10 | 1 |

**Tableau II.41** : Résultat de la requête avec filtre flou TAILLE **NFLEQ** [180,220]

Le tableau II.41 montre 13 joueurs qui sont nécessairement plus petit ou égal à 2m20. Le filtre flou TAILLE NFLEQ [180,220] cherche les tuples à partir de l'équation I.10 (voir paragraphe I.10). Nous utilisons alors les règles de cette équation dans la condition WHERE de la requête. La règle $D_R \leq C_S$ avec $C_S = 220$ cherche les tuples avec une mesure de nécessité égale à 1. Il s'agit des valeurs crisp 170, 177, 192 et 198, et des étiquettes linguistiques BAS, NORMAL et GRAND comme le montre le tableau II.2. Ensuite, la règle $D_R > C_S$ avec $C_S = 220$ et $C_R < D_S$ avec $D_S = 220$ cherche les tuples avec une mesure de nécessité entre 0 et 1. D'après le tableau II.2 il n'y a aucune valeur $C_R$ et $D_R$ qui satisfont cette règle. Le comparateur NFLEQ assigne par définition la valeur 0 à la mesure de nécessité de UNKNOWN.

## II.5.4 Filtres flous MGT(MLT)/NMGT(NMLT) : Much Greater (Less) Than, Possiblement/nécessairement beaucoup plus grand (petit) que

Soient les requêtes MGT/NMGT et MLT/NMLT « Sélectionner toutes les informations des joueurs qui sont possiblement/nécessairement beaucoup plus grande (petite) que l'intervalle [180,220] ». Les figures II.18 et II.19 montrent la syntaxe FSQL associé aux modèles MGT/NMGT et MLT/NMLT respectivement, tandis que les tableaux II.42, II.43, II.44 et II.45 montrent les résultats de ces requêtes floues FSQL.

| SELECT JOUEURS. %, CDEG (TAILLE) FROM JOUEURS WHERE TAILLE **MGT** [180,220] THOLD 0.05; | SELECT JOUEURS. %, CDEG (TAILLE) FROM JOUEURS WHERE TAILLE **NMGT** [180,220] THOLD 0.05; |
|---|---|

**Figure II.18 :** Requêtes avec filtres flous **MGT/NMGT** et intervalle

| SELECT JOUEURS. %, CDEG (TAILLE) FROM JOUEURS WHERE TAILLE **MLT** [180,220] THOLD 0.05; | SELECT JOUEURS. %, CDEG (TAILLE) FROM JOUEURS WHERE TAILLE **NMLT** [180,220] THOLD 0.05; |
|---|---|

**Figure II.19 :** Requêtes avec filtres flous **MLT/NMLT** et intervalle

| Nº Fila | JOUEUR | ÉQUIPE | TAILLE | QUALITÉ | CDEG(TAILLE) |
|---|---|---|---|---|---|
| 1 | P2 | Córdoba | TRÈS_GRAND | [2,7,10,15] | 1 |
| 2 | P7 | Málaga | TRÈS_GRAND | 35±10 | 1 |
| 3 | P11 | Cádiz | TRÈS_GRAND | 25±10 | 1 |
| 4 | P14 | Almería | TRÈS_GRAND | 8±10 | 1 |
| 5 | P17 | Huelva | UNKNOWN | UNKNOWN | 1 |
| 6 | P18 | Jaén | UNKNOWN | [8,12,15,25] | 1 |

**Tableau II.42 :** Résultat de la requête avec filtre flou TAILLE **MGT** [180,220]

Le tableau II.42 montre 6 joueurs qui sont possiblement beaucoup plus grand que 2m20. Le filtre flou TAILLE MGT [180, 220] cherche les tuples à partir de l'équation I.11 (voir paragraphe I.11). Nous utilisons alors les règles de cette équation dans la condition WHERE de la requête. La règle $C_R \geq D_S + M$ avec $D_S = 220$ et $M = 6$ cherche les tuples avec une mesure de possibilité égale à 1. Ceci correspond à l'étiquette linguistique TRÈS_GRAND comme le montre le tableau II.2. Ensuite, la règle $C_R < D_S + M$ avec $D_S = 220$ et $D_R > C_S + M$ avec $C_S = 220$ cherche les tuples avec une mesure de possibilité entre 0 et 1. Or, d'après le tableau II.2 il n'y a aucune valeur pour $C_R$ et $D_R$ qui satisfont cette règle. La mesure de possibilité vaut 0 dans les autres cas, c'est-à-dire où $D_R \leq C_S + M$ avec $C_S = 220$ et $M = 6$. Il s'agit des valeurs crisp 170, 177, 192 et 198, et des étiquettes linguistiques BAS, NORMAL et GRAND comme le montre le tableau II.2. Le comparateur MGT assigne par définition la valeur 1 à la mesure de possibilité de UNKNOWN.

| Nº Fila | JOUEUR | ÉQUIPE | TAILLE | QUALITÉ | CDEG(TAILLE) |
|---|---|---|---|---|---|
| | | | | | |

**Tableau II.43 :** Résultat de la requête avec filtre flou TAILLE **NMGT** [180,220]

Le tableau II.43 montre une requête vide, cela signifie qu'il n' y a aucun joueur nécessairement beaucoup plus grand que 2m20. En effet, le filtre flou TAILLE NMGT [180,220] cherche les tuples à partir de l'équation I.12 (voir paragraphe I.12). Nous utilisons alors les règles de cette équation dans la condition WHERE de la requête. La règle $A_R \geq D_S + M$ avec $D_S = 220$ et $M = 6$ cherche les tuples avec une mesure de nécessité égale à 1. Nous avons alors, d'après le tableau II.2, qu'il n'y a aucune valeur $A_R$ qui satisfait cette règle. Ensuite, la règle $A_R < D_S + M$ avec $D_S = 220$ et $B_R > C_S + M$ avec $C_S = 220$ cherche les tuples avec une mesure de nécessité entre 0 et 1. Or, d'après le tableau II.2 il n'y a aucun valeur $A_R$ et $B_R$ qui satisfont cette règle. En revanche, la règle $B_R \leq C_S + M$ avec $C_S = 220$ et $M = 2$ est vérifie dans tous les cas, c'est-à-dire pour les valeurs crisp 170, 177, 192 et 198, et les étiquettes linguistiques BAS, NORMAL, GRAND et TRÈS_GRAND, cela signifie que la mesure de nécessité vaut 0. Enfin, le comparateur NMGT assigne par définition la valeur 0 à la mesure de nécessité de UNKNOWN.

| Nº Fila | JOUEUR | ÉQUIPE | TAILLE | QUALITÉ | CDEG(TAILLE) |
|---|---|---|---|---|---|
| 1 | P1 | Córdoba | BAS | [30,38,40,45] | 1 |
| 2 | P8 | Málaga | 170 | [31,34,35,38] | 0,33 |
| 3 | P9 | Sevilla | BAS | 15±10 | 1 |
| 4 | P12 | Cádiz | BAS | TRÈS_BON | 1 |
| 5 | P17 | Huelva | UNKNOWN | UNKNOWN | 1 |
| 6 | P18 | Jaén | UNKNOWN | [8,12,15,25] | 1 |

**Tableau II.44 :** Résultat de la requête avec filtre flou TAILLE **MLT** [180,220]

| TAILLE | $p = B_S - M - A_R$ | $r = B_R - A_R$ | $s = A_S - B_S$ | $q = r - s$ | CDEG (TAILLE) $p/q$ |
|--------|------|------|------|------|------|
| 170 | 180 – 12 -164 | 170 - 164 | 174 - 180 | 6 + 6 | 0.33 |

Le tableau II.44 montre 6 joueurs qui sont possiblement beaucoup plus petit que 1m80. Le filtre flou TAILLE MLT [180,220] cherche les tuples à partir de l'équation I.13 (voir paragraphe I.13). Nous utilisons alors les règles de cette équation dans la condition WHERE de la requête. La règle $B_R \le A_S - M$ avec $A_S = 180$ et $M = 6$ cherche les tuples avec une mesure de possibilité égale à 1. Ceci correspond à la valeur crisp **170** et à l'étiquette linguistique BAS comme le montre le tableau II.2. Ensuite, la règle $B_R > A_S - M$ avec $A_S = 180$ et $A_R < B_S - M$ avec $B_S = 180$ cherche les tuples avec une mesure de possibilité entre 0 et 1. Or, d'après le tableau II.2 il n'y a aucun valeur $A_R$ et $B_R$ qui satisfont cette règle ; la mesure de possibilité vaut 0 dans les autres cas, c'est-à-dire où $A_R \ge B_S - M$ avec $B_S = 180$ et $M = 6$. Il s'agit des valeurs crisp 177, 192 et 198, et des étiquettes linguistiques NORMAL, GRAND et TRÈS_GRAND comme le montre le tableau II.2. Le comparateur MLT assigne par définition la valeur 1 à la mesure de possibilité de UNKNOWN.

| Nº File | JOUEUR | ÉQUIPE | TAILLE | QUALITÉ | CDEG(TAILLE) |
|---------|--------|--------|--------|---------|--------------|
| 1 | P17 | Huelva | UNKNOWN | UNKNOWN | 1 |
| 2 | P18 | Jaén | UNKNOWN | [8,12,15,25] | 1 |

**Tableau II.45 :** Résultat de la requête avec filtre flou TAILLE **NMLT** [180,220]

Le tableau II.45 montre 2 joueurs qui sont nécessairement beaucoup plus petit que 1m80. Le filtre flou TAILLE NMLT [180,220] cherche les tuples à partir de l'équation I.14 (voir paragraphe I.14). Nous utilisons alors les règles de cette équation dans la condition WHERE de la requête. La règle $D_R \le A_S - M$ avec $A_S = 180$ et $M = 6$ cherche les tuples avec une mesure de nécessité égale à 1. Ceci correspond au valeur crisp 170 comme le montre le tableau II.2. Ensuite, la règle $D_R > A_S - M$ avec $A_S = 180$ et $C_R < B_S - M$ avec $B_S = 180$ cherche les tuples avec une mesure de nécessité entre 0 et 1. Or, d'après le tableau II.2 il n'y a aucune valeur $A_R$ et $B_R$ qui satisfont cette règle. La mesure de nécessité donne 0 dans les autres cas, c'est-à-dire où $C_R \ge B_S - M$ avec $B_S = 180$ et $M = 6$. Il s'agit des valeurs crisp 177, 192 et 198, et des étiquettes linguistiques BAS, NORMAL, GRAND et TRÈS_GRAND comme le montre le tableau II.2. Le comparateur NMLT assigne par définition la valeur 1 à la mesure de nécessité de UNKNOWN.

## II.6   Modèle comportant un nombre autour-de-*n* FSQL comme constante

Soit la valeur autour #197 comme intervalle donc d'après le tableau II.4, on a : $A_S = 187$, $B_S = C_S = 197$ et $D_S = 207$.

### II.6.1      Filtres flous FEQ/NFEQ : Fuzzy/Necessarily EQual, Possiblement/nécessairement égale

Soit la requête FEQ/NFEQ « Sélectionner toutes les informations des joueurs qui sont possiblement/nécessairement égale autour de 197 ». La figure II.20 montre la syntaxe FSQL associé au modèle FEQ/NFEQ, tandis que les tableaux II.46 et II.47 montrent les résultats de ces requêtes floues FSQL.

| SELECT JOUEURS. %, CDEG (TAILLE) FROM JOUEURS | SELECT JOUEURS. %, CDEG (TAILLE) FROM JOUEURS |
|---|---|
| WHERE TAILLE **FEQ** #197 THOLD 0.05; | WHERE TAILLE **NFEQ** #197 THOLD 0.05; |

**Figure II.20 :** Requêtes avec filtres flous **FEQ**/**NFEQ** et autour-de-*n*

| Nº Fila | JOUEUR | ÉQUIPE | TAILLE | QUALITÉ | CDEG(TAILLE) |
|---|---|---|---|---|---|
| 1 | P2 | Córdoba | TRÈS_GRAND | [2,7,10,15] | 0,19 |
| 2 | P3 | Granada | NORMAL | RÉGULIER | 0,25 |
| 3 | P4 | Granada | 192 | RÉGULIER | 0,17 |
| 4 | P5 | Granada | GRAND | 10±10 | 1 |
| 5 | P6 | Málaga | 198 | MAUVAIS | 0,83 |
| 6 | P7 | Málaga | TRÈS_GRAND | 35±10 | 0,19 |
| 7 | P10 | Sevilla | NORMAL | BON | 0,25 |
| 8 | P11 | Cádiz | TRÈS_GRAND | 25±10 | 0,19 |
| 9 | P13 | Almería | GRAND | TRÈS_BON | 1 |
| 10 | P14 | Almería | TRÈS_GRAND | 8±10 | 0,19 |
| 11 | P16 | Huelva | GRAND | TRÈS_BON | 1 |
| 12 | P17 | Huelva | UNKNOWN | UNKNOWN | 1 |
| 13 | P18 | Jaén | UNKNOWN | [8,12,15,25] | 1 |
| 14 | P19 | Jaén | NORMAL | 25±10 | 0,25 |

**Tableau II.46 :** Résultat de la requête avec filtre flou **FEQ** et autour-de-*n*

| Nº Fila | JOUEUR | ÉQUIPE | TAILLE | QUALITÉ | CDEG(TAILLE) |
|---|---|---|---|---|---|
| 1 | P4 | Granada | 192 | RÉGULIER | 0,17 |
| 2 | P5 | Granada | GRAND | 10±10 | 0,19 |
| 3 | P6 | Málaga | 198 | MAUVAIS | 0,83 |
| 4 | P13 | Almería | GRAND | TRÈS_BON | 0,19 |
| 5 | P16 | Huelva | GRAND | TRÈS_BON | 0,19 |

**Tableau II.47 :** Résultat de la requête avec filtre flou **NFEQ** et autour-de-*n*

## II.6.2 Filtres flous FGT (FGEQ)/NFGT(NFGEQ) : Fuzzy/Necessarily Greater (Equal) Than, possiblement/nécessairement plus grand (égal) que

Soient les requêtes FGT/NFGT et FGEQ/NFGEQ « Sélectionner toutes les informations des joueurs qui sont possiblement/nécessairement plus grand (égale) qu'autour de #197 ». Les figures II.21 et II.22 montrent la syntaxe FSQL associé aux modèles FGT/NFGT et FGEQ/NFGEQ respectivement, tandis que les tableaux II.48, II.49, II.50 et II.51 montrent les résultats de ces requêtes floues FSQL.

| SELECT JOUEURS. %, CDEG (TAILLE) FROM JOUEURS WHERE TAILLE **FGT** #197 THOLD 0.05; | SELECT JOUEURS. %, CDEG (TAILLE) FROM JOUEURS WHERE TAILLE **NFGT** #197 THOLD 0.05; |
|---|---|

**Figure II.21 :** Requêtes avec filtres flous FGT/NFGT et autour-de-*n*

| SELECT JOUEURS. %, CDEG (TAILLE) FROM JOUEURS WHERE TAILLE **FGEQ** #197 THOLD 0.05; | SELECT JOUEURS. %, CDEG (TAILLE) FROM JOUEURS WHERE TAILLE **NFGEQ** #197 THOLD 0.05; |
|---|---|

**Figure II.22 :** Requêtes avec filtres flous **FGEQ/NFGEQ** et autour-de-*n*

| Nº Fila | JOUEUR | ÉQUIPE | TAILLE | QUALITÉ | CDEG(TAILLE) |
|---|---|---|---|---|---|
| 1 | P2 | Córdoba | TRÈS_GRAND | [2,7,10,15] | 1 |
| 2 | P5 | Granada | GRAND | 10±10 | 0,81 |
| 3 | P6 | Málaga | 198 | MAUVAIS | 0,17 |
| 4 | P7 | Málaga | TRÈS_GRAND | 35±10 | 1 |
| 5 | P11 | Cádiz | TRÈS_GRAND | 25±10 | 1 |
| 6 | P13 | Almería | GRAND | TRÈS_BON | 0,81 |
| 7 | P14 | Almería | TRÈS_GRAND | 8±10 | 1 |
| 8 | P16 | Huelva | GRAND | TRÈS_BON | 0,81 |
| 9 | P17 | Huelva | UNKNOWN | UNKNOWN | 1 |
| 10 | P18 | Jaén | UNKNOWN | [8,12,15,25] | 1 |

**Tableau II.48 :** Résultat de la requête avec filtre flou TAILLE **FGT** #197

Le tableau II.48 montre 10 joueurs qui sont possiblement plus grand que 2m 07. Le filtre flou TAILLE FGT #197 cherche les tuples à partir de l'équation I.3 (voir paragraphe I.3). Nous utilisons alors les règles de cette équation dans la condition WHERE de la requête. La règle $C_R \geq D_S$ avec $D_S = 207$ cherche les tuples avec une mesure de possibilité égale à 1. Nous avons alors, d'après le tableau II.2, que la valeur $C_R$ qui satisfait cette règle vaut 1000, ceci correspond à l'étiquette linguistique TRÈS_GRAND. Ensuite, la règle $C_R < D_S$ avec $D_S = 207$ et $D_R > C_S$ avec $C_S = 197$ cherche les tuples avec une mesure de possibilité entre 0 et 1. Or, d'après le tableau II.2 les valeurs $C_R$ et $D_R$ qui satisfont cette règle correspondent à la valeur crisp 198 et aux valeurs 200 et 210 pour l'étiquette linguistique GRAND. Nous avons résumé les calculs de la mesure de possibilité dans le tableau suivant :

| TAILLE | $p = D_R - C_S$ | $r = D_S - C_S$ | $s = C_R - D_R$ | $q = r - s$ | CDEG (TAILLE) p/q |
|---|---|---|---|---|---|
| **GRAND** | **210 – 197** | **203 – 197** | **200 – 210** | **6 + 10** | **0.81** |
| **198** | **198 - 197** | **203 – 197** | **198 – 198** | **6** | **0.17** |

La mesure de possibilité vaut 0 dans les autres cas, c'est-à-dire oú $D_R \leq C_S$ avec $C_S = 197$. Il s'agit des valeurs crisp 170, 177 et 192, et des étiquettes linguistiques BAS et NORMAL selon le montre le tableau II.2. Le comparateur FGT assigne par définition la valeur 1 à la mesure de possibilité de UNKNOWN.

| Nº Fila | JOUEUR | ÉQUIPE | TAILLE | QUALITÉ | CDEG(TAILLE) |
|---|---|---|---|---|---|
| 1 | P2 | Córdoba | TRÈS_GRAND | [2,7,10,15] | 0,81 |
| 2 | P6 | Málaga | 198 | MAUVAIS | 0,17 |
| 3 | P7 | Málaga | TRÈS_GRAND | 35±10 | 0,81 |
| 4 | P11 | Cádiz | TRÈS_GRAND | 25±10 | 0,81 |
| 5 | P14 | Almería | TRÈS_GRAND | 8±10 | 0,81 |

**Tableau II.49** : Résultat de la requête avec filtre flou TAILLE **NFGT** #197

Le tableau II.49 montre 5 joueurs qui sont nécessairement plus grand que 2m 07. Le filtre flou TAILLE NFGT #197 cherche les tuples à partir de l'équation I.4 (voir paragraphe I.4). Nous utilisons alors les règles de cette équation dans la condition WHERE de la requête. La règle $A_R \geq D_S$ avec $D_S = 207$ cherche les tuples avec une mesure de nécessité égale à 1. Nous avons alors, d'après le tableau II.2, qu'il n'y a aucune valeur $A_R$ qui satisfait cette règle. Ensuite, la règle $A_R < D_S$ avec $D_S = 207$ et $B_R > C_S$ avec $C_S = 197$ cherche les tuples avec une mesure de nécessité entre 0 et 1. Or, d'après le tableau II.2 les valeurs $A_R$ et $B_R$ qui satisfont cette règle correspondent à la valeur crisp 198 et aux valeurs 200 et 210 pour l'étiquette linguistique TRÈS_GRAND. Nous avons résumé les calculs de la mesure de possibilité dans le tableau suivant :

| TAILLE | $p = B_R - C_S$ | $r = D_S - C_S$ | $s = A_R - B_R$ | $q = r - s$ | CDEG (TAILLE) p/q |
|---|---|---|---|---|---|
| **TRÈS_GRAND** | **210 – 197** | **203 – 197** | **200 – 210** | **6 + 10** | **0.81** |
| **198** | **198 - 197** | **203 – 197** | **198 – 198** | **6** | **0.17** |

La mesure de nécessité vaut 0 dans les autres cas, c'est-à-dire oú $B_R \leq C_S$ avec $C_S = 197$. Il s'agit des valeurs crisp 170, 177 et 192, et des étiquettes linguistiques BAS, NORMAL et GRAND comme le montre le tableau II.2. Le comparateur NFGT assigne par définition la valeur 0 à la mesure de nécessité de UNKNOWN.

| Nº Fila | JOUEUR | ÉQUIPE | TAILLE | QUALITÉ | CDEG(TAILLE) |
|---|---|---|---|---|---|
| 1 | P2 | Córdoba | TRÈS_GRAND | [2,7,10,15] | 1 |
| 2 | P3 | Granada | NORMAL | RÉGULIER | 0,25 |
| 3 | P4 | Granada | 192 | RÉGULIER | 0,17 |
| 4 | P5 | Granada | GRAND | 10±10 | 1 |
| 5 | P6 | Málaga | 198 | MAUVAIS | 1 |
| 6 | P7 | Málaga | TRÈS_GRAND | 35±10 | 1 |
| 7 | P10 | Sevilla | NORMAL | BON | 0,25 |
| 8 | P11 | Cádiz | TRÈS_GRAND | 25±10 | 1 |
| 9 | P13 | Almería | GRAND | TRÈS_BON | 1 |
| 10 | P14 | Almería | TRÈS_GRAND | 8±10 | 1 |
| 11 | P16 | Huelva | GRAND | TRÈS_BON | 1 |
| 12 | P17 | Huelva | UNKNOWN | UNKNOWN | 1 |
| 13 | P18 | Jaén | UNKNOWN | [8,12,15,25] | 1 |
| 14 | P19 | Jaén | NORMAL | 25±10 | 0,25 |

**Tableau II.50 :** Résultat de la requête avec filtre flou TAILLE **FGEQ** #197

Le tableau II.50 montre 14 joueurs qui sont possiblement plus grand ou égal à 1m97. Le filtre flou TAILLE FGEQ #197 cherche les tuples à partir de l'équation I.5 (voir paragraphe I.5). Nous utilisons alors les règles de cette équation dans la condition WHERE de la requête. La règle $C_R \geq B_S$ avec $B_S = 197$ cherche les tuples avec une mesure de possibilité égale à 1. Il s'agit de crisp 198, et des étiquettes linguistiques GRAND et TRÈS_GRAND comme le montre le tableau II.2. Ensuite, la règle $C_R < B_S$ avec $B_S = 197$ et $D_R > A_S$ avec $A_S = 187$ cherche les tuples avec une mesure de possibilité entre 0 et 1. Or, d'après le tableau II.2 les valeurs $C_R$ et $D_R$ qui satisfont cette règle correspondent à la valeur crisp 192 et aux valeurs 185 et 195 pour l'étiquette linguistique NORMAL. Nous avons résumé les calculs de la mesure de possibilité dans le tableau suivant :

| TAILLE | $p = D_R - A_S$ | $r = B_S - A_S$ | $s = C_R - D_R$ | $q = r - s$ | CDEG (TAILLE) p/q |
|---|---|---|---|---|---|
| NORMAL | 195 – 191 | 197 – 191 | 185 – 195 | 6 + 10 | 0.25 |
| 192 | 192 – 191 | 197 – 191 | 192 – 192 | 6 | 0.17 |

La mesure de possibilité donne 0 dans les autres cas, c'est-à-dire oú $D_R \leq A_S$ avec $A_S = 187$. Il s'agit des valeurs crisp 170 et 177, et de l'étiquette linguistique BAS comme le montre le tableau II.2. Le comparateur FGEQ assigne par définition la valeur 1 à la mesure de possibilité de UNKNOWN.

| N° Fila | JOUEUR | ÉQUIPE | TAILLE | QUALITÉ | CDEG(TAILLE) |
|---|---|---|---|---|---|
| 1 | P2 | Córdoba | TRÈS_GRAND | [2,7,10,15] | 1 |
| 2 | P4 | Granada | 192 | RÉGULIER | 0.17 |
| 3 | P5 | Granada | GRAND | 10±10 | 0.25 |
| 4 | P6 | Málaga | 198 | MAUVAIS | 1 |
| 5 | P7 | Málaga | TRÈS_GRAND | 35±10 | 1 |
| 6 | P11 | Cádiz | TRÈS_GRAND | 25±10 | 1 |
| 7 | P13 | Almería | GRAND | TRÈS_BON | 0.25 |
| 8 | P14 | Almería | TRÈS_GRAND | 8±10 | 1 |
| 9 | P16 | Huelva | GRAND | TRÈS_BON | 0.25 |

**Tableau II.51** : Résultat de la requête avec filtre flou TAILLE **NFGEQ** #197

Le tableau II.51 montre 9 joueurs qui sont nécessairement plus grand ou égal au 1m97. Le filtre flou TAILLE NFGEQ #197 cherche les tuples à partir de l'équation I.6 (voir paragraphe I.6). Nous utilisons alors les règles de cette équation dans la condition WHERE de la requête. La règle $A_R \geq B_S$ avec $B_S = 197$ cherche les tuples avec une mesure de nécessité égale à 1. Il s'agit de crisp 198, et de l'étiquette linguistique TRÈS_GRAND comme le montre le tableau II.2. Ensuite, la règle $A_R < B_S$ avec $B_S = 197$ et $B_R > A_S$ avec $A_S = 187$ cherche les tuples avec une mesure de nécessité entre 0 et 1. Or, d'après le tableau II.2 les valeurs $A_R$ et $B_R$ qui satisfont cette règle correspondent à la valeur crisp 192 et aux valeurs 185 et 195 pour l'étiquette linguistique GRAND. Nous avons résumé les calculs de la mesure de nécessité dans le tableau suivant :

| TAILLE | $p = B_R - A_S$ | $r = B_S - A_S$ | $s = A_R - B_R$ | $q = r - s$ | CDEG (TAILLE) p/q |
|---|---|---|---|---|---|
| GRAND | 195 – 191 | 197 – 191 | 185 – 195 | 6 + 10 | 0.25 |
| 192 | 192 – 191 | 197 – 191 | 192 – 192 | 6 | 0.17 |

La mesure de nécessité vaut 0 dans les autres cas, c'est-à-dire oú $B_R \leq A_S$ avec $A_S = 187$. Il s'agit des valeurs crisp 170 et 177, et des étiquettes linguistiques BAS et NORMAL selon le montre le tableau II.2. Le comparateur NFGEQ assigne par définition la valeur 0 à la mesure de nécessité de UNKNOWN.

## II.6.3 Filtres flous FLT (FLEQ)/NFLT (NFLEQ) : Fuzzy/Necessarily Less (Equal) than, possiblement/nécessairement plus petit (égale) que

Soient les requêtes FLT/NFLT et FLEQ/NFLEQ « Sélectionner toutes les informations des joueurs qui sont possiblement/nécessairement plus petite (égale) qu'autour de 197 ». Les figures II.23 et II.24 montrent la syntaxe FSQL associé aux modèles FGT/NFGT et FGEQ/NFGEQ respectivement, tandis que les tableaux II.52, II.53, II.54 et II.55 montrent les résultats de ces requêtes floues FSQL.

| SELECT JOUEURS. %, CDEG (TAILLE) FROM JOUEURS WHERE TAILLE **FLT** #197 THOLD 0.05; | SELECT JOUEURS. %, CDEG (TAILLE) FROM JOUEURS WHERE TAILLE **NFLT** #197 THOLD 0.05; |
|---|---|

**Figure II.23 :** Requêtes avec filtres flous FLT/NFLT et autour-de-*n*

| SELECT JOUEURS. %, CDEG (TAILLE) FROM JOUEURS WHERE TAILLE **FLEQ** #197 THOLD 0.05 ; | SELECT JOUEURS. %, CDEG (TAILLE) FROM JOUEURS WHERE TAILLE **NFLEQ** #197 THOLD 0.05; |
|---|---|

**Figure II.24 :** Requêtes avec filtres flous **FLEQ/NFLEQ** et autour-de-*n*

| Nº Fila | JOUEUR | ÉQUIPE | TAILLE | QUALITÉ | CDEG(TAILLE)] |
|---|---|---|---|---|---|
| 1 | P1 | Córdoba | BAS | [30,38,40,45] | 1 |
| 2 | P3 | Granada | NORMAL | RÉGULIER | 1 |
| 3 | P4 | Granada | 192 | RÉGULIER | 0,83 |
| 4 | P5 | Granada | GRAND | 10±10 | 0,75 |
| 5 | P8 | Málaga | 170 | [31,34,35,38] | 1 |
| 6 | P9 | Sevilla | BAS | 15±10 | 1 |
| 7 | P10 | Sevilla | NORMAL | BON | 1 |
| 8 | P12 | Cádiz | BAS | TRÈS_BON | 1 |
| 9 | P13 | Almería | GRAND | TRÈS_BON | 0,75 |
| 10 | P15 | Almería | 177 | 6±10 | 1 |
| 11 | P16 | Huelva | GRAND | TRÈS_BON | 0,75 |
| 12 | P17 | Huelva | UNKNOWN | UNKNOWN | 1 |
| 13 | P18 | Jaén | UNKNOWN | [8,12,15,25] | 1 |
| 14 | P19 | Jaén | NORMAL | 25±10 | 1 |

**Tableau II.52 :** Résultat de la requête avec filtre flou TAILLE **FLT** #197

Le tableau II.52 montre 14 joueurs qui sont possiblement plus petit que 1m91. Le filtre flou TAILLE FLT #197 cherche les tuples à partir de l'équation I.7 (voir paragraphe I.7). Nous utilisons alors les règles de cette équation dans la condition WHERE de la requête. La règle $B_R \leq A_S$ avec $A_S = 191$ cherche les tuples avec une mesure de possibilité égale à 1. Il s'agit des valeurs crisp 170 et 177, et des étiquettes linguistiques BAS et NORMAL selon le montre le tableau II.2. Ensuite, la règle $B_R > A_S$ avec $A_S = 191$ et $A_R < B_S$ avec $B_S = 197$ cherche les tuples avec une mesure de possibilité entre 0 et 1. Or, d'après le tableau II.2 les valeurs $A_R$ et $B_R$ qui satisfont cette règle correspondent à la valeur crisp 192 et aux valeurs 185 et 195 pour l'étiquette linguistique GRAND. Nous avons résumé les calculs de la mesure de nécessité dans le tableau suivant :

| **TAILLE** | **p = A$_R$ − B$_S$** | **r = A$_S$ − B$_S$** | **s = B$_R$ − A$_R$** | **q = r − s** | **CDEG (TAILLE) p/q** |
|---|---|---|---|---|---|
| **GRAND** | **185 − 197** | **191 − 197** | **195 − 185** | **-6 − 10** | **0.75** |
| **192** | **192 − 197** | **191 − 197** | **192 − 192** | **-6** | **0.83** |

La mesure de possibilité donne 0 dans les autres cas, c'est-à-dire où $A_R \geq B_S$ avec $B_S = 197$. Il s'agit de crisp 198, et de l'étiquette linguistique TRÈS_GRAND selon le montre le tableau II.2. Le comparateur FLT assigne par définition la valeur 1 à la mesure de possibilité de UNKNOWN.

| Nº Fila | JOUEUR | ÉQUIPE | TAILLE | QUALITÉ | CDEG(TAILLE)] |
|---|---|---|---|---|---|
| 1 | P1 | Córdoba | BAS | [30,38,40,45] | |
| 2 | P3 | Granada | NORMAL | RÉGULIER | 0,75 |
| 3 | P4 | Granada | 192 | RÉGULIER | 0,83 |
| 4 | P8 | Málaga | 170 | [31,34,35,38] | 1 |
| 5 | P9 | Sevilla | BAS | 15±10 | 1 |
| 6 | P10 | Sevilla | NORMAL | BON | 0,75 |
| 7 | P12 | Cádiz | BAS | TRÈS_BON | 1 |
| 8 | P15 | Almería | 177 | 6±10 | 1 |
| 9 | P19 | Jaén | NORMAL | 25±10 | 0,75 |

**Tableau II.53 :** Résultat de la requête avec filtre flou TAILLE **NFLT** #197

Le tableau II.53 montre 9 joueurs qui sont nécessairement plus petit que 1m91. Le filtre flou TAILLE NFLT #197 cherche les tuples à partir de l'équation I.8 (voir paragraphe I.8). Nous utilisons alors les règles de cette équation dans la condition WHERE de la requête. La règle $D_R \leq A_S$ avec $A_S = 191$ cherche les tuples avec une mesure de nécessité égale à 1. Il s'agit des valeurs crisp 170 et 177, et de l'étiquette linguistique BAS comme le montre le tableau II.2. Ensuite, la règle $D_R > A_S$ avec $A_S = 191$ et $C_R < B_S$ avec $B_S = 197$ cherche les tuples avec une mesure de nécessité entre 0 et 1. Or, d'après le tableau II.2 les valeurs $C_R$ et $D_R$ qui satisfont cette règle correspondent à la valeur crisp 192 et aux valeurs 185 et 195 pour l'étiquette linguistique NORMAL. Nous avons résumé les calculs de la mesure de nécessité dans le tableau suivant :

| TAILLE | $p = C_R - B_S$ | $r = A_S - B_S$ | $s = D_R - C_R$ | $q = r - s$ | CDEG (TAILLE) p/q |
|---|---|---|---|---|---|
| NORMAL | 185 – 197 | 191 – 197 | 195 – 185 | -6 – 10 | 0.75 |
| 192 | 192 – 197 | 191 – 197 | 192 – 192 | -6 | 0.83 |

La mesure de nécessité vaut 0 dans les autres cas, c'est-à-dire où $C_R \geq B_S$ avec $B_S = 197$. Il s'agit de crisp 198 et des étiquettes linguistiques GRAND et TRÈS_GRAND comme le montre le tableau II.2. Le comparateur NFLT assigne par définition la valeur 0 à la mesure de nécessité de UNKNOWN.

| Nº Fila | JOUEUR | ÉQUIPE | TAILLE | QUALITÉ | CDEG(TAILLE) |
|---|---|---|---|---|---|
| 1 | P1 | Córdoba | BAS | [30,38,40,45] | 1 |
| 2 | P2 | Córdoba | TRÈS_GRAND | [2,7,10,15] | 0,19 |
| 3 | P3 | Granada | NORMAL | RÉGULIER | 1 |
| 4 | P4 | Granada | 192 | RÉGULIER | 1 |
| 5 | P5 | Granada | GRAND | 10±10 | 1 |
| 6 | P6 | Málaga | 198 | MAUVAIS | 0,83 |
| 7 | P7 | Málaga | TRÈS_GRAND | 35±10 | 0,19 |
| 8 | P8 | Málaga | 170 | [31,34,35,38] | 1 |
| 9 | P9 | Sevilla | BAS | 15±10 | 1 |
| 10 | P10 | Sevilla | NORMAL | BON | 1 |
| 11 | P11 | Cádiz | TRÈS_GRAND | 25±10 | 0,19 |
| 12 | P12 | Cádiz | BAS | TRÈS_BON | 1 |
| 13 | P13 | Almería | GRAND | TRÈS_BON | 1 |
| 14 | P14 | Almería | TRÈS_GRAND | 8±10 | 0,19 |
| 15 | P15 | Almería | 177 | 6±10 | 1 |
| 16 | P16 | Huelva | GRAND | TRÈS_BON | 1 |
| 17 | P17 | Huelva | UNKNOWN | UNKNOWN | 1 |
| 18 | P18 | Jaén | UNKNOWN | [8,12,15,25] | 1 |
| 19 | P19 | Jaén | NORMAL | 25±10 | 1 |

**Tableau II.54 :** Résultat de la requête avec filtre flou TAILLE **FLEQ** #197

Le tableau II.54 montre 19 joueurs qui sont possiblement plus petite ou égal à 1m97. Le filtre flou TAILLE FLEQ #197 cherche les tuples à partir de l'équation I.9 (voir paragraphe I.9). Nous utilisons alors les règles de cette équation dans la condition WHERE de la requête. La règle $B_R \leq C_S$ avec $C_S = 197$ cherche les tuples avec une mesure de possibilité égale à 1. Il s'agit des valeurs crisp 170, 177 et 192, et des étiquettes linguistiques BAS, NORMAL et GRAND comme le montre le tableau II.2. Ensuite, la règle $B_R > C_S$ avec $C_S = 197$ et $A_R < D_S$ avec $D_S = 203$ cherche les tuples avec une mesure de possibilité entre 0 et 1. Or, d'après le tableau II.2 les valeurs $A_R$ et $B_R$ qui satisfont cette règle correspondent à la valeur crisp 199 et aux valeurs 200 et 210 pour l'étiquette linguistique TRÈS_GRAND. Nous avons résumé les calculs de la mesure de nécessité dans le tableau suivant :

| TAILLE | $p = D_S - A_R$ | $r = B_R - A_R$ | $s = C_S - D_S$ | $q = r - s$ | CDEG (TAILLE) p/q |
|---|---|---|---|---|---|
| TRÈS_GRAND | 203 – 200 | 210 – 200 | 197 – 203 | 10 + 6 | 0.19 |
| 198 | 203 – 198 | 198 – 198 | 197 – 203 | +6 | 0.83 |

Le comparateur FLEQ assigne par définition la valeur 1 à la mesure de possibilité de UNKNOWN.

| Nº Fila | JOUEUR | ÉQUIPE | TAILLE | QUALITÉ | CDEG(TAILLE) |
|---|---|---|---|---|---|
| 1 | P1 | Córdoba | BAS | [30,38,40,45] | 1 |
| 2 | P3 | Granada | NORMAL | RÉGULIER | 1 |
| 3 | P4 | Granada | 192 | RÉGULIER | 1 |
| 4 | P5 | Granada | GRAND | 10±10 | 1 |
| 5 | P6 | Málaga | 198 | MAUVAIS | 0,83 |
| 6 | P8 | Málaga | 170 | [31,34,35,38] | 1 |
| 7 | P9 | Sevilla | BAS | 15±10 | 1 |
| 8 | P10 | Sevilla | NORMAL | BON | 1 |
| 9 | P12 | Cádiz | BAS | TRÈS_BON | 1 |
| 10 | P13 | Almería | GRAND | TRÈS_BON | 0,19 |
| 11 | P15 | Almería | 177 | 6±10 | 1 |
| 12 | P16 | Huelva | GRAND | TRÈS_BON | 0,19 |
| 13 | P19 | Jaén | NORMAL | 25±10 | 1 |

**Tableau II.55 :** Résultat de la requête avec filtre flou flou TAILLE **NFLEQ** #197

Le tableau II.55 montre 13 joueurs qui sont nécessairement plus petit ou égal à 1m 97. Le filtre flou TAILLE NFLEQ #197 cherche les tuples à partir de l'équation I.10 (voir paragraphe I.10). Nous utilisons alors les règles de cette équation dans la condition WHERE de la requête. La règle $D_R \leq C_S$ avec $C_S = 197$ cherche les tuples avec une mesure de nécessité égale à 1. Il s'agit des valeurs crisp 170, 177 et 192, et des étiquettes linguistiques BAS et

---

NORMAL selon le montre le tableau II.2. Ensuite, la règle $D_R > C_S$ avec $C_S = 197$ et $C_R < D_S$ avec $D_S = 203$ cherche les tuples avec une mesure de nécessité entre 0 et 1. Or, d'après le tableau II.2 les valeurs $C_R$ et $D_R$ qui satisfont cette règle correspondent à la valeur crisp 198 et aux valeurs 200 et 210 pour l'étiquette linguistique GRAND. Nous avons résumé les calculs de la mesure de nécessité dans le tableau suivant :

| TAILLE | p = $C_R$ – $D_S$ | r = $C_S$ – $D_S$ | s = $D_R$ – $C_R$ | q = r – s | CDEG (TAILLE) p/q |
|--------|------|------|------|------|------|
| GRAND | 200 – 203 | 197 – 203 | 210 – 200 | -6 – 10 | 0.19 |
| 198 | 198 – 203 | 197 – 203 | 198 – 198 | -6 | 0.83 |

La mesure de nécessité vaut 0 dans les autres cas, c'est-à-dire oú $C_R \geq D_S$ avec $D_S = 203$. Il s'agit de l'étiquette linguistique TRÈS_GRAND comme le montre le tableau II.2. Le comparateur NFLEQ assigne par définition la valeur 0 à la mesure de nécessité de UNKNOWN.

## II.6.4 Filtres flous MGT(MLT)/NMGT(NMLT) :Much Greater (Less) Than, Possiblement/nécessairement beaucoup plus grand (petit) que

Soient les requêtes MGT/NMGT et MLT/NMLT « Sélectionner toutes les informations des joueurs qui sont possiblement/nécessairement beaucoup plus grande (petite) qu'autour de 197 ». Les figures II.25 et II.26 montrent la syntaxe FSQL associé aux modèles MGT/NMGT et MLT/NMLT respectivement, tandis que les tableaux II.56, II.57, II.58 et II.59 montrent les résultats de ces requêtes floues FSQL.

| SELECT JOUEURS. %, CDEG (TAILLE) FROM JOUEURS WHERE TAILLE **MGT** #197 THOLD 0.05; | SELECT JOUEURS. %, CDEG (TAILLE) FROM JOUEURS WHERE TAILLE **NMGT** #197 THOLD 0.05; |
|---|---|

**Figure II.25 :** Requêtes avec filtres flous MGT/NMGT et autour-de-*n*

| SELECT JOUEURS. %, CDEG (TAILLE) FROM JOUEURS WHERE TAILLE **MLT** #197 THOLD 0.05; | SELECT JOUEURS. %, CDEG (TAILLE) FROM JOUEURS WHERE TAILLE **NMLT** #197 THOLD 0.05; |
|---|---|

**Figure II.26 :** Requêtes avec filtres flous **MLT**/**NMLT** et autour-de-*n*

| Nº Fila | JOUEUR | ÉQUIPE | TAILLE | QUALITÉ | CDEG[(TAILLE)] |
|---|---|---|---|---|---|
| 1 | P2 | Córdoba | TRÈS_GRAND | [2,7,10,15] | 1 |
| 2 | P5 | Granada | GRAND | 10±10 | 0.06 |
| 3 | P7 | Málaga | TRÈS_GRAND | 35±10 | 1 |
| 4 | P11 | Cádiz | TRÈS_GRAND | 25±10 | 1 |
| 5 | P13 | Almería | GRAND | TRÈS_BON | 0.06 |
| 6 | P14 | Almería | TRÈS_GRAND | 8±10 | 1 |
| 7 | P16 | Huelva | GRAND | TRÈS_BON | 0.06 |
| 8 | P17 | Huelva | UNKNOWN | UNKNOWN | 1 |
| 9 | P18 | Jaén | UNKNOWN | [8,12,15,25] | 1 |

**Tableau II.56 :** Résultat de la requête avec filtre flou TAILLE **MGT** #197

Le tableau II.56 montre 9 joueurs qui sont possiblement beaucoup plus grand que 2m 03. Le filtre flou TAILLE MGT #197 cherche les tuples à partir de l'équation I.11 (voir paragraphe I.11). Nous utilisons alors les règles de cette équation dans la condition WHERE de la requête. La règle $C_R \geq D_S + M$ avec $D_S = 203$ et M = 6 cherche les tuples avec une mesure de possibilité égale à 1. Ceci correspond à l'étiquette linguistique TRÈS_GRAND comme le montre le tableau II.2. Ensuite, la règle $C_R < D_S + M$ avec $D_S = 203$ et $D_R > C_S + M$ avec $C_S = 197$ cherche les tuples avec une mesure de possibilité entre 0 et 1. Or, d'après le tableau II.2 les valeurs $C_R$ et $D_R$ qui satisfont cette condition valent 200 et 210 pour l'étiquette linguistique GRAND. La mesure de possibilité est donnée dans le tableau suivant :

| TAILLE | $p = C_S + M - D_R$ | $r = C_R - D_R$ | $s = D_S - C_S$ | $q = r - s$ | CDEG (TAILLE) p/q |
|---|---|---|---|---|---|
| GRAND | 197 + 12 – 210 | 200 – 210 | 203 – 197 | -10 – 6 | 0.06 |

La mesure de possibilité donne 0 dans les autres cas, c'est-à-dire où $D_R \leq C_S + M$ avec $C_S = 197$ et M = 6. Il s'agit des valeurs crisp 170, 177, 192 et 198, et des étiquettes linguistiques BAS et NORMAL comme le montre le tableau II.2. Le comparateur MGT assigne par définition la valeur 1 à la mesure de possibilité de UNKNOWN.

| N° Fila | JOUEUR | ÉQUIPE | TAILLE | QUALITÉ | CDEG(TAILLE) |
|---|---|---|---|---|---|
| 1 | P2 | Córdoba | TRÈS_GRAND | [2,7,10,15] | 0,06 |
| 2 | P7 | Málaga | TRÈS_GRAND | 35±10 | 0,06 |
| 3 | P11 | Cádiz | TRÈS_GRAND | 25±10 | 0,06 |
| 4 | P14 | Almería | TRÈS_GRAND | 8±10 | 0,06 |

**Tableau II.57 :** Résultat de la requête avec filtre flou TAILLE **NMGT** #197

Le tableau II.57 montre une requête vide, cela signifie qu'il n' y a aucun joueur nécessairement beaucoup plus grand que 2m 03. En effet, le filtre flou TAILLE NMGT #197 cherche les tuples à partir de l'équation I.12 (voir paragraphe I.12). Nous utilisons alors les règles de cette équation dans la condition WHERE de la requête. La règle $A_R \geq D_S + M$ avec $D_S = 203$ et M = 6 cherche les tuples avec une mesure de nécessité égale à 1. Nous avons alors, d'après le tableau II.2, qu'il n'y a aucune valeur $A_R$ qui satisfait cette règle. Ensuite, la règle $A_R < D_S + M$ avec $D_S = 203$ et $B_R > C_S + M$ avec $C_S = 197$ cherche les tuples avec une mesure de nécessité entre 0 et 1. Or, d'après le tableau II.2 les valeurs $C_R$ et $D_R$ qui satisfont cette condition vaux 200 et 210 pour l'étiquette linguistique TRÈS_GRAND. La mesure de possibilité est donnée dans le tableau suivant :

| TAILLE | $p = C_S + M - B_R$ | $r = A_R - B_R$ | $s = D_S - C_S$ | $q = r - s$ | CDEG (TAILLE) p/q |
|---|---|---|---|---|---|
| TRÈS_GRAND | 197 + 12 – 210 | 200 – 210 | 203 – 197 | -10 – 6 | 0.06 |

La mesure de possibilité vaut 0 dans les autres cas, c'est-à-dire où $D_R \leq C_S + M$ avec $C_S = 197$ et M = 6. Il s'agit des valeurs crisp 170, 177, 192 et 198, et des étiquettes linguistiques BAS, NORMAL et GRAND comme le montre le tableau II.2. Le comparateur NMGT assigne par définition la valeur 0 à la mesure de possibilité de UNKNOWN.

| N° Fila | JOUEUR | ÉQUIPE | TAILLE | QUALITÉ | CDEG(TAILLE) |
|---|---|---|---|---|---|
| 1 | P1 | Córdoba | BAS | [30,38,40,45] | 1 |
| 2 | P3 | Granada | NORMAL | RÉGULIER | 0,91 |
| 3 | P8 | Málaga | 170 | [31,34,35,38] | 1 |
| 4 | P9 | Sevilla | BAS | 15±10 | 1 |
| 5 | P10 | Sevilla | NORMAL | BON | 0,91 |
| 6 | P12 | Cádiz | BAS | TRÈS_BON | 1 |
| 7 | P15 | Almería | 177 | 6±10 | 1 |
| 8 | P17 | Huelva | UNKNOWN | UNKNOWN | 1 |
| 9 | P18 | Jaén | UNKNOWN | [8,12,15,25] | 1 |
| 10 | P19 | Jaén | NORMAL | 25±10 | 0,91 |

**Tableau II.58 :** Résultat de la requête avec filtre flou TAILLE **MLT** #197

Le tableau II.58 montre 10 joueurs qui sont possiblement beaucoup plus petite que 1m 91. Le filtre flou TAILLE MLT #197 cherche les tuples à partir de l'équation I.13 (voir paragraphe I.13). Nous utilisons alors les règles de cette équation dans la condition WHERE de la requête. La règle $B_R \leq A_S - M$ avec $A_S = 191$ et M = 6 cherche les tuples avec une mesure de possibilité égale à 1. Ceci correspond aux valeurs crisp 170 et 177, et aux étiquettes linguistiques BAS et NORMAL selon le tableau II.2. Ensuite, la règle $B_R > A_S - M$ avec $A_S = 191$ et $A_R < B_S - M$ avec $B_S = 197$ cherche les tuples avec une mesure de possibilité entre 0 et 1. D'après le tableau II.2 les valeurs $A_R$ et $B_R$ qui satisfont cette règle correspondent à la valeur crisp 198 et aux valeurs 175 et 180 pour l'étiquette linguistique GRAND. Nous avons résumé les calculs de la mesure de nécessité dans le tableau suivant :

| TAILLE | $p = B_S - M - A_R$ | $r = B_R - A_R$ | $s = A_S - B_S$ | $q = r - s$ | CDEG |
|---|---|---|---|---|---|

| | | | | | (TAILLE)<br>p/q |
|---|---|---|---|---|---|
| NORMAL | 197 – 12 – 175 | 180 – 175 | 191 – 197 | 5 + 6 | 0.91 |

La mesure de possibilité donne 0 dans les autres cas, c'est-à-dire oú $A_R \geq B_S - M$ avec $B_S = 197$ et $M = 6$. Il s'agit des valeurs crisp 192 et 198, et de l'étiquette linguistique TRÈS_GRAND comme le montre le tableau II.2. Le comparateur MLT assigne par définition la valeur 1 à la mesure de possibilité de UNKNOWN.

| Nº Fila | JOUEUR | ÉQUIPE | TAILLE | QUALITÉ | CDEG(TAILLE) |
|---|---|---|---|---|---|
| 1 | P1 | Córdoba | BAS | [30,38,40,45] | 0,91 |
| 2 | P8 | Málaga | 170 | [31,34,35,38] | 1 |
| 3 | P9 | Sevilla | BAS | 15±10 | 0,91 |
| 4 | P12 | Cádiz | BAS | TRÈS_BON | 0,91 |
| 5 | P15 | Almería | 177 | 6±10 | 0,67 |
| 6 | P17 | Huelva | UNKNOWN | UNKNOWN | 1 |
| 7 | P18 | Jaén | UNKNOWN | [8,12,15,25] | 1 |

**Tableau II.59 :** Résultat de la requête avec filtre flou TAILLE **NMLT** #197

Le tableau II.59 montre 7 joueurs qui sont nécessairement beaucoup plus petite que 1m 91. Le filtre flou TAILLE NMLT #197 cherche les tuples à partir de l'équation I.14 (voir paragraphe I.14). Nous utilisons alors les règles de cette équation dans la condition WHERE de la requête. La règle $D_R \leq A_S - M$ avec $A_S = 191$ et $M = 6$ cherche les tuples avec une mesure de nécessité égale à 1. Ceci correspond aux valeurs crisp 170 et 177, et à l'étiquette linguistique **BAS** selon le montre le tableau II.2. Ensuite, la règle $D_R > A_S - M$ avec $A_S = 191$ et $C_R < B_S - M$ avec $B_S = 197$ cherche les tuples avec une mesure de nécessité entre 0 et 1. Or, d'après le tableau II.2 les valeurs $C_R$ et $D_R$ qui satisfont cette règle correspondent aux valeurs 185 et 195 pour l'étiquette linguistique NORMAL. Nous avons résumé les calculs de la mesure de nécessité dans le tableau suivant :

| TAILLE | $p = B_S - M - C_R$ | $r = D_R - C_R$ | $s = A_S - B_S$ | $q = r - s$ | CDEG<br>(TAILLE)<br>p/q |
|---|---|---|---|---|---|
| BAS | 197 – 12 – 175 | 180 – 175 | 191 – 197 | 5 + 6 | 0.91 |
| 177 | 197 – 12 – 177 | 183 – 177 | 191 – 197 | 6 + 6 | 0.67 |

La mesure de nécessité donne 0 dans les autres cas, c'est-à-dire oú $C_R \geq B_S - M$ avec $B_S = 197$ et $M = 6$. Il s'agit des valeurs crisp 192 et 198, et des étiquettes linguistiques GRAND et TRÈS_GRAND comme le montre le tableau II.2. Le comparateur NMLT assigne par définition la valeur 1 à la mesure de nécessité de UNKNOWN.

**RESUME**

Cette thèse s'intéresse à une réalité de gestion des connaissances imparfaites, réalité dans laquelle la connaissance existe simultanément et nécessairement dans un domaine : flou et non flou. Cette thèse propose un modèle autopoïétique pour la gestion des connaissances imparfaites. Il est composé de deux sous-systèmes, à savoir :

- le système de connaissance. Ce système est conçu d'une part, dans le domaine de l'organisation, par la description de l'unité à travers le modèle autopoïétique de la marguerite, et d'autre part, dans le domaine de la structure, par les composants de l'unité conçus par l'approche de l'enaction de Maturna et Varela ;

- le système opérationnel. Ce système est conçu, d'une part, dans le domaine de l'organisation, par la description de l'unité à travers le modèle autopoïétique OIDC, et d'autre part, dans le domaine de la structure, par les composants de l'unité conçus par l'approche cognitiviste selon une représentation imparfaite de la connaissance.

L'imperfection de la connaissance se réfère à la prise en charge du flou sous l'angle des bases de données relationnelles floues. Le point de vue génie industriel se traduit finalement dans un modèle sociotechnique. Ce modèle permet d'expliquer un phénomène de gestion des connaissances à la fois d'un point de vue social et technique. La richesse du modèle, réside dans le fait que l'unité existe (simultanément et nécessairement) dans une dualité organisation/structure. Notre champ d'expérimentation a été une industrie manufacturière de carton dans la région du Maulé au Chili. Cette expérience a permis de décrire les produits, procédés, processus, autour d'un noyau invariant, qui a permis de valider l'organisation et la structure du modèle proposé dans un domaine industriel.

**Mots clés :** connaissances, gestion des connaissances, autopoïèse, apprentissage, évolution.

**SUMMARY**

In order to conceive of 'living systems' in terms of the processes that realized them, rather in terms of their relationships with an environment (as in system theory), Maturana and Varela at the University of Chile (1969) coined the term 'autopoiesis' ($\alpha \upsilon \tau \sigma \sigma$ = self, $\pi \sigma \iota \epsilon \nu \iota \nu$ = creation, production) to denote the central tenet of their organization; *autonomy*. The inherent meanings of this term attempt to covey the perceived nature of living systems as systems that maintain their *identity* through their own operations of continuous self-renewal to preserve their *unity*. Moreover, these systems can only be characterized and understood with reference to themselves and whatever activity takes place within them is necessarily and constitutively determined in relation to themselves, i.e., self-production or self-referentiality.

This thesis proposes a model – Fuzzy Autopoietic Knowledge Management (FAKM) – in order to integrate the system theory of living systems, the cybernetic theory of viable systems, and the autopoiesis theory of the autopoietic systems, with the hope of going beyond the knowledge management models that are based, among other things, on Cartesian dualism cognition/action (i.e., on a model of symbolic cognition as the processing of representational information in a knowledge management system). Instead, the FAKM model, proposed here, uses a dualism organization/structure to define an autopoietic system in a socio-technical approach.

Our experimental source for this model is a manufacturing company located within the Maule Region south of Santiago in Chile.

**Key words:** knowledge, knowledge management, autopoiesis, learning and evolution.